\RequirePackage{fix-cm}
\documentclass[smallextended]{svjour3}       
\smartqed  

\usepackage{natbib}
\usepackage{amssymb,amsmath}
\usepackage{graphicx}
\usepackage{mathptmx}      
\usepackage{booktabs}
\usepackage{ulem,color,siunitx,physics}
\usepackage[colorlinks=true,citecolor=blue,urlcolor=blue,breaklinks]{hyperref}
\newcommand{\prl}{Phys.~Rev.~Lett.}
\newcommand{\prc}{Phys.~Rev.~C}
\newcommand{\prd}{Phys.~Rev.~D}

\newcommand{\apj}{Astrophys.~J.}
\newcommand{\apjl}{Astrophys.~J.~Lett.}
\newcommand{\apjs}{Astrophys.~J.~Suppl.}
\newcommand{\mnras}{Mon.~Not.~R.~Astron.~Soc.}
\newcommand{\aap}{Astron.~Astrophys.}
\newcommand{\pasj}{Publ.~Astron.~Soc.~Jpn}
\newcommand{\pasp}{Publ.~Astron.~Soc.~Pac}
\newcommand{\actaa}{Acta~Astron.}
\newcommand{\nphysa}{Nucl.~Phys.~A}
\newcommand{\nat}{Nature}
\newcommand{\araa}{Annu.~Rev.~Astron.~Astrophys.}
\newcommand{\ssr}{Space~Sci.~Rev.}
\newcommand{\physrep}{Phys.~Rep.}
\newcommand{\jcap}{J.~Cosmol.~Astropart.~Phys.}
\newcommand{\jqsrt}{J.~Quant.~Spectrosc.~Radiat.~Transf.}
\journalname{Living Reviews in Relativity}
\begin{document}

\title{Coalescence of black hole--neutron star binaries%
\thanks{This article is a revised version of \url{https://doi.org/10.12942/lrr-2011-6}.\\\textbf{Change summary:} Major revision, updated and expanded.\\\textbf{Change details:} Section~1 has been updated reflecting the current status of gravitational-wave and multimessenger astronomy. Section~3 now includes the review of dynamical mass ejection and postmerger activity such as the disk outflow and neutrino emission. A lot of figures are added. Section~4 is newly added to discuss electromagnetic signals from black hole--neutron star binaries and observational distinguishability of binary types. Appendices~A and B review formulation for quasieqiulibrium states and dynamical simulations, respectively. Appendix C is devoted to some relevant analytic estimation. The number of references has increased from 232 to 651.}}

\author{Koutarou Kyutoku \and Masaru Shibata \and Keisuke Taniguchi}
\authorrunning{K. Kyutoku, M. Shibata, and K. Taniguchi} 

\institute{
K. Kyutoku \at Department of Physics, Kyoto University \\ Kyoto
606-8502, Japan \\ Center for Gravitational Physics, Yukawa Institute
for Theoretical Physics, Kyoto University \\ Kyoto 606-8502, Japan \\
Interdisciplinary Theoretical and Mathematical Sciences Program
(iTHEMS), RIKEN, \\ Wako, Saitama 351-0198, Japan \\
\email{kyutoku@tap.scphys.kyoto-u.ac.jp} \\
\url{http://www-tap.scphys.kyoto-u.ac.jp/~kyutoku/index.html}
\and
M. Shibata \at Max Planck Institute for Gravitational Physics (Albert
Einstein Institute) \\ Am M{\"u}hlenberg 1, 14476 Potsdam-Golm,
Germany \\ Center for Gravitational Physics, Yukawa Institute for
Theoretical Physics, Kyoto University \\ Kyoto 606-8502, Japan \\
\email{mshibata@aei.mpg.de} \\
\url{http://www2.yukawa.kyoto-u.ac.jp/~mshibata/}
\and
K. Taniguchi \at Department of Physics, University of the Ryukyus \\
Nishihara, Okinawa 903-0213, Japan \\ \email{ktngc@sci.u-ryukyu.ac.jp}
\\ \url{http://www.phys.u-ryukyu.ac.jp/~keisuke/}}


\maketitle

\begin{abstract}
 We review the current status of general relativistic studies for
 coalescences of black hole--neutron star binaries. First,
 high-precision computations of black hole--neutron star binaries in
 quasiequilibrium circular orbits are summarized, focusing on the
 quasiequilibrium sequences and the mass-shedding limit. Next, the
 current status of numerical-relativity simulations for the merger of
 black hole--neutron star binaries is described. We summarize our
 understanding for the merger process, tidal disruption and its
 criterion, properties of the merger remnant and ejected material,
 gravitational waveforms, and gravitational-wave spectra. We also
 discuss expected electromagnetic counterparts to black hole--neutron
 star coalescences.
\end{abstract}

\setcounter{tocdepth}{3}
\tableofcontents


\section{Introduction} \label{sec:intro}

\subsection{Why is the black hole--neutron star binary merger
  important?} \label{sec:intro_bg}

After the first release of this review article in 2011 \citep{1st}, the
research environment for compact binary coalescences has changed
completely. The turning point was the first gravitational-wave event
GW150914 from a binary-black-hole merger detected by the LIGO and Virgo
Collaboration \citep{GW150914}. We obtained the strongest evidence for
the existence of black holes and mergers of their binaries within the
Hubble time. We learned that some stellar-mass black holes are
significantly more massive than those found in our Galaxy by X-ray
observations \citep{GW150914_2}. Furthermore, we confirmed that general
relativity is consistent with observations even for this dynamical and
strongly gravitating phenomenon \citep[see also
\citealt{GWTC1GR,GWTC2GR} for the
update]{GW150914GR,Yunes_Yagi_Pretorius2016}. After further
observations, the number of stellar-mass black holes detected by
gravitational waves has already exceeded those by electromagnetic
radiation (see \citealt{GWTC1,GWTC2} for reported detections as of
2020). Gravitational-wave astronomy of binary black holes is rapidly
becoming an established branch of astrophysics.

Subsequently, the first binary-neutron-star merger, GW170817, was
detected with not only gravitational waves \citep{GW170817,GW170817_2}
but also electromagnetic waves by multiband instruments all over the
world \citep{EM170817,GRB170817A}. Tidal deformability of the neutron
star was constrained from gravitational-wave data analysis, and
extremely stiff equations of state are no longer favored
\citep{GW170817,De_FLBBB2018,GW170817EOS,GW170817_2,Narikawa_UKKKST2020}. Binary-neutron-star
mergers were strongly suggested to be central engines of short gamma-ray
bursts by the detection of a weak GRB 170817A
\citep{GRB170817A,Goldstein_etal2017,Savchenko_etal2017} and by longterm
observations of its off-axis afterglow
\citep{Mooley_DGNHBFHCH2018,Mooley_etal2018,Alexander_etal2018,Lamb_etal2019}.
The host galaxy NGC 4993 was identified by the kilonova/macronova AT
2017gfo
\citep{Coulter_etal2017,Arcavi_etal2017,Lipunov_etal2017,SoaresSantos_etal2017,Tanvir_etal2017,Valenti_SYCTCJRHK2017},
and Hubble's constant was inferred in a novel manner by combining the
cosmological redshift of NGC 4993 and the luminosity distance estimated
from GW170817 \citep{GW170817Hubble}. Furthermore, binary-neutron-star
mergers were indicated to be a site of \textit{r}-process
nucleosynthesis
\citep{Tanaka_etal2017,Kasen_MBQR2017,Watson_etal2019}. This event
heralded a new era of multimessenger astronomy with gravitational and
electromagnetic radiation.

Finally, during the review process of this article, detections of black
hole--neutron star binaries, GW200105 and GW200115, are reported
\citep{GW200105200115}. Together with another candidate of a black
hole--neutron star binary merger GW190426\_152155 \citep{GWTC2}, we may
now safely consider that black hole--neutron star binaries are actually
merging in our Universe. The merger rate is currently inferred to be
$\approx 12$--$\SI{240}{Gpc^{-3}.yr^{-1}}$, which is largely consistent
with previous theoretical estimation \citep[see also
\citealt{Narayan_Piran_Shemi1991,Phinney1991} for pioneering rate
estimation]{Dominik_BOMBFHBP2015,Kruckow_TLKI2018,Neijssel_VSBGBMSVM2019,Zevin_SBK2020,Santoilquido_MGBA2021}. Unfortunately,
despite the presence of neutron stars, no associated electromagnetic
counterpart was detected. This is consistent with the current
theoretical understanding reviewed throughout this article, because the
most likely mass ratios of these binaries are as high as $4$--$5$ and
the spin of the black holes are likely to be zero or retrograde. Tidal
disruption do not occur for these parameters, and thus neither mass
ejection nor electromagnetic emission is expected to occur. We also note
that some other gravitational-wave events reported in LIGO-Virgo O3 are
also consistent with black hole--neutron star binaries under generous
assumptions on the mass of compact objects
\citep{GW190425,GW190814},\footnote{``NSBH'' of the LIGO-Virgo
classification scheme does not necessarily mean that the lighter
component is a neutron star, because this label only indicates that the
mass is smaller than $3\,M_\odot$. ``MassGap'' means that at least one
member of the binary has the mass between $3\,M_\odot$ and $5\,M_\odot$. The
chirp mass, the best-determined parameter for inspiral-dominated events,
of a $3\,M_\odot$--$3\,M_\odot$ binary is identical to that of
$7.6\,M_\odot$--$1.35\,M_\odot$. A finite probability of ``BNS'' for the
source classification derived in the real-time data analysis is
consistent with a black hole--neutron star binary with the black-hole
mass being $\lesssim 7.6\,M_\odot$ for the case in which the neutron-star
mass is $1.35\,M_\odot$. Similarly, the chirp mass of a
$5\,M_\odot$--$5\,M_\odot$ binary is identical to that of
$8.8\,M_\odot$--$3\,M_\odot$ as well as that of
$25.9\,M_\odot$--$1.35\,M_\odot$, and a finite probability of ``BBH''
indicates a black hole heavier than this for a black hole--neutron star
binary.} partly because no electromagnetic counterpart was detected
\citep{Kyutoku_FHKKST2020,Han_THLJJFW2020,Kawaguchi_Shibata_Tanaka2020-2}.

\begin{figure}[tbp]
 \centering \includegraphics[width=0.95\linewidth,clip]{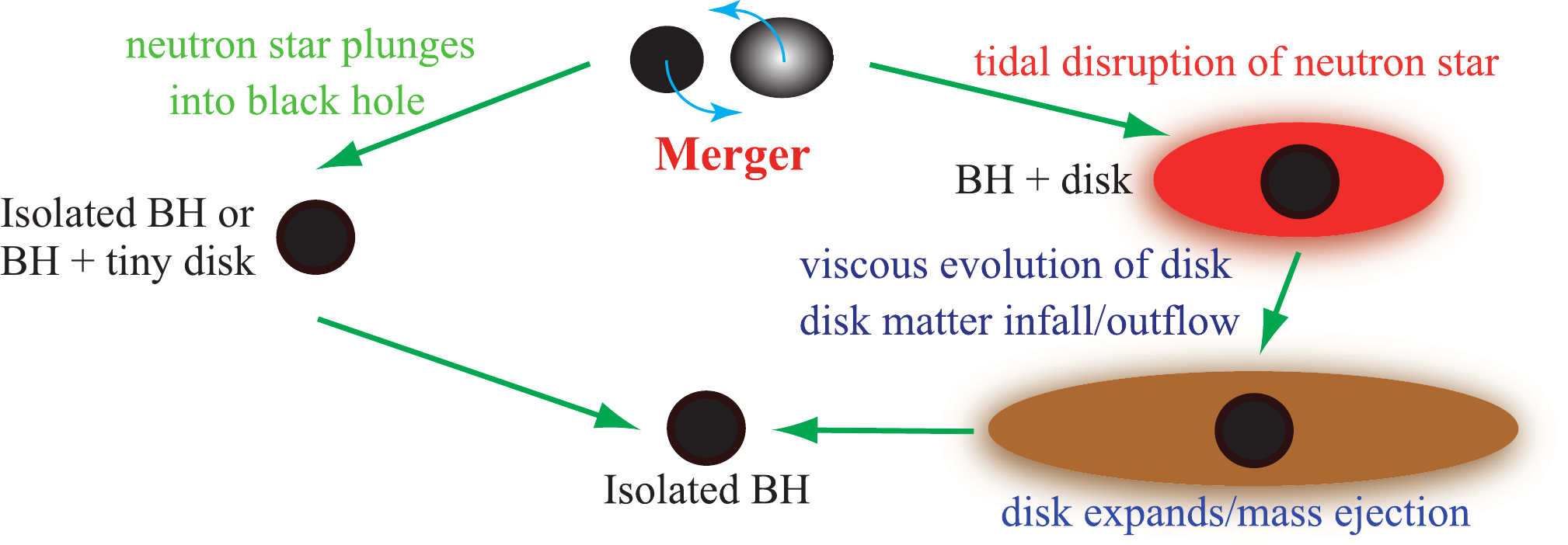}
 \caption{Summary for the merger and postmerger evolution of black
 hole--neutron star binaries. The fate is classified into two categories
 according to whether the neutron star is tidally disrupted (right) or
 not (left). Image adapted from \citet{Shibata_Hotokezaka2019}, copyright by Annual Reviews.} \label{fig:fate}
\end{figure}

One of the remaining issues for ground-based gravitational-wave
detectors is to discover coalescences of black hole--neutron star
binaries accompanied by tidal disruption and hence electromagnetic
emission. Indeed, among the mergers of black hole--neutron star
binaries, those resulting in tidal disruption of the neutron star by the
black hole are of physical and astrophysical interest and deserve
detailed investigations. Specifically, the tidal disruption is required
to occur outside the innermost stable circular orbit of the black hole
for inducing astrophysically interesting outcomes. If the neutron star
is not disrupted, as is likely the case of GW200105 and GW200115, it
behaves like a point particle throughout the coalescence, and the merger
process will be indistinguishable from that of (highly asymmetric)
binary black holes \citep{Foucart_BDGKMMPSS2013} except for possible
electromagnetic emission associated with crust shattering
\citep{Tsang_RHPB2012}, magnetospheric activities \citep[see also
\citealt{Ioka_Taniguchi2000} for earlier work on binary neutron
stars]{Hansen_Lyutikov2001,McWilliams_Levin2011,Lai2012,Paschalidis_Etienne_Shapiro2013,DOrazio_LMP2016,Carrasco_Shibata2020,Wada_Shibata_Ioka2020,East_LLP2021,Carrasco_Shibata_Reula2021},
or charged black holes
\citep{Levin_DOrazio_GarciaSaenz2018,Zhang2019,Dai2019,Pan_Yang2019,Zhong_Dai_Deng2019}. These
two possibilities for the fate of merger are summarized schematically in
Fig.~\ref{fig:fate}.\footnote{There may, in principle, be a third
possibility that the binary initiates stable mass transfer after the
onset of mass shedding. This might seem possible from the Newtonian
intuition, because the heavier component (black hole) accretes material
from the lighter component (neutron star). Although (pseudo-)Newtonian
simulations have indeed witnessed episodic mass transfer
\citep{Janka_ERF1999,Rosswog_Speith_Wynn2004,Ruffert_Janka2010}, this
process has never been identified in numerical-relativity simulations of
quasicircular inspirals as we discuss in
Sect.~\ref{sec:intro_history}. Readers interested in the stable mass
transfer should be referred to Appendix~\ref{app:ae_smt} for detailed
discussions.}

Focusing on the cases in which tidal disruption occurs, many researchers
have vigorously studied the following three aspects. Accordingly, most
parts of this review will be devoted to their detailed discussions.
\begin{itemize}
 \item Gravitational waves will enable us to study the finite-size
       properties and hence the equation of state of neutron
       stars. First, tidal deformability, $\Lambda$ (see also
       Sect.~\ref{sec:intro_tidal}), of neutron stars will be extracted
       from the phase evolution in the inspiral phase
       \citep{Flanagan_Hinderer2008} along with the masses and the spins
       of binary components
       \citep{Finn_Chernoff1993,Jaranowski_Krolak1994,Cutler_Flanagan1994,Poisson_Will1995}. Although
       the tidal deformability could be inferred even if tidal
       disruption does not occur, realistic measurements will be
       possible only when the finite-size effect is so sizable that the
       neutron star is disrupted
       \citep{Lackey_KSBF2012,Lackey_KSBF2014}. Second, the orbital
       frequency at tidal disruption depends on the compactness of the
       neutron star, $\mathcal{C}$
       \citep{Vallisneri2000,Shibata_KYT2009}. Because the mass can be
       extracted or constrained from inspiral signals along with the
       spin as stated above, gravitational waveforms from tidal
       disruption of a neutron star may bring us invaluable information
       about its radius, which is strongly but not perfectly correlated
       with the tidal deformability
       \citep{Hotokezaka_KSS2016,De_FLBBB2018}. The measurement of these
       quantities with black hole--neutron star binaries could serve as
       an additional tool for exploring supranuclear-density matter
       \citep{Lindblom1992,Harada2001}. For this purpose, it is crucial
       to understand the dependence of gravitational waveforms,
       including characteristic observable features associated with
       tidal disruption, on possible equations of state by theoretical
       calculations.
 \item The remnant disk formed from the disrupted neutron star is a
       promising central engine of short-hard gamma-ray bursts
       \citep[see also \citealt{Blinnikov_NPP1984} for an earlier idea
       and \citealt{Paczynski1986,Goodman1986,Eichler_LPS1989} for
       binary-neutron-star
       scenarios]{Paczynski1991,Narayan_Paczynski_Piran1992,Mochkovitch_HIM1993}. A
       typical beaming-corrected energy of the jet, $\sim \SI{e50}{erg}$
       \citep{Fong_BMZ2015}, can be explained if, for example, $\sim
       0.1\%$ of the rest-mass energy is converted from a $\sim 0.1
       \,M_\odot$ accretion disk. This could be realized via neutrino pair
       annihilation \citep{Rees_Meszaros1992}, which is effective when
       the disk is sufficiently hot and dense to cool via neutrino
       radiation, called the neutrino-dominated accretion flow
       \citep{Popham_Woosley_Fryer1999,Narayan_Piran_Kumar2001,Kohri_Mineshige2002,DiMatteo_Perna_Narayan2002,Kohri_Narayan_Piran2005,Chen_Beloborodov2007,Kawanaka_Mineshige2007}. Another
       possible energy source is the rotational energy of a spinning
       black hole extracted by magnetic fields, i.e., the
       Blandford-Znajek mechanism
       \citep{Blandford_Znajek1977,Meszaros_Rees1997}. For this
       mechanism to work, magnetic-field strength in the disk needs to
       be amplified by turbulent motion resulting from
       magnetohydrodynamic instabilities such as the magnetorotational
       instability \citep{Balbus_Hawley1991}, and subsequently, strong
       magnetic fields threading the spinning black hole need to be
       developed to form a surrounding magnetosphere. One of the
       ultimate goals for numerical simulations of compact binary
       coalescences may be to clarify how, if possible, the
       ultrarelativistic jet is launched from the merger
       remnant. Theoretical investigations should also clarify whether
       longterm activity of short-hard gamma-ray bursts, e.g., the
       extended and plateau emission
       \citep{Norris_Bonnell2006,Rowlinson_OMTL2013,Gompertz_OWR2013,Kisaka_Ioka_Sakamoto2017},
       can really be explained by the merger remnant of black
       hole--neutron star binaries. Because of the diversity associated
       with stellar-mass black holes, black hole--neutron star binaries
       might naturally explain the variety observed in short-hard
       gamma-ray bursts (see \citealt{Nakar2007,Berger2014} for
       reviews).
 \item A substantial amount of neutron-rich material will be ejected and
       synthesize \textit{r}-process elements \citep[see also
       \citealt{Lattimer2019} for retrospection by an
       originator]{Lattimer_Schramm1974}, i.e., about half of the
       elements heavier than iron in the universe, whose origin has not
       yet been fully understood
       \citep{Burbidge_BFH1957,Cameron1957}. Subsequently, radioactive
       decays of unstable nuclei will heat up the ejecta, resulting in
       quasithermal emission in UV-optical-IR bands on a time scale of
       $\order{10}$ days \citep{Li_Paczynski1998}. This transient,
       called the kilonova \citep{Metzger_MDQAKTNPZ2010} or macronova
       \citep{Kulkarni2005}, serves as the most promising
       omnidirectional electromagnetic counterparts to gravitational
       waves (see \citealt{Metzger2019} for reviews). The ejecta are
       eventually mixed with the interstellar medium and contribute to
       the chemical evolution of galaxies, and this interaction may
       drive another electromagnetic counterpart such as synchrotron
       radiation from nonthermal electrons \citep{Nakar_Piran2011} and
       possibly inverse Compton emission
       \citep{Takami_Kyutoku_Ioka2014}. To derive nucleosynthetic yields
       and characteristics of electromagnetic counterparts, we need to
       understand properties of the ejecta such as the mass, the
       velocity, and the electron fraction that characterizes the
       neutron richness. In particular, the electron fraction primarily
       determines the nucleosynthetic yield, which controls features of
       the kilonova/macronova via the opacity
       \citep{Kasen_Badnell_Barnes2013,Tanaka_Hotokezaka2013,Tanaka_etal2018,Tanaka_KGK2020,Banerjee_TKKG2020}
       and the heating rate
       \citep{Hotokezaka_WTBTP2016,Barnes_KWM2016,Kasen_Barnes2019,Waxman_Ofek_Kushnir2019,Hotokezaka_Nakar2020}. If
       a significant fraction of the ejecta keeps extreme neutron
       richness of the neutron star, ultraheavy elements may be produced
       in abundance, and the associated fission and/or $\alpha$-decay
       will power the kilonova/macronova at late times
       \citep{Wanajo_SNKKS2014,Zhu_etal2018,Wu_BMM2019}. They could also
       be the origin of exceptionally \textit{r}-process enhanced
       metal-poor stars, so-called actinide-boost stars \citep[see,
       e.g.,][]{Mashonkina_Christlieb_Eriksson2014}. Last but not least,
       the geometrical shape of the ejecta could be important for
       understanding the diversity of electromagnetic counterparts to
       black hole--neutron star binaries
       \citep{Kyutoku_Ioka_Shibata2013,Tanaka_HKWKSS2014}.
\end{itemize}

\subsection{Life of black hole--neutron star binaries}
\label{sec:intro_orbit}

We first overview the entire evolution of black hole--neutron star
binaries from their birth. Binaries consisting of a black hole and/or a
neutron star, hereafter collectively called compact object binaries, are
generally born after evolution of isolated massive binaries (see, e.g.,
\citealt{Postnov_Yungelson2014} for reviews) or via dynamical processes
in dense environments (see, e.g., \citealt{Benacquista_Downing2013} for
reviews). Relative contributions of these two paths to black
hole--neutron star binaries have not been understood yet, as well as for
compact object binaries of other types. We do not go into details of the
formation path in this article, commenting only that the evolution of
isolated binaries is usually regarded as the dominant channel for black
hole--neutron star binaries \citep[see, e.g., discussions
in][]{GW200105200115}.

After the formation of black hole--neutron star binaries, their orbital
separation decreases gradually due to longterm gravitational radiation
reaction. If we would like to observe their coalescences, the binaries
are required to merge within the Hubble time of $\approx
\SI{1.4e10}{yr}$. This condition is also a prerequisite for them to
drive short-hard gamma-ray bursts and to produce \textit{r}-process
elements. The lifetime of a black hole--neutron star binary in a
circular orbit for a given orbital separation $r$ is given by
\begin{align}
 t_\mathrm{GW} & = \frac{5c^5}{256G^3} \frac{r^4}{( M_\mathrm{BH} +
 M_\mathrm{NS} ) M_\mathrm{BH} M_\mathrm{NS}} \notag \\
 & = \SI{1.01e10}{yr} \pqty{\frac{r}{\SI{6e6}{\km}}}^4
 \pqty{\frac{M_\mathrm{BH}}{7\,M_\odot}}^{-1}
 \pqty{\frac{M_\mathrm{NS}}{1.4\,M_\odot}}^{-1}
 \pqty{\frac{m_0}{8.4\,M_\odot}}^{-1} \label{eq:tmerge}
\end{align}
in the adiabatic approximation, which is appropriate when the radiation
reaction time scale is much longer than the orbital period. Here, $G$,
$c$, $M_\mathrm{BH}$, $M_\mathrm{NS}$, and $m_0$ are the gravitational
constant, the speed of light, the gravitational mass of the black hole,
the gravitational mass of the neutron star, and the total mass of the
binary $m_0 := M_\mathrm{BH} + M_\mathrm{NS}$, respectively (cf., Table
\ref{table:para}). The orbital eccentricity only reduces the time to
merger for a given value of the semimajor axis
\citep{Peters_Mathews1963,Peters1964}. Thus, a black hole--neutron star
binary merges within the Hubble time if its initial semimajor axis is
smaller than $\sim \SI{e7}{\km}$ with the precise value depending on the
masses of the objects and the initial eccentricity. Because the spin and
finite-size properties of the objects come into play only as
higher-order corrections in terms of the orbital velocity or other
appropriate parameters (see, e.g., \citealt{Blanchet2014} for reviews of
the post-Newtonian formalism), Eq.~\eqref{eq:tmerge} with eccentricity
corrections is adequate for judging whether a binary merges within the
Hubble time.

Two remarks should be made regarding the longterm evolution. First, the
orbital eccentricity decreases rapidly, specifically $e \propto
a^{-19/12}$ in an asymptotically circular regime with $a$ being the
semimajor axis, due to gravitational radiation reaction
\citep{Peters1964}. Accordingly, black hole--neutron star binaries right
before merger (e.g., when gravitational waves are detected by
ground-based detectors) may safely be approximated as circular. Second,
the neutron star is unlikely to be tidally-locked, because the effects
of viscosity are likely to be insufficient
\citep{Kochanek1992,Bildsten_Cutler1992}. Thus, the spin of the neutron
star can affect the merger dynamics significantly only if the rotational
period is extremely short at the outset and the spin-down is not
severe. Quantitatively, the dimensionless spin parameter of the neutron
star is approximately written as
\begin{align}
 \chi_\mathrm{NS} & = \frac{c I_\mathrm{NS} \Omega_\mathrm{rot}}{G
 M^2_\mathrm{NS}} = \num{4.9e-4} \pqty{\frac{I_\mathrm{NS} / (
 M_\mathrm{NS} R_\mathrm{NS}^2 )}{1/3}}
 \pqty{\frac{M_\mathrm{NS}}{1.4\,M_\odot}}^{-1}
 \pqty{\frac{R_\mathrm{NS}}{\SI{12}{\km}}}^2
 \pqty{\frac{P_\mathrm{rot}}{\SI{1}{\second}}}^{-1} ,
\end{align}
where $I_\mathrm{NS}$, $R_\mathrm{NS}$, and $P_\mathrm{rot}$ are the
moment of inertia, the radius, and the rotational period, respectively,
of the neutron star. Observationally, the shortest rotational period of
known pulsars in Galactic binary neutron stars that merge within the
Hubble time is $\approx \SI{17}{\milli\second}$, which is equivalent to
only $\chi_\mathrm{NS} \approx 0.03$ \citep{Stovall_etal2018}. Moreover,
black hole--neutron star binaries are unlikely to harbor recycled
pulsars, because the neutron star is expected to be formed after the
black hole, having no chance for mass accretion. Hence, it is reasonable
to approximate neutron stars as nonspinning in the merger of black
hole--neutron star binaries. Exceptions to these remarks might arise
from dynamical formation in dense environments such as galactic centers
and globular clusters, e.g., exchange interactions involving recycled
pulsars \citep[see, e.g.,][]{Fragione_GLPP2019,Ye_FKRCFR2020}, and/or
black-hole formation from the secondary caused by mass transfer in
isolated massive binaries \citep[see, e.g.,][]{Kruckow_TLKI2018}.

The late inspiral and merger phases of black hole--neutron star binaries
are promising targets of gravitational waves for ground-based detectors
irrespective of the degree of tidal disruption. The frequency $f$ and
the amplitude $h$ of gravitational waves from black hole--neutron star
binaries with the orbital separation $r$ at the luminosity distance $D$
are estimated in the quadrupole approximation for two point particles as
\begin{align}
 f & \approx \frac{\Omega}{\pi} = \SI{523}{\hertz} \pqty{\frac{r}{6Gm_0
 / c^2}}^{-3/2} \pqty{\frac{m_0}{8.4\,M_\odot}}^{-1} , \\
 h & \approx \frac{4G^2 M_\mathrm{BH} M_\mathrm{NS}}{c^4 r D} =
 \num{3.7e-22} \pqty{\frac{\mu}{1.17\,M_\odot}} \pqty{\frac{r}{6Gm_0 /
 c^2}}^{-1} \pqty{\frac{D}{\SI{100}{Mpc}}}^{-1} ,
\end{align}
where $\mu := M_\mathrm{BH} M_\mathrm{NS} / m_0$ is the reduced
mass. Here, the most favorable direction and orientation are assumed for
evaluating $h$. These values indicate that black hole--neutron star
binaries near the end of their lives fall within the observable window
of ground-based gravitational-wave detectors as far as the distance is
sufficiently close.

However, the quadrupole approximation for point particles is not
sufficiently accurate for describing the evolution of black
hole--neutron star binaries in the late inspiral, merger, and postmerger
phases. As the orbital separation gradually approaches the radius of the
object, spins and finite-size effects such as tidal deformation begin to
modify the gravitational interaction between the binary in a noticeable
manner. The adiabatic approximation also breaks down for the very close
orbit, because the radiation reaction time scale and the orbital period
become comparable near an approximate innermost stable
orbit\footnote{Definition of the innermost stable circular orbit is
subtle for comparable mass binaries \citep[see,
e.g.,][]{Blanchet_Iyer2003}. In this article, we basically refer to the
minimum energy circular orbit as the innermost stable circular orbit.}
as
\begin{equation}
 \frac{t_\mathrm{GW}}{P_\mathrm{orb}} = 2.0 \pqty{\frac{r}{6Gm_0 /
  c^2}}^{5/2} \pqty{\frac{M_\mathrm{BH}}{7\,M_\odot}}^{-1}
  \pqty{\frac{M_\mathrm{NS}}{1.4\,M_\odot}}^{-1}
  \pqty{\frac{m_0}{8.4\,M_\odot}}^2 . \label{eq:gwvsorb}
\end{equation}
Thus, dynamics in the late inspiral and merger phases depends crucially
on complicated hydrodynamics associated with neutron stars, whose
properties are controlled by the equation of state, and on nonlinear
gravity of general relativity. Furthermore, the evolution of the remnant
disk in the postmerger phase is governed by neutrino emission triggered
by shock-induced heating and turbulence associated with
magnetohydrodynamic instabilities
\citep{Lee_RamirezRuiz_Page2004,Setiawan_Ruffert_Janka2004,Lee_RamirezRuiz_Page2005,Setiawan_Ruffert_Janka2006,Shibata_Sekiguchi_Takahashi2007}. All
these facts make fully general-relativistic numerical studies the unique
tool to clarify the final evolution of black hole--neutron star
coalescences in a quantitative manner.

\subsection{Tidal problem around a black hole} \label{sec:intro_tidal}

As we stated in Sect.~\ref{sec:intro_bg}, this review will focus
primarily on numerical studies of black hole--neutron star binaries for
which finite-size effects play a significant role. To set the stage for
understanding numerical results, in this Sect.~\ref{sec:intro_tidal}, we
will discuss requirement for the binary to cause significant tidal
disruption, which starts with the mass shedding from the inner edge of
the neutron star.

\subsubsection{Mass-shedding condition} \label{sec:intro_tidal_ms}

The orbital separation at which the mass shedding sets in is determined
primarily by the mass ratio of the binary and the radius of the neutron
star. The orbit at which the mass shedding sets in, the so-called
mass-shedding limit, can be estimated semiquantitatively by Newtonian
calculations as follows. Mass shedding from the neutron star occurs when
the tidal force exerted by the black hole overcomes the self-gravity of
the neutron star at the inner edge of the stellar surface. This
condition is approximately given by
\begin{equation}
 \frac{2G M_\mathrm{BH} ( c_\mathrm{R} R_\mathrm{NS} )}{r^3} \gtrsim
  \frac{G M_\mathrm{NS}}{( c_\mathrm{R} R_\mathrm{NS} )^2} ,
  \label{eq:mscond}
\end{equation}
where the factor $c_\mathrm{R} \ge 1$ represents the degree of tidal
(and rotational if the neutron star is rapidly spinning) elongation of
the stellar radius. The precise value of this factor depends on the
neutron-star properties and the orbital separation. The mass-shedding
limit may be defined as the orbit at which this inequality is
approximately saturated,
\begin{equation}
 r_\mathrm{ms} := 2^{1/3} c_\mathrm{R}
  \pqty{\frac{M_\mathrm{BH}}{M_\mathrm{NS}}}^{1/3} R_\mathrm{NS}
  . \label{eq:msrad}
\end{equation}

We emphasize here that Eq.~\eqref{eq:mscond} is a necessary condition
for the onset of mass shedding. Tidal disruption occurs only after
substantial mass is stripped from the surface of the neutron star, while
the orbital separation decreases continuously due to gravitational
radiation reaction during this process. Thus, the tidal disruption
should occur at a smaller orbital separation than
Eq.~\eqref{eq:msrad}. We also note that the neutron star will be
disrupted immediately after the onset of mass shedding if its radius
increases rapidly in response to the mass loss, although typical
equations of state predict that the radius in equilibrium depends only
weakly on the mass (see, e.g.,
\citealt{Lattimer_Prakash2016,Ozel_Freire2016,Oertel_HKT2017} for
reviews).

Tidal disruption induces observable astrophysical consequences only if
it occurs outside the innermost stable circular orbit of the black hole,
inside which stable circular motion is prohibited by strong gravity of
general relativity; If the tidal disruption fails to occur outside this
orbit, the material is rapidly swallowed by the black hole and does not
leave a remnant disk or unbound ejecta in an appreciable manner. This
implies that observable tidal disruption requires, at least, the
mass-shedding limit to be located outside the innermost stable circular
orbit. The radius of the innermost stable circular orbit depends
sensitively on the dimensionless spin parameter of the black hole,
$\chi$. Specifically, it is given in terms of a dimensionless decreasing
function $\hat{r}_\mathrm{ISCO} ( \chi )$ of $\chi$ for an orbit on the
equatorial plane of the black hole by
\citep{Bardeen_Press_Teukolsky1972}
\begin{equation}
 r_\mathrm{ISCO} = \hat{r}_\mathrm{ISCO} ( \chi )
  \frac{GM_\mathrm{BH}}{c^2} .
\end{equation}
Here, we adopt the convention that the positive and negative values of
$\chi$ indicate the prograde and retrograde orbits, i.e., the orbits
with their angular momenta aligned and anti-aligned with the black-hole
spin, respectively. Specifically, the value of $\hat{r}_\mathrm{ISCO}$
is $9$ for a retrograde orbit around an extremally-spinning black hole
($\chi = -1$), $6$ for an orbit around a nonspinning black hole ($\chi =
0$), and $1$ for a prograde orbit around an extremally-spinning black
hole ($\chi = 1$). If the spin of the black hole is inclined with
respect to the orbital angular momentum, the spin effect described here
is not determined by the magnitude of the spin angular momentum but by
that of the component parallel to the orbital angular momentum. Thus,
even if the black-hole spin is high, its effect can be minor in the
presence of spin misalignment.

To sum up, the final fate of a black hole--neutron star binary is
determined primarily by the mass ratio of the binary, the compactness of
the neutron star, and the dimensionless spin parameter of the black
hole. The ratio of the radius of the mass-shedding limit and that of the
innermost stable circular orbit is given by
\begin{equation}
 \frac{r_\mathrm{ms}}{r_\mathrm{ISCO}} = \frac{2^{1/3}
  c_\mathrm{R}}{\hat{r}_\mathrm{ISCO} ( \chi )}
  \pqty{\frac{M_\mathrm{BH}}{M_\mathrm{NS}}}^{-2/3}
  \pqty{\frac{GM_\mathrm{NS}}{c^2 R_\mathrm{NS}}}^{-1}
  . \label{eq:iscovsms}
\end{equation}
This semiquantitative estimate suggests that tidal disruption of a
neutron star could occur if one or more of the following conditions are
satisfied:
\begin{enumerate}
 \item the mass ratio of the binary, $Q := M_\mathrm{BH} /
       M_\mathrm{NS}$, is low,
 \item the compactness of the neutron star, $\mathcal{C} :=
       GM_\mathrm{NS} / ( c^2 R_\mathrm{NS} )$, is small,
 \item the dimensionless spin parameter of the black hole, $\chi$, is
       high with the definition of signature stated above.
\end{enumerate}
If we presume that the neutron-star mass is fixed, the conditions 1 and
2 may be restated as
\begin{enumerate}
 \item[1'.] the black-hole mass is small,
 \item[2'.] the neutron-star radius is large,
\end{enumerate}
respectively.

Quantitative discussions have to take the general-relativistic nature of
black hole--neutron star binaries into account. For this purpose, it is
advantageous to rewrite Eq.~\eqref{eq:mscond} in terms of the orbital
angular velocity as
\begin{equation}
 \Omega^2 \ge \frac{1}{2c_\mathrm{R}^3}
  \frac{GM_\mathrm{NS}}{R_\mathrm{NS}^3} \pqty{1 + Q^{-1}} ,
\end{equation}
because $\Omega$ can be defined in a gauge-invariant manner even for a
comparable-mass binary in general relativity. It should be remarked that
the orbital frequency at the onset of mass shedding is determined
primarily by the average density of the neutron star, $\propto
\sqrt{M_\mathrm{NS} / R_\mathrm{NS}^3}$. According to the results of
fully general-relativistic numerical studies for quasiequilibrium states
\citep[see Sect.~\ref{sec:eq} for the
details]{Taniguchi_BFS2007,Taniguchi_BFS2008}, the mass-shedding
condition is given by
\begin{equation}
 \Omega^2 \ge \Omega_\mathrm{ms}^2 := C_\Omega^2 \frac{G
  M_\mathrm{NS}}{R_\mathrm{NS}^3} \pqty{1 + Q^{-1}} , \label{eq:msomega}
\end{equation}
where $C_\Omega \lesssim 0.3$ for binaries of a nonspinning black hole
and a neutron star with the irrotational velocity field. The smallness
of $C_\Omega < 1/\sqrt{2}$ indicates that the mass shedding is helped by
significant tidal deformation, i.e., $c_\mathrm{R}>1$, and/or by
relativistic gravity. This condition also indicates that the
gravitational-wave frequency at the onset of mass shedding is given by
\begin{equation}
 f_\mathrm{ms} = \frac{\Omega_\mathrm{ms}}{\pi} \gtrsim
  \SI{1.0}{\kilo\hertz} \pqty{\frac{C_\Omega}{0.3}}
  \pqty{\frac{M_\mathrm{NS}}{1.4\,M_\odot}}^{1/2}
  \pqty{\frac{R_\mathrm{NS}}{\SI{12}{\km}}}^{-3/2} \pqty{1 +
  Q^{-1}}^{1/2} . \label{eq:msfgw}
\end{equation}
This value might be encouraging for ground-based gravitational-wave
detectors, which have high sensitivity up to $\sim
\SI{1}{\kilo\hertz}$. However, we again caution that the mass shedding
is merely a necessary condition for tidal disruption, and thus the
frequency at tidal disruption should be higher than this value.

\subsubsection{Tidal interaction in the orbital evolution}
\label{sec:intro_tidal_orb}

The discussion in Sect.~\ref{sec:intro_tidal_ms} did not take the effect
of tidal deformation of a neutron star on the orbital motion into
account except for a fudge factor $c_\mathrm{R}$. Tidally-induced higher
multipole moments of the neutron star modify the gravitational
interaction between the binary components \citep[see,
e.g.,][]{Poisson_Will}, so are the orbital evolution and the criterion
for tidal disruption. This problem has thoroughly been investigated in
Newtonian gravity with the ellipsoidal approximation, in which the
isodensity contours are assumed to be self-similar ellipsoids
\citep{Lai_Rasio_Shapiro1993,Lai_Rasio_Shapiro1993_2,Lai_Rasio_Shapiro1994,Lai_Rasio_Shapiro1994_2}. They
find that the tidal interaction acts as additional attractive force and
accordingly the radius of the innermost stable circular orbit is
increased \citep[see
also][]{Rasio_Shapiro1992,Rasio_Shapiro1994,Lai_Wiseman1996,Shibata1996}.
Because (i) the tidally-deformed neutron star develops a reduced
quadrupole moment with the magnitude of components being $\propto
r^{-3}$ associated with the tidal field of the black hole and (ii) the
reduced quadrupole moment produces potential of the form $\propto
r^{-3}$, the gravitational potential in the binary develops an $r^{-6}$
term in addition to the usual $r^{-1}$ term of the monopolar (i.e.,
mass) interaction. The reason that this interaction works as the
attraction is that the neutron star is stretched along the line
connecting the binary components and the enhancement of the pull at the
inner edge dominates over the reduction at the outer edge. The steep
dependence of the potential on the orbital separation indicates that the
tidal interaction is especially important for determining properties of
the close orbit.

These discussions about the tidal effects on the orbital motion have
been revived in the context of gravitational-wave modeling and data
analysis \citep{Flanagan_Hinderer2008}. Specifically, it has been
pointed out that the finite-size effect of a star on the orbital
evolution and hence gravitational waveforms are characterized
quantitatively by the tidal deformability as far as the deformation is
perturbative
\citep{Hinderer2008,Binnington_Poisson2009,Damour_Nagar2009}. Because
the additional attractive force increases the orbital angular velocity
required to maintain a circular orbit for a given orbital separation,
the gravitational-wave luminosity is also increased. In addition, the
coupling of the quadrupole moments between the binary and the deformed
star also enhances the luminosity. These effects accelerate the orbital
decay particularly in the late inspiral phase to the extent that the
difference of gravitational waveforms may be used to extract tidal
deformability of neutron stars. This extraction has been realized in
GW170817
\citep{GW170817,GW170817EOS,De_FLBBB2018,GW170817_2,Narikawa_UKKKST2020}
and GW190425 \citep{GW190425}, whereas the statistical errors are
large. It should also be cautioned that the effect of tidal
deformability is not very large compared to various other effects, e.g.,
the spin and the eccentricity
\citep{Yagi_Yunes2014,Favata2014,Wade_COLFLR2014}. In particular, the
tidal effect comes into play effectively at the fifth post-Newtonian
order ($r^{-6} / r^{-1} = r^{-5}$), but the point-particle terms at this
order have not yet been derived in the post-Newtonian
approximation. Thus, accurate extraction of tidal deformability requires
sophistication not only in the description of tidal effects but also in
the higher-order post-Newtonian corrections to point-particle, monopolar
interactions. This fact has motivated gravitational-wave modeling in the
effective-one-body formalism
\citep{Buonanno_Damour1999,Buonanno_Damour2000} and numerical
relativity.

Tidal interaction and criteria for mass shedding in general relativity
have long been explored for a circular orbit of a ``test'' Newtonian
fluid star around a Kerr (or Schwarzschild) black hole as follows
\citep{Fishbone1973,Mashhoon1975,Lattimer_Schramm1976,Shibata1996,Wiggins_Lai2000,Ishii_Shibata_Mino2005}. The
center of mass of the star is assumed to obey the geodesic equation in
the background spacetime, and the stellar structure is computed with a
model based on the Newtonian Euler's equation of the form
\begin{equation}
 \dv{u_i}{\tau} = - \frac{1}{\rho} \pdv{P}{x^i} - \pdv{\phi}{x^i} -
  C_{ij} x^j ,
\end{equation}
where $\tau$, $x^i$, $u_i$, $\rho$, $P$, $\phi$, and $C_{ij}$ denote the
proper time of the stellar center, spatial coordinates orthogonal to the
geodesic, the internal velocity, the rest-mass density, the pressure,
the gravitational potential associated with the star itself, and the
tidal tensor associated with the black hole, respectively. The
self-gravity of the fluid star is computed in a Newtonian manner from
Poisson's equation sourced by $4\pi G\rho$. The tidal force of the black
hole is incorporated up to the quadrupole order via the tidal tensor
derived from the fully relativistic Riemann tensor \citep[see also
\citealt{vandeMeent2020}]{Marck1983}. Because the gravity of the fluid
star is assumed not to affect the orbital motion and general relativity
is not taken into account for describing its self-gravity, the analysis
based on this model is valid quantitatively only for the cases in which
the black hole is much heavier than the fluid star ($Q \gg 1$) and the
fluid star is not compact ($\mathcal{C} \ll 1$). In addition, the tidal
force of the black hole beyond the quadrupole order, $C_{ij}$, is
neglected \citep{Marck1983}, and this model is valid only if the stellar
radius is much smaller than the curvature scale of the background
spacetime (again, $Q \gg 1$ is assumed). Regarding this point,
higher-order tidal interactions have also been incorporated
\citep{Ishii_Shibata_Mino2005} via the tidal potential computed in the
Fermi normal coordinates \citep{Manasse_Misner1963}.

A series of analysis described above confirms the qualitative dependence
of the mass-shedding and tidal-disruption conditions inferred from
Eq.~\eqref{eq:iscovsms} on binary parameters in a semiquantitative
manner. Specifically, the mass shedding from an incompressible star is
found to occur for the mass and the spin of the black hole satisfying
\begin{equation}
 M_\mathrm{BH} \lesssim C_M ( \chi ) \,M_\odot
  \pqty{\frac{M_\mathrm{NS}}{1.4\,M_\odot}}^{-1/2}
  \pqty{\frac{R_\mathrm{NS}}{\SI{10}{\km}}}^{3/2} ,
\end{equation}
where $C_M (0) \approx 4.6$, $C_M (0.5) \approx 7.8$, $C_M (0.75)
\approx 12$, $C_M (0.9) \approx 19$, and $C_M (1) \approx 68$
\citep{Shibata1996}. This condition tells us that tidal disruption of a
neutron star by a nonspinning black hole is possible only if the
black-hole mass is small compared to astrophysically typical values
\citep[see, e.g.,][]{Ozel_PNM2010,Kreidberg_BFK2012,GWTC1,GWTC2}. At the
same time, the increase in the threshold mass by a factor of $\approx
15$ for extremal black holes is impressive particularly in light of many
massive black holes discovered by gravitational-wave observations.

The threshold mass of the black hole for mass shedding and thus tidal
disruption also depends on the neutron-star equation of state even if
the mass and the radius are identical
\citep{Wiggins_Lai2000,Ishii_Shibata_Mino2005}. If we focus on
polytropes, stiffer equations of state characterized by a larger
adiabatic index are more susceptible to tidal deformation due to the
flatter, less centrally condensed density profile. Conversely, neutron
stars with a soft equation of state are less subject to tidal disruption
than those with a stiff one. These features are also reflected in the
tidal Love number and tidal deformability \citep{Hinderer2008}. Note
that the incompressible model corresponds to the stiffest possible
equation of state. According to the computations performed adopting
compressible stellar models
\citep{Wiggins_Lai2000,Ishii_Shibata_Mino2005}, the threshold mass of
the black hole may be reduced by 10\%--20\% for a soft equation of state
characterized by a small adiabatic index.

In reality, the self-gravity of the neutron star needs to be treated in
a general-relativistic manner. General-relativistic effects associated
with the neutron star have been investigated by a series of work in a
phenomenological manner based on the ellipsoidal approximation
\citep{Ferrari_Gualtieri_Pannarale2009,Ferrari_Gualtieri_Pannarale2010,Pannarale_Tonita_Rezzolla2011,Ferrari_Gualtieri_Maselli2012,Maselli_GPF2012}. However,
quantitative understanding of the mass shedding and tidal disruption
ultimately requires numerical computations of quasiequilibrium states
and dynamical simulations of the merger process in full general
relativity.

\subsection{Brief history of studies on black hole--neutron
  star binaries} \label{sec:intro_history}

Here, we briefly review studies on black hole--neutron star binaries
from the historical perspectives in an approximate chronological
order. We also introduce pioneering studies that are not fully
relativistic, e.g., Newtonian computations of equilibrium states and
partially-relativistic simulations of the coalescences. In the main part
of this article, Sect.~\ref{sec:eq} and Sect.~\ref{sec:sim}, we will
review fully general-relativistic results, i.e., quasiequilibrium states
satisfying the Einstein constraint equations and dynamical evolution
derived by solving the full Einstein equation, from the physical
perspectives.

\subsubsection{Nonrelativistic equilibrium computation}
\label{sec:intro_history_nreq}

Equilibrium configurations of a neutron star governed by Newtonian
self-gravity in general-relativistic gravitational fields of a
background black hole was first studied in \citet{Fishbone1973} for
incompressible fluids in the corotational motion (i.e., the fluid is at
rest in the corotating frame of the binary).  The criterion for mass
shedding was investigated and qualitative results were obtained. This
type of studies has been generalized to accommodate irrotational
velocity fields \citep[i.e., the vorticity is absent;][]{Shibata1996},
compressible, polytropic equations of state \citep{Wiggins_Lai2000}, and
higher-order tidal potential of the black hole
\citep{Ishii_Shibata_Mino2005}. Another direction of extension was to
remove the assumption of the extreme mass ratio of the binary. This
extension was done in \citet{Taniguchi_Nakamura1996} by adopting
modified pseudo-Newtonian potential for the black hole based on the
so-called Paczy{\'n}ski--Wiita potential \citep{Paczynski_Wiita1980} to
determine the location of the innermost stable circular orbit.

However, all these studies have limitation even if we accept the
Newtonian self-gravity of neutron stars. The ellipsoidal approximation
is strictly valid only if the fluid is incompressible and the tidal
field beyond the quadrupole order can be neglected
\citep{Chandrasekhar}. Thus, the internal structure of compressible
neutron stars in a close orbit can be investigated only qualitatively.

The hydrostationary equilibrium of black hole--neutron star binaries was
derived in \citet{Uryu_Eriguchi1998} assuming that the black hole was a
point source of Newtonian gravity and that the neutron star with
irrotational velocity fields obeyed a polytropic equation of state
(irrotational Roche--Riemann problem). The center-of-mass motion of the
neutron star was computed fully accounting for its self-gravity, and the
tidal field of the Newtonian point source was incorporated to the full
order in the ratio of the stellar radius to the orbital
separation. Their subsequent work, \citet{Uryu_Eriguchi1999}, considered
both the corotational and irrotational velocity fields, and differences
from the ellipsoidal approximation have been analyzed.

\subsubsection{Relativistic quasiequilibrium computation}
\label{sec:intro_history_req}

One of the essential features of general relativity is the existence of
gravitational radiation, whose reaction prohibits exactly stationary
equilibria of binaries. Still, an approximately stationary solution to
the Einstein equation may be obtained by solving the constraint
equations, quasiequilibrium conditions derived by some of the evolution
equations, and hydrostationary equations. Such solutions are called
quasiequilibrium states, and Eq.~\eqref{eq:gwvsorb} suggests that they
are reasonable approximations to inspiraling binaries except near merger
\citep[see also][]{Blackburn_Detweiler1992,Detweiler1994}. The
quasiequilibrium states are important not only by their own but also as
initial data of realistic numerical-relativity simulations.

Quasiequilibrium states and sequences of black hole--neutron star
binaries in full general relativity were first studied in
\citet{Miller2001} with preliminary formulation. Approximate
quasiequilibrium states in the extreme mass ratio limit were obtained
for the corotational velocity field in
\citet{Baumgarte_Skoge_Shapiro2004} and later for the irrotational
velocity field in \citet{Taniguchi_BFS2005}. Because gravitational
fields around the black hole are not required to be solved in the
extreme mass ratio limit, these computations were performed only around
(relativistic) neutron stars.

General-relativistic quasiequilibrium states for comparable-mass
binaries were obtained in 2006 by various groups both in the excision
\citep{Grandclement2006,Taniguchi_BFS2006} and the puncture frameworks
\citep{Shibata_Uryu2006,Shibata_Uryu2007}. A general issue in the
numerical computation of black-hole spacetimes is how to handle the
associated physical or coordinate singularity. The excision framework
handles the black hole by removing the interior of a suitably-defined
horizon \citep[see,
e.g.,][]{Dreyer_KSS2003,Ashtekar_Krishnan2004,Gourgoulhon_Jaramillo2006}
from the computational domains and by imposing appropriate boundary
conditions \citep{Cook2002,Cook_Pfeiffer2004}. The puncture framework
separates the singular and regular components in an analytic manner so
that only the latter terms are solved numerically
\citep{Bowen_York1980,Brandt_Brugmann1997}. The details are presented in
Appendix~\ref{app:init}.

\citet{Taniguchi_BFS2007} derived accurate quasiequilibrium sequences in
the excision framework by adopting the conformally-flat background and
investigated properties of close black hole--irrotational neutron star
binaries such as the mass-shedding limit \citep[see
also][]{Grandclement2006e}. \citet{Taniguchi_BFS2008} further improved
the sequences by enforcing nonspinning conditions for the black hole in
a sophisticated manner via the boundary condition at the
horizon. Quasiequilibrium states with spinning black holes were computed
with the same code as initial data for numerical simulations
\citep{Etienne_LSB2009}.

Quasiequilibrium states in the puncture framework were also derived for
irrotational velocity fields in \citet{Shibata_Taniguchi2008} by
extending the formulation for corotating neutron stars
\citep{Shibata_Uryu2006,Shibata_Uryu2007}. \citet{Kyutoku_Shibata_Taniguchi2009}
obtained quasiequilibrium sequences of nonspinning black holes with
varying the method for determining the center of mass of the binary,
which is not uniquely defined in the puncture
framework. Quasiequilibrium states in the puncture framework were
extended to black holes with aligned and inclined spins
\citep{Kyutoku_OST2011,Kawaguchi_KNOST2015}, and the same formulation
has also been adopted to perform merger simulations in the
conformal-flatness approximation \citep{Just_BAGJ2015}. Recently, the
eccentricity reduction method has been implemented in this framework
\citep[see below for preceding work in the excision
framework]{Kyutoku_KKST2021}.

Except for early work in the puncture framework
\citep{Shibata_Uryu2006,Shibata_Uryu2007}, all the computations
described above were performed with the spectral-method library, LORENE,
which enables us to achieve very high precision (see
\citealt{Grandclement_Novak2009} for reviews). Note that
Grandcl{\'e}ment \citep{Grandclement2006} and Taniguchi
\citep{Taniguchi_BFS2006} are two of the main developers of LORENE.

\citet{Foucart_KPT2008} also computed quasiequilibrium sequences in the
excision framework with an independent code, SPELLS
\citep{Pfeiffer_KST2003}. This code implemented a modified Kerr-Schild
background metric for computing highly-spinning black holes
\citep{Lovelace_OPC2008} and the eccentricity reduction method for
performing realistic inspiral simulations
\citep{Pfeiffer_BKLLS2007}. Initial data with inclined black-hole spins
were also derived \citep{Foucart_DKT2011}. The computations of
quasiequilibrium states have now been extended to high-compactness
\citep{Henriksson_FKT2016} and/or spinning neutron stars
\citep{Tacik_FPMKSS2016}. \citet{Papenfort_TGMR2021} have also derived
quasiequilibrium sequences by using another spectral-method library,
KADATH \citep{Grandclement2010}.

\subsubsection{Non/partially-relativistic merger simulations}
\label{sec:intro_history_nrsim}

The merger process of black hole--neutron star binaries was first
studied in Newtonian gravity primarily with the aim of assessing the
potentiality for the central engine of gamma-ray bursts. In the early
work, the black hole was modeled by a point source of Newtonian gravity
with (artificial) absorbing boundary conditions. A series of simulations
with a smoothed-particle-hydrodynamics code explored influences of the
rotational states of the fluids and stiffness of the (polytropic)
equations of state for neutron-star matter
\citep{Kluzniak_Lee1998,Lee_Kluzniak1999,Lee_Kluzniak1999-2,Lee2000,Lee2001}. They
studied the process of tidal disruption, subsequent formation of a black
hole--disk system and mass ejection, properties of the remnant disk and
unbound material, and gravitational waveforms emitted during merger.

Around the same time, \citet{Janka_ERF1999} performed simulations
incorporating detailed microphysics with a mesh-based hydrodynamics
code. Specifically, their code had implemented a temperature- and
composition-dependent equation of state \citep{Lattimer_Swesty1991} and
neutrino emission modeled in terms of the leakage scheme
\citep{Ruffert_Janka_Schafer1996}. By performing simulations for various
configurations of binaries, a hot and massive remnant disk with $\gtrsim
\SI{10}{\mega\eV}$ and $0.2$--$0.7 \,M_\odot$ was suggested to be formed,
and the neutrino luminosity was found to reach
\num{e53}--\SI{e54}{erg.s^{-1}} during 10--\SI{20}{\milli\second} after
formation of a massive disk. Pair annihilation of neutrinos was also
investigated by post-process calculations \citep{Ruffert_JTS1997} and
was found to be capable of explaining the total energy of gamma-ray
bursts. Although these Newtonian results were still highly qualitative
and both the disk mass and temperature can be overestimated for given
binary parameters (see below), it was first suggested by dynamical
simulations that black hole--neutron star binary coalescences could be
promising candidates for the central engine of short-hard gamma-ray
bursts if the massive accretion disk was indeed
formed. \citet{Rosswog_Speith_Wynn2004} also performed simulations in
similar setups with a smoothed-particle-hydrodynamics code adopting a
different equation of state \citep{Shen_TOS1998}.

The gravity in the vicinity of a black hole modeled by a Newtonian point
source is qualitatively different from that in reality. In particular,
the innermost stable circular orbit is absent in the Newtonian
point-particle model. Because of this difference, early Newtonian work
concluded that the neutron star might be disrupted without an immediate
plunge even in an orbit closer to the black hole than $\lesssim
6GM_\mathrm{BH} / c^2$. Consequently, they indicated that a massive
remnant disk with $\gtrsim 0.1\,M_\odot$ might be formed around the black
hole irrespective of the mass ratio and the rotational states of
fluids. It should also be mentioned that the orbital evolution within
Newtonian gravity frequently exhibited episodic mass transfer (see
\citealt{Clark_Eardley1977,Cameron_Iben1986,Benz_BCP1990} for relevant
discussions). That is, the neutron star is only partially disrupted via
the stable mass transfer during the close encounter with the black hole,
becomes a ``mini-neutron star'' \citep{Rosswog2005} with increasing the
binary separation, and continues the orbital motion. This has never been
found in fully relativistic simulations (although not completely
rejected throughout the possible parameter space) and may be regarded as
another qualitative difference associated with the realism of
gravitation (see also Appendix~\ref{app:ae_smt}).

To overcome these drawbacks, \citet{Rosswog2005} performed
smoothed-particle-hydrodynamics simulations by modeling the black-hole
gravity in terms of a pseudo-Newtonian potential
\citep{Paczynski_Wiita1980}. A potential for modeling the gravity of a
spinning black hole \citep{Artemova_Bjornsson_Novikov1996} was also
adopted in later mesh-based simulations \citep{Ruffert_Janka2010}, and
the episodic mass transfer was still observed for some parameters of
binaries. These work found that the massive disk with $\gtrsim
0.1\,M_\odot$ was formed only for binaries with low-mass and/or spinning
black holes. Because this feature agrees qualitatively with the fully
relativistic results, simulations with a pseudo-Newtonian potential
might be helpful to understand the nature of black hole--neutron star
binary mergers qualitatively or even semiquantitatively.\footnote{It
should be cautioned that, in the Newtonian and pseudo-Newtonian
simulations in which the black hole is modeled by a point particle,
numerical results can depend significantly on the treatment for the
accreted material. For example, it is not clear how the angular momentum
of the infalling material is distributed to the spin and the orbital
angular momentum of the black hole in this treatment. Thus, a rule has
to be artificially determined. By contrast, in
fully-general-relativistic simulations, the evolution of the black hole
is computed in an unambiguous manner.}

While numerical-relativity simulations have been feasible since 2006
(see Sect.~\ref{sec:intro_history_rsim}), approximately
general-relativistic simulations have also been performed without
simplifying the black holes by point sources of gravity. This is
particularly the case of smoothed-particle-hydrodynamics codes, which
are especially useful to track the motion of the material ejected from
the system but have not been available in numerical relativity (see
\citealt{Rosswog_Diener2021} for recent development). One of the popular
approaches to incorporate general relativity is the conformal-flatness
approximation \citep{Faber_BST2006,Faber_BSTR2006,Just_BAGJ2015}. In
these work, the gravity of neutron stars was also treated in a
general-relativistic manner. Another work adopted the Kerr background
for modeling the black hole, while the neutron star is modeled as a
Newtonian self-gravitating object, and studied dependence of the merger
process on the inclination angle of the black-hole spin with respect to
the orbital angular momentum \citep{Rantsiou_KLR2008}. Results of these
simulations agree qualitatively with those from pseudo-Newtonian and
numerical-relativity simulations.

\subsubsection{Fully-relativistic merger simulations}
\label{sec:intro_history_rsim}

General-relativistic effects play a crucial role in the dynamics of
close black hole--neutron star binaries. First of all, the inspiral is
driven by gravitational radiation reaction. Dynamics right before merger
is affected crucially by further general-relativistic effects, which
include strong attractive force between two bodies, associated presence
of the innermost stable circular orbit, spin-orbit coupling, and
relativistic self-gravity of neutron stars. Accordingly, the orbital
evolution, the merger process, the criterion for tidal disruption, and
the evolution of disrupted material are all affected substantially by
the general-relativistic effects. Although non-general-relativistic work
has provided qualitative insights, numerical simulations in full general
relativity are obviously required for accurately and quantitatively
understanding the nature of black hole--neutron star binary
coalescences. This is particularly the case in development of an
accurate gravitational-wave template for the data analysis.

Soon after the breakthrough success in simulating binary-black-hole
mergers \citep{Pretorius2005,Campanelli_LMZ2006,Baker_CCKV2006},
\citet{Shibata_Uryu2006,Shibata_Uryu2007,Shibata_Taniguchi2008} started
exploration of black hole--neutron star binary coalescences in full
general relativity extending their early work on binary neutron stars
\citep{Shibata1999,Shibata_Uryu2000,Shibata_Uryu2002,Shibata_Taniguchi_Uryu2003,Shibata_Taniguchi_Uryu2005,Shibata_Taniguchi2006}.
While these work adopted initial data computed in the puncture framework
for moving-puncture simulations, \citet{Etienne_FLSTB2008} independently
performed moving-puncture simulations with excision-based initial data
by extending their early work on binary neutron stars \citep[see also
\citealt{Loffler_Rezzolla_Ansorg2006} for early work on a head-on
collision in a similar setup]{Duez_MSB2003}. \citet{Duez_FKPST2008} also
performed simulations for black hole--neutron star binaries based on the
excision method by introducing hydrodynamics equation solvers to a
spectral-method code, SpEC, for binary black holes
\citep{Boyle_BKMPSCT2007,Boyle_BKMPPS2008,Scheel_BCKMP2009}. All these
studies were performed for nonspinning black holes and neutron stars
modeled by a polytropic equation of state with $\Gamma = 2$.

To derive accurate gravitational waveforms, longterm simulations need to
be performed. Effort in this direction was made with the aid of an
adaptive-mesh-refinement code (see Appendix~\ref{app:sim_mr}), SACRA
\citep{Yamamoto_Shibata_Taniguchi2008}, by the authors
\citep{Shibata_KYT2009,Shibata_KYT2009e}. Around the same time,
\citet{Etienne_LSB2009} independently studied the effect of black-hole
spins by another adaptive-mesh-refinement code. Systematic longterm
studies were started employing nuclear-theory based equations of state
approximated by piecewise-polytropic equations of state
\citep{Read_LOF2009} for both nonspinning
\citep{Kyutoku_Shibata_Taniguchi2010,Kyutoku_Shibata_Taniguchi2010e} and
spinning black holes \citep{Kyutoku_OST2011}. A tabulated, temperature-
and composition-dependent equation of state \citep{Shen_TOS1998} was
also incorporated in simulations with SpEC around the same time
\citep{Duez_FKOT2010}, while neutrino reactions were not considered in
this early work. SpEC was also used to simulate systems with inclined
black-hole spins \citep{Foucart_DKT2011} or an increased mass ratio of
$Q=7$ \citep{Foucart_DKSST2012}. These simulations clarified
quantitatively the criterion for tidal disruption, properties of the
remnant disk and black hole, and emitted gravitational waves in the
merger phase.

Mass ejection from black hole--neutron star binaries and associated
fallback of material began to be explored in numerical relativity at the
beginning of 2010's \citep{Chawla_ABLLMN2010,Kyutoku_OST2011}. Actually,
these studies predate the corresponding investigations for binary
neutron stars in numerical relativity
\citep{Hotokezaka_KKOSST2013}. Although the authors of this article
suggested that $\ge 0.01\,M_\odot$ may be ejected dynamically from black
hole--neutron star binaries in this early work \citep{Kyutoku_OST2011},
they were unable to show the ejection of material with confidence,
partly because the artificial atmosphere was not tenuous enough and the
computational domains were not large enough. By contrast, the mass
ejection from hyperbolic encounters was discussed clearly
\citep{Stephens_East_Pretorius2011,East_Pretorius_Stephens2012,East_Paschalidis_Pretorius2015}.

Serious investigations of the dynamical mass ejection were initiated in
2013
\citep{Foucart_DDKMOPSST2013,Lovelace_DFKPSS2013,Kyutoku_Ioka_Shibata2013},
right after the first version of this review article was released,
stimulated by the importance of electromagnetic counterparts to
gravitational waves \citep[the preprint version of which appeared on
arXiv in 2013]{KLV2020}. \citet{Kyutoku_IOST2015} systematically studied
kinematic properties such as the mass and the velocity as well as the
morphology of dynamical ejecta with reducing the density of artificial
atmospheres and enlarging the computational
domains. \citet{Kawaguchi_KNOST2015} also investigated the impact of the
inclination angle of the black-hole spin for both the remnant disk and
the dynamical ejecta \citep[see also][]{Foucart_DDKMOPSST2013}. The
study of mass ejection has now become routine in numerical-relativity
simulations of black hole--neutron star binary
coalescences. Accordingly, a lot of discussions about mass ejection are
newly added in this update.

The current frontier of numerical relativity for neutron-star mergers is
the incorporation of magnetohydrodynamics and neutrino-radiation
hydrodynamics as accurately as possible. These physical processes play
essentially no role during the inspiral phase
\citep{Chawla_ABLLMN2010,Etienne_LPS2012,Etienne_Paschalidis_Shapiro2012,Deaton_DFOOKMSS2013},
and thus gravitational radiation, disk formation, and dynamical mass
ejection are safely studied by pure hydrodynamics simulations except for
the chemical composition of the dynamical ejecta. However, both
neutrinos and magnetic fields are key agents for driving postmerger
evolution including disk outflows and ultrarelativistic jets. In
addition, the electron fraction of the ejected material, either
dynamical or postmerger, can be quantified only by simulations
implementing composition-dependent equations of state with appropriate
schemes for neutrino transport.

Although numerical-relativity codes for magnetohydrodynamics were
already available in the late 2000's and a preliminary simulation was
performed by \citet{Chawla_ABLLMN2010}, magnetohydrodynamics simulations
are destined to struggle with the need to resolve short-wavelength modes
of instability (see, e.g., \citealt{Balbus_Hawley1998} for
reviews). Various simulations have been performed aiming at clarifying
the launch of an ultrarelativistic jet, generally finding magnetic-field
amplification difficult to resolve
\citep{Etienne_LPS2012,Etienne_Paschalidis_Shapiro2012}. The situation
was much improved by high-resolution simulations performed in \citet[see
also \citealt{Wan2017} for a follow-up]{Kiuchi_SKSTW2015}. At around the
same time, magnetohydrodynamics simulations with a presumed strong
dipolar field have been performed to clarify the potentiality of black
hole--neutron star binaries as a central engine of short-hard gamma-ray
bursts
\citep{Paschalidis_Ruiz_Shapiro2015,Ruiz_Shapiro_Tsokaros2018,Ruiz_PTS2020}. Simulations
beyond ideal magnetohydrodynamics have recently been performed aiming at
clarifying magnetospheric activities right before merger
\citep{East_LLP2021}.

As the electron fraction of the ejected material is crucial to determine
the nucleosynthetic yield and the feature of associated
kilonovae/macronovae, numerical-relativity simulations with neutrino
transport have been performed extensively following those for
binary-neutron-star mergers
\citep{Sekiguchi_KKS2011,Sekiguchi_KKS2011-2}. Neutrino-radiation-hydrodynamics
simulations of black hole--neutron star binaries were first performed in
\citet{Deaton_DFOOKMSS2013,Foucart_DDOOHKPSS2014,Foucart_DBDKHKPS2017,Brege_DFDCHKOPS2018}
with neutrino emission based on a leakage scheme \citep[see][for the
description in Newtonian
cases]{Ruffert_Janka_Schafer1996,Rosswog_Liebendorfer2003} and with
composition-dependent equations of state. These simulations are also
capable of predicting neutrino emission from the postmerger
system. \citet{Kyutoku_KSST2018} incorporated neutrino absorption by the
material based on the two moment formalism
\citep{Thorne1981,Shibata_KKS2011} again following work on binary
neutron stars \citep{Wanajo_SNKKS2014,Sekiguchi_KKS2015}.

The longterm postmerger evolution of the remnant accretion disk also
requires numerical investigations. These simulations need to be
performed with sophisticated microphysics, because the evolution of the
disk is governed by weak interaction processes such as the neutrino
emission and absorption and magnetohydrodynamical processes. The
liberated gravitational binding energy may eventually be tapped to
launch a postmerger wind as well as an ultrarelativistic
jet. Simulations focusing on the postmerger evolution with detailed
neutrino transport are initially performed without incorporating sources
of viscosity for a short term \citep{Foucart_ORDHKOPSS2015}. This work
has been extended in the Cowling approximation to incorporate magnetic
fields that provide effective viscosity and induce magnetohydrodynamical
effects \citep{Nouri_etal2018}.
\citet{Fujibayashi_SWKKS2020,Fujibayashi_SWKKS2020-2} performed
fully-general-relativistic viscous-hydrodynamics simulations for
postmerger systems with detailed neutrino transport. Finally,
\citet{Most_PTR2021-2} reported results of postmerger simulations for
nearly-equal-mass black hole--neutron star binaries with neutrino
transport and magnetohydrodynamics in full general relativity. Still, it
is not feasible to perform fully-general-relativistic simulations
incorporating both detailed neutrino transport and well-resolved
magnetohydrodynamics. Because the postmerger mass ejection is now widely
recognized as an essential source of nucleosynthesis and electromagnetic
emission for binary-neutron-star mergers \citep{Shibata_FHKKST2017},
sophisticated simulations for this problem will remain the central topic
in the future study of black hole--neutron star binary coalescences.

After the detection of gravitational waves from binary neutron stars
GW170817 \citep{GW170817} and GW190425 \citep{GW190425}, it became
apparent that robust characterization of source properties requires us
to distinguish binary neutron stars from black hole--neutron star
binaries (see Sect.~\ref{sec:dis_dis_ns}). This situation motivated
studies on mergers of very-low-mass black hole--neutron star binaries to
clarify disk formation, mass ejection, and gravitational-wave emission
in a more precise manner than what was done before
\citep{Foucart_DKNPS2019,Hayashi_KKKS2021,Most_PTR2021}. Here, ``very
low mass'' means that it is consistent with being a neutron star and
that observationally distinguishing binary types is not
straightforward. Interestingly, these studies are revealing overlooked
features of very-low-mass ratio systems. Finally, longterm simulations
with $\gtrsim 10$--$15$ inspiral orbits have recently been performed
aiming at deriving accurate gravitational waveforms
\citep{Foucart_etal2019,Foucart_etal2021}.

\subsection{Outline and notation} \label{sec:intro_outline}

This review article is organized as follows. In Sect.~\ref{sec:eq}, we
review the current status of the study on quasiequilibrium states of
black hole--neutron star binaries in general relativity. First, we
summarize physical parameters characterizing a binary in
Sect.~\ref{sec:eq_param}. Next, the parameter space surveyed is
summarized in Sect.~\ref{sec:eq_space}. Results and implications are
reviewed in Sect.~\ref{sec:eq_res} and Sect.~\ref{sec:eq_end},
respectively. In Sect.~\ref{sec:sim}, we review the current status of
the study on coalescences of black hole--neutron star binaries in
numerical relativity. Methods for numerical simulations are briefly
described in Sect.~\ref{sec:sim_meth}, and the parameter space
investigated is summarized in Sect.~\ref{sec:sim_space}. The remainder
of Sect.~\ref{sec:sim} is devoted to reviewing numerical results, and
readers interested only in them should jump into
Sect.~\ref{sec:sim_mrg}, in which we start from reviewing the overall
process of the binary coalescence and tidal disruption in the late
inspiral and merger phases. Properties of the remnant disk, remnant
black hole, fallback material, and dynamical ejecta are summarized in
Sect.~\ref{sec:sim_rem}. Postmerger evolution of the remnant disk is
reviewed in Sect.~\ref{sec:sim_pm}. Gravitational waveforms and spectra
are reviewed in Sect.~\ref{sec:sim_gw}. Finally in Sect.~\ref{sec:dis},
we discuss implications of numerical results to electromagnetic emission
and characterization of observed astrophysical sources. Formalisms to
derive quasiequilibrium states and to simulate dynamical evolution are
reviewed in Appendix~\ref{app:init} and Appendix~\ref{app:sim},
respectively. Appendix~\ref{app:ae} presents analytic estimates related
to discussions made in this article.

\begin{table} \caption{List of symbols}
 \centering
 \begin{tabular}{cc}
  \toprule
  symbol & content \\
  \midrule
  & geometric quantity \\ \midrule
  $g_{\mu\nu}$ & spacetime metric \\
  $g$ & determinant of $g_{\mu \nu}$ \\
  $\nabla_\mu$ & covariant derivative associated with $g_{\mu \nu}$ \\
  $\Sigma_t$ & three-dimensional hypersurface of a constant time $t$ \\
  $n^{\mu}$ & future-directed timelike unit vector normal to
      $\Sigma_t$ \\
  $\alpha$  & lapse function \\
  $\beta^i$  & shift vector \\
  $\gamma_{\mu\nu}$ & induced metric $\gamma_{\mu \nu} := g_{\mu \nu} +
      n_\mu n_\nu$ on $\Sigma_t$ \\
  $D_i$ & covariant derivative associated with $\gamma_{ij}$ \\
  $\gamma$ & determinant of $\gamma_{ij}$; $\sqrt{-g} = \alpha
      \sqrt{\gamma}$ \\
  $K_{ij}$ & extrinsic curvature on $\Sigma_t$ \\
  $K$ & trace of the extrinsic curvature $K := \gamma^{ij} K_{ij} $ \\
  \midrule
  & hydrodynamical quantity \\
  \midrule
  $T^{\mu \nu}$ & energy--momentum tensor \\
  $u^\mu$ & four velocity of the fluid \\
  $v^i$ & three velocity of the fluid $u^i / u^t$ \\
  $w$ & Lorentz factor of the fluid $\alpha u^t$ \\
  $\rho$ & baryon rest-mass density \\
  $\varepsilon$ & specific internal energy \\
  $P$ & pressure \\
  $h$ & specific enthalpy $h := c^2 + \varepsilon + P/ \rho$ \\
  $T$ & temperature \\
  $Y_\mathrm{e}$ & electron fraction \\
  $\kappa$ & polytropic constant \\
  $\Gamma$ & adiabatic index \\
  $\alpha_\nu$ & alpha parameter for the viscosity \textit{\`{a} la}
      \citet{Shakura_Sunyaev1973} \\
  \midrule
  & parameter of the black hole \\
  \midrule
  $M_\mathrm{BH}$ & gravitational mass of the black hole in isolation \\
  $S_\mathrm{BH}$ & spin angular momentum of the black hole \\
  $\chi$ & dimensionless spin parameter of the black hole $\chi :=
      c S_\mathrm{BH} / (G M_\mathrm{BH}^2)$ \\
  $\iota$ & inclination angle of the black-hole spin with respect to the
      orbital plane \\
  \midrule
  & parameter of the neutron star \\
  \midrule
  $M_\mathrm{NS}$ & gravitational mass of the neutron star in
      isolation \\
  $M_\mathrm{B}$ & baryon rest mass of the neutron star \\
  $R_\mathrm{NS}$ & circumferential radius of a spherical neutron star
      in isolation \\
  $\mathcal{C}$ & compactness of the neutron star $\mathcal{C} :=
      GM_\mathrm{NS} / (c^2 R_\mathrm{NS}$) \\
  \midrule
  & binary parameter \\
  \midrule
  $m_0$ & gravitational mass of the binary at infinite separation $m_0
      := M_\mathrm{BH} + M_\mathrm{NS}$ \\
  $Q$ & mass ratio of the binary $Q := M_\mathrm{BH} / M_\mathrm{NS}$ \\
  $\Omega$ & orbital angular velocity of the binary \\
  $f$ & gravitational-wave frequency \\
  \bottomrule
 \end{tabular} \label{table:para}
\end{table}

The notation adopted in this article is summarized in Table
\ref{table:para}. Among the parameters shown in this table,
$M_\mathrm{BH}$, $\chi$, $\iota$, $M_\mathrm{NS}$, and $R_\mathrm{NS}$
are frequently used to characterize binary systems in this article. The
negative value of $\chi$ is allowed for describing anti-aligned spins,
meaning that the dimensionless spin parameter and the inclination angle
are given by $\abs{\chi}$ and $\iota = \ang{180}$,
respectively. Hereafter, the dimensionless spin parameter is referred to
also by the spin parameter for simplicity. Greek and Latin indices
denote the spacetime and space components, respectively. We adopt
geometrical units in which $G=c=1$ in Sect.~\ref{sec:eq},
Sect.~\ref{sec:sim_gw}, Appendix~\ref{app:init}, and Appendix
\ref{app:sim}.

This article focuses on fully general-relativistic studies of black
hole--neutron star binaries, and other types of compact object binaries
are not covered in a comprehensive manner. Numerical-relativity
simulations of compact object binaries in general are reviewed in, e.g.,
\citet{Lehner_Pretorius2014,Duez_Zlochower2019}. Simulations of compact
object binaries involving neutron stars and their implications for
electromagnetic counterparts are reviewed in, e.g.,
\citet{Paschalidis2017,Baiotti_Rezzolla2017,Shibata_Hotokezaka2019}. Black
hole--neutron star binaries are also reviewed briefly in
\citet{Foucart2020}.

\newpage

\section{Quasiequilibrium state and sequence} \label{sec:eq}

Quasiequilibrium states of compact object binaries in close orbits are
important from two perspectives. First, they enable us to understand
deeply the tidal interaction of comparable-mass binaries in general
relativity. Second, they serve as realistic initial conditions for
dynamical simulations in numerical relativity.

For the purpose of the former, a sequence of quasiequilibrium states
parametrized by the orbital separation or angular velocity, i.e.,
quasiequilibrium sequences, should be investigated as an approximate
model for the evolution path of the binary. In this section, we review
representative numerical results of quasiequilibrium sequences derived
to date. The formulation to construct black hole--neutron star binaries
in quasiequilibrium is summarized in Appendix~\ref{app:init}. Because
the differential equations to be solved are typically of elliptic type
(see also the end of this section), most numerical computations adopt
spectral methods for achieving high precision (see
\citealt{Grandclement_Novak2009} for reviews). In this section,
geometrical units in which $G=c=1$ is adopted.

\subsection{Physical parameters of the binary} \label{sec:eq_param}

In this Sect.~\ref{sec:eq_param}, we present physical quantities
required for quantitative analysis of quasiequilibrium sequences. Each
sequence is specified by physical quantities conserved at least
approximately along the sequences, and these quantities also serve as
labels of binary models in dynamical simulations. We also need physical
quantities that characterize each quasiequilibrium state to study its
property.

To begin with, we introduce a helical Killing vector used in modeling
quasiequilibrium states of black hole--neutron star binaries (see also
Appendix~\ref{app:init}). Because the time scale of gravitational
radiation reaction is much longer than the orbital period except for
binaries in a very close orbit as we discussed in
Sect.~\ref{sec:intro_orbit}, the binary system appears approximately
stationary in the comoving frame. Such a system is considered to be in
quasiequilibrium and is usually modeled by assuming the existence of a
helical Killing vector with the form of
\begin{equation}
 \xi^\mu = ( \partial_t )^\mu + \Omega ( \partial_\varphi )^\mu ,
  \label{eq:helical}
\end{equation}
where $\Omega$ denotes the orbital angular velocity of the system. The
helical Killing vector is timelike and spacelike in the near zone of
$\varpi \lesssim c/\Omega$ and the far zone of $\varpi \gtrsim
c/\Omega$, respectively, where $\varpi$ denotes the distance from the
rotational axis and we inserted $c$ for clarity. Thus, if we focus only
on quasiequilibrium configurations in the near zone, we may assume the
existence of a timelike Killing vector, which allows us to define
several physical quantities in a meaningful manner.

Here, it is necessary to keep the following two (not independent)
caveats in mind if we consider helically symmetric spacetimes. First,
spacetimes of binaries cannot be completely helically symmetric in full
general relativity. That is, it is not realistic to assume that a
helical Killing vector exists in the entire spacetime. The reason is
that the helical symmetry holds throughout the spacetime only if
standing gravitational waves are present everywhere. However, the total
energy of the system diverges for such a case. Thus, the helical Killing
vector can be supposed to exist only in a limited region of the
spacetime, e.g., in the local wave zone. A simpler strategy for studying
quasiequilibrium states of a binary is to neglect the presence of
gravitational waves. Although this is an overly simplified assumption,
this strategy has been employed in the study of quasiequilibrium states
of compact object binaries. The results introduced in this section are
derived by assuming that gravitational waves are absent. Moreover, the
induced metric is assumed to be conformally flat (see Appendix
\ref{app:init} for the details.)

Second, gravitational radiation reaction violates the helical symmetry
in full general relativity. To compute realistic quasiequilibrium states
of compact object binaries, we need to take radiation reaction into
account. Procedures for this are described in Appendix
\ref{app:init_beyond_ecc}.

\subsubsection{Parameters of the black hole} \label{sec:eq_param_bh}

It is reasonable to assume that the irreducible mass (i.e., the area of
the event horizon) and the magnitude of the spin angular momentum of the
black hole are conserved along a quasiequilibrium sequence, because the
absorption of gravitational waves by the black hole is only a tiny
effect \citep[see also \citealt{Poisson_Sasaki1995,Poisson2004} for
relevant work in black-hole perturbation
theory]{Alvi2001,Chatziioannou_Poisson_Yunes2013}. The irreducible mass
of the black hole is defined by \citep{Christodoulou1970}
\begin{equation}
 M_\mathrm{irr} := \sqrt{\frac{A_\mathrm{EH}}{16\pi}} ,
\end{equation}
where $A_\mathrm{EH}$ is the proper area of the event horizon. Because
the event horizon cannot be identified in quasiequilibrium
configurations, its area, $A_\mathrm{EH}$, is usually approximated by
that of the apparent horizon, $A_\mathrm{AH}$. It is reasonable to
consider that $A_\mathrm{AH}$ agrees at least approximately with
$A_\mathrm{EH}$ in the current context, because a timelike Killing
vector is assumed to exist in the vicinity of the black hole
\citep[Chap.~9]{Hawking_Ellis}. The magnitude of the spin angular
momentum is determined in terms of an approximate Killing vector
$\xi_\mathcal{S}^i$ for axisymmetry on the horizon as
\citep{Dreyer_KSS2003,Caudill_CGP2006}
\begin{equation}
 S_\mathrm{BH} := \frac{1}{8\pi} \oint_\mathcal{S} \pqty{K_{ij} - K
  \gamma_{ij}} \xi_\mathcal{S}^j \dd{S}^i . \label{eq:approxspin}
\end{equation}
An approximate Killing vector, $\xi_\mathcal{S}^i$, may be determined on
the horizon by requiring some properties satisfied by genuine Killing
vectors to hold (see Appendix~\ref{app:init_grav_ex}).

The angle of the black-hole spin angular momentum is usually evaluated
in terms of coordinate-dependent quantities specific to individual
formulations, because no geometric definition is known. It should be
cautioned that the direction of the black-hole spin seen from a distant
observer changes for the case in which the precession motion occurs
\citep{Apostolatos_CST1994}. Still, the angle between the black-hole
spin and the orbital angular momentum of the binary is approximately
conserved during the evolution, and hence, can be employed to
characterize the system.

The gravitational mass \citep[sometimes called the Christodoulou
mass;][]{Christodoulou1970} of the black hole in isolation is given by
\begin{equation}
 M_\mathrm{BH}^2 = M_\mathrm{irr}^2 + \frac{S_\mathrm{BH}^2}{4
  M_\mathrm{irr}^2} .
\end{equation}
By introducing a dimensionless spin parameter of the black hole,
\begin{equation}
 \chi := \frac{S_\mathrm{BH}}{M_\mathrm{BH}^2} ,
\end{equation}
the gravitational mass is also written by
\begin{equation}
 M_\mathrm{BH} = M_\mathrm{irr} \bqty{\frac{2}{1 + ( 1 - \chi^2
  )^{1/2}}}^{1/2} .
\end{equation}
Because $M_\mathrm{BH}$ is directly measured in actual observations,
this quantity rather than $M_\mathrm{irr}$ is usually adopted to label
quasiequilibrium sequences and the models of binary systems in dynamical
simulations for spinning black holes, along with the spin parameter,
$\chi$.

\subsubsection{Parameters of the neutron star} \label{sec:eq_param_ns}

The baryon rest mass of the neutron star given by
\begin{equation}
 M_\mathrm{B} = \int \rho u^t \sqrt{-g} \dd[3]{x} = \int \rho \alpha u^t
  \sqrt{\gamma} \dd[3]{x}
\end{equation}
is conserved along quasiequilibrium sequences assuming that the
continuity equation holds and that mass ejection from the neutron star
does not occur prior to merger. The spin angular momentum of the neutron
star may be evaluated on the stellar surface by
Eq.~\eqref{eq:approxspin} \citep{Tacik_FPMKSS2016}, although it is not
conserved on a long time scale due to the spin-down. The orientation of
the spin is also affected by the precession motion.

In contrast to black holes, an equation of state needs to be provided to
specify finite-size properties of neutron stars such as the radius and
the tidal deformability, although the realistic equation of state at
supranuclear density is still uncertain (see, e.g.,
\citealt{Lattimer_Prakash2016,Oertel_HKT2017,Baym_HKPST2018} for
reviews). Because of rapid cooling by neutrino emission in the initial
stage and subsequent photon emission (see, e.g.,
\citealt{Potekhin_Pons_Page2015} for reviews), temperature of
not-so-young neutron stars relevant to coalescing compact object
binaries is likely to be much lower than the Fermi energy of constituent
particles. Thus, it is reasonable to adopt a fixed zero-temperature
equation of state throughout the quasiequilibrium sequence. The
zero-temperature equation of state allows us to express all the
thermodynamic quantities as functions of a single variable, e.g., the
rest-mass density.

As a qualitative model, the polytrope of the form
\begin{equation}
 P ( \rho ) = \kappa \rho^{\Gamma} ,
\end{equation}
where $\kappa$ and $\Gamma$ are the polytropic constant and the
adiabatic index, respectively, has often been adopted in the study of
quasiequilibrium sequences. The neutron-star matter is frequently
approximated by a polytrope with $\Gamma \approx 2$--3. More
sophisticated models include piecewise polytropes \citep{Read_LOF2009}
and generalization thereof \citep{OBoyle_MSR2020}, spectral
representations \citep{Lindblom2010}, and various nuclear-theory-based
tabulated equations of state. We will come back to this topic later in
Sect.~\ref{sec:sim_meth_eos}.

Once a hypothetical equation of state is given, the gravitational mass
of a neutron star in isolation, $M_\mathrm{NS}$, is determined for a
given value of the baryon rest mass, $M_\mathrm{B}$, via the
Tolman-Oppenheimer-Volkoff equation
\citep{Tolman1939,Oppenheimer_Volkoff1939} [if the neutron star is
spinning, the magnitude of the spin also comes into play
\citep{Hartle1967,Friedman_Stergioulas}]. The gravitational mass rather
than the baryon rest mass is usually adopted to label the models of
dynamical simulations, primarily because the gravitational mass is
directly measured in actual observations. The equation of state also
determines the radius and the compactness for a given mass of the
neutron star. By imposing perturbative tidal fields on a background
spherical configuration, the tidal deformability as a function of the
neutron-star mass is computed from the ratio of the tidally-induced
multipole moment and the exerted tidal field
\citep{Hinderer2008,Binnington_Poisson2009,Damour_Nagar2009}.

\subsubsection{Parameters of the binary system}

The Arnowitt--Deser--Misner (ADM) mass of the system
\citep{Arnowitt_Deser_Misner2008} is evaluated in isotropic Cartesian
coordinates (see, e.g., \citealt[Chap.~8]{York1979,Gourgoulhon} for
further details) as
\begin{equation}
 M_\mathrm{ADM} = - \frac{1}{2\pi} \oint_{r \to \infty} \partial_i \psi
  \dd{S}^i ,
\end{equation}
where $\psi$ is the conformal factor, which is given by $\psi =
\gamma^{1/12}$ in Cartesian coordinates for a conformally-flat case (see
Appendix~\ref{app:init}). This quantity should decrease as the orbital
separation decreases along a quasiequilibrium sequence because of the
strengthening of gravitational binding. The binding energy of a binary
system is often defined by
\begin{equation}
 E_\mathrm{b} := M_\mathrm{ADM} - m_0 ,
\end{equation}
where the total mass $m_0 := M_\mathrm{BH} + M_\mathrm{NS}$ corresponds
to the ADM mass of the binary system at infinite orbital separation.

The Komar mass is originally defined as a charge associated with a
timelike Killing vector \citep{Komar1959} and is evaluated in the $3+1$
formulation by \citep[see, e.g.,][Sect.~5]{Shibata}
\begin{equation}
 M_\mathrm{K} = \frac{1}{4\pi} \oint_{r \to \infty} \pqty{\partial_i
  \alpha - K_{ij} \beta^j} \dd{S}^i .
\end{equation}
Since quasiequilibrium states of black hole--neutron star binaries are
computed assuming the existence of a helical Killing vector which is
timelike in the near zone, the Komar mass may be considered as a
physical quantity, at least approximately. If the second term in the
integral falls off sufficiently rapidly, as is typical for the case in
which the linear momentum of the system vanishes, the Komar mass may be
evaluated only from the first term, i.e., the derivative of the lapse
function. Because the ADM and Komar masses should agree if a timelike
Killing vector exists \citep[see also
\citealt{Beig1978,Ashtekar_MagnonAshtekar1979}]{Friedman_Uryu_Shibata2002,Shibata_Uryu_Friedman2004},
their fractional difference,
\begin{equation}
 \delta M := \abs{\frac{M_\mathrm{ADM} - M_\mathrm{K}}{M_\mathrm{ADM}}}
  ,
\end{equation}
measures the global error in the numerical computation. This quantity is
sometimes called the virial error.

An ADM-like angular momentum of the system may be defined by
\citep{York1979}
\begin{equation}
 J_i := \frac{1}{16\pi} \underline{\epsilon}_{ijk} \oint_{r \to \infty}
  \pqty{X^j K^{kl} - X^k K^{jl}} \dd{S}_l ,
\end{equation}
where $X^i$ and $\underline{\epsilon}_{ijk}$ denote, respectively,
Cartesian coordinates relative to the center of mass of the binary and
the Levi-Civita tensor for the flat space. It should be cautioned that
this quantity is well-defined only in restricted coordinate systems
\citep[see, e.g.,][Chap.~8]{York1979,Gourgoulhon}. This subtlety is
irrelevant to the results reviewed in this article. For binary systems
with the reflection symmetry about the orbital plane, only the component
normal to the plane is nonvanishing and will be denoted by $J$.

\subsection{Current parameter space surveyed} \label{sec:eq_space}

\begin{table}
 \caption{Summary of the study on quasiequilibrium sequences. The first
 column points to the references. The other columns are explained in the
 body text. We do not list references which present individual
 quasiequilibrium states, including initial data for dynamical
 simulations.} \label{table:qe} \centering
 \begin{tabular}{llllllll}
  \toprule
  Reference & Metric & Hole & Spin & Flow & EOS & Compactness & Mass
  Ratio \\
  \midrule
  \citet{Taniguchi_BFS2006} & KS  & Ex & $0$ & Ir & Poly & $0.088$ &
                              $5$ \\
  \citet{Taniguchi_BFS2007} & CFMS & Ex & $0$ & Ir & Poly & $0.088,
                          0.145$ & $1$--$10$ \\
  \citet{Taniguchi_BFS2008} & CFMS & Ex & $0$ & Ir & Poly &
                          $0.109$--$0.132$ & $1$--$9$ \\
  \citet{Grandclement2006,Grandclement2006e} & CFMS & Ex & 0 & Ir & Poly
                      & $0.075$--$0.150$ & $5$ \\
  \citet{Shibata_Uryu2006} & CFMS & Pu & $0$ & Co & Poly & $0.139$ &
                              $2.47$ \\
  \citet{Shibata_Uryu2007} & CFMS & Pu & $0$ & Co & Poly & $0.139,
                          0.148$ & $2.47, 3.08$ \\
  \citet{Kyutoku_Shibata_Taniguchi2009} & CFMS & Pu & $0$ & Ir & Poly &
                          $0.132, 0.145, 0.160$ & $1, 3, 5$ \\
  \citet{Foucart_KPT2008} & CFMS & Ex & $0$ & Ir & Poly & $0.144$ &
                              $1$ \\
  ~ & MKS & Ex & $-0.5, 0$ & Ir & Poly & $0.144$ & $1$ \\
  \citet{Henriksson_FKT2016} & MKS & Ex & $0$ & Ir & Tab & $0.23, 0.25$
                          & $6$ \\
  \citet{Papenfort_TGMR2021} & CFMS & Ex & $0$ & Ir & Poly & $0.138$ &
                              $1$ \\
  \bottomrule
 \end{tabular}
\end{table}

Although a decade has passed since the release of the first version of
this article, the parameter space surveyed for the study of
quasiequilibrium sequences remains narrow. The main progress achieved
during this period may be the computations of sequences involving
high-compactness neutron stars with a tabulated equation of state
\citep{Henriksson_FKT2016}. We classify the current study shown in
Table~\ref{table:qe} according to the following seven items (see
Appendix~\ref{app:init} for the details).
\begin{enumerate}
 \item Metric: Choice of the spatial background metric
       $\hat{\gamma}_{ij}$ and the extrinsic curvature $K$. The
       abbreviations ``CFMS,'' ``KS,'' and ``MKS'' mean, respectively,
       the conformally-flat and maximal-slicing condition ($K=0$; see
       Appendix~\ref{app:init_grav}), Kerr-Schild, and modified
       Kerr-Schild.
 \item Hole: Method to handle the singularity associated with the black
       hole. The abbreviations ``Ex'' and ``Pu'' indicate the excision
       and the puncture approaches, respectively.
 \item Spin: Dimensionless spin parameter of the black hole $\chi$. Note
       that the spin is zero or anti-aligned with the orbital angular
       momentum in these work.
 \item Flow: State of the fluid flow in the neutron star. The
       abbreviations ``Ir'' and ``Co'' indicate irrotational and
       corotational flows, respectively. Arbitrary spins of neutron
       stars are not considered in these computations.
 \item EOS: Equation of state for neutron-star matter. Here, ``Poly''
       means the polytrope (specifically, that with $\Gamma = 2$) and
       ``Tab'' means a tabulated equation of state. Specifically, the
       SLy equation of state \citep{Douchin_Haensel2001} is adopted for
       computing quasiequilibrium sequences in
       \citet{Henriksson_FKT2016}.
 \item Compactness: Compactness of the neutron star $\mathcal{C}$.
 \item Mass Ratio: Mass ratio $Q$.
\end{enumerate}

Quasiequilibrium sequences with high and/or misaligned spins of either
component have not been derived. Meanwhile, computations of individual
quasiequilibrium states have been extended to a wide range of the
parameter space including spin vectors for both black holes and neutron
stars \citep{Tacik_FPMKSS2016}. Additionally, many quasiequilibrium
configurations, including those with a radial approaching velocity to
reduce the orbital eccentricity
\citep{Foucart_KPT2008,Kyutoku_KKST2021}, have been computed as initial
conditions of dynamical simulations, whose results will be discussed in
Sect.~\ref{sec:sim}.

\subsection{Numerical results} \label{sec:eq_res}

Hereafter in this section, we focus on the results reported in
\citet{Taniguchi_BFS2008}, because a systematic survey for a wide range
of the parameter space was performed only there. Accordingly, all the
results are derived in the excision approach, the conformal-flatness and
maximal-slicing condition, nonspinning black holes, the irrotational
flow, and the $\Gamma = 2$ polytrope. Still, this work reasonably
captures the properties of quasiequilibrium sequences. As the polytropic
equation of state has only a single dimensional parameter, $\kappa$, we
may normalize various quantities such as the length, mass, and time by
the polytropic length scale,
\begin{equation}
 R_\mathrm{poly} := \kappa^{1/(2\Gamma - 2)} \bqty{= \kappa^{1/(2\Gamma
  - 2)} G^{-1/2} c^{(\Gamma - 2)/(\Gamma - 1)}} .
\end{equation}
We will put an overbar above a symbol for indicating quantities in the
polytropic unit, e.g., $\bar{M}_\mathrm{B} := M_\mathrm{B} /
R_\mathrm{Poly} [= G M_\mathrm{B}/ (c^2 R_\mathrm{poly}) ]$.

\begin{figure}[htbp]
 \centering \includegraphics[width=0.8\linewidth,clip]{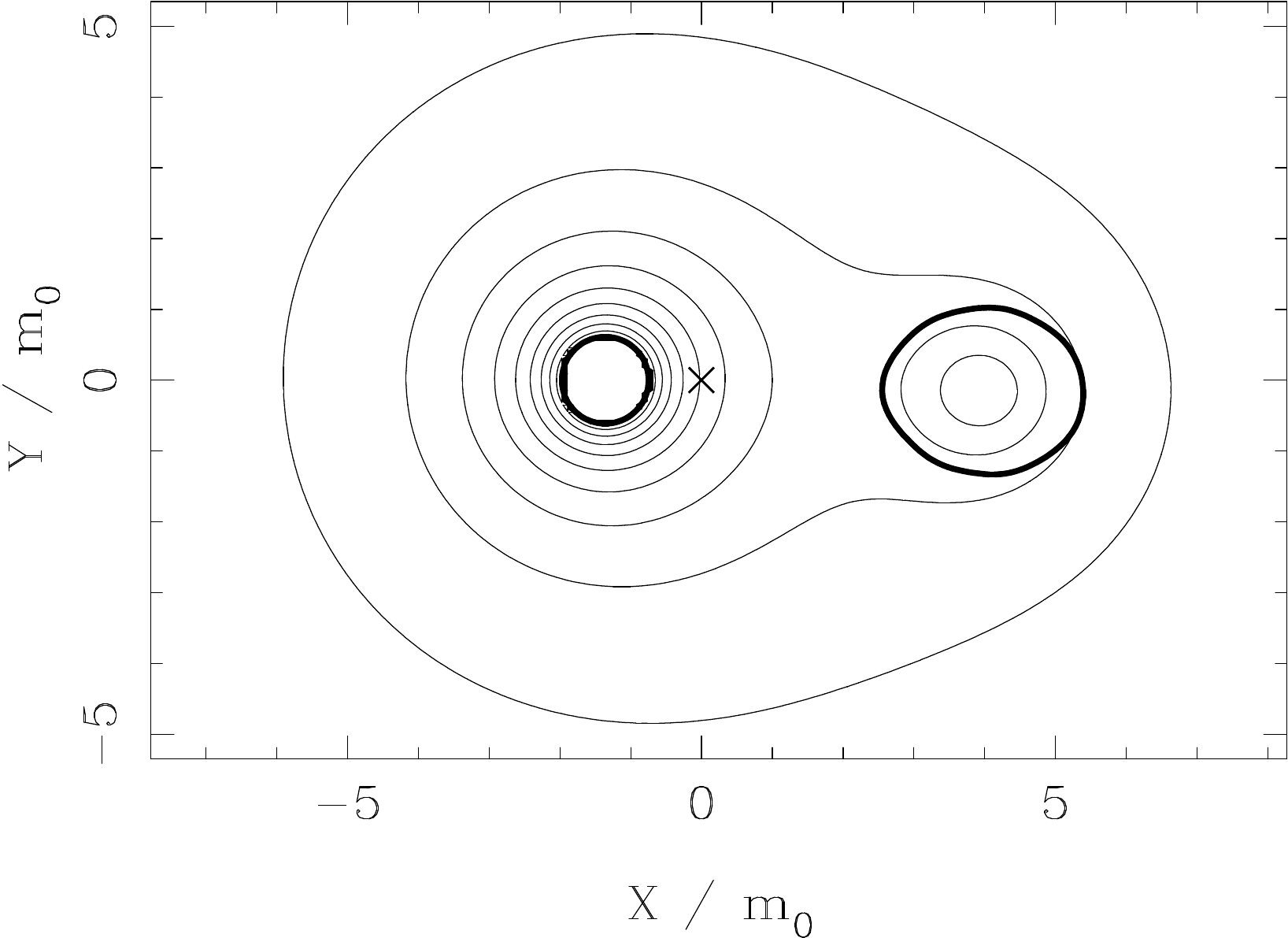}
 \caption{Contour of the conformal factor, $\psi$, on the orbital plane
 for the innermost configuration with the mass ratio $Q=3$ and the
 baryon rest mass of the neutron star $\bar{M}_{\mathrm{B}}=0.15$
 ($\mathcal{C} = 0.145$). The cross symbol indicates the position of the
 rotational axis. Image reproduced with permission from \citet{Taniguchi_BFS2008}, copyright by APS.}
 \label{fig:isoBHNS}
\end{figure}

Figure~\ref{fig:isoBHNS} displays contours of the conformal factor,
$\psi$, for a black hole--neutron star binary with the mass ratio $Q=3$
and the baryon rest mass of the neutron star $\bar{M}_{\mathrm{B}}=0.15$
($\mathcal{C} = 0.145$). This contour plot shows the configuration at
the smallest orbital separation for which the code used in
\citet{Taniguchi_BFS2008} successfully achieved a converged
solution. The thick solid circle for $X<0$ (left) denotes the excised
surface, i.e., the apparent horizon, while that for $X>0$ (right)
denotes the surface of the neutron star. A saddle point exists between
the black hole and the neutron star, and for this close orbit, it is
located in the vicinity of the inner edge of the neutron star. This fact
suggests that the orbit of the binary is close to the mass-shedding
limit. The value of $\psi$ on the excised surface is not constant,
because a Neumann boundary condition is imposed (see Appendix
\ref{app:init_grav_ex}).

\subsubsection{Binding energy and total angular momentum}

\begin{figure}[htbp]
 \centering
 \begin{tabular}{cc}
  \includegraphics[width=.48\linewidth,clip]{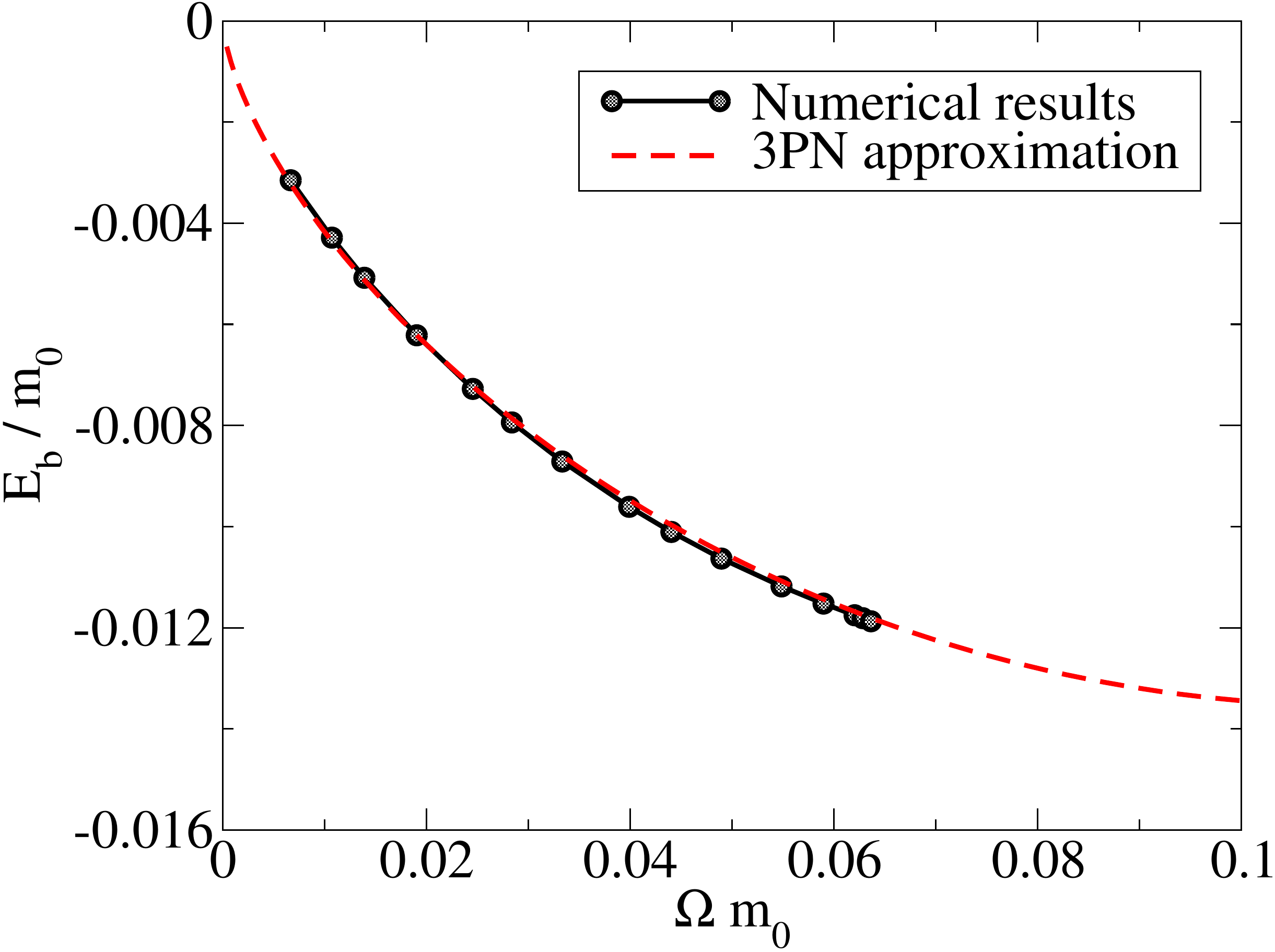} &
  \includegraphics[width=.46\linewidth,clip]{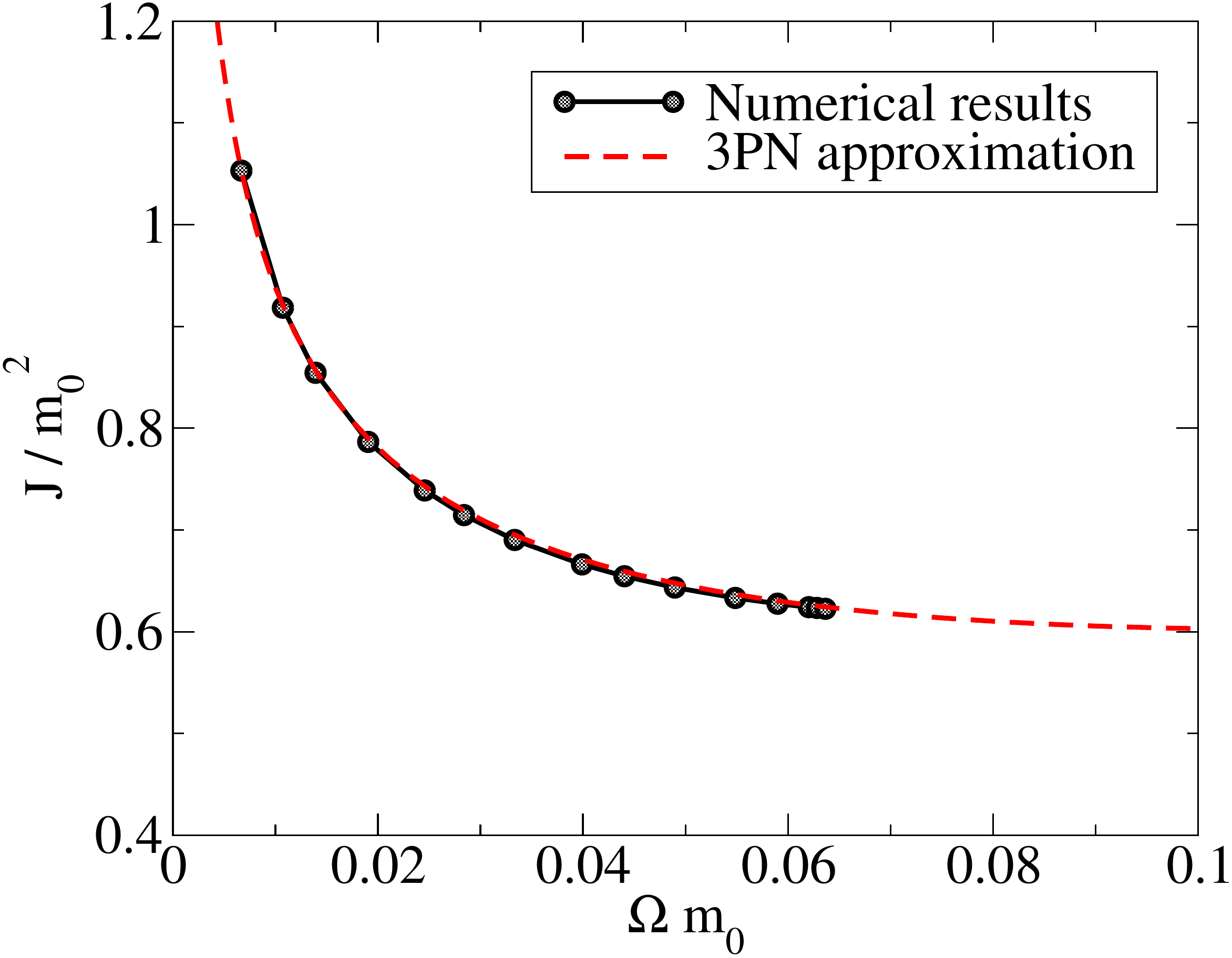}
 \end{tabular}
 \caption{Binding energy $E_{\mathrm{b}}/m_0 [=E_\mathrm{b}/(m_0 c^2)]$
 (left) and total angular momentum $J/m_0^2 [=cJ/(Gm_0^2)]$ (right) as
 functions of $\Omega m_0 (=G \Omega m_0 /c^3)$ for binaries with $Q=3$
 and $\bar{M}_{\mathrm{B}}=0.15$ ($\mathcal{C} = 0.145$). The solid
 curve with the filled circles shows numerical results, and the dashed
 curve denotes the results derived in the third-order post-Newtonian
 approximation for point particles \citep{Blanchet2002}. Image reproduced with permission from \citet{Taniguchi_BFS2008}, copyright by APS.} \label{fig:eneangmom}
\end{figure}

Figure~\ref{fig:eneangmom} shows the binding energy, $E_{\mathrm{b}}/m_0
[= E_\mathrm{b} / (m_0 c^2)]$, and the total angular momentum, $J/m_0^2
[= cJ/(Gm_0^2)]$, as functions of the orbital angular velocity, $\Omega
m_0 (= G \Omega m_0 /c^3)$, for a binary with $Q=3$ and
$\bar{M}_{\mathrm{B}}=0.15$ ($\mathcal{C} = 0.145$). All the quantities
of the binary are expressed as dimensionless quantities normalized by
the total mass, $m_0$. This figure shows that the numerical results
agree quantitatively with the third-order post-Newtonian approximation
\citep{Blanchet2002}. The results also agree with the up-to-date,
fourth-order post-Newtonian approximations \citep{Blanchet2014}, which
differ only by $< 1\%$ and $< 3\%$ for the binding energy and the total
angular momentum, respectively, from the third-order ones in the range
of $\Omega m_0$ considered here. For this parameter set, the numerical
sequence terminates at the mass-shedding limit, i.e., at an orbit for
which a cusp is formed at the inner edge of the neutron star and the
material begins to flow out, before the innermost stable circular orbit
is encountered, i.e., before the minimum of the binding energy appears.

If the binary separation at the mass-shedding limit is substantially
larger than the radius of the innermost stable circular orbit, the
neutron star is expected not only to start shedding mass but also to be
tidally disrupted before being swallowed by the black
hole. Equation~\eqref{eq:iscovsms} presented in
Sect.~\ref{sec:intro_tidal_ms} suggests that the ratio of the binary
separation at the mass-shedding limit to the radius of the innermost
stable circular orbit decreases with increasing the mass ratio, $Q$,
and/or the compactness of the neutron star, $\mathcal{C}$. Thus, it is
naturally expected that quasiequilibrium sequences encounter minima in
the binding energy and the total angular momentum as in the
two-point-particle problem in general relativity, if the mass ratio is
sufficiently high and/or the compactness is sufficiently large.

\begin{figure}[htbp]
 \centering
 \begin{tabular}{cc}
  \includegraphics[width=.48\linewidth,clip]{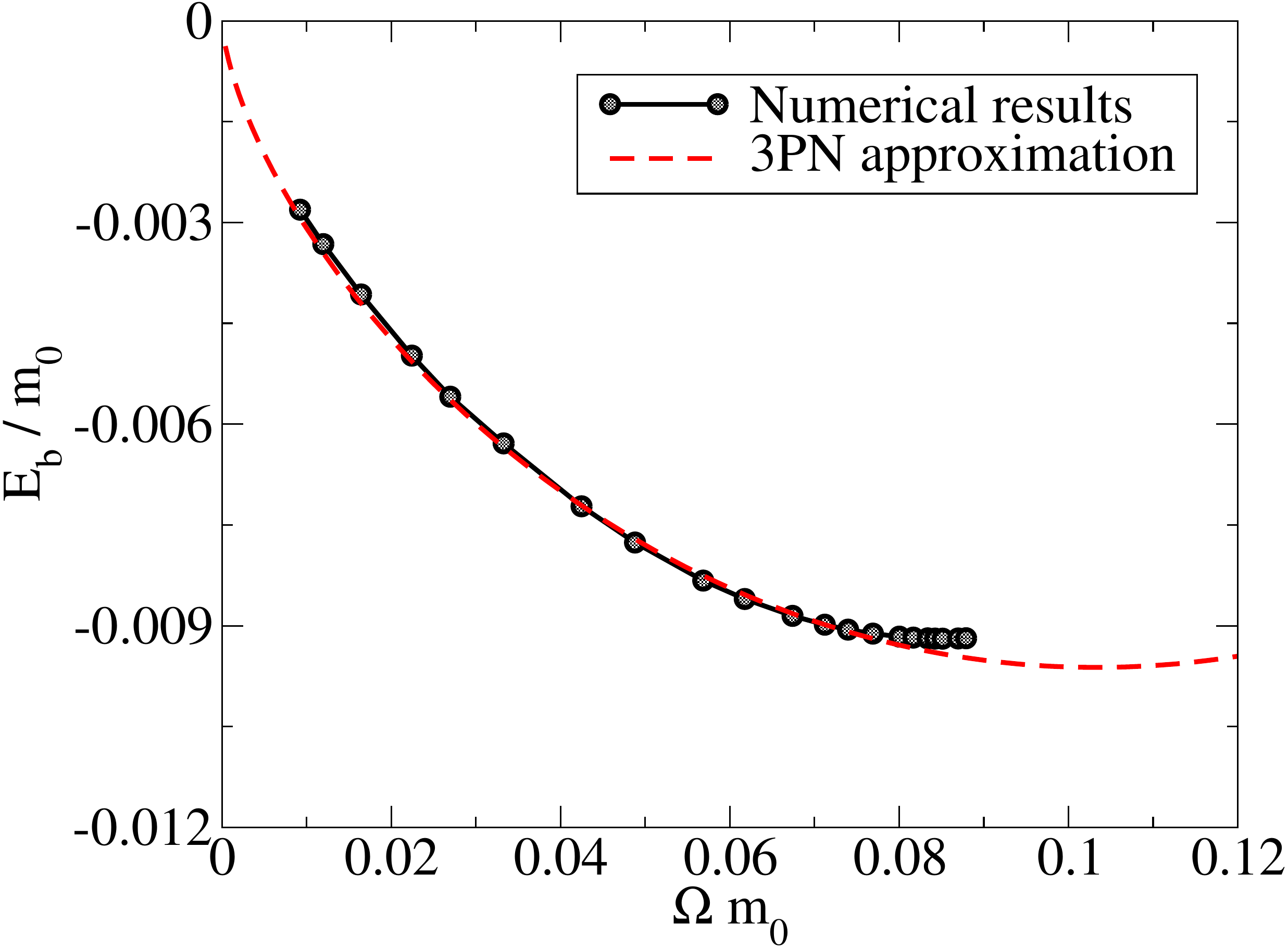} &
  \includegraphics[width=.46\linewidth,clip]{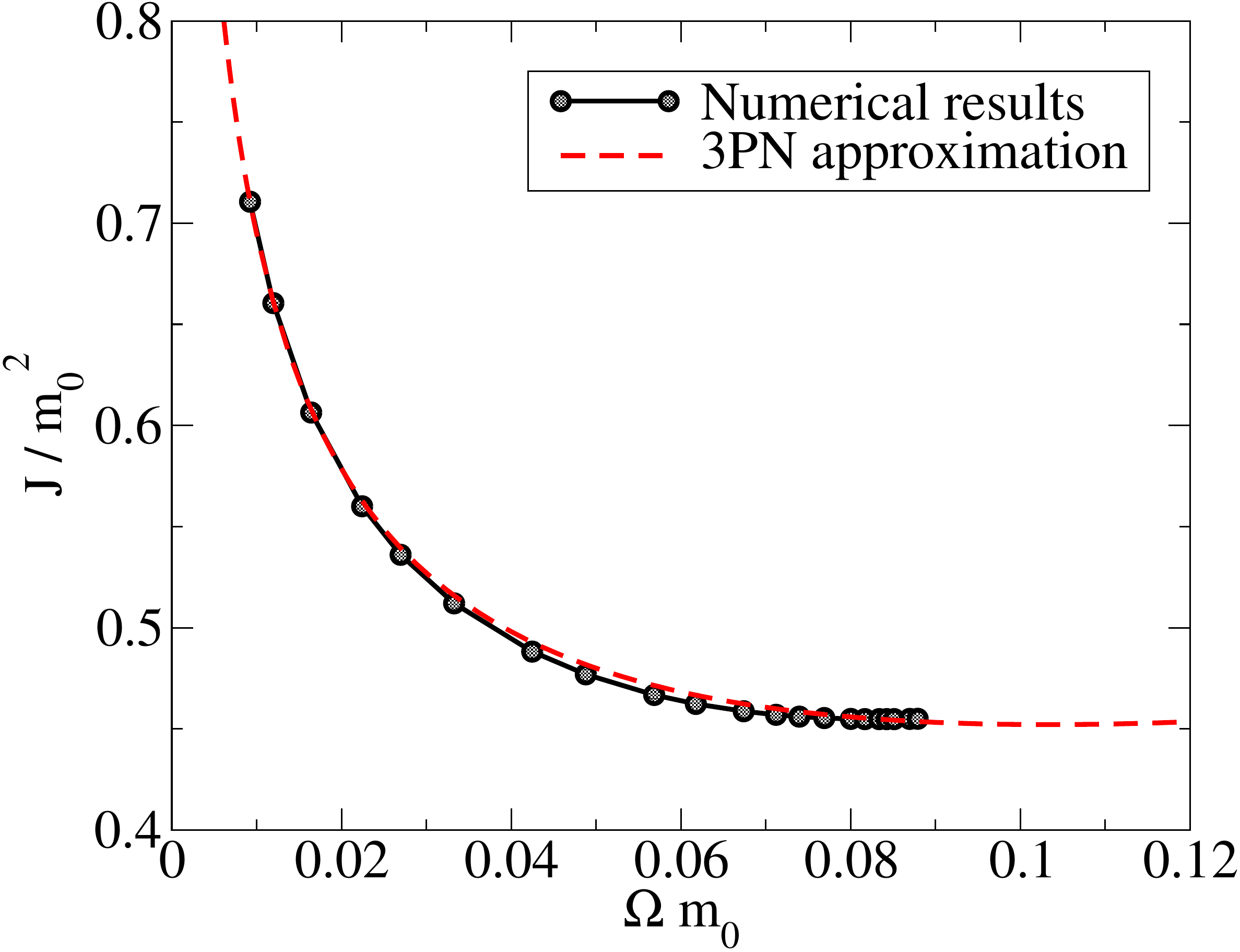}
 \end{tabular}
 \caption{Same as Fig.~\ref{fig:eneangmom} but for the sequence with
 $Q=5$. Image reproduced with permission from \citet{Taniguchi_BFS2008}, copyright by APS.}
 \label{fig:eneangmom2}
\end{figure}

Figure~\ref{fig:eneangmom2} shows the binding energy,
$E_\mathrm{b}/m_0$, and the total angular momentum, $J/m_0^2$, as
functions of $\Omega m_0$ for a binary with $Q=5$ and
$\bar{M}_\mathrm{B}=0.15$ ($\mathcal{C}=0.145$). While the compactness
of the neutron star is the same as that shown in
Fig.~\ref{fig:eneangmom}, the mass ratio is higher. In this sequence, an
innermost stable circular orbit is encountered before the onset of mass
shedding, i.e., we see minima in the binding energy and the angular
momentum just before the end of the sequence.

\begin{figure}[htbp]
 \centering \includegraphics[width=.7\linewidth,clip]{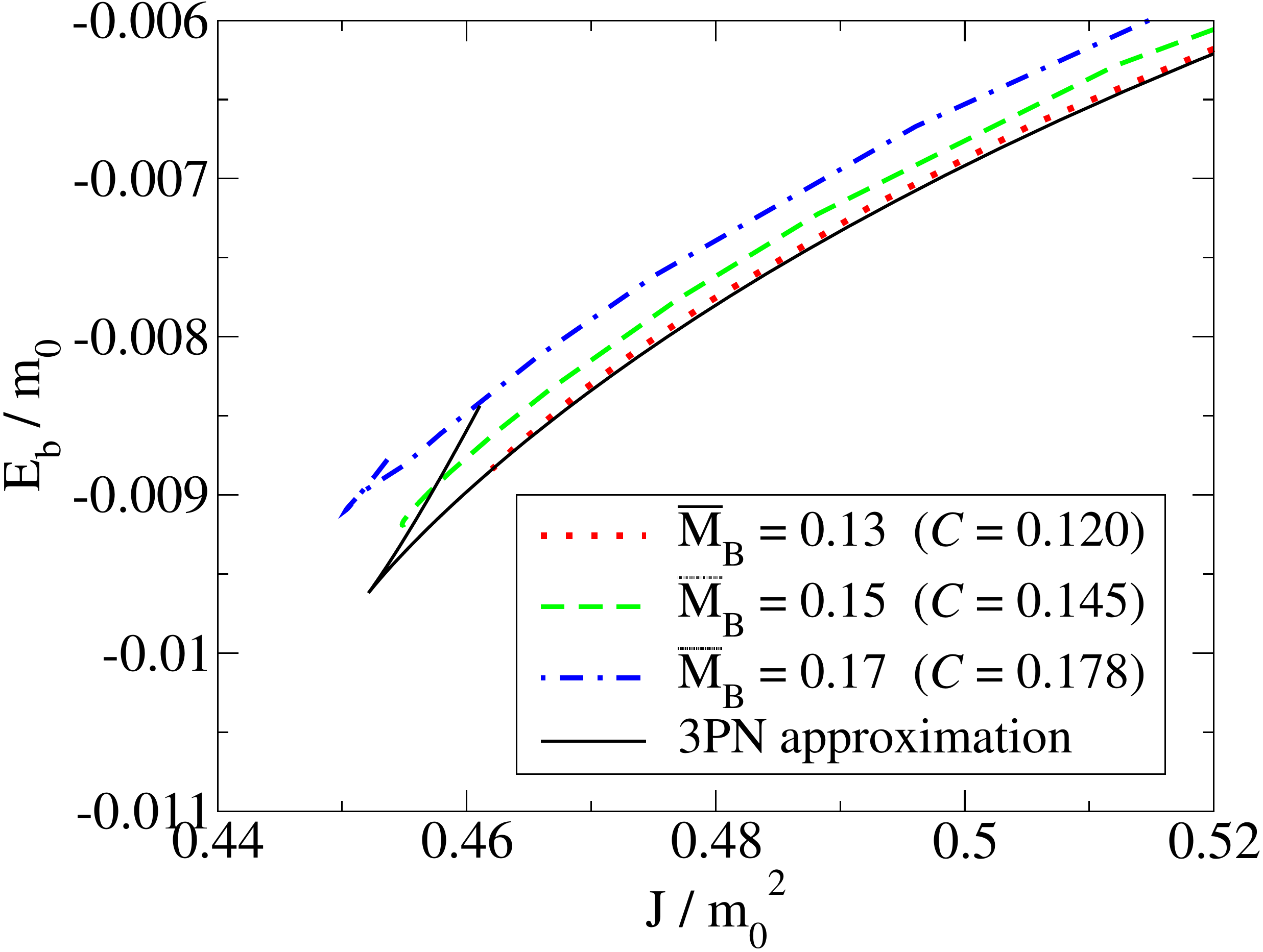}
 \caption{Binding energy as a function of the total angular momentum for
 binaries with $Q=5$ and different compactnesses of neutron stars. The
 solid curve denotes the results derived in the third-order
 post-Newtonian approximation for point particles
 \citep{Blanchet2002}. Image reproduced with permission from \citet{Taniguchi_BFS2008}, copyright by APS.} \label{fig:evsjqe}
\end{figure}

Figure~\ref{fig:eneangmom2} shows that the turning points in the binding
energy and the total angular momentum appear simultaneously within the
numerical accuracy as described by theory of binary thermodynamics
\citep{Friedman_Uryu_Shibata2002}. A simultaneous turning point appears
as a cusp in the curve representing the relation between the binding
energy and the total angular momentum. This fact is clearly seen in
Fig.~\ref{fig:evsjqe}, which shows the relations for sequences with
$Q=5$ but with different compactnesses of neutron stars. As suggested by
Eq.~\eqref{eq:iscovsms}, turning points are not found for small
compactnesses such as $\mathcal{C}=0.120$, since the sequences terminate
at mass shedding before encountering an innermost stable circular
orbit. By contrast, for large compactnesses, these curves indeed form a
cusp. Note that sequences derived in post-Newtonian approximations for
point particles cannot identify mass shedding and therefore always
exhibit turning points. The difference of $E_\mathrm{b} / m_0$ as a
function of $J / m_0^2$ between the third- and fourth-order
post-Newtonian approximations is less than 0.1\% before the binary
reaches an innermost stable circular orbit for the binary parameters
chosen here. Although it may appear from Fig.~\ref{fig:evsjqe} that the
numerical results deviate from the post-Newtonian approximation as the
compactness increases, i.e., approaching a point-particle limit,
quasiequilibrium sequences of binary black holes also deviate from the
analytic computation based on the post-Newtonian approximation in a
similar manner to the high-compactness sequence of black hole--neutron
star binaries \citep{Cook_Pfeiffer2004,Caudill_CGP2006}.

\subsection{Endpoint of the sequence} \label{sec:eq_end}

One of the most important questions in the study of black hole--neutron
star binaries is whether the coalescence leads to mass shedding of the
neutron star before reaching the innermost stable circular orbit or to
the plunge of the neutron star into the black hole before the onset of
mass shedding. The answer to this question is essential for the topics
raised in Sect.~\ref{sec:intro_bg}: Orbital dynamics and gravitational
waves in the late inspiral and merger phases are affected strongly by
this issue, and hence, its precise understanding is necessary for
developing theoretical templates. For driving a short-hard gamma-ray
burst, formation of an accretion disk surrounding the black hole is the
most promising model. The \textit{r}-process nucleosynthesis and
electromagnetic emission occur if the material of the neutron star is
ejected from the system. Both the disk formation and the mass ejection
can result only if the neutron star is disrupted prior to reaching the
innermost stable circular orbit.

Quantitative exploration of this issue ultimately requires dynamical
simulations, and their results will be reviewed in
Sect.~\ref{sec:sim}. However, in-depth studies of quasiequilibrium
sequences also provide a guide to the binary parameters that separate
mass shedding and the dynamical plunge. In the following, we summarize
quantitative insights obtained from the study of quasiequilibrium
sequences.  Specifically, we will review semiquantitative expressions
that may be used to predict whether a black hole--neutron star binary of
arbitrary values of $Q$ and $\mathcal{C}$ encounters an innermost stable
circular orbit before the onset of mass shedding or not. Since the
detailed study was performed only for nonspinning black holes, we do not
consider the effect of black-hole spins in the following. However, as
discussed in Sect.~\ref{sec:intro_tidal}, we have to keep in mind that
the spin of the black hole is crucial for determining the fate of
general black hole--neutron star binaries.

\subsubsection{Mass-shedding limit} \label{sec:eq_end_ms}

First, we summarize the results for the orbital angular velocity at
which mass shedding sets in, i.e., the mass-shedding limit. For this
purpose, a mass-shedding indicator\footnote{This quantity is originally
denoted by $\chi$, but in this article we keep $\chi$ for the
dimensionless spin parameter, which is now widely used in studies of
compact binary coalescences.}  introduced in studies of binary neutron
stars
\citep{Gourgoulhon_GTMB2001,Taniguchi_Gourgoulhon2002,Taniguchi_Gourgoulhon2003}
\begin{equation}
 \mathcal{X} := \frac{\eval{\pdv*{(\ln
  h)}{r}}_\mathrm{eq}}{\eval{\pdv*{(\ln h)}{r}}_\mathrm{pole}}
\end{equation}
is used to determine the mass-shedding limit
\citep{Taniguchi_BFS2006,Taniguchi_BFS2007,Taniguchi_BFS2008}. The
numerator is the radial derivative of the log-enthalpy at the surface
toward the black hole on the orbital plane. The denominator is that on
the pole. Here, the radial coordinate is defined with respect to the
center of the neutron star. The mass-shedding indicator, $\mathcal{X}$,
is unity for a spherical neutron star at infinite orbital separation,
while $\mathcal{X} = 0$ indicates the formation of a cusp on the stellar
surface, and hence, the onset of mass shedding. Note that the
quasiequilibrium sequences have been analyzed only for binary systems
with the reflection symmetry about the orbital plane. Thus, ``pole'' and
``eq'' are well-defined.

In Newtonian gravity and partially-relativistic approaches, simple
formulae may be introduced to fit the effective radius of a Roche lobe
\citep{Paczynski1971,Eggleton1983,Wiggins_Lai2000,Ishii_Shibata_Mino2005}.
In \citet{Shibata_Uryu2006,Shibata_Uryu2007}, a fitting formula is
introduced for binaries composed of a nonspinning black hole and a
corotating neutron star in general relativity. In this
Sect.~\ref{sec:eq_end_ms}, we review how to derive a fitting formula
from data of \citet{Taniguchi_BFS2008} for a nonspinning black hole and
an irrotational neutron star.

To derive the fitting formula, we need to determine the orbital angular
velocity at the mass-shedding limit. However, stellar configurations
with cusps cannot be constructed by the numerical code used in
\citet{Taniguchi_BFS2008}, because it is based on a spectral method and
accompanied by the Gibbs phenomena in the presence of a nonsmooth
stellar surface \citep[but see
also][]{Ansorg_KleinWachter_Meinel2003,Grandclement2010}. This is also
the case for a configuration with smaller values of $\mathcal{X} \le
0.5$, even though a configuration with a cusp does not appear. Thus,
data for the mass-shedding limits have to be determined by
extrapolation.

\begin{figure}[htbp]
 \centering \includegraphics[width=.7\linewidth,clip]{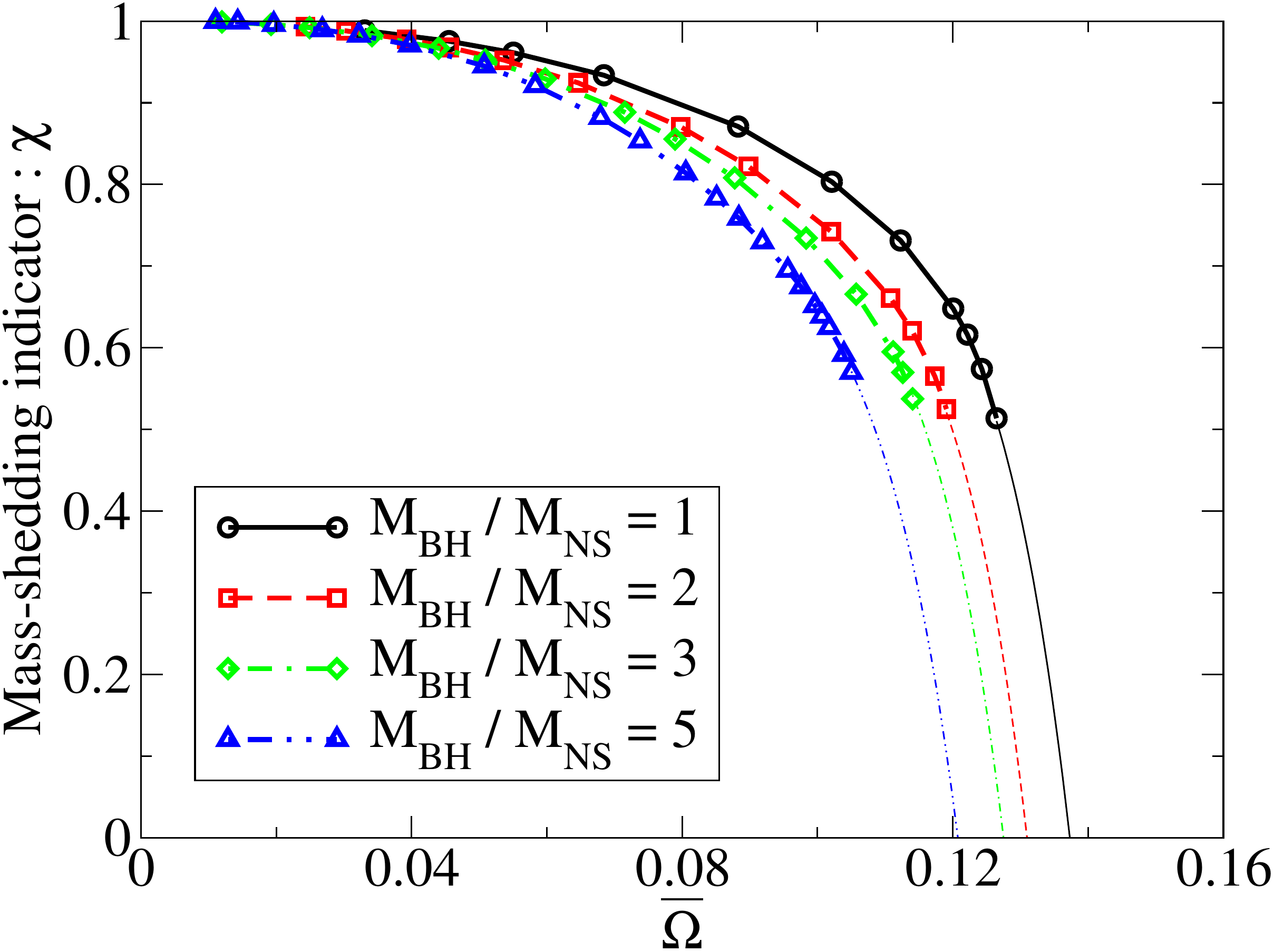}
 \caption{Extrapolation of sequences with $\mathcal{C} = 0.145$ to the
 mass-shedding limit, $\mathcal{X} = 0$. The thick curves are sequences
 constructed using numerical data, and the thin curves are their
 extrapolations. The horizontal axis is the orbital angular velocity in
 the polytropic unit, $\bar{\Omega} = \Omega R_{\mathrm{poly}} [= \Omega
 R_\mathrm{poly} / c]$. Image reproduced with permission from \citet{Taniguchi_BFS2008}, copyright by APS.} \label{fig:msind}
\end{figure}

\citet{Taniguchi_BFS2008} tabulated the values of $\mathcal{X}$ as a
function of the orbital angular velocity and their sequence is
extrapolated to $\mathcal{X} = 0$ by using polynomial functions to find
the orbits at the onset of mass shedding. Figure~\ref{fig:msind} shows
an example of such extrapolations for sequences with $\mathcal{C} =
0.145$ and $Q=1,2,3,$ and $5$. By extrapolating results toward
$\mathcal{X} = 0$, the orbital angular velocity at the mass-shedding
limit, $\Omega_\mathrm{ms}$, is approximately determined for each set of
$Q$ and $\mathcal{C}$.

\begin{figure}[htbp]
 \centering \includegraphics[width=.7\linewidth,clip]{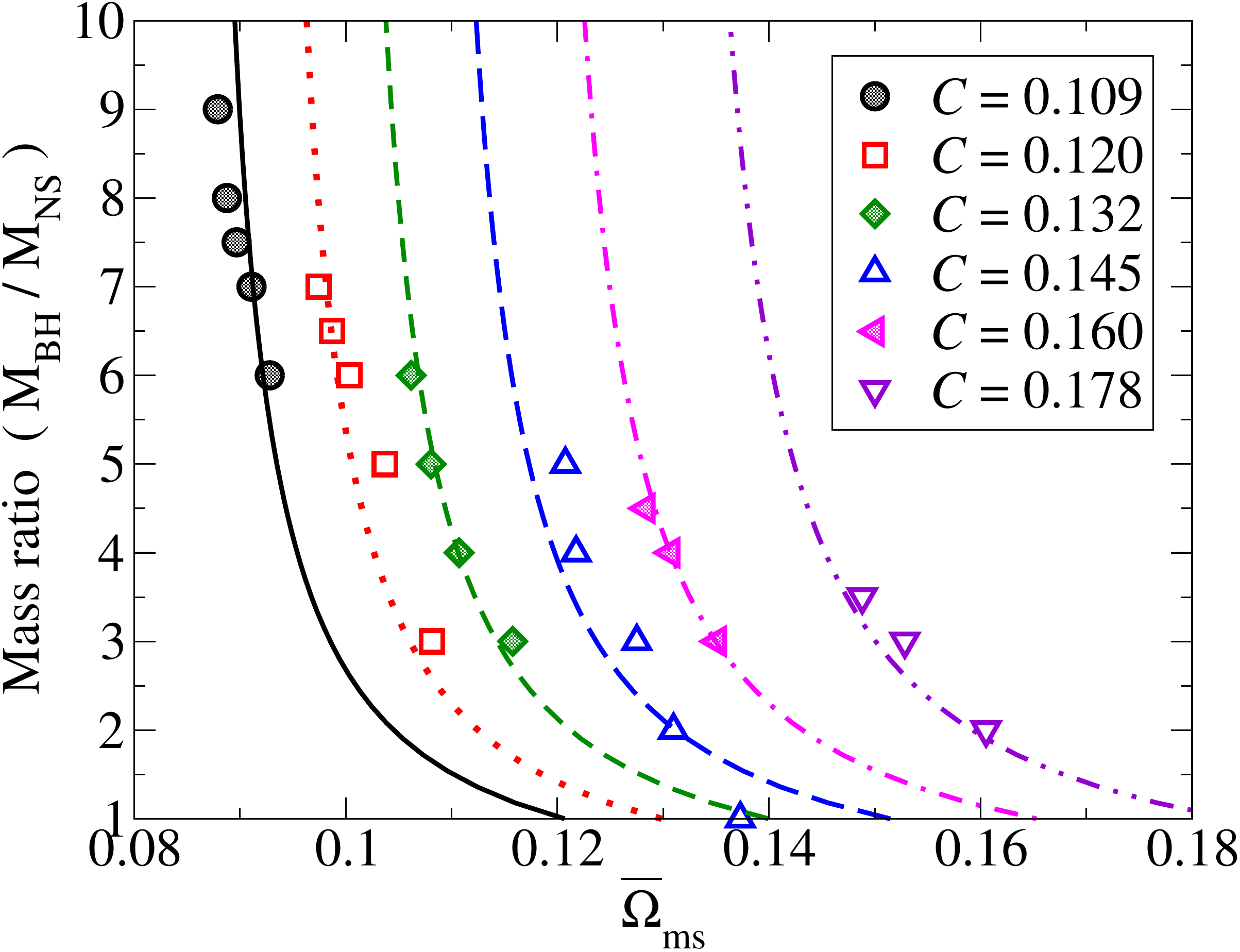}
 \caption{Fit of the mass-shedding limit by the analytic expression,
 Eq.~\eqref{eq:msfit}. The mass-shedding limit for each compactness of
 the neutron star and mass ratio is computed by extrapolating the
 numerical data. Image reproduced with permission from \citet{Taniguchi_BFS2008}, copyright by APS.}
 \label{fig:extrapall}
\end{figure}

To derive a fitting formula for $\Omega_\mathrm{ms}$ as a function of
$Q$ and $\mathcal{C}$, the Newtonian expression of
Eq.~\eqref{eq:msomega} is useful, although it is semiquantitative. By
fitting the sequence of data with respect to this expression,
\citet{Taniguchi_BFS2008} determined the value of $C_{\Omega}$ for the
$\Gamma = 2$ polytrope to be $0.270$, i.e.,
\begin{equation}
 \bar{\Omega}_\mathrm{ms} = 0.270
  \frac{\mathcal{C}^{3/2}}{\bar{M}_{\mathrm{NS}}} \pqty{1 +
  Q^{-1}}^{1/2} , \label{eq:msfit}
\end{equation}
or equivalently,
\begin{equation}
 \Omega_{\mathrm{ms}} m_0 = 0.270 \mathcal{C}^{3/2} (1 +Q) \pqty{1 +
  Q^{-1}}^{1/2}. \label{eq:msfit2}
\end{equation}
Figure~\ref{fig:extrapall} shows the results of the fitting for the
mass-shedding limit. The agreement is not perfect but fairly good for $Q
\ge 2$.

It may be interesting to note that the value of $C_{\Omega} = 0.270$ is
the same as that found for quasiequilibrium sequences in general
relativity of binary neutron stars \citep{Taniguchi_Shibata2010} and of
black hole--corotating neutron star binaries
\citep{Shibata_Uryu2006,Shibata_Uryu2007}. Thus, the value of
$C_{\Omega} = 0.270$ could be widely used for an estimation of the
orbital angular velocity at the mass-shedding limit of a neutron star in
a relativistic binary system with $\Gamma = 2$. We also note that
$C_\Omega = 0.270$ corresponds to $c_\mathrm{R} = 1.90$, which is
defined in Sect.~\ref{sec:intro_tidal_ms}.

\subsubsection{Innermost stable circular orbit}

\begin{figure}[htbp]
 \centering \includegraphics[width=.7\linewidth,clip]{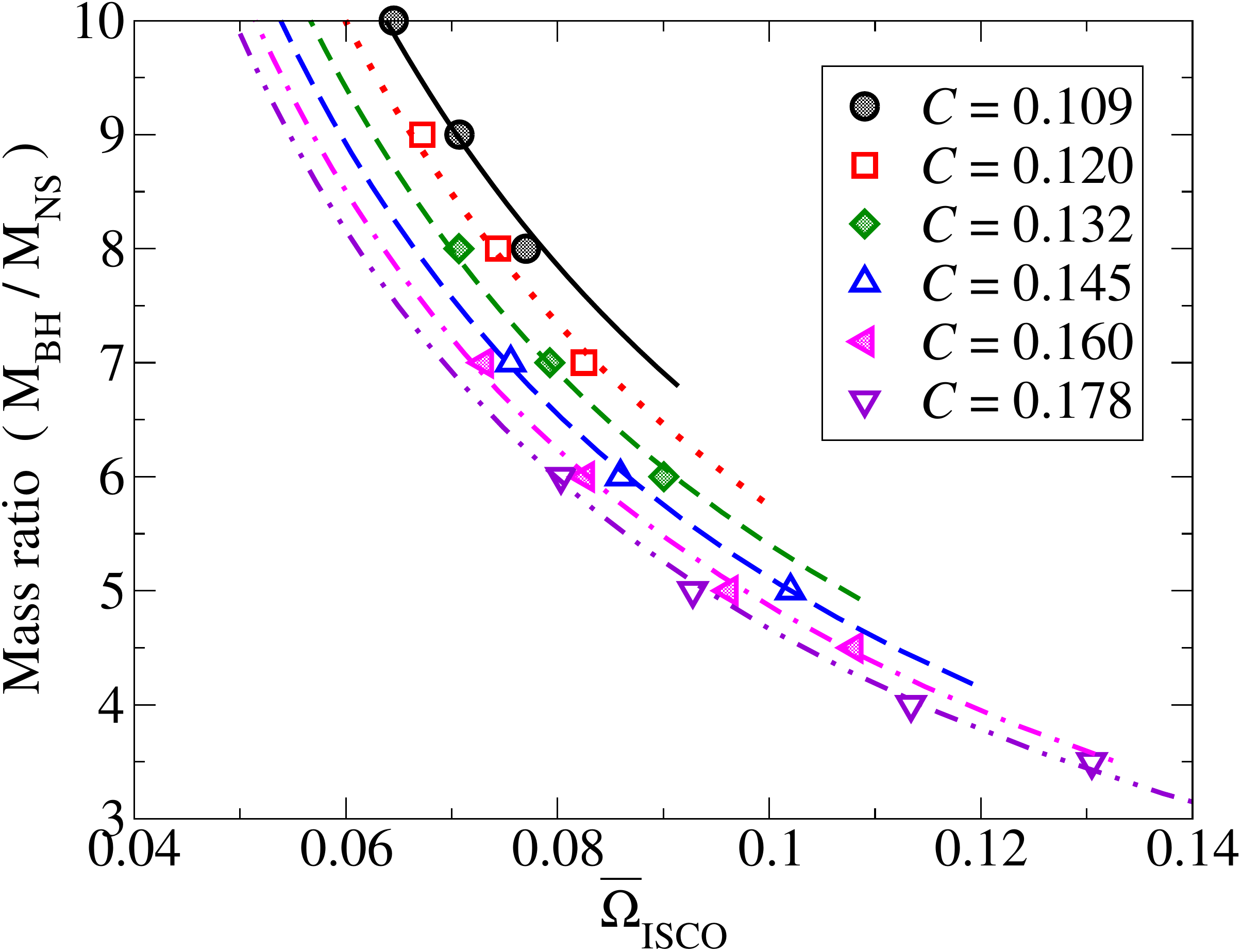}
 \caption{Fit of the minimum point of the binding energy curve by
 Eq.~\eqref{eq:iscofit}. Image reproduced with permission from \citet{Taniguchi_BFS2008}, copyright by APS.} \label{fig:isco}
\end{figure}

Next, we summarize the results for the orbital angular velocity at which
the minimum of the binding energy, i.e., the innermost stable circular
orbit appears. Because the numerical data are discrete and do not
necessarily give the exact minimum, the minimum point may be located
approximately by fitting three nearby points of the sequence to a
second-order polynomial.

A simple empirical fitting that predicts the angular velocity
$\Omega_\mathrm{ISCO}$ at the innermost stable circular orbit for an
arbitrary companion orbiting a black hole may be derived in the manner
of \citet{Taniguchi_BFS2008}. They search for an expression of
$\Omega_\mathrm{ISCO}$ as a function of the mass ratio $Q$ and the
compactness $\mathcal{C}$ of the companion. Specifically, they assume a
functional form of
\begin{equation}
 \Omega_\mathrm{ISCO} m_0 = c_1 \bqty{1 - \frac{c_2}{Q^{p_1}} \pqty{1 -
  c_3 \mathcal{C}^{p_2}}} ,
\end{equation}
where $c_1$, $c_2$, $c_3$, $p_1$, and $p_2$ are parameters to be
determined in the following manner. The coefficients $c_1$, $c_2$, and
$c_3$ are determined for given values of $p_1$ and $p_2$ by requiring
that three known values of $\Omega_\mathrm{ISCO}$ are recovered: (1)
that of a test particle orbiting a Schwarzschild black hole,
$\Omega_\mathrm{ISCO} m_0 =6^{-3/2}$ (for $Q \to \infty$), (2) that of
an equal-mass binary-black-hole system, $\Omega_\mathrm{ISCO} m_0
=0.1227$ \citep[for $Q=1$ and $\mathcal{C} = 0.5$;][]{Caudill_CGP2006},
and finally (3) that of a black hole--neutron star configuration with
$Q=5$ and $\mathcal{C}=0.1452$, $\Omega_\mathrm{ISCO} m_0 = 0.0854$
\citep{Taniguchi_BFS2008}. The exponents $p_1$ and $p_2$ are determined
by requiring the fitted curves to lie near the data points for all the
systems.

As demonstrated in Fig.~\ref{fig:isco}, the numerical data are fitted
nicely by a function
\begin{equation}
 \Omega_\mathrm{ISCO} m_0 = 0.0680 \bqty{1 - \frac{0.444}{Q^{0.25}}
  \pqty{1 - 3.54 \mathcal{C}^{1/3}}} . \label{eq:iscofit}
\end{equation}
The agreement is sufficient for finding the orbital angular velocity at
the innermost stable circular orbit within the error of $\sim 10\%$.

\subsubsection{Critical mass ratio}

\begin{figure}[htbp]
 \centering \includegraphics[width=.7\linewidth,clip]{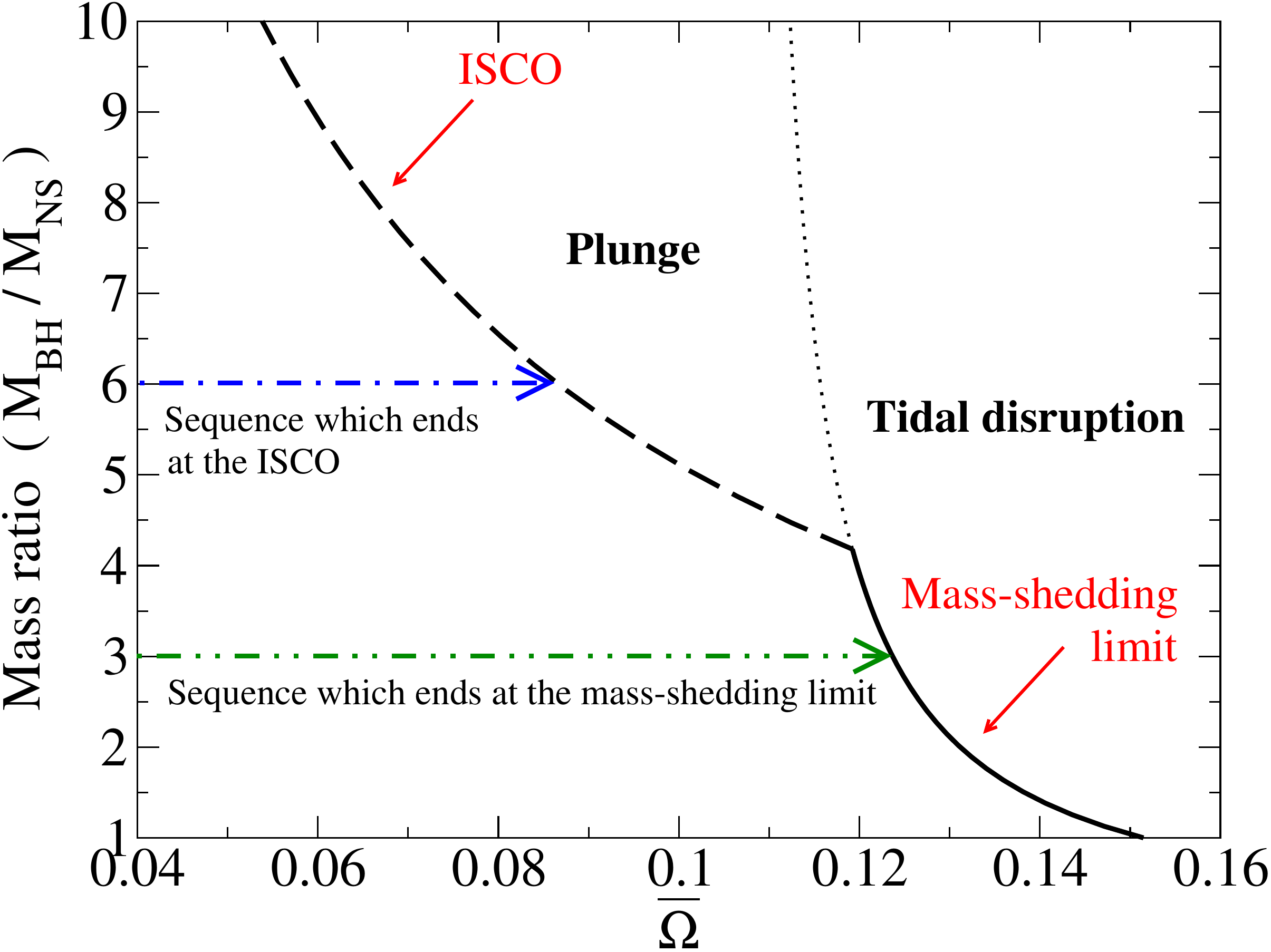}
 \caption{Example of the boundary between the mass-shedding limit and
 the innermost stable circular orbit for $\mathcal{C} = 0.145$. The
 solid and dashed curves denote the mass ratio that gives the
 mass-shedding limit and the innermost stable circular orbit (ISCO),
 respectively, for a given value of the orbital angular velocity in the
 polytropic unit. The thin dotted curve denotes the mass-shedding limit
 for unstable quasiequilibrium sequences. Image reproduced with permission from \citet{Taniguchi_BFS2008}, copyright by APS.} \label{fig:endpoint}
\end{figure}

Combining Eqs.~\eqref{eq:msfit2} and \eqref{eq:iscofit}, we can identify
the critical binary parameters which separate two final fates that the
binary encounters an innermost stable circular orbit before initiating
mass shedding or that the neutron star reaches the mass-shedding limit
before plunging into the black hole. Figure~\ref{fig:endpoint}
illustrates the final fate of black hole--neutron star binaries with
$\mathcal{C} = 0.145$. Because the orbital angular velocities at the
mass-shedding limit [Eq.~\eqref{eq:msfit2}] and the innermost stable
circular orbit [Eq.~\eqref{eq:iscofit}] depend differently on the mass
ratio, $Q$, they intersect in this figure. An inspiraling binary evolves
along horizontal lines toward increasing $\bar{\Omega}$, i.e., from the
left to the right, until it reaches either the innermost stable circular
orbit or the mass-shedding limit. For a sufficiently high mass ratio,
the binary reaches an innermost stable circular orbit. Quasiequilibrium
sequences cannot predict the fate of the neutron star after that,
because it enters a dynamical plunge phase (see
Sect.~\ref{sec:sim}). Thus, the mass-shedding limit for unstable
quasiequilibrium sequences included in Fig.~\ref{fig:endpoint} should be
regarded as only indicative. As shown in Fig.~\ref{fig:endpoint}, the
sequence with $Q=6$ (dot-dashed line) encounters the innermost stable
circular orbit, while that with $Q=3$ (dot-dot-dashed line) ends at the
mass-shedding limit. The intersection of the curve for the mass-shedding
limit and that for the innermost stable circular orbit marks a critical
point that separates the two distinct fates of the binary
inspiral. Specifically, the critical mass ratio is found to be $Q
\approx 4.2$ for this case, i.e., binaries of a nonspinning black hole
and an irrotational neutron star with $\mathcal{C} = 0.145$ modeled by a
$\Gamma = 2$ polytrope.

\begin{figure}[htbp]
 \centering \includegraphics[width=.7\linewidth,clip]{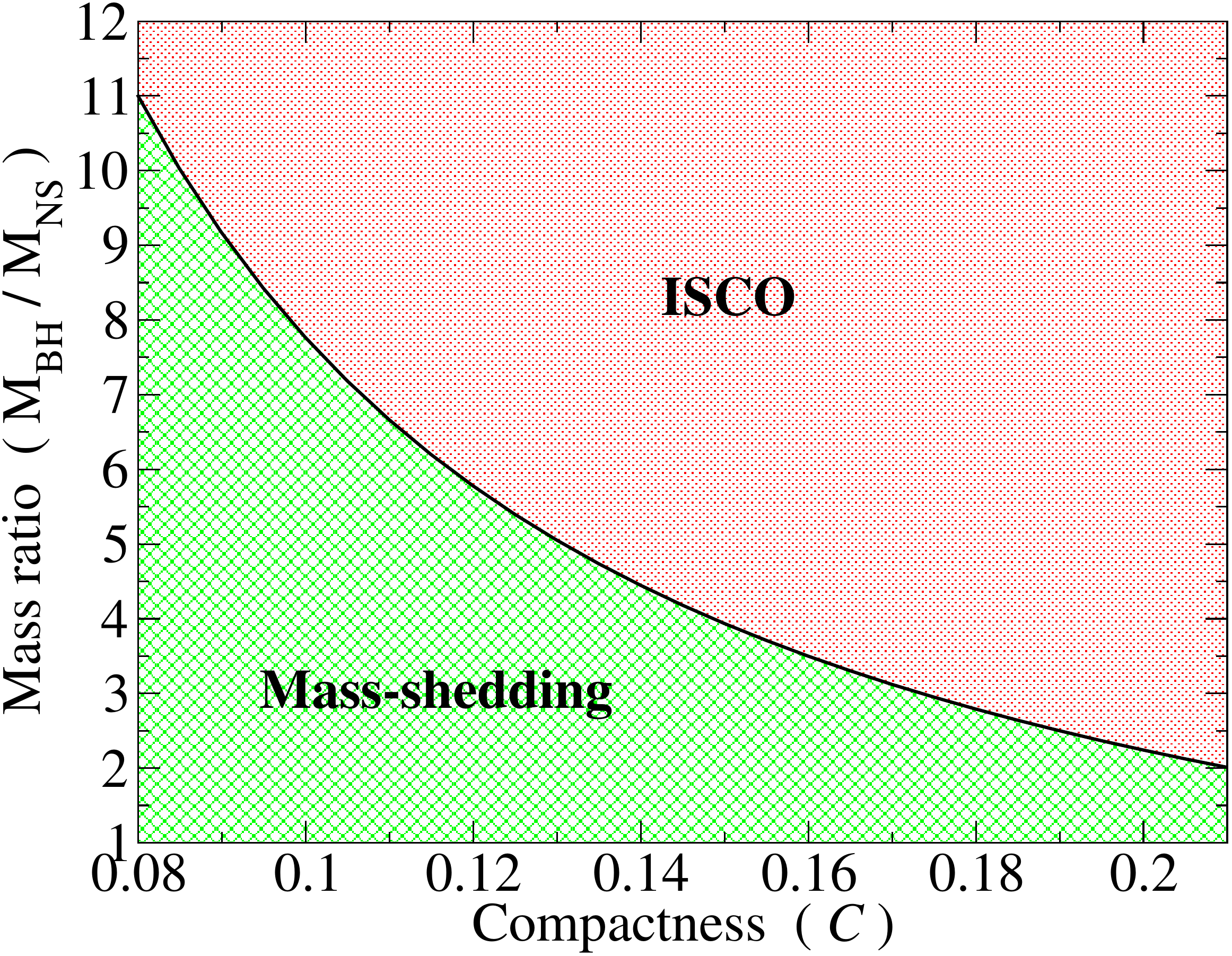}
 \caption{Critical mass ratio which separates black hole--neutron star
 binaries that encounter an innermost stable circular orbit before
 initiating mass shedding and those undergoing tidal disruption, as a
 function of the compactness of the neutron star modeled by the $\Gamma
 = 2$ polytrope. As we discuss in Sect.~\ref{sec:sim}, while
 configurations in ``ISCO'' nearly certainly end up in the plunge
 without tidal disruption, only a part of configurations in
 ``Mass-shedding'' leads to appreciable tidal disruption. Image reproduced with permission from \citet{Taniguchi_BFS2008}, copyright by APS.} \label{fig:cqdiag}
\end{figure}

The critical mass ratio which separates black hole--neutron star
binaries that encounter an innermost stable circular orbit before
initiating mass shedding and those reach the mass-shedding limit before
the plunge is obtained as a function of the compactness of the neutron
star by equating Eqs.~\eqref{eq:msfit2} and
\eqref{eq:iscofit}. Specifically, the critical mass ratio is determined
by
\begin{equation}
 0.270 \mathcal{C}^{3/2} (1 + Q) \pqty{1 + Q^{-1}}^{1/2} =  0.0680
  \bqty{1 - \frac{0.444}{Q^{0.25}} \pqty{1 - 3.54 \mathcal{C}^{1/3}}} ,
  \label{eq:critical}
\end{equation}
and the curve that separates those two regions on the $Q$-$\mathcal{C}$
plane is shown in Fig.~\ref{fig:cqdiag}. If the mass ratio of a black
hole--neutron star binary is higher than the critical value, the
quasiequilibrium sequence terminates by encountering the innermost
stable circular orbit, while if lower, it ends at the mass-shedding
limit. We emphasize that the mass shedding is only a necessary condition
for tidal disruption. In Sect.~\ref{sec:sim}, we will see that dynamical
simulations tend to predict that tidal disruption occurs for a more
restricted range of parameters than that shown in Fig.~\ref{fig:cqdiag}.

We again caution that the classification shown in Fig.~\ref{fig:cqdiag}
is appropriate only for nonspinning black hole--neutron star
binaries. The spin of the black hole significantly modifies the critical
mass ratio. For example, if the spin of the black hole is high and
prograde, i.e., aligned with the orbital angular momentum of the binary,
the region of ``Mass-shedding'' will be enlarged
significantly. Systematic studies on the effect of black-hole spins have
not been performed yet, and further investigations are awaited.

\subsection{Summary and issues for the future} \label{sec:eq_summary}

In this section, we have reviewed the current status of the studies on
quasiequilibrium sequences of black hole--neutron star binaries in
general relativity, focusing mainly on \citet{Taniguchi_BFS2008}. In
particular, we highlighted a curve of the critical mass ratio, which
separates black hole--neutron star binaries that encounter an innermost
stable circular orbit before initiating mass shedding and vice versa, as
a function of the compactness of the neutron star. The result is shown
in Eq.~(\ref{eq:critical}) and in Fig.~\ref{fig:cqdiag}. Such a critical
curve clearly classifies the possible final fate of black hole--neutron
star binaries, which depends on the mass ratio and the compactness of
the neutron star for a given equation of state. The final fate depends
also on the spin of the black hole (see Sect.~\ref{sec:sim}), although
only nonspinning configurations are discussed here.

As seen in Table~\ref{table:qe}, the parameter space surveyed is still
quite narrow, partly because the community has been devoting effort to
dynamical simulations. A systematic study of quasiequilibrium sequences
has been done only for binaries composed of a nonspinning black hole and
an irrotational neutron star with the $\Gamma = 2$ polytrope. It may be
useful to survey the remaining parameter space in a systematic manner in
the future. Systematic numerical results for such a study will be
helpful for predicting the final fate of black hole--neutron star
binaries and for checking the results derived in numerical simulations
in light of the high computational precision of quasiequilibrium
states. Specifically, quasiequilibrium sequences of binaries composed of
a spinning black hole and a neutron star with an equation of state other
than the $\Gamma = 2$ polytrope remain to be studied in detail.

The formulation for quasiequilibrium states should be improved further
for more rigorous studies. To date, all the computations adopted
formulations which solve only five out of ten components of the Einstein
equation: constraint equations and the slicing condition (see Appendix
\ref{app:init}). Employing an improved formulation in which full
components of the Einstein equation are solved is left for the future
(see \citealt{Shibata_Uryu_Friedman2004,Cook_Baumgarte2008} for proposed
formulations and \citealt{Uryu_LFGS2006,Uryu_LFGS2009} for studies of
binary neutron stars).

\newpage

\section{Merger and postmerger simulation} \label{sec:sim}

Dynamical simulations for coalescences of black hole--neutron star
binaries have been performed in full general relativity by several
groups since \citet{Shibata_Uryu2006}. These studies have explored the
merger process, the criterion for tidal disruption of a neutron star,
the mass and the spin parameter of the remnant black hole, properties of
the remnant disk and the ejected material, and gravitational
waveforms. Results from different groups agree with each other
quantitatively when they can be compared. Longterm simulations of the
accretion disks around black holes for $\sim \SI{10}{\second}$ are also
becoming available in recent years. These studies begin to clarify the
evolution of the dense and hot accretion disk, neutrino luminosity, the
role of neutrino emission, properties of the disk outflow, and possible
launch of an ultrarelativistic jet. In this section, we review our
current understanding about these topics.

\subsection{Numerical method for coalescence simulations}
\label{sec:sim_meth}

Numerical-relativity simulations for black hole--neutron star binary
coalescences are performed by solving the Einstein evolution equations
with appropriate gauge conditions and hydrodynamics equations, which may
involve neutrino-radiation transfer and magnetohydrodynamics. General
formulation and numerical techniques are summarized in Appendix
\ref{app:sim}. In this Sect.~\ref{sec:sim_meth}, we summarize general
aspects of initial data and equations of state adopted in numerical
simulations of black hole--neutron star binary coalescences.

Numerical-relativity simulations for black hole--neutron star binaries
throughout the inspiral-merger-postmerger phases have typically been
performed only for $\lesssim \SI{100}{\ms}$ and are still in the early
stage for studying the postmerger evolution. To explore the longterm
evolution of the merger remnant, simulations of black hole--accretion
disk systems have also been performed in full general relativity for
$\gtrsim \SI{1}{\second}$ \citep[see also
\citealt{Most_PTR2021-2}]{Fujibayashi_SWKKS2020,Fujibayashi_SWKKS2020-2}. The
setup of these simulations will be described separately in
Sect.~\ref{sec:sim_pm}.

\subsubsection{Initial condition} \label{sec:sim_meth_ini}

Realistic simulations of black hole--neutron star binary coalescences
always adopt quasiequilibrium states reviewed in Sect.~\ref{sec:eq} as
their initial conditions. Typically, simulations based on the
generalized harmonic formalism with the excision method adopt
quasiequilibrium states computed in the excision framework. While many
simulations based on the BSSN formalism (or its extension) with the
moving-puncture method adopt quasiequilibrium states computed in the
puncture framework, quasiequilibrium states computed in the excision
framework have also been adopted
\citep{Etienne_FLSTB2008,Etienne_LSB2009,Etienne_LPS2012,Etienne_Paschalidis_Shapiro2012,Paschalidis_Ruiz_Shapiro2015,Ruiz_Shapiro_Tsokaros2018,Ruiz_PTS2020,Most_PTR2021}. Because
the moving-puncture method needs data of the metric inside the excision
surface in the initial configurations, the interior needs to be filled
artificially by extrapolating the data outside the excision
surface. This extrapolation generally produces constraint-violating
initial data, and care must be taken so that this violation does not
affect significantly the evolution outside the excision surface
\citep[see also \citealt{Brown_SSTDHP2007}]{Etienne_FLSB2007}.

Strictly speaking, quasiequilibrium states derived under the assumption
of the helical symmetry cannot be realistic, because the radial
approaching velocity induced by gravitational radiation reaction is not
taken into account. This drawback gives rise to the inspiral motion with
the residual eccentricity of $e \gtrsim 0.01$ for typical initial
data. Because the eccentricity of $e \approx 0.01$ introduces a phase
shift larger than the tidal effect \citep{Favata2014}, numerical
simulations used for developing theoretical templates are required to
adopt initial data with the eccentricity as low as $e \lesssim
\num{e-3}$ in order not to bias estimation of tidal deformability in the
analysis of gravitational waves from realistic circular
binaries. Although the residual eccentricity may be reduced if we could
start simulations from a distant orbit at which radiation reaction is
sufficiently weak, this is not practical with current and near-future
computational resources. To obtain low-eccentricity inspirals with a
reasonable initial separation, iterative eccentricity reduction is
routinely applied in the excision-based simulations \citep[see Appendix
\ref{app:init_beyond_ecc} for details]{Foucart_KPT2008}. Essentially the
same technique has recently been developed for and applied to
puncture-based initial data \citep{Kyutoku_KKST2021}.

Because all the quasiequilibrium states have been computed in the
framework of pure ideal hydrodynamics assuming that the neutron star is
cold, additional variables need to be specified if we perform
simulations with detailed microphysics. For evolving neutron stars with
composition-dependent equations of state, we need to give the electron
fraction in the initial condition. These variables are usually
determined by the condition of a(n approximate) zero-temperature
$\beta$-equilibrium \citep[see, e.g.,][]{Duez_FKOT2010}. This step is in
particular necessary for neutrino-radiation-hydrodynamics simulations
\citep{Deaton_DFOOKMSS2013,Foucart_DDOOHKPSS2014,Foucart_DBDKHKPS2017,Brege_DFDCHKOPS2018,Kyutoku_KSST2018,Foucart_DKNPS2019,Most_PTR2021}.

For magnetohydrodynamics simulations, magnetic fields are superposed on
the initial configuration with arbitrary strength and arbitrary
geometry. Their magnitude should not be very large, because too strong
magnetic fields destroy the hydrostationary equilibrium. Still, this
condition admits an astrophysically strong magnetic fields of $\lesssim
\SI{e17}{G}$ in the neutron star, for which the gravitational binding
energy is larger by orders of magnitude. To resolve short-wavelength
modes associated with the magnetorotational instability in the
postmerger phase \citep{Balbus_Hawley1991}, it is customary to impose
magnetar-level magnetic fields inside neutron stars. This may be
justified, because the magnetic fields are likely to be amplified on a
dynamical time scale in the accretion disk formed after merger in the
real world. However, it has not been clarified yet whether a strong and
coherent poloidal field is really established by some mechanism, e.g.,
the dynamo process. While many magnetohydrodynamics simulations have
adopted poloidal magnetic fields initially confined in the neutron star
to avoid difficulty in handling force-free magnetospheres
\citep{Chawla_ABLLMN2010,Etienne_LPS2012,Etienne_Paschalidis_Shapiro2012,Kiuchi_SKSTW2015,Most_PTR2021},
pulsar-like dipolar magnetic fields are also adopted with a non-tenuous
artificial atmosphere outside the neutron star in simulations that focus
on the possible jet launch
\citep{Paschalidis_Ruiz_Shapiro2015,Ruiz_Shapiro_Tsokaros2018,Ruiz_PTS2020}. We
note that an artificial atmosphere itself is always required by
hydrodynamics simulations performed in a conservative scheme (see also
Appendix~\ref{app:sim_hydro_m}).

\subsubsection{Equation of state} \label{sec:sim_meth_eos}

The equation of state for neutron-star matter is a key ingredient for
deriving realistic outcomes of and multimessenger signals from black
hole--neutron star binary coalescences. The primary reason for this is
that the equation of state determines the density distribution and hence
the radius of the neutron star for a given value of the mass. Whether
tidal disruption occurs or not during the coalescence and, if it occurs,
its degree are determined primarily by the radius or the compactness of
the neutron star for given masses and spins of binary components [see
Eq.~\eqref{eq:critical}]. Thus, the gravitational waveform, the
properties of the remnant disk, and the properties of the ejecta are
governed crucially by the equation of state. The equation of state also
determines the tidal deformability of the neutron star, which affects
the late inspiral phase of compact binary coalescences.

However, as mentioned in Sect.~\ref{sec:eq_param_ns}, the equation of
state for supranuclear-density matter is still uncertain (see, e.g.,
\citealt{Lattimer_Prakash2016,Oertel_HKT2017,Baym_HKPST2018} for
reviews). In the study of compact binary coalescences involving neutron
stars, it is more beneficial to explore the possibility of determining
the equation of state via gravitational-wave observations
\citep{Lindblom1992,Vallisneri2000,Read_MSUCF2009,Ferrari_Gualtieri_Pannarale2010,Lackey_KSBF2012,Maselli_Gualtieri_Ferrari2013,Lackey_KSBF2014,Pannarale_BKLS2015}
than to derive observable signals relying on a single candidate of the
realistic equation of state. For this purpose, it is necessary to
prepare theoretical templates of gravitational waveforms by performing
simulations systematically over the parameter space of black
hole--neutron star binaries with a wide variety of hypothetical
equations of state. Such a systematic survey is also indispensable for
predicting electromagnetic counterparts (see Sect.~\ref{sec:sim_rem} for
the dependence of the disk and ejecta properties on the equation of
state). Thus, all the equations of state adopted in simulations reviewed
in this article should be understood as hypothetical.

Due to the reason described in Sect.~\ref{sec:eq_param_ns}, we may
safely adopt zero-temperature equations of state during the inspiral and
early merger phases before the shock heating begins to play a
role. Moreover, the zero-temperature equations of state are sufficient
for simulating black hole--neutron star binary coalescences which do not
result in tidal disruption of neutron stars, because essentially no
heating process is involved. However, because polytropes are not
quantitative models of neutron stars, it is desirable to adopt
nuclear-theory-based equations of state for the purpose of investigating
realistic black hole--neutron star binary coalescences.

Sophisticated zero-temperature equations of state are implemented in
numerical simulations by various means. A straightforward method is to
adopt numerical tables calculated based on hypothetical models of
nuclear physics \citep[see,
e.g.,][]{Glendenning_Moszkowski1991,Muller_Serot1996,Akmal_Pandharipande_Ravehnall1998,Douchin_Haensel2001,Alford_BPR2005}. The
drawback of this method is that the capability of systematic studies is
limited by available tables. A popular tool for conducting a systematic
study is an analytic, piecewise-polytropic equation of state, with which
the pressure is given by a broken power-law function of the rest-mass
density as
\begin{equation}
 P ( \rho ) = \kappa_i \rho^{\Gamma_i} \; ( \rho_i \le \rho < \rho_{i+1}
  ) ,
\end{equation}
where $i \in [0:n-1]$, $\rho_0 = 0$, and $\rho_n \to \infty$ \citep[see
also \citealt{OBoyle_MSR2020} for generalization and
\citealt{Haensel_Potekhin2004,Lindblom2010,Potekhin_FCPG2013} for other
analytic approaches]{Read_MSUCF2009,Read_LOF2009,Ozel_Psaltis2009}. It
has been shown that most of the nuclear-theory-based equations of state
for neutron-star matter can be approximated to reasonable accuracy up to
the rest-mass density of $\gtrsim \SI{e15}{\gram\per\cubic\cm}$ by
piecewise polytropes consisting of one for the crust region and three
for the core region if we choose $\rho_2 =
\SI{e14.7}{\gram\per\cubic\cm}$ and $\rho_3 =
\SI{e15}{\gram\per\cubic\cm}$ \citep{Read_LOF2009}.

It is remarkable that the maximum density in the system only decreases
in time (except for possible minor fluctuations) during the black
hole--neutron star binary coalescences. This is a striking difference
from the binary-neutron-star merger, after which a massive or collapsing
neutron star with increased central density is formed \citep[see,
e.g.,][]{Hotokezaka_KOSK2011,Takami_Rezzolla_Baiotti2015,Dietrich_BUB2015,Foucart_HDOORKLPS2016}. This
property indicates that black hole--neutron star binaries are not
influenced by the equation of state at very high density, e.g., several
times the nuclear saturation density, unless the neutron star is close
to the maximum-mass configuration. Thus, numerical simulations for
binaries with plausibly canonical $\sim 1.4\,M_\odot$ neutron stars may
safely adopt simplified models of nuclear-matter equations of state. For
example, if the central density is lower than
$\SI{e15}{\gram\per\cubic\cm}$, we may adopt piecewise polytropes with a
reduced number of pieces for the core focusing only on its low-density
part
\citep{Read_MSUCF2009,Kyutoku_Shibata_Taniguchi2010,Kyutoku_OST2011}. This
feature also indicates a weak point that we will not be able to
investigate properties of ultrahigh-density matter from observations of
black hole--neutron star binaries without extrapolation relying on
theoretical models \citep[see,
e.g.,][]{GW170817EOS,Raaijmakers_etal2019}.

Once the heating process is activated in the merger phase, particularly
via the shock associated with self-crossing of the tidal tail, the
zero-temperature approximation is no longer valid. Finite-temperature
effects become increasingly important in the remnant disk, because
temperature increases due to the shock interaction and presumably to
subsequent viscous heating associated with magnetohydrodynamical
turbulence, while the Fermi energy decreases due to the decreased
rest-mass density compared to neutron stars.

One popular and qualitative approach for incorporating
finite-temperature effects is to supplement zero-temperature equations
of state with an approximate correction. A simple prescription for this
purpose is to add an ideal-gas-like term
\citep{Janka_Zwerger_Moenchmeyer1993},
\begin{equation}
 P_\mathrm{th} = ( \Gamma_\mathrm{th} - 1 ) \rho \varepsilon_\mathrm{th}
  , \label{eq:thideal}
\end{equation}
where $\varepsilon_\mathrm{th} ( \varepsilon , \rho ) := \varepsilon -
\varepsilon_\mathrm{cold} ( \rho )$ is the finite-temperature part of
the specific internal energy with $\varepsilon_\mathrm{cold} ( \rho )$
being the specific internal energy given by a zero-temperature equation
of state. Indeed, the ideal-gas equation of state, $P = ( \Gamma - 1 )
\rho \varepsilon$, which reduces to a polytrope with the adiabatic index
$\Gamma$ for the isentropic fluid and/or at zero temperature assumed in
computations of quasiequilibrium, is occasionally adopted in dynamical
simulations as a qualitative model of neutron-star matter. A parameter
$\Gamma_\mathrm{th}$ represents the strength of the thermal effect, and
its appropriate value may be estimated by calibration with simulations
performed adopting genuinely finite-temperature equations of state,
which are usually given by numerical tables
\citep{Bauswein_Janka_Oechslin2010,Figura_LBLS2020}. It should be
cautioned that, however, the constant value of $\Gamma_\mathrm{th}$ is
not faithful to nuclear-theory-based calculations
\citep{Constantinou_MPL2015} even though uncertain thermal effects at
supranuclear density may not be relevant to black hole--neutron star
binary coalescences \citep{Carbone_Schwenk2019}. In addition, the use of
zero-temperature equations of state is not fully justified after tidal
disruption even if the temperature is not increased. This is because,
although zero-temperature equations of state are derived as a function
of a single variable, e.g., rest-mass density, assuming the
$\beta$-equilibrium, the rapid decompression of the disrupted material
preserves the composition on a dynamical time scale and violates the
$\beta$-equilibrium condition \citep{Foucart_DBDKHKPS2017}.

To investigate the entire merger and postmerger phases in a
self-consistent manner, it is necessary to adopt temperature- and
composition-dependent equations of state with an appropriate scheme for
neutrino transport (see Appendix~\ref{app:sim_hydro_r}). These equations
of state are usually given in a tabulated form as e.g.,
\begin{align}
 P & = P ( \rho , T , Y_\mathrm{e} ) , \\
 \varepsilon & = \varepsilon ( \rho , T , Y_\mathrm{e} ) ,
\end{align}
where $T$ and $Y_\mathrm{e}$ are the temperature and the electron
fraction, respectively \citep[see, e.g.,][see also
\citealt{Raithel_Ozel_Psaltis2019} for a detailed analytic approach to
augment zero-temperature equations of state in a similar manner to
Eq.~\eqref{eq:thideal}]{Lattimer_Swesty1991,Shen_TOS1998,Hempel_FSL2012,Steiner_Hempel_Fisher2013,Banik_Hempel_Bandyopadhyay2014}. Because
these variables are tightly related to neutrino emission and absorption,
neutrino transport is an essential ingredient for accurately determining
thermal properties of material in the postmerger phase. Multidimensional
neutrino-radiation-hydrodynamics simulations in full general relativity
have been developed in the stellar-core collapse
\citep{Sekiguchi2010,Sekiguchi_Shibata2011} and are later applied to
binary neutron stars \citep{Sekiguchi_KKS2011,Sekiguchi_KKS2011-2} as
well as black hole--neutron star binaries
\citep{Deaton_DFOOKMSS2013,Foucart_DDOOHKPSS2014,Kyutoku_KSST2018}.

\subsection{Current parameter space surveyed} \label{sec:sim_space}

As reviewed in Sect.~\ref{sec:eq_param}, models of black hole--neutron
star binaries are characterized by various parameters. In this section,
we focus only on the models in which neutron stars are initially in the
irrotational state, which is presumably realistic for the majority of
compact object binaries as we discussed in Sect.~\ref{sec:intro_orbit}
(see \citealt{Foucart_etal2019,Ruiz_PTS2020} for studies on spinning
neutron stars). Then, the properties of binaries are characterized by
the mass of the black hole, $M_\mathrm{BH}$, the spin parameter of the
black hole, $\chi$, its orientation, $\iota$, and the mass of the
neutron star, $M_\mathrm{NS}$. Furthermore, hypothetical equations of
state for neutron-star matter should also be regarded as a free
parameter (or function) characterizing the models. Properties of an
equation of state may be usefully represented by the radius of the
neutron star, $R_\mathrm{NS}$, particularly when we focus on tidal
disruption.

To date, numerical-relativity simulations of black hole-neutron star
binaries have been performed focusing mainly on neutron stars with
typical masses in our Galaxy of $M_\mathrm{NS} \approx
1.2$--$1.5\,M_\odot$
\citep{Tauris_etal2017,Farrow_Zhu_Thrane2019}. Accordingly, the results
are reviewed below focusing on the models with these typical
values. Because the mass of the neutron star does not vary much among
the models, it is useful in many occasions to characterize a binary
model by quantities directly related to the criterion for mass shedding
to occur outside the innermost stable circular orbit,
Eq.~\eqref{eq:iscovsms}, namely the mass ratio $Q$ and the compactness
$\mathcal{C}$ instead of $M_\mathrm{BH}$ and $R_\mathrm{NS}$,
respectively. This parametrization is also sufficient for qualitative
but scale-free polytropic equations of state (see
Sect.~\ref{sec:eq_res}). It will be worthwhile in the future to simulate
mergers of black hole--neutron star binaries with $M_\mathrm{NS} \sim
2\,M_\odot$ employing nuclear-theory-based equations of state,
particularly in light of possible detections of such neutron stars with
gravitational waves \citep{GW190425,GW200105200115}, although tidal
disruption is unlikely to be common due to the large compactness (see
also Sect.~\ref{sec:dis_dis_bh}).

The range of the mass ratio covered by numerical-relativity simulations
has been enlarged to $1 \le Q \lesssim 8.3$ in the last decade. The
small value in this range is adopted to clarify the difference between
black hole--neutron star binaries and binary neutron stars in the
feature of the mergers, and we will discuss this topic in
Sect.~\ref{sec:dis_dis_ns}. The large values of $Q$ are directly related
to realistic black hole--neutron star binaries, taking into account the
fact that the observed stellar-mass black holes typically have
$M_\mathrm{BH} \gtrsim 5$--$7\,M_\odot$ \citep[see also
\citealt{Thompson_etal2019} for a low-mass black-hole candidate with
$\sim 3.3^{+2.8}_{-0.7}
\,M_\odot$]{Ozel_PNM2010,Kreidberg_BFK2012,GWTC1,GWTC2}.

A wide range of black-hole spins, both in terms of the magnitude and the
orientation, have been adopted in simulations of black hole--neutron
star binaries. Most of the recent simulations have focused on the
prograde spin, because it is required for tidal disruption by massive
black holes with $M_\mathrm{BH} \gtrsim 5\,M_\odot$ or $Q \gtrsim
4$. Notably, the largest value of the spin parameter simulated is
increased to $\chi = 0.97$ \citep{Lovelace_DFKPSS2013}. Although this is
only the case for a single system with $Q=3$, the capability of
simulating nearly-extremal black holes is important for future
investigations of tidal disruption in high mass-ratio systems. Inclined
spins of the black holes are also handled in many simulations.
Essentially all the orientations of the spin have already been handled,
although covering the parameter space becomes computationally demanding
simply because of the increased degree of freedom.

\begin{figure}[htbp]
 \centering \includegraphics[width=0.9\linewidth,clip]{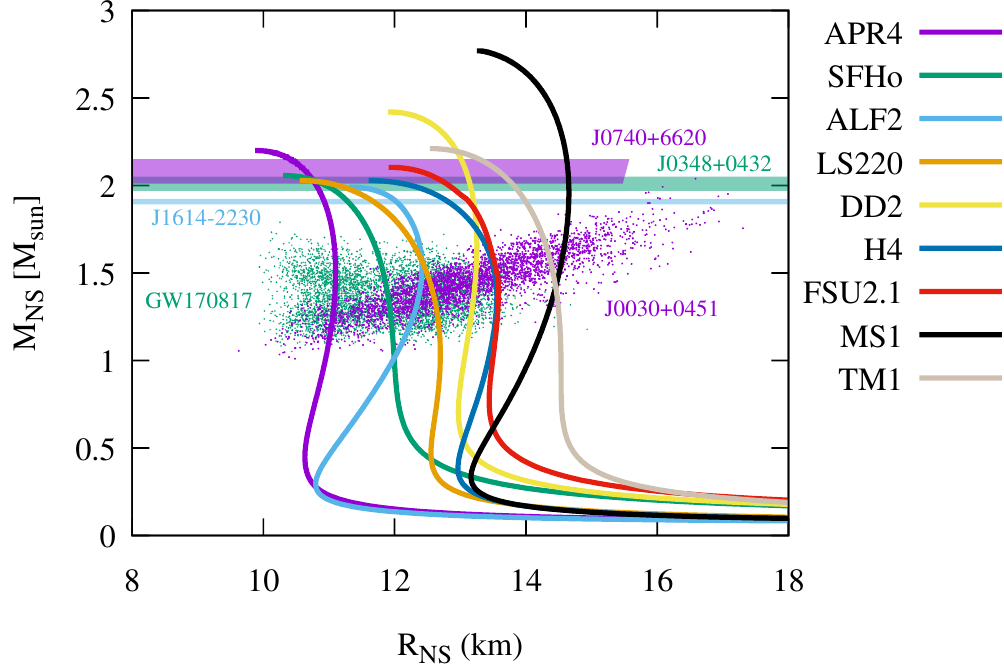}
 \caption{Mass-to-radius relation of cold, spherical neutron stars for
 various equations of state. Unstable configurations with small radii
 are not shown in this plot. We also display the measured mass (68.3\%
 credibility) of a pulsar in J1614-2230 \citep[cyan
 band:][]{Demorest_PRRH2010,Arzoumanian_etal2018}, that in J0348+0432
 \citep[green band:][]{Antoniadis_etal2013}, that in J0740+6620
 \citep[magenta band:][]{Cromartie_etal2020,Fonseca_etal2021}, posterior
 samples obtained by analysis of J0030+0451 \citep[magenta dot:][thinned
 out from provided samples]{Miller_etal2019}, and those by analysis of
 GW170817 \citep[green dot:][]{GW170817EOS}.} \label{fig:MR}
\end{figure}

\begin{table}
 \caption{List of representative equations of state adopted in
 simulations of black hole--neutron star binary coalescences and
 characteristic quantities of neutron stars. $M_\mathrm{max}$ denotes
 the maximum mass of cold, spherical neutron
 stars. $M_{\mathrm{B},1.35}$, $R_{1.35}$, $\mathcal{C}_{1.35}$, and
 $\Lambda_{1.35}$ are, respectively, the baryon rest mass, the
 circumferential radius, the compactness, and the $l=2$ tidal
 deformability of a $1.35\,M_\odot$ neutron star.} \label{table:eos}
 \centering
 \begin{tabular}{lllllll}
  \toprule
  Name & Reference & $M_\mathrm{max} [\,M_\odot]$ & $M_{\mathrm{B},1.35}
  [\,M_\odot]$ & $R_{1.35} (\si{\km})$ & $\mathcal{C}_{1.35}$ &
  $\Lambda_{1.35}$ \\
  \midrule
  APR4 & \citet{Akmal_Pandharipande_Ravehnall1998} & $2.20$ & $1.50$ &
                  $11.1$ & $0.180$ & $322$ \\
  SFHo & \citet{Steiner_Hempel_Fisher2013} & $2.06$ & $1.48$ & $11.9$ &
                      $0.167$ & $420$ \\
  ALF2 & \citet{Alford_BPR2005} & $1.99$ & $1.49$ & $12.4$ & $0.161$ &
                          $733$ \\
  LS220 & \citet{Lattimer_Swesty1991}& $2.03$ & $1.48$ & $12.6$ &
                      $0.158$ & $653$ \\
  DD2 & \citet{Banik_Hempel_Bandyopadhyay2014} & $2.42$ & $1.47$ &
                  $13.2$ & $0.151$ & $854$ \\
  H4 & \citet{Lackey_Nayyar_Owen2006} & $2.03$ & $1.47$ & $13.6$ &
                      $0.147$ & $1110$ \\
  FSU2.1 & \citet{Shen_Horowitz_OConner2011} & $2.10$ & $1.46$ & $13.6$
                  & $0.147$ & $980$ \\
  MS1 & \citet{Muller_Serot1996} & $2.77$ & $1.46$ & $14.4$ & $0.138$ &
                          $1740$ \\
  TM1 & \citet{Hempel_FSL2012} & $2.21$ & $1.46$ & $14.5$ & $0.138$ &
                          $1430$ \\
  \bottomrule
 \end{tabular}
\end{table}

After the early days of adopting qualitative ideal-gas (polytropic at
zero temperature) equations of state, many simulations have been
performed with nuclear-theory-based equations of state. In particular,
temperature- and composition-dependent equations of state are routinely
adopted in neutrino-radiation-hydrodynamics numerical-relativity
simulations. Figure \ref{fig:MR} shows the mass-to-radius relations of
neutron stars for various equations of state which are popular in
numerical-relativity simulations, along with constraints on the
neutron-star properties derived by observations of Galactic massive
pulsars
\citep{Demorest_PRRH2010,Antoniadis_etal2013,Arzoumanian_etal2018,Cromartie_etal2020,Fonseca_etal2021},
J0030+0451 by NICER \citep{Miller_etal2019}, and GW170817 by the
LIGO-Virgo collaboration \citep{GW170817EOS}. Table \ref{table:eos}
summarizes characteristic quantities of neutron stars modeled by these
equations of state. Because the maximum mass of the neutron star is
widely accepted to exceed $\sim 2\,M_\odot$ from the observations of
massive pulsars, recent numerical simulations seldom employ soft
equations of state which are incompatible with these measurements.

Last but not least, various magnetohydrodynamics simulations have been
performed. It should be cautioned that the initial strength and geometry
of magnetic fields need to be chosen somewhat arbitrarily in current
simulations (see Sect.~\ref{sec:sim_pm_mag}). This limitation might not
affect the numerical evolution of the remnant accretion disk if the grid
resolution is sufficiently high. This is because the magnetic fields
inside the accretion disk are expected to be amplified by the
magnetorotational instability and a turbulent state is expected to be
developed (see, e.g., \citealt{Balbus_Hawley1998} for
reviews). Consequently, the relaxed quasisteady state should not depend
on the initial conditions. However, the grid resolution is usually not
high enough for guaranteeing numerical convergence due to the limited
computational resources. Thus, results obtained by current
magnetohydrodynamics simulations should be carefully interpreted. We
also caution that the global configuration of the magnetic field in the
final state could be impacted by the initial choice of a large-scale
poloidal field.

\subsection{Merger process} \label{sec:sim_mrg}

We begin with the review of the merger process focusing on tidal
disruption, subsequent disk formation, and dynamical mass ejection (or
absence thereof). In particular, this Sect.~\ref{sec:sim_mrg} focuses on
the dynamics until $\sim \SI{10}{\ms}$ after the onset of
merger. Effects of the magnetic field
\citep{Chawla_ABLLMN2010,Etienne_LPS2012,Etienne_Paschalidis_Shapiro2012,Kiuchi_SKSTW2015}
and/or neutrino transport
\citep{Deaton_DFOOKMSS2013,Foucart_DDOOHKPSS2014,Kyutoku_KSST2018} play
a significant role in the dynamical evolution of the system only after
the disrupted material winds around the black hole and collides itself
to form a circularized disk. Thus, properties of the neutron star are
characterized only by the mass and zero-temperature equations of state
during the processes discussed here.

\begin{figure}[htbp]
 \centering
 \begin{tabular}{cc}
  \includegraphics[width=.47\linewidth,clip]{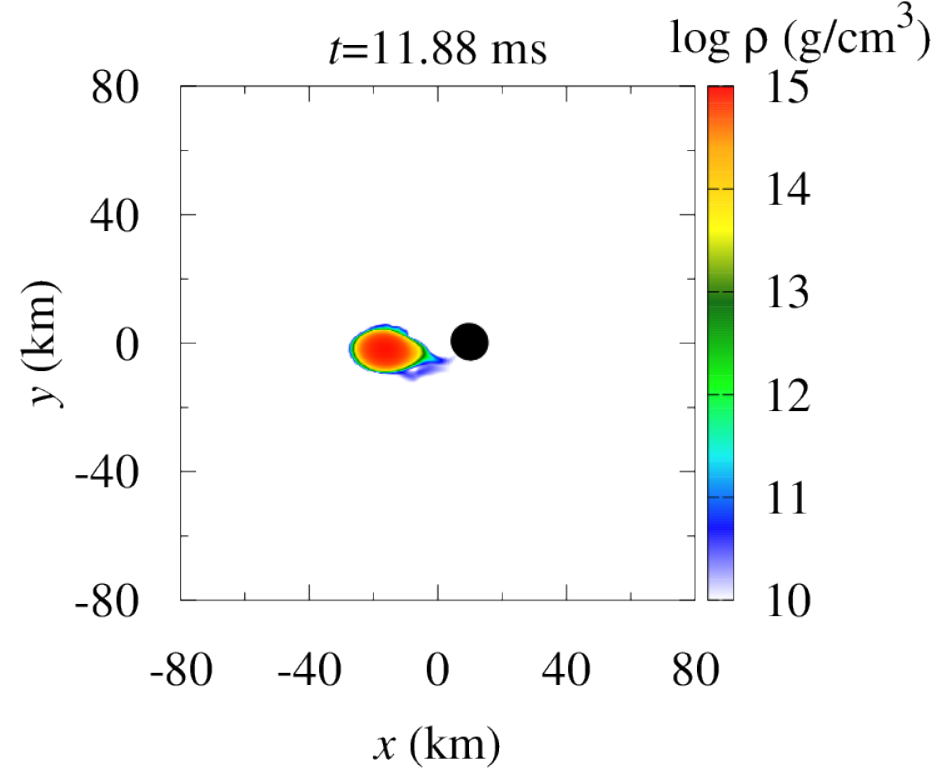} &
  \includegraphics[width=.47\linewidth,clip]{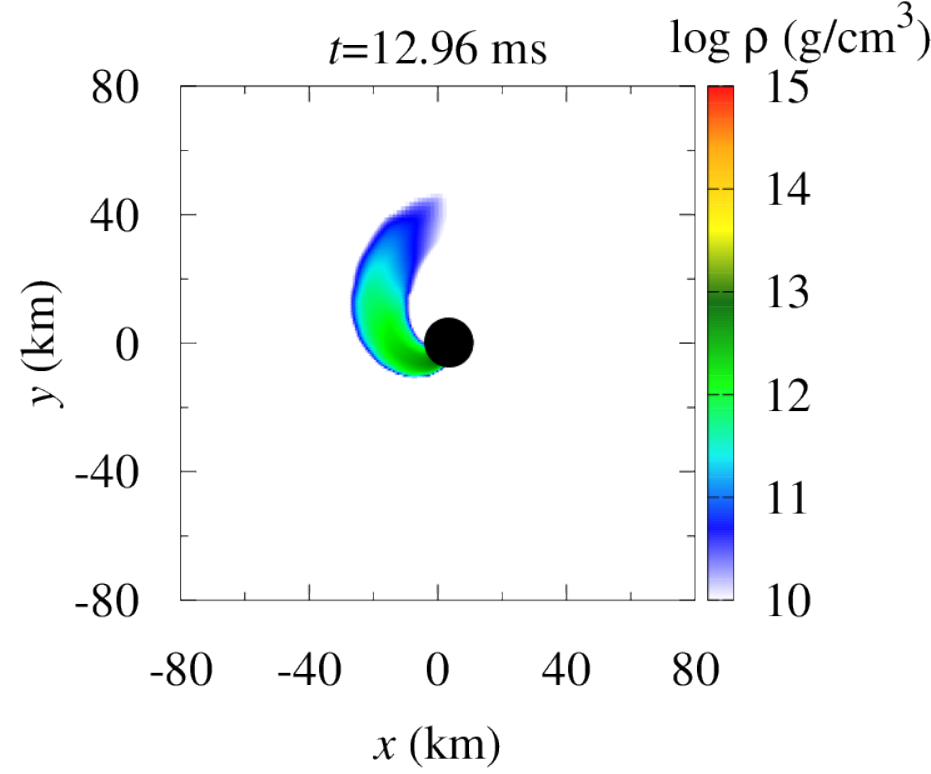} \\
  \includegraphics[width=.47\linewidth,clip]{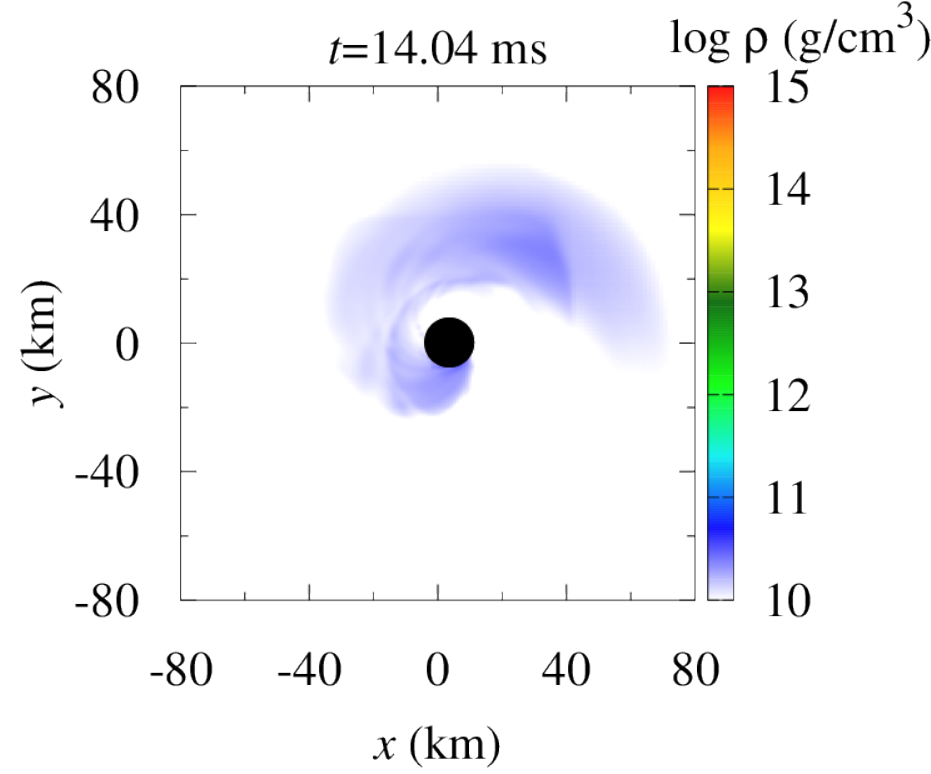} &
  \includegraphics[width=.47\linewidth,clip]{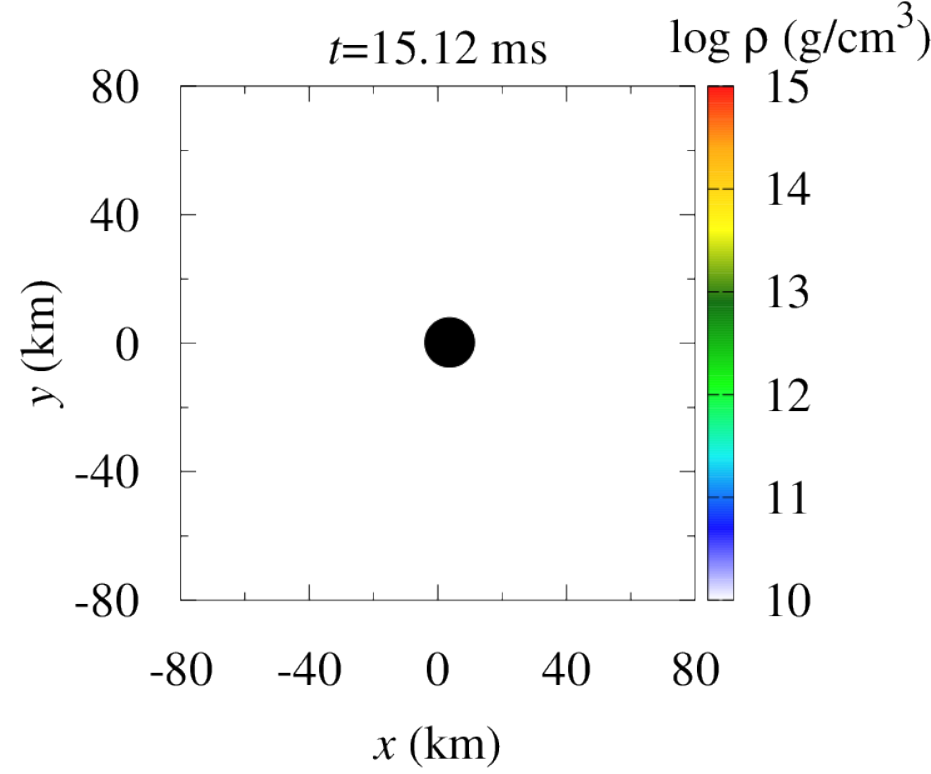}
 \end{tabular}
 \caption{Evolution of the rest-mass density profile and the location of
 the apparent horizon on the equatorial plane for a binary with
 $M_\mathrm{BH} = 4.05\,M_\odot$, $\chi = 0$, $M_\mathrm{NS} =
 1.35M_{\odot}$, and $R_\mathrm{NS} = \SI{11.1}{\km}$ ($Q=3$,
 $\mathcal{C}=0.180$) modeled by a piecewise-polytropic approximation of
 the APR4 equation of state
 \citep{Akmal_Pandharipande_Ravehnall1998}. The black filled circles
 denote the regions inside the apparent horizon of the black hole. The
 color map of each figure shows $\log_{10} ( \rho [
 \si{\gram\per\cubic\cm} ])$. This figure is generated from data of
 \citet{Kyutoku_IOST2015}.} \label{fig:snap_notd}
\end{figure}

\begin{figure}[htbp]
 \centering
 \begin{tabular}{cc}
  \includegraphics[width=.47\linewidth,clip]{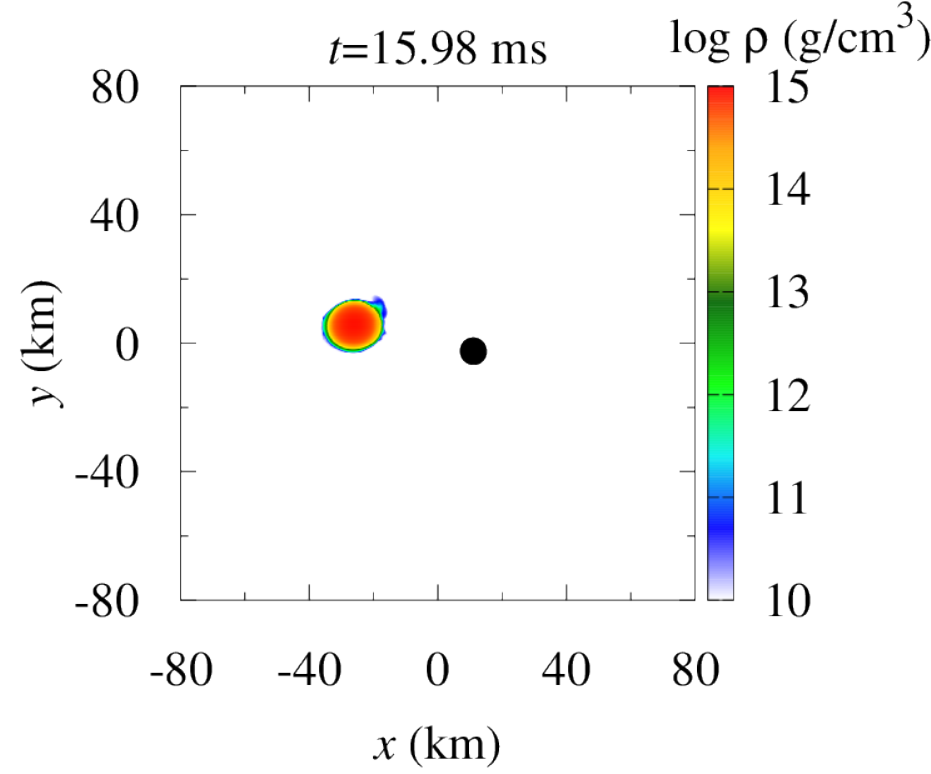} &
  \includegraphics[width=.47\linewidth,clip]{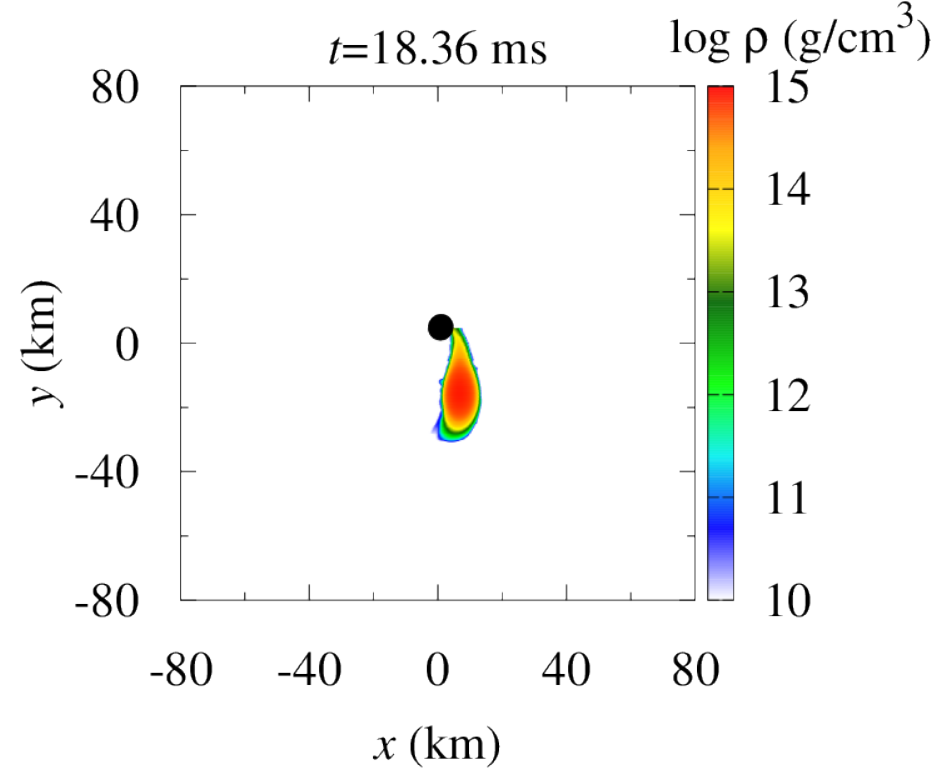} \\
  \includegraphics[width=.47\linewidth,clip]{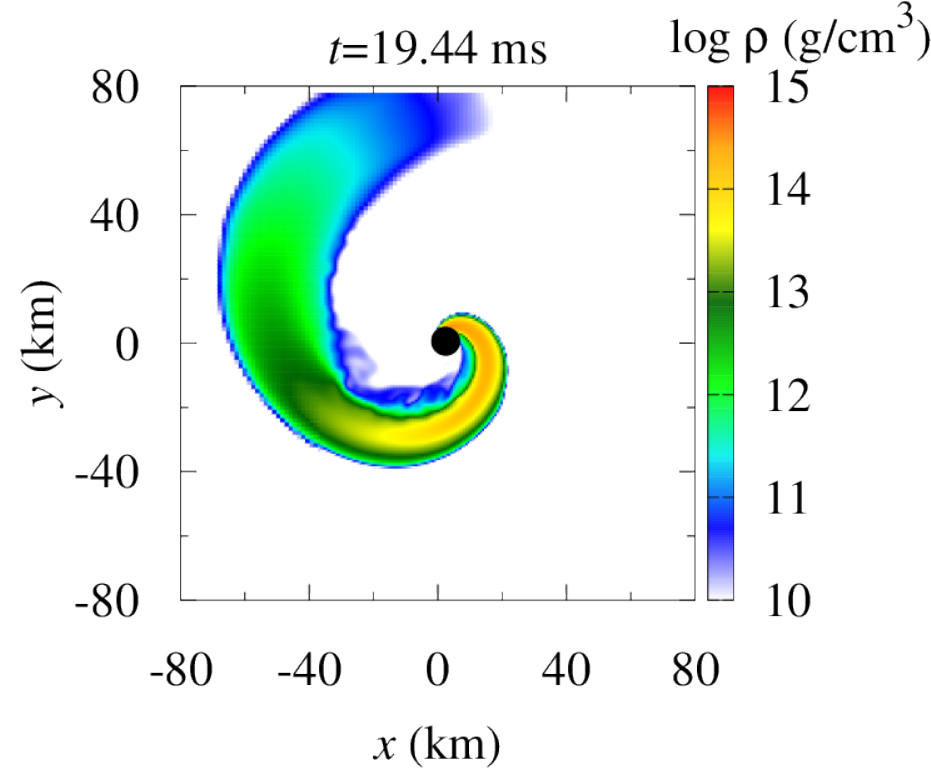} &
  \includegraphics[width=.47\linewidth,clip]{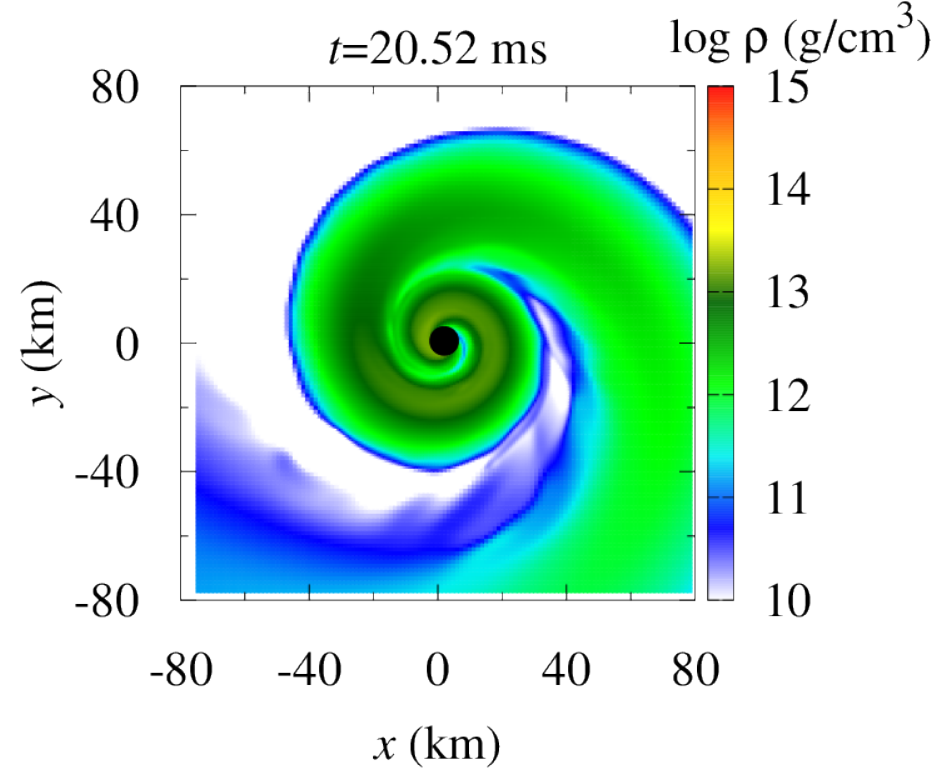} \\
  \includegraphics[width=.47\linewidth,clip]{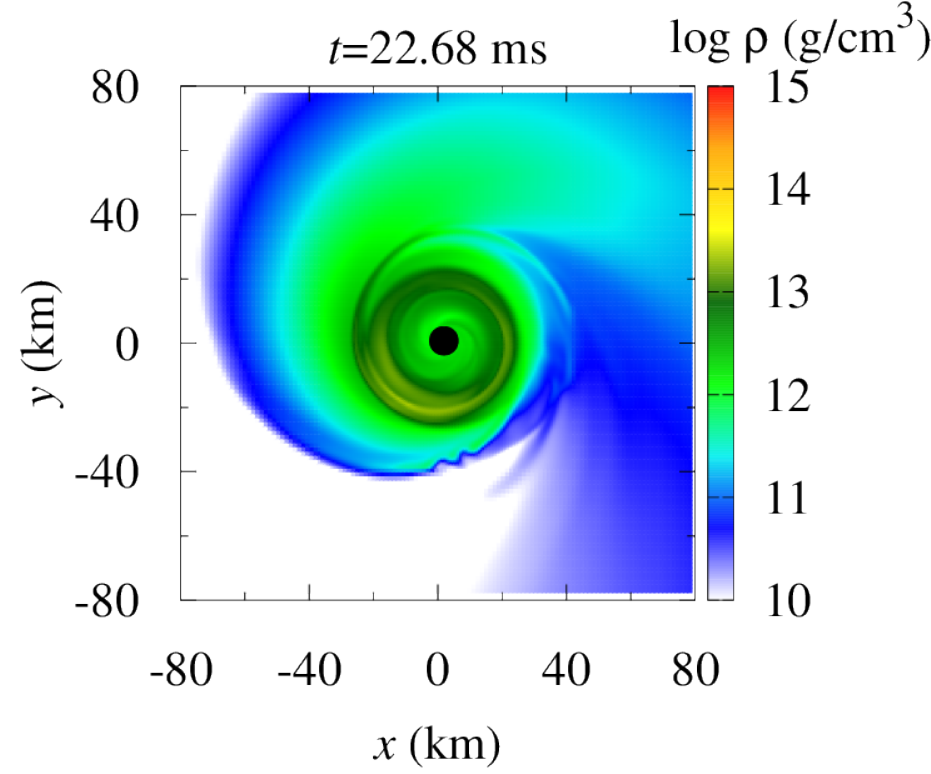} &
  \includegraphics[width=.47\linewidth,clip]{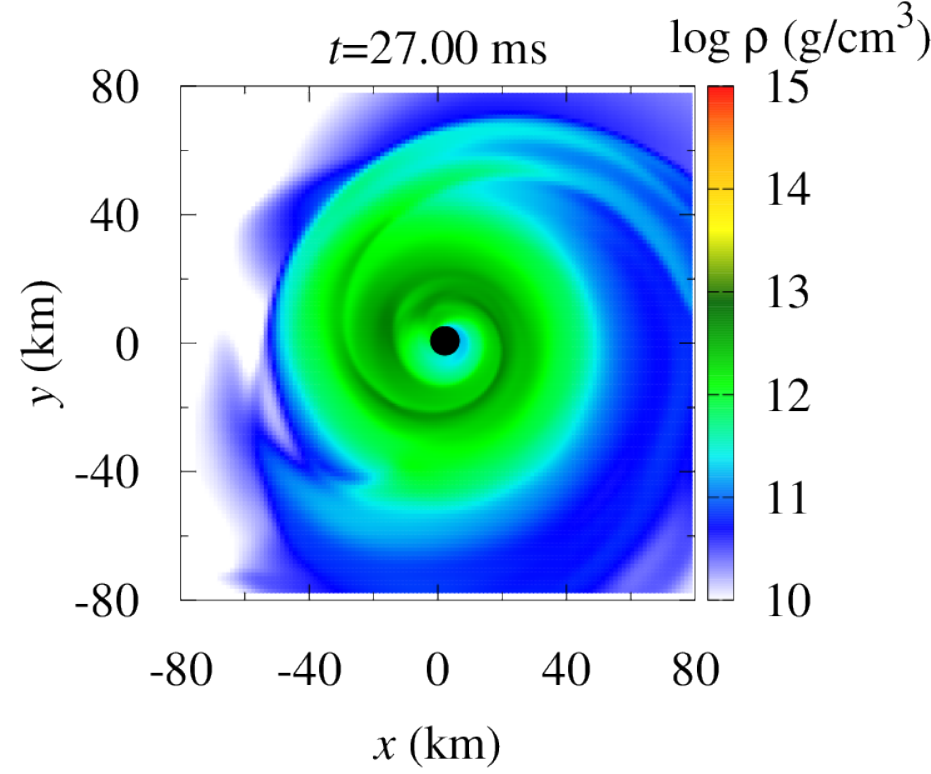}
 \end{tabular}
 \caption{Same as Fig.~\ref{fig:snap_notd} but for a binary with
 $M_\mathrm{BH} = 4.05M_{\odot}$, $\chi = 0.75$, $M_\mathrm{NS} =
 1.35\,M_\odot$, and $R_\mathrm{NS} = \SI{11.1}{\km}$ ($Q=3$,
 $\mathcal{C}=0.180$) modeled by a piecewise-polytropic approximation of
 the APR4 equation of state
 \citep{Akmal_Pandharipande_Ravehnall1998}. This figure is generated
 from data of \citet{Kyutoku_IOST2015}.} \label{fig:snap_td}
 \end{figure}

The orbital separation of a black hole--neutron star binary decreases
due to dissipation of the energy and the angular momentum via
gravitational radiation reaction, and eventually two objects merge. As
we discussed in Sect.~\ref{sec:intro}, the final fate of black
hole--neutron star binary coalescences is classified into two categories
(we will later refine this dichotomy). One is the case in which the
neutron star is swallowed by the black hole without experiencing tidal
disruption. The other is the case in which the neutron star is tidally
disrupted outside the innermost stable circular orbit of the black
hole. As we described in Sect.~\ref{sec:intro_tidal_ms}, which of these
two possibilities is realized is determined primarily by competition
between the orbital separation at which the tidal disruption occurs and
the radius of the innermost stable circular orbit. Figures
\ref{fig:snap_notd} and \ref{fig:snap_td} display the snapshots of the
rest-mass density and the region inside the apparent horizon on the
equatorial plane at selected time slices for typical examples of these
two categories \citep{Kyutoku_IOST2015}.

Figure \ref{fig:snap_notd} illustrates the case in which the neutron
star is not tidally disrupted before it is swallowed by the black
hole. This system is characterized by $M_\mathrm{BH} = 4.05\,M_\odot$,
$\chi = 0$, $M_\mathrm{NS} = 1.35\,M_\odot$, and $R_\mathrm{NS} =
\SI{11.1}{\km}$ ($Q=3$, $\mathcal{C} = 0.180$) modeled by a
piecewise-polytropic approximation of the APR4 equation of state
\citep{Akmal_Pandharipande_Ravehnall1998}. Because the neutron star is
tidally deformed significantly only after it comes very close to the
black hole, mass shedding sets in for an orbit too close to induce
subsequent disruption outside the innermost stable circular orbit. This
is consistent with the expectation from Fig.~\ref{fig:cqdiag} presented
in Sect.~\ref{sec:eq_end}. Accordingly, the masses of the remnant disk
and the dynamical ejecta are negligible, say, $\ll 0.01\,M_\odot$. At the
same time, most of the neutron-star material falls into the black hole
approximately simultaneously through a narrow region of the
horizon. This coherent infall efficiently excites quasinormal-mode
oscillations of the remnant black hole. We discuss gravitational waves
later in Sect.~\ref{sec:sim_gw}. Overall, the behavior of the system in
this category universally resembles that of binary-black-hole
coalescences with the same masses and spins of components, because the
finite-size effect of the neutron star does not play a role
\citep{Foucart_BDGKMMPSS2013}. This seems to be the case for all the
black hole--neutron star binaries and their candidates reported as of
2021 \citep{GWTC2,GW200105200115}.

Figure \ref{fig:snap_td} illustrates the case in which the neutron star
is disrupted before the binary reaches the innermost stable circular
orbit. This system is characterized by $M_\mathrm{BH} = 4.05\,M_\odot$,
$\chi = 0.75$, $M_\mathrm{NS} = 1.35\,M_\odot$, and $R_\mathrm{NS} =
\SI{11.1}{\km}$ ($Q=3$, $\mathcal{C} = 0.180$) modeled by a
piecewise-polytropic approximation of the APR4 equation of state
\citep{Akmal_Pandharipande_Ravehnall1998}. In this case, mass shedding
from an inner cusp of the deformed neutron star sets in at an orbital
separation much larger than that of the innermost stable circular
orbit. After a substantial amount of material is removed from the inner
cusp, the neutron star is tidally disrupted outside the innermost stable
circular orbit. It should be emphasized that tidal disruption does not
occur immediately after the onset of mass shedding but occurs for an
orbital separation smaller than that for the onset of mass shedding as
illustrated by Fig.~\ref{fig:snap_td}. Thus, conditions such as
Eq.~\eqref{eq:critical} are not a sufficient condition but a necessary
condition for tidal disruption.

Once the neutron star is disrupted, the material spreads around the
black hole and forms a one-armed spiral structure, so-called tidal
tail. As a result of the angular momentum transport from the inner to
the outer parts of the tidal tail, a large amount of material in the
outer part avoids being swallowed immediately by the black hole. Because
the tidal tail is in differential rotation, it eventually winds around
the black hole and collides with itself (the right middle panel of
Fig.~\ref{fig:snap_td}). This results in circularization and thus
formation of an approximately axisymmetric disk surrounding the remnant
black hole. The disk material can no longer be treated as zero
temperature because of shock heating, and longterm simulations for the
disk require appropriate implementations of finite-temperature
effects. Still, the disk does not become completely axisymmetric in the
typical rotational period of $\sim \SI{5}{\ms} (m_0 / 10\,M_\odot)$ (see
also Sect.~\ref{sec:sim_rem_bh}). A one-armed spiral structure with a
small amplitude persists for a long time and gradually transports the
angular momentum outward. Hence, mass accretion by the black hole
continues even if viscous or magnetohydrodynamical processes do not set
in. Because the accretion time scale is much longer than the rotational
period, the disk is in a quasisteady state on a time scale of $\gg
\SI{10}{\ms}$. This evolution process agrees qualitatively with that
found for longterm evolution of black hole--accretion disk systems
\citep{Hawley1991,Korobkin_ASSZ2011,Kiuchi_SMF2011,Wessel_PTRS2021}. In
reality, longterm evolution of the disk will be driven by
magnetically-induced turbulent viscosity \citep{Balbus_Hawley1991}. We
will defer discussions about this stage to Sect.~\ref{sec:sim_pm}.

\begin{figure}[htbp]
 \centering
 \begin{tabular}{cc}
  \includegraphics[width=.45\linewidth,clip]{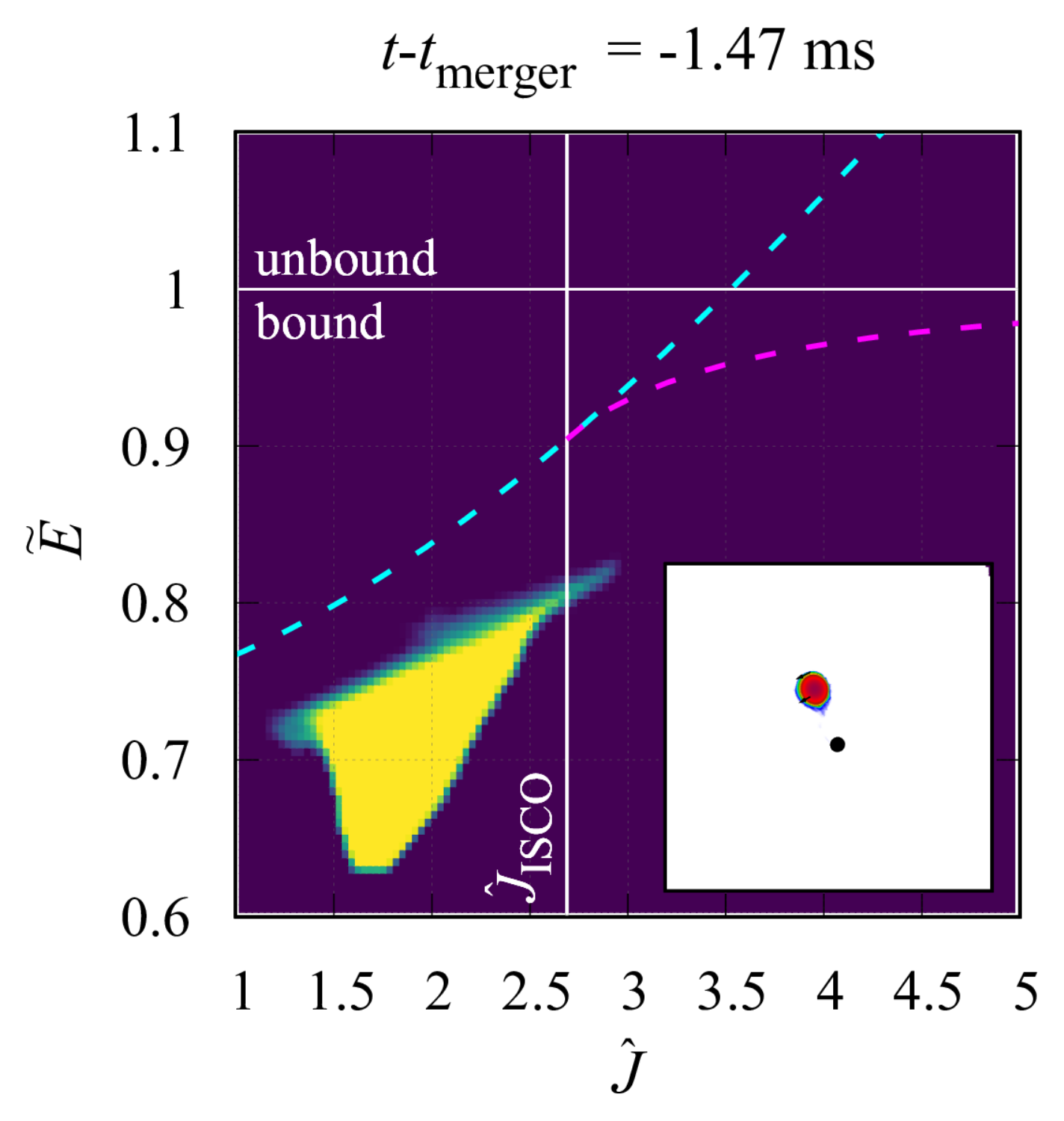} &
  \includegraphics[width=.48\linewidth,clip]{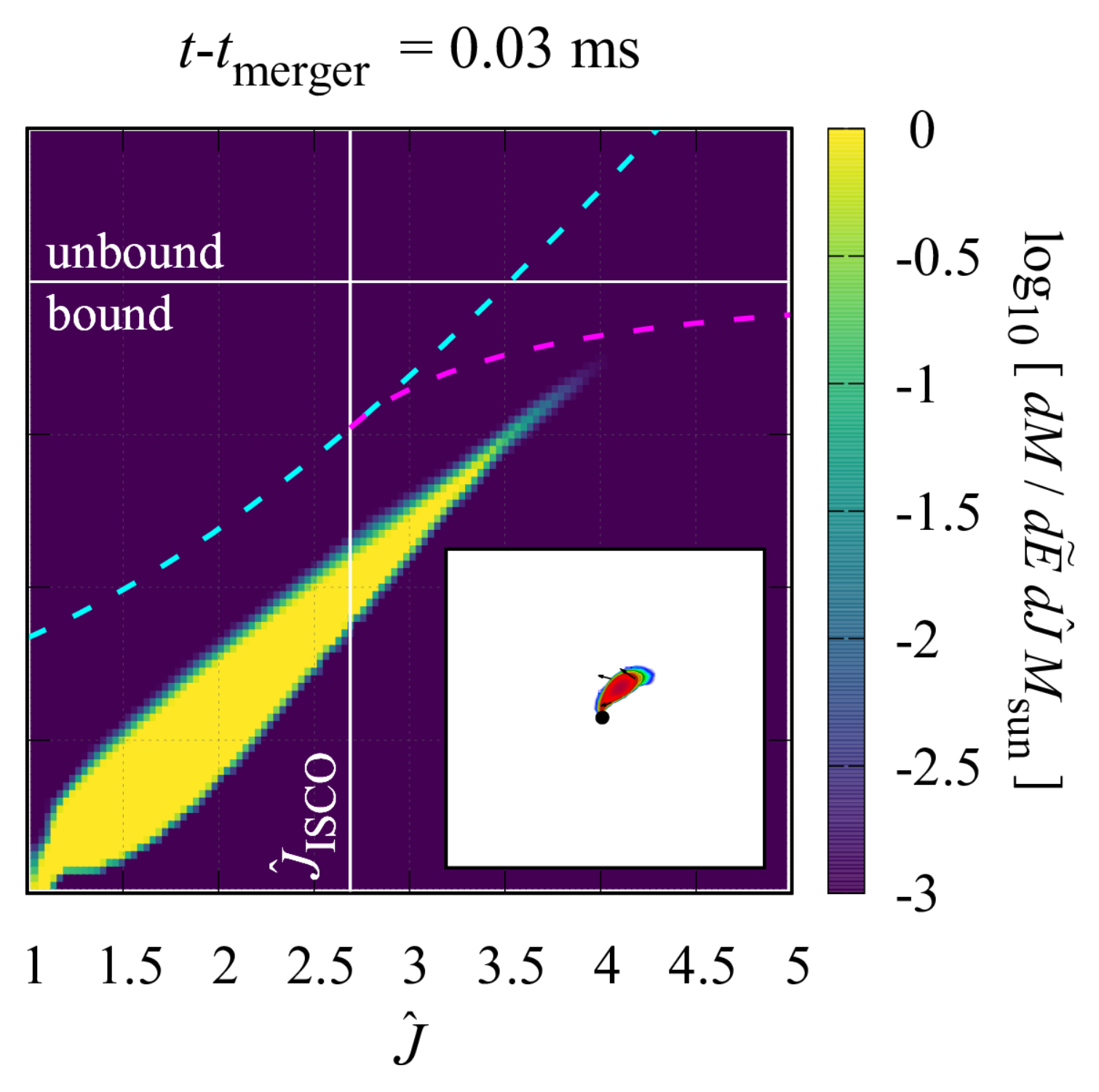} \\
  \includegraphics[width=.45\linewidth,clip]{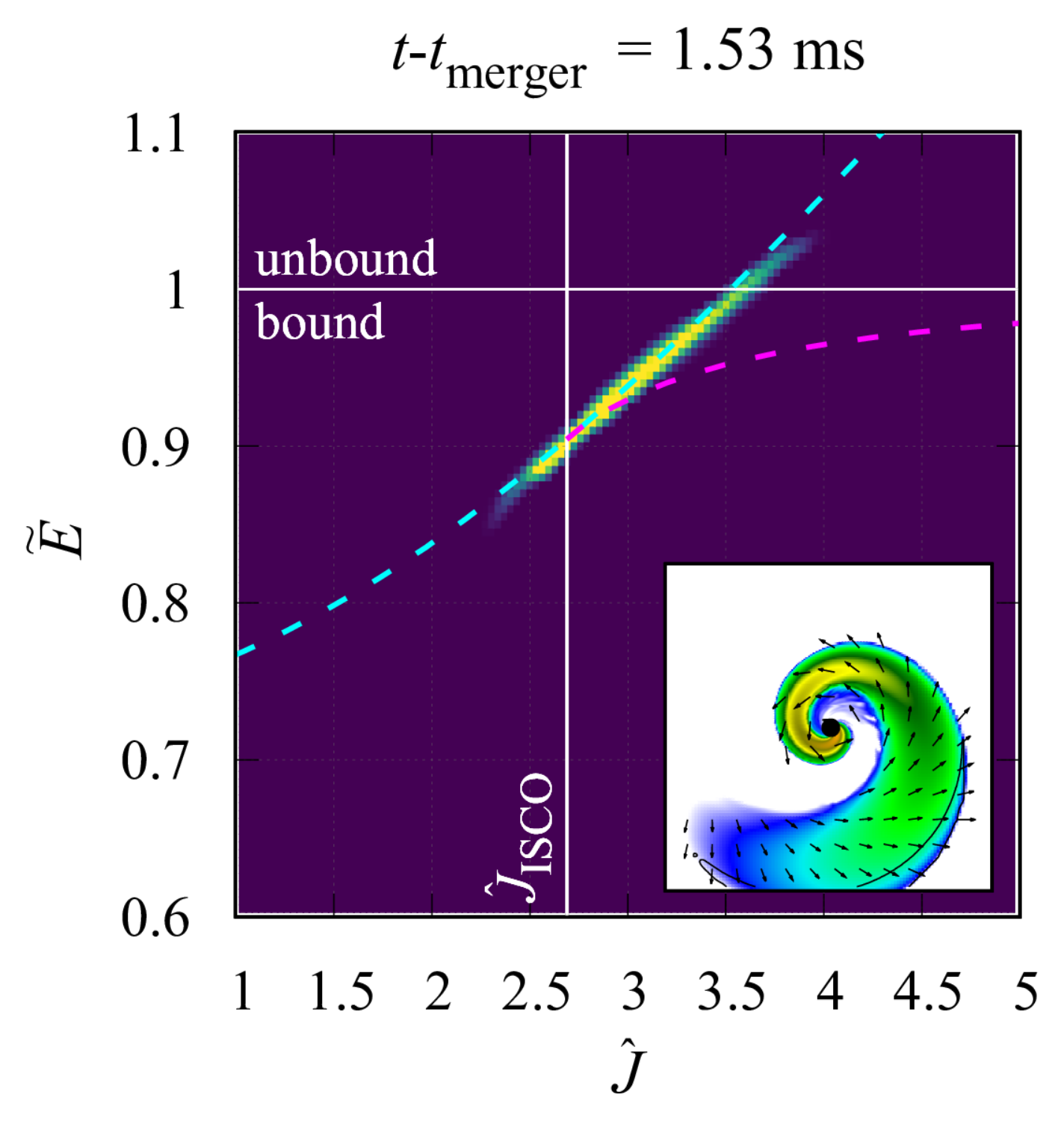} &
  \includegraphics[width=.48\linewidth,clip]{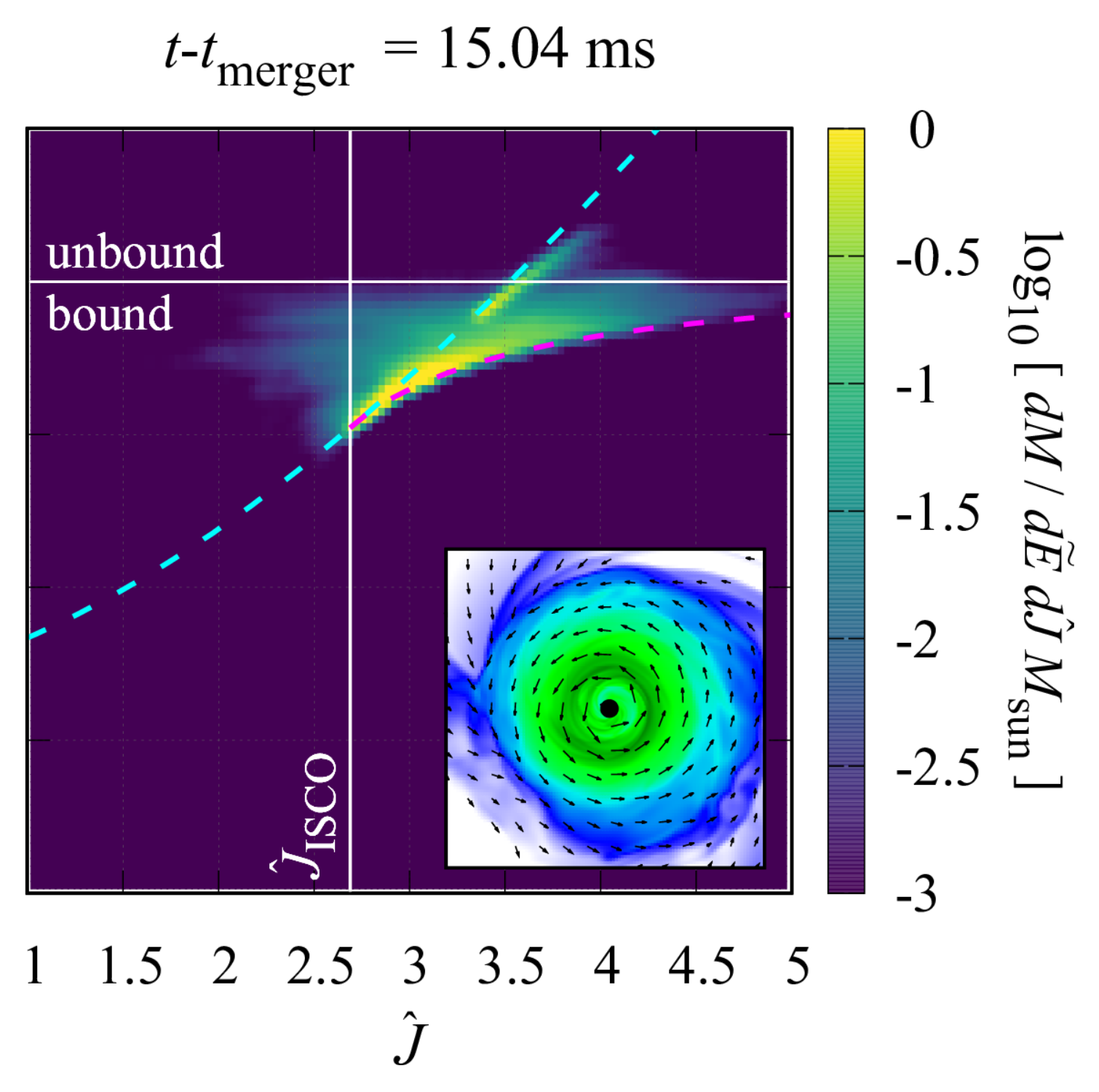}
 \end{tabular}
 \caption{Distribution of material on the phase space of specific energy
 $\tilde{E}$ and specific angular momentum normalized by the mass of the
 remnant black hole $\hat{J}$ for a binary with $M_\mathrm{BH} =
 3\,M_\odot$, $\chi = 0$, $M_\mathrm{NS} = 1.35\,M_\odot$, and
 $R_\mathrm{NS} = \SI{12.3}{\km}$ ($Q \approx 2.2$, $\mathcal{C}=0.162$)
 modeled by a piecewise polytrope called H \citep{Read_MSUCF2009}. The
 left top panel shows the distribution in the final stage of the
 inspiral. The right top panel shows the distribution at the onset of
 merger. The distribution is broadened and spans a wide range of
 $\tilde{E}$ and $\hat{J}$ due to the angular momentum transport. The
 left bottom panel shows the state after the infall of the material with
 low $\hat{J}$. Only the material with the angular momentum exceeding
 that for the innermost stable circular orbit, $\hat{J}_\mathrm{ISCO}$,
 remains outside the black hole. This material has gained exclusively
 the energy during the coalescence of the black hole and the major part
 of the neutron star. The right bottom panel shows the stage in which
 the remnant disk establishes a quasisteady state. The purple dashed
 curves denote the relation for stable circular orbits
 \citep{Bardeen_Press_Teukolsky1972}, and the material along this curve
 is the remnant disk. The cyan dashed curves denote the relation for
 material with a fixed periastron distance, and the material along this
 curve consists of the dynamical ejecta and fallback material. Image reproduced with permission from \citet{Hayashi_KKKS2021}, copyright by APS.} \label{fig:phasespace}
\end{figure}

\begin{figure}[htbp]
 \centering
 \begin{tabular}{cc}
  \includegraphics[width=.47\linewidth,clip]{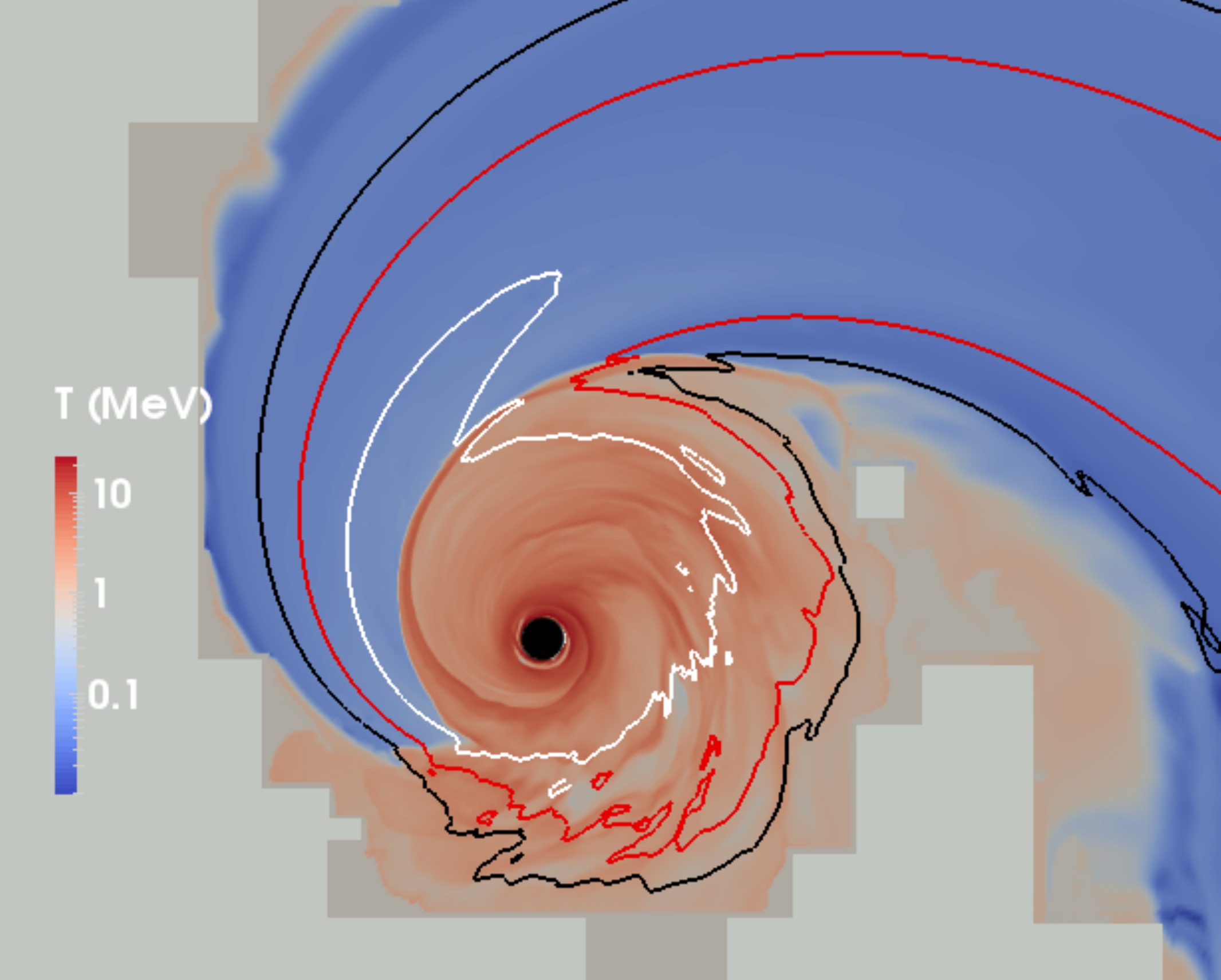} &
  \includegraphics[width=.47\linewidth,clip]{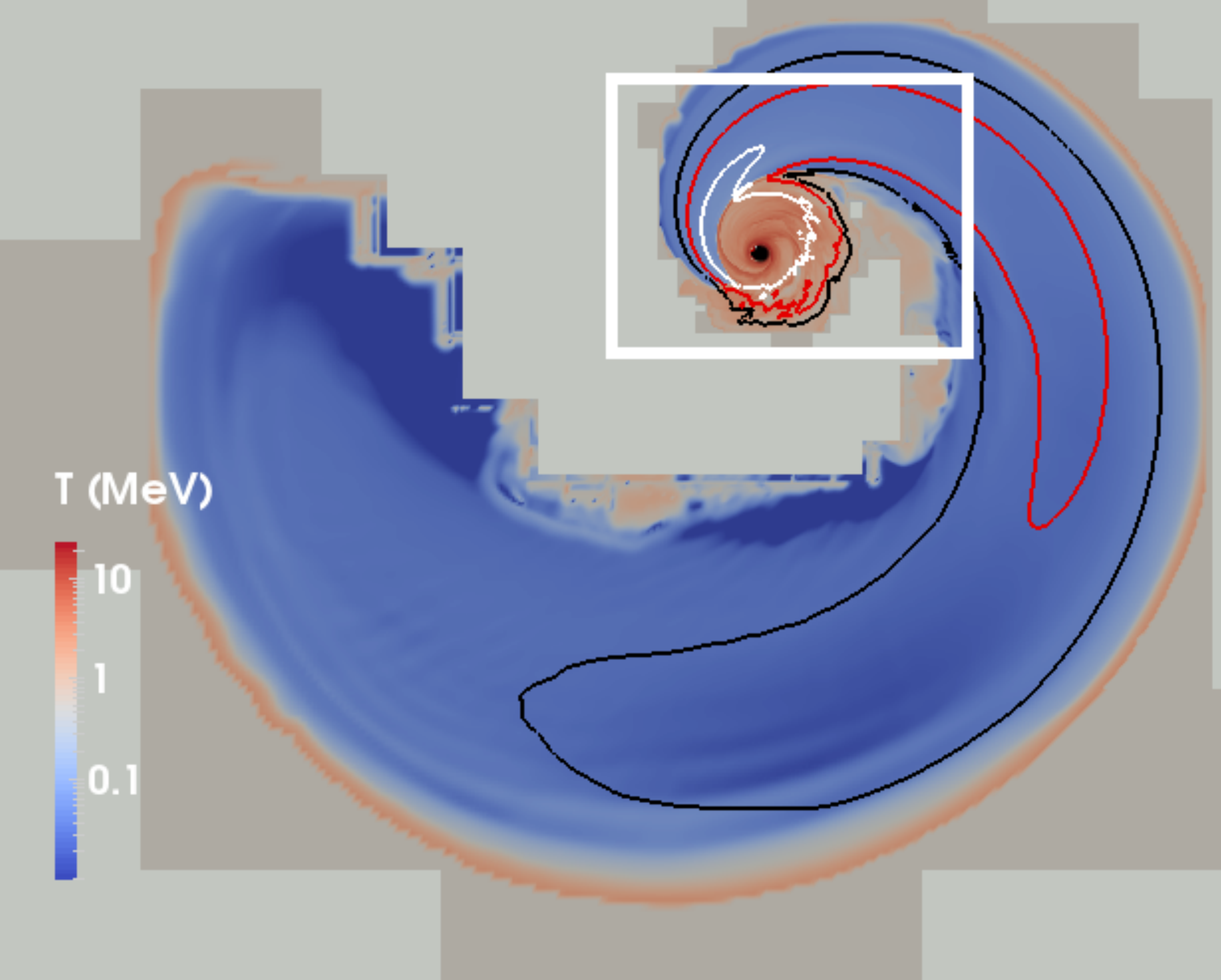}
 \end{tabular}
 \caption{Temperature profile and the location of the apparent horizon
 (black filled circle) at \SI{7}{\ms} after the onset of merger for a
 binary with $M_\mathrm{BH} = 7\,M_\odot$, $\chi = 0.9$, $M_\mathrm{NS} =
 1.2\,M_\odot$, and $R_\mathrm{NS} = \SI{13.5}{\km}$ ($Q \approx 5.8$,
 $\mathcal{C}=0.130$) modeled by the FSU2.1 equation of state
 \citep{ToddRutel_Piekarewicz2005,Shen_Horowitz_OConner2011}. The left
 panel is magnification of the white box in the right panel. The white,
 red, and black curves denote \num{e11}, \num{e10}, and
 \SI{e9}{\gram\per\cubic\cm}, respectively. Image reproduced with permission from
 \citet{Brege_DFDCHKOPS2018}, copyright by APS.} \label{fig:snap_temp}
\end{figure}

The outermost part of the tidal tail obtains energy sufficient to become
unbound from the remnant black hole. Figure \ref{fig:phasespace}
visualizes the process of dynamical mass ejection on the phase space of
the specific energy and the specific angular momentum for a system with
$M_\mathrm{BH} = 3\,M_\odot$, $\chi = 0$, $M_\mathrm{NS} = 1.35\,M_\odot$,
and $R_\mathrm{NS} = \SI{12.3}{\km}$ ($Q \approx 2.2$,
$\mathcal{C}=0.162$) modeled by a piecewise polytrope called H
\citep{Hayashi_KKKS2021}. Here, unboundedness is identified by the
criterion $-u_t > 1$, which is suitable for dynamical mass ejection as
far as the spacetime is approximated as stationary, because the shock
heating, and hence, the contribution of the internal energy, does not
play a role. If the internal energy contributes significantly, a
reasonable criterion may be defined based on $-hu_t$ taking the offset
associated with the composition into account \citep[see,
e.g.,][]{Fujibayashi_SWKKS2020}. First, the outer part acquires the
angular momentum and also the energy via the tidal torque (from the left
top to the right top panels). Next, it gains exclusively the energy via
work done by impulsive outward radial force, which is likely to be
associated with the infall of the major part of the neutron star to the
black hole (from the right top to the left bottom panels). If the energy
of a fluid element exceeds the gravitational binding energy, the element
escapes from the system as the dynamical ejecta.

Material behind the dynamical ejecta also acquires some energy but
remains bound to the remnant black hole, and thus it eventually falls
back onto the disk. The disk and fallback components may approximately
be distinguished by temperature higher and lower than
$0.1$--\SI{1}{\mega\eV}, respectively, as shown in
Fig.~\ref{fig:snap_temp} generated by \citet{Brege_DFDCHKOPS2018},
because the shock interaction sets in when the fallback material hits
the outer edge of the disk. Although the longterm fallback dynamics
cannot be fully tracked in current simulations of black hole--neutron
star binary coalescences, estimated fallback rates of the mass are found
to coincide with the well-known $t^{-5/3}$ law for tidal disruption
events \citep[see also \citealt{Rosswog2007} for early Newtonian
work]{Chawla_ABLLMN2010,Kyutoku_IOST2015,Brege_DFDCHKOPS2018}.

The left bottom panel of Fig.~\ref{fig:phasespace} implies that the
fallback material and the dynamical ejecta may be considered to be
launched from an approximately common periastron. Taking the $t^{-5/3}$
fallback behavior into account, the process depicted here might seem
similar to tidal disruption of stars by supermassive black holes in
slightly unbound, hyperbolic encounters
\citep{Rees1988,Phinney1989}. However, it should be remarked that
gravitational-wave-driven mergers of compact object binaries occur in a
strongly bound, quasicircular orbit. Therefore, it is not trivial
\textit{a priori} that even a finite amount of material could be ejected
by tidal disruption in black hole--neutron star binary coalescences.

The process of tidal disruption described in Fig.~\ref{fig:snap_td} is
qualitatively common for systems with a large neutron-star radius, a
small black-hole mass, and/or a high black-hole spin. However,
quantitative details depend on the parameters of the binary. The
orientation of the black-hole spin also introduces qualitative
differences in the merger dynamics and morphology of the remnant. In the
following, we review the dependence on these parameters.

\subsubsection{Dependence on the equation of state}
\label{sec:sim_mrg_eos}

As found from the analysis of Sect.~\ref{sec:intro_tidal} and
Sect.~\ref{sec:eq}, the merger process depends on the compactness of the
neutron star, which is determined by the equation of state. Systematic
studies performed employing a variety of piecewise polytropes clearly
show that neutron stars with smaller compactnesses are tidally disrupted
more easily
\citep{Kyutoku_Shibata_Taniguchi2010,Kyutoku_OST2011,Kyutoku_IOST2015}. This
tendency also holds for nuclear-theory-based tabulated equations of
state \citep{Kyutoku_KSST2018,Brege_DFDCHKOPS2018}.

Even if the compactness and the mass are identical, the density profiles
generally differ among neutron stars modeled by different equations of
state. If the density profile is more centrally condensed, the neutron
star is less subject to tidal disruption as discussed in
Sect.~\ref{sec:intro_tidal_orb}. This tendency is demonstrated by a
study employing two-piecewise polytropes with different adiabatic
indices for the core region
\citep{Kyutoku_Shibata_Taniguchi2010}. Specifically, if the adiabatic
index for the core region is smaller, the neutron star with a given
compactness is more centrally condensed and less subject to tidal
disruption.

\subsubsection{Dependence on the mass ratio} \label{sec:sim_mrg_Q}

As indicated by the analysis of Sect.~\ref{sec:intro_tidal} and
Sect.~\ref{sec:eq}, the possibility of tidal disruption increases as the
mass ratio decreases. For example, the amount of the mass remaining
outside the black hole is likely to be larger for the lower mass
ratio. This dependence is particularly important for nonspinning black
holes, because significant tidal disruption occurs only for low-mass
black holes such as those in the putative mass gap
\citep{Shibata_KYT2009,Shibata_KYT2009e}. Stated differently, for a
plausibly realistic mass ratio of $Q \gtrsim 4$, the neutron star can be
tidally disrupted only if the black hole has a high prograde spin, as we
discuss in Sect.~\ref{sec:sim_mrg_s1}.

After the discovery of binary neutron stars by gravitational waves,
very-low-mass black hole--neutron star coalescences acquire renewed
interest \citep[see Sect.~\ref{sec:intro_history_rsim} for definition of
``very low
mass'']{Foucart_etal2019,Foucart_DKNPS2019,Hayashi_KKKS2021,Most_PTR2021}. The
primary reason for this is that they could be potential mimickers of
binary-neutron-star coalescences, rendering astrophysical interpretation
of gravitational-wave sources ambiguous
\citep{Hinderer_etal2019,Kyutoku_FHKKST2020}. Distinguishing
very-low-mass black hole--neutron star binaries and binary neutron stars
would be invaluable for gaining knowledge about the maximum mass of
neutron stars, the mass gap between black holes and neutron stars
\citep[see, e.g.,][]{Kreidberg_BFK2012}, and the formation mechanism of
these compact objects, i.e., stellar core collapse and supernova
explosions.

Recent numerical simulations have shown that susceptibility to tidal
disruption is not reflected monotonically in the remnant material for
very-low-mass-ratio systems
\citep{Foucart_DKNPS2019,Hayashi_KKKS2021,Most_PTR2021}. Rather, the
mass of the remnant disk saturates to a value of $\sim \order{0.1}
\,M_\odot$ for the case in which the black hole is nonspinning (see also
\citealt{Brege_DFDCHKOPS2018} and Appendix of
\citealt{Hayashi_KKKS2021}). Quantitatively, the saturated value of the
disk mass depends on the equation of state. Moreover, the mass of the
dynamical ejecta, which increases with decreasing $Q$ down to a
moderately large values of $Q = Q_\mathrm{peak} \sim 3$, begins to
decrease as the mass ratio decreases for $Q < Q_\mathrm{peak}$
\citep{Hayashi_KKKS2021}. The precise value of $Q_\mathrm{peak}$ again
depends on the equation of state. We will discuss quantitative
dependence of the remnant disk and the dynamical ejecta on the mass
ratio later in this section.

We note that, while the physical reason of the behavior at very low-mass
ratios described above is not fully understood yet, it may not be
unexpected taking the fact that an extremely low-mass black hole with
$M_\mathrm{BH} \ll M_\mathrm{NS} (Q \ll 1)$ cannot make the neutron star
unbound because of the tiny contribution of such a minute black hole to
the dynamics of the entire system. Related simulations have been
performed in the context of consumption of a neutron star by an
endoparasitic black hole at the center
\citep{East_Lehner2019,Richards_Baumgarte_Shapiro2021}.

\subsubsection{Dependence on the black-hole spin (I) aligned spin}
\label{sec:sim_mrg_s1}

\begin{figure}[htbp]
 \centering
 \begin{tabular}{cc}
  \includegraphics[width=.47\linewidth,clip]{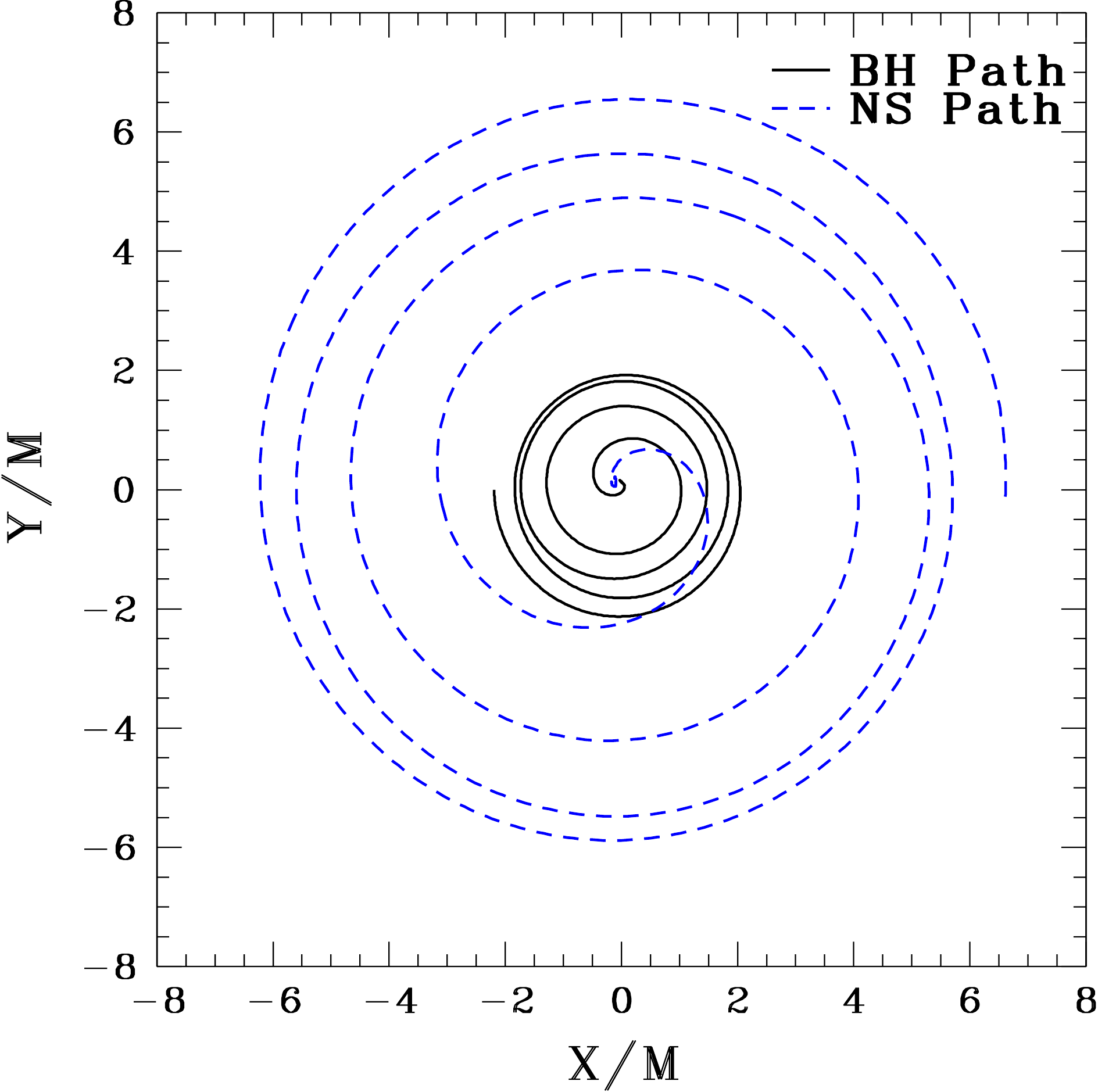} &
  \includegraphics[width=.47\linewidth,clip]{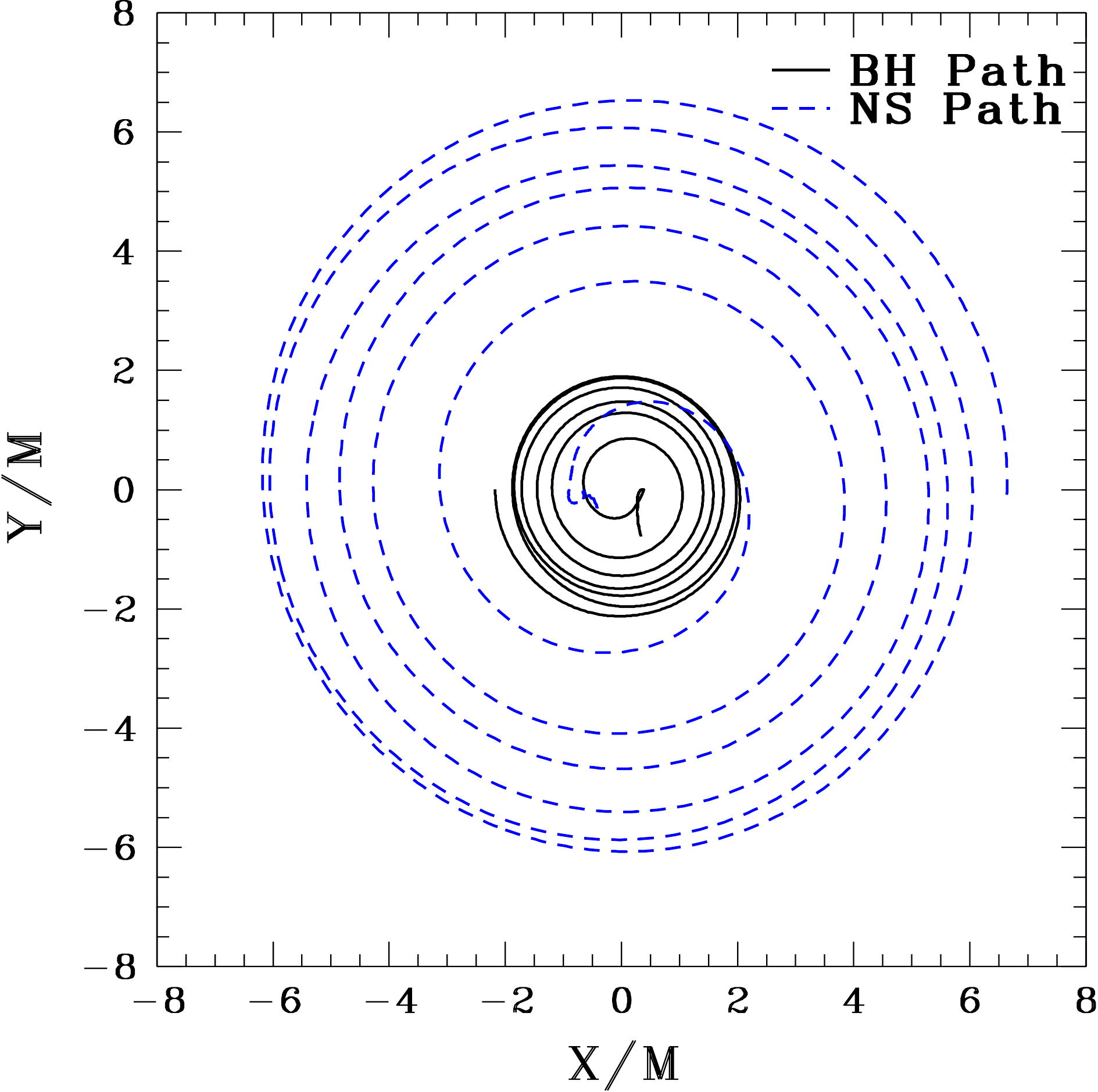}
 \end{tabular}
 \caption{Coordinate trajectory of the black hole (black solid curve)
 and the neutron star (blue dashed curve) on the orbital plane for
 binaries with $Q=3$, $\mathcal{C}=0.145$ modeled by a $\Gamma = 2$
 polytrope, and $\chi = 0$ (left) and $\chi = 0.75$ (right). Image reproduced with permission from \citet{Etienne_LSB2009}, copyright by APS.} \label{fig:orb_spin}
\end{figure}

The spin of the black hole quantitatively modifies the orbital evolution
in the late inspiral phase and the merger dynamics. First, we focus on
the cases in which the black-hole spin is (anti-)aligned with respect to
the orbital angular momentum of the binary. Figure \ref{fig:orb_spin}
generated by \citet{Etienne_LSB2009} shows the trajectories of the black
hole and the neutron star for systems characterized by $Q=3$,
$\mathcal{C}=0.145$ modeled by a $\Gamma = 2$ polytrope, and two
different values of the spin parameter $\chi = 0$ (left) and $\chi =
0.75$ (right). Because a polytropic equation of state is adopted, the
mass of the binary components can be scaled arbitrarily as discussed in
Sect.~\ref{sec:eq_res}. Both systems have the same values of initial
orbital angular velocity normalized by the total mass, $G m_0 \Omega /
c^3 \approx 0.033$. The nonspinning ($\chi = 0$) and spinning ($\chi =
0.75$) systems merge after $\sim 4$ and $6$ orbits, respectively. The
difference in the number of orbits is ascribed mainly to the spin-orbit
interaction. Specifically, this interaction serves as repulsive force
for a prograde spin of the black hole, $\chi > 0$, and vice versa (see,
e.g., \citealt{Kidder_Will_Wiseman1993,Kidder1995} for two-body
equations of motion in the post-Newtonian approximation). The repulsion
for $\chi > 0$ counteracts the gravitational pull between the binary
components and reduces the orbital angular velocity to maintain a
circular orbit for a given orbital separation (or a given
circumferential radius of the orbit). Because the gravitational-wave
luminosity is as sensitive to the orbital angular velocity as $\propto
\Omega^{10/3}$, the approaching velocity associated with the radiation
reaction is also decreased. This effect increases the lifetime of the
binary. In addition, the spin-orbit repulsion decreases the radius of
the innermost stable orbit and strengthens gravitational binding there
\citep{Bardeen_Press_Teukolsky1972}. This further helps to increase the
lifetime of a progradely-spinning black hole--neutron star binary,
because it needs to emit a larger amount of energy to reach the
innermost stable circular orbit than that for a nonspinning black
hole. These effects increase the number of inspiral orbits.

The higher the prograde spin of the black hole, the neutron star is
disrupted more easily, and thus, the disk formation and the mass
ejection are more pronounced. This is clearly shown by comparing
Figs.~\ref{fig:snap_notd} and \ref{fig:snap_td}, between which the only
difference is the spin of the black hole. Quantitatively, while the mass
of the disk for $\chi = 0$ is less than $\num{e-3} \,M_\odot$, it
increases to $0.19 \,M_\odot$ for $\chi = 0.75$ in these examples. The
mass of the dynamical ejecta also increases as the black-hole spin
increases, specifically from $\ll \num{e-3} \,M_\odot$ to $0.01\,M_\odot$ in
these examples \citep{Kyutoku_IOST2015}. These increases are ascribed
primarily to the small radius of the innermost stable circular orbit
with the prograde spin \citep{Bardeen_Press_Teukolsky1972}. As an
extreme, it has been shown that about more than a half of the
neutron-star material remains outside the remnant black hole right after
the onset of merger for a binary with $Q=3$, $\mathcal{C}=0.144$ modeled
by a $\Gamma = 2$ polytrope, and $\chi = 0.97$, which is the largest
value of the spin parameter simulated for black hole--neutron star
binaries to date \citep{Lovelace_DFKPSS2013}. In light of the
astrophysically plausible range of $Q$ and $\mathcal{C}$ (e.g., $Q
\gtrsim 4$ and $\mathcal{C} \gtrsim 0.16$), it is remarkable that the
prograde spin enables tidal disruption to occur for a binary which
results in the plunge if the black hole is nonspinning. By contrast, if
the spin of the black hole is retrograde, the neutron star is swallowed
by the black hole without tidal disruption even if $Q<3$ for a wide
range of equations of state. We defer further quantitative discussions
to Sect.~\ref{sec:sim_rem_disk}.

If tidal disruption occurs in a binary with a spinning black hole and a
realistic mass ratio of $Q \gtrsim 4$, the elongated neutron star can be
swallowed by the black hole through a narrow region of its large surface
\citep{Kyutoku_OST2011}. This feature is advantageous for exciting
nonaxisymmetric, fundamental quasinormal modes of the remnant black hole
as we discuss in Sect.~\ref{sec:sim_gw}. This does not occur for
nonspinning black holes, because the tidal disruption is possible only
for a binary with a low mass ratio of $Q \lesssim (3\mathcal{C})^{-3/2}$
[cf., Eq.~\eqref{eq:iscovsms} and Fig.~\ref{fig:cqdiag}] and thus for a
black hole with a small surface. For such a case, the tidally-disrupted
material is swallowed through a wide region of the black-hole surface,
and the quasinormal-mode excitation is suppressed. These differences are
reflected in both gravitational waveforms and spectra as predicted by a
black-hole perturbation study
\citep{Saijo_Nakamura2000,Saijo_Nakamura2001}.

\begin{figure}[htbp]
 \centering
 \begin{tabular}{ccc}
  \includegraphics[width=.30\linewidth,clip]{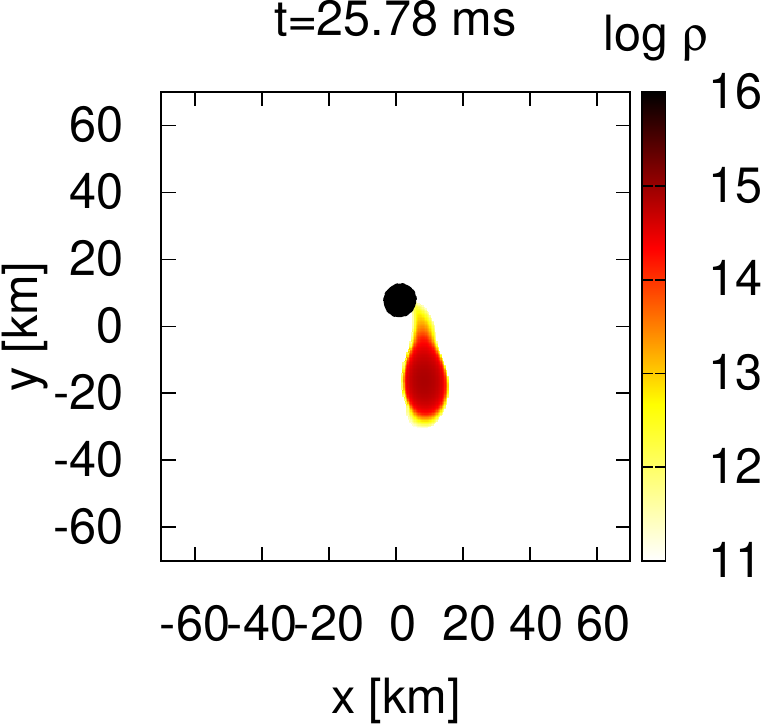} &
  \includegraphics[width=.30\linewidth,clip]{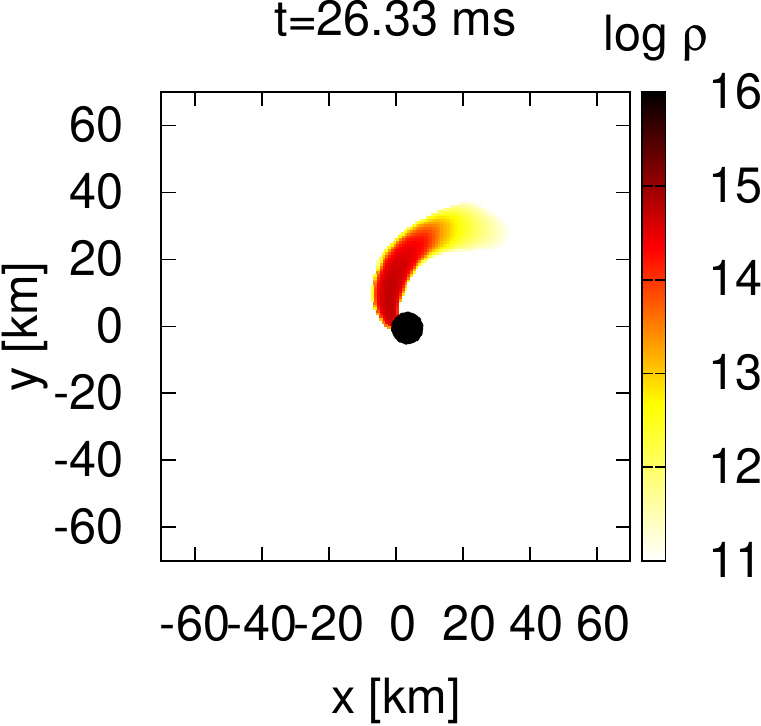} &
  \includegraphics[width=.30\linewidth,clip]{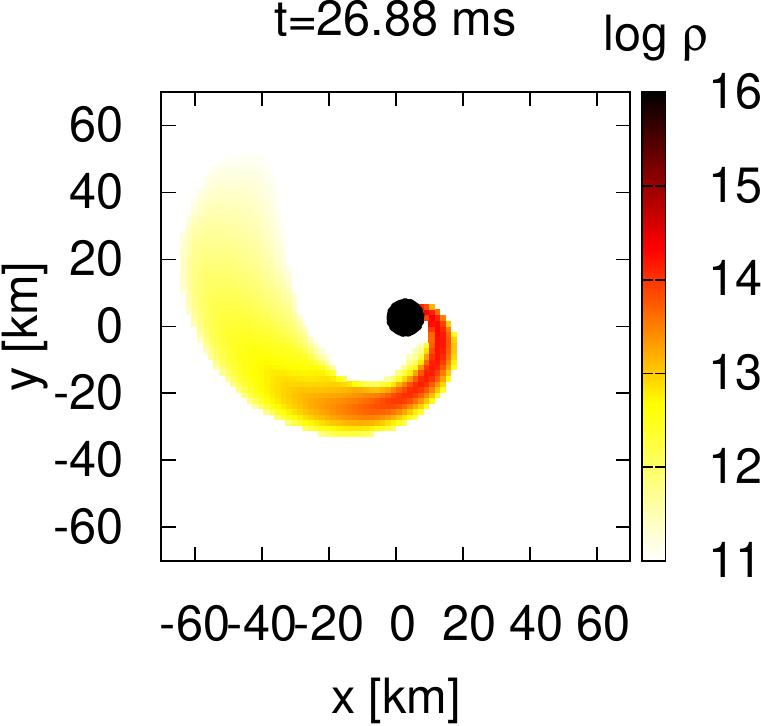} \\
  \includegraphics[width=.30\linewidth,clip]{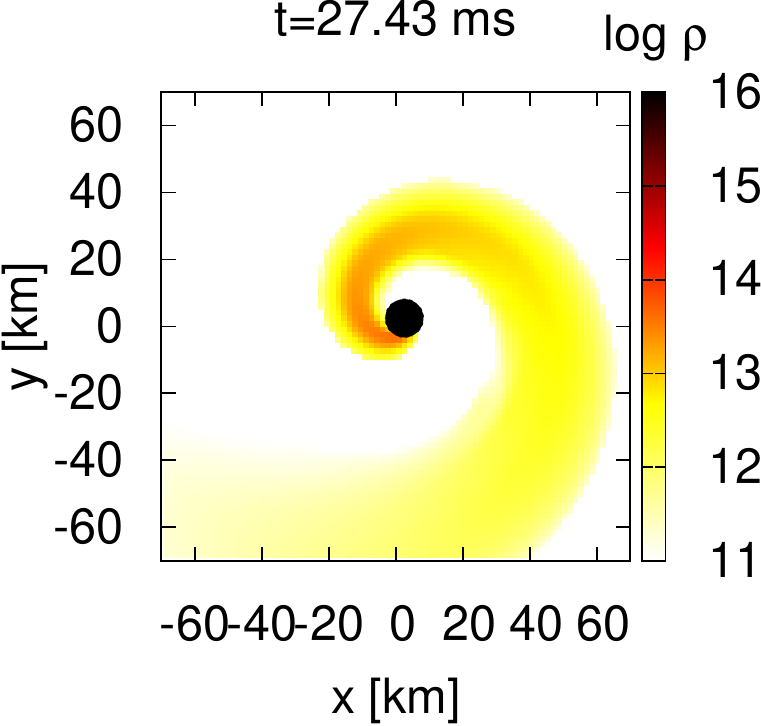} &
  \includegraphics[width=.30\linewidth,clip]{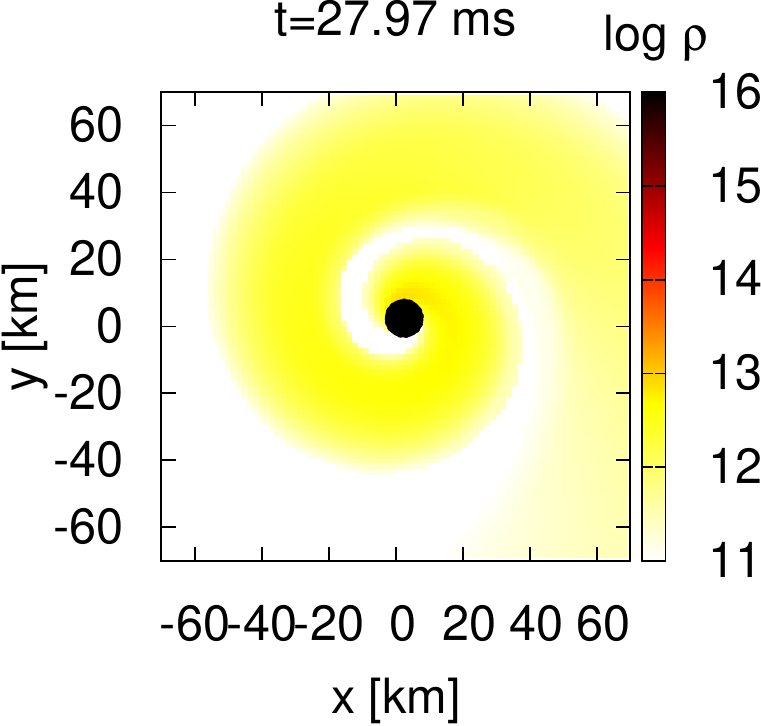} &
  \includegraphics[width=.30\linewidth,clip]{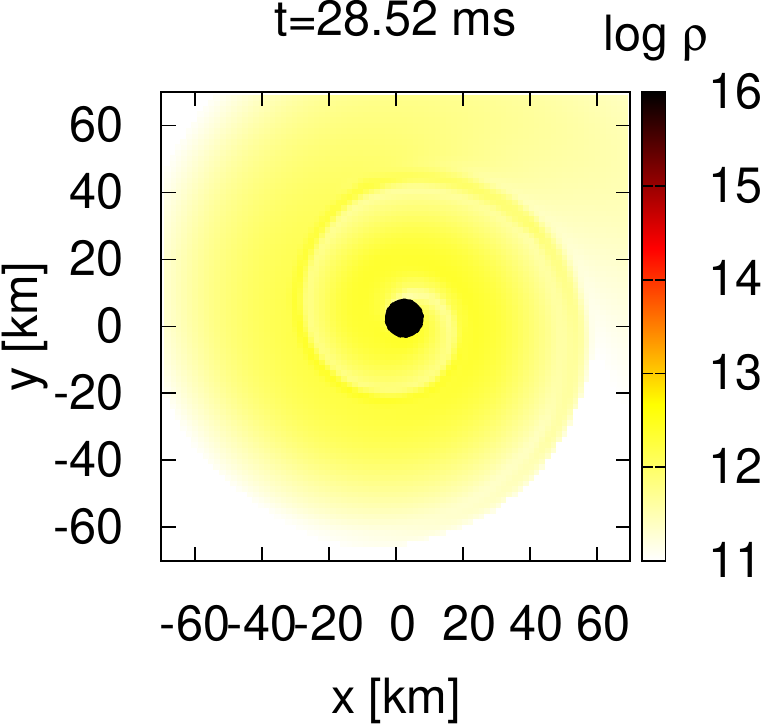}
 \end{tabular}
 \caption{Same as Fig.~\ref{fig:snap_notd} but for a binary with
 $M_\mathrm{BH} = 4.05\,M_\odot$, $\chi = 0.5$, $M_\mathrm{NS} =
 1.35\,M_\odot$, and $R_\mathrm{NS} = \SI{11.6}{\km}$ ($Q=3$,
 $\mathcal{C}=0.172$) modeled by a piecewise polytrope called HB
 \citep{Read_MSUCF2009}. Image reproduced with permission from
 \citet{Kyutoku_OST2011}, copyright by APS.} \label{fig:snap_asmtd}
\end{figure}

Figure \ref{fig:snap_asmtd} illustrates the case described above, i.e.,
the tidally-elongated neutron star is swallowed through a narrow region
of the black-hole surface \citep{Kyutoku_OST2011}. This system is
characterized by $M_\mathrm{BH} = 4.05\,M_\odot$, $M_\mathrm{NS} =
1.35\,M_\odot$, $R_\mathrm{NS} = \SI{11.6}{\km}$ ($Q=3$, $\mathcal{C} =
0.172$) modeled by a piecewise polytrope called HB, and a moderately
high and prograde spin of $\chi = 0.5$. Tidal disruption occurs at an
orbit outside but close to the innermost stable circular orbit. The
dense part of disrupted material does not have a sufficient time for
winding around the black hole before the infall. Thus, it falls into the
black hole in a significantly nonaxisymmetric manner and excites
quasinormal-mode oscillations. This behavior is frequently found for a
binary with a high mass ratio and a high black-hole spin. Conversely,
for a retrograde spin, tidal disruption becomes insignificant even for a
small value of $Q=2$--$3$ \citep{Kyutoku_OST2011}. An alternative
interpretation of this finding is that the orientation of the black-hole
spin plays an important role. We will discuss this viewpoint for a
general inclination angle in Sect.~\ref{sec:sim_mrg_s2}.

\begin{figure}[htbp]
 \centering \includegraphics[width=0.95\linewidth,clip]{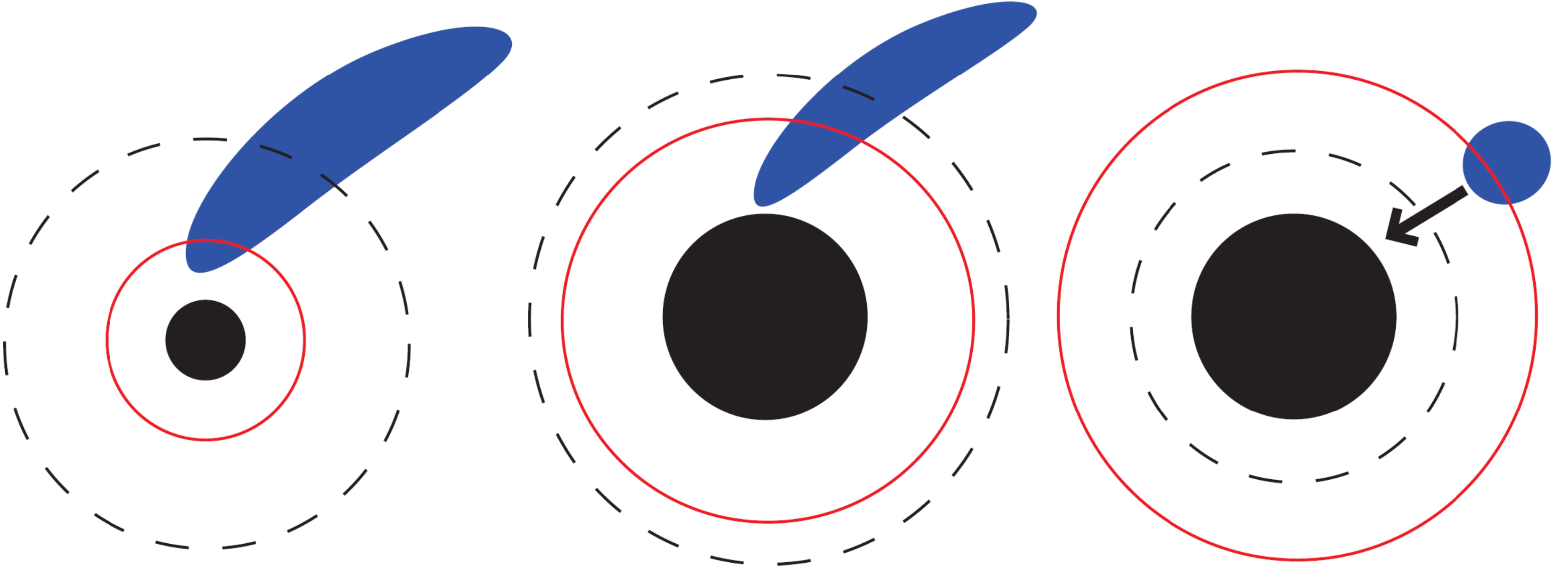}
 \caption{Schematic picture for three types of merger processes. The
 filled circle, the solid red circle, and the black dashed circle denote
 the black hole, the innermost stable circular orbit, and the radius at
 which tidal disruption occurs, respectively. The deformed ellipsoid
 denotes the neutron star. The left, middle, and right panels correspond
 to the types 1, 2, and 3, described in the body text,
 respectively. Image adapted from \citet{Kyutoku_OST2011}, copyright by APS.} \label{fig:schematic}
\end{figure}

To summarize, numerical-relativity simulations have revealed that the
merger process may be classified into three types according to the mass
and the spin of black holes for a given equation of state (see also
\citealt{Pannarale_BKS2013,Pannarale_BKLS2015_2} for relevant
classifications):
\begin{enumerate}
 \item The neutron star is tidally disrupted at an orbit far from the
       innermost stable circular orbit. This occurs for the cases in
       which the black-hole mass is small and/or the black-hole spin is
       prograde and sufficiently high.
 \item The neutron star is tidally disrupted at an orbit close to the
       innermost stable circular orbit. This occurs for the cases in
       which the black-hole mass is not small and the black-hole spin is
       prograde and high.
 \item The neutron star is not tidally disrupted. This occurs for the
       cases in which the black-hole mass is not small and/or the
       black-hole spin is retrograde or prograde but not high. An
       approximate criterion for tidal disruption is found in
       Eq.~\eqref{eq:iscovsms}.
\end{enumerate}
These three types are displayed schematically in
Fig.~\ref{fig:schematic}. The differences of merger processes for these
types, particularly that between 1 and 2, are imprinted in gravitational
waveforms and spectra described in Sect.~\ref{sec:sim_gw}.

\subsubsection{Dependence on the black-hole spin (II) inclined spin}
\label{sec:sim_mrg_s2}

\begin{figure}[htbp]
 \centering
 \begin{tabular}{cc}
  \includegraphics[width=.47\linewidth,clip]{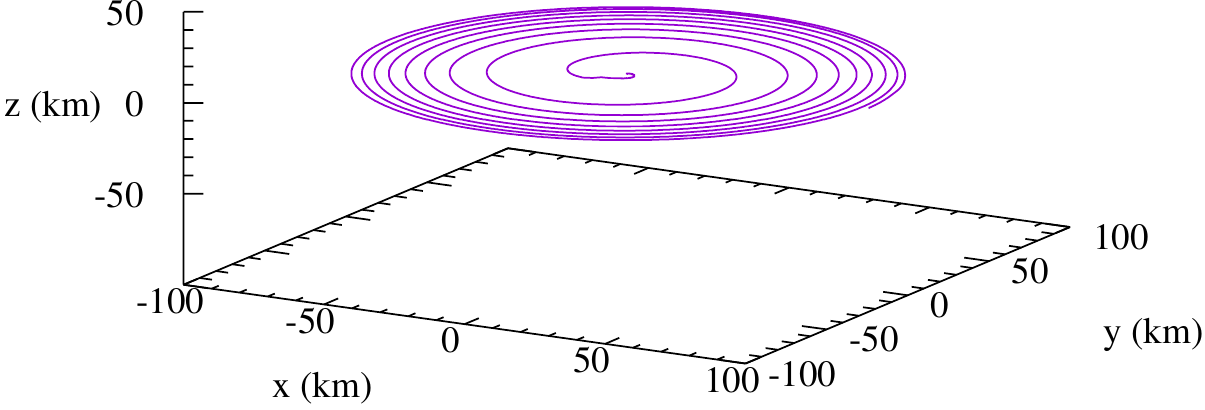} &
  \includegraphics[width=.47\linewidth,clip]{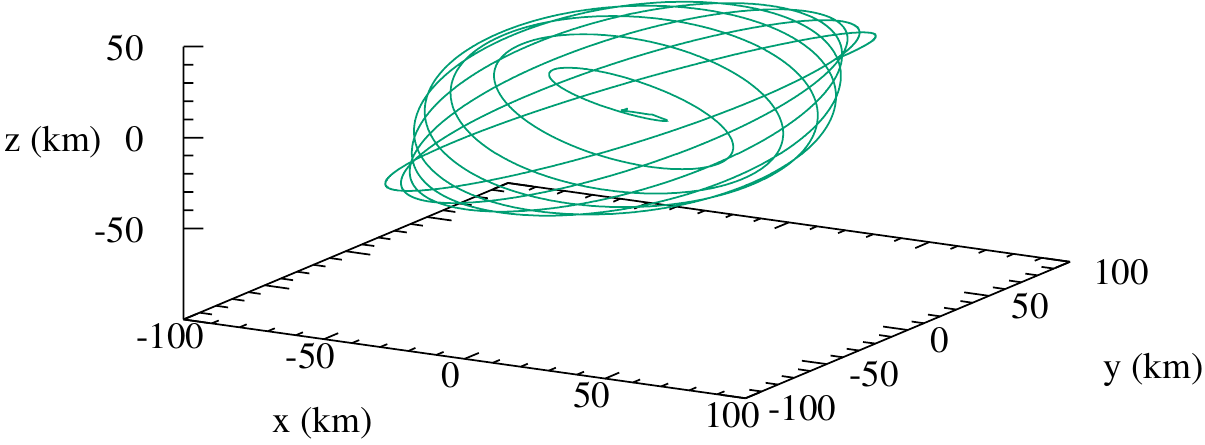}
 \end{tabular}
 \caption{Evolution of the coordinate separation between the black hole
 and the neutron star for binaries with $M_\mathrm{BH} = 4.05\,M_\odot$,
 $M_\mathrm{NS} = 1.35\,M_\odot$, and $R_\mathrm{NS} = \SI{11.1}{\km}$
 ($Q=3$, $\mathcal{C}=0.180$) modeled by a piecewise-polytropic
 approximation of the APR4 equation of state
 \citep{Akmal_Pandharipande_Ravehnall1998}. The spin of the black hole
 is zero for the left panel and $\chi = 0.75$ with the inclination angle
 $\iota \approx \ang{90}$ for the right panel. The $z$-axis is taken to
 be the direction of the total angular momentum at the initial
 instant. This figure is generated from data of
 \citet{Kyutoku_KKST2021}.} \label{fig:orb_incl}
\end{figure}

The inclination angle has a qualitative impact on the inspiral and
merger dynamics
\citep{Foucart_DKT2011,Foucart_DDKMOPSST2013,Kawaguchi_KNOST2015,Foucart_DBDKHKPS2017,Foucart_etal2021}. Figure
\ref{fig:orb_incl} compares typical orbital evolution of black
hole--neutron star binaries for which the spin of the black hole is
absent or (anti-)aligned with respect to the orbital angular momentum of
the binary (left: $\chi = 0$) and inclined (right: $\chi = 0.75$ and
$\iota \approx \ang{90}$). The reflection symmetry about the orbital
plane is lost in the presence of spin misalignment. Because the orbital
angular velocity vector is inclined with respect to the total angular
momentum of the system, the vector normal to the orbital plane precesses
approximately around the total angular momentum during the inspiral
phase \citep{Apostolatos_CST1994,Kidder1995,Racine2008}.

The spin misalignment also reduces the degree of tidal disruption, as
well as the masses of the remnant disk and the dynamical ejecta, for the
same magnitude of the spin. This is because the spin-orbit coupling is
proportional to, in the post-Newtonian terminology, the inner product
$\vb{S} \vdot \vb{L}$ of the spin angular momentum $\vb{S}$ and the
orbital angular momentum $\vb{L}$. The effect of the black-hole spin on
the radius of the innermost stable circular orbit, or an innermost
stable spherical orbit
\citep{Hughes2001,Buonanno_Chen_Damour2006,Fragile_BAS2007,Stone_Loeb_Berger2013},
is also determined primarily by this inner product. Thus, the spin-orbit
repulsion for a given magnitude of the spin becomes weak as the
inclination angle increases. For systems with the black-hole spin being
confined in the orbital plane, $\vb{S} \vdot \vb{L} \approx 0$, the spin
of the black hole is likely to play only a minor role in tidal
disruption of the neutron star irrespective of its magnitude, while
systematic surveys have not yet been performed.

\begin{figure}[htbp]
 \centering
 \begin{tabular}{cc}
  \includegraphics[width=.47\linewidth,clip]{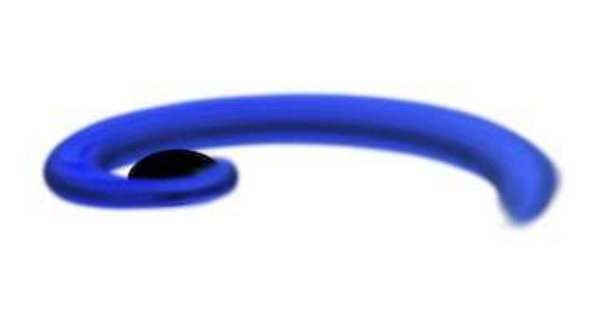} &
  \includegraphics[width=.47\linewidth,clip]{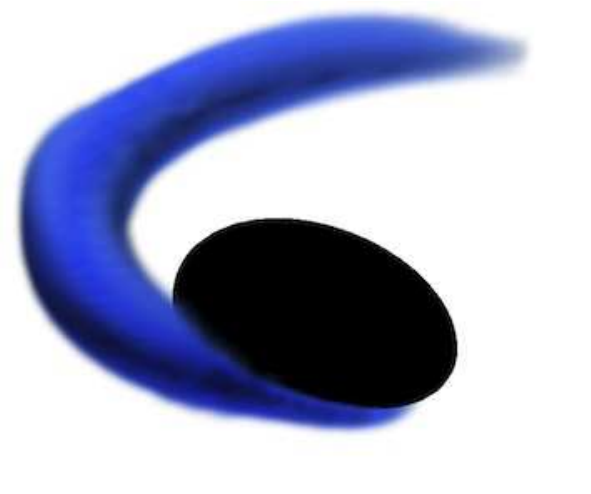} \\
  \includegraphics[width=.47\linewidth,clip]{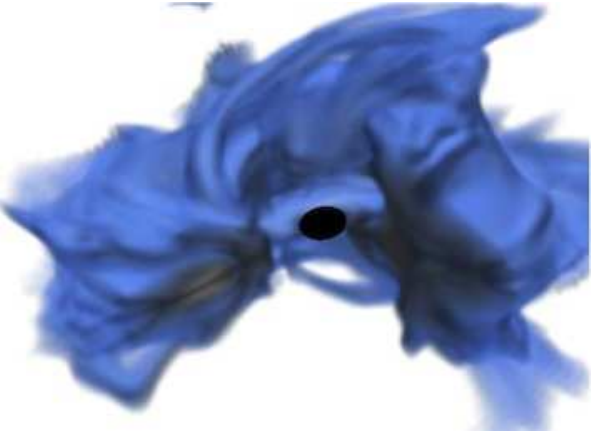} &
  \includegraphics[width=.47\linewidth,clip]{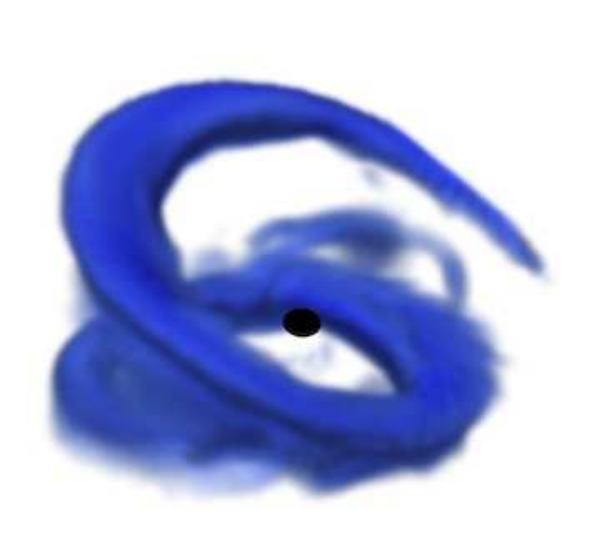}
 \end{tabular}
 \caption{Three-dimensional plot of the rest-mass density and the
 location of the apparent horizon (denoted by the filled black region)
 for binaries with $Q=7$, $\chi = 0.9$, and $\mathcal{C}=0.144$
 ($R_\mathrm{NS} = \SI{14.4}{\km}$ if we suppose a $1.4\,M_\odot$ neutron
 star) modeled by a $\Gamma = 2$ polytrope. The angles between the spin
 angular momentum of the black hole and the orbital angular momentum of
 the binary are \ang{0} and \ang{40} for the left and right panels,
 respectively. The top and bottom panels show snapshots at different
 times. Image reproduced with permission from \citet{Foucart_DDKMOPSST2013}, copyright by APS.}
 \label{fig:snap_incl}
\end{figure}

Orbital precession caused by the spin misalignment introduces
qualitative differences also in the morphology of the disrupted material
\citep{Foucart_DDKMOPSST2013,Kawaguchi_KNOST2015}. Because the
misalignment breaks the reflection symmetry, the remnant also exhibits a
reflection-asymmetric structure. Figure \ref{fig:snap_incl} generated by
\citet{Foucart_DDKMOPSST2013} shows three-dimensional plots of the
rest-mass density and the region inside the apparent horizon. The system
on the left column is characterized by $Q=7$, $\mathcal{C}=0.144$
modeled by a $\Gamma = 2$ polytrope, and $\chi = 0.9$ aligned with the
orbital angular momentum of the binary. The merger dynamics and the
morphology of the remnant are essentially the same as those described in
Fig.~\ref{fig:snap_td}. The system on the right column of
Fig.~\ref{fig:snap_incl} has the same parameters as those on the left
column, except that the black-hole spin has an inclination angle of
$\iota = \ang{40}$ with respect to the orbital angular momentum. The
tidal tail of this system inherits precessing motion of the inspiral
phase. Accordingly, it does not form a circularized disk immediately
after single orbital revolution. Instead, the tidal tail eventually
collides with itself from various directions and forms a thick
torus. The inclination angle between the angular momentum of the remnant
torus and the spin angular momentum of the remnant black hole becomes
smaller than the inclination angle during the inspiral phase, $\iota$,
because a substantial fraction of the orbital angular momentum is
brought into the black hole by the infalling neutron-star material
\citep{Foucart_DKT2011,Kawaguchi_KNOST2015}. This effect is more
significant for lower mass-ratio systems, for which the initial spin
angular momentum of the black hole accounts for a smaller fraction of
the total angular momentum.

In the long run, a tilted disk will evolve in a manner different from an
aligned one via the Lense-Thirring precession
\citep{Bardeen_Petterson1975,Papaloizou_Pringle1983},
magnetically-induced turbulent viscosity \citep{Fragile_BAS2007} and/or
magnetic coupling with the remnant black hole
\citep{McKinney_Tchekhovskoy_Blandford2013}. Longterm evolution of a
remnant torus of precessing black hole--neutron star binaries with
realistic microphysics is a subject for future studies in numerical
relativity (but see \citealt{Mewes_FGMS2016} for pure hydrodynamics).

\subsection{Remnant} \label{sec:sim_rem}

In this Sect.~\ref{sec:sim_rem}, we present quantitative details of the
remnant black hole, disk, fallback material, and dynamical ejecta
derived by merger simulations. Unless explicitly stated, we discuss the
cases in which the spin of the black hole is aligned with respect to the
orbital angular momentum and the system possesses reflection symmetry
about the orbital plane. We will make it explicit when we consider the
effects of the spin misalignment. The longterm evolution of the remnant
disk and associated outflows are discussed separately in
Sect.~\ref{sec:sim_pm}.

\subsubsection{Black hole} \label{sec:sim_rem_bh}

The mass and the spin of the black hole change during merger, because it
swallows the material of the neutron star. The mass of the remnant black
hole $M_\mathrm{BH,f}$ is approximately estimated by
\citep{Shibata_Uryu2007}
\begin{equation}
 M_\mathrm{BH,f} \approx M_\mathrm{BH} + M_\mathrm{NS} -
  M_{r>{r_\mathrm{AH}}} - E_\mathrm{GW} ,
\end{equation}
where $M_{r>{r_\mathrm{AH}}}$ denotes the mass of the material remaining
outside the black hole, which is composed of the remnant disk, the
fallback material, and the dynamical ejecta, and $E_\mathrm{GW}$ denotes
the energy carried away by gravitational radiation (see
\citealt{Duez_FKPST2008,Shibata_KYT2009} for other methods of
estimation). Because a large fraction of the neutron-star material falls
into the black hole for most cases and also $E_\mathrm{GW}$ is much
smaller than the total rest-mass energy of the system, $M_\mathrm{BH,f}$
is larger than $0.9 m_0$ except for nearly-extremal spins
\citep{Lovelace_DFKPSS2013}. Because both $M_{r>{r_\mathrm{AH}}}$ and
$E_\mathrm{GW}$ decrease as the spin of the black hole decreases,
$M_\mathrm{BH,f}$ becomes close to $m_0$ for small values of $\chi$
and/or large inclination angles. Specifically, the difference between
$M_\mathrm{BH,f}$ and $m_0$ is only a few percent for anti-aligned,
retrograde spins of $\chi = -0.5$ \citep[stated differently, $\chi =
0.5$ and $\iota = \ang{180}$;][]{Kyutoku_OST2011}.

\begin{figure}[htbp]
 \centering \includegraphics[width=0.7\linewidth,clip]{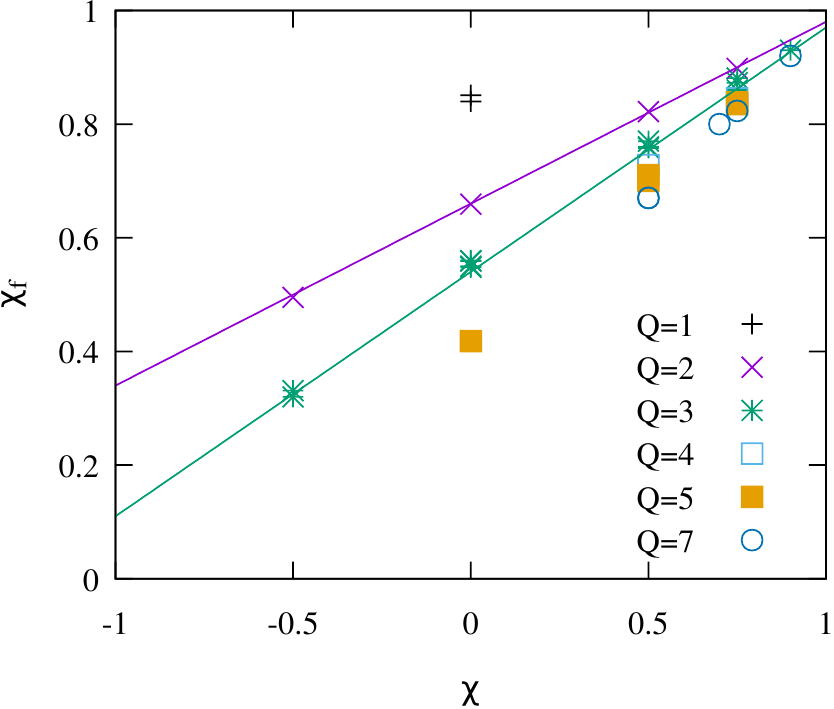}
 \caption{Spin parameter of the remnant black hole, $\chi_\mathrm{f}$,
 for a range of binary parameters as a function of the initial spin
 parameter, $\chi$. The solid lines denote the fitting for the results
 of $Q=2$ and $3$ \citep{Kyutoku_OST2011}. This figure is generated from
 data of
 \citet{Etienne_LSB2009,Kyutoku_Shibata_Taniguchi2010,Foucart_DKT2011,Kyutoku_OST2011,Foucart_DKSST2012,Kyutoku_IOST2015}. If
 the neutron-star mass and/or equation of state are varied in a single
 paper, the results are averaged.} \label{fig:remspin}
\end{figure}

The spin parameter of the remnant black hole $\chi_\mathrm{f}$ is
determined primarily by the mass ratio of the binary, $Q$, and the
initial spin parameter of the black hole, $\chi$
\citep{Kyutoku_OST2011}. Figure \ref{fig:remspin} shows the spin
parameter of the remnant black hole as a function of the initial spin
parameter $\chi$ for various mass ratios obtained by pure hydrodynamics
simulations performed by several groups with various numerical
implementations. Some particular combinations of $Q$ and $\chi$ are
simulated by several groups independently, and results are
overplotted. Taking inherent variation associated with the neutron-star
mass, the neutron-star equation of state, and the methods for evaluating
$\chi_\mathrm{f}$ \citep{Duez_FKPST2008,Shibata_KYT2009} into account,
all the results are consistent among independent groups. This figure
shows that the dependence on the initial spin parameter is more
pronounced for higher mass ratios. The reason for this is that, as we
describe below, the spin angular momentum becomes dominant and the
orbital angular momentum gives a minor contribution in such
systems. Meanwhile, the remnant spin parameter is found to depend only
weakly on the equation of state (not shown in this figure), and this
fact indicates that the material remaining outside the remnant black
hole takes only a minor fraction of the mass and the angular momentum
for the systems considered. It is also found that the remnant spin
parameter is larger than that for binary black holes with the same
values of $Q$ and $\chi$ for the case in which tidal disruption
occurs. For example, the coalescence of equal-mass, nonspinning binary
black holes is known to form a Kerr black hole with $\chi_\mathrm{f}
\approx 0.686$ \citep{Scheel_BCKMP2009}, which is significantly smaller
than $\chi_\mathrm{f} \approx 0.84$ for black hole--neutron star
binaries \citep{Etienne_LSB2009,Foucart_DKNPS2019}. This difference
stems from the fact that black hole--neutron star binaries which result
in tidal disruption do not experience orbits as close as those in binary
black holes. Accordingly, black hole--neutron star binaries do not emit
gravitational waves as strongly as binary black holes do.

Qualitative dependence of $\chi_\mathrm{f}$ on $Q$ and $\chi$ can be
understood by the following analysis. The total angular momentum of two
point particles in a circular orbit with the orbital angular velocity
$\Omega$ is given in Newtonian gravity by
\begin{equation}
 J_\mathrm{orb} = \frac{G^{2/3} M_\mathrm{BH} M_\mathrm{NS}}{(\Omega
  m_0)^{1/3}} .
\end{equation}
The spin parameter of the system, which is also denoted by
$\chi_\mathrm{f}$ here, may be given approximately by
\begin{align}
 \chi_\mathrm{f} & = \frac{c J_\mathrm{orb} / G + \chi
 M_\mathrm{BH}^2}{m_0^2} \notag \\
 & = \frac{\pqty{G m_0 \Omega / c^3}^{-1/3} Q + \chi Q^2}{(Q+1)^2} ,
\end{align}
where we assumed that the spin of the black hole is aligned with the
orbital angular momentum. Because the orbital angular velocity at the
onset of merger or at tidal disruption is given by $G m_0 \Omega / c^3
\sim 0.05$--$0.1$ for a wide range of binary parameters, $( G m_0 \Omega
/ c^3 )^{-1/3}$ takes a narrow range of $2.2$--$2.7$. Thus, as far as we
may neglect the mass and the angular momentum of the remnant material
and the energy carried away by gravitational radiation,
$\chi_\mathrm{f}$ gives the spin parameter of the remnant black
hole. This expression depends primarily on the mass ratio and the
initial spin parameter of the black hole, and furthermore, explains the
pronounced dependence on the initial spin parameter for a high
mass-ratio system.

Numerical simulations suggest that the remnant black hole is not
overspun by the infall of the neutron star, respecting the cosmic
censorship conjecture \citep{Penrose1969,Penrose2002}. Rather, fitting
formulae derived in \citet{Kyutoku_OST2011} predict that the spin
parameter should decrease during merger for nearly-extremal spins as
shown at the right edge of Fig.~\ref{fig:remspin}. Simulations of a
system with $Q=3$ and $\chi = 0.97$ illustrate that the spin parameter
indeed decreases below $0.97$ \citep{Lovelace_DFKPSS2013}, consistently
with the prediction of the fitting formulae. Detailed phenomenological
models of both the mass and the spin of the remnant black hole are
provided by \citet{Pannarale2013,Pannarale2014}.

The magnitude of the spin parameter of the remnant black hole decreases
as the inclination angle increases for a given value of the magnitude of
the spin parameter of the initial black hole
\citep{Foucart_DKT2011,Kawaguchi_KNOST2015,Foucart_DBDKHKPS2017}. This
is because the magnitude of the spin angular momentum, which is a
vectorial quantity, does not increase as sizably as the mass of the
remnant black hole in the presence of the inclination. The magnitude of
the spin parameter can even decrease from the initial value for a large
inclination angle \citep{Kawaguchi_KNOST2015,Foucart_DBDKHKPS2017} in
the same manner as the cases with anti-aligned spins shown in
Fig.~\ref{fig:remspin}. The direction of the spin of the remnant black
hole is approximately aligned with the total angular momentum of the
system right before merger, as most of the orbital angular momentum is
swallowed by the black hole.

\subsubsection{Accretion disk, or material remaining outside the black
   hole} \label{sec:sim_rem_disk}

We separate discussions about the remnant disk into two parts. Here in
Sect.~\ref{sec:sim_rem_disk}, we describe dependence of the disk mass,
or the mass of the remnant (see below), on binary
parameters. Thermodynamic variables such as the rest-mass density of the
remnant disk will be discussed later in Sect.~\ref{sec:sim_rem_disk2}.

Regarding the disk mass, early systematic surveys have rather focused on
the total mass of the material remaining outside the black hole,
$M_{r>r_\mathrm{AH}}$, without distinguishing bound and unbound
components. This quantity can be derived more accurately than the disk
or bound mass, which suffers from a subtle task of determining the
boundary between bound and unbound components. Furthermore,
$M_{r>r_\mathrm{AH}}$ is found to show clearer correlations with the
neutron-star compactness than the disk or bound mass is. Thus, we base
our discussions on $M_{r>r_\mathrm{AH}}$.

\begin{figure}[htbp]
 \centering \includegraphics[width=0.7\linewidth,clip]{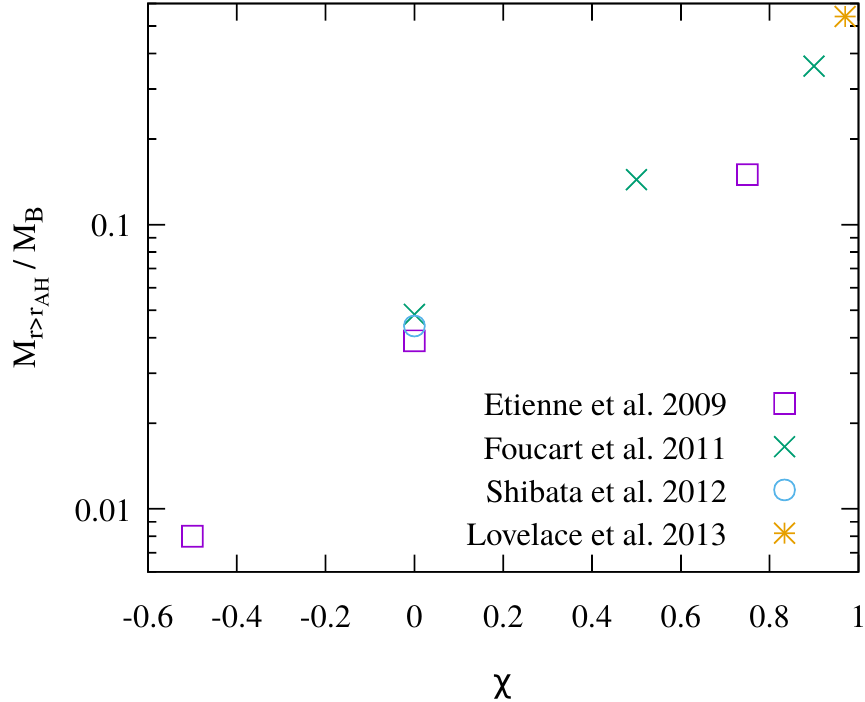}
 \caption{Summary of the mass of the material remaining outside the
 black hole, $M_{r>r_\mathrm{AH}}$, as a function of the spin parameter
 of the black hole for binaries with $Q=3$ and $\mathcal{C} \approx
 0.145$ modeled by a $\Gamma = 2$ polytrope computed by three
 independent groups. The vertical axis shows the fraction of
 $M_{r>r_\mathrm{AH}}$ to the baryon rest mass of the neutron star,
 $M_\mathrm{B}$. Note that these values are measured at different times
 in each simulation \citep[see also][]{Foucart2012}. This figure is
 generated from data of
 \citet{Shibata_KYT2009e,Etienne_LSB2009,Foucart_DKT2011,Lovelace_DFKPSS2013}.}
 \label{fig:diskG2}
\end{figure}

To date, dependence of $M_\mathrm{r>r_\mathrm{AH}}$ on the spin
parameter has been studied most extensively for a qualitative $\Gamma =
2$ polytrope by several independent groups. Figure \ref{fig:diskG2}
plots the fractional baryonic mass of the material remaining outside the
black hole obtained by three independent groups as a function of the
spin parameter of the black hole for systems with $Q=3$ and $\mathcal{C}
\approx 0.145$ modeled by the $\Gamma = 2$ polytrope. Care must be taken
for the fact that these quantities are measured at different times in
each simulation, and particularly the result for $\chi = 0.75$ is
inferred as late as $\approx \SI{25}{\ms}$ after the onset of merger for
a hypothetical value of $M_\mathrm{NS} = 1.35\,M_\odot$
\citep[Fig.~13]{Etienne_LSB2009}. Taking this limitation into account,
the results agree approximately for $\chi = 0$
\citep{Etienne_LSB2009,Foucart_DKT2011,Shibata_KYT2009e}. Figure
\ref{fig:diskG2} also shows that $M_{r>r_\mathrm{AH}}$ increases as the
value of the spin parameter increases for $Q=3$ and this equation of
state. Specifically, the values of $M_{r>r_\mathrm{AH}} / M_\mathrm{B}$
are approximately proportional to $\exp ( b \chi )$ with $b \approx
2.5$--$3$ for a range displayed in Fig.~\ref{fig:diskG2}. This enables
us to reconfirm that the degree of tidal disruption depends strongly on
the spin of the black hole.

\begin{figure}[htbp]
 \centering \includegraphics[width=0.7\linewidth,clip]{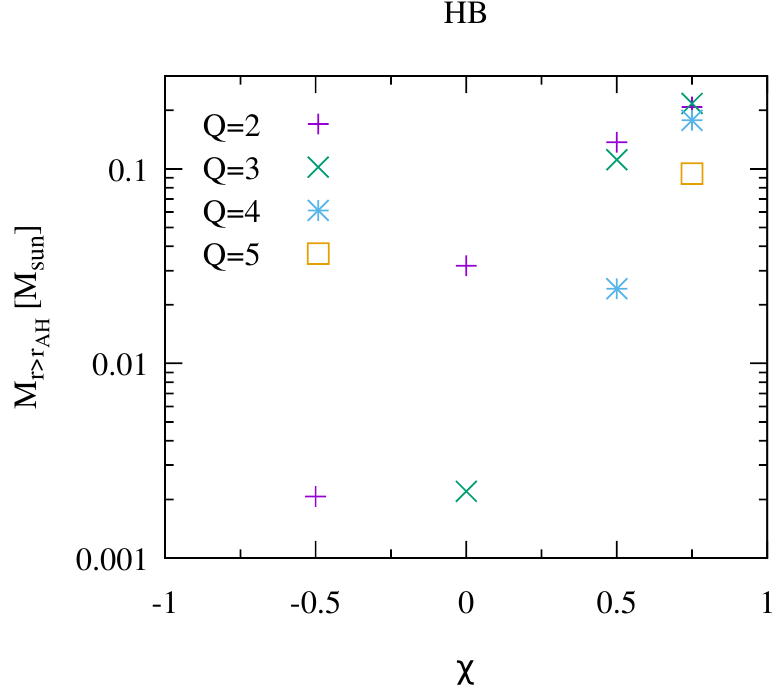}
 \caption{Mass of the material remaining outside the black hole at
 \SI{10}{\ms} after the onset of merger as a function of the spin
 parameter of the black hole for a variety of the mass ratio. The mass
 and the radius of the neutron star are fixed, respectively, to
 $M_\mathrm{NS} = 1.35\,M_\odot$ and $R_\mathrm{NS} = \SI{11.6}{\km}$
 ($\mathcal{C} = 0.172$) modeled by a piecewise polytrope called HB
 \citep{Read_MSUCF2009}. This figure is generated from data of
 \citet{Kyutoku_OST2011}.} \label{fig:diskHB}
\end{figure}

Figure \ref{fig:diskHB} plots $M_{r>r_\mathrm{AH}}$ as a function of the
spin parameter of the black hole for systems with $M_\mathrm{NS} =
1.35\,M_\odot$ and $R_\mathrm{NS} = \SI{11.6}{\km}$ ($\mathcal{C} =
0.172$) modeled by a piecewise polytrope called HB, which approximates
soft nuclear-theory-based equations of state. This figure again
illustrates the importance of the black-hole spin depicted in
Fig.~\ref{fig:diskG2}. Quantitatively, $M_{r>r_\mathrm{AH}}$ becomes as
large as $\approx 0.1\,M_\odot$ for $\chi = 0.75$ even if the compactness
is realistically large with this soft equation of state and the mass
ratio $Q$ is as high as $5$. Various subsequent studies have further
shown that large values of the spin parameter allow such significant
tidal disruption to occur even for $Q \approx 7$, which corresponds to
typical masses of Galactic black holes, $\approx 10\,M_\odot$
\citep{Foucart_DDKMOPSST2013,Foucart_DDOOHKPSS2014,Kyutoku_IOST2015}. Conversely,
it is highly unlikely that the neutron star with $\mathcal{C} \gtrsim
0.17$ is disrupted by a black hole with $M_\mathrm{BH} \gtrsim
10\,M_\odot$ to leave material of $\gtrsim 0.1\,M_\odot$ unless the spin
parameter is as high as $\chi \gtrsim 0.9$.

\begin{figure}[htbp]
 \centering \includegraphics[width=0.7\linewidth,clip]{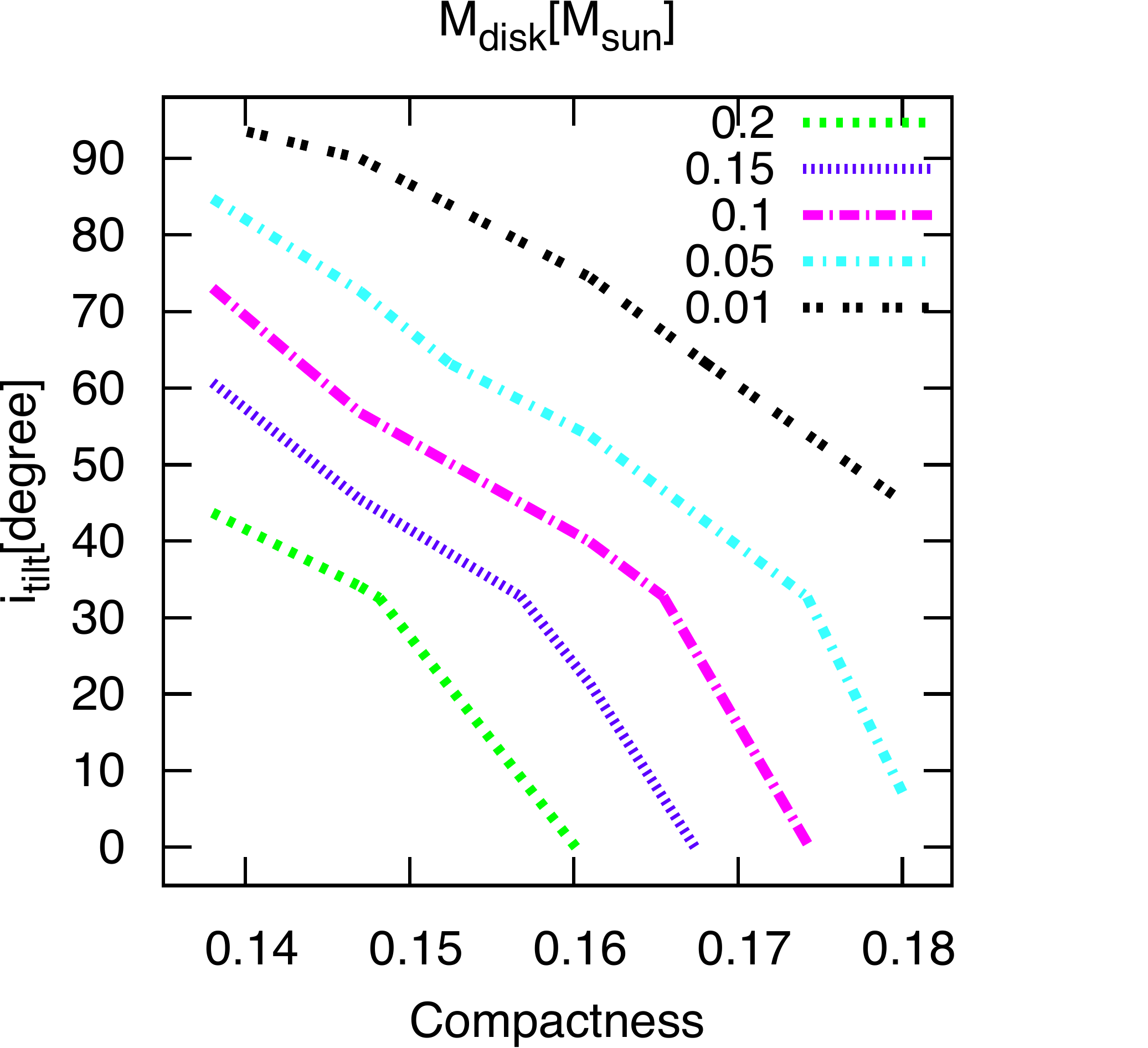}
 \caption{Contour for the mass of the bound material (denoted by
 $M_\mathrm{disk}$ in this figure) at \SI{10}{\ms} after the onset of
 merger on the compactness-inclination angle (denoted by
 $i_\mathrm{tilt}$ in this figure) plane for binaries with
 $M_\mathrm{BH} = 6.75\,M_\odot$, $\chi = 0.75$, and $M_\mathrm{NS} =
 1.35\,M_\odot$ ($Q=5$). The mass of the unbound material is
 excluded. Image reproduced with permission from \citet{Kawaguchi_KNOST2015}, copyright by APS.}
 \label{fig:disk_incl}
\end{figure}

As the inclination angle of the black-hole spin increases, the degree of
tidal disruption and thus the mass of the remnant disk decrease
\citep{Foucart_DKT2011,Foucart_DDKMOPSST2013,Kawaguchi_KNOST2015}. Figure
\ref{fig:disk_incl} shows contours for the mass of the bound material
only (not $M_{r>r_\mathrm{AH}}$). The masses of the black hole and the
neutron star are fixed to be $M_\mathrm{BH} = 6.75\,M_\odot$ and
$M_\mathrm{NS} = 1.35\,M_\odot$, respectively, and the magnitude of the
black-hole spin is fixed to be $\chi = 0.75$. This figure shows that the
mass of the bound material decreases as the inclination angle
increases. A similar trend holds for the unbound material and thus for
the total mass remaining outside the black hole. Quantitatively, the
mass of the bound material decreases from $0.1\,M_\odot$ (magenta) to
$0.01\,M_\odot$ (black) with the increase of the inclination angle by only
$\approx \ang{20}$--\ang{30} for a given value of the compactness at
$\mathcal{C} \lesssim 0.17$. It has been pointed out that the mass of
the material remaining outside the black hole with inclined spins is
approximately reproduced by a model with an aligned black-hole spin
whose magnitude derives the same radius of the innermost stable
(circular or spherical) orbits as that of the original configuration
\citep{Foucart_DDKMOPSST2013,Stone_Loeb_Berger2013} or, more simply, by
a model with an aligned spin whose dimensionless magnitude is $\chi \cos
\iota$ \citep{Kawaguchi_KNOST2015}. The effect of the spin orientation
has not been explored systematically over the parameter space, and this
is a subject for future study. Figure \ref{fig:disk_incl} also shows
that the mass of the bound material decreases as the compactness of the
neutron star increases. We discuss the dependence of
$M_{r>r_\mathrm{AH}}$ on the neutron-star compactness below, going back
to the aligned-spin systems.

\begin{figure}[htbp]
 \centering
 \begin{tabular}{cc}
  \includegraphics[width=.47\linewidth,clip]{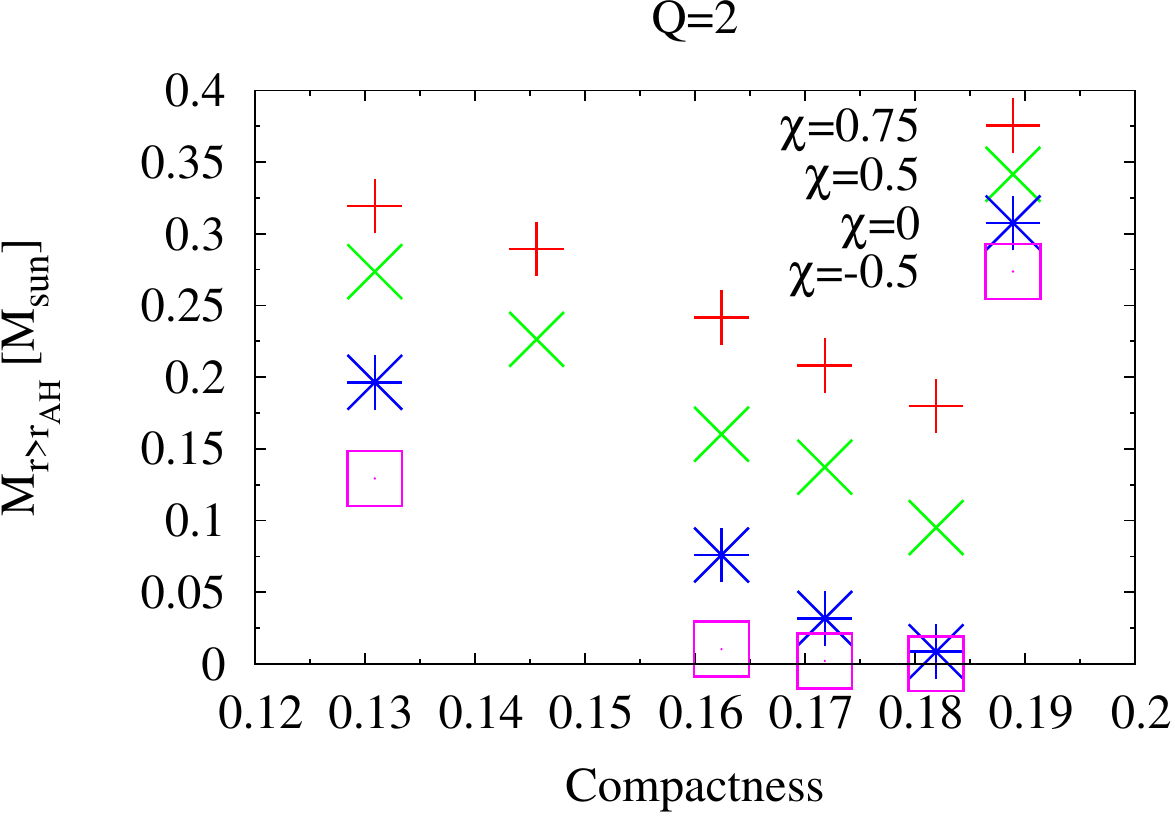} &
  \includegraphics[width=.47\linewidth,clip]{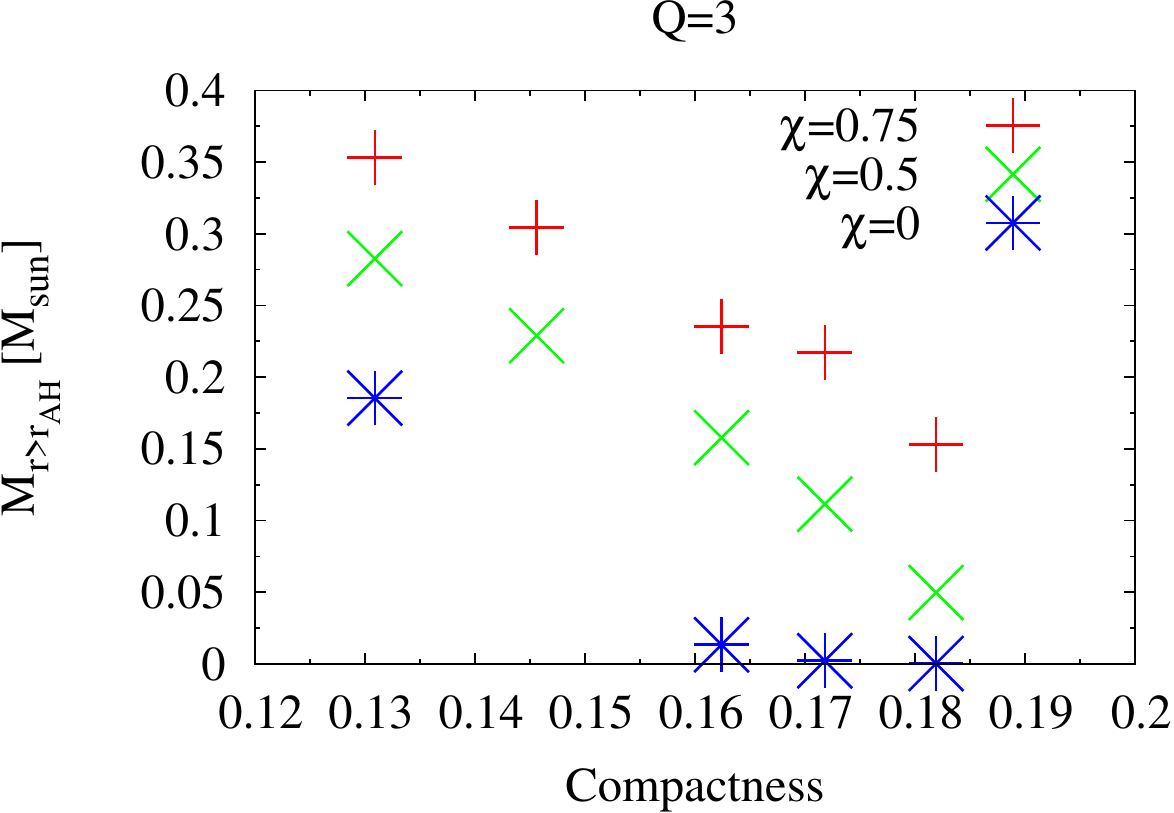}
 \end{tabular}
 \caption{Mass of the material remaining outside the black hole at
 \SI{10}{\ms} after the onset of merger for binaries with $M_\mathrm{NS}
 = 1.35\,M_\odot$ as a function of the compactness, which is varied by
 adopting a one-parameter family of piecewise polytropes. The spin
 parameter of the black hole is also varied. The left and right panels
 show the results of systems with $Q=2$ and $3$, respectively. Image adapted from \citet{Kyutoku_OST2011}, copyright by APS.} \label{fig:diskQ}
\end{figure}

Figure \ref{fig:diskQ} plots $M_{r>r_\mathrm{AH}}$ as a function of the
compactness for $Q=2$ (left) and $Q=3$ (right). The mass of the neutron
star is fixed to be $M_\mathrm{NS} = 1.35\,M_\odot$, and the compactness
is varied by adopting a one-parameter family of piecewise polytropes
(see \citealt{Kyutoku_OST2011} for the details). The values of the spin
parameter are also varied systematically. This figure shows that
$M_{r>r_\mathrm{AH}}$ decreases approximately linearly as the
compactness increases until the value decreases to $\lesssim
0.01\,M_\odot$. This trend holds irrespective of the values of $Q$ or
$\chi$. Figure \ref{fig:diskQ} also shows that $M_{r>r_\mathrm{AH}}$
increases steeply as the spin parameter of the black hole increases as
already seen in Figs.~\ref{fig:diskG2} and
\ref{fig:diskHB}. Quantitatively, $M_{r>r_\mathrm{AH}} \gtrsim
0.1\,M_\odot$ is achieved for a wide range of compactness, $\mathcal{C}
\lesssim 0.18$, if the black-hole spin is as high as $0.5$ for $Q=2$ and
$0.75$ for $Q=3$.

Although the primary effect of the equation of state is captured by the
compactness of the neutron star, the density profile also affects the
susceptibility to tidal disruption. A comparison performed using a
two-parameter family of piecewise polytropes shows that the mass of the
material remaining outside the black hole decreases by more than a
factor of $2$ for a centrally-condensed profile of the neutron star even
if the compactness is approximately the same
\citep{Kyutoku_Shibata_Taniguchi2010,Kyutoku_Shibata_Taniguchi2010e}. This
is because the central condensation tends to suppress the degree of
tidal deformation and thus tidal disruption is appreciably delayed from
the onset of mass shedding. We note that it has been pointed out that
the correlation of $M_{r>r_\mathrm{AH}}$ with the tidal deformability,
$\Lambda$, is not stronger than that with the compactness
\citep{Foucart2012,Foucart_Hinderer_Nissanke2018}. Thus, neither the
compactness nor the tidal deformability is fully appropriate to
determine the amount of material remaining outside the black hole.

\begin{figure}[htbp]
 \centering
 \begin{tabular}{cc}
  \includegraphics[width=.47\linewidth,clip]{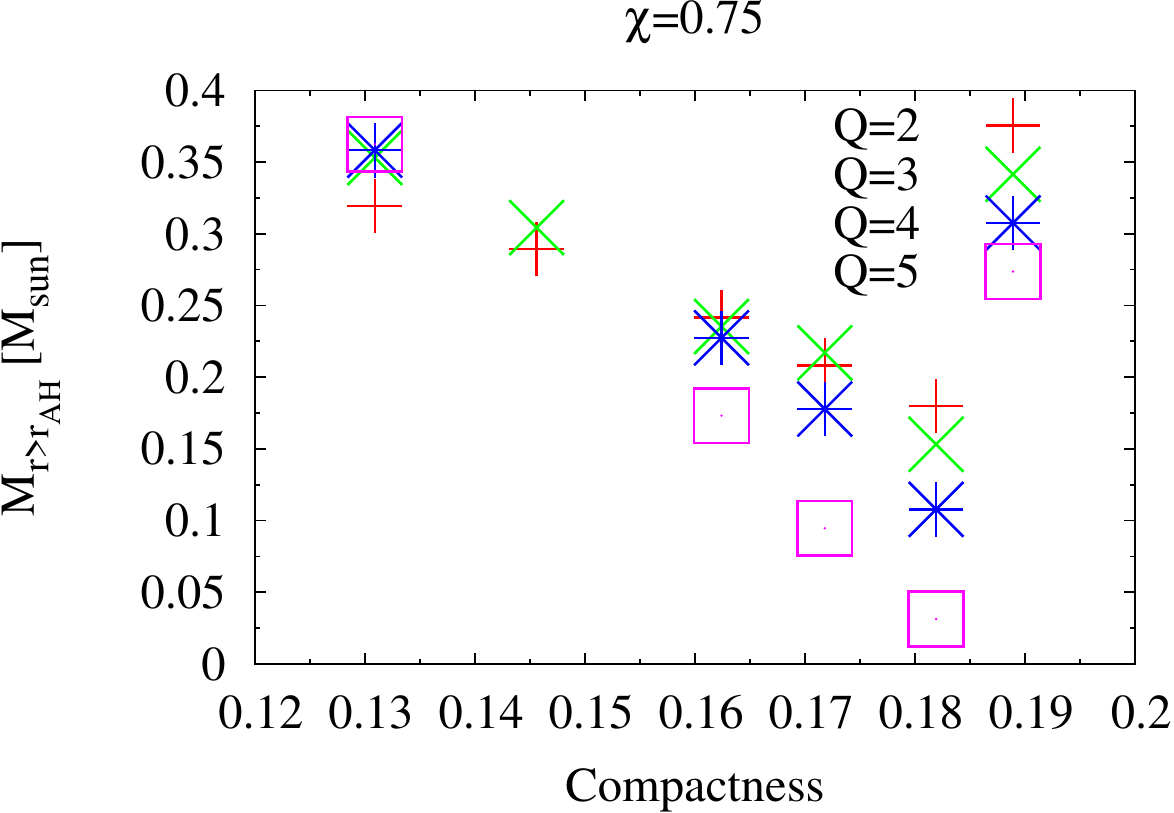} &
  \includegraphics[width=.47\linewidth,clip]{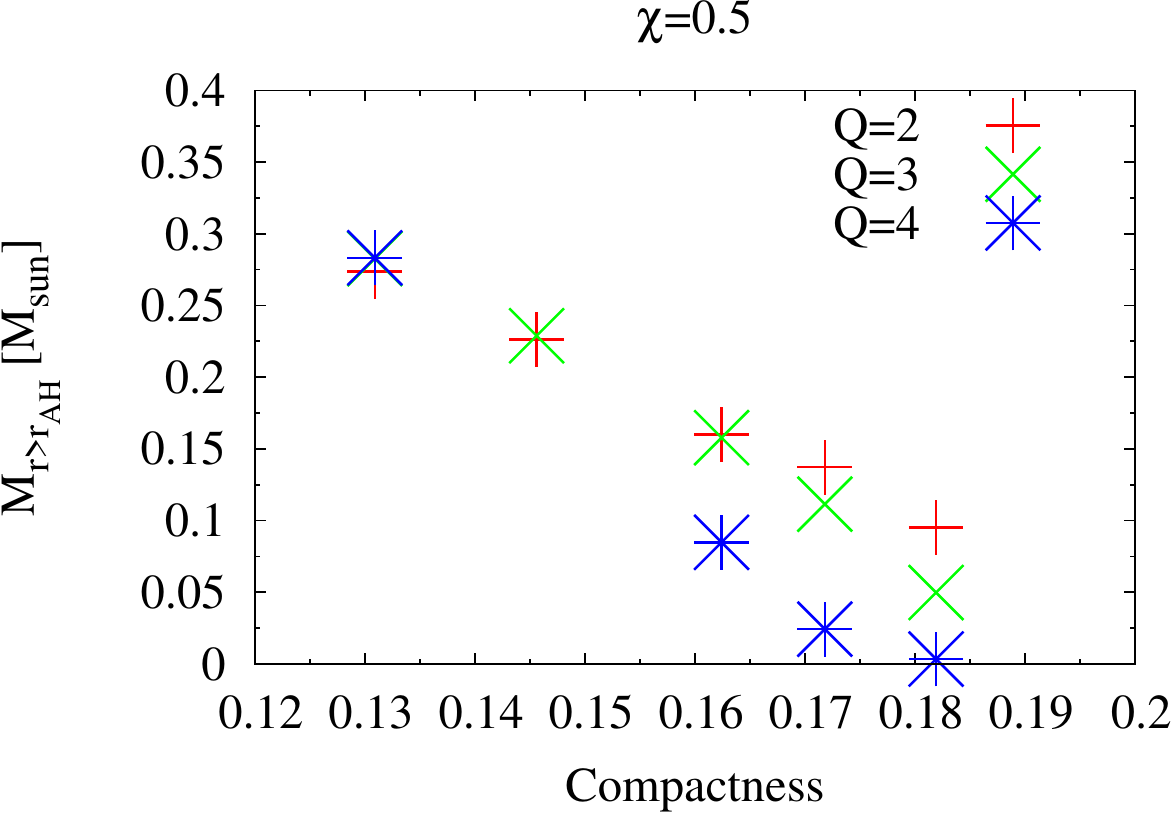}
 \end{tabular}
 \caption{Same as Fig.~\ref{fig:diskQ} but for $\chi = 0.75$ (left) and
 $\chi = 0.5$ (right). The mass ratio is varied in each plot. Image adapted from \citet{Kyutoku_OST2011}, copyright by APS.} \label{fig:diskchi}
\end{figure}

Figure \ref{fig:diskchi} shows the dependence of $M_{r>r_\mathrm{AH}}$
on the neutron-star compactness from another perspective for fixed
values of $\chi = 0.75$ (left) and $\chi = 0.5$ (right). The mass of the
neutron star is fixed to be $M_\mathrm{NS} = 1.35\,M_\odot$. Again, it is
found that $M_{r>r_\mathrm{AH}}$ increases approximately linearly as the
compactness decreases irrespective of $Q$ or $\chi$. One notable feature
that becomes clearly visible in this plot, although presaged in previous
figures, is that the dependence of $M_{r>r_\mathrm{AH}}$ on the mass
ratio, $Q$, becomes weak or even inverted from naive expectations at the
small compactness of $\mathcal{C} \lesssim 0.15$. Analytic estimation in
Sect.~\ref{sec:intro_tidal} suggests that the degree of tidal disruption
and thus $M_{r>r_\mathrm{AH}}$ is likely to decrease as the black hole
becomes massive, i.e., as $Q$ becomes high. This indeed holds for
systems with a large neutron-star compactness, $\mathcal{C} \gtrsim
0.15$, but does not for $\mathcal{C} \lesssim 0.15$. This fact suggests
that the dependence on the mass ratio is worth investigating in detail.

\begin{figure}[htbp]
 \centering \includegraphics[width=.7\linewidth,clip]{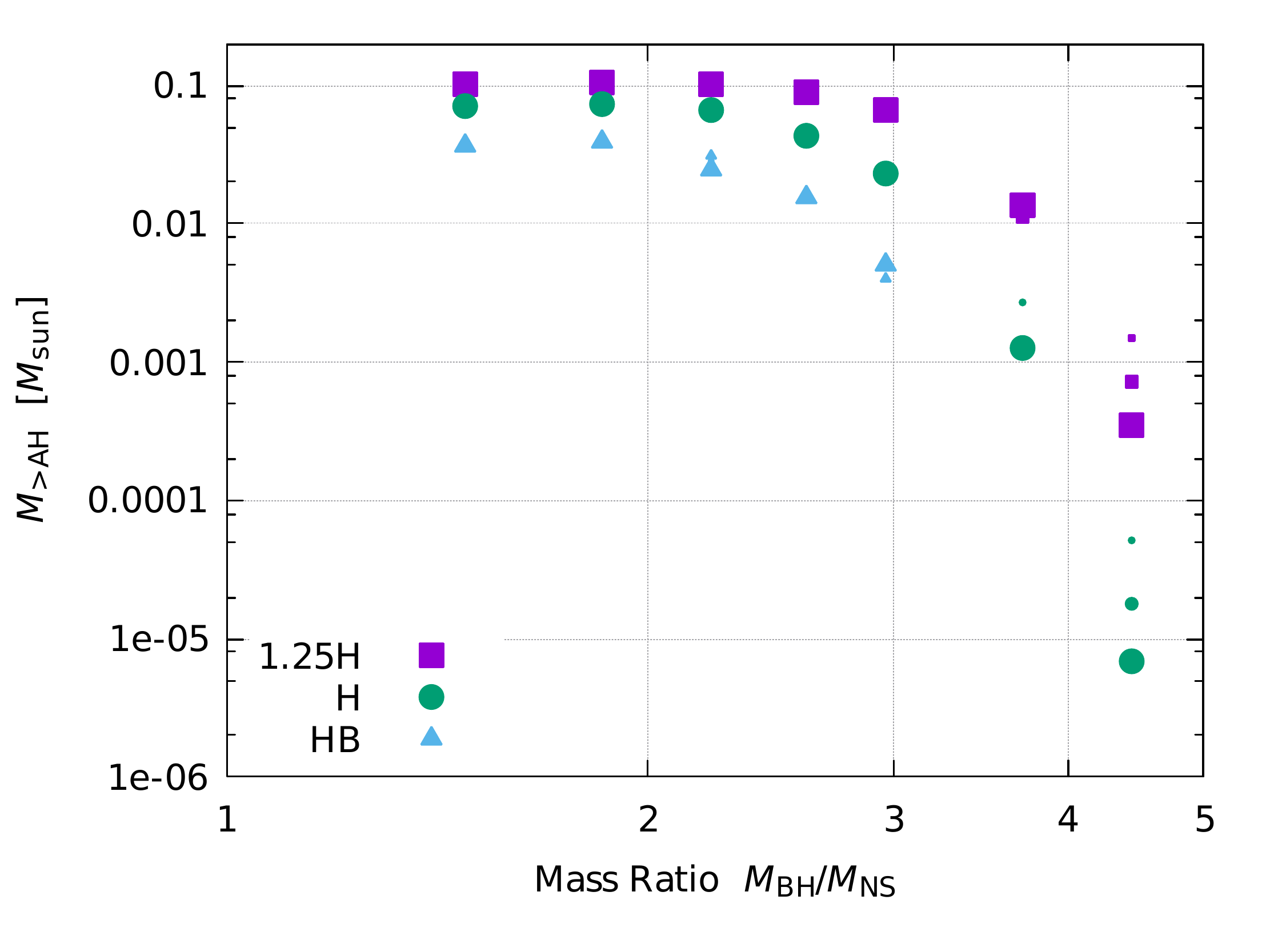}
 \caption{Mass of the material remaining outside the black hole at
 \SI{12}{\ms} after the onset of merger as a function of the mass ratio
 for binaries with $\chi = 0$ and $M_\mathrm{NS} =
 1.35\,M_\odot$. Equations of state are varied by adopting a one-parameter
 family of piecewise polytropes called HB, H, and 1.25H
 \citep{Read_MSUCF2009,Lackey_KSBF2012}. Symbols with different sizes
 show the result obtained with different grid resolutions and are not
 visible for $M_{r>r_\mathrm{AH}} \gtrsim 0.01\,M_\odot$ on the scale of
 this figure. Image reproduced with permission from \citet{Hayashi_KKKS2021}, copyright by APS.}
 \label{fig:diskQlow}
\end{figure}

The dependence of $M_{r>r_\mathrm{AH}}$ on the mass ratio, $Q$, has
recently been found to become very weak for very-low-mass and
nonspinning black holes \citep{Hayashi_KKKS2021}. Figure
\ref{fig:diskQlow} shows $M_{r>r_\mathrm{AH}}$ as a function of $Q$ for
nonspinning black hole--neutron star binaries with various equations of
state. The mass of the neutron star is fixed to be $M_\mathrm{NS} =
1.35\,M_\odot$. It is clearly shown that the value of
$M_{r>r_\mathrm{AH}}$ levels off at $0.05$--$0.1\,M_\odot$ for $Q \lesssim
2$--$3$, where their precise values depend on the equation of state (and
presumably the black-hole spin). We note that, as we discuss later in
Sect.~\ref{sec:sim_rem_dyn}, $M_{r>r_\mathrm{AH}}$ is approximately
identical to the mass of the remnant disk at this low-$Q$ saturation
regime. In fact, such disappearance of the dependence on binary
parameters may be observed in various regions of the parameter space for
which the degree of tidal disruption increases
\citep{Brege_DFDCHKOPS2018,Hayashi_KKKS2021} as is also suggested by
Fig.~\ref{fig:diskchi}. By what and how the saturation values of
$M_{r>r_\mathrm{AH}}$ are determined have not been understood yet and
may be counted as one of the unsolved problems about black hole--neutron
star binaries.

\begin{figure}[htbp]
 \centering \includegraphics[width=.7\linewidth,clip]{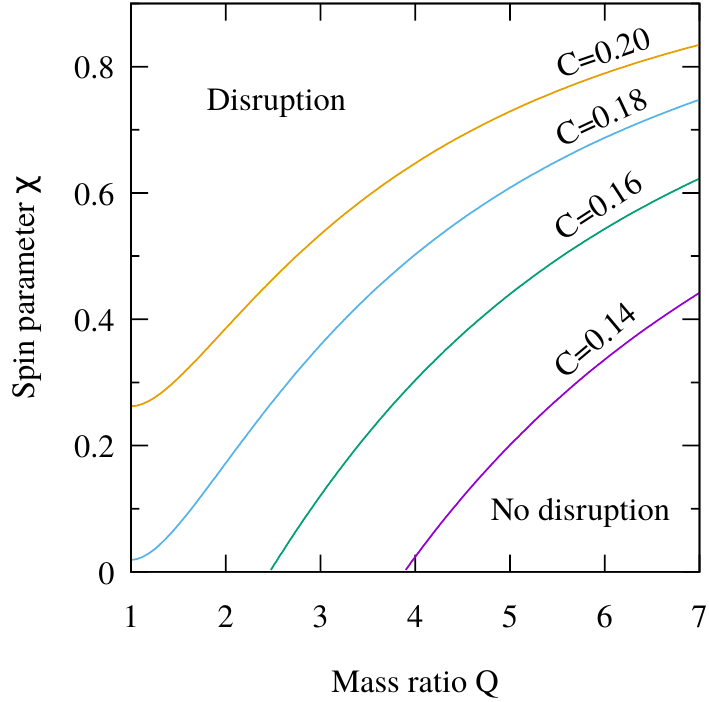}
 \caption{Criterion for tidal disruption predicted by the fitting
 formula of \citet{Foucart_Hinderer_Nissanke2018} for various
 compactnesses of neutron stars denoted by $C$. Here, the criterion is
 defined by the condition that more than 1\% of the baryon rest mass of
 the neutron star is left outside the apparent horizon at \SI{10}{\ms}
 after the onset of merger. The baryon rest mass increases toward the
 left top region of this $Q$-$\chi$ plane for a given value of the
 compactness.} \label{fig:criterion}
\end{figure}

Fitting formulae for the mass of the material remaining outside the
black hole are provided by
\citet{Foucart2012,Foucart_Hinderer_Nissanke2018}. These fitting
formulae are especially useful for deriving an approximate criterion for
tidal disruption. Figure \ref{fig:criterion} displays the contour above
which more than $1\%$ of the baryon rest mass of the neutron star is
left outside the apparent horizon at \SI{10}{\milli\second} after the
onset of merger for a given value of the neutron-star compactness
(denoted by $C$ in this figure) on the $Q$-$\chi$ plane adopting a
formula of \citet{Foucart_Hinderer_Nissanke2018}. This formula is
claimed to be accurate within $\sim 15\%$ for $1 \le Q \le 7$ and $-0.5
\le \chi \le 0.9$ for the case in which less than 30\% of the baryon
rest mass remains outside the apparent horizon. Because the equation of
state is uncertain and the mass of the neutron star is different among
realistic binaries, we draw contours for various values of the
compactness, $\mathcal{C}$, which is the only finite-size effect of the
neutron star taken into account in the formula adopted here (see
\citealt{Foucart_Hinderer_Nissanke2018} for another formula based on
tidal deformability). This figure clearly shows that high prograde spins
enable tidal disruption to occur outside the innermost stable circular
orbit of massive black holes for a wide range of the neutron-star
compactness. It should be cautioned that, however, all the previous
simulations of black hole--neutron star binaries have not systematically
varied masses of the neutron stars except for a scale-free, qualitative
$\Gamma=2$ polytrope. Thus, predictions for $M_\mathrm{NS} \lesssim
1.2\,M_\odot$ or $M_\mathrm{NS} \gtrsim 1.5\,M_\odot$ need to be taken with
particular care.

\subsubsection{Thermodynamic properties of the disk}
\label{sec:sim_rem_disk2}

Continuing the discussions in Sect.~\ref{sec:sim_rem_disk}, we summarize
thermodynamic properties of the remnant disk. For a given value of the
mass, the structure and time evolution of the remnant disk depend
primarily on the total mass of the system, $m_0$. This is because the
length scale of the system after merger is proportional to the mass of
the remnant black hole, which agrees approximately with the total mass
as we discussed in Sect.~\ref{sec:sim_rem_bh}. If we focus on systems
with a fixed value of $M_\mathrm{NS}$, the total mass of the system is
controlled by the mass ratio as $m_0 = (1+Q) M_\mathrm{NS}$.

\begin{figure}[htbp]
 \centering
 \begin{tabular}{cc}
  \includegraphics[width=.47\linewidth,clip]{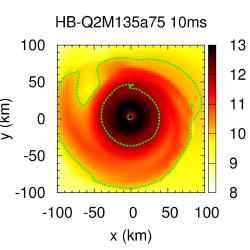} &
  \includegraphics[width=.47\linewidth,clip]{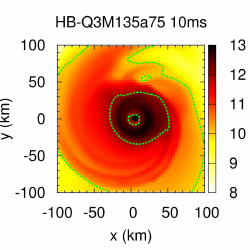} \\
  \includegraphics[width=.47\linewidth,clip]{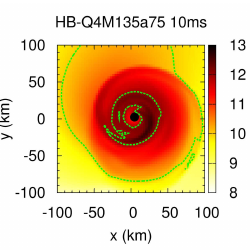} &
  \includegraphics[width=.47\linewidth,clip]{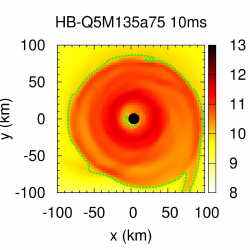}
 \end{tabular}
 \caption{Profile of the rest-mass density and the location of the
 apparent horizon on the equatorial plane at \SI{10}{\ms} after the
 onset of merger for binaries with $\chi = 0.75$, $M_\mathrm{NS} =
 1.35\,M_\odot$, and $R_\mathrm{NS} = \SI{11.6}{\km}$
 ($\mathcal{C}=0.172$) modeled by a piecewise polytrope called HB
 \citep{Read_MSUCF2009}. The mass of the black hole, $M_\mathrm{BH}$, is
 varied. The total mass of the system is $4.05\,M_\odot$ (left top,
 $Q=2$), $5.4\,M_\odot$ (right top, $Q=3$), $6.75\,M_\odot$ (left bottom,
 $Q=4$), and $8.1\,M_\odot$ (right bottom, $Q=5$). The color map shows
 $\log_{10} ( \rho [\si{\gram\per\cubic\cm}] )$. The dashed curves
 denote isodensity contours of \num{e10} and
 \SI{e12}{\gram\per\cubic\cm}. Image reproduced with permission from
 \citet{Kyutoku_OST2011}, copyright by APS.} \label{fig:snap_rho}
\end{figure}

Figure \ref{fig:snap_rho} displays the profiles of the rest-mass density
of the remnant disk on the equatorial plane for systems with $\chi =
0.75$, $M_\mathrm{NS} = 1.35\,M_\odot$, $R_\mathrm{NS} = \SI{11.6}{\km}$
($\mathcal{C} = 0.172$) modeled by a piecewise polytrope called HB but
with different values of $M_\mathrm{BH}$ \citep{Kyutoku_OST2011}. This
figure shows that the rest-mass density in the inner region is
systematically higher for a smaller value of the total
mass. Quantitatively, the maximum rest-mass density is $\approx
\SI{e13}{\gram\per\cubic\cm}$ for $m_0 = 4.05\,M_\odot$ ($Q=2$) and
$\approx \SI{e11}{\gram\per\cubic\cm}$ for $m_0 = 8.1\,M_\odot$
($Q=5$). This difference cannot be ascribed to the difference in the
mass of the material, because $M_{r>r_\mathrm{AH}}$ varies only by a
factor of $\approx 2$ among the systems presented in this
figure. Instead, reflecting the fact that the typical length scale such
as the radius of the innermost stable circular orbit is smaller for a
smaller value of $m_0$, the remnant disk is spatially more concentrated
and the rest-mass density is increased. Conversely, low-density regions
with $\rho \lesssim \SI{e10}{\gram\per\cubic\cm}$ extend to a more
distant region for larger values of $m_0$. Thus, the density gradient
becomes shallow as the total mass increases.

The angular momentum profile after circularization is slightly
sub-Keplerian by $\sim \order{10\%}$ due to pressure support,
particularly in the outer part
\citep{Foucart_DDKMOPSST2013,Lovelace_DFKPSS2013,Foucart_DDOOHKPSS2014}. In
addition, the disk is not geometrically thick at its formation. These
features are qualitatively different from those of a geometrically-thick
torus with constant specific angular momentum, which is typically
adopted in longterm simulations for black hole--accretion disk systems
\citep{Fernandez_Metzger2013,Just_BAGJ2015,Siegel_Metzger2017,Siegel_Metzger2018,Fernandez_TQFK2019,Miller_RDBFFKLMW2019,Fernandez_Foucart_Lippuner2020}.

\begin{figure}[htbp]
 \centering
 \begin{tabular}{ccc}
  \includegraphics[width=0.3\linewidth,clip]{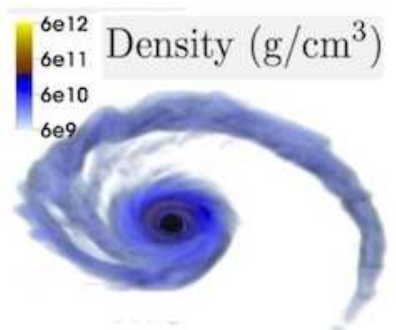} &
  \includegraphics[width=0.3\linewidth,clip]{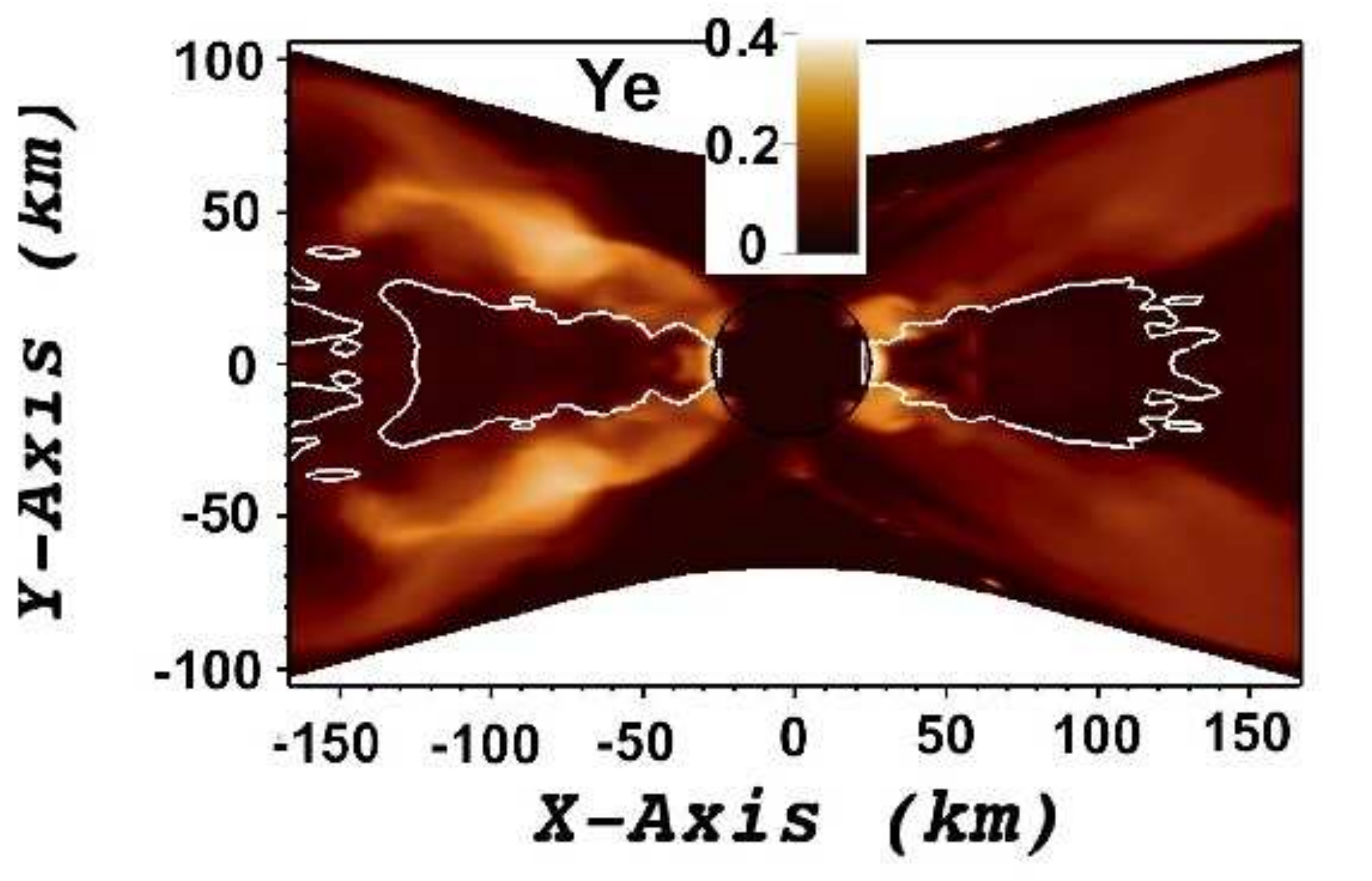} &
  \includegraphics[width=0.3\linewidth,clip]{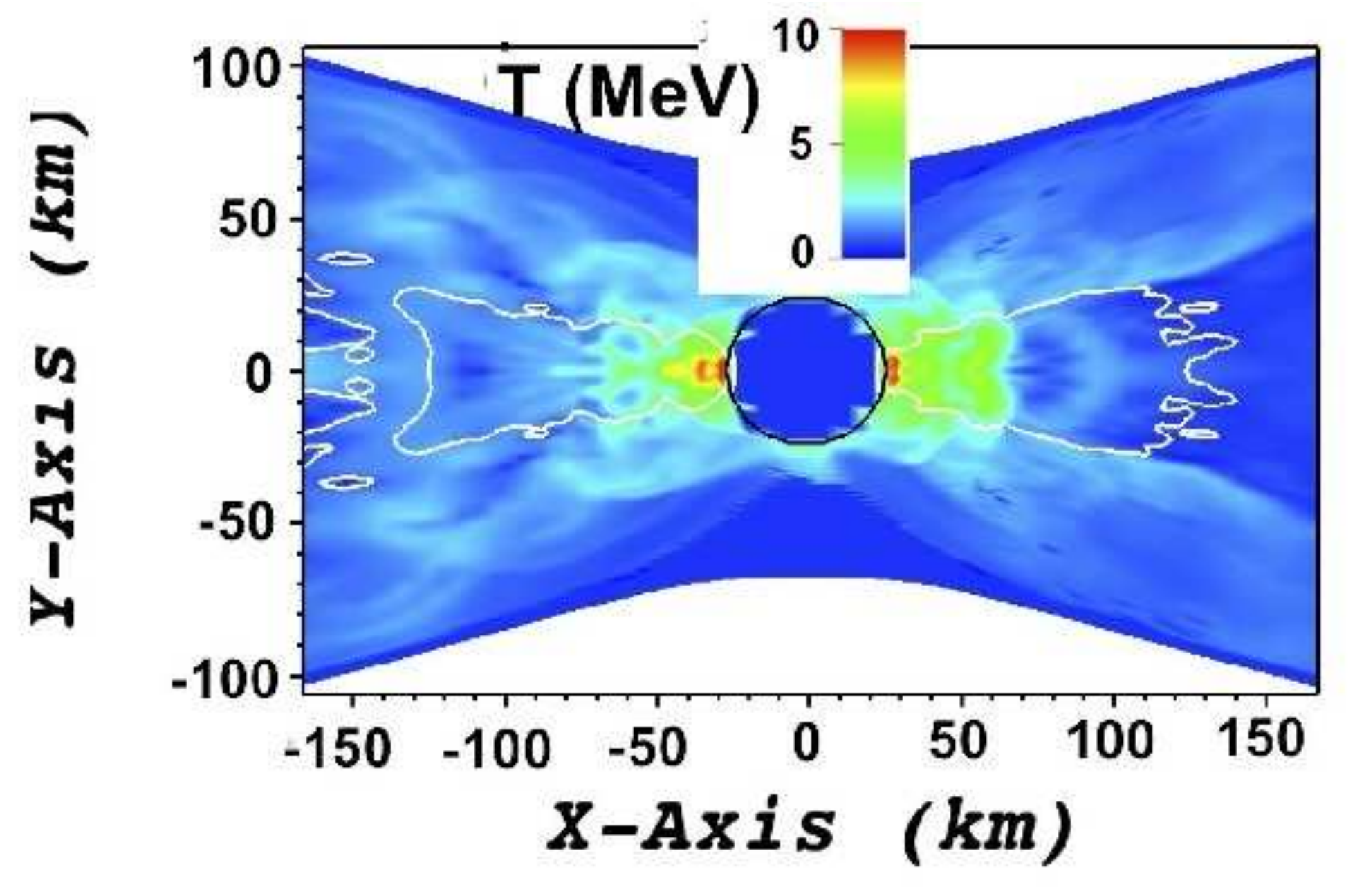}
 \end{tabular}
 \caption{Three-dimensional plot of the rest-mass density (left),
 profile of the electron fraction on the meridional plane (middle), and
 that of the temperature in units of \si{\mega\eV} (right) at
 \SI{5}{\ms} after the onset of merger for a binary with $M_\mathrm{BH}
 = 7\,M_\odot$, $\chi = 0.8$, $M_\mathrm{NS} = 1.4\,M_\odot$, and
 $R_\mathrm{NS} = \SI{12.7}{\km}$ ($Q=5$, $\mathcal{C}=0.163$) modeled
 by the LS220 equation of state \citep{Lattimer_Swesty1991}. Images repoduced with permission from \citet{Foucart_DDOOHKPSS2014}, copyright by APS.} \label{fig:snap_mer}
\end{figure}

Although the profiles of the rest-mass density and the angular momentum
are derived reasonably by pure hydrodynamics simulations, determining
temperature and chemical composition of the disk requires
implementations of a temperature- and composition-dependent equation of
state and a scheme for neutrino transport. Such simulations are now
becoming available along with gradual sophistication of the neutrino
transport scheme
\citep{Deaton_DFOOKMSS2013,Foucart_DDOOHKPSS2014,Foucart_DBDKHKPS2017,Kyutoku_KSST2018,Brege_DFDCHKOPS2018,Most_PTR2021}. Figure
\ref{fig:snap_mer} generated by \citet{Foucart_DDOOHKPSS2014}
illustrates a three-dimensional plot of the rest-mass density as well as
the profiles of the electron fraction and the temperature on the
meridional cross section at \SI{5}{\ms} after the onset of merger for a
typical model. This simulation is performed with the LS220 equation of
state \citep{Lattimer_Swesty1991} and a leakage scheme for neutrino
emission (but without neutrino absorption).

Maximum temperature in the accretion disk reaches $\gtrsim
\SI{10}{\mega\eV}$, i.e., $\gtrsim 1\%$ of the rest-mass energy of
nucleons. For example, the system shown in Fig.~\ref{fig:snap_mer}
realizes $\sim \SI{15}{\mega\eV}$ \citep{Foucart_DDOOHKPSS2014}. Other
simulations by an independent group show that another set of systems
with $M_\mathrm{BH} = 5.4\,M_\odot$ and $M_\mathrm{NS} = 1.35\,M_\odot$
($Q=4$) also realizes $10$--$\SI{20}{\mega\eV}$
\citep{Kyutoku_KSST2018}. These values are reasonably expected, because
the temperature of the disk is increased by liberating the kinetic
energy of the radial motion, whose velocity is $\sim \order{0.1c}$ at
tidal disruption. Closer inspection reveals that the temperature is
higher for a more compact neutron star, which is disrupted at a closer
orbit to the black hole with higher velocity
\citep{Kyutoku_KSST2018}. Although the maximum temperature exceeds
\SI{10}{\mega\eV}, the average temperature of the disk is typically
$3$--\SI{5}{\mega\eV} at the disk formation, because the shock heating
plays a major role only in a limited region of the disk.

The electron fraction of the remnant disk is $Y_\mathrm{e} \approx 0.1$
in most of the region, and it is increased to $0.2$--$0.3$ from the
cold, $\beta$-equilibrium value of the neutron star for the region with
high temperature of $\gtrsim \SI{10}{\mega\eV}$
\citep{Deaton_DFOOKMSS2013,Foucart_DDOOHKPSS2014,Foucart_DBDKHKPS2017,Kyutoku_KSST2018}. The
reason for this increase is that the originally neutron-rich material of
the remnant disk formed from the neutron star is protonized by capture
reactions of thermally-produced electron/positron pairs onto
nucleons. The profile of the electron fraction has been found to depend
on binary parameters including equations of state
\citep{Foucart_DDOOHKPSS2014,Kyutoku_KSST2018}. Systematic studies for
the temperature and the chemical composition of the remnant disk have
not been performed vigorously throughout the parameter space compared to
studies in pure hydrodynamics for the mass and the rest-mass
density. These may be one of the future directions.

In this Sect.~\ref{sec:sim_rem_disk2}, we have reviewed thermodynamic
properties of the disk right after the onset of merger, e.g., at
$\approx \SI{10}{\ms}$ after the disk formation. Longterm evolution and
states at late times are determined by ensuing neutrino cooling and
magnetic-field amplification \citep[see,
e.g.,][]{Siegel_Metzger2017,Nouri_etal2018,Fernandez_TQFK2019,Miller_RDBFFKLMW2019,Christie_LTFFQK2019}. After
the initial circularization stage, the disk temperature will gradually
decrease due to the accretion and neutrino emission unless heating due
to the effective viscosity associated with magnetohydrodynamical
turbulence is highly efficient. As the disk expands and the rest-mass
density decreases, the electron fraction will gradually increase due to
weak interactions such as electron/positron captures and neutrino
irradiation
\citep{Fujibayashi_SWKKS2020,Fujibayashi_SWKKS2020-2,2021arXiv210208387J}. In
fact, these longterm evolution processes are partially observed in the
simulations introduced above. However, these simulations are not
realistic for $\gtrsim \SI{10}{\ms}$ due to the lack of the angular
momentum transport expected to be driven by magnetohydrodynamical
effects. These topics will be discussed in Sect.~\ref{sec:sim_pm}.

\subsubsection{Fallback material} \label{sec:sim_rem_fb}

When a part of the material outside the black hole becomes unbound as we
discuss in Sect.~\ref{sec:sim_rem_dyn}, some material behind them
simultaneously obtains energy but remains bound to the remnant black
hole. The latter material travels to a distant region, turns around,
joins the remnant disk with shock heating, and may energize
electromagnetic transients such as extended and plateau emission of
short-hard gamma-ray bursts
\citep{Rossi_Begelman2009,Lee_RamirezRuiz_LopezCamara2009,Metzger_AQM2010,Cannizzo_Troja_Gehrels2011,Kisaka_Ioka2015,Desai_Metzger_Foucart2019,2021arXiv210404708I}. Although
current numerical-relativity simulations are not capable of tracking
very longterm evolution of the temporally-ejected material for $\gg
\SI{1}{\second}$, the fallback rate may be estimated from instantaneous
profiles of the fluid
\citep{Chawla_ABLLMN2010,Kyutoku_IOST2015,Brege_DFDCHKOPS2018} with the
aid of extrapolation by analytic models \citep{Rosswog2007}.

\begin{figure}[htbp]
 \centering
 \begin{tabular}{cc}
  \includegraphics[width=.5\linewidth,clip]{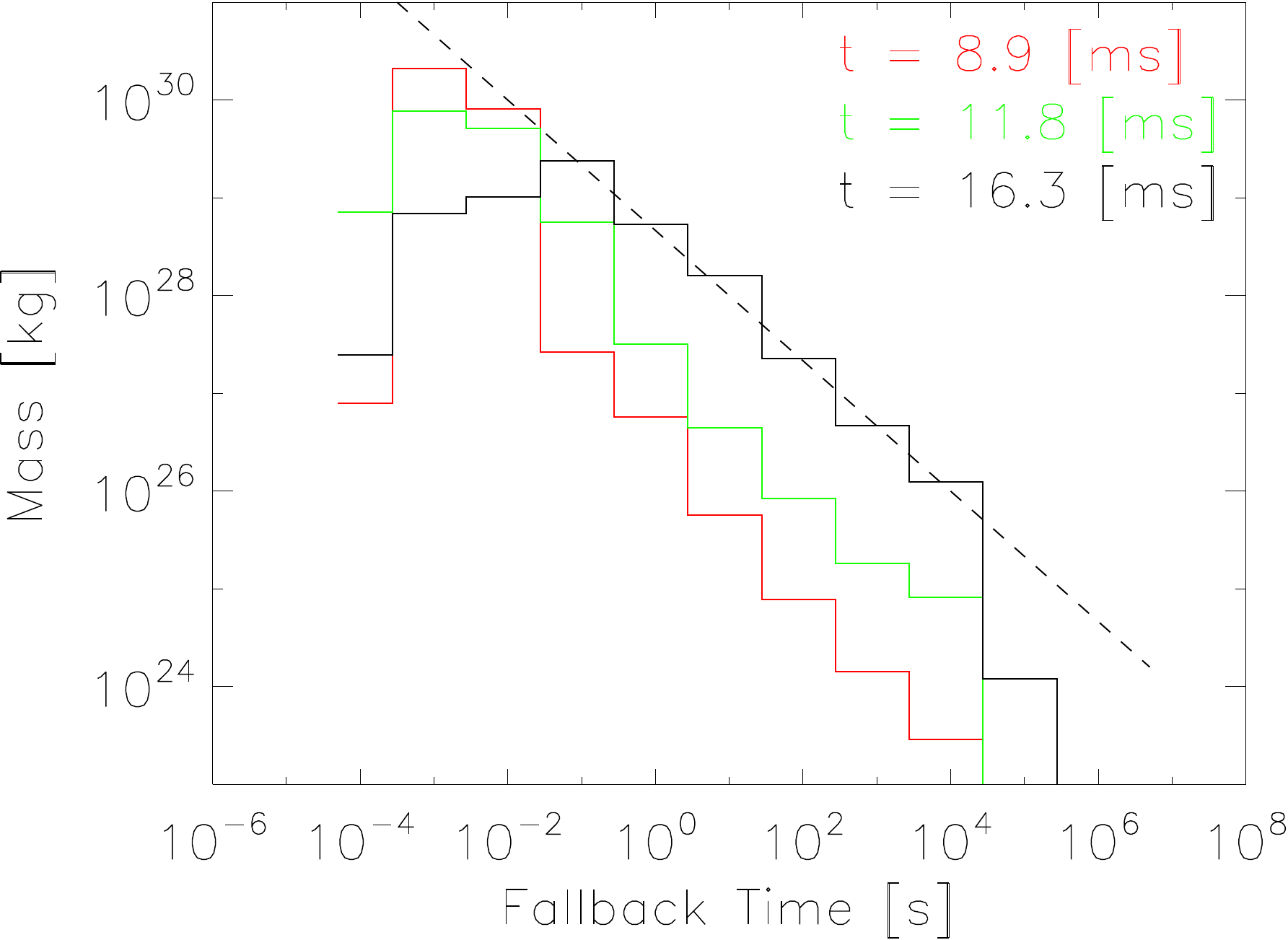} &
  \includegraphics[width=.42\linewidth,clip]{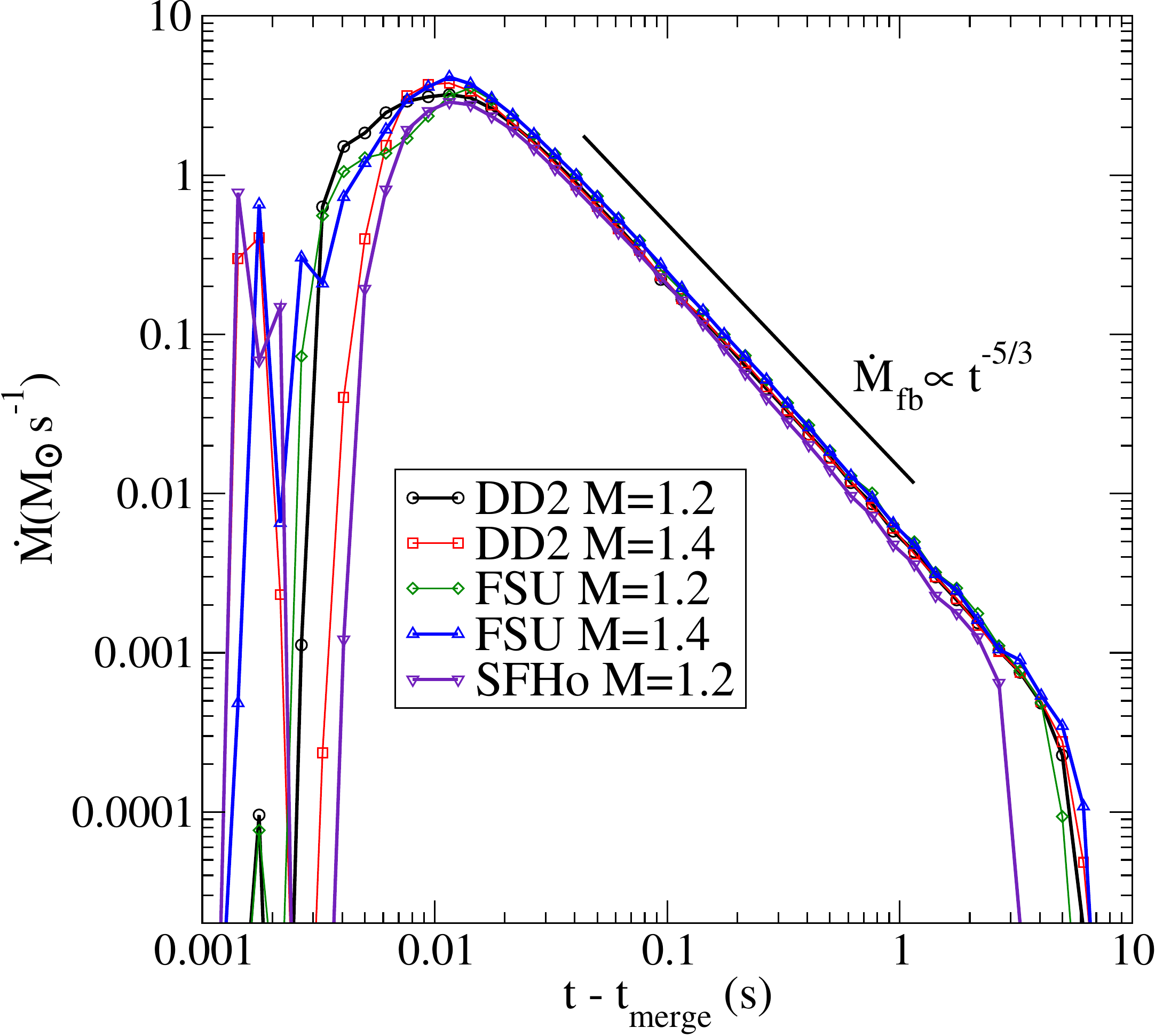}
 \end{tabular}
 \caption{Left: Mass distribution as a function of the fallback time
 predicted from data of a simulation for a binary with $Q=5$, $\chi =
 0.5$ and $\mathcal{C}=0.1$ ($R_\mathrm{NS} = \SI{20.7}{\km}$ for a
 $1.4\,M_\odot$ neutron star) modeled by a $\Gamma = 2$ polytrope at
 different times. The dashed line shows expected $t^{-2/3}$
 behavior. Right: Mass fallback rate predicted from data of simulations
 for various black hole--neutron star binaries. Images repoduced with permission from [left] \citet{Chawla_ABLLMN2010}; and [right] from \citet{Brege_DFDCHKOPS2018}, copyright by APS.} \label{fig:fallback}
\end{figure}

Figure \ref{fig:fallback} generated by \citet{Chawla_ABLLMN2010} (left)
and \citet{Brege_DFDCHKOPS2018} (right) shows the mass fallback rate
right after tidal disruption in black hole--neutron star binaries for
various systems. The fallback rate is found to obey the power law in
time of $\propto t^{-5/3}$ irrespective of the models. This time
dependence coincides with that of tidal disruption events of stars in
marginally unbound orbits around supermassive black holes
\citep{Rees1988,Phinney1989}. Normalization of the fallback rate depends
significantly on the binary system, and $\sim
0.001$--$0.01\,M_\odot\,\si{\per\second} (t/ \SI{1}{\second} )^{-5/3}$ may
reasonably be expected for the case in which significant tidal
disruption occurs \citep{Kyutoku_IOST2015,Brege_DFDCHKOPS2018}.

\begin{figure}[htbp]
 \centering \includegraphics[width=.7\linewidth,clip]{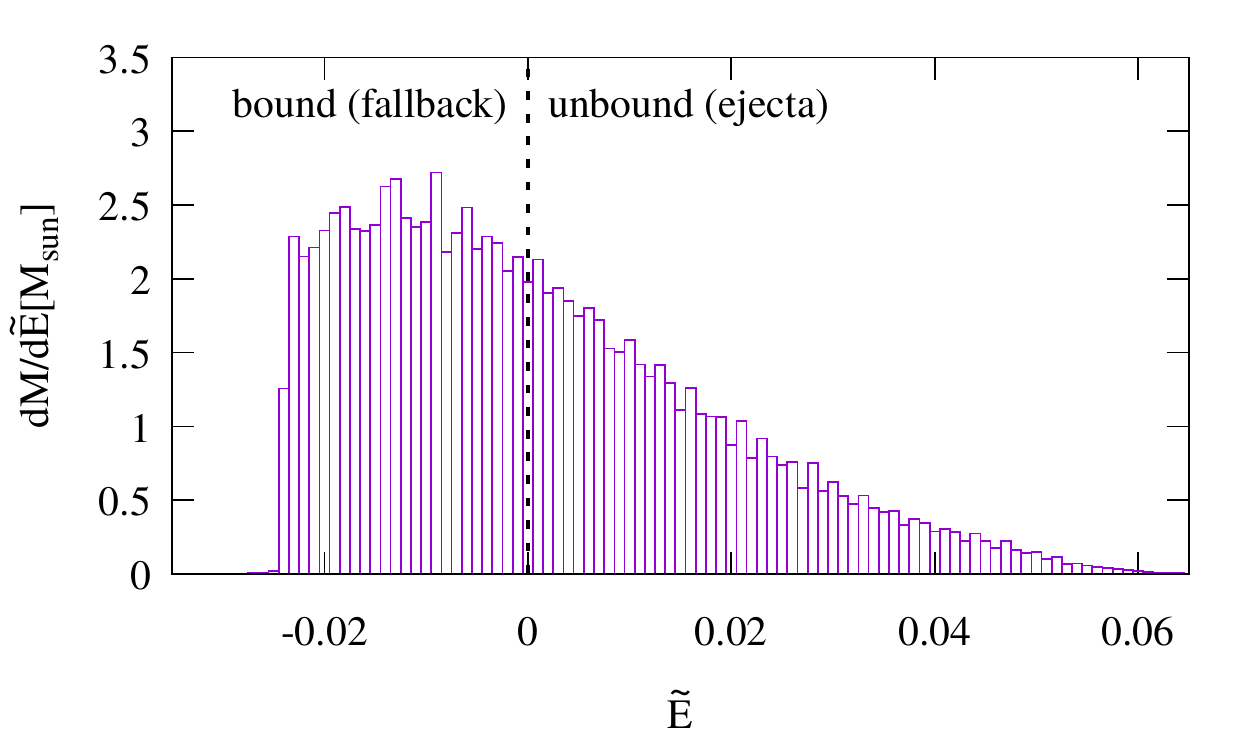}
 \caption{Mass distribution as a function of the specific energy of
 material remaining outside the black hole at \SI{10}{\ms} after the
 onset of merger for a binary with $M_\mathrm{BH} = 4.05\,M_\odot$, $\chi
 = 0.75$, $M_\mathrm{NS} = 1.35\,M_\odot$ and $R_\mathrm{NS} =
 \SI{13.6}{\km}$ ($Q=3$, $\mathcal{C}=0.147$) modeled by a
 piecewise-polytropic approximation of the H4 equation of state
 \citep{Glendenning_Moszkowski1991,Lackey_Nayyar_Owen2006}. Components
 with $\tilde{E}<0$ and $\tilde{E}>0$ represent bound and unbound
 material, respectively. The material within \SI{200}{\km} around the
 central black hole is removed from the plot assuming that it represents
 an accretion disk. This figure is generated from data of
 \citet{Kyutoku_IOST2015}} \label{fig:fbspec}
\end{figure}

As we stated in Sect.~\ref{sec:sim_mrg}, coalescences of circular
compact binaries in bound orbits are fundamentally different from tidal
disruption events associated with hyperbolic encounters. Thus, it is not
expected \textit{a priori} that the fallback rate is given by a power
law with the index of $-5/3$. A key ingredient for realizing this index
is that the mass of fallback material is distributed uniformly with
respect to the specific energy, $\tilde{E}$ \citep{Rees1988}. Actually,
the uniform distribution is found to be approximately realized for bound
material with $\tilde{E} < 0$ in black hole--neutron star
binaries. Figure \ref{fig:fbspec} shows the distribution at \SI{10}{\ms}
after the onset of merger for a typical system. We exclude material
within \SI{200}{\km} around the remnant black hole from this plot
considering that they approximately correspond to disk components, and
the disk outflow is also absent in this simulation. While the amount of
dynamical ejecta with $\tilde{E}>0$ decreases as $\tilde{E}$ increases,
the fallback material with $\tilde{E}<0$ exhibits an approximately flat
profile, resulting in the approximate $t^{-5/3}$ fallback rate. The
mechanism that realizes this distribution has not been clarified yet,
and it would be worthwhile to study this topic in more detail (see,
e.g., \citealt{Lodato_King_Pringle2009} for a study on tidal disruption
events).

Because the disk outflow is launched at some point after disk formation
(see Sect.~\ref{sec:sim_pm_wind}), a part of it will prevent fallback of
the material at late times. Furthermore, mass ejection from the remnant
disk itself could produce additional fallback material. The realistic
fallback rate and relevance to observable features will be determined by
interaction of various components. In addition, \textit{r}-process
heating also modifies the fallback rate
\citep{Metzger_AQM2010,Desai_Metzger_Foucart2019,2021arXiv210404708I}. They
need further investigations.

\subsubsection{Dynamical ejecta} \label{sec:sim_rem_dyn}

\begin{figure}[htbp]
 \centering
 \begin{tabular}{c}
  \includegraphics[width=0.7\linewidth,clip]{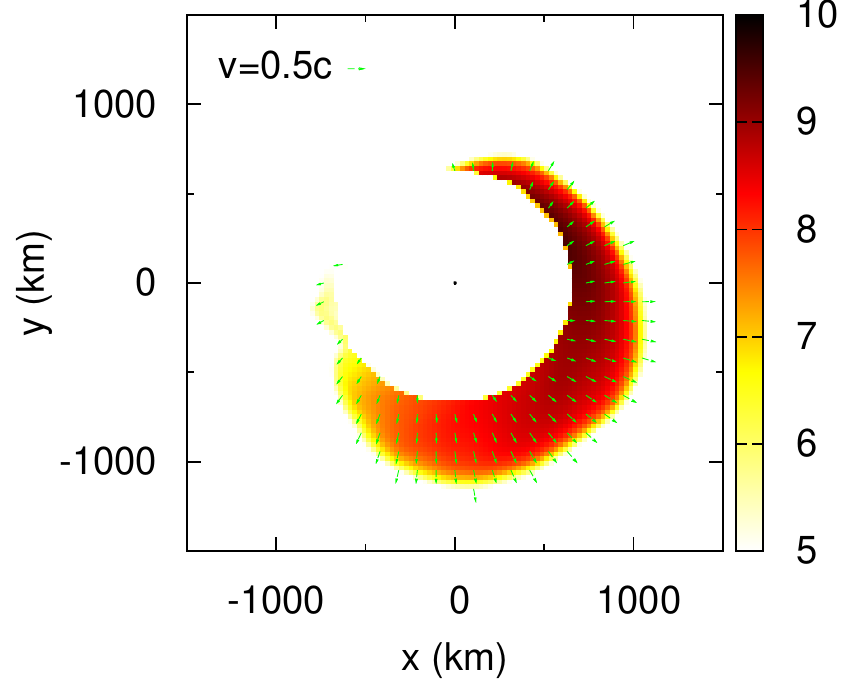} \\
  \hspace{2mm}
  \includegraphics[width=0.65\linewidth,clip]{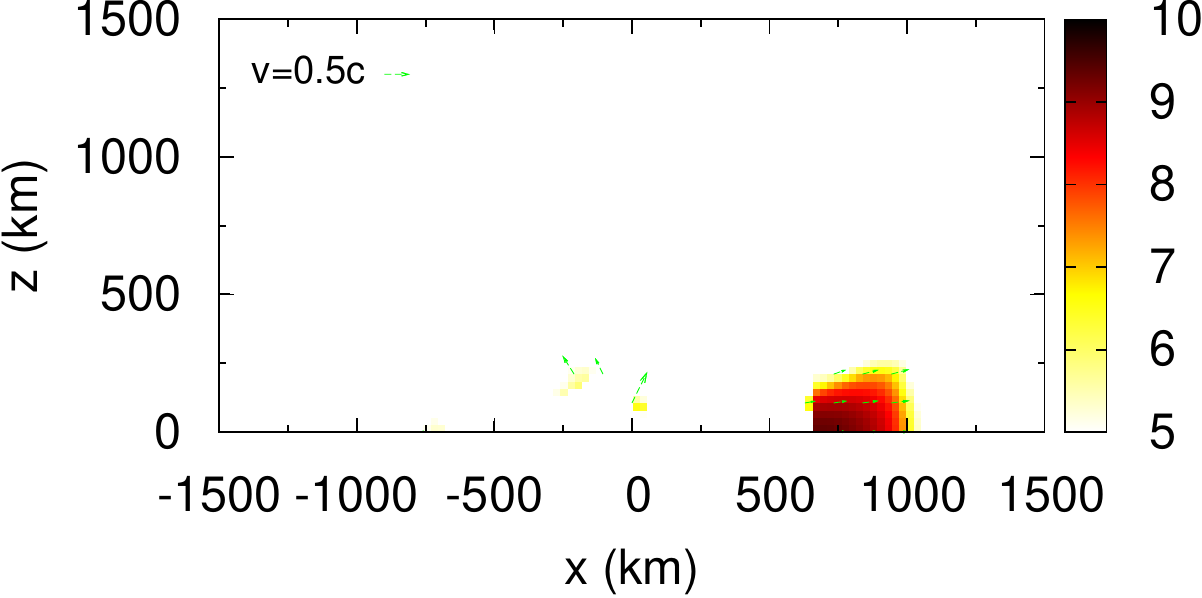}
 \end{tabular}
 \caption{Profile of the rest-mass density and the velocity of the
 unbound material on the equatorial plane (top) and the meridional plane
 (bottom) at \SI{10}{\ms} after the onset of merger for a binary with
 $M_\mathrm{BH} = 4.05 \,M_\odot$, $\chi = 0.75$, $M_\mathrm{NS} = 1.35
 \,M_\odot$, and $R_\mathrm{NS} = \SI{13.6}{\km}$ ($Q=3$,
 $\mathcal{C}=0.147$) modeled by a piecewise-polytropic approximation of
 the H4 equation of state
 \citep{Glendenning_Moszkowski1991,Lackey_Nayyar_Owen2006}. Image reproduced with permission from \citet{Kyutoku_Ioka_Shibata2013}, copyright by APS.}
 \label{fig:morphology}
\end{figure}

Dynamical ejecta from black hole--neutron star binaries are generated
from the outermost part of the tidal tail and characterized by the high
degree of nonsphericity \citep{Kyutoku_Ioka_Shibata2013}. Figure
\ref{fig:morphology} illustrates the rest-mass density of only unbound
material at \SI{10}{\ms} after the onset of merger for a representative
case. The dynamical ejecta typically take a crescent-like shape on the
equatorial plane at this instant as long as the degree of tidal
disruption is significant. Quantitatively, they sweep out only about a
half of the orbital plane and are concentrated around the plane with a
half opening angle of $\sim \ang{10}$--\ang{20}
\citep{Kyutoku_Ioka_Shibata2013,Foucart_DBDKHKPS2017,Brege_DFDCHKOPS2018}. Anisotropy
of mass ejection is ascribed to the (not independent) facts that (i)
black hole--neutron star binaries are completely asymmetric systems in
terms of the profile of material and (ii) dynamical mass ejection in
black hole--neutron star binaries is driven primarily by tidal torque,
which is most efficient in the direction along the orbital plane. This
morphology should be contrasted with more spherical dynamical ejecta
from binary neutron stars, especially equal-mass systems \citep[see,
e.g.,][]{Hotokezaka_KKOSST2013}. The opening angles depend only weakly
on the equation of state except for the cases in which tidal disruption
occurs only marginally \citep{Kyutoku_IOST2015}, and we present its
simplified explanation in Appendix~\ref{app:ae_ej}.

Anisotropic mass ejection may induce various potentially-observable
features \citep{Kyutoku_Ioka_Shibata2013,Kyutoku_IOST2015}. Possible
effects on electromagnetic emission are discussed later in
Sect.~\ref{sec:dis_impl}. One kinematic consequence is recoil motion of
the remnant black hole (see \citealt{Rosswog_DTP2000} for early
discussions in the case of asymmetric binary neutron stars). Because the
dynamical ejecta escape from the system carrying a net linear momentum
with a typical center-of-mass velocity of $0.1$--$0.2c$, the remnant
black hole obtains the recoil velocity in the direction opposite to the
ejecta. The center-of-mass velocity of the remnant black hole may become
as high as $\sim \SI{1000}{\km\per\second}$, particularly if the system
experiences significant tidal disruption. We note that significant tidal
disruption suppresses gravitational-wave recoil \citep{Shibata_KYT2009},
which might be efficient in the final plunge phase in the absence of
tidal disruption \citep{Wiseman1992,Blanchet_Qusailah_Will2005}.

As they escape from the gravitational binding of the remnant black hole,
the expansion of the dynamical ejecta gradually becomes homologous
\citep{Kyutoku_IOST2015}. In the course of this transition, the
crescent-like shape should change to a half-disk-like shape, because the
inner region is decelerated further by the influence of a deep
gravitational potential. The morphology of the ejecta may also be
modified by the radioactive decay heat of \textit{r}-process elements
\citep[see also \citealt{Rosswog_KATP2014} for binary neutron stars in
Newtonian gravity]{Darbha_KFP2021}. To derive precise observational
features related to the ejecta morphology, (not necessarily fully
general relativistic) longterm evolution for $1$--\SI{10}{\second} is
necessary. On a long time scale of $10$--$\SI{100}{yr}$ which depends
not only on the ejecta property but also on the density of the
interstellar medium, the dynamical ejecta as well as the disk outflow
(see Sect.~\ref{sec:sim_pm_wind}) are decelerated by the interstellar
medium and eventually dissolve into it \citep{Nakar_Piran2011}.

\begin{figure}[htbp]
 \centering \includegraphics[width=.7\linewidth,clip]{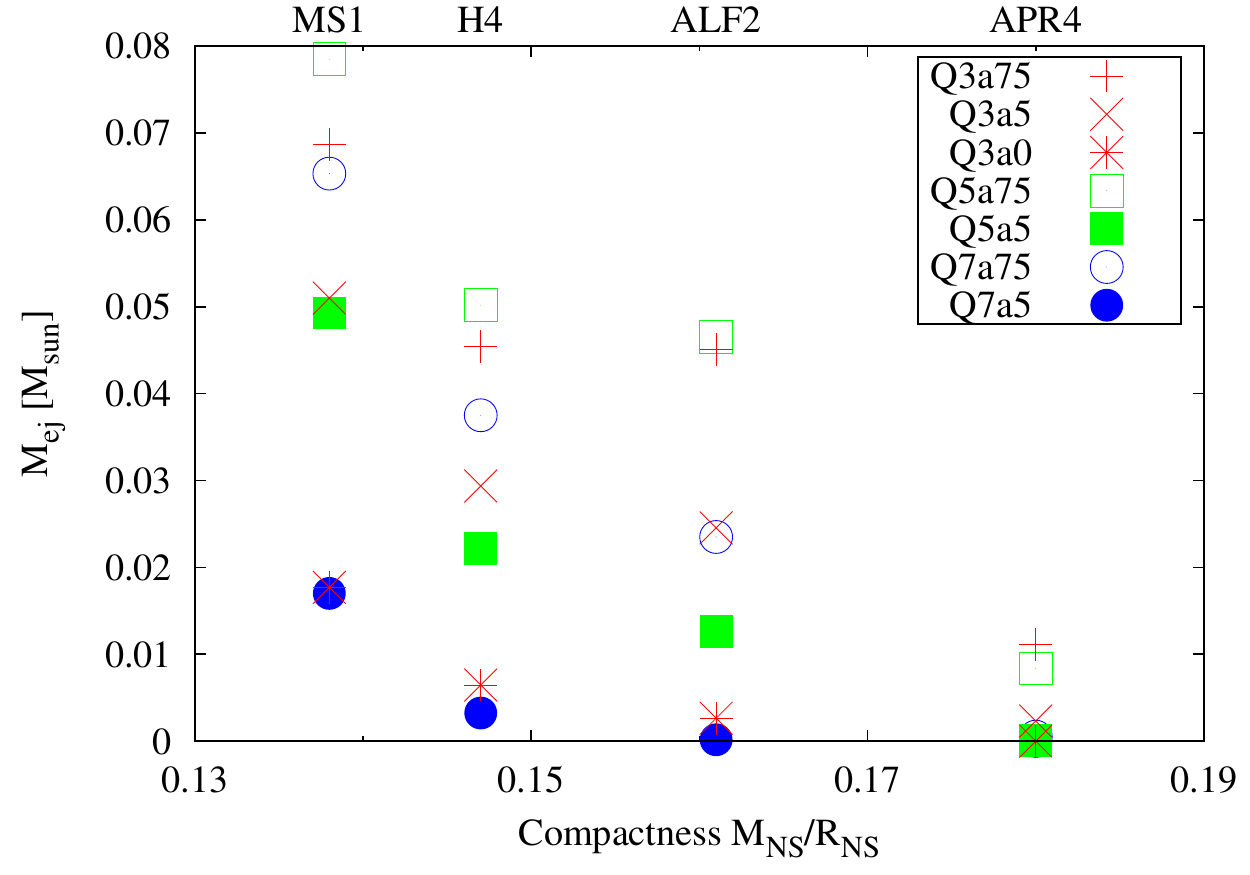}
 \caption{Mass of the dynamical ejecta for various black hole--neutron
 star binaries with $M_\mathrm{NS} = 1.35\,M_\odot$. The horizontal axis
 is the compactness of the neutron star, and corresponding equations of
 state are written above the top axis. The legend indicates the mass
 ratio and the spin parameter of the black hole, e.g., Q5a75 means $Q=5$
 and $\chi = 0.75$. Image reproduced with permission from \citet{Kyutoku_IOST2015}, copyright by APS.}
 \label{fig:mdynej}
\end{figure}

Dynamical ejecta from black hole--neutron star binaries can take a wide
range of mass depending on binary parameters
\citep{Kyutoku_IOST2015}. Figure \ref{fig:mdynej} displays the mass of
the dynamical ejecta, $M_\mathrm{ej}$, for various black hole--neutron
star binaries. If tidal disruption does not occur, the ejecta mass is
essentially zero. This is a typical outcome for binaries with a high
mass ratio, a low black-hole spin, and/or a large neutron-star
compactness such as those detected by and/or reported with gravitational
waves during the LIGO-Virgo O3 \citep{GWTC2,GW200105200115}. By
contrast, the mass can easily exceed $0.01\,M_\odot$ if tidal disruption
is significant. Notably, \citet{Lovelace_DFKPSS2013} reported that
$(0.26 \pm 0.16) \,M_\odot ( 1.4\,M_\odot / M_\mathrm{NS} )$ may be ejected
from a system with $Q=3$, $\mathcal{C} = 0.144$ modeled by a $\Gamma =
2$ polytrope, and a very high black-hole spin of $\chi = 0.97$. These
large values should be compared with dynamical ejecta from binary
neutron stars, whose mass is limited to $\lesssim 0.02 \,M_\odot$
\citep[see][for reports of $\gtrsim
0.01\,M_\odot$]{Hotokezaka_KKOSST2013,Sekiguchi_KKS2015,Sekiguchi_KKST2016,Kiuchi_KST2019,Vincent_FDHKPS2020}
unless the mass ratio is higher than an extreme value (as binary neutron
stars) of $\approx 1.5$ \citep{Dietrich_UTBB2017}.

\begin{figure}[htbp]
 \centering
 \begin{tabular}{cc}
  \includegraphics[width=.52\linewidth,clip]{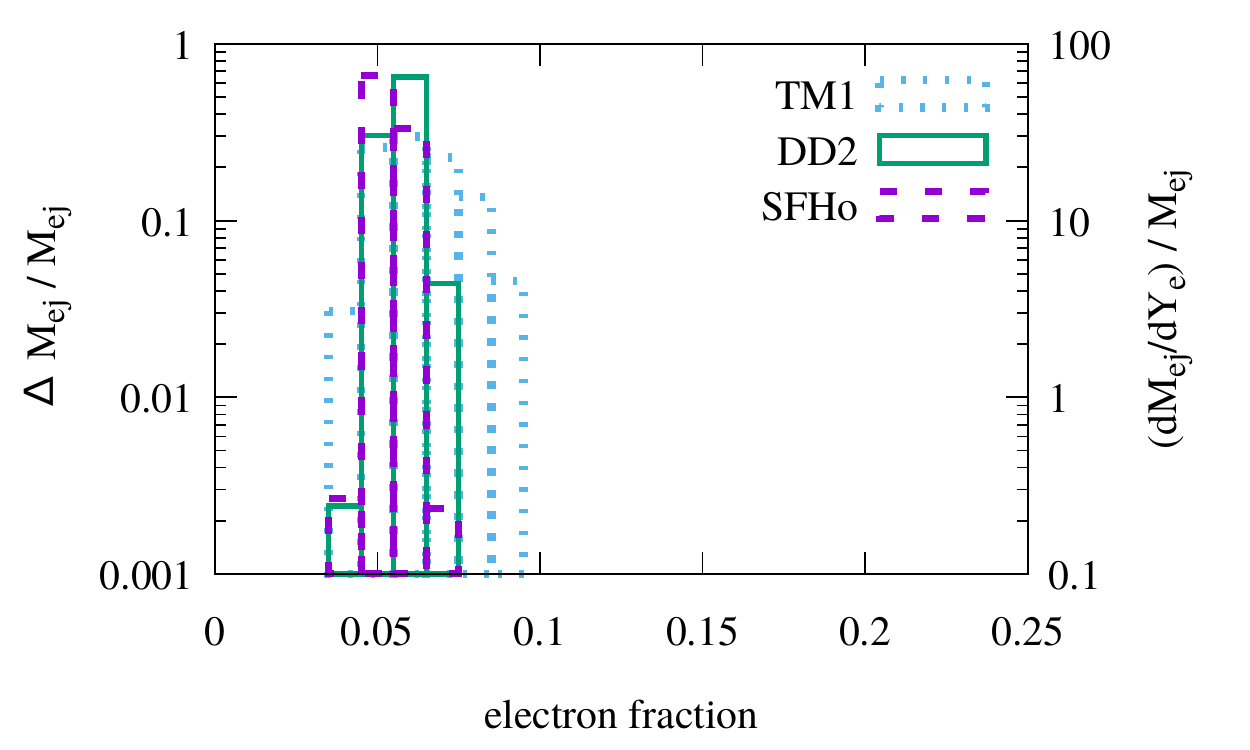} &
  \includegraphics[width=.42\linewidth,clip]{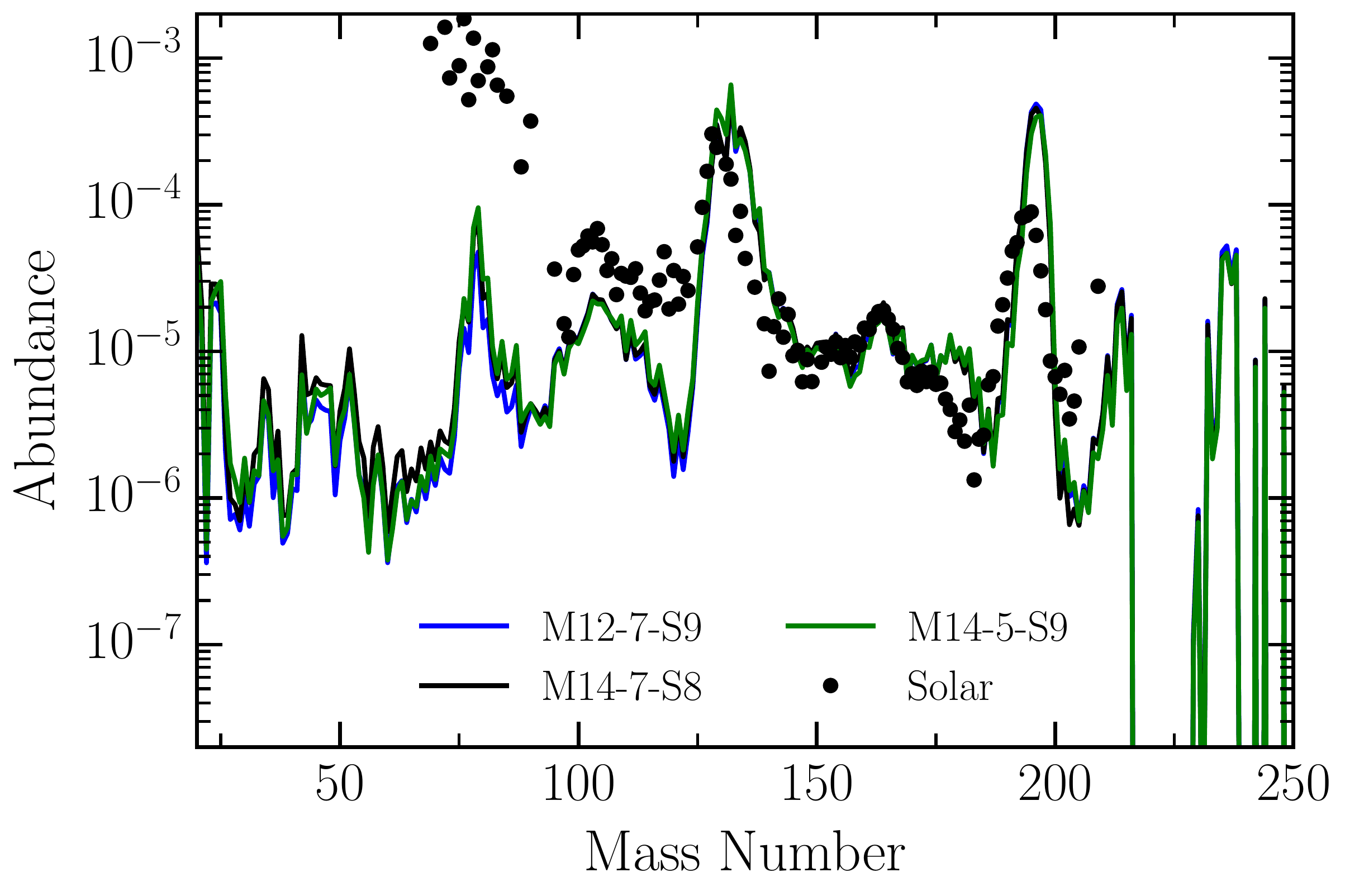}
 \end{tabular}
 \caption{Left: Mass distribution as a function of the electron fraction
 in the dynamical ejecta for binaries with $M_\mathrm{BH}=5.4\,M_\odot$,
 $\chi = 0.75$, and $M_\mathrm{NS}=1.35\,M_\odot$ ($Q=4$). Adopted
 equations of state are SFHo \citep{Steiner_Hempel_Fisher2013}, DD2
 \citep{Banik_Hempel_Bandyopadhyay2014}, and TM1 \citep{Hempel_FSL2012},
 with which the radii of $1.35\,M_\odot$ neutron stars are $11.9$, $13.2$,
 and \SI{14.5}{\km}, respectively. The simulations are performed
 incorporating neutrino irradiation. Right: Abundance pattern of the
 \text{r}-process nucleosynthesis in dynamical ejecta from various
 models of black hole--neutron star binaries simulated in
 \citet{Deaton_DFOOKMSS2013,Foucart_DDOOHKPSS2014}. Neutrino irradiation
 is taken into account by a post process, and it does not change the
 results appreciably. Images reproduced with permission from [left] \citet{Kyutoku_KSST2018}, copyright by APS; and [right] \citet{Roberts_LDFFLNOP2017}, copyright by the authors.} \label{fig:dynej_ye}
\end{figure}

Because the shock interaction plays only a minor role in the ejection
process, the dynamical ejecta keep extreme neutron richness of the cold
neutron star
\citep{Deaton_DFOOKMSS2013,Foucart_DDOOHKPSS2014,Foucart_DBDKHKPS2017,Kyutoku_KSST2018,Brege_DFDCHKOPS2018,Foucart_DKNPS2019}. The
left panel of Fig.~\ref{fig:dynej_ye} shows the mass distribution as a
function of the electron fraction for simulations with neutrino
irradiation \citep{Kyutoku_KSST2018}. This figure shows that the
electron fraction is as low as $Y_\mathrm{e} \approx 0.05$--$0.1$
preserving original values of the neutron-star material even if the
neutrino irradiation is taken into account in numerical
simulations. This should again be compared with the dynamical ejecta
from binary neutron stars, the range of whose electron fraction can be
extended up to $Y_\mathrm{e} \approx 0.4$ by the shock heating and
associated weak interactions, i.e., electron/positron captures and
neutrino irradiation
\citep{Sekiguchi_KKS2015,Palenzuela_LNLCOA2015,Foucart_HDOORKLPS2016,Sekiguchi_KKST2016,Radice_GLROR2016,Lehner_LPCOAN2016,Bovard_MGARK2017,Radice_PHFBR2018,Vincent_FDHKPS2020}. There
are two reasons for this significant difference. First, because the
dynamical ejecta from black hole--neutron star binaries do not
experience significant shock heating, the temperature is kept so low
that the electron/positron pairs are not efficiently produced. Second,
the dynamical ejecta are located at the distant region when the
circularized accretion disk starts to emit a copious amount of
neutrinos. Thus, the neutrino flux is low when neutrinos catch up the
dynamical ejecta. These two facts do not allow the dynamical ejecta to
increase the electron fraction. As a minor effect, equations of state
that predict small neutron-star radii tend to give slightly low electron
fraction. This correlation may stem from the correlation between the
symmetry energy of the nuclear matter and the radius of the neutron star
\citep{Lattimer_Prakash2001}.

The low electron fraction of the dynamical ejecta indicates that the
yield of subsequent nucleosynthesis is dominated by heavy
\textit{r}-process elements with the mass number $\gtrsim 130$, i.e.,
beyond the second peak. This expectation is confirmed in the right panel
of Fig.~\ref{fig:dynej_ye} generated by \citet{Roberts_LDFFLNOP2017},
which shows the abundance pattern of the nucleosynthesis for selected
models of black hole--neutron star binaries. The abundance pattern is
robust against variation of binary parameters, although the detailed
features depend on nuclear physics inputs \citep[see,
e.g.,][]{MendozaTemis_WLMBJ2015,Mumpower_SMA2016,Wu_FMM2016,Zhu_LBSVMMS2021}.

\begin{figure}[htbp]
 \centering \includegraphics[width=0.7\linewidth,clip]{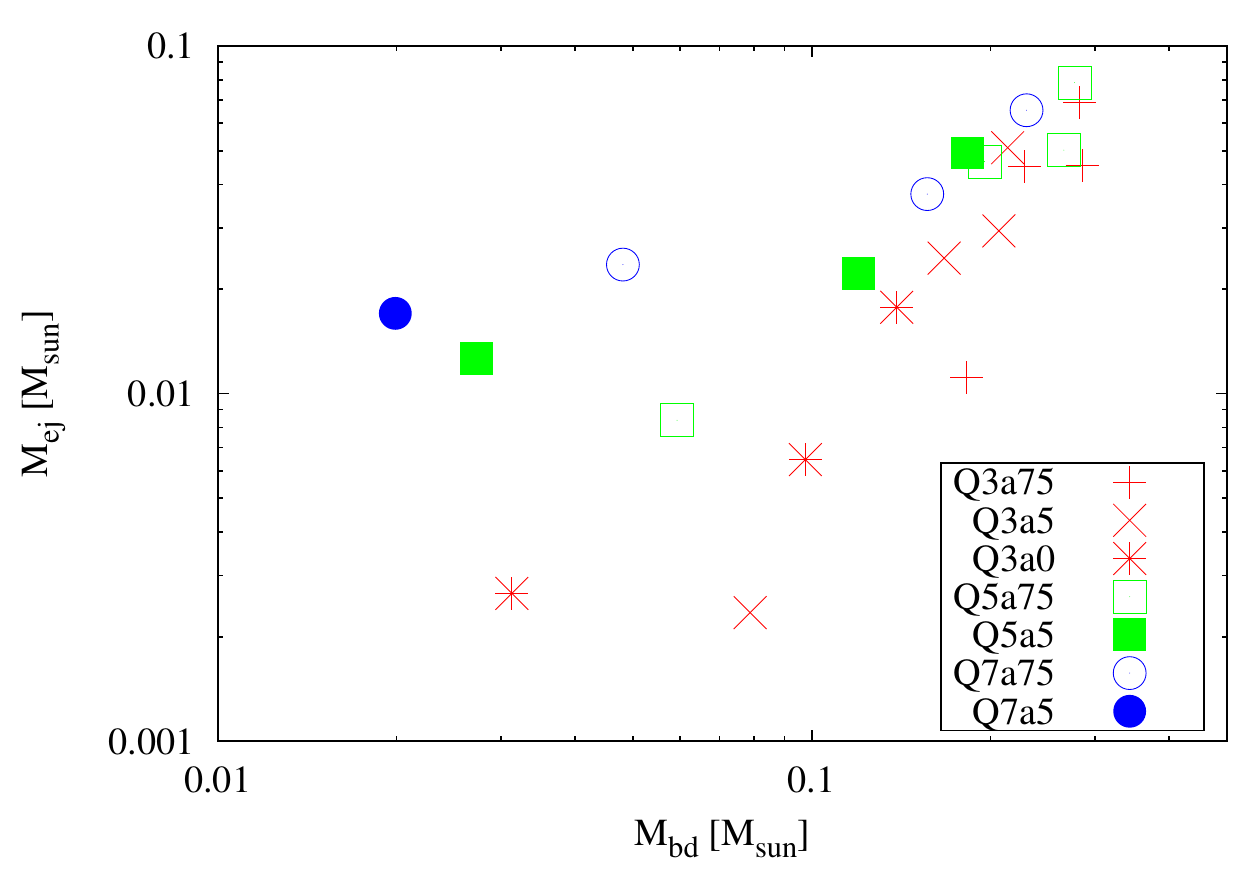}
 \caption{Comparison between the mass of the dynamical ejecta,
 $M_\mathrm{ej}$, and that of the bound material, $M_\mathrm{bd}$, which
 may be regarded as the disk mass. The mass of the neutron star
 $M_\mathrm{NS}$ is fixed to be $1.35\,M_\odot$. The legend indicates the
 mass ratio and the spin parameter of the black hole, e.g., Q5a75 means
 $Q=5$ and $\chi = 0.75$. Equations of state are varied for given values
 of $Q$ and $\chi$. Image reproduced with permission from \citet{Kyutoku_IOST2015}, copyright by APS.}
 \label{fig:mej_mdisk}
\end{figure}

One intriguing question may be as follows: ``Do the dynamical ejecta or
the disk outflow dominate the mass of the ejected material?''  This is
important for understanding abundance of \textit{r}-process elements and
emission features of the kilonova/macronova for black hole--neutron star
binaries (see also Sect.~\ref{sec:sim_pm}). In the case of binary
neutron stars, it is strongly believed that the disk outflow dominates
the dynamical ejecta from theoretical calculations and observations of
AT 2017gfo \citep[see,
e.g.,][]{Kasen_MBQR2017,Shibata_FHKKST2017,Perego_Radice_Bernuzzi2017,Villar_etal2017}.

The answer for black hole--neutron star binaries may be that it depends
on the mass ratio. Figure \ref{fig:mej_mdisk} compares the mass of the
dynamical ejecta with that of the remnant disk for various systems with
$M_\mathrm{NS} = 1.35\,M_\odot$. This figure indicates that the mass of
the dynamical ejecta is correlated with that of the bound material in a
manner dependent on the mass ratio for a parameter range depicted here,
specifically $3 \le Q \le 7$. It should be cautioned that, however, the
neutron-star mass is not varied in Fig.~\ref{fig:mej_mdisk}. In fact,
most simulations of black hole--neutron star binaries have focused on
the plausibly typical value of $M_\mathrm{NS} \lesssim
1.35$--$1.4\,M_\odot$ to date. Further investigations are required to
reveal precise dependence on binary parameters
\citep{Foucart_DDOOHKPSS2014}. In particular, high mass-ratio systems
with $Q>7$ should also be investigated systematically. Accordingly, the
following discussions need to be understood with this caveat in mind.

For high mass-ratio systems $Q \gtrsim 5$, the mass of the dynamical
ejecta depends relatively weakly on the mass of the disk. For example,
they become comparable at $M_\mathrm{bd} \approx M_\mathrm{ej} \approx
0.02\,M_\odot$ for $Q=7$. Presuming that $\sim 30\%$ could be ejected from
the remnant disk (see Sect.~\ref{sec:sim_pm_wind}), dynamical mass
ejection could be a dominant mechanism for black hole--neutron star
binaries with $Q \gtrsim 7$, particularly for the case in which the
tidal disruption is moderate (that is, the amount of material remaining
outside the black hole is not very large). As we will discuss later in
Sect.~\ref{sec:sim_pm_wind}, the realistic fraction of the material
ejected from the remnant disk is estimated to be 15\%--30\% from various
simulations.

As the mass ratio decreases, the fraction of the disk outflow
increases. Figure \ref{fig:mej_mdisk} shows that the mass of the
dynamical ejecta is typically only $\sim 5\%$--20\% of the disk mass for
$Q \sim 3$--$5$. Thus, it is likely that the amount of the disk outflow
becomes comparable to that of the dynamical ejecta. This trend is
enhanced for the regime of very-low-mass black holes, or equivalently,
very low mass ratios. The mass of the dynamical ejecta decreases to
$\lesssim \num{e-3}\,M_\odot$ for nonspinning black holes with $Q \lesssim
3$ \citep{Foucart_DKNPS2019,Hayashi_KKKS2021,Most_PTR2021}, although the
remnant disk can be as massive as $\sim 0.05$--$0.1\,M_\odot$. For these
systems, the disk outflow will dominate the amount of the ejected
material.

The mass ratio also governs the average velocity of the dynamical
ejecta, although significant dispersion is found associated with
variations of other binary parameters
\citep{Kyutoku_IOST2015,Foucart_DBDKHKPS2017,Hayashi_KKKS2021}. Here,
the average velocity is defined from the kinetic energy,
$T_\mathrm{kin}$, and the mass, $M_\mathrm{ej}$, by
$\sqrt{2T_\mathrm{kin} / M_\mathrm{ej}}$. Quantitatively, the asymptotic
velocity is as high as $\sim 0.2c$ for high mass-ratio systems with $Q
\sim 5$--$7$, for which corresponding kinetic energy reaches $\sim
\SI{e52}{erg}$. As the mass ratio decreases to $Q \lesssim 1.5$, the
asymptotic velocity decreases to $\sim 0.1c$ and the kinetic energy
decreases by orders of magnitude due to the small value of the ejecta
mass \citep{Foucart_DKNPS2019,Hayashi_KKKS2021}.

\begin{figure}[htbp]
 \centering \includegraphics[width=0.7\linewidth,clip]{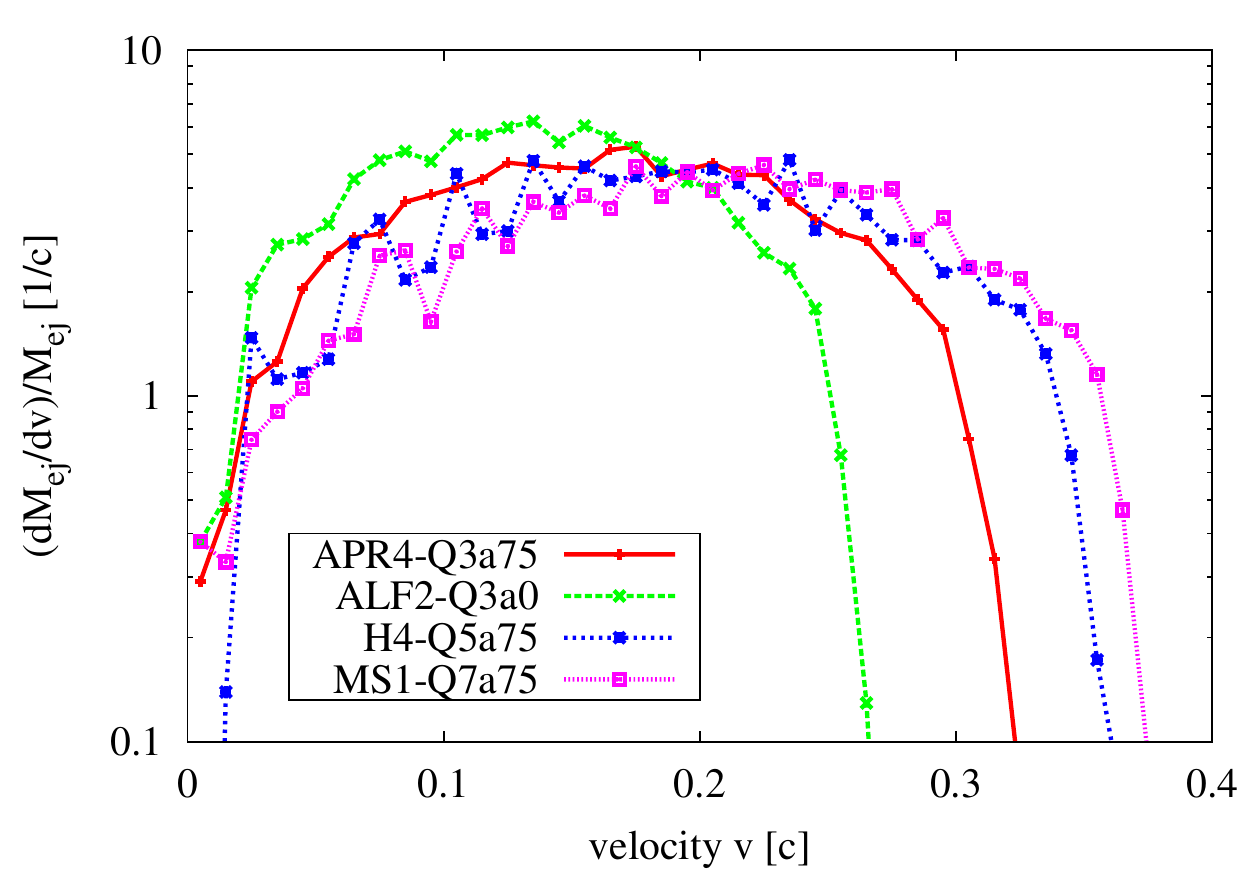}
 \caption{Mass distribution as a function of the velocity of the
 dynamical ejecta measured at \SI{10}{\ms} after the onset of merger for
 various black hole--neutron star binaries. The distribution is
 normalized by the total mass of the ejecta. It should be cautioned that
 Fig.~\ref{fig:fbspec} is drawn with respect to the specific energy
 instead of the velocity used here. Although this figure is derived by
 analyzing only the material on the equatorial plane, the conclusion
 remains the same even if all the dynamical ejecta are taken into
 account \citep{Brege_DFDCHKOPS2018,Most_PTR2021}. Image reproduced with permission from \citet{Kyutoku_IOST2015}, copyright by APS.} \label{fig:dynej_v}
\end{figure}

The velocity of the dynamical ejecta is distributed approximately
symmetrically about the averaged value and does not extend to high
velocity of $\gtrsim 0.5c$ \citep[see also
\citealt{Rosswog_Piran_Nakar2013} for early Newtonian
work]{Kyutoku_IOST2015,Brege_DFDCHKOPS2018,Most_PTR2021}. Figure
\ref{fig:dynej_v} shows the mass distribution as a function of the
velocity for various black hole--neutron star binaries
\citep{Kyutoku_IOST2015}, which is directly related to the $\tilde{E} >
0$ side of Fig.~\ref{fig:fbspec}. The cutoff features on the
highest-velocity side in these distributions are distinct from the
so-called fast tail found in binary-neutron-star mergers, which may
extend to $\gtrsim 0.8c$
\citep{Hotokezaka_KKOSST2013,Hotokezaka_KSNP2018,Radice_PHFBR2018}. This
difference may be reflected in electromagnetic counterparts,
particularly in the early part of kilonova/macronova remnants as we
discuss in Sect.~\ref{sec:dis}. We note that, although
Fig.~\ref{fig:dynej_v} is derived by analyzing only the material on the
equatorial plane, the conclusion remains the same even if all the
dynamical ejecta are taken into account
\citep{Brege_DFDCHKOPS2018,Most_PTR2021}.

Fitting formulae for the mass of the dynamical ejecta are provided by
\citet{Kawaguchi_KST2016,Kruger_Foucart2020}, where
\citet{Kawaguchi_KST2016} also derive a formula for the velocity. A
fitting formula for the disk outflow is also proposed based on the
fitting formulae for the mass of the material remaining outside the
black hole and for the dynamical ejecta \citep{2021arXiv210211569R},
while the efficiency of the ejection depends on the mass and the
compactness of the disk and thus is highly uncertain (see
Sect.~\ref{sec:sim_pm_wind}).

\subsection{Postmerger activity} \label{sec:sim_pm}

The remnant disk is considered to evolve via neutrino emission and
magnetohydrodynamical turbulence on a long viscous time scale. In a wide
range of astrophysics studies, the kinematic shear viscosity $\nu$ in
the accretion disk is often parametrized by the so-called alpha
parameter $\alpha_\nu$ as \citep{Shakura_Sunyaev1973}
\begin{equation}
 \nu = \alpha_\nu c_s H ,
\end{equation}
where $c_s$ and $H$ are the sound speed and the scale height of the
disk, respectively. The value of $\alpha_\nu$ is believed to be
determined by magnetohydrodynamical processes as we describe in
Sect.~\ref{sec:sim_pm_mag}, and recent studies suggest $\alpha_\nu
\approx 0.01$--$0.1$ is reasonable
\citep{Fernandez_TQFK2019,Christie_LTFFQK2019}. If we presume such
values of $\alpha_\nu$, the viscous time scale is estimated by
\begin{equation}
 t_\mathrm{vis} := \frac{R^2}{\nu} \approx \SI{0.33}{s}
  \pqty{\frac{\alpha_\nu}{0.03}}^{-1} \pqty{\frac{c_s}{0.1c}}^{-1}
  \pqty{\frac{H/R}{1/3}}^{-1} \pqty{\frac{R}{\SI{100}{\km}}} ,
  \label{eq:tvis}
\end{equation}
where $R$ is the cylindrical radius of the disk. This is much longer
than the dynamical time scale of the disk,
\begin{equation}
 t_\mathrm{dyn} := 2\pi \sqrt{\frac{R^3}{GM_\mathrm{BH,f}}} =
  \SI{5.5}{\ms} \pqty{\frac{R}{\SI{100}{\km}}}^{3/2}
  \pqty{\frac{M_\mathrm{BH,f}}{10\,M_\odot}}^{-1/2} .
\end{equation}
Furthermore, $t_\mathrm{vis}$ becomes even longer for an outer region of
the disk. This comparison indicates that a longterm simulation of
$\order{1}\si{\second} \gg t_\mathrm{dyn}$ is required for understanding
the evolution of the accretion disk. Because the evolution is also
governed by neutrino emission, implementation of neutrino transfer is
another key ingredient for quantitative exploration. These facts make
the physical postmerger simulation computationally challenging.

To date, only a few work have reported longterm simulations of an
accretion disk surrounding a black hole in full general relativity i.e.,
numerical relativity,
\citep{Shibata_Sekiguchi2012,Foucart_ORDHKOPSS2015,Fujibayashi_SWKKS2020,Fujibayashi_SWKKS2020-2,Most_PTR2021-2},
although many work have been done in (pseudo-)Newtonian gravity
\citep{Setiawan_Ruffert_Janka2004,Lee_RamirezRuiz_Page2005,Setiawan_Ruffert_Janka2006,Fernandez_Metzger2013,Just_BAGJ2015,Fernandez_Foucart_Lippuner2020}
or with adopting a fixed background spacetime
\citep{Shibata_Sekiguchi_Takahashi2007,Siegel_Metzger2017,Siegel_Metzger2018,Fernandez_TQFK2019,Miller_RDBFFKLMW2019,Christie_LTFFQK2019},
in some cases starting with initial conditions taken from merger
simulations \citep[Cowling approximation;][]{Nouri_etal2018}. In this
Sect.~\ref{sec:sim_pm}, we review the current status of our
understanding for neutrino emission, the disk outflow powered by the
viscous heating, and the effect of magnetic fields, focusing on the
results obtained by fully general-relativistic simulations. We caution
that magnetohydrodynamics simulations in numerical relativity as long as
\SI{1}{\second} have not been reported and are needed for robust
understanding of the postmerger evolution in the future \citep[see][for
a recent simulation up to $\sim
\SI{350}{\milli\second}$]{Most_PTR2021-2}.

\subsubsection{Neutrino emission} \label{sec:sim_pm_nu}

The hot and massive remnant accretion disk emits a copious amount of
neutrinos primarily via electron/positron captures onto nucleons
\citep[see also \citealt{Janka_ERF1999} for pioneering study in
Newtonian
gravity]{Deaton_DFOOKMSS2013,Foucart_DDOOHKPSS2014,Kyutoku_KSST2018}. If
a massive accretion disk of $\gtrsim 0.1\,M_\odot$ is formed around a
rapidly spinning black hole with $\lesssim 10\,M_\odot$, the luminosity
reaches $\gtrsim \SI{e53}{erg.s^{-1}}$ at $\approx \SI{10}{\ms}$ after
the onset of merger, when the tidal tail collides with itself and forms
a remnant disk. The order of the peak luminosity is the same as that
found for stellar core collapse and binary-neutron-star mergers. The
emission efficiency is a few to several percent of the accretion rate
onto the black hole for a disk with the maximum rest-mass density of
$\gtrsim \SI{e11}{\gram\per\cubic\cm}$ and increases as the spin
parameter of the remnant black hole increases
\citep{Fujibayashi_SWKKS2020,Fujibayashi_SWKKS2020-2}. Because the
optical depth of the disk to neutrinos can exceed unity (see, e.g.,
\citealt{Deaton_DFOOKMSS2013,Foucart_ORDHKOPSS2015,Nouri_etal2018}), the
neutrino luminosity is saturated to $\sim \SI{e53}{erg.s^{-1}}$ in the
early stage of the postmerger evolution until the disk becomes optically
thin to neutrinos as a result of the accretion onto the black hole
\citep[see also][]{Lee_RamirezRuiz_Page2005,Setiawan_Ruffert_Janka2006}.

\begin{figure}[htbp]
 \centering \includegraphics[width=0.7\linewidth,clip]{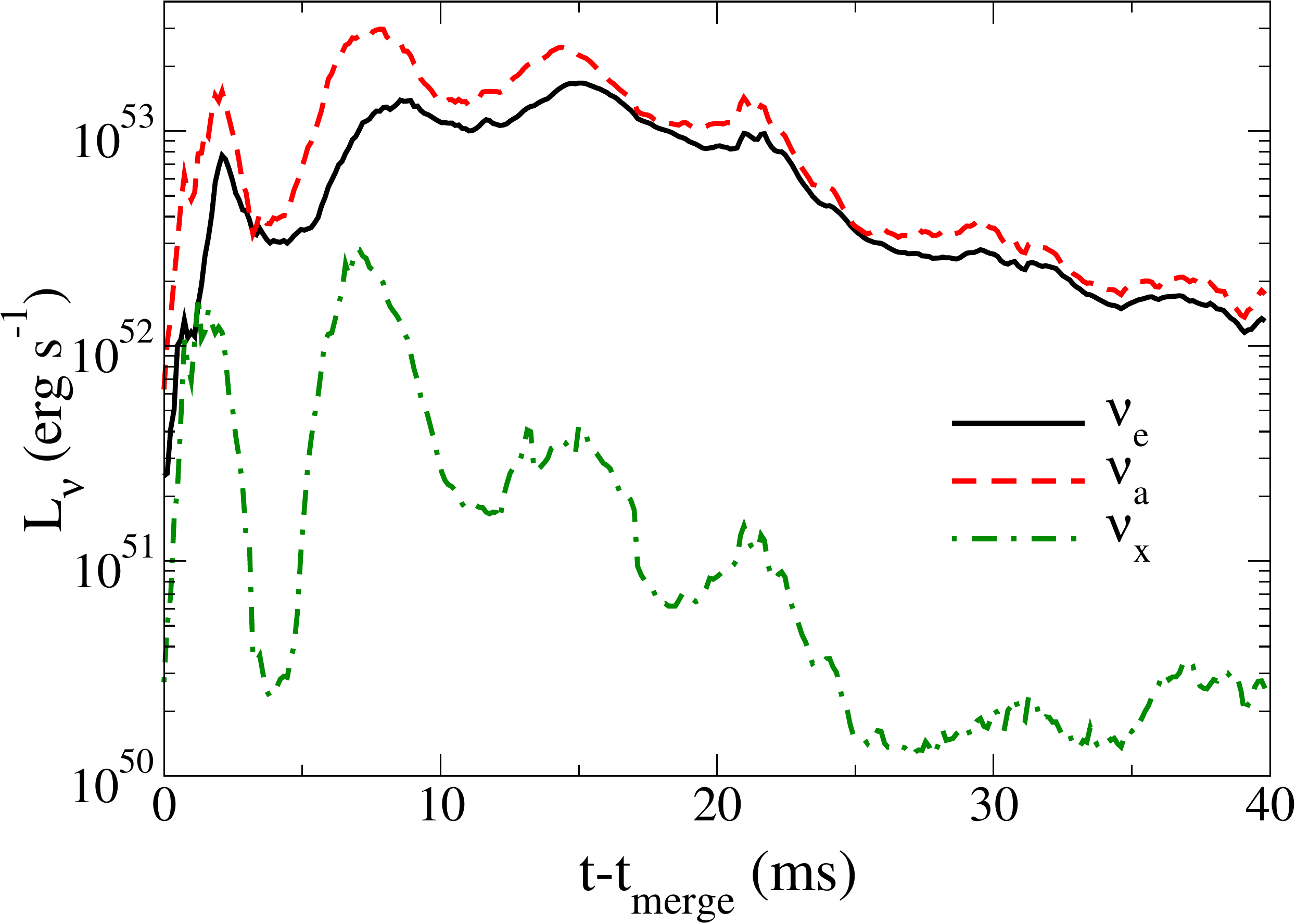}
 \caption{Time evolution of the luminosity for electron neutrinos
 ($\nu_\mathrm{e}$: black solid), electron antineutrinos
 ($\nu_\mathrm{a}$: red dashed), and one of the other neutrinos
 ($\nu_\mathrm{x}$: green dot-dashed), for the same model as that shown
 in Fig.~\ref{fig:snap_mer}. The disk mass at \SI{20}{\ms} after the
 onset of merger is reported to be $0.08\,M_\odot$. The mass and the spin
 parameter of the remnant black hole are $8.00\,M_\odot$ and $0.87$,
 respectively. Image reproduced with permission from \citet{Foucart_DDOOHKPSS2014}, copyright by APS.}
 \label{fig:nuluminosity}
\end{figure}

Among the six species of neutrinos, electron antineutrinos always carry
away the largest amount of energy. Figure \ref{fig:nuluminosity}
generated by \citet{Foucart_DDOOHKPSS2014} shows a typical example of
luminosity evolution in an early stage of the postmerger accretion disk
\citep{Foucart_DDOOHKPSS2014}. Quantitatively, the peak luminosity of
electron antineutrinos typically reaches $\gtrsim \SI{e53}{erg.s^{-1}}$
if the accretion disk with $\gtrsim 0.1\,M_\odot$ is formed, and it is
higher than the luminosity of electron neutrinos by a factor of $\sim
2$.

The reasons for the dominance of electron antineutrinos are twofold, and
both are ascribed to the neutron-rich composition of the remnant disk
inherited from the neutron star. First, the number of capture reactions
is larger for positrons than for electrons, because the remnant disk
equilibrates toward a protonized state from a neutron-rich
state. Second, the neutron-rich disk is optically thicker to electron
neutrinos than to electron antineutrinos. This feature puts the
neutrinosphere (an analog of the photosphere for photons) for electron
antineutrinos at high-temperature regions close to the midplane of the
disk. Thus, the number of emitted neutrinos is larger and the energy of
individual neutrinos is higher for electron antineutrinos than for
electron neutrinos. The dominance in number should be contrasted with
neutrino emission from core-collapse supernovae, in which electron
neutrinos are larger in number than electron antineutrinos due to
continuous deleptonization. This difference may introduce differences in
neutrino oscillations such as matter-neutrino resonances \citep[see also
the end of this Sect.~\ref{sec:sim_pm_nu}]{Malkus_KMS2012}. Muon and tau
neutrinos and their antineutrinos, denoted collectively by ``x''
neutrinos in the context of stellar core collapse and compact binary
mergers, are typically dimmer by more than an order of magnitude than
two dominant species. This is because the x neutrinos are emitted only
via neutral-current processes and the temperature of the disk is
relatively low as described in Sect.~\ref{sec:sim_rem_disk2}.

\begin{figure}[htbp]
 \centering \includegraphics[width=0.7\linewidth,clip]{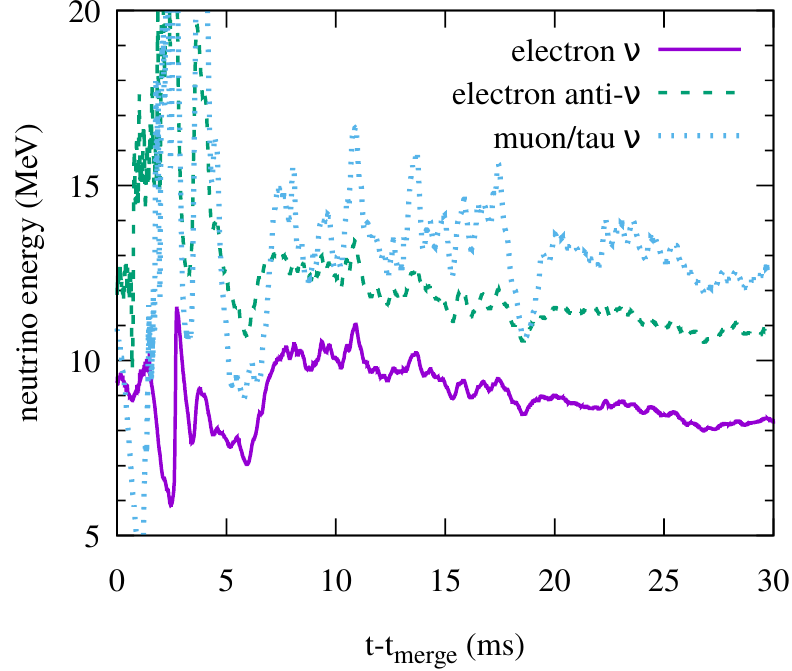}
 \caption{Time evolution of the average energy of electron neutrinos
 (purple solid), electron antineutrinos (green dashed), and other
 neutrinos (cyan dotted) for a binary with $M_\mathrm{BH} = 5.4\,M_\odot$,
 $\chi = 0.75$, $M_\mathrm{NS} = 1.35\,M_\odot$,
 $R_\mathrm{NS}=\SI{13.2}{\km}$ ($Q=4$, $\mathcal{C}=0.151$) modeled by
 the DD2 equation of state \citep{Banik_Hempel_Bandyopadhyay2014}. This
 figure is generated from data of \citet{Kyutoku_KSST2018}.}
 \label{fig:nuenergy}
\end{figure}

More precisely, the energy of individual neutrinos is the lowest for
electron neutrinos, the middle for electron antineutrinos, and the
highest for x neutrinos. Quantitatively, for example in the case of
$M_\mathrm{BH} = 5.4\,M_\odot$, $\chi = 0.75$, and $M_\mathrm{NS} =
1.35\,M_\odot$ ($Q=4$) shown in Fig.~\ref{fig:nuenergy}, the average
values of the energy for electron neutrinos, electron antineutrinos, and
x neutrinos are $8$, $11$, and \SI{13}{\mega\eV}, respectively, at
\SI{20}{\ms} after the onset of merger, depending weakly on the equation
of state \citep{Kyutoku_KSST2018}. The average values of neutrino energy
owe their hierarchy to the optical depth of the disk to each species of
neutrinos described above. That is, the average values reflect the
location and the temperature of the neutrinosphere for each species. We
caution that the energy of neutrinos is derived only approximately by
the total energy and number of neutrinos in these
simulations. Quantitative estimation of the neutrino energy requires a
multienergy transport scheme, which has never been adopted in
simulations of black hole--neutron star binaries. Studies such as the
one conducted by \citet{Foucart_DHKPS2020} for binary neutron stars are
necessary.

The peak neutrino luminosity does not always increase with the increase
of the disk mass, because the temperature can be lower for a more
massive disk \citep{Kyutoku_KSST2018}. As we discussed in
Sect.~\ref{sec:sim_rem_disk}, the disk mass is smaller for a more
compact neutron star if the other binary parameters are fixed. However,
as we discussed in Sect.~\ref{sec:sim_rem_disk2}, the temperature tends
to be higher for a more compact neutron star, because tidal disruption
occurs at an orbit closer to the black hole, and reflecting higher
velocity at the closer orbit, the shock interaction results in higher
temperature. As the emission rate of energy via the electron/positron
capture is approximately proportional to $T^6$
\citep{Fuller_Fowler_Newman1985}, the temperature, $T$, plays a more
decisive role in neutrino emission than the disk mass or the rest-mass
density.

To date, no inspiral-merger-postmerger simulation in neutrino-radiation
hydrodynamics has incorporated well-resolved magnetohydrodynamical or
viscous heating of the disk, and thus the luminosity for $\gtrsim
\SI{10}{\ms}$ after the onset of merger is likely to be underestimated
\citep[but see also][for the effort]{Most_PTR2021,Most_PTR2021-2}. In
reality, longterm neutrino emission is controlled by
magnetically-induced viscous heating, which compensates the neutrino
cooling
\citep{Lee_RamirezRuiz_Page2005,Setiawan_Ruffert_Janka2006}. Magnetohydrodynamics
simulations in the Cowling approximation found that magnetically-induced
turbulent viscosity enhances the neutrino luminosity by a factor of
$\sim 2$ for initial $\sim \SI{50}{\ms}$ if the magnetic-field strength
is $\gtrsim \SI{e15}{G}$ at the maximum inside the accretion disk
\citep{Nouri_etal2018}. Further longterm simulations have been performed
in viscous-hydrodynamics numerical-relativity simulations in
axisymmetry, which are described in Sect.~\ref{sec:sim_pm_wind} with a
particular emphasis on the disk outflow
\citep{Fujibayashi_SWKKS2020,Fujibayashi_SWKKS2020-2}. Generally
speaking, the neutrino luminosity is higher than $\SI{e53}{erg.s^{-1}}$
only in the initial $10$--\SI{100}{\ms} and decreases to $\lesssim
\SI{e50}{erg.s^{-1}}$ at $\sim \SI{1}{\second}$ after the disk
formation, because the time scale of the weak interaction becomes too
long due to the viscous disk expansion. Thus, it is not very likely that
neutrino pair annihilation \citep[see also the next
paragraph]{Rees_Meszaros1992} can drive an ultrarelativistic jet for the
entire duration of short-hard gamma-ray bursts, which is typically
$0.1$--$\SI{1}{\second}$ (see, e.g., \citealt{Nakar2007,Berger2014} for
reviews), in black hole--neutron star binary coalescences. It should
also be noted that neutrino-driven winds have not been observed in
numerical-relativity simulations without viscosity
\citep{Kyutoku_KSST2018}. Neutrino absorption is not found to enhance
the mass of the viscous disk outflow, either, in numerical-relativity
simulations with viscosity \citep{Fujibayashi_SWKKS2020}.

We should mention two neutrino processes that have never been
incorporated in numerical-relativity simulations of black hole--neutron
star binaries as a topic for future investigations. One is the neutrino
pair annihilation, which could power an ultrarelativistic jet
\citep{Mochkovitch_HIM1993,Janka_ERF1999,Birkl_AJM2007,Zalamea_Beloborodov2011}. Although
this process may be handled semiquantitatively within the moment
formalism \citep{Fujibayashi_SKS2017,Fujibayashi_KNSS2018}, solving
Boltzmann's equation directly will be valuable for incorporating precise
angular dependence of neutrinos (see, e.g.,
\citealt{Cardall_Endeve_Mezzacappa2013,Shibata_NSY2014} for approaches
in numerical relativity). Monte-Carlo neutrino-radiation transport is
also useful to incorporate pair annihilation
\citep{Foucart_DHKPS2020}. The other is the neutrino oscillation, which
could modify nucleosynthetic yields for some part of the ejecta
\citep{Malkus_KMS2012,Malkus_McLaughlin_Surman2016,Wu_Tamborra2017,Wu_TJJ2017,PadillaGay_Shalgar_Tamborra2021,Li_Siegel2021}. Its
modeling will require us to address the challenging task of solving
quantum kinetic equations in a dynamical spacetime (see, e.g.,
\citealt{Richers_MKV2019} for relevant work). The so-called fast flavor
conversion is partially incorporated in a disk simulation in a fixed,
Kerr background \citep{Li_Siegel2021}.

\subsubsection{Disk outflow} \label{sec:sim_pm_wind}

The long time scale of $\gtrsim \SI{1}{\second}$ for the evolution of
the remnant disk makes its accurate simulation a challenging task. In
particular, because magnetohydrodynamic instabilities are often
characterized by a short wavelength and magnetohydrodynamical turbulence
can be maintained only in three-dimensional simulations (see
\citealt{Balbus_Hawley1998} for reviews), the computational cost is
extremely high if we try to study the longterm evolution of the remnant
disk by magnetohydrodynamics simulations irrespective of whether the
gravity is Newtonian or relativistic. To save computational costs, it is
customary to adopt viscous hydrodynamics with a value of $\alpha_\nu$
chosen phenomenologically to reproduce results of high-resolution
magnetohydrodynamics simulations. Here, we need to keep in mind that
this prescription does not reproduce all of the magnetohydrodynamical
effects in a faithful manner (see also the final paragraph of
Sect.~\ref{sec:sim_pm_wind}).

Numerical-relativity simulations of black hole--accretion disk systems
have recently been performed in two-dimensional, axisymmetric
neutrino-radiation viscous hydrodynamics
\citep{Fujibayashi_SWKKS2020,Fujibayashi_SWKKS2020-2}. Initial
conditions of these work are given by an axisymmetric equilibrium disk
surrounding a black hole \citep{Shibata2007}. While the entropy per
baryon is taken to be constant throughout the disk, the profiles of the
angular momentum and the electron fraction are modeled by functions
motivated by results of inspiral-merger-postmerger simulations of black
hole--neutron star binaries in neutrino-radiation hydrodynamics
\citep{Kyutoku_KSST2018}. The equation of state is given by the DD2
equation of state for high density
\citep{Banik_Hempel_Bandyopadhyay2014} and by the Helmholtz equation of
state for low density \citep{Timmes_Swesty2000}. In the following, we
describe properties of the viscous disk outflow investigated in
\citet[see also \citealt{Fernandez_Metzger2013,Just_BAGJ2015} for
pioneering pseudo-Newtonian
simulations]{Fujibayashi_SWKKS2020,Fujibayashi_SWKKS2020-2}. We focus
mainly on the results for the fiducial value of $\alpha_\nu = 0.05$.

\begin{figure}[htbp]
 \centering
 \begin{tabular}{cc}
  \includegraphics[width=0.47\linewidth,clip]{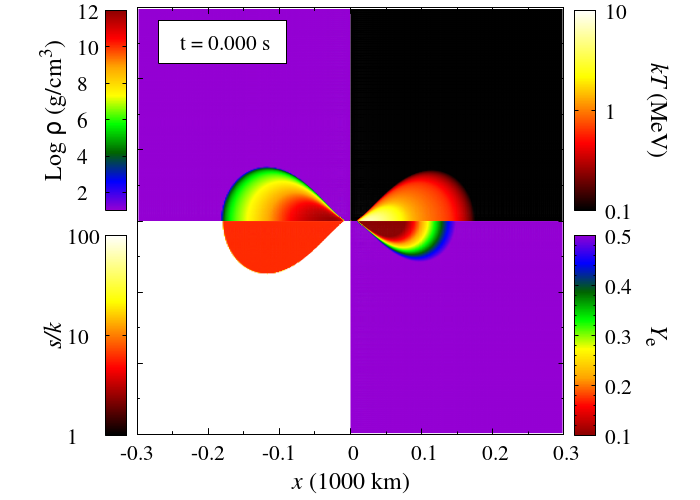} &
  \includegraphics[width=0.47\linewidth,clip]{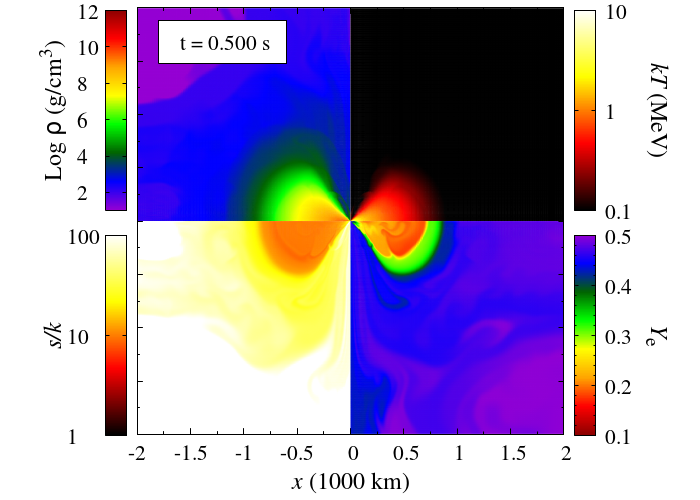} \\
  \includegraphics[width=0.47\linewidth,clip]{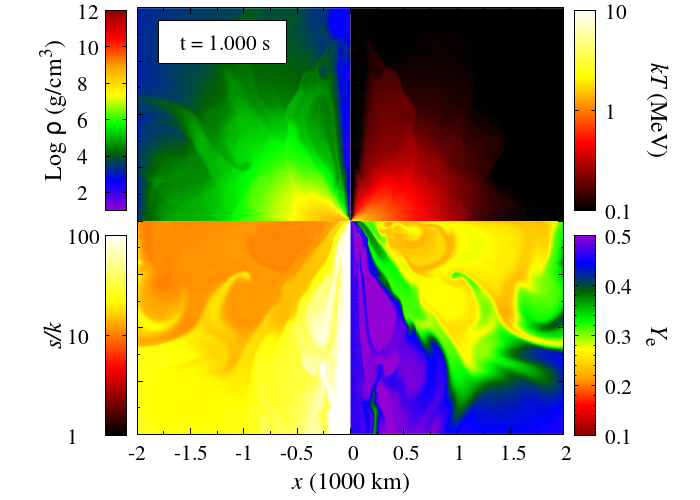} &
  \includegraphics[width=0.47\linewidth,clip]{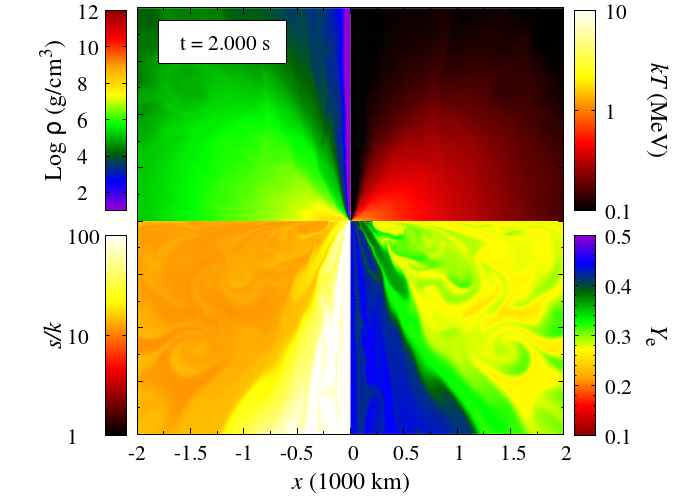}
 \end{tabular}
 \caption{Profile of the rest-mass density (left top), the temperature
 in units of \si{\mega\eV} (right top), the entropy per baryon in units
 of the Boltzmann constant (left bottom), and the electron fraction
 (right bottom) at \SI{1}{\second} (left) and \SI{2}{\second} (right)
 after the start of viscous evolution for a $0.1\,M_\odot$ disk
 surrounding a $3\,M_\odot$ black hole with $\chi = 0.8$. The size of the
 drawing area is $\SI{300}{\meter}$ for the left top panel and
 $\SI{2}{\km}$ for the others. The alpha parameter $\alpha_\nu$ is taken
 to be $0.05$. Image reproduced with permission from \citet{Fujibayashi_SWKKS2020}, copyright by the authors.}
 \label{fig:snap_visc}
\end{figure}

In the early stage of the postmerger evolution, most of the internal
energy generated by the viscous heating is consumed by the neutrino
emission from the accretion disk. The viscous disk outflow is
insignificant during the stage in which the neutrino cooling is
efficient. As the temperature and the rest-mass density decrease due to
the viscous disk expansion and the mass accretion onto the black hole,
the time scale for weak interactions becomes long. Once it exceeds the
viscous time scale of $\order{1}\si{\second}$, the internal energy
generated by the viscous heating increases the entropy in the innermost
region, because the neutrino emission becomes inefficient. As shown in
Fig.~\ref{fig:snap_visc}, the entropy gradient activates convective
motion so that the disk begins to expand significantly at $\approx
\SI{0.5}{\second}$ after the start of viscous evolution. The viscous
disk outflow sets in as well due to the energy transport associated with
the convection. Because the time scale of the convective motion, $\sim
\SI{10}{\ms}$, is much shorter than the viscous time scale, the outer
part of the disk immediately gains energy once the neutrino cooling
becomes inefficient.

\begin{figure}[htbp]
 \centering
 \begin{tabular}{c}
  \includegraphics[width=0.7\linewidth,clip]{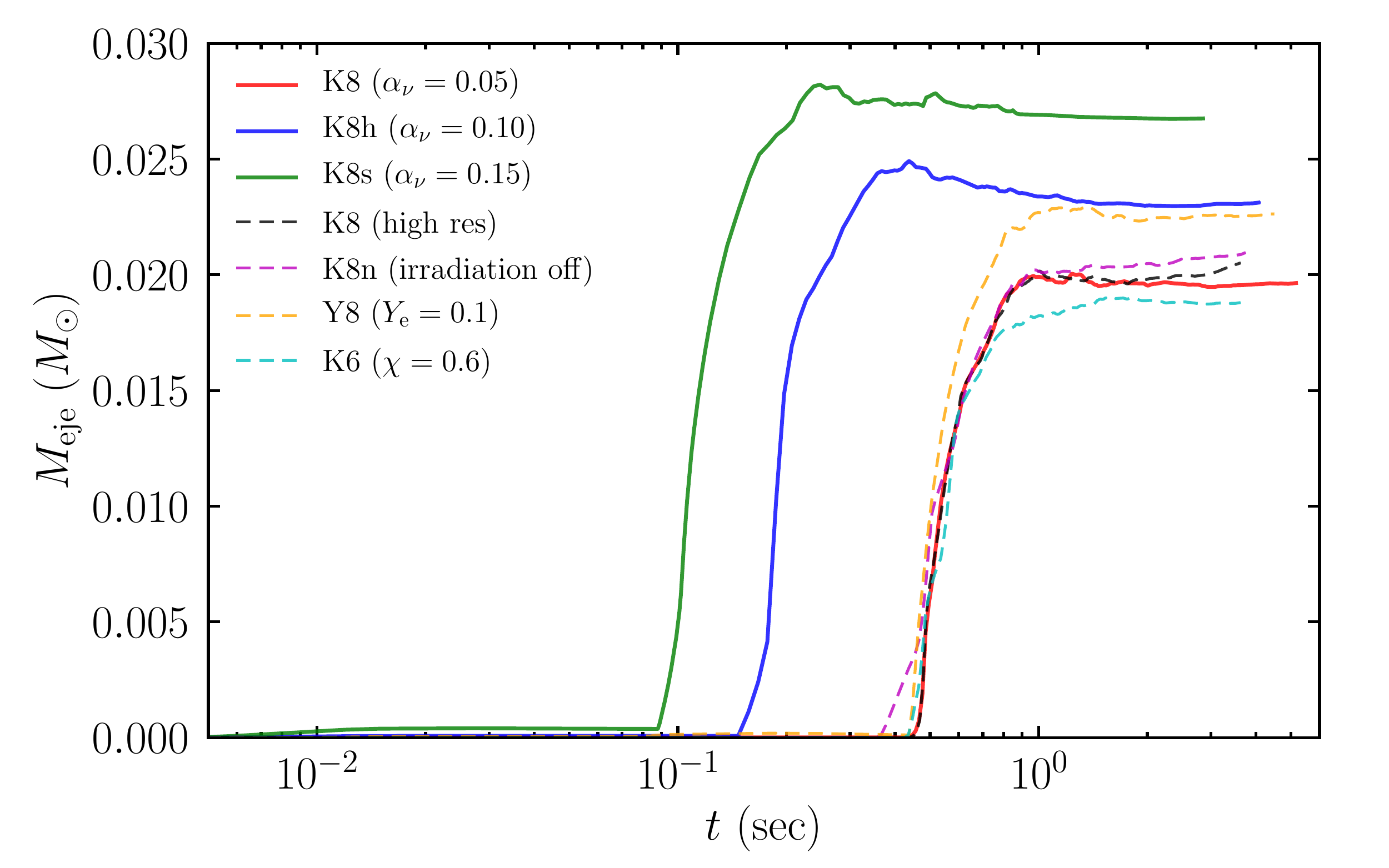} \\
  \includegraphics[width=0.7\linewidth,clip]{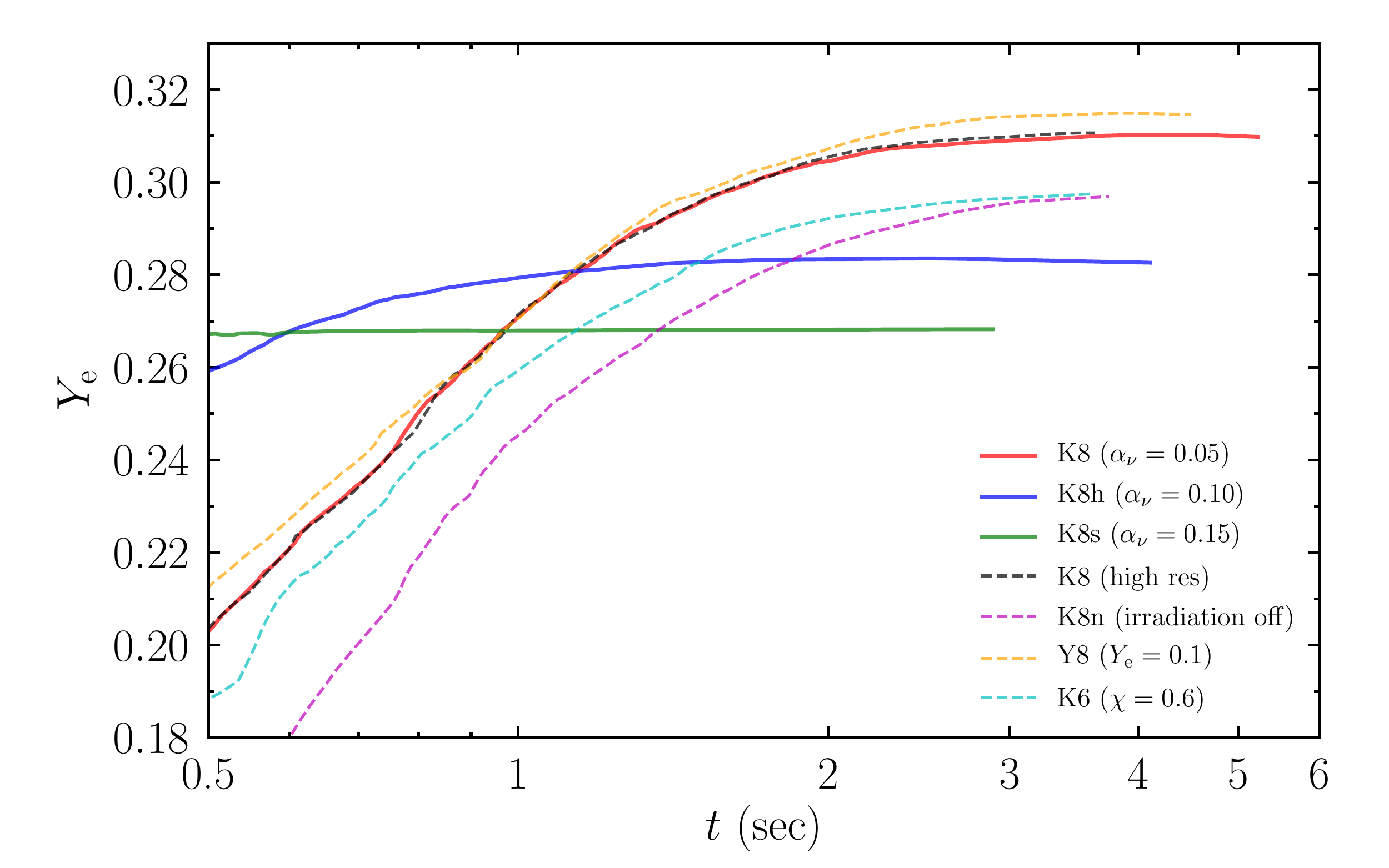}
 \end{tabular}
 \caption{Time evolution of the mass (top) and the electron fraction
 (bottom) of the material ejected from a $0.1\,M_\odot$ disk surrounding a
 $3\,M_\odot$ black hole for several initial configurations and three
 values of $\alpha_\nu$. The initial radius of the disk is $\sim
 \SI{180}{\km}$, and the spin parameter of the black hole $\chi$ is
 $0.8$ except for the model denoted by K6 with $\chi = 0.6$. Note that
 the scales of the horizontal axes are different between two
 panels. Image reproduced with permission from \citet{Fujibayashi_SWKKS2020}, copyright by the authors.}
 \label{fig:wind}
\end{figure}

The viscous outflow from the remnant disk of black hole--neutron star
binary mergers typically ejects 15\%--30\% of the initial disk material
\citep[see also, e.g.,][for a wider range of
predictions]{Siegel_Metzger2018,Fernandez_TQFK2019,Christie_LTFFQK2019,Fernandez_Foucart_Lippuner2020}. The
precise fraction of the ejected mass depends on the physical condition
of the black hole--accretion disk system such as the mass and the
compactness of the disk as well as the effective viscosity
\citep{Fujibayashi_SWKKS2020,Fujibayashi_SWKKS2020-2}. In particular,
the higher viscosity, or equivalently the larger value of the alpha
parameter, results in the larger fraction. The top panel of
Fig.~\ref{fig:wind} shows the time evolution of the mass ejected from
the system \citep{Fujibayashi_SWKKS2020}. This figure shows that the
amount of ejected material increases appreciably as the value of
$\alpha_\nu$ increases (compare K8, K8h, and K8s). Thus, accurate
determination of the magnitude of the kinematic viscosity is important
to predict quantitatively the amount of the disk outflow. Physically, it
is necessary to clarify how the magnetorotational instability enhances
the turbulence (see Sect.~\ref{sec:sim_pm_mag}).

The velocity of the viscous disk outflow is typically $0.05$--$0.1c$,
depending only weakly on the physical condition of the system. This
value is smaller than that for the dynamical ejecta, because this
outflow is driven primarily from the outer region of the disk, where the
typical velocity scale is smaller than the orbital velocity at tidal
disruption during merger. The viscous disk outflow is driven nearly
isotropically except for a narrow polar region, into which the angular
momentum barrier prohibits the penetration of the material.

Although the black hole--accretion disk system does not host a strong
neutrino emitter such as the hot and massive neutron star, the average
electron fraction of the disk increases to $\expval{Y_\mathrm{e}}
\approx 0.3$ in $t_\mathrm{vis} \sim \SI{0.5}{\second}$ from the onset
of viscous disk evolution, and the viscous disk outflow is also
characterized by relatively high electron fraction if it is launched
later than $t_\mathrm{vis}$ (see below for the discussion about the case
of rapid launch). The increase of the electron fraction is ascribed to
the relaxation to the equilibrium of electron/positron captures onto
nucleons. Because the rest-mass density of the disk decreases to $\rho
\lesssim \SI{e9}{\gram\per\cubic\cm}$ prior to the launch of the outflow
in this case, the degeneracy of electrons is not strong with $kT \sim
\SI{2}{\mega\eV}$ where $k$ is the Boltzmann constant. The bottom panel
of Fig.~\ref{fig:wind} displays the time evolution of the average
electron fraction of the ejected material and shows that the average
electron fraction is higher than $0.25$ for all the cases studied in
\citet{Fujibayashi_SWKKS2020}. Although neutrino irradiation from the
disk itself helps to increase the electron fraction by $\approx 0.05$,
this works only in an early evolution stage of $\lesssim
\SI{0.1}{\second}$ and is not a key ingredient for increasing the
electron fraction of the disk outflow. This fact is found by comparing
models K8 (neutrino irradiation is taken into account) and K8n (not
taken into account) in the bottom panel of Fig.~\ref{fig:wind}. Because
the rest-mass density is lower for disks surrounding more massive black
holes as we have discussed in Sect.~\ref{sec:sim_rem_disk2}, the
electron fraction is even higher due to the weaker degeneracy of
electrons \citep{Fujibayashi_SWKKS2020-2}. An extensive study of the
dependence of the electron fraction on various physics inputs is
performed within pseudo-Newtonian gravity by
\citet{2021arXiv210208387J}.

\begin{figure}[htbp]
 \centering
 \begin{tabular}{cc}
  \includegraphics[width=.47\linewidth,clip]{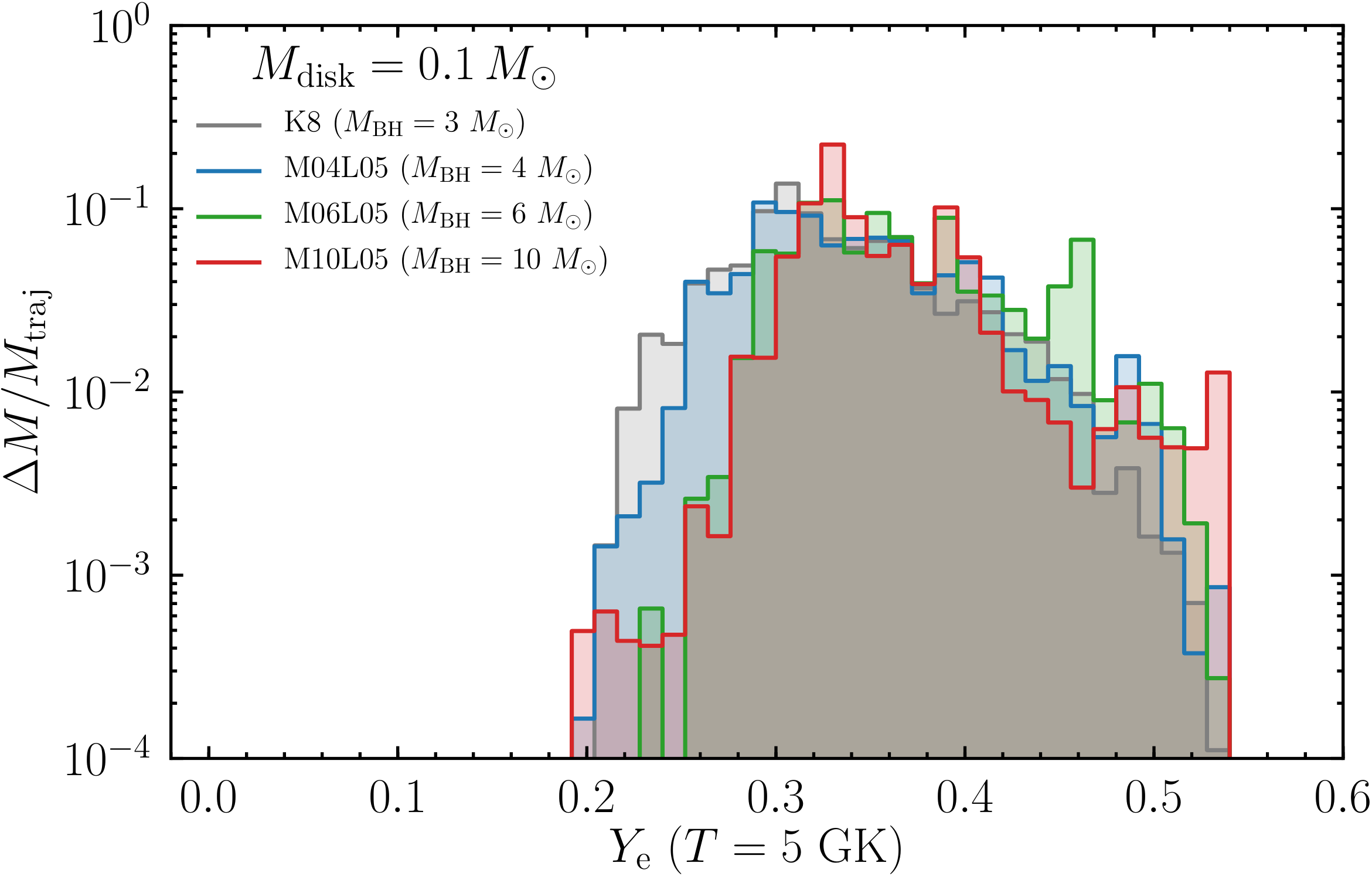} &
  \includegraphics[width=.47\linewidth,clip]{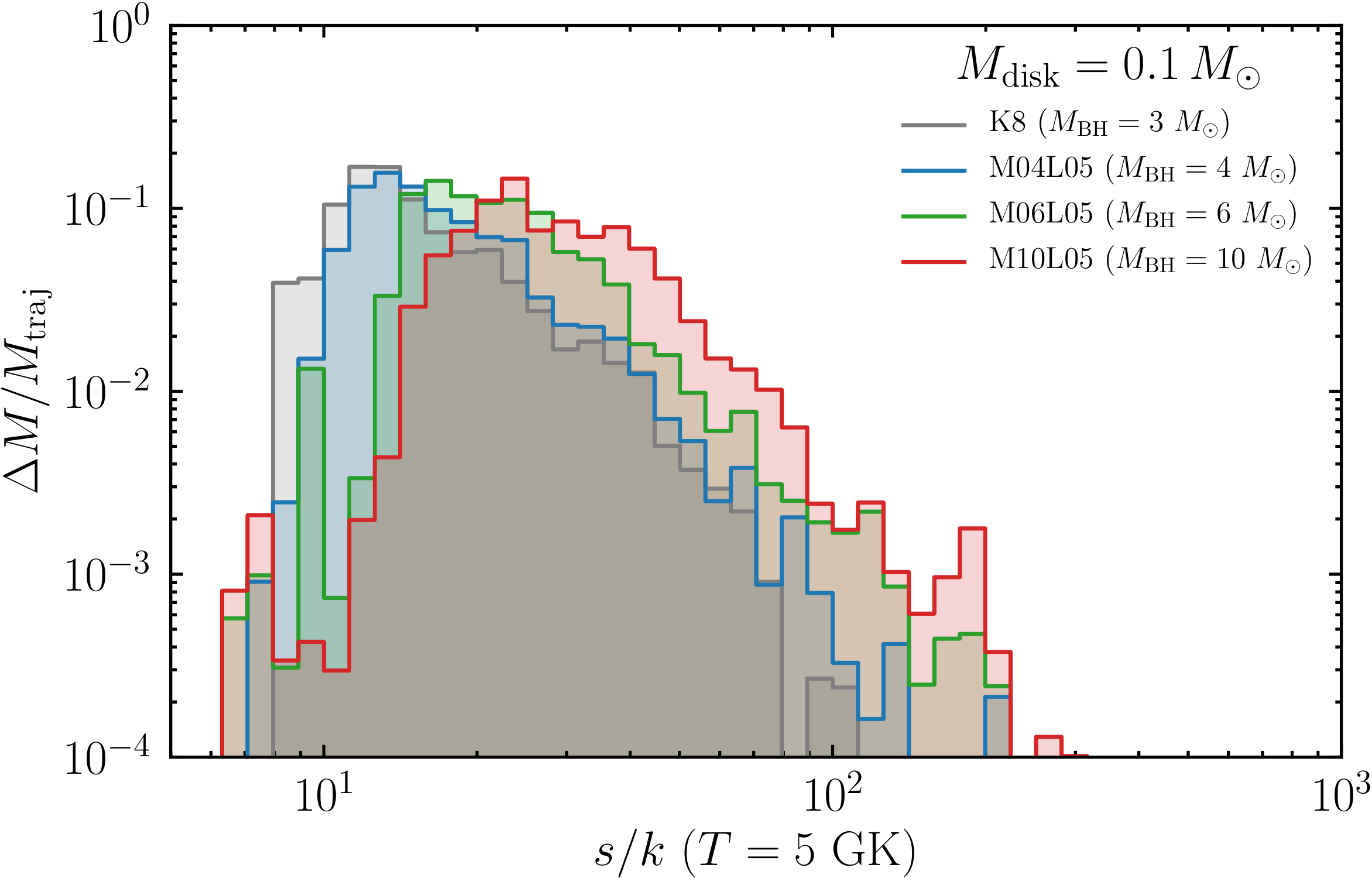} \\
  \includegraphics[width=.47\linewidth,clip]{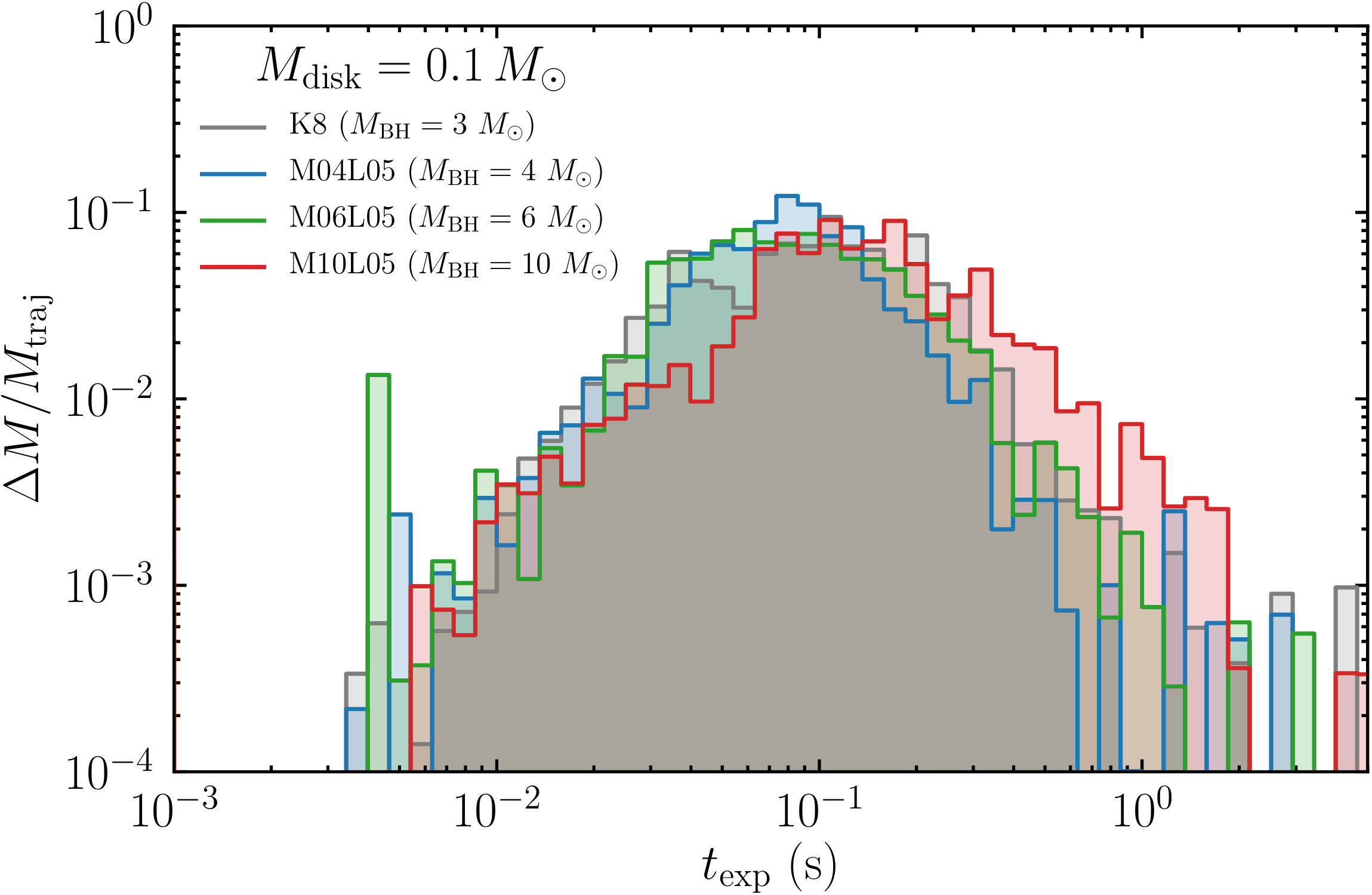} &
  \includegraphics[width=.47\linewidth,clip]{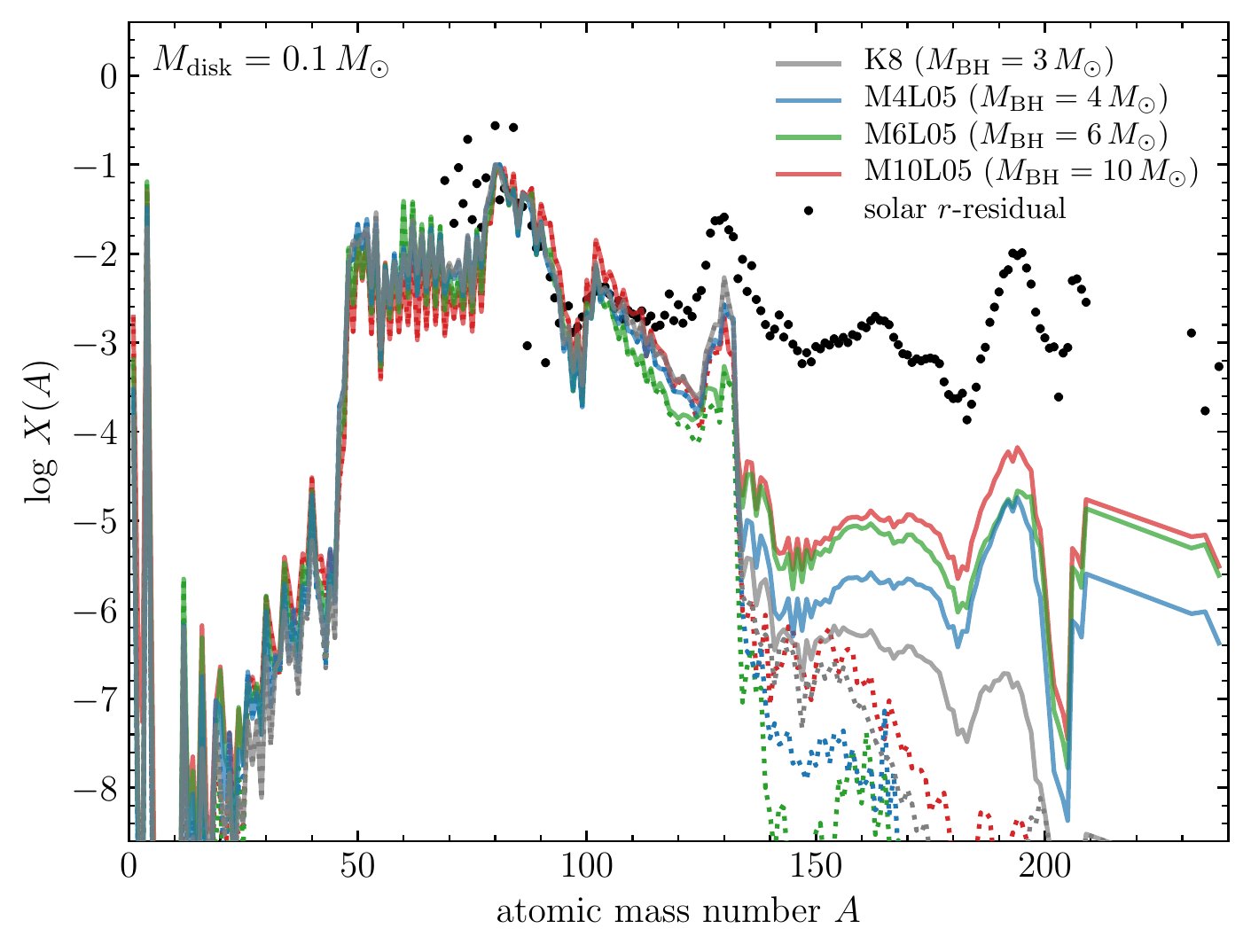}
 \end{tabular}
 \caption{Mass distribution as a function of the electron fraction (left
 top), the entropy per baryon (right top), and the expansion time scale
 (left bottom) for the viscous outflow from a $0.1\,M_\odot$ disk with the
 initial radius $\sim \SI{180}{\km}$ surrounding black holes with $\chi
 \approx 0.8$ and various masses. The alpha parameter $\alpha_\nu$ is
 taken to be $0.05$. All the quantities are measured when temperature of
 fluid elements decreases to \SI{5e9}{\kelvin} by a tracer-particle
 approach. The right bottom panel shows corresponding abundance patterns
 of the \textit{r}-process nucleosynthesis. Image reproduced with permission from \citet{Fujibayashi_SWKKS2020-2}, copyright by the authors.} \label{fig:visc_yield}
\end{figure}

Reflecting the relatively high value of the electron fraction, the
\textit{r}-process nucleosynthesis in the viscous disk outflow does not
efficiently produce heavy \textit{r}-process elements with the mass
number $\gtrsim 130$, i.e., beyond the second peak. Figure
\ref{fig:visc_yield} shows the mass distribution as a function of three
important quantities which control the production of the heavy
\textit{r}-process elements, namely the electron fraction, the entropy
per baryon, and the expansion timescale of the ejected
material. According to the criterion proposed in
\citet{Hoffman_Woosley_Qian1997}, the viscous disk outflow considered
here is not capable of producing abundant heavy \textit{r}-process
elements with the mass number $\gtrsim 130$. Indeed, results of
nucleosynthesis calculations shown also in Fig.~\ref{fig:visc_yield}
indicate that the viscous disk outflow dominantly produces transiron and
light \textit{r}-process elements with the mass number $\sim 50$--$130$,
and the fraction of heavy \textit{r}-process elements with the mass
number $\gtrsim 130$ is limited.

This yield indicates that, if the viscous disk outflow is the dominant
source of the mass ejection for a black hole--neutron star binary, the
resultant abundance may not resemble the solar \textit{r}-process
pattern. This particularly applies to black hole--neutron star binaries
with $Q \lesssim 3$, for which dynamical mass ejection is inefficient in
comparison with the disk formation (see
Sect.~\ref{sec:sim_rem_dyn}). For black hole--neutron star binaries with
$Q \gtrsim 5$, the neutron-rich dynamical ejecta may contribute
substantially to the yield, and thus the abundance pattern may be
inclined to heavy \textit{r}-process elements. Suitable superposition of
the dynamical ejecta and the viscous disk outflow might reproduce the
solar \textit{r}-process pattern \citep{Just_BAGJ2015}, and it would be
worthwhile to investigate whether this can be realized by realistic
black hole--neutron star binaries.

A word of caution is necessary here. Although the fractional mass and
the velocity of the disk outflow agree semiquantitatively among various
simulations, the electron fraction is vigorously debated \citep[see
\citealt{2021arXiv210208387J} for a detailed
investigation]{Fernandez_Metzger2013,Just_BAGJ2015,Siegel_Metzger2018,Fernandez_TQFK2019,Fernandez_Foucart_Lippuner2020}. In
particular, if the material is ejected as early as $\lesssim
\SI{0.1}{\second}$ after merger, the electron fraction of the disk
outflow is unlikely to be increased to $\expval{Y_\mathrm{e}} \gtrsim
0.25$, because the equilibrium of electron/positron captures is not
achieved for this short time scale. In the framework of viscous
hydrodynamics, the rapid mass ejection can be realized by adopting a
large value of the alpha parameter of $\alpha_\nu \gtrsim 0.1$. Indeed,
the average electron fraction of the viscous disk outflow is found to be
as low as $\expval{Y_\mathrm{e}} \approx 0.25$ if such a large value of
$\alpha_\nu$ is adopted
\citep{Fujibayashi_SWKKS2020,Fujibayashi_SWKKS2020-2}. Thus, the
electron fraction is likely to be determined by the realistic process of
mass ejection, which is still not well-understood. The electron fraction
of the disk outflow also decreases if the disk mass is small, because
the outflow is launched earlier and the material does not spend a long
time for increasing the electron fraction.

Precise strength of the effective viscosity can be understood only by
high-resolution magnetohydrodynamics simulations performed with
realistic magnetic-field geometry. The rapid ejection may occur in the
presence of hypothetical, coherent magnetic fields right after merger
via the magnetic winding and the Lorentz force
\citep{Fernandez_TQFK2019}. Indeed, fully-relativistic
magnetohydrodynamics simulations of black hole--neutron star binary
mergers throughout the coalescence indicated enhancement of the
effective viscosity associated with magnetically-induced turbulence
\citep{Kiuchi_SKSTW2015}. We discuss the current status of
magnetohydrodynamics simulations in numerical relativity in
Sect.~\ref{sec:sim_pm_mag}.

\subsubsection{Magnetic activity} \label{sec:sim_pm_mag}

Magnetohydrodynamics simulations (with a sufficiently high resolution)
are indispensable for clarifying not only the disk outflow and
accompanying \textit{r}-process nucleosynthesis but also the mechanism
of short-hard gamma-ray bursts. It is strongly believed that the
effective viscosity in the accretion disks stems from turbulence induced
by the magnetorotational instability \citep{Balbus_Hawley1991}. Magnetic
winding and subsequent magnetic braking can also contribute to
transporting the angular momentum inside the differentially-rotating
remnant disk and may drive the mass accretion. These magnetic effects
are always active in accretion disks and govern their longterm
evolution. If the magnetic field amplified in the accretion disks by the
magnetohydrodynamic instabilities gives rise to a strong and
globally-coherent configuration penetrating the black hole, the
Blandford--Znajek mechanism could extract rotational kinetic energy of
the black hole and launch an ultrarelativistic jet
\citep{Blandford_Znajek1977,Meszaros_Rees1997}. Furthermore, as we
discussed in the end of Sect.~\ref{sec:sim_pm_wind}, magnetic-field
configurations may be the key for determining the timing at which the
disk outflow is launched. For example, the magnetocentrifugal effect
\citep{Blandford_Payne1982} with the field lines anchored in the inner
region of the disk can contribute to the disk outflow.

\begin{figure}[htbp]
 \centering
 \begin{tabular}{ccc}
  \includegraphics[width=0.3\linewidth,clip]{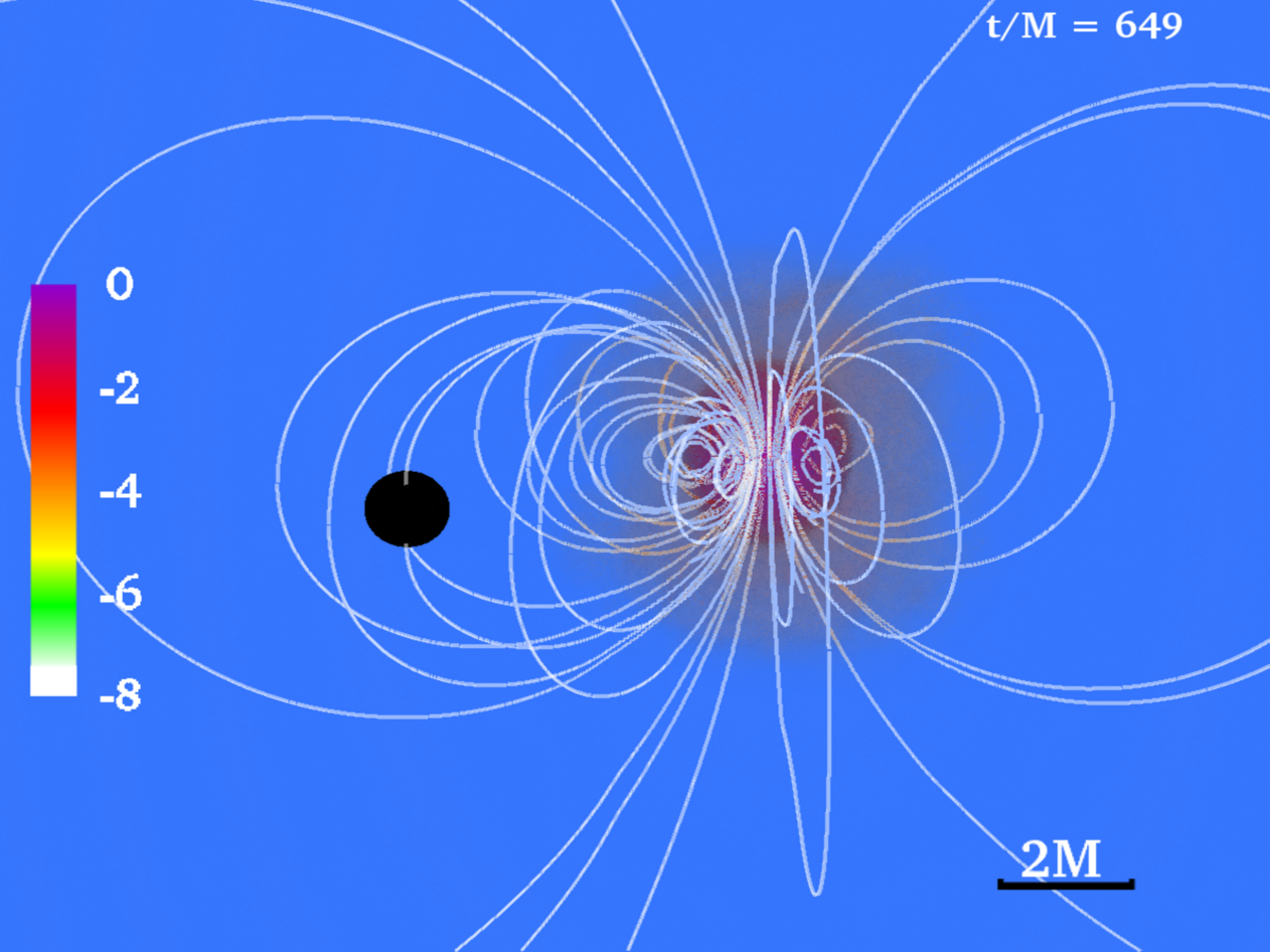} &
  \includegraphics[width=0.3\linewidth,clip]{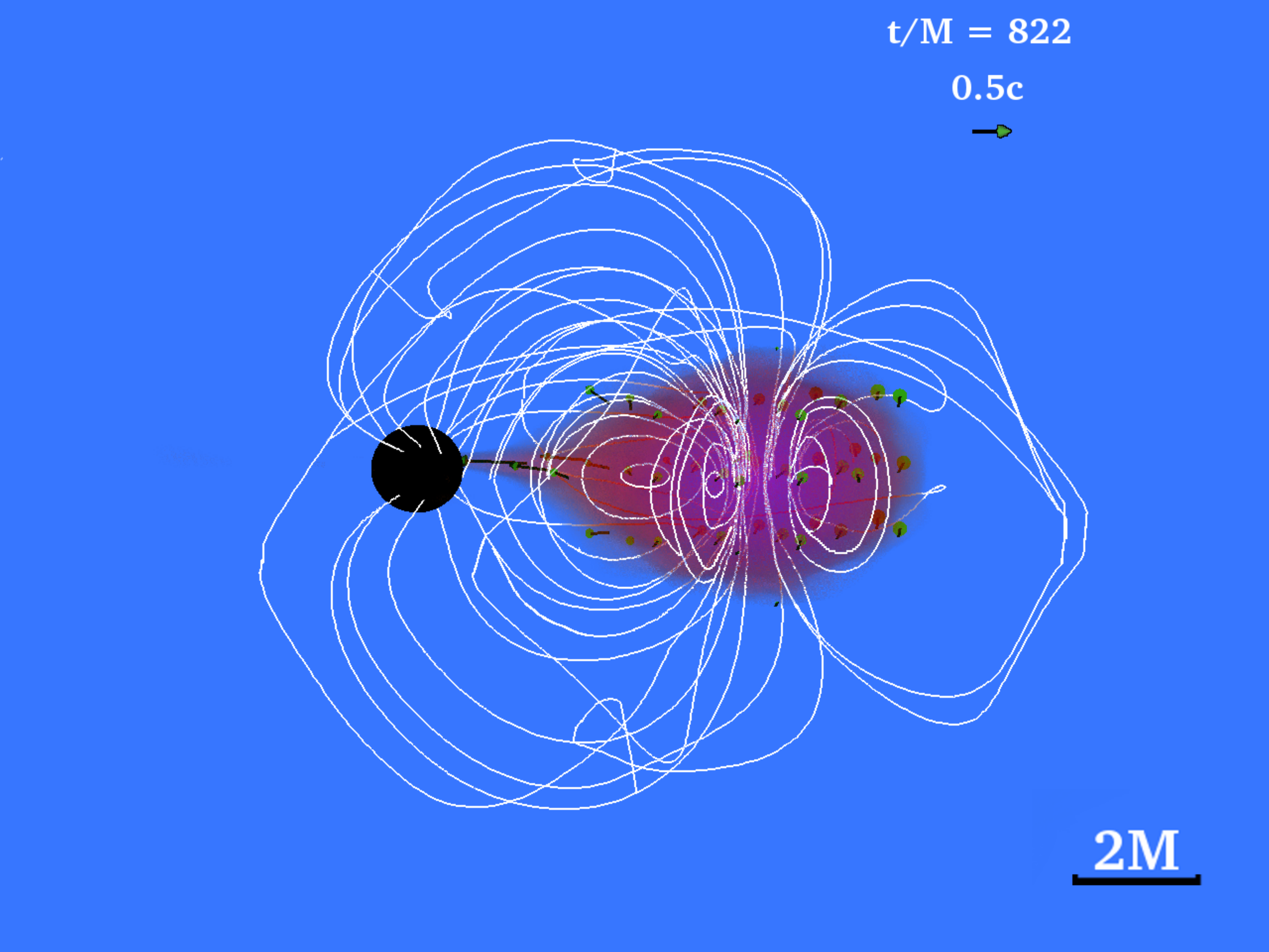} &
  \includegraphics[width=0.3\linewidth,clip]{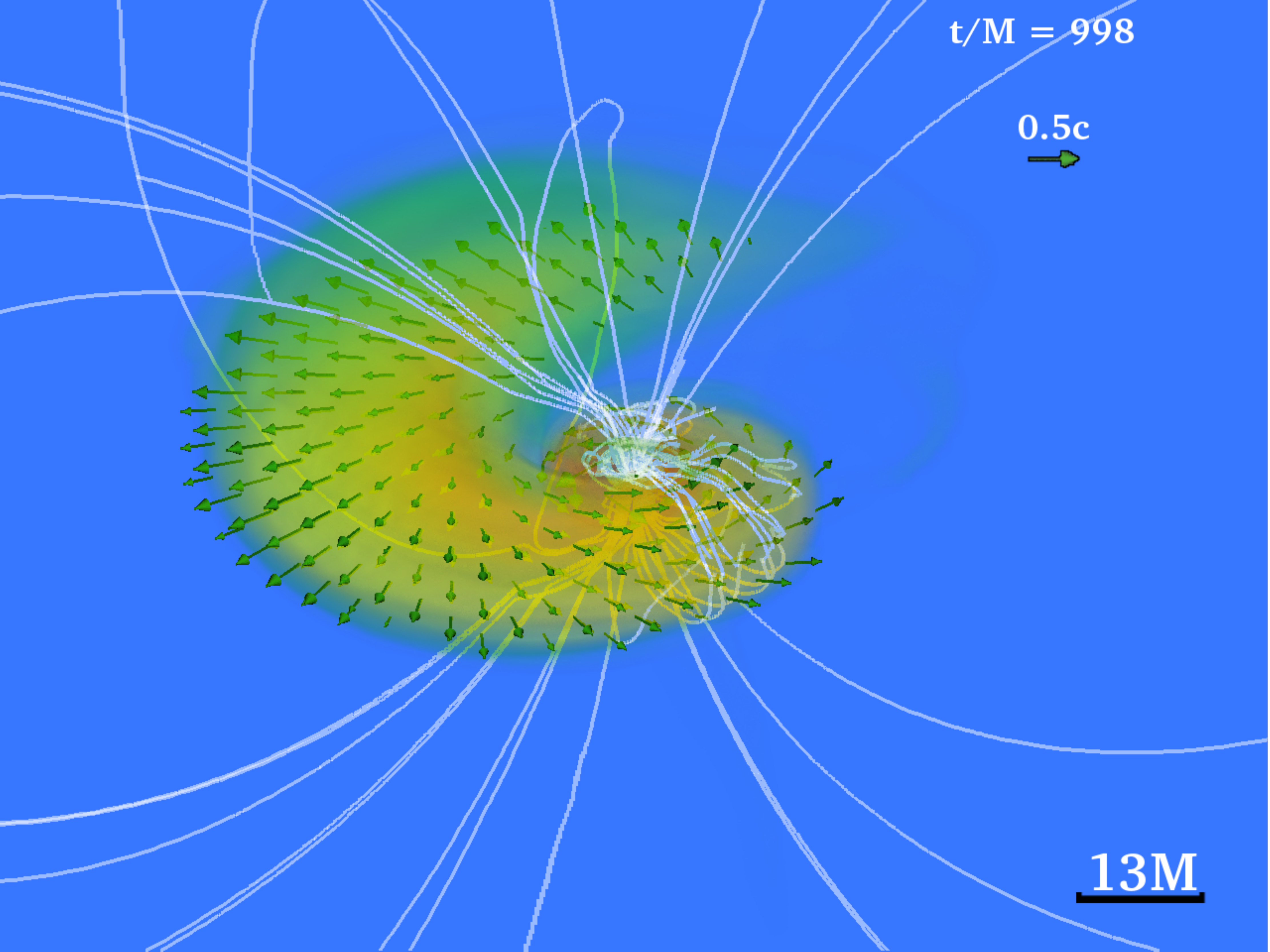} \\
  \includegraphics[width=0.3\linewidth,clip]{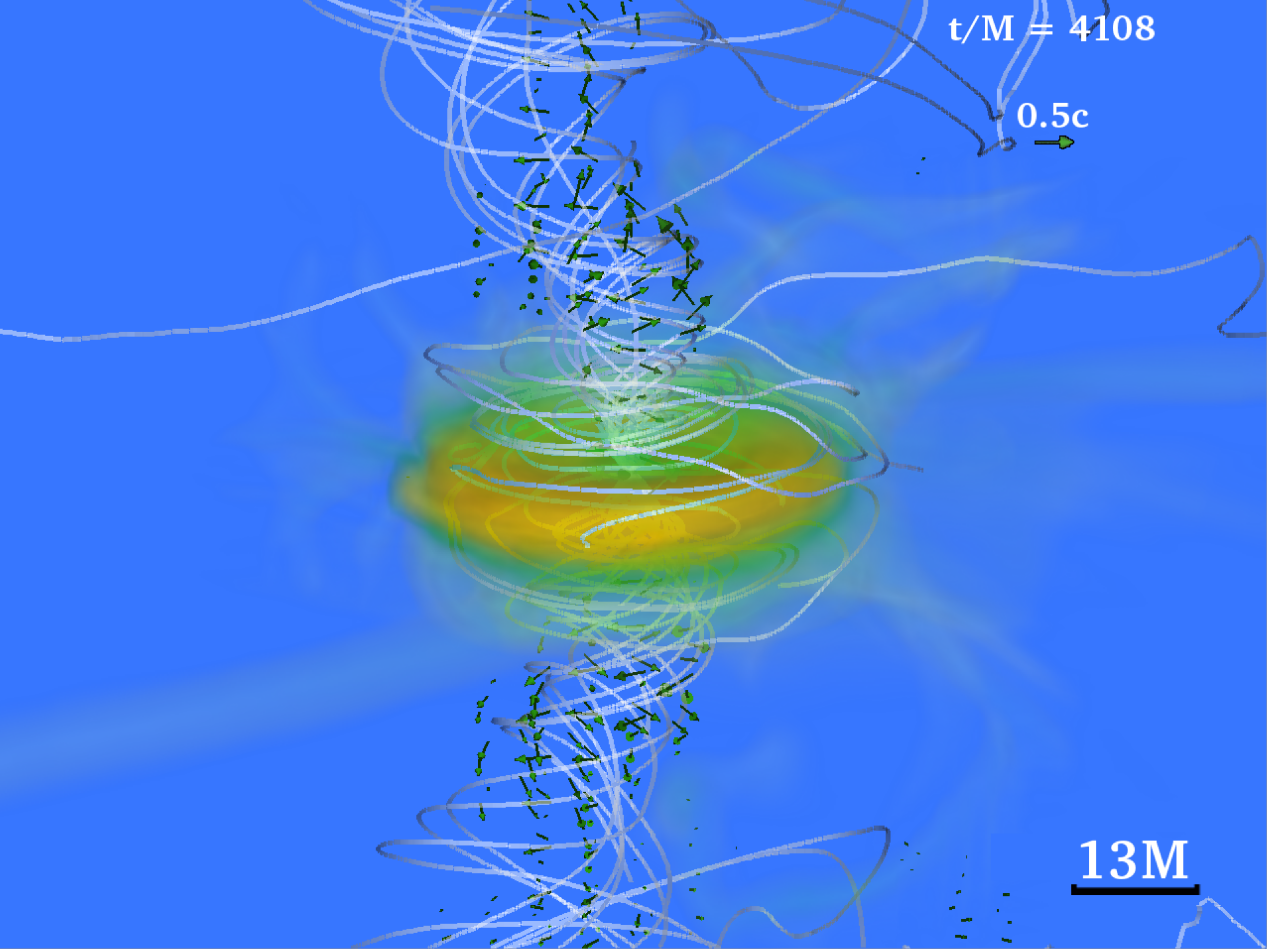} &
  \includegraphics[width=0.3\linewidth,clip]{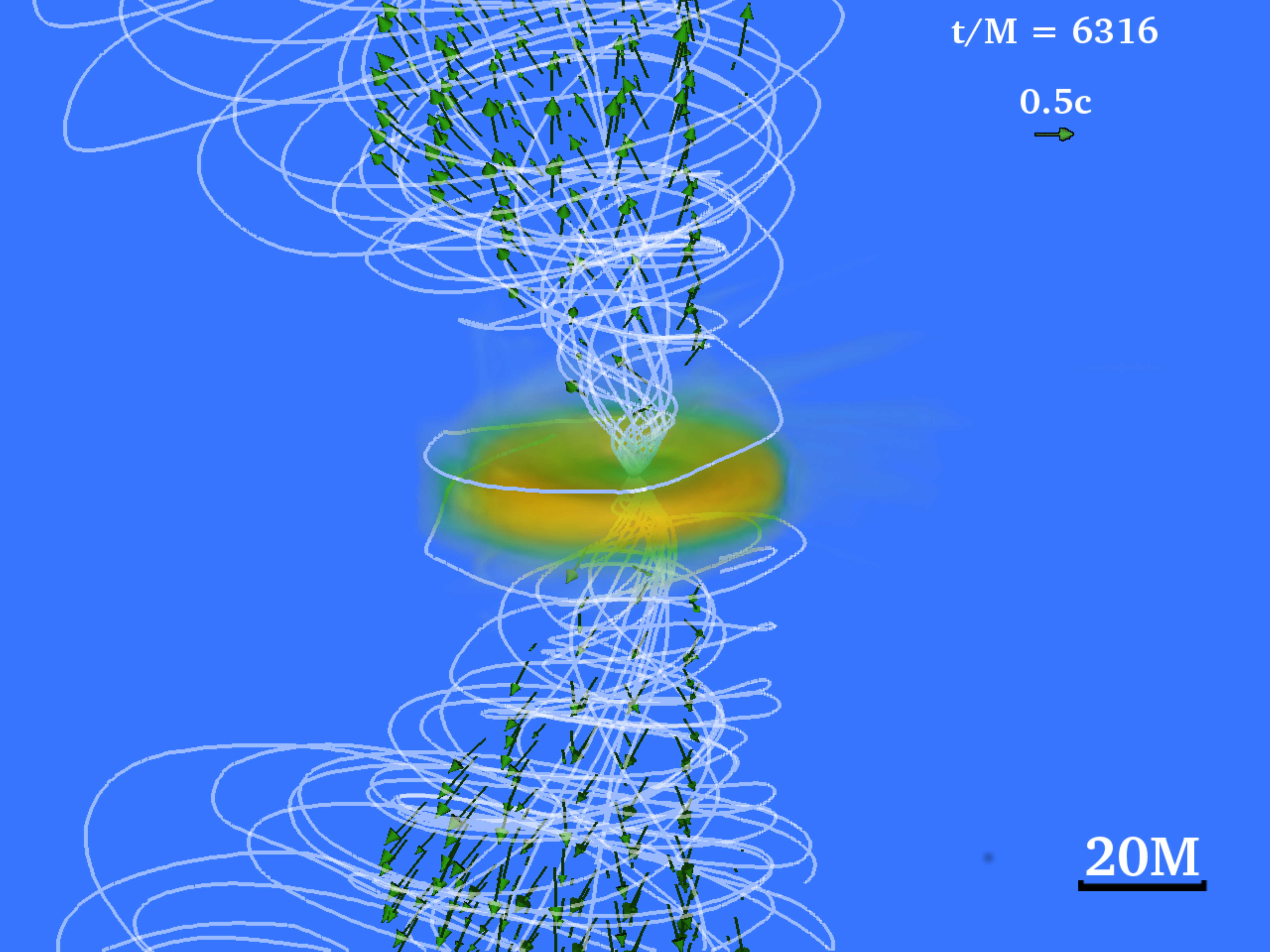} &
  \includegraphics[width=0.3\linewidth,clip]{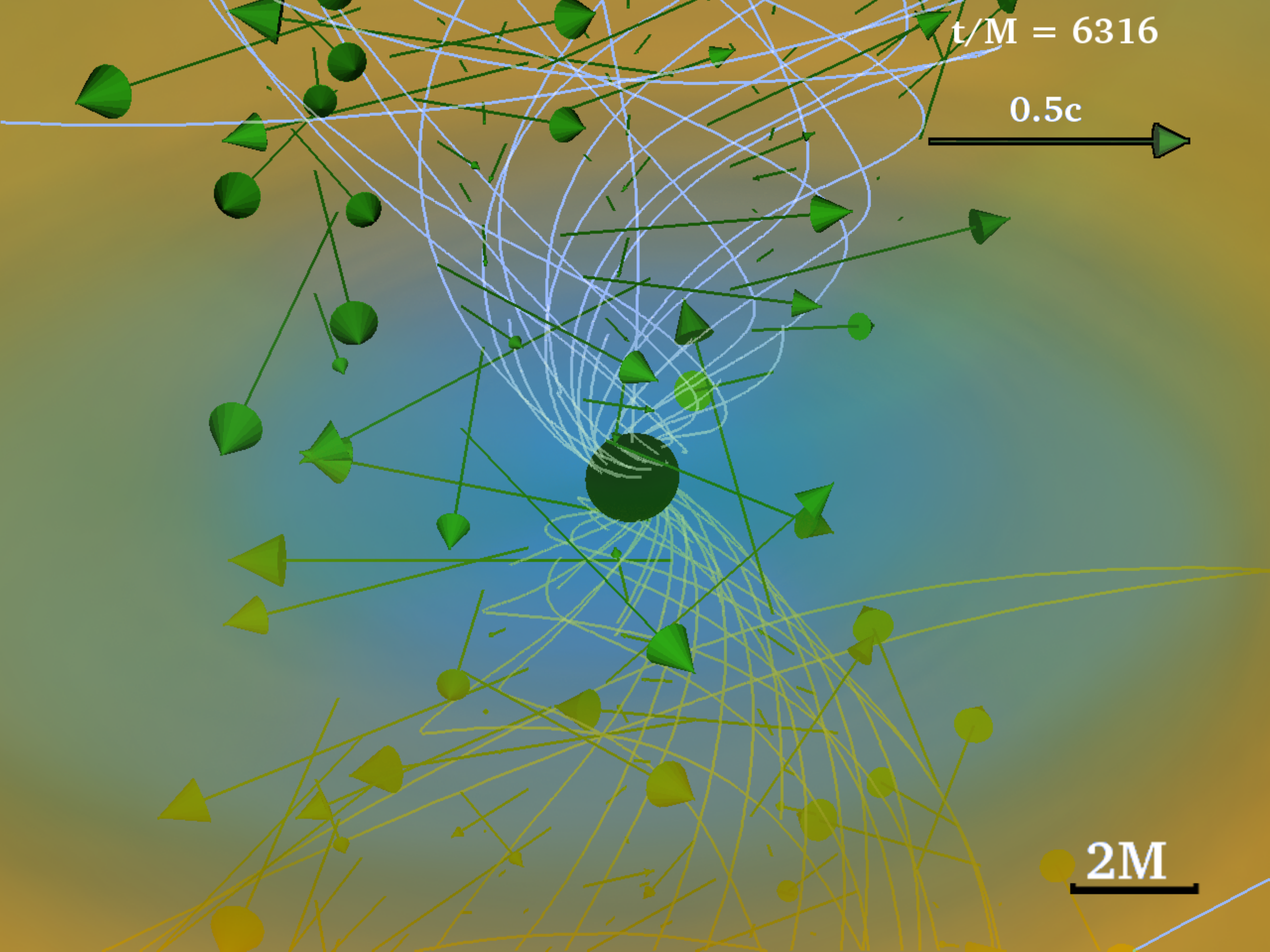}
 \end{tabular}
 \caption{Evolution of the profile of the rest-mass density (normalized
 by its initial maximum value), the location of the apparent horizon
 (black filled sphere), and the magnetic-field lines (white solid) for a
 binary with $Q = 3$, $\chi = 0.75$, and $\mathcal{C} = 0.145$ modeled
 by a $\Gamma = 2$ polytrope. The maximum value of the initial
 magnetic-field strength is $\approx \SI{e17}{G} ( 1.4\,M_\odot /
 M_\mathrm{NS} )$. Image reproduced with permission from
 \citet{Paschalidis_Ruiz_Shapiro2015}, copyright by AAS.} \label{fig:snap_mag}
\end{figure}

If the neutron star is seeded with strong dipolar magnetic fields
aligned with the orbital angular momentum of the binary at the outset, a
collimated outflow with possibly ultrarelativistic terminal velocity may
be launched from the remnant black hole--disk system
\citep{Paschalidis_Ruiz_Shapiro2015}. Figure \ref{fig:snap_mag}
generated by \citet{Paschalidis_Ruiz_Shapiro2015} illustrates the
rest-mass density and the structure of magnetic fields for a system with
$Q=3$, $\chi = 0.75$, $\mathcal{C} = 0.145$ modeled by a $\Gamma = 2$
polytrope, and the initial dipolar magnetic field reaching up to
$\approx \SI{e17}{G} ( 1.4\,M_\odot / M_\mathrm{NS} )$ in the stellar
interior. Strong magnetic fields are helpful for resolving the
magnetorotational instability in the accretion disk with a sufficient
number of grids, and the magnetic stress in this system is found to
correspond to $\alpha_\nu \approx 0.01$--0.04. Although this value is
broadly consistent with results obtained by numerical studies of other
similar systems \citep[see, e.g.,][]{Kiuchi_KSS2018,Fernandez_TQFK2019},
the value of $\alpha_\nu$ has not yet settled due to the difficulty in
achieving convergence in three-dimensional magnetohydrodynamics
simulations. Actually, it is not claimed that the magnetorotational
instability is resolved in \citet{Paschalidis_Ruiz_Shapiro2015}.

The results of \citet{Paschalidis_Ruiz_Shapiro2015} suggest that strong
poloidal magnetic fields on a large scale may be essential for driving
an ultrarelativistic jet in the low-density polar region, which may
become force-free in the real world. While differential rotation of the
tidal tail amplifies the toroidal field after tidal disruption, the
poloidal component still persists and connects distinct portions of the
remnant disk. The differential rotation of the disk continues to twist
the poloidal component in the polar region, and the amplified toroidal
field eventually overcomes the ram pressure and establishes a
funnel-like, low-density environment. A collimated and
mildly-relativistic outflow is driven in this region with the Lorentz
factor of $\approx 1.2$--$1.3$ (equivalently, $\sim 0.6c$) and the
opening angle of $\sim \ang{20}$. However, the acceleration cannot be
fully tracked to the terminal velocity, because a force-free environment
cannot be resolved and, as a result, the acceleration to
ultrarelativistic velocity close to the speed of light is prohibited in
magnetohydrodynamics simulations. Despite this limitation, the terminal
Lorentz factor in \citet{Paschalidis_Ruiz_Shapiro2015} is suggested to
be as large as $\sim 100$ if the baryon loading is not severe. In
addition, the large terminal Lorentz factor is realized only if the
magnetic-field energy is converted efficiently to the kinetic energy of
the material (see, e.g., \citealt{Rees_Gunn1974,Kennel_Coroniti1984} for
the so-called sigma problem in the pulsar wind nebula). Because the
Poynting luminosity reaches \SI{e51}{erg.s^{-1}} and the accretion time
scale is estimated to be $\sim \SI{0.5}{\second}$, their results support
the idea that the Blandford-Znajek mechanism drives a short-hard
gamma-ray burst in black hole--neutron star binaries if the assumed
magnetic-field configuration is realistic.

However, all this depends on whether strong poloidal magnetic fields
presumed in \citet{Paschalidis_Ruiz_Shapiro2015} at the outset are
developed on a large scale within the lifetime of the accretion
disk. Actually, their own follow-up study revealed that the dipolar
field tilted by \ang{90} did not allow an outflow to be launched
\citep{Ruiz_Shapiro_Tsokaros2018}. This may be reasonable, because the
tilt angle of \ang{90} eliminates net vertical magnetic fields threading
the orbital plane. This finding indicates the importance of deriving the
strength and geometry of magnetic fields at the disk formation in a
self-consistent manner by merger simulations, even if it is a
computationally challenging task
\citep{Etienne_LPS2012,Etienne_Paschalidis_Shapiro2012}. Although the
Kelvin-Helmholtz instability at disk circularization and subsequent
magnetorotational instability rapidly amplify small-scale magnetic
fields \citep[see also \citealt{Kiuchi_CKSS2015} for magnetic-field
amplification at the contact surface of binary neutron
stars]{Kiuchi_SKSTW2015}, it is not obvious whether and, if yes, how
globally coherent fields stronger than the initial fields are developed.

\begin{figure}[htbp]
 \centering \includegraphics[width=0.8\linewidth,clip]{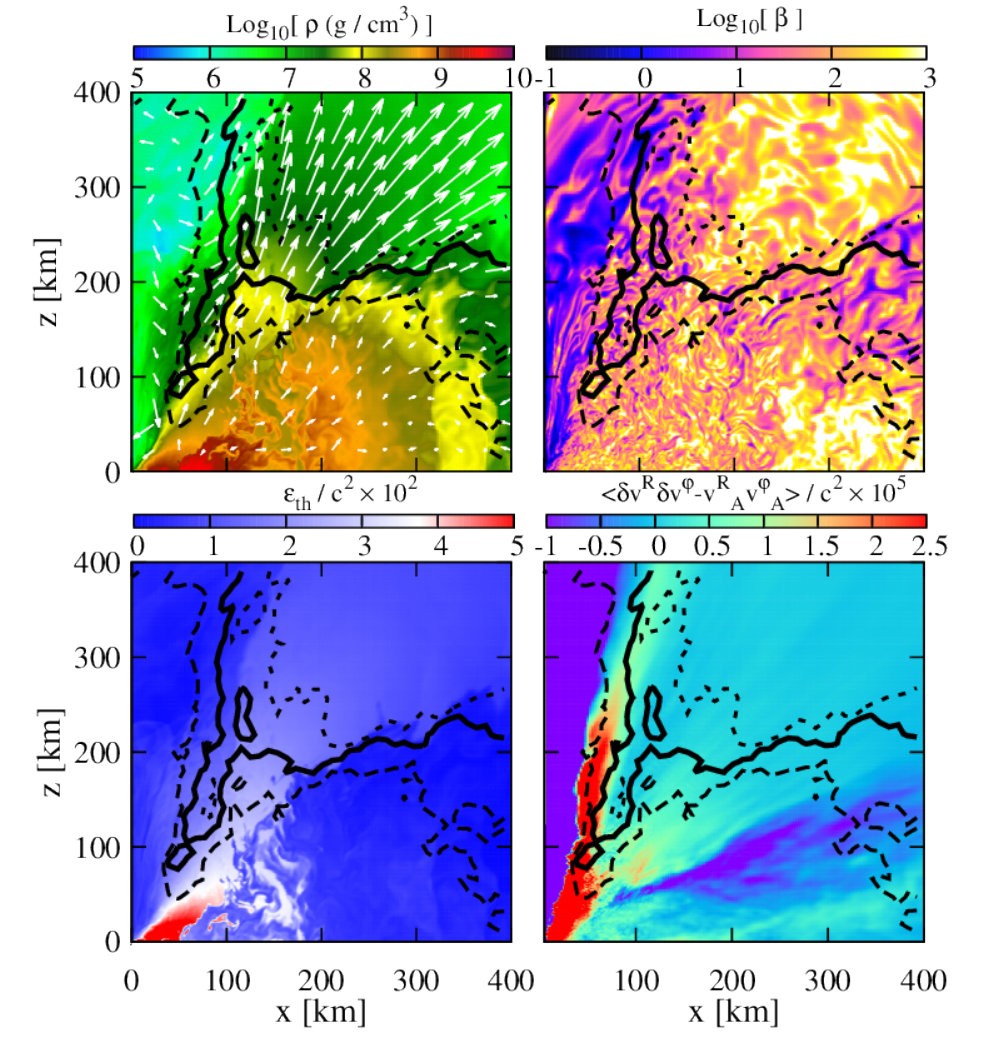}
 \caption{Profile of the rest-mass density (left top), the plasma
 $\beta$, i.e., the ratio of the gas pressure to the magnetic pressure,
 (right top), the thermal part of the specific internal energy (left
 bottom), and the sum of Maxwell and Reynolds stress (right bottom) on
 the meridional plane at $\approx \SI{50}{\ms}$ after the onset of
 merger for a binary with $M_\mathrm{BH} = 5.4\,M_\odot$, $\chi = 0.75$,
 $M_\mathrm{NS} = 1.35\,M_\odot$, and $R_\mathrm{NS} = \SI{13.6}{\km}$
 ($Q=4$, $\mathcal{C}=0.147$) modeled by a piecewise-polytropic
 approximation of the H4 equation of state
 \citep{Glendenning_Moszkowski1991,Lackey_Nayyar_Owen2006}. The maximum
 value of the initial magnetic-field strength is $\approx
 \SI{e15}{G}$. The white arrows in the left top panel denote the
 velocity field on this plane. The black contours indicate $u_t = -0.98$
 (thick dashed), $-1$ (solid), and $-1.02$ (thin dashed). Image reproduced with permission from \citet{Kiuchi_SKSTW2015}, copyright by APS.} \label{fig:mhdwind}
\end{figure}

If a magnetohydrodynamical disk outflow is launched, global poloidal
magnetic fields are likely to be developed due to the flux
freezing. Actually, fully-relativistic magnetohydrodynamics simulations
have witnessed the launch of a disk outflow from the innermost region of
the accretion disk with high velocity as shown in Fig.~\ref{fig:mhdwind}
\citep{Kiuchi_SKSTW2015}. As a result, the wind generates strong and
coherent magnetic fields, which may subsequently extract rotational
energy of the black hole via the Blandford-Znajek mechanism with the
luminosity of \num{e49}--\SI{e50}{erg.s^{-1}}. To achieve this
luminosity, a coherent magnetic field with $\gtrsim \SI{e14}{G}$ needs
to penetrate the black hole. Although numerical convergence is not fully
confirmed, the magnetorotational instability is likely to be resolved in
the highest-resolution simulation with the grid spacing of
\SI{120}{\meter} in \citet{Kiuchi_SKSTW2015}.

It is worthwhile to note that \citet{Liska_Tchekhovskoy_QuataertQ2020}
demonstrated that poloidal magnetic fields can be generated from purely
toroidal fields in very-high-resolution simulations for an accretion
disk in a fixed, Kerr black-hole background. This global poloidal field
is generated by a runaway growth of one of flux loops generated locally
in a stochastic manner. The realistic magnetic fields and subsequent
outflows may be clarified by future numerical-relativity simulations if
the computational resources allow us to achieve sufficiently high
resolutions.

One caveat in most magnetohydrodynamics simulations of black
hole--neutron star binary coalescences performed to date is that
neutrino transport is not taken into account. It is unclear whether
magnetohydrodynamical disk outflows like that found in
\citet{Kiuchi_SKSTW2015} are driven in the presence of neutrino
cooling. Actually, this outflow is not observed in neutrino-radiation
magnetohydrodynamics simulations performed by \citet{Most_PTR2021-2},
although this difference may simply be ascribed to the insufficient grid
resolution. A sufficiently-high-resolution numerical-relativity
simulation of the whole coalescence process with detailed microphysics
will become a milestone for clarifying realistic magnetic-field
configurations, postmerger evolution, jet launch, and mass
ejection. Last but not least, for clarifying the jet launch in
particular, we need to develop novel numerical techniques for accurately
resolving the force-free environment with low density and strong
magnetic fields.

\subsection{Gravitational waves} \label{sec:sim_gw}

The last topic of Sect.~\ref{sec:sim} is gravitational waves from the
merger of black hole--neutron star binaries. After discussing general
features of the time- and frequency-domain waveforms in
Sect.~\ref{sec:sim_gw_time} and Sect.~\ref{sec:sim_gw_spec},
respectively, we quantify the correlation between the hypothetical
neutron-star equation of state and the cutoff frequency of the spectrum
observed in the final inspiral and merger phases in
Sect.~\ref{sec:sim_gw_fcut}. Our discussions are based mainly on
systematic surveys performed in
\citet{Kyutoku_Shibata_Taniguchi2010,Kyutoku_OST2011}, because the
features of gravitational waves are qualitatively the same among the
results obtained by independent groups. We also focus on the dominant
$l=\abs{m}=2$ modes of gravitational waves from nonprecessing binaries,
because they have been studied systematically in previous work. Higher
harmonic modes and precession-induced modulation are discussed briefly
in Sect.~\ref{sec:sim_gw_high} while keeping in mind that further
investigations are required to understand these topics in black
hole--neutron star binaries. In this Sect.~\ref{sec:sim_gw}, we adopt
geometrical units in which $G=c=1$.

We note that the remnant disk does not emit gravitational waves
significantly unless instabilities such as the Papaloizou-Pringle
instability set in \citep{Papaloizou_Pringle1977}. While it has been
suggested that nearly-extremally-spinning black hole--neutron star
binary coalescences may be accompanied by gravitational waves from
massive remnant disks, this issue has not yet been settled
\citep{Lovelace_DFKPSS2013}. Thus, postmerger gravitational waves from
black hole--neutron star binary coalescences are considered to be
composed either of a ringdown signal from the remnant black hole or of a
steep shutdown of the amplitude due to tidal disruption (see
below). This point is in stark contrast with postmerger gravitational
waves from binary-neutron-star coalescences, which can be highly diverse
depending on masses and equations of state of neutron stars \citep[see,
e.g.,][]{Hotokezaka_KKMSST2013}.

\subsubsection{Waveform} \label{sec:sim_gw_time}

In preparation for discussing gravitational waveforms from black
hole--neutron star binaries, it would be useful to review those from
binary black holes, which have been detected more often than on a weekly
basis in the LIGO-Virgo O3 \citep[see also
\citealt{Jani_HCLLS2016,Healy_LLOZC2019,Boyle_etal2019} for catalogs of
gravitational waveforms derived in numerical
relativity]{GWTC1,GWTC2}. In the inspiral phase, the orbital frequency
continues to increase due to gravitational radiation reaction. Both the
amplitude and the frequency of gravitational waves increase accordingly,
and the waveform with this feature is observed as the so-called chirp
signal. As the merger approaches, general-relativistic effects on the
orbital motion such as the spin-orbit and spin-spin couplings gradually
become significant. While the waveform remains to be of the chirp type,
an accurate description requires increasingly higher-order
post-Newtonian corrections in this phase (see \citealt{Blanchet2014} for
reviews). The merger time may be identified by the time at which the
amplitude becomes maximum, and the waveform during the merger phase is
smoothly connected with the chirp waveform in the inspiral phase and the
ringdown waveform in the postmerger phase. A reliable merger waveform
can be derived only by numerical-relativity simulations (see, e.g.,
\citealt{Centrella_BKM2010,Duez_Zlochower2019} for
reviews). Gravitational waves in the postmerger phase are dominated by
the ringdown waveform, which is a superposition of quasinormal modes of
the remnant Kerr black holes (see, e.g.,
\citealt{Berti_Cardoso_Starinets2009} for reviews). It should be pointed
out that gravitational waves associated with the quasinormal modes are
emitted efficiently in the coalescence of binary black holes, because
the merging process is highly nonaxisymmetric and dynamical.

\begin{figure}[htbp]
 \centering \includegraphics[width=0.7\linewidth,clip]{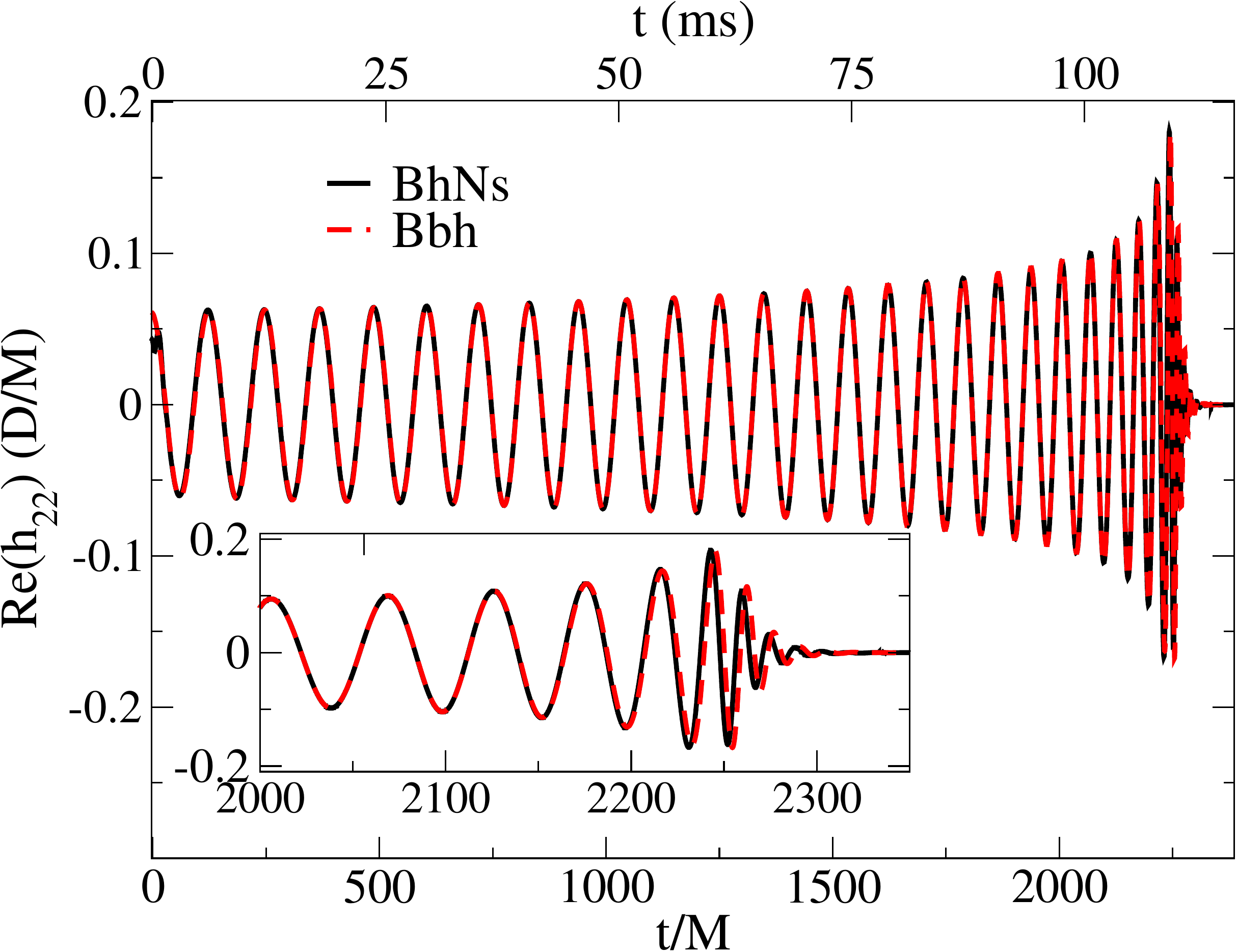}
 \caption{Comparison of the $l=\abs{m}=2$ mode of gravitational
 waveforms for black hole--neutron star binaries (black solid) and
 binary black holes (red dashed) for systems with $Q=6$, $\chi = 0$, and
 $\mathcal{C} = 0.156$ (for the black hole--neutron star binary) modeled
 by a $\Gamma = 2$ polytrope. Image reproduced with permission from
 \citet{Foucart_BDGKMMPSS2013}, copyright by APS.} \label{fig:gwBHNSvsBHBH}
\end{figure}

If the neutron star is not significantly deformed (and thus not
disrupted) by the black hole, gravitational waveforms from black
hole--neutron star binaries are essentially the same as those from
binary black holes with the same values of the masses and the
spins. Figure \ref{fig:gwBHNSvsBHBH} generated by
\citet{Foucart_BDGKMMPSS2013} is a comparison of the two waveforms for
systems with $Q=6$, $\chi = 0$, and $\mathcal{C} = 0.156$ (for the black
hole--neutron star binary) derived by numerical-relativity
simulations. Because tidal disruption is absent for this nonspinning
black hole--neutron star binary with a high mass ratio, the two
waveforms are practically indistinguishable. While a slight deviation is
found during and after merger, this is within estimated numerical
errors. This indistinguishability generally holds for nondisruptive
systems, which are typical for high mass ratios, zero or retrograde
spins of the black hole, and/or large compactnesses of the neutron star.

\begin{figure}[htbp]
 \centering
 \begin{tabular}{cc}
  \includegraphics[width=0.47\linewidth,clip]{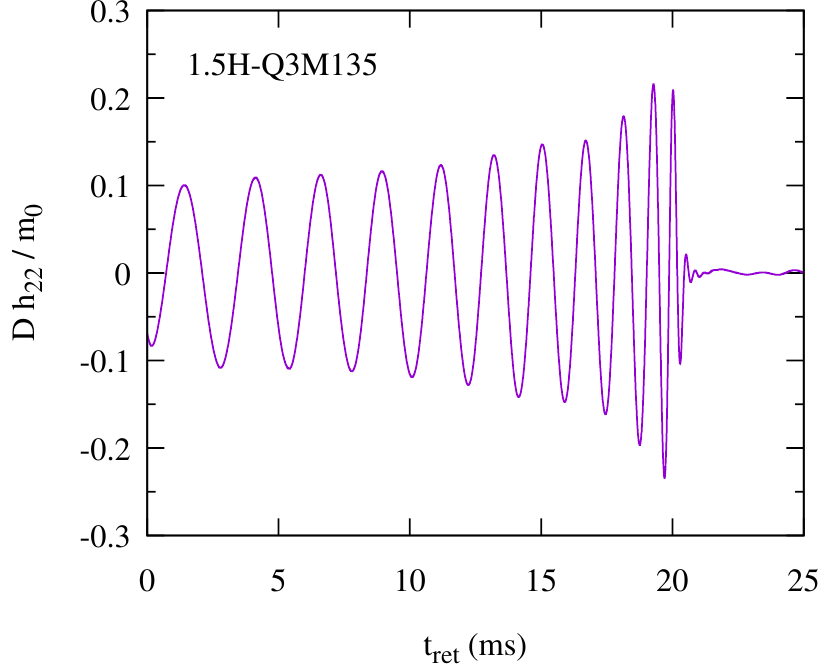} &
  \includegraphics[width=0.47\linewidth,clip]{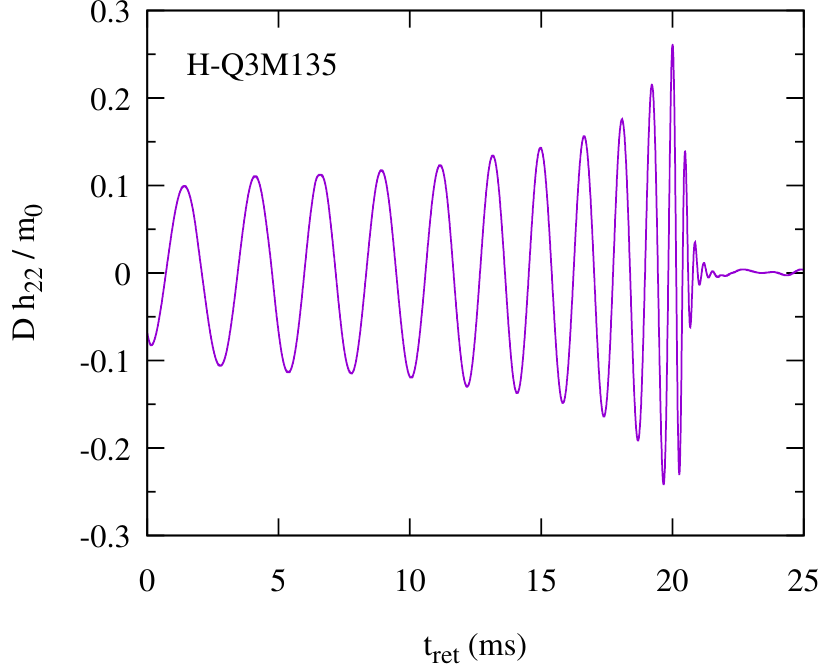} \\
  \includegraphics[width=0.47\linewidth,clip]{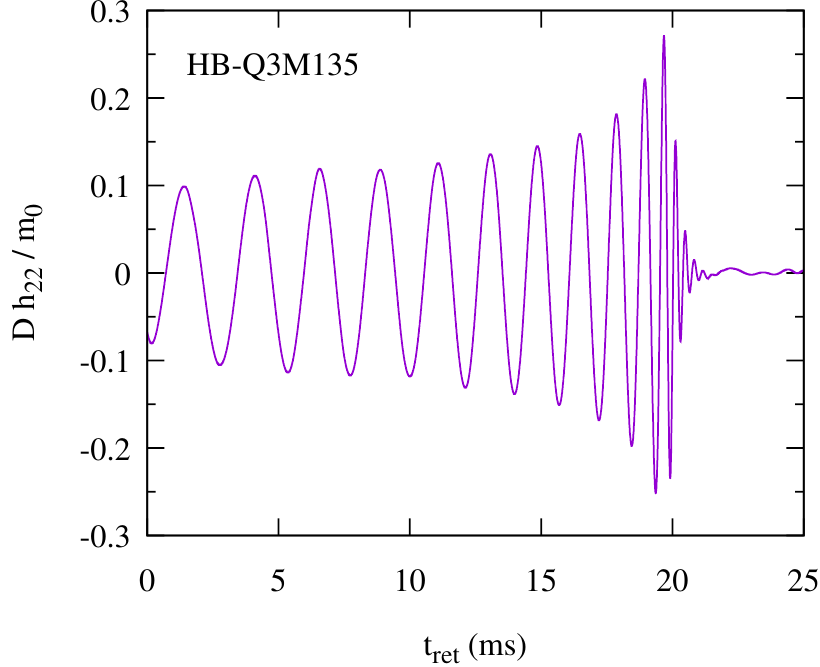} &
  \includegraphics[width=0.47\linewidth,clip]{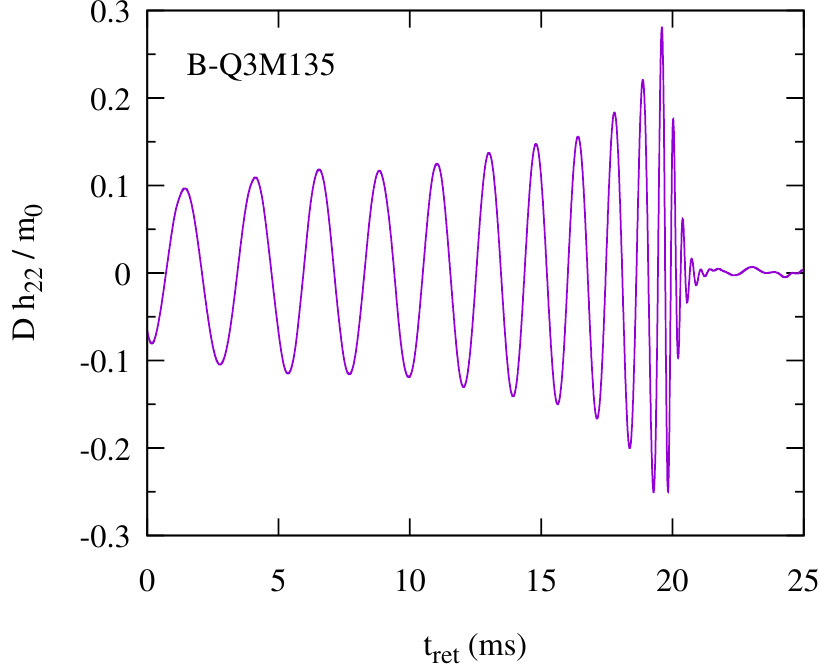}
 \end{tabular}
 \caption{$l=\abs{m}=2$ mode of gravitational waveforms for black
 hole--neutron star binaries modeled by various equations of state. The
 left top, right top, left bottom, and right bottom panels show the
 results for a very stiff (1.5H: $R_\mathrm{NS} = \SI{13.7}{\km}$,
 $\mathcal{C}=0.146$), stiff (H: $R_\mathrm{NS} = \SI{12.3}{\km}$,
 $\mathcal{C}=0.162$), moderate (HB: $R_\mathrm{NS} = \SI{11.6}{\km}$,
 $\mathcal{C} = 0.172$), and soft (B: $R_\mathrm{NS} = \SI{11.0}{\km}$,
 $\mathcal{C}=0.182$) equations of state, respectively
 \citep{Read_MSUCF2009,Lackey_KSBF2012}. Other parameters are fixed to
 be $M_\mathrm{BH} = 4.05\,M_\odot$, $\chi = 0$, and $M_\mathrm{NS} =
 1.35\,M_\odot$ ($Q=3$). This figure is generated from data of
 \citet{Kyutoku_Shibata_Taniguchi2010,Lackey_KSBF2012}.}
 \label{fig:gw_eos}
\end{figure}

Presence of the neutron-star matter influences gravitational waves from
black hole--neutron star binaries for the cases in which tidal effects
are strong, particularly in the merger and postmerger phases. Figure
\ref{fig:gw_eos} displays gravitational waves from nonspinning, $\chi =
0$ black hole--neutron star binaries with $M_\mathrm{BH} = 4.05\,M_\odot$,
$M_\mathrm{NS} = 1.35\,M_\odot$ ($Q=3$) and various equations of state
\citep{Kyutoku_Shibata_Taniguchi2010}. The most prominent imprint of
tidal disruption is the abrupt shutdown of gravitational-wave emission
and suppression of the ringdown waveform for low mass-ratio binaries
with low-compactness neutron stars, e.g., for 1.5H-Q3M135 in
Fig.~\ref{fig:gw_eos}. The primary mechanism for the suppression of the
ringdown waveform is phase cancellation
\citep{Nakamura_Sasaki1981,Shapiro_Wasserman1982,Nakamura_Oohara1983}. That
is, quasinormal modes are excited only incoherently by the infall of
widely-spread material as depicted in the left panel of
Fig.~\ref{fig:schematic}, and thus their amplitude is reduced by the
interference. This effect is enhanced more appreciably for a neutron
star with a larger radius, which is disrupted at a more distant orbit
from the innermost stable circular orbit. Here, it is appropriate to
point out that the incoherent infall of the widely-spread material is
realized because of the small radius of a low-mass black hole compared
to that of a neutron star. We recall that, if the black hole is
nonspinning, tidal disruption is possible only for low mass-ratio
systems discussed here. If the spin parameter of the black hole is
large, tidal disruption becomes possible for high mass-ratio systems,
and the gravitational waveform may exhibit different features as we
discuss later.

The gravitational-wave frequency at which this shutdown occurs is
determined approximately by the orbital frequency at which tidal
disruption occurs. One important remark is that this is in general
distinct from the orbital frequency at the onset of mass shedding. Even
after the mass shedding sets in, the neutron star remains to be
approximately gravitationally self-bound and preserves its orbital
evolution as a two-body system for a while. Tidal disruption occurs only
after the orbital separation decreases further as a result of continued
emission of inspiral-like gravitational waves. Because the
gravitational-wave amplitude suddenly decreases only at tidal
disruption, the orbital frequency at the onset of mass shedding may not
leave noticeable imprints in the waveform. By contrast, if the neutron
star is not disrupted, e.g., Fig.~\ref{fig:gwBHNSvsBHBH} and B-Q3M135 in
Fig.~\ref{fig:gw_eos}, the maximum frequency is determined universally
by the quasinormal mode of the remnant black hole. This is because most
of the neutron-star material falls into the black hole simultaneously
through a narrow region (see the middle panel of
Fig.~\ref{fig:schematic}) and the quasinormal modes are excited
efficiently. The maximum amplitude of gravitational waves also becomes
as large as that of binary black holes in the absence of tidal
disruption.

\begin{figure}[htbp]
 \centering
 \begin{tabular}{cc}
  \includegraphics[width=0.47\linewidth,clip]{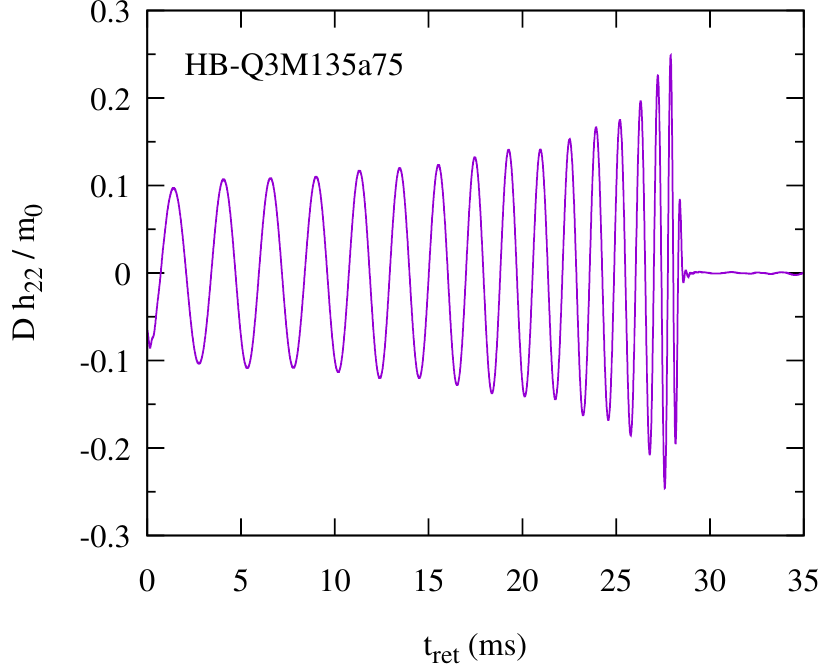} &
  \includegraphics[width=0.47\linewidth,clip]{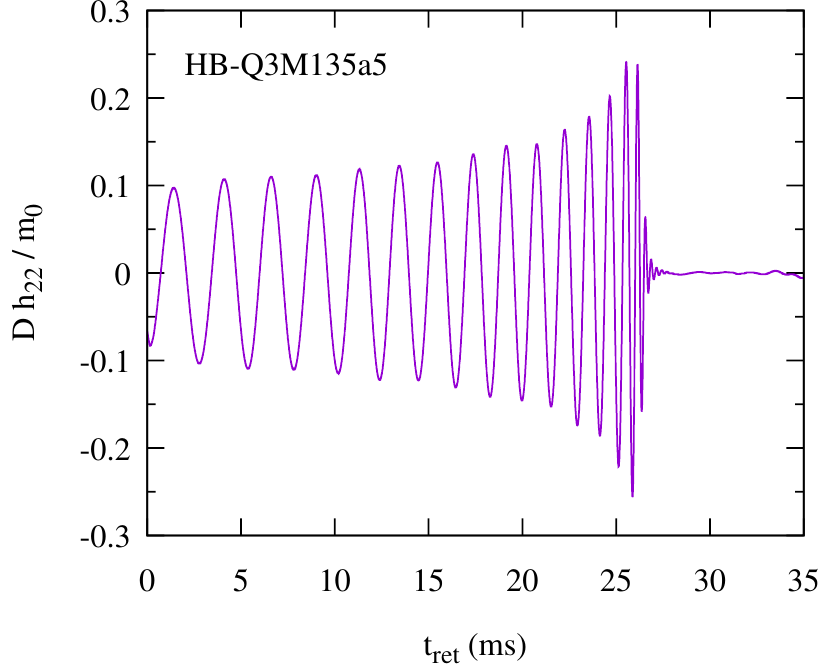} \\
  \includegraphics[width=0.47\linewidth,clip]{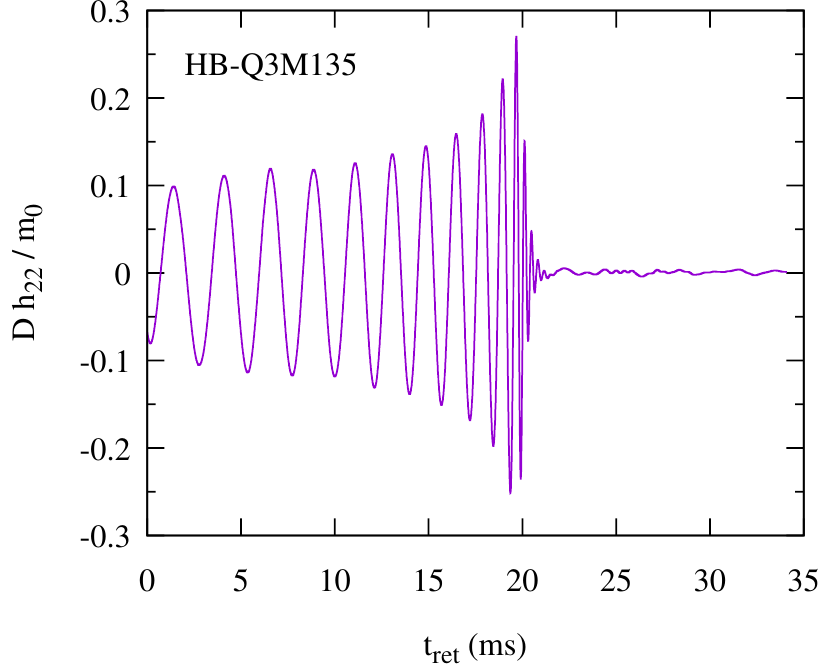} &
  \includegraphics[width=0.47\linewidth,clip]{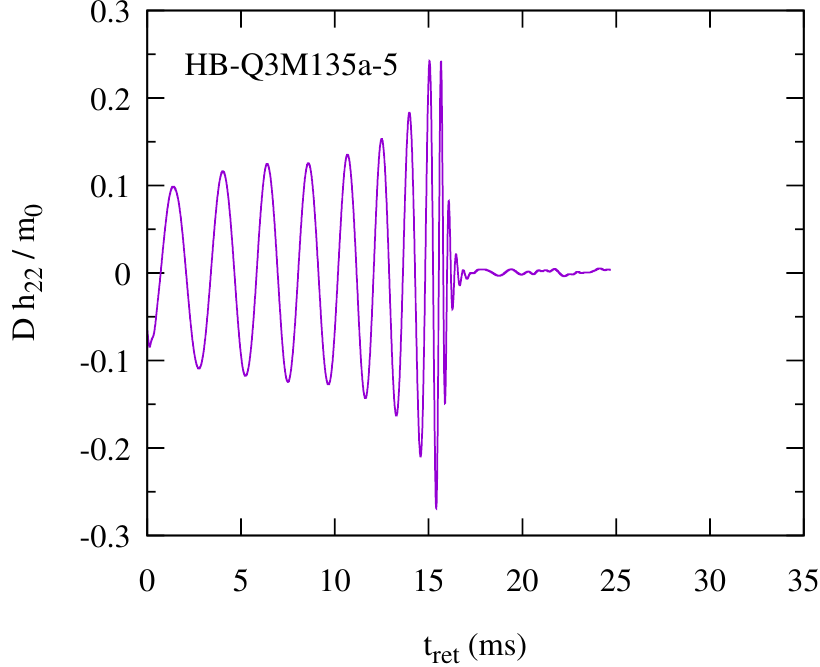}
 \end{tabular}
 \caption{Same as Fig.~\ref{fig:gw_eos} but for various values of the
 black-hole spin. The left top, right top, left bottom, and right bottom
 panels show the results for $\chi = 0.75$, $0.5$, $0$, and $-0.5$,
 respectively. Other parameters are fixed to be $M_\mathrm{BH} =
 4.05\,M_\odot$, $M_\mathrm{NS} = 1.35\,M_\odot$, and $R_\mathrm{NS} =
 \SI{11.6}{\km}$ ($Q=3$, $\mathcal{C}=0.172$) modeled by a piecewise
 polytrope called HB \citep{Read_MSUCF2009}. All the models have the
 same values of the initial orbital angular velocity, $m_0 \Omega =
 0.030$. The result for HB-Q3M135 has already been shown in
 Fig.~\ref{fig:gw_eos}, but we replot it for comparison. This figure is
 generated from data of
 \citet{Kyutoku_Shibata_Taniguchi2010,Kyutoku_OST2011}.}
 \label{fig:gw_spin}
\end{figure}

The spin of the black hole also modifies the gravitational waveform
qualitatively as it affects the merger process (Sect.~\ref{sec:sim_mrg})
and properties of the remnant (Sect.~\ref{sec:sim_rem}). Figure
\ref{fig:gw_spin} displays gravitational waves from black hole--neutron
star binaries with $M_\mathrm{BH} = 4.05\,M_\odot$, $M_\mathrm{NS} =
1.35\,M_\odot$, $R_\mathrm{NS} = \SI{11.6}{\km}$ ($Q=3$, $\mathcal{C} =
0.172$) modeled by a piecewise polytrope called HB for various values of
the (anti-)aligned spin of the black hole. This figure clearly shows
that the lifetime of the binary and the number of gravitational-wave
cycles in the inspiral phase increase as the spin of the black hole
increases. This is caused mainly by the spin-orbit coupling as described
in Sect.~\ref{sec:sim_mrg_s1}.

Figure \ref{fig:gw_spin} also illustrates that gravitational waveforms
in the merger phase are modified qualitatively by the spin of the black
hole, because it critically affects the final fate of the companion
neutron star. In particular, the quasinormal-mode excitation is
suppressed for HB-Q3M135a75 with $\chi = 0.75$ compared to HB-Q3M135
with $\chi = 0$. This is because the radius of the innermost stable
circular orbit is decreased and tidal disruption occurs far outside it
if the black hole has a prograde spin. Conversely, quasinormal modes are
clearly visible for zero and retrograde spins of the black hole, $\chi
\le 0$, because tidal disruption does not occur. Remarkably, HB-Q3M135a5
with $\chi = 0.5$ shows the ringdown signal, although the neutron star
is tidally disrupted and leaves the material as much as $0.11\,M_\odot$
outside the apparent horizon at \SI{10}{\ms} after the onset of
merger. The reason that the quasinormal modes are excited even if the
neutron star is tidally disrupted is that the tidal disruption occurs
only near the innermost stable circular orbit and thus the infall of the
disrupted material proceeds from a narrow region of the black-hole
surface without spreading around it (see the right panel of
Fig.~\ref{fig:schematic}).

\begin{figure}[htbp]
 \centering
 \begin{tabular}{cc}
  \includegraphics[width=0.47\linewidth,clip]{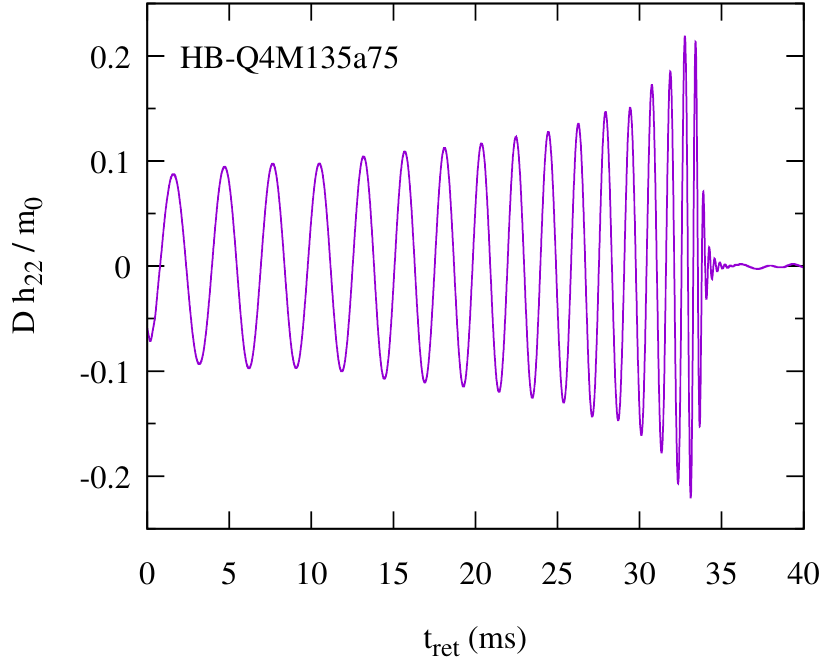} &
  \includegraphics[width=0.47\linewidth,clip]{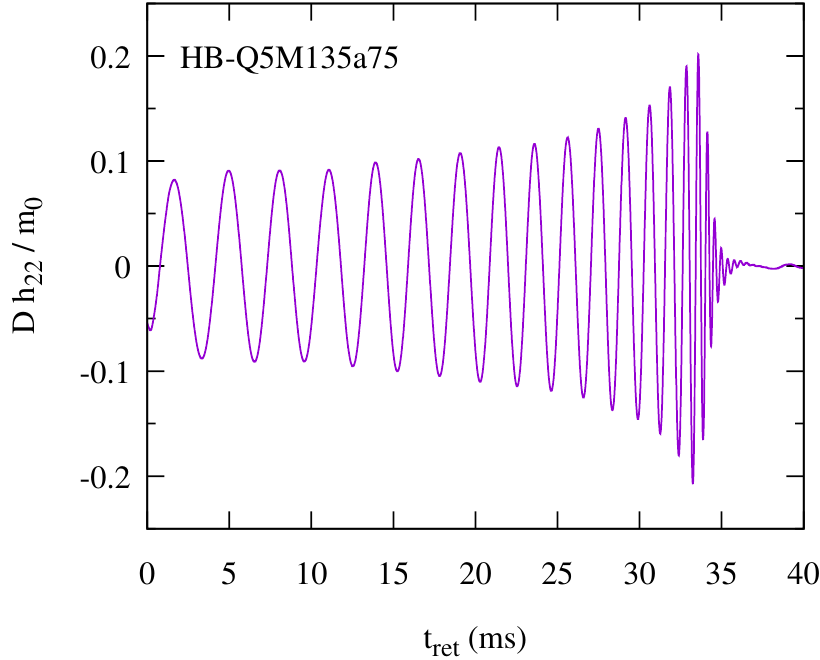}
 \end{tabular}
 \caption{Same as Fig.~\ref{fig:gw_eos} but for two values of the mass
 ratio. The left and right panels show the results for $M_\mathrm{BH} =
 5.4\,M_\odot$ ($Q=4$) and $6.75\,M_\odot$ ($Q=5$), respectively (see also
 the left top panel of Fig.~\ref{fig:gw_spin} for $Q=3$). Other
 parameters are fixed to be $\chi = 0.75$, $M_\mathrm{NS} =
 1.35\,M_\odot$, and $R_\mathrm{NS} = \SI{11.6}{\km}$
 ($\mathcal{C}=0.172$) modeled by a piecewise polytrope called HB
 \citep{Read_MSUCF2009}. This figure is generated from data of
 \citet{Kyutoku_OST2011}.} \label{fig:gw_Q}
\end{figure}

Tidal disruption of the neutron star in high mass-ratio systems is
typically compatible with the quasinormal-mode excitation of the remnant
black hole. Figure \ref{fig:gw_Q} displays the gravitational waveforms
for black hole--neutron star binaries with $\chi = 0.75$, $M_\mathrm{NS}
= 1.35\,M_\odot$, $R_\mathrm{NS} = \SI{11.6}{\km}$ ($\mathcal{C} = 0.172$)
modeled by a piecewise polytrope called HB, and two different values of
the black-hole mass $M_\mathrm{BH} = 5.4\,M_\odot$ ($Q=4$, left) and
$6.75\,M_\odot$ ($Q=5$, right). The ringdown waveform is clearly visible
for both systems due to the same reason as that described above for
HB-Q3M135a5. Although tidal disruption is less significant for higher
mass-ratio systems, HB-Q5M135a75 with $Q=5$ still leaves $0.095\,M_\odot$
outside the apparent horizon at \SI{10}{\ms} after the onset of
merger. Thus, the excitation of quasinormal modes cannot be ascribed to
the absence of tidal disruption. The existence of the ringdown waveform,
however, does not imply that the entire waveform is identical to that of
binary black holes. We discuss this point further using the spectrum in
Sect.~\ref{sec:sim_gw_spec}.

\subsubsection{Spectrum} \label{sec:sim_gw_spec}

Gravitational waves in the frequency domain clearly exhibit rich
information about their sources. Frequency-domain gravitational
waveforms are also useful for applications to gravitational-wave data
analysis, which is usually performed in the frequency domain. For the
early stage of the inspiral phase, during which the orbital frequency is
$\lesssim \SI{1}{\kilo\hertz}$, the spectral amplitude and phase of the
frequency-domain waveform agree approximately with those derived by the
post-Newtonian calculations. The effective spectral amplitude defined by
$h_\mathrm{eff} := f \abs*{\tilde{h} (f)}$, where $\tilde{h} (f)$ is the
Fourier transform of the time-domain waveform, behaves universally as
$\propto f^{-1/6}$ at low frequency in accordance with the prediction of
the quadrupole formula \citep[see,
e.g.,][]{Thorne,Sathyaprakash_Dhurandhar1991}. The gravitational-wave
phase is determined primarily by the chirp mass, $\mathcal{M} :=
M_\mathrm{BH}^{3/5} M_\mathrm{NS}^{3/5} / m_0^{1/5}$, and thus
$\mathcal{M}$ is determined with high accuracy by analyzing
gravitational waves in this stage
\citep{Sathyaprakash_Dhurandhar1991,Finn_Chernoff1993}. In closer
inspiral orbits with higher frequency, the spectral index increases as
the nonlinearity of general relativity and finite-size effects of the
neutron star become important at the small orbital separation. In this
stage, we may be able to extract the mass ratio, which allows us to
determine the masses of individual components, and spin parameters of
the binary \citep{Cutler_Flanagan1994,Poisson_Will1995}. These
measurements are indeed realized for black hole--neutron star binaries
in LIGO-Virgo O3 \citep{GWTC2,GW200105200115}. Tidal deformability of
the neutron star may also be extracted if the finite-size effect is
appreciable and the signal-to-noise ratio is large
\citep{Flanagan_Hinderer2008}.

The final fate of the neutron star in black hole--neutron star binaries
is clearly reflected in the damping of the gravitational-wave spectrum
at high frequency of $f \gtrsim \SI{1}{\kilo\hertz}$. If tidal
disruption (not the mass shedding) occurs at a distant orbit from the
innermost stable circular orbit, the spectral amplitude of gravitational
waves is characterized by damping above cutoff frequency,
$f_\mathrm{cut}$. Its precise value is given by the frequency of
gravitational waves in the last stage of the inspiral phase, $\sim
1$--\SI{2}{\kilo\hertz}, and depends primarily on the compactness of the
neutron star (see Sect.~\ref{sec:intro_tidal_ms}). If tidal deformation
(and thus disruption) is not significant, the spectrum in the
high-frequency range agrees approximately with that for binary black
holes. Even if the neutron star is tidally disrupted near the innermost
stable circular orbit, the inspiral-like motion continues inside it on a
dynamical time scale and gravitational waves with a large amplitude are
emitted. Thus, the effective amplitude, $h_\mathrm{eff} (f)$, is larger
than the case in which tidal effects play a significant role in
determining the final fate of the binary coalescence. If the tidal
effect is very weak, the effective amplitude even increases for $f
\gtrsim \SI{1}{\kilo\hertz}$ until it damps exponentially above the
cutoff frequency determined by quasinormal modes of the remnant black
hole in a similar manner to the case of binary-black-hole
coalescences. A proposal for quantitative definition of the cutoff
frequency will be reviewed later in Sect.~\ref{sec:sim_gw_fcut}.

\begin{figure}[htbp]
 \centering \includegraphics[width=0.7\linewidth,clip]{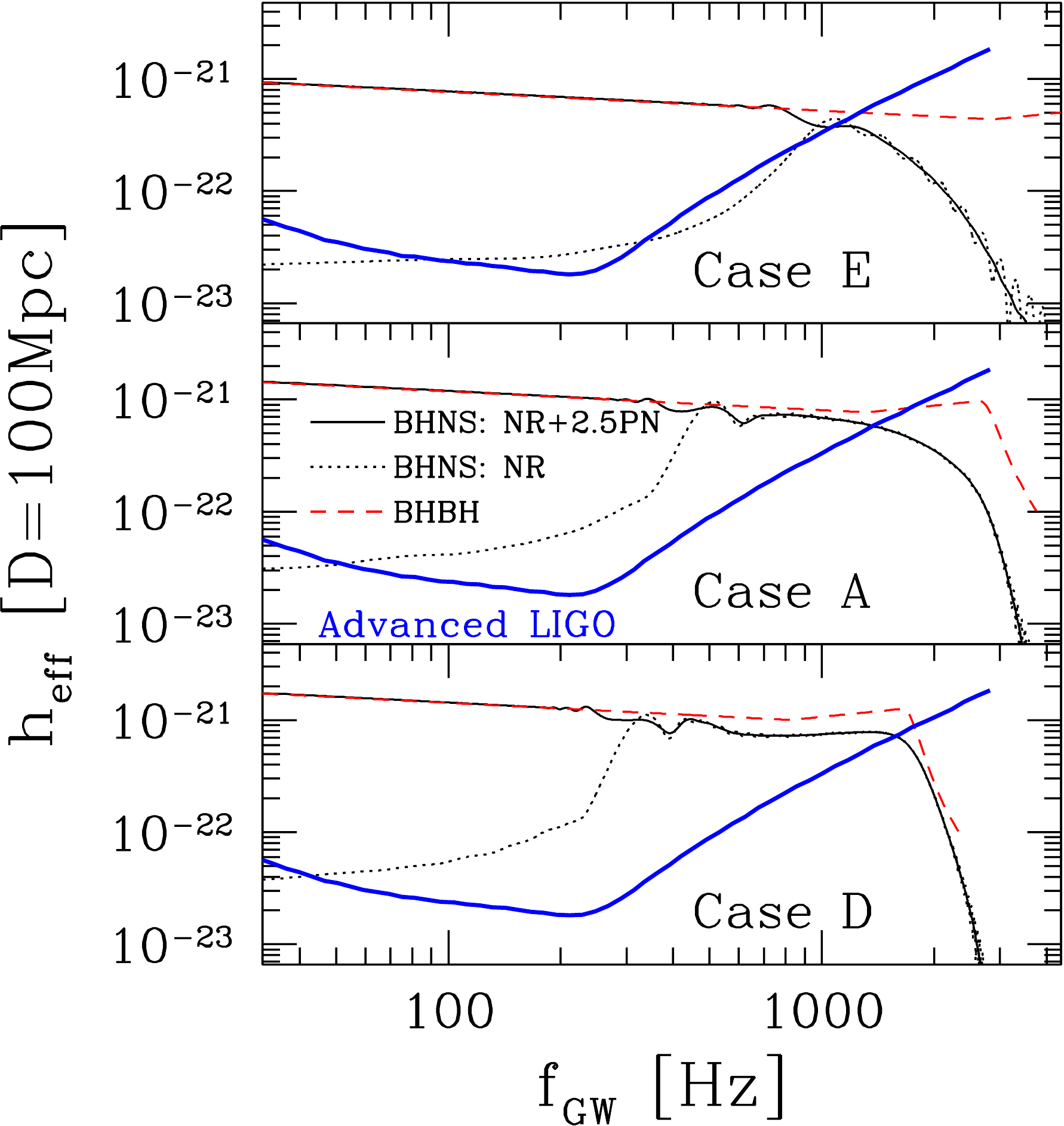}
 \caption{Gravitational-wave spectrum for black hole--neutron star
 binaries with $Q=1$ (top: Case E), $3$ (middle: Case A), and $5$
 (bottom: Case D). Other parameters are fixed to be $\chi = 0$ and
 $\mathcal{C} = 0.145$ modeled by a $\Gamma = 2$ polytrope. To plot the
 curves in physical units, the baryon rest mass of the neutron star is
 set to be $1.4\,M_\odot$ and the distance of $D=\SI{100}{Mpc}$ is
 assumed. The black solid curves show the spectra of hybrid waveforms
 constructed by a second-and-a-half-order post-Newtonian approximation
 and numerical-relativity simulations, while the black dotted curves
 show only the latter. The red dashed curve shows the so-called PhenomA
 model for binary black holes \citep{Ajith_etal2008}. The blue solid
 curve is the noise spectral density of the Advanced LIGO planned as of
 2009. Image reproduced with permission from \citet{Etienne_LSB2009}.}
 \label{fig:spec_Q}
\end{figure}

Figure \ref{fig:spec_Q} generated by \citet{Etienne_LSB2009} displays
the variation of the spectral cutoff with respect to the degree of tidal
disruption by comparing results for nonspinning black hole--neutron star
binaries with different mass ratios (equivalently, different masses of
the black holes with the same mass and radius of the neutron stars). The
top panel for $Q=1$ shows that the spectrum damps at $f \sim
\SI{1}{\kilo\hertz}$, which is far below the quasinormal-mode frequency
for the remnant black hole of $\sim \SI{6}{\kilo\hertz}$. The cutoff of
this spectrum reflects tidal disruption at a distant orbit from the
innermost stable circular orbit. The bottom panel for $Q=5$ shows that
the cutoff frequency of this spectrum agrees approximately with the
quasinormal-mode frequency, because tidal disruption is
insignificant. The middle panel for $Q=3$ shows intermediate
behavior. Specifically, the cutoff frequency is determined by tidal
disruption which occurs near the innermost stable circular orbit.

\begin{figure}[htbp]
 \centering \includegraphics[width=0.7\linewidth,clip]{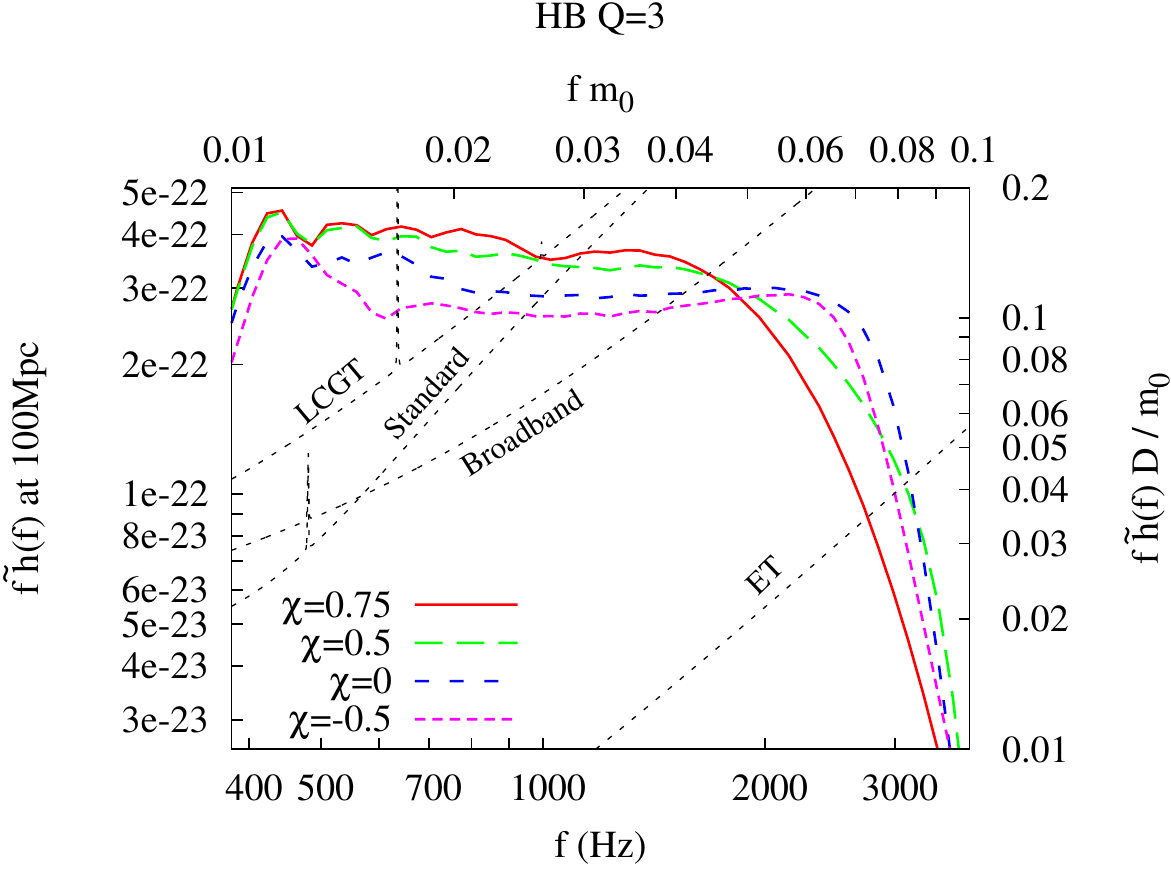}
 \caption{Gravitational-wave spectrum of the $l=\abs{m}=2$ mode for
 black hole--neutron star binaries with various values of the black-hole
 spin. Other parameters are fixed to be $M_\mathrm{BH} = 4.05\,M_\odot$,
 $M_\mathrm{NS} = 1.35\,M_\odot$, and $R_\mathrm{NS} = \SI{11.6}{\km}$
 ($Q=3$, $\mathcal{C}=0.172$) modeled by a piecewise polytrope called HB
 \citep{Read_MSUCF2009}. These spectra are derived from the waveforms
 shown in Fig.~\ref{fig:gw_spin}, and the cutoff at $f \lesssim
 \SI{400}{\hertz}$ results from the fact that numerical simulations
 started there. The left and bottom axes show these quantities in
 physical units assuming that the observer is located at \SI{100}{Mpc}
 along the direction perpendicular to the orbital plane. The right and
 top axes show normalized, dimensionless amplitude and frequency,
 respectively. The dashed curves labeled by ``Standard'' and
 ``Broadband'' are two options for the noise spectral density of the
 Advanced LIGO planned as of 2011. Those labeled by ``LCGT'' and ``ET''
 are the planned noise spectral density of KAGRA, formerly called LCGT,
 and the Einstein Telescope, respectively. Image adapted from
 \citet{Kyutoku_OST2011}; copyright by APS.}
 \label{fig:spec_spin}
\end{figure}

\begin{figure}[htbp]
 \centering
 \includegraphics[width=0.8\linewidth,clip]{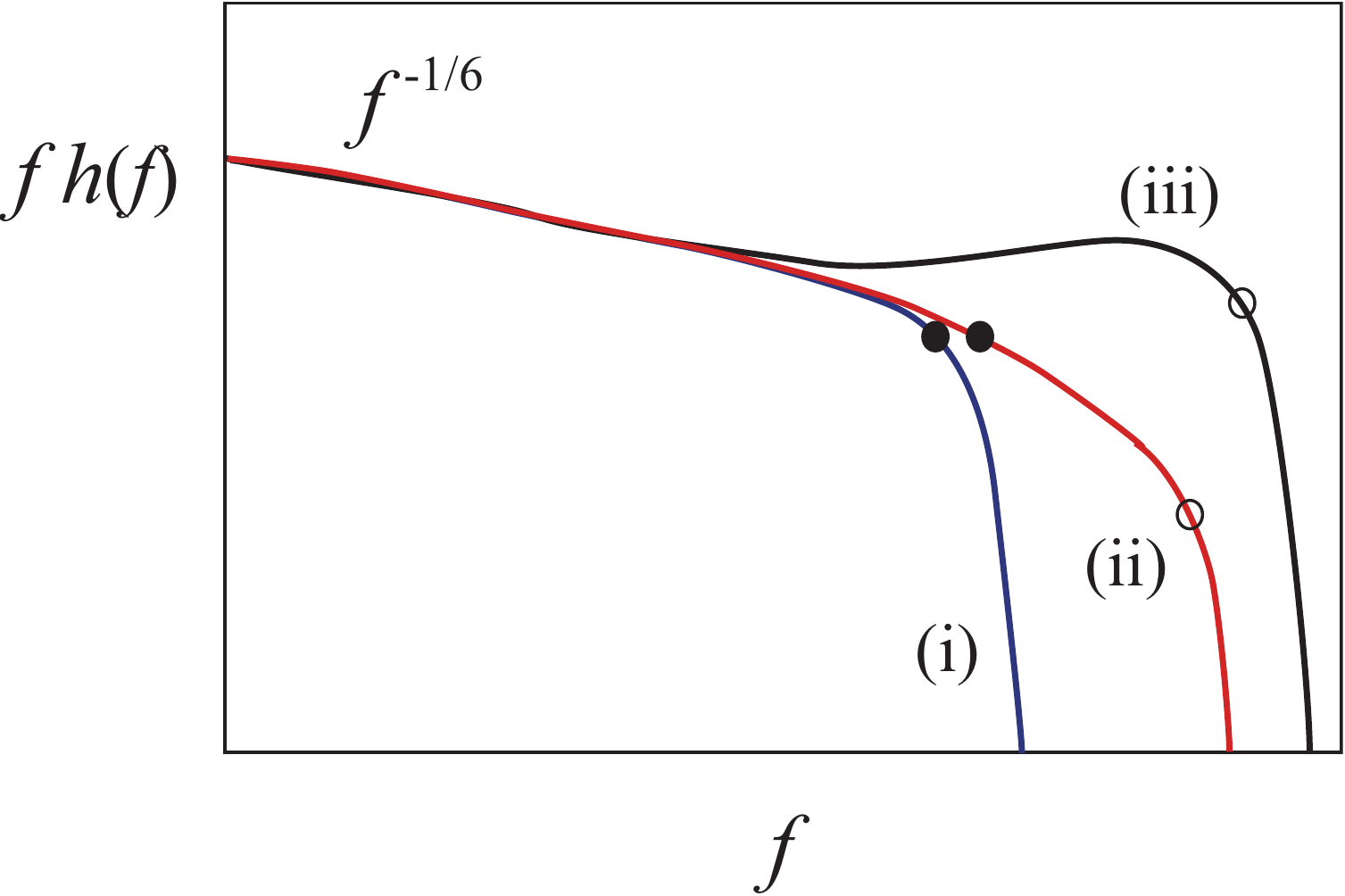}
 \caption{Schematic description of three types of gravitational-wave
 spectra for black hole--neutron star binaries. The filled and open
 circles denote frequency associated with tidal disruption and a
 quasinormal mode of the remnant black hole, respectively. The spectrum
 (i) corresponds to the case depicted in the left panel of
 Fig.~\ref{fig:schematic}, for which tidal disruption is induced by a
 black hole with a small mass. The spectrum (ii) corresponds to the case
 depicted in the middle panel of Fig.~\ref{fig:schematic}, for which
 tidal disruption is induced by a black hole with a large mass and a
 high spin. The spectrum (iii) corresponds to the case depicted in the
 right panel of Fig.~\ref{fig:schematic}, for which tidal disruption
 does not occur as the black hole is massive and does not have a high
 prograde spin.} \label{fig:specschematic}
\end{figure}

The spin of the black hole is the key for observing cutoff behavior
associated with tidal disruption
\citep{Etienne_LSB2009,Kyutoku_OST2011}. Figure \ref{fig:spec_spin}
shows that the spectrum depends quantitatively and also qualitatively on
the black-hole spin. Because tidal disruption does not occur
significantly for systems with $\chi \le 0$ shown in this figure, the
cutoff frequency, $f_\mathrm{cut}$, is $\sim 2.5$--\SI{3}{\kilo\hertz}
determined by the quasinormal modes of the remnant black holes with
$\sim 5.3\,M_\odot$. For the system with $\chi = 0.75$, the cutoff
frequency becomes as low as $f_\mathrm{cut} \sim
1.5$--$\SI{2}{\kilo\hertz}$, because tidal disruption occurs outside the
innermost stable circular orbit. Notably, the system with $\chi = 0.5$
shows both of these features. That is, the spectrum for $\chi = 0.5$
first exhibits softening of the spectrum at $f \sim \SI{2}{\kilo\hertz}$
associated with tidal disruption and next damps exponentially above $f
\sim \SI{3}{\kilo\hertz}$ associated with the quasinormal mode,
consistently with the gravitational waveform of HB-Q3M135a5 (exactly the
same model) shown in the left top panel of
Fig.~\ref{fig:gw_spin}. Hereafter, we refer to the former as the cutoff
frequency, because we are primarily interested in tidal disruption. This
type of the spectrum is produced for the case depicted in the right
panel of Fig.~\ref{fig:schematic}, and schematic gravitational-wave
spectra for these three types of the merger process are shown in
Fig.~\ref{fig:specschematic}.

The spin-orbit coupling described in Sect.~\ref{sec:sim_mrg_s1} induces
two features preferable for observing the cutoff frequency by
gravitational-wave detectors if the spin is prograde. First, because the
prograde spin reduces the orbital angular velocity to maintain a
circular orbit for a given orbital separation, the cutoff frequency of
gravitational waves at tidal disruption is reduced. Next, the effective
amplitude before tidal disruption increases for given frequency due to
the reason explained by the following post-Newtonian argument
\citep{Kyutoku_OST2011}. By retaining only the first-and-a-half-order
post-Newtonian term related to the spin-orbit coupling, the binding
energy of the system, $E$, as a function of the gravitational-wave
frequency behaves as \citep{Kidder1995}
\begin{equation}
 \dv{E}{f} = \frac{Q}{3(1+Q)^2} \frac{(\pi m_0 f)^{5/3}}{\pi f^2}
  \bqty{1 + \chi \vu{S} \vdot \vu{L} \frac{5(4Q+3)}{3(1+Q)^2} ( \pi m_0
  f )} ,
\end{equation}
where $\vu{S}$ and $\vu{L}$ are the unit vectors in the direction of the
black-hole spin and the orbital angular momentum, respectively. Note
that we have also neglected the spin of the neutron star. Here,
$\dv*{E}{f}$ is related to the effective amplitude via
\begin{equation}
 \dv{E}{f} \propto \bqty{f \abs{\tilde{h}(f)}}^2 = h_\mathrm{eff}^2 (f)
  .
\end{equation}
Thus, for given frequency, the effective amplitude increases as the spin
of the black hole increases in the prograde direction, reflecting the
fact that the binary has to emit larger energy via gravitational waves
for increasing the orbital angular velocity against the spin-orbit
repulsive force. These two effects are clearly shown in
Fig.~\ref{fig:spec_spin}.

Arguably the most important feature of the cutoff frequency is that it
depends on the equation of state of the neutron star, in particular the
average density (see Sect.~\ref{sec:intro_tidal_ms}). Thus, the cutoff
frequency will give us information about the equation of state for
supranuclear-density matter, particularly if it is stiff
\citep{Vallisneri2000}.

\begin{figure}[htbp]
 \centering
 \begin{tabular}{cc}
  \includegraphics[width=0.47\linewidth,clip]{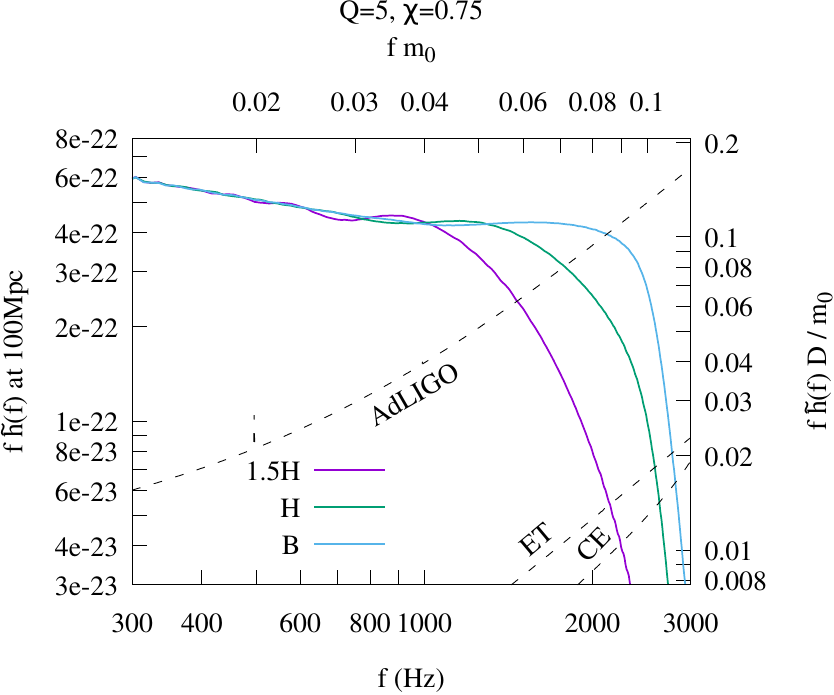} &
  \includegraphics[width=0.47\linewidth,clip]{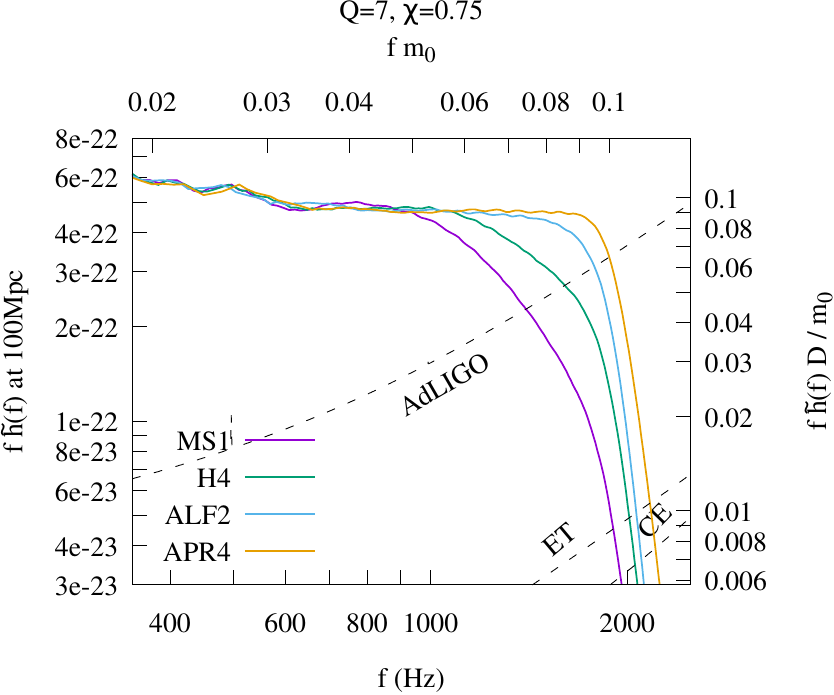}
 \end{tabular}
 \caption{Same as Fig.~\ref{fig:spec_spin} but for binaries with $\chi =
 0.75$ and $M_\mathrm{NS} = 1.35\,M_\odot$ modeled by various equations of
 state. The left panel is for $M_\mathrm{BH} = 6.75\,M_\odot$ ($Q=5$), and
 the equations of state are taken to be piecewise polytropes called 1.5H
 ($R_\mathrm{NS} = \SI{13.7}{\km}$, $\mathcal{C}=0.146$), H
 ($R_\mathrm{NS} = \SI{12.3}{\km}$, $\mathcal{C}=0.162$), and B
 ($R_\mathrm{NS} = \SI{11.0}{\km}$, $\mathcal{C}=0.182$)
 \citep{Read_MSUCF2009,Lackey_KSBF2012}. The right panel is for
 $M_\mathrm{BH} = 9.45\,M_\odot$ ($Q=7$), and the equations of state are
 taken to be piecewise polytropic approximations of MS1
 \citep[$R_\mathrm{NS}=\SI{14.4}{\km}$,
 $\mathcal{C}=0.138$]{Muller_Serot1996}, H4
 \citep[$R_\mathrm{NS}=\SI{13.6}{\km}$,
 $\mathcal{C}=0.147$]{Glendenning_Moszkowski1991,Lackey_Nayyar_Owen2006},
 ALF2 \citep[$R_\mathrm{NS}=\SI{12.4}{\km}$,
 $\mathcal{C}=0.161$]{Alford_BPR2005}, and APR4 \citep[$R_\mathrm{NS} =
 \SI{11.1}{\km}$,
 $\mathcal{C}=0.180$]{Akmal_Pandharipande_Ravehnall1998}. The dashed
 curves labeled by ``AdLIGO,'' ``ET,'' and ``CE'' are the noise spectral
 density of the Advanced LIGO at the design sensitivity, Einstein
 Telescope, and Cosmic Explorer, respectively, taken from
 \citet{CosmicExplorer}. The left and right panels are generated from
 data of unpublished work and \citet{Kyutoku_IOST2015}, respectively.}
 \label{fig:spec_EOS}
\end{figure}

Figure \ref{fig:spec_EOS} plots the gravitational-wave spectrum for
black hole--neutron star binaries with $\chi = 0.75$ and $M_\mathrm{NS}
= 1.35\,M_\odot$ modeled by a variety of equations of state. In this
figure, $M_\mathrm{BH}$ is taken to be $6.75\,M_\odot$ (left: $Q=5$,
unpublished) and $9.45\,M_\odot$ \citep[right:
$Q=7$,][]{Kyutoku_IOST2015}. This figure clearly shows that the cutoff
frequency at which the effective amplitude begins to damp depends on the
equation of state for given values of the mass and the spin of the
components. Specifically, the cutoff frequency is in the range of $\sim
1$--$\SI{2.5}{\kilo\hertz}$ and becomes lower as the equation of state
becomes stiffer (i.e., the compactness of the neutron star becomes
smaller) for the systems considered here, because tidal disruption
occurs at a more distant orbit with lower frequency. At the same time,
the effective amplitude for given frequency at $f \gtrsim
\SI{1}{\kilo\hertz}$ increases as the equation of state becomes soft
(i.e., the compactness becomes large). For a softer equation of state,
the cutoff frequency becomes higher and saturates approximately to the
quasinormal-mode frequency as tidal disruption becomes
insignificant. These features are discussed in more detail in
Sect.~\ref{sec:sim_gw_fcut}.

A remarkable feature of high mass-ratio systems with the high prograde
spin of the black hole is that the neutron star can be tidally disrupted
outside the innermost stable circular orbit for a wide range of
equations of state. This feature can make variation of the spectra and
the cutoff frequency visible in the sensitivity band of current
ground-based detectors such as the Advanced LIGO. Thus, binaries of a
high-mass, high-spin black hole and a neutron star is a promising target
for studying the equation of state via the cutoff frequency. For even
larger values of the black-hole spin parameter, say $\chi \ge 0.9$,
tidal disruption of a neutron star with the typical mass in our Galaxy
of $1.3$--$1.4\,M_\odot$ will also be possible for even larger values of
the mass ratio, $Q \gtrsim 10$ or equivalently more massive black holes
with $M_\mathrm{BH} \gtrsim 13$--$14\,M_\odot$. Exploration of tidal
disruption at such a high mass-ratio regime is an important topic for a
future study in numerical relativity.

Conversely, the cutoff frequency is higher than $\sim
\SI{2}{\kilo\hertz}$ if the mass ratio of the system is $Q \lesssim
3$. We recall that the cutoff frequency is determined not by mass
shedding but by tidal disruption, which occurs in a closer orbit with
higher frequency, in particular for the small values of $Q$. Thus, the
cutoff frequency of low mass-ratio systems allows us to investigate
neutron-star equations of state only if we are capable of detecting
signals at $f \gtrsim \SI{2}{\kilo\hertz}$, for which the sensitivity of
detectors is degraded. Moreover, the amplitude is not very large because
of the small mass. These facts imply that the study of the cutoff
frequency with nonspinning black hole--neutron star binaries will await
third-generation detectors such as the Einstein Telescope
\citep{EinsteinTelescope} and the Cosmic Explorer
\citep{CosmicExplorer}.

Before concluding Sect.~\ref{sec:sim_gw_spec}, we comment on the current
status of the development of phenomenological frequency-domain waveform
templates with various methods utilizing results of numerical
relativity. Preliminary models, one of which is called LEA, were
constructed with the aid of a phenomenological model called PhenomC
\citep{Santamaria_etal2010} or the effective-one-body formalism
\citep{Taracchini_PBBBCLPS2012} by calibrating free parameters for both
the amplitude and the phase with numerical-relativity simulations of
black hole--neutron star binaries \citep{Lackey_KSBF2014}. One of the
up-to-date models, PhenomNSBH \citep{Thompson_FKNPDH2020}, is
constructed by combining an amplitude model calibrated with
numerical-relativity simulations of black hole--neutron star binaries
\citep{Pannarale_BKLS2015_2} with a phase model consisting of a model
called PhenomD for the point-particle baseline \citep{Khan_HHOPFB2016}
and the NRTidal model for the tidal effect
\citep{Dietrich_Bernuzzi_Tichy2017,Dietrich_SKJDT2019}, which is derived
also in a phenomenological manner by calibrating free parameters with
numerical-relativity simulations of binary neutron stars. Another model,
SEOBNR\_NSBH \citep{Matas_etal2020}, is constructed in a similar manner
except that the point-particle baseline is taken to be an
effective-one-body model called SEOBNR \citep{Bohe_etal2017} and new
numerical-relativity waveform models are used to recalibrate the
amplitude model. These phenomenological models are available in LSC
Algorithm Library. A common drawback among these waveform models may be
that no higher harmonic mode or no precession has been modeled despite
their relevance for black hole--neutron star binaries. These issues are
discussed in Sect.~\ref{sec:sim_gw_high}. In addition, the ringdown
phase is not incorporated appropriately in the phase model.

\subsubsection{Correlation of the cutoff frequency and the compactness}
\label{sec:sim_gw_fcut}

To quantify the cutoff frequency and contained information,
\citealt{Shibata_KYT2009,Kyutoku_Shibata_Taniguchi2010,Kyutoku_OST2011}
have fitted the gravitational-wave spectra by a function of the form
\begin{align}
 \frac{f \tilde{h}_\mathrm{fit} (f) D}{m_0} = \frac{f
 \tilde{h}_\mathrm{3PN} (f) D}{m_0}
 e^{-(f/f_\mathrm{ins})^{\sigma_\mathrm{ins}}} + A
 e^{-(f/f_\mathrm{dam})^{\sigma_\mathrm{dam}}} \bqty{1 -
 e^{-(f/f_\mathrm{ins2})^{\sigma_\mathrm{ins2}}}} ,
\end{align}
where $\tilde{h}_\mathrm{3PN}$ is obtained by Fourier-transforming the
so-called TaylorT4 approximant \citep{Buonanno_Chen_Vallisneri2003},
which is one variant of post-Newtonian approximants
\citep{Damour_Iyer_Sathyaprakash2001,Buonanno_IOPS2009}, and
$f_\mathrm{ins}$, $f_\mathrm{ins2}$, $f_\mathrm{dam}$,
$\sigma_\mathrm{ins}$, $\sigma_\mathrm{ins2}$, $\sigma_\mathrm{dam}$,
and $A$ are free parameters to be determined. Then,
\citet{Kyutoku_OST2011} determined the cutoff frequency $f_\mathrm{cut}$
by the higher of two frequency at which the second term of this function
becomes a half of the maximum value. Note that earlier work have simply
defined $f_\mathrm{cut}$ by $f_\mathrm{dam}$
\citep{Shibata_KYT2009,Kyutoku_Shibata_Taniguchi2010}, but this
definition turned out to be not useful for spinning black hole--neutron
star binaries, primarily because the functional form is not flexible
enough to model the spectrum (ii) in
Fig.~\ref{fig:specschematic}. Robust and useful definition of the cutoff
frequency, $f_\mathrm{cut}$, is still to be explored.

\begin{figure}[htbp]
 \centering
 \begin{tabular}{cc}
  \includegraphics[width=0.47\linewidth,clip]{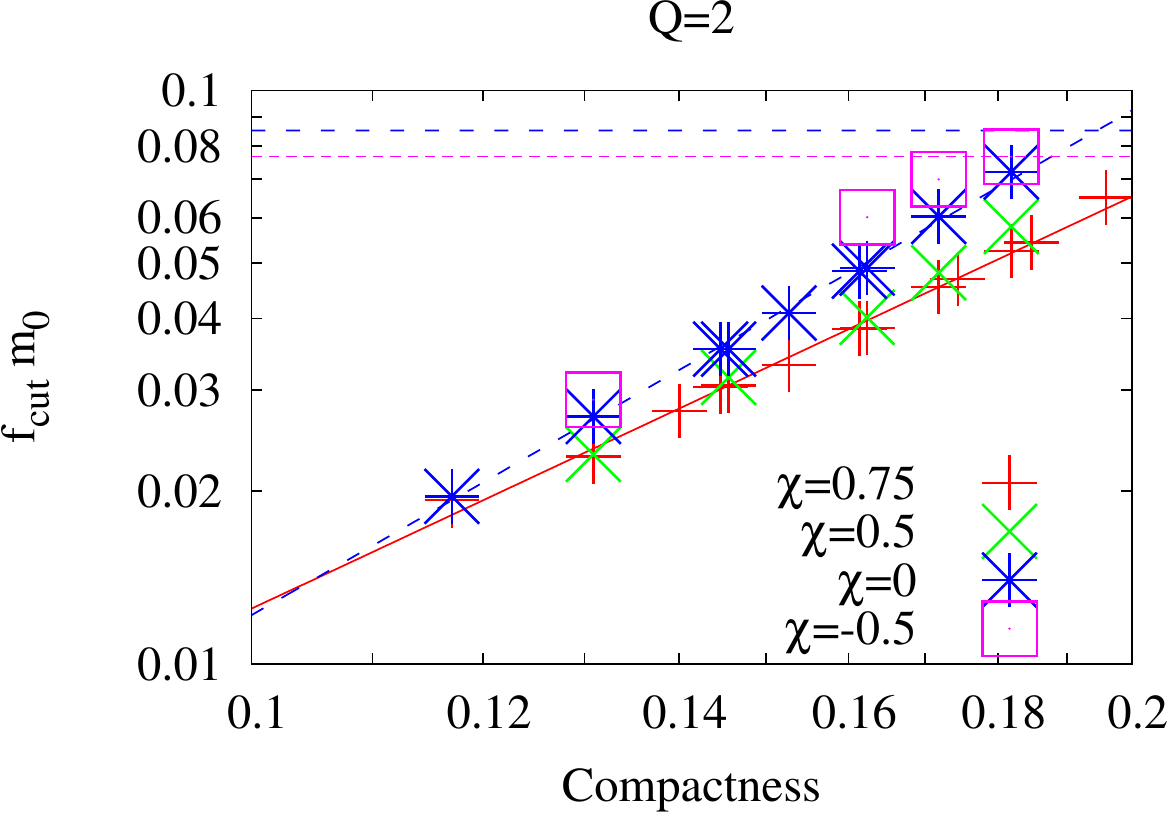} &
  \includegraphics[width=0.47\linewidth,clip]{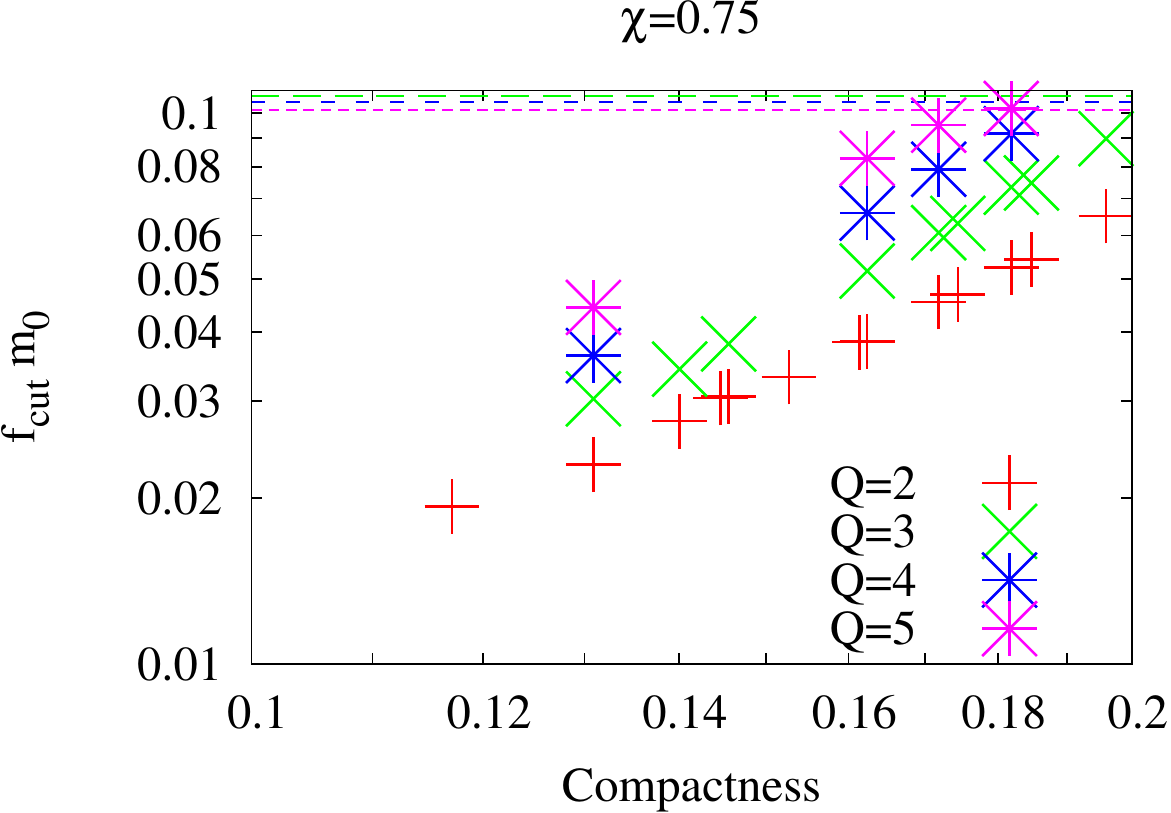}
 \end{tabular}
 \caption{Relation between the normalized, dimensionless cutoff
 frequency $f_\mathrm{cut} m_0$ and the compactness of the neutron star
 $\mathcal{C}$ derived by a suite of simulations with piecewise
 polytropes. The left panel show the results for $Q=2$ with $\chi =
 0.75$, $0.5$, $0$, and $-0.5$. The red solid and blue dashed lines are
 results of linear fittings for $\chi = 0.75$ ($f_\mathrm{cut} m_0
 \propto \mathcal{C}^{2.39}$) and $\chi=0$ ($f_\mathrm{cut} m_0 \propto
 \mathcal{C}^{2.92}$), respectively. Notice that the power-law indices
 are larger than $3/2$ [cf., Eq.~\eqref{eq:msfgw}]. The right panel
 shows the results for $\chi = 0.75$ with $Q=2$, $3$, $4$, and $5$. The
 dashed horizontal lines in the upper parts of the plots denote typical
 values of the quasinormal-mode frequency for each binary
 parameter. Image adapted from \citet{Kyutoku_OST2011}, copyright by APS.} \label{fig:fcut}
\end{figure}

Figure \ref{fig:fcut} plots the dimensionless cutoff frequency
$f_\mathrm{cut} m_0$ for various systems. The left panel is for $Q=2$
with various values of $\chi$, and the right panel is for $\chi = 0.75$
with various values of $Q$. This figure shows that the cutoff frequency
normalized by the inverse of the total mass has a tight correlation with
the compactness of the neutron star for given values of the mass ratio
and the spin parameter of the black hole. Thus, if we are able to
determine masses and spin parameters from gravitational-wave data
analysis \citep[see, e.g.,][for degeneracy of the mass and the
spin]{Cutler_Flanagan1994,Hannam_BFFH2013}, the cutoff frequency may
enable us to constrain the compactness of the neutron star and hence the
equation of state. Although it is not clear from this figure, for the
case in which tidal disruption does not occur significantly, which is
typical for a large compactness, the cutoff frequency settles
approximately to the quasinormal-mode frequency shown in the upper part
of the panels \citep{Shibata_KYT2009,Kyutoku_Shibata_Taniguchi2010}. It
has also been found that, by adopting piecewise polytropes with
different values of the adiabatic index for the core region, the cutoff
frequency also depends weakly on the density profile of the neutron star
\citep{Kyutoku_Shibata_Taniguchi2010}. This trend is consistent with
that for the mass of the material remaining outside the apparent horizon
described in Sect.~\ref{sec:sim_rem_disk}.

An important finding is that the dimensionless cutoff frequency,
$f_\mathrm{cut} m_0$, depends more strongly on the neutron-star
compactness than expected from the mass-shedding condition,
Eq.~\eqref{eq:msomega}. Specifically, although the mass-shedding
condition suggests $f_\mathrm{cut} m_0 \propto \mathcal{C}^{3/2}$, the
results of nonspinning black hole--neutron star binaries with $Q=2$ are
found to be approximated by \citep{Kyutoku_OST2011}
\begin{equation}
 \ln( f_\mathrm{cut} m_0 ) = ( 2.92 \pm 0.06 ) \ln \mathcal{C} + ( 2.32
  \pm 0.12 )
\end{equation}
up to the compactness above which tidal disruption becomes
insignificant. That is, $f_\mathrm{cut} m_0$ is approximately
proportional to $\mathcal{C}^3$. It should be cautioned that this
power-law index depends on the definition of $f_\mathrm{cut}$ (e.g.,
\citealt{Kyutoku_Shibata_Taniguchi2010} derived a larger value of
$\approx 4$) and also decreases for a prograde spin as shown in
Fig.~\ref{fig:fcut}, specifically $f_\mathrm{cut} m_0 \propto
\mathcal{C}^{2.4}$ for $Q=2$ and $\chi = 0.75$. Still, the index is
universally larger than $3/2$ derived by the mass-shedding
condition. This fact illustrates that the cutoff frequency is not
determined simply by mass shedding. Actually, the steep dependence is
reasonably expected, because the survival time of a neutron star against
tidal disruption after the onset of mass shedding is generally longer
for a more compact neutron star with a more centrally-condensed density
profile.

\subsubsection{Higher harmonic mode and spin-induced precession}
\label{sec:sim_gw_high}

Up to here, we have discussed only the dominant $l=\abs{m}=2$ modes of
gravitational waves. They approximately represent gravitational waves
observed from the direction perpendicular to the orbital plane of
nonprecessing binaries. If we observe nonprecessing binaries from
inclined directions, various harmonic modes contribute significantly to
the observed signal, particularly because black hole--neutron star
binaries are highly asymmetric in terms of the masses of components and
emit odd modes such as $l=\abs{m}=3$ efficiently
\citep{GWTC2,GW200105200115}. The mode mixing induces modulation of
gravitational-wave amplitude and phase, which cannot be mimicked only by
the dominant mode \citep{Foucart_etal2021}. Thus, accurate modeling of
gravitational waves particularly in the strong-field regime requires us
to take into account higher harmonic modes. It should also be mentioned
that numerical-relativity simulations of compact binary coalescences
have only recently begun to extract $m=0$ modes relevant to
gravitational-wave memory in a reliable and accurate manner
\citep{Mitman_MSTBDKT2020}.

\begin{figure}[htbp]
 \centering
 \begin{tabular}{cc}
  \includegraphics[width=0.48\linewidth,clip]{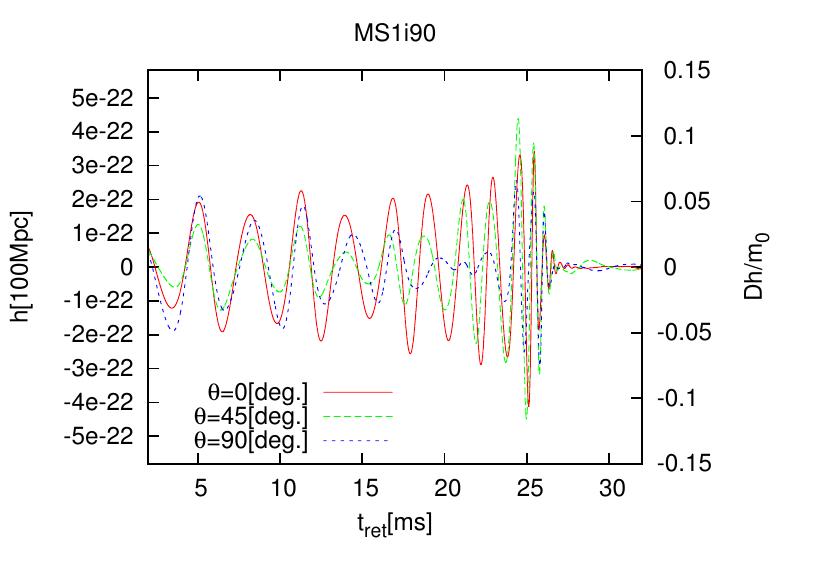} &
  \includegraphics[width=0.48\linewidth,clip]{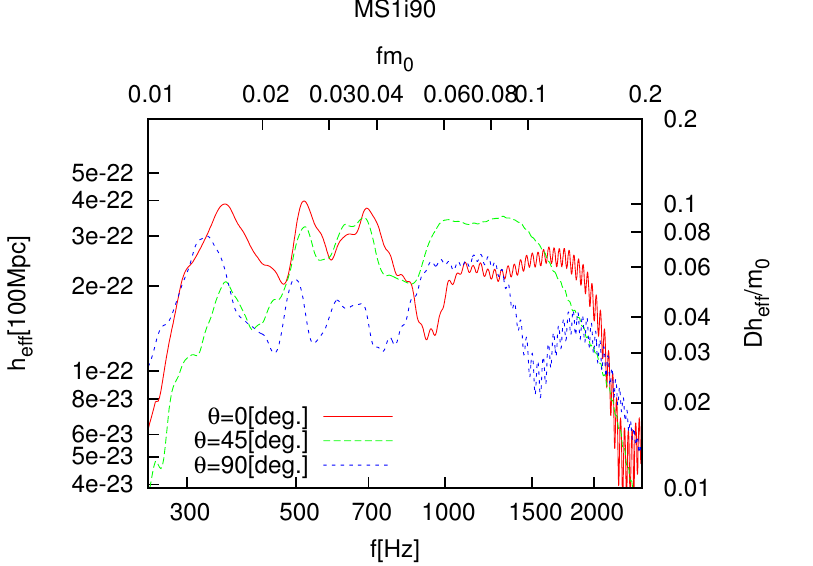}
 \end{tabular}
 \caption{Plus-mode gravitational waveform (left) and spectrum (right)
 for a binary with $M_\mathrm{BH} = 6.75\,M_\odot$, $\chi = 0.75$,
 $M_\mathrm{NS} = 1.35\,M_\odot$, and $R_\mathrm{NS} = \SI{14.4}{\km}$
 ($Q=5$, $\mathcal{C}=0.138$) modeled by a piecewise-polytropic
 approximation of the MS1 equation of state \citep{Muller_Serot1996}
 observed from different directions with respect to the total angular
 momentum at the initial instant of simulations. All the modes with
 $l=2,3,$ and $4$ are included. In the left panel, the left axis shows
 the amplitude in physical units assuming the distance between the
 source and the observer to be $D=\SI{100}{Mpc}$, and the right axis
 shows the dimensionless, normalized amplitude. In the right panel, the
 left and bottom axes show the effective amplitude and the frequency in
 physical units, respectively, and the right and top axes show those in
 the dimensionless form. The cutoff of the spectrum at $f \lesssim
 \SI{300}{\hertz}$ results from the fact that numerical simulations
 started there. Image reproduced with permission from \citet{Kawaguchi_KNOST2015}, copyright by APS.}
 \label{fig:gw_incl}
\end{figure}

If the spin of the black hole is inclined with respect to the orbital
angular momentum, the orbital plane precesses as shown in
Fig.~\ref{fig:orb_incl}, and the amplitude and phase of gravitational
waveforms are modulated due to the coupling of orbital and precessional
frequency
\citep{Foucart_DKT2011,Foucart_DDKMOPSST2013,Kawaguchi_KNOST2015,Foucart_etal2021}. The
left panel of Fig.~\ref{fig:gw_incl} displays the plus-mode
gravitational waveforms observed from various directions for a system
with $M_\mathrm{BH} = 6.75\,M_\odot$, $\chi = 0.75$, $\iota \approx
\ang{90}$, $M_\mathrm{NS} = 1.35\,M_\odot$, and $R_\mathrm{NS} =
\SI{14.4}{\km}$ ($Q=5$, $\mathcal{C} = 0.138$) modeled by a
piecewise-polytropic approximation of the MS1 equation of state. In this
figure, the viewing angle, $\theta$, is measured from the direction of
the total angular momentum at the initial instant of the numerical
simulations. Even if we observe the binary from $\theta = 0$, the
amplitude and phase of the waveform are modulated by the precession of
the orbital plane. The modulation of the amplitude becomes stronger as
the viewing angle becomes larger, because higher harmonic modes
contribute more significantly to the waveform.

Mode couplings and modulation make the extraction of cutoff frequency,
$f_\mathrm{cut}$, from the spectrum challenging in realistic
observations. The right panel of Fig.~\ref{fig:gw_incl} shows that the
spectra, including their dampings for $f \gtrsim \SI{2}{\kilo\hertz}$,
depend significantly on the viewing angle. Because the mass ratio of
this system is as high as $5$ and also tidal disruption is not very
significant with leaving only $\approx 0.02 \,M_\odot$ outside the
apparent horizon due to the large inclination angle of $\approx
\ang{90}$, the spectrum is expected to damp around the quasinormal-mode
frequency of $\sim \SI{2}{\kilo\hertz}$. However, the different amount
of mode mixings prohibits straightforward extraction of the cutoff
frequency independent of the viewing angle. It is a topic for future
investigations how to identify the cutoff frequency which is tightly
correlated with properties of neutron stars in actual observations (see
\citealt{Kawaguchi_KNS2017} for a preliminary study). Fortunately, the
impact of higher harmonic modes on estimation of parameters common to
binary black holes such as the mass and spin is likely to be marginal
for black hole--neutron star binaries with moderate signal-to-noise
ratios of $\lesssim 20$
\citep{OShaughnessy_FOCKL2014,OShaughnessy_FOCRKL2014}.

\subsection{Summary and issues for the future}

In this section, we have reviewed the current status of the studies on
dynamical simulations of black hole--neutron star binary coalescences in
full general relativity. Black hole--neutron star binaries are studied
focusing mainly on their potentiality as a source of gravitational
waves, a central engine of short-hard gamma-ray bursts, a site of
\textit{r}-process nucleosynthesis, and a progenitor of resulting
kilonovae/macronovae. In the last decade, not only the parameter space
surveyed has been expanded, but also qualitatively new directions have
been opened by the improvement in numerical techniques and
implementations. It has become popular to perform
neutrino-radiation-hydrodynamics simulations with temperature- and
composition-dependent equations of state. Magnetohydrodynamics
simulations have also been performed frequently, although the grid
resolution is still not high enough to derive robust conclusions
starting from realistic profiles of the magnetic fields. In addition,
qualitatively new directions of research, namely mass ejection, have
been explored in numerical relativity. This direction is benefited from
the development of fully-relativistic viscous hydrodynamics simulations
in axisymmetry.

The final fate of the black hole-neutron star binaries is classified
into two categories. One is tidal disruption of the neutron star before
the plunge, and the other is the plunge of the neutron star into the
black hole before tidal disruption. Tidal disruption is a more likely
outcome for systems with a lower mass ratio, a higher prograde spin of
the black hole, and/or a smaller compactness of the neutron star. If the
black hole is nonspinning, tidal disruption of a neutron star with $\sim
1.35\,M_\odot$ is possible only for a low mass-ratio system with $Q
\lesssim 3$ if the compactness is in a plausibly realistic range of
$0.14 \lesssim \mathcal{C} \lesssim 0.20$ (cf.,
Fig.~\ref{fig:criterion}). If the black hole has a high and prograde
spin, tidal disruption becomes possible for a higher mass-ratio system
with $Q \gtrsim 5$ because of the reduced radius of the innermost stable
circular orbit.

If tidal disruption occurs sufficiently outside the innermost stable
circular orbit, an accretion disk is formed and a fraction of material
is ejected dynamically. The masses of the disk and the ejecta depend on
the degree of tidal disruption. If the black hole is nonspinning, the
mass of the remnant disk is limited to $\lesssim 0.1\,M_\odot$ even for a
very low mass ratio. By contrast, if the spin is nearly extremal, the
mass of the disk could increase to $\gtrsim 0.5\,M_\odot$. The mass of the
dynamical ejecta is more sensitive to the mass ratio of the binary than
that of the disk. For very low mass-ratio systems with $Q \lesssim 2$,
the mass of the dynamical ejecta is limited to $\lesssim
\num{e-3}\,M_\odot$ even if the disk as massive as $0.1\,M_\odot$ is
formed. For high mass-ratio systems, the mass of the dynamical ejecta
may become comparable to the mass of the remnant disk and could increase
to $\gtrsim 0.2\,M_\odot$. Because the dynamical mass ejection in black
hole--neutron star binaries is accompanied by neither significant shock
heating nor neutrino irradiation, the dynamical ejecta preserve an
extremely low electron fraction of the original neutron-star matter,
$Y_\mathrm{e} \sim 0.05$--$0.1$. Thus, the dynamical ejecta mainly
produce heavy \textit{r}-process elements with the mass number $\gtrsim
130$, i.e., beyond the second peak.

The remnant disk ejects 15\%--30\% of its mass via viscous and/or
magnetohydrodynamical processes for a wide range of physical
conditions. In the early stage of the postmerger evolution, the internal
energy generated by viscous or magnetohydrodynamical heating is consumed
by neutrino cooling with the peak luminosity of $\gtrsim
\SI{e53}{erg.s^{-1}}$. At $\sim 0.1$--$\SI{1}{\second}$ after the disk
formation, the time scale of weak interactions becomes longer than the
heating time scale due to the disk expansion. Because of this
transition, the material in the outer part of the disk is gradually
ejected in an approximately isotropic manner. The disk outflow is likely
to dominate the dynamical ejecta in terms of the mass for low mass-ratio
systems with $Q \lesssim 3$, while the dynamical ejecta become important
for high mass-ratio systems with $Q \gtrsim 5$.

The electron fraction of the disk outflow is still uncertain. If the
launch is delayed until $\gtrsim \SI{0.5}{\second}$ after the disk
formation, the disk outflow is characterized by an increased electron
fraction of $\expval{Y_\mathrm{e}} \approx 0.3$ for typical cases,
because the disk relaxes to an equilibrium of electron/positron captures
onto nucleons during its longterm viscous or magnetohydrodynamical
evolution. Neutrino irradiation also plays a role in increasing the
electron fraction of the disk material in its early evolution stage of
$\lesssim \SI{0.1}{\second}$ as the electron antineutrinos are always
the brightest. Thus, for the cases with late mass ejection, the disk
outflow mainly produces light \textit{r}-process elements with the mass
number $80$--$130$ around the first peak. However, if the outflow is
launched earlier by some mechanisms, e.g., magnetohydrodynamical
activity, the electron fraction of the disk outflow could be lower. If
strong and coherent poloidal magnetic fields penetrating the remnant
black hole are developed as a result of binary coalescence, the merger
remnant may even launch an ultrarelativistic jet via the
Blandford-Znajek mechanism. However, this possibility has not been
deeply explored in numerical relativity.

The final fate of the black hole--neutron star binaries is reflected in
gravitational waveforms in the final inspiral and merger phases. In
particular, whether tidal disruption occurs or not has a significant
impact on the cutoff frequency, $f_\mathrm{cut}$, above which the
gravitational-wave spectrum damps. If tidal disruption does not occur,
quasinormal modes of the remnant black hole are excited and determine
the cutoff frequency. If tidal disruption occurs around a low-mass black
hole whose areal radius is comparable to or smaller than the
neutron-star radius, tidal disruption suddenly suppresses
gravitational-wave emission. Accordingly, the spectrum damps at the
cutoff frequency corresponding to the orbital frequency at which tidal
disruption occurs---not the onset of mass shedding. If tidal disruption
occurs around a high-mass black hole, which has to be spinning in a
prograde direction, tidal disruption does not prevent excitation of
quasinormal modes. In this case, the spectrum exhibits two
characteristic frequencies, one for tidal disruption and the other for
quasinormal modes. These three types of merger processes and
gravitational-wave spectra are presented schematically in
Fig.~\ref{fig:schematic} and Fig.~\ref{fig:specschematic}, respectively.

Despite the remarkable progress made in the last decade, there still
remain various issues to be explored. One important issue is to
investigate tidal disruption in high mass-ratio binaries with $Q \gtrsim
10$ and/or nearly-extremally spinning black holes of $\chi \gtrsim
0.99$. Although the spin parameter of the black hole has been limited to
$\chi \le 0.97$ in the previous study of black hole--neutron star
binaries, a larger value of the spin parameter can be handled in
numerical relativity
\citep{Lovelace_OPC2008,Liu_Etienne_Shapiro2009,Ruchlin_Healy_Lousto_Zlochower2017},
and simulations of binary black holes with $\chi > 0.97$ have been
performed \citep{Lovelace_BSS2012,Scheel_GHLKBSK2015}. Regarding the
inspiral phase, systematic derivations of longterm gravitational
waveforms, particularly for precessing black-hole spins (see
\citealt{Foucart_etal2019,Foucart_etal2021} for effort in this
direction), are required to develop accurate theoretical templates
toward the era of third-generation detectors. Dependence of the cutoff
frequency in the gravitational-wave spectrum on neutron-star properties
should be revealed thoroughly with adopting a variety of equations of
state, along with developing methods for the extraction from observed
gravitational-wave data. Another topic for future study in the inspiral
phase is the time evolution of neutron-star magnetospheres. Because it
is extremely difficult to evolve force-free magnetospheres directly by
equations of the usual magnetohydrodynamics, this topic likely requires
a novel method. Such simulations are also necessary for deriving
realistic magnetic-field configurations at the formation of the remnant
accretion disk and tracking acceleration to ultrarelativistic velocity
of a possible jet. Neutrino-radiation-magnetohydrodynamics simulations
in numerical relativity for the evolution of the postmerger system with
the duration of $\gtrsim \SI{1}{\second}$ are ultimately required to
clarify the whole picture of mass ejection, \textit{r}-process
nucleosynthesis, and electromagnetic emission such as short-hard
gamma-ray bursts and kilonovae/macronovae (see also
Sect.~\ref{sec:dis}). For this purpose, improved schemes of neutrino
transport, e.g., solving full Boltzmann's equation \citep[see][for
viable
approaches]{Cardall_Endeve_Mezzacappa2013,Shibata_NSY2014,Foucart_DHKPS2020},
and sophisticated methods of magnetohydrodynamics will play a vital role
in the foreseeable future.

\newpage

\section{Discussion} \label{sec:dis}

\subsection{Implication for electromagnetic counterparts}
\label{sec:dis_impl}

Characteristics of various postmerger electromagnetic signals are
derived from the properties of the dynamical ejecta and the disk outflow
reviewed in Sect.~\ref{sec:sim_rem_dyn} and Sect.~\ref{sec:sim_pm_wind},
respectively. In this Sect.~\ref{sec:dis_impl}, we briefly discuss
possible electromagnetic counterparts to black hole--neutron star binary
mergers focusing on the relation to the ejected material. Discussions
about gravitational-wave memory and cosmic-ray acceleration are also
given in \citet{Kyutoku_Ioka_Shibata2013}.

\subsubsection{Kilonova/macronova} \label{sec:dis_impl_knmn}

Features of the kilonova/macronova, a quasithermal transient powered by
radioactive decays of \textit{r}-process nuclei, are governed primarily
by (i) the mass of the ejecta $M$, (ii) the velocity of the ejecta $V$,
(iii) the opacity of the ejecta $\kappa$, and (iv) the radioactive
heating rate of the material. A simplified spherical model with
homologous expansion, uniform density, and grey opacity predicts the
peak time and the peak bolometric luminosity to be
\citep{Li_Paczynski1998}
\begin{align}
 t_\mathrm{peak,s} & \approx \SI{13}{\day}
 \pqty{\frac{\kappa}{\SI{10}{\square\cm\per\gram}}}^{1/2}
 \pqty{\frac{M}{0.01\,M_\odot}}^{1/2} \pqty{\frac{V}{0.1c}}^{-1/2} , \\
 L_\mathrm{peak,s} & \approx \SI{1.6e40}{erg.s^{-1}}
 \pqty{\frac{f_\mathrm{heat}}{\num{e-6}}}
 \pqty{\frac{\kappa}{\SI{10}{\square\cm\per\gram}}}^{-1/2}
 \pqty{\frac{M}{0.01\,M_\odot}}^{1/2} \pqty{\frac{V}{0.1c}}^{1/2} ,
\end{align}
respectively, for the case in which the specific heating rate of the
material is given by $f_\mathrm{heat} c^2 / t$ with $t$ being the time
after mass ejection. In this model, $V$ denotes the surface velocity of
the sphere. On this time scale, the homologous expansion is safely
justified (see Sect.~\ref{sec:dis_impl_rem}). In reality, however, the
density profile of the ejecta is neither uniform nor spherical. The
opacity depends on the wavelength, the ionization degree, which varies
in time, and moreover, the composition of the \textit{r}-process
elements in the ejecta, because the complexity of the level structure
depends significantly on the atomic species
\citep{Kasen_Badnell_Barnes2013,Tanaka_Hotokezaka2013}. The radioactive
heating rate also depends on the composition and evolves in time in a
complicated manner, because the majority of the heating is given by the
sum of $\beta$-decays of various nuclides near the $\beta$-stable line
with different lifetimes
\citep{Metzger_MDQAKTNPZ2010,Wanajo_SNKKS2014,Wanajo2018}. In typical
situations, the total heating rate may be approximated by $\propto
t^{-(1.2 \mathchar`- \mathchar`- 1.4)}$ \citep[see also
\citealt{Kasen_Barnes2019,Waxman_Ofek_Kushnir2019,Hotokezaka_Nakar2020}
for the late-time
behavior]{Metzger_MDQAKTNPZ2010,Hotokezaka_Sari_Piran2017}. Quantitative
calculations of (iii) and (iv) require information about the abundance
pattern in the ejecta.

Numerical-relativity simulations are the unique tool to derive
information required for computing the light curve and the spectrum of
the kilonova/macronova in a quantitative and accurate
manner. Macroscopic properties (i) and (ii) along with the density
profile are provided by solving hydrodynamical evolution. Although the
average velocity of the ejecta does not vary much from $0.1$--$0.2c$,
the mass depends significantly on the binary parameters as we have
already discussed in Sect.~\ref{sec:sim}. Furthermore, simulations with
weak interactions including neutrino-radiation transfer give us the
distributions of thermodynamic quantities such as the electron fraction
in the ejecta, which determine the nucleosynthetic yield and hence
microscopic properties (iii) and (iv). Precise multiband light curves
have been calculated by photon-radiation-transfer simulations based on
the ejecta properties derived by numerical-relativity simulations of
black hole--neutron star binaries
\citep{Tanaka_HKWKSS2014,Fernandez_FKLDR2017,Kawaguchi_Shibata_Tanaka2020,Bulla_KTCBMMTW2021,Darbha_KFP2021}. In
the following, we briefly review current understanding of the
kilonova/macronova with particular emphasis on features specific to
black hole--neutron star binaries.

First, we focus on the emission from the dynamical ejecta. The
lanthanide-rich abundance pattern of the dynamical ejecta implies that
the associated (Planck mean) opacity becomes as high as $\sim
\order{10}\si{\square\cm\per\gram}$ at $\sim
\num{2000}$--\SI{20000}{\kelvin}
\citep{Kasen_Badnell_Barnes2013,Tanaka_Hotokezaka2013,Tanaka_etal2018,Tanaka_KGK2020}. Then,
the kilonova/macronova is expected to shine in red-optical and infrared
bands on a time scale of $\sim \SI{10}{\day}$ if the spherical model is
valid. Because the dynamical ejecta from black hole--neutron star
binaries are extremely neutron rich, ultraheavy elements such as
${}^{254}\mathrm{Cf}$ might be produced in abundance. If this is the
case, their fission and/or $\alpha$-decay will become an important
source of the heating and emission at late times, for which the
$\beta$-decay heating becomes inefficient due to the low density
\citep{Wanajo_SNKKS2014,Zhu_etal2018,Wu_BMM2019}.

The nonspherical morphology of dynamical ejecta described in
Sect.~\ref{sec:sim_rem_dyn} induces various differences from spherical
cases, including directional dependence of the kilonova/macronova
\citep{Kyutoku_Ioka_Shibata2013,Kyutoku_IOST2015}. For simplicity, we
assume that the binary has the reflection symmetry about the orbital
plane (i.e., the spin misalignment is not taken into account). We
further parametrize the opening angles in the polar and azimuthal
directions by $2\theta$ and $\varphi$, respectively, where the prefactor
$2$ of the former comes from the reflection symmetry. Because photons
diffuse preferentially toward the direction perpendicular to the
equatorial plane, for which the length scale is smaller by $\approx
\theta$ than for the radial direction, the kilonova/macronova may be
brighter and bluer from an earlier epoch if an observer is located along
the polar direction than along the equatorial direction. This
qualitative prediction is confirmed by photon-radiation-transfer
simulations \citep{Tanaka_HKWKSS2014}. Because the gravitational-wave
observation is biased toward finding binary coalescences from the polar
direction \citep{Schutz2011}, this feature is advantageous for the
follow-up detection of the kilonovae/macronovae. If the dynamical ejecta
becomes quasispherical during the expansion due to the
\textit{r}-process heating, however, this directional dependence weakens
\citep{Darbha_KFP2021}. Instead, the Doppler effect associated with the
center-of-mass motion dominates the directional dependence of the
emission, making the emission brighter and bluer toward the direction of
the motion, and vice versa \citep{Fernandez_FKLDR2017,Darbha_KFP2021}.

The nonsphericity also modifies the peak time and the peak bolometric
luminosity. The preferential diffusion toward the polar direction
associated with a small value of $\theta$ shortens the peak time
compared to the spherical case. At the same time, if the opening angle
in the azimuthal direction, $\varphi$, is small, the peak time is
delayed due to the increased density and hence the increased optical
depth for given values of the mass and the velocity of the
ejecta. Dependence of the peak time and the peak bolometric luminosity
on the opening angles may be derived by approximating that photons
diffuse only toward the polar direction as
\citep{Kyutoku_Ioka_Shibata2013,Kyutoku_IOST2015}
\begin{equation}
 t_\mathrm{peak} \approx \SI{11}{\day}
  \pqty{\frac{\kappa}{\SI{10}{\square\cm\per\gram}}}^{1/2}
  \pqty{\frac{M}{0.01\,M_\odot}}^{1/2} \pqty{\frac{V}{0.1c}}^{-1/2}
  \pqty{\frac{\theta}{1/5}}^{1/2} \pqty{\frac{\varphi}{\pi}}^{-1/2} ,
\end{equation}
\begin{align}
 L_\mathrm{peak} \approx \SI{1.8e40}{erg.s^{-1}} &
 \pqty{\frac{f_\mathrm{heat}}{\num{e-6}}}
 \pqty{\frac{\kappa}{\SI{10}{\square\cm\per\gram}}}^{-1/2}
 \pqty{\frac{M}{0.01\,M_\odot}}^{1/2} \notag \\
 \times & \pqty{\frac{V}{0.1c}}^{1/2} \pqty{\frac{\theta}{1/5}}^{-1/2}
 \pqty{\frac{\varphi}{\pi}}^{1/2} .
\end{align}
The dependence of $L_\mathrm{peak}$ on various quantities may be
understood from Arnett's rule \citep{Arnett1982}, $L_\mathrm{peak} \sim
f_\mathrm{heat} M c^2 / t_\mathrm{peak}$ in our case. For typical values
of the opening angles found in numerical-relativity simulations, the
peak time is earlier and the peak bolometric luminosity is higher by
10\%--20\% than predictions for spherical
ejecta. Photon-radiation-transfer simulations also find that the
temperature is enhanced for nonspherical ejecta with given values of the
mass and the velocity, because the reduced volume increases the heating
efficiency \citep{Tanaka_HKWKSS2014}. Although the realistic spectrum
can be derived only by simulations with a detailed line list, the
diffusion approximation suggests that the effective temperature is
increased by 30\%--50\%, which may have an impact on multiband light
curves \citep{Kyutoku_IOST2015}. Detailed photon-radiation-transfer
simulations predict that the peak absolute AB magnitude in the $H$ band
is $\approx -(16$--$16.5)$ at $\approx \SI{5}{\day}$ after merger
depending weakly on the viewing angle, if the mass of the dynamical
ejecta is $0.02\,M_\odot$ \citep[see also
\citealt{Darbha_KFP2021}]{Kawaguchi_Shibata_Tanaka2020}.

Another possible characteristic observable feature is a relatively high
degree of polarization induced by significant deformation of the
photosphere in an early epoch
\citep{Kyutoku_Ioka_Shibata2013,Kyutoku_IOST2015}. If Thomson scattering
contributes appreciably to the opacity, a deformed photosphere produces
net linear polarization along the short principal axis. Although the
ejecta dominated by \textit{r}-process elements with large atomic
numbers tend to reduce the polarization due to the small number of
scattering electrons and depolarization via bound-bound transitions
compared to those consisting of elements up to the iron peak
\citep{Kyutoku_IOST2015}, photon-radiation-transfer simulations have
suggested that the linear polarization of a few percent may be expected
at 1--\SI{2}{\day} after merger in near-infrared bands
\citep{Bulla_KTCBMMTW2021}.

Next, we discuss the emission from the disk outflow. Because the disk
outflow is likely to be associated with moderate lanthanide fraction,
its (Planck mean) opacity is expected to be as low as $\sim
\SI{1}{\square\cm\per\gram}$
\citep{Tanaka_Hotokezaka2013,Tanaka_KGK2020}. If this is the case,
depending on the mass and the velocity, the kilonova/macronova from the
disk outflow could be bright in blue-optical bands from the early epoch
of $\sim \SI{1}{\day}$ after merger. This type of emission is sometimes
called the blue kilonova. For example, a photon-radiation-transfer
simulation shows that the peak absolute AB magnitude is $\approx -16$ in
the $r$ band at $\approx \SI{1}{\day}$ and $\approx -15$ in the $H$ band
at $\approx \SI{3}{\day}$, if the mass of the disk outflow is
$0.03\,M_\odot$ and the average velocity is $0.06c$
\citep{Kawaguchi_Shibata_Tanaka2020}. Instead, if the disk outflow is
lanthanide rich due to rapid ejection, the opacity becomes $\sim
\SI{10}{\square\cm\per\gram}$ and the emission will peak in red-optical
and infrared bands on a time scale of $\sim \SI{10}{\day}$. Again, if
the mass is $0.03\,M_\odot$ and the average velocity is $0.06c$, the peak
absolute AB magnitude is $\approx -15$ for both the $r$ and $H$ bands at
$\approx \SI{1}{\day}$ and $\approx \SI{7}{\day}$, respectively
\citep{Kawaguchi_Shibata_Tanaka2020}. In this case, the overall emission
feature is similar to that from the dynamical ejecta. Because the disk
outflow is characterized by quasispherical morphology with mild
concentration toward the equatorial plane, the emission is expected to
be approximately isotropic irrespective of the lanthanide fraction.

As is frequently discussed in the case of AT 2017gfo, which was
associated with a binary-neutron-star merger
\citep{Tanaka_etal2017,Kasen_MBQR2017,Perego_Radice_Bernuzzi2017,Kawaguchi_Shibata_Tanaka2018,Waxman_OKG2018},
realistic features of the kilonova/macronova from black hole--neutron
star binaries are likely to be determined by the combination of the
dynamical ejecta and the disk outflow
\citep{Kawaguchi_Shibata_Tanaka2020,Kyutoku_FHKKST2020}. The relation
between the disk mass and the ejecta mass discussed in
Sect.~\ref{sec:sim_rem_disk} suggests that the disk outflow tends to be
the dominant source of the kilonova/macronova for a system with a low
mass ratio. In the presence of lanthanide-rich dynamical ejecta with
high velocity, they can reprocess the blue kilonova from the disk
outflow and emit later in redder bands, called the lanthanide-curtain
effect \citep{Kasen_Fernandez_Metzger2015}. Although the blocking by the
dynamical ejecta may occur for a wide range of the viewing angle in the
case of binary neutron stars \citep{Kawaguchi_Shibata_Tanaka2018}, it
occurs only in a limited range for black hole--neutron star binaries
because of the aforementioned nonsphericity. If an observer is located
along the direction in which the disk outflow is not covered by the
preceding dynamical ejecta, the blue kilonova can be observed without
being concealed by the dynamical ejecta. In this case, additional
emission from the dynamical ejecta, including the reprocessed emission,
enhances the luminosity of the kilonova/macronova. If an observer is in
the direction covered by the dynamical ejecta, the lanthanide-curtain
effect reduces the luminosity in blue bands. This occurs even if the
disk outflow is lanthanide rich and has a composition similar to the
dynamical ejecta, because the opacity characterizing the expanding
medium \citep[so-called expansion
opacity:][]{Karp_LCS1977,Eastman_Pinto1993} is inversely proportional to
the rest-mass density \citep[here we presume that the dynamical ejecta
have a small mass and high velocity;][]{Kyutoku_FHKKST2020}. For a
system with a high mass ratio, the dynamical ejecta tend to become
dominant sources of the kilonova/macronova. Because the expansion occurs
rapidly due to the high velocity, the emission from the dynamical ejecta
is likely to be observed directly irrespective of the disk outflow. The
polarization degree of $\sim 1\%$ for the wavelength of $\gtrsim
\SI{0.7}{\micro\meter}$ might still be expected in a wide range of the
viewing angle at 1--\SI{2}{\day} after merger for a combination of
ejecta components \citep[see also \citealt{Li_Shen2019} for an
early-time polarization with hypothetical existence of free
neutrons]{Bulla_KTCBMMTW2021}.

Various phenomenological models of the kilonova/macronova from dynamical
ejecta \citep{Kawaguchi_KST2016} and the sum of them and the disk
outflow are available in the literature. For a reader who is interested
in such models, we recommend to refer to
\citet{Barbieri_SPCG2019,Barbieri_SPCG2020,Zhu_YLHZLYG2020}.

\subsubsection{Kilonova/macronova remnant} \label{sec:dis_impl_rem}

Another promising electromagnetic counterpart is nonthermal emission
from the blast wave formed between the ejecta and the interstellar
medium \citep{Nakar_Piran2011}. After a period of the homologous
expansion, the ejecta are eventually decelerated by accumulating the
interstellar medium. Blast waves develop between the ejecta and the
interstellar medium in the course of this collision, and a part of the
kinetic energy of the ejecta is converted to the internal energy of the
postshock material behind the forward shock. A fraction of this
postshock internal energy is devoted to accelerating electrons to
nonthermal velocity distribution \citep[see,
e.g.,][]{Bell1978,Blandford_Ostriker1978} and to amplifying magnetic
fields via plasma instabilities \citep[see,
e.g.,][]{Weibel1959,Bell2004}. Then, the accelerated electrons will emit
synchrotron radiation in the magnetized environment.

The flux density of this emission is highly sensitive to the velocity of
the material, which primarily determines the available energy in the
postshock region. Thus, the dynamical ejecta, which can have a component
of the velocity up to $\sim 0.4c$, will play a more important role than
the disk outflow compared to the case of the kilonova/macronova. For
example, if the power-law index of nonthermal electron distribution with
respect to the Lorentz factor is given by $-2.5$, the peak flux density
in \si{\giga\hertz} bands is proportional to $v^{2.75}$ for a
single-velocity outflow with a fixed kinetic energy
\citep{Piran_Nakar_Rosswog2013}. Here, the velocity may be identified
with the average velocity of the ejecta described in the end of
Sect.~\ref{sec:sim_rem_dyn}, i.e., the value defined from the mass and
the kinetic energy.

The emission is expected to be brightest when the deceleration becomes
significant as the accumulated mass becomes comparable to that of the
ejecta. Thus, the time at the onset of significant deceleration
(hereafter, simply referred to as the deceleration time) is critically
important to assess detectability. By assuming that the interstellar
medium is composed of hydrogen of mass $m_\mathrm{p}$ with the number
density $n$, the deceleration time (also called the Sedov time) for
spherical ejecta is estimated from $M = (4\pi/3) (vt_\mathrm{dec,s})^3
m_\mathrm{p} n$ to be
\begin{equation}
 t_\mathrm{dec,s} \approx \SI{15}{yr} \pqty{\frac{M}{0.01\,M_\odot}}^{1/3}
  \pqty{\frac{v}{0.1c}}^{-1}
  \pqty{\frac{n}{\SI{1}{\per\cubic\cm}}}^{-1/3} , \label{eq:sedov}
\end{equation}
where $M$ again denotes the mass of the ejecta. This estimate may be
applicable to the quasispherical disk outflow. After this time, the
ejecta profile evolves approximately according to Sedov-Taylor's
self-similar solution under the assumption that nonrelativistic
treatment is valid \citep{Piran_Nakar_Rosswog2013}. The flux density as
a function of time and frequency is derived by assuming the fraction of
postshock internal energy that goes into the accelerated electrons
(typically assumed to be $\sim 0.1$), the fraction that goes into the
amplified magnetic field ($\sim 0.01$--$0.1$), and the power-law index
of the electron distribution ($\sim 2.5$--$3$)
\citep{Nakar_Piran2011}. The fraction of electrons involved in the
particle acceleration may also be introduced
\citep{Takami_Kyutoku_Ioka2014,Kyutoku_IOST2015}.

The deceleration time of nonspherical dynamical ejecta becomes long for
given mass and velocity, because they sweep only a limited solid
angle. Quantitatively, it is given in terms of the opening angles
defined in Sect.~\ref{sec:dis_impl_knmn} by
\citep{Kyutoku_Ioka_Shibata2013,Kyutoku_IOST2015}
\begin{equation}
 t_\mathrm{dec} \approx \SI{38}{yr} \pqty{\frac{M}{0.01\,M_\odot}}^{1/3}
  \pqty{\frac{v}{0.1c}}^{-1}
  \pqty{\frac{n}{\SI{1}{\per\cubic\cm}}}^{-1/3}
  \pqty{\frac{\theta}{1/5}}^{-1/3} \pqty{\frac{\varphi}{\pi}}^{-1/3} .
\end{equation}
We again note that the morphology and hence the opening angles may be
modified by the \textit{r}-process heating in the early part of the
evolution. Because the peak flux density is expected to be independent
of the opening angles except for low frequency at which self-absorption
plays a role, this estimate suggests that the nonthermal emission from
the nonspherical dynamical ejecta of black hole--neutron star binaries
will be long lasting. After the peak time given by $\approx
t_\mathrm{dec}$, evolution of the flux density is more complicated for
nonspherical ejecta than for spherical ejecta, because the
hydrodynamical evolution becomes nontrivial due to the lateral
expansion. To derive quantitative predictions, longterm numerical
simulations are necessary for modeling geometrical evolution (see
\citealt{Margalit_Piran2015} for a piecewise-spherical approximation).

We finally mention possible proper motion of the radio image caused by
the motion of the ejecta
\citep{Kyutoku_Ioka_Shibata2013,Kyutoku_IOST2015}. As the dynamical
ejecta carry a net linear momentum, their center-of-mass velocity
$v_\mathrm{COM}$ typically reaches $0.1$--$0.2c$ as described in
Sect.~\ref{sec:sim_rem_dyn}. A characteristic distance scale for the
movement of the ejecta may be estimated by $v_\mathrm{COM}
t_\mathrm{dec}$. For a hypothetical event at $\order{100}\si{Mpc}$, the
radio image could move on the sky by $\order{1}
\si{milliarcsecond}$. Taking the fact that superluminal motion of the
radio image by $2.7 \pm \SI{0.3}{milliarcsecond}$ has been detected in
the afterglow of GRB 170817A into account \citep{Mooley_DGNHBFHCH2018},
there may be a chance for detecting radio proper motion associated with
the dynamical ejecta from black hole--neutron star binaries in
forthcoming observations. The proper motion may be detectable until the
slow, quasispherical disk outflow begins to dominate the entire radio
emission.

\subsubsection{Influence on the jet propagation}
\label{sec:dis_impl_jet}

The material ejected to the polar region crucially affects propagation
and collimation of a hypothetical jet launched from the black
hole--accretion disk system. It is not clear whether the ejecta are
helpful for realizing observed gamma-ray bursts. On the one hand, if the
jet accumulates too much material from the ejecta during its
propagation, the terminal velocity cannot be ultrarelativistic
\citep[the so-called baryon-loading
problem;][]{Meszaros_Rees2000}. Massive ejecta may even choke the jet
before it breaks out \citep{Gottlieb_NPH2018,Nakar_GPKH2018}. On the
other hand, gas pressure of the surrounding material composed of the
ejecta may be necessary for collimating the jet to a narrow opening
angle \citep[see also
\citealt{Matzner2003,Bromberg_NPS2011,Mizuta_Ioka2013} for related work
on the stellar
envelope]{Nagakura_HSSI2014,Harrison_Gottlieb_Nakar2018,Hamidani_Ioka2021}.

The current numerical simulations of black hole--neutron star binaries
are not fully successful in driving an ultrarelativistic jet. Thus, for
a theoretical study on the jet propagation, it is customary to inject
internal energy artificially in the central region for investigating the
interaction of an expanding fireball and the surrounding material
\citep[see also, e.g.,
\citealt{Nagakura_HSSI2014,MurguiaBerthier_MRDL2014,Duffell_Quataert_MacFadyen2015,MurguiaBerthier_RMDRRTPL2017,Lazzati_LCMPW2017}
for the case of binary neutron stars]{Just_OJBS2016}. While the robust
conclusion cannot be drawn at this stage, an ultrarelativistic jet may
be able to penetrate through the ejecta with reasonable collimation
\citep{Just_OJBS2016}. One difference from the case of
binary-neutron-star mergers is that the dynamical ejecta do not cover
the polar region. Thus, the hypothetical jet is likely to interact only
with the slow disk outflow. Indeed, this is the reason that
\citet{Just_OJBS2016} find black hole--neutron star binaries are more
advantageous for launching a successful jet with avoiding the
baryon-loading problem than binary neutron stars.

In terms of the energy injection, the Blandford-Znajek mechanism is
likely to be required for explaining observed properties of the
short-hard gamma-ray bursts such as the duration of $\sim
\SI{1}{\second}$, because other mechanisms, e.g., neutrino pair
annihilation, are unlikely to operate for such a long time
scale. Although fully consistent, end-to-end simulations for the
ultrarelativistic jet may be prohibitive, jet propagation and
collimation for black hole--neutron star binaries require further
investigations. For example, it is uncertain whether the
Blandford-Znajek jet is launched before or after the substantial
ejection of the disk outflow. Because the fate of ultrarelativistic jets
after propagation depends strongly on the timing difference of the
launch \citep{Nagakura_HSSI2014}, this point needs to be explored
systematically.

\subsection{Distinguishability of binary types} \label{sec:dis_dis}

One of the lessons from the LIGO-Virgo O3 is that realistic compact
object binaries sometimes challenge our ability to distinguish black
holes and neutron stars \citep{GW190425,GW190814}. Because the
sensitivity of the current detectors are not high enough, some
parameters of gravitational-wave sources such as the masses of
individual objects are not determined precisely. Moreover, because we do
not know the precise upper limit of the neutron-star mass, an object
with $\approx 2$--$3\,M_\odot$ cannot be concluded as either a black hole
or a neutron star with confidence, even if its mass is determined
without uncertainty. In this Sect.~\ref{sec:dis_dis}, we discuss
prospects for observationally distinguishing binary types for the case
in which one of the components has $2$--$3\,M_\odot$.

\subsubsection{Can we distinguish black hole--neutron star
   binaries from binary black holes?} \label{sec:dis_dis_bh}

If the neutron star is disrupted during merger, both gravitational waves
and electromagnetic counterparts for black hole--neutron star binaries
are significantly different from those for binary black holes. Thus,
disruptive black hole--neutron star binaries may be distinguishable from
binary black holes if the distance to the event is small enough that
gravitational waves and electromagnetic counterparts are detected. For
example, numerical-relativity-calibrated waveform models suggest that
gravitational waves with the signal-to-noise ratio of $\gtrsim 30$--$40$
enable us to distinguish these two types of binaries
\citep{Thompson_FKNPDH2020} even if we do not consider cutoff behavior
of the waveform or electromagnetic signals. The distance to which
electromagnetic counterparts can be detected depends on both the
properties of the ejecta and available observational instruments. For
example, events at $\sim \SI{400}{Mpc}$ give us a fair chance for
detections of the kilonova/macronova for a wide range of binary
parameters if rapid follow-up is performed by \SI{8}{\meter}-class
optical telescopes or sensitive space-borne infrared telescopes such as
the Nancy Grace Roman Space Telescope (formerly called WFIRST) become
available \citep{Tanaka_HKWKSS2014}. Because the typical distance at
which black hole--neutron star binaries are observed to merge is
expected to be larger than $\sim \SI{200}{Mpc}$ [e.g., the distances to
GW200105 and GW200115 are estimated to be $\sim \SI{300}{Mpc}$
\citep{GW200105200115}], large-scale telescopes will play a vital role
for detecting associated electromagnetic counterparts. If a short-hard
gamma-ray burst is launched toward us, the detectable distance will
exceed a cosmological distance of $z \sim 1$ \citep[see,
e.g.,][]{Fong_BMZ2015}.

As discussed in Sect.~\ref{sec:intro}, distinguishing nondisruptive
black hole--neutron star binaries and binary black holes seems a
formidable challenge \citep{Foucart_BDGKMMPSS2013}. This is demonstrated
by actual detections of black hole--neutron star binaries GW200105 and
GW200115 \citep{GW200105200115} and a possible candidate
GW190426\_152155 \citep{GWTC2}. A highly asymmetric binary-black-hole
candidate GW190814 may also fall into this category
\citep{GW190814}. Numerical-relativity-calibrated waveform models
suggest that the tidal effect in gravitational waves will not be useful
for identifying the binary type unless the signal-to-noise ratio exceeds
$\approx 100$ \citep{Thompson_FKNPDH2020}, which is not likely to be
realized in the foreseeable future. The only signal that might allow us
to distinguish these two types of compact object binaries is an
electromagnetic precursor associated with magnetospheric activities,
although its understanding is currently highly uncertain \citep[see also
\citealt{Ioka_Taniguchi2000} for binary neutron
stars]{Hansen_Lyutikov2001,McWilliams_Levin2011,Lai2012,Paschalidis_Etienne_Shapiro2013,DOrazio_LMP2016,Carrasco_Shibata2020,Wada_Shibata_Ioka2020,East_LLP2021,Carrasco_Shibata_Reula2021}. Because
the precursor emission needs to be detected without gravitational-wave
informed localization, a wide-field monitoring is required on the
observational side. Sophistication of theoretical predictions is also
important on the theoretical side for distinguishing the signal from
other transients.

\subsubsection{Can we distinguish black hole--neutron star binaries from
   binary neutron stars?} \label{sec:dis_dis_ns}

Distinguishing binary neutron stars from a black hole--neutron star
binary with a small total mass is also not an easy task. Very strictly
speaking, even GW170817 is not inconsistent with a very-low-mass black
hole--neutron star binary if we accept existence of black holes with
$M_\mathrm{BH} < 2\,M_\odot$ \citep{Hinderer_etal2019}. A more subtle case
is the second binary-neutron-star candidate, GW190425
\citep{GW190425}. Because the total mass of this system,
$3.4^{+0.3}_{-0.1}\,M_\odot$ for the high-spin prior (i.e., the spin
parameter of each component is assumed to be $<0.89$ in the data
analysis), is an outlier compared to the mass distribution of Galactic
binary neutron stars \citep{Tauris_etal2017,Farrow_Zhu_Thrane2019}, this
system may indicate the existence of a novel path for compact binary
formation. However, a possibility that GW190425 is a very-low-mass black
hole--neutron star binary with $M_\mathrm{BH} > 2\,M_\odot$ is not
rejected, partly because the localization was poor and no
electromagnetic counterpart was detected
\citep{Kyutoku_FHKKST2020,Han_THLJJFW2020}. Because either scenario
involves astrophysically unexpected populations of compact object
binaries, it is worthwhile to explore how we can distinguish the type of
binaries in a reliable manner in the foreseeable future \citep[see
also][for a similar discussion]{Yang_East_Lehner2018}.

Although gravitational waves from the inspiral phase give us a plenty of
information about the source, it is unlikely to be powerful enough to
distinguish low-mass black hole--neutron star binaries and binary
neutron stars. Because the mass ratio and the spin parameter are tightly
correlated \citep{Cutler_Flanagan1994,Poisson_Will1995}, the mass of
each component is determined only with a limited precision
\citep{Hannam_BFFH2013}. Although this degeneracy may be mitigated by
taking the precession into account \citep{Chatziioannou_CKY2015}, this
is likely to be effective only for edge-on systems, against which
gravitational-wave detectors are biased \citep{Littenberg_FCKH2015}.
Even if the degeneracy is resolved, complete distinction is hindered
from the uncertainty in the maximum mass of the neutron star. It is also
difficult to separate tidal effects of individual components to claim
that one is devoid of tidal deformability \citep[see,
e.g.,][]{Wade_COLFLR2014}. If the sensitivity at high frequency of
$\gtrsim \SI{1}{\kilo\hertz}$ is improved with third-generation
detectors \citep{CosmicExplorer,EinsteinTelescope}, detections of
postmerger signals from the remnant massive neutron star will allow us
to identify the system to be binary neutron stars without doubt,
particularly for binary neutron stars less massive than a canonical mass
of $\sim 2 \times 1.35\,M_\odot$ \citep{Shibata2005,Hotokezaka_KOSK2011}.

Electromagnetic counterparts will give us useful information about the
nature of sources, but in general it is not easy to distinguish low-mass
black hole--neutron star binaries from asymmetric binary neutron stars
\citep{Kyutoku_FHKKST2020}. If the total mass is large, nearly-symmetric
binary neutron stars will collapse promptly into a black hole. Because
prompt collapse is unlikely to leave substantial material outside the
formed black hole \citep{Hotokezaka_KKOSST2013}, detections of
electromagnetic counterparts such as the gamma-ray burst and the
kilonova/macronova are not compatible with symmetric, massive binary
neutron stars. However, the outcomes of black hole--neutron star
binaries and asymmetric binary neutron stars are similar if the masses
of each component are identical. A possible difference may be induced by
the mass of the dynamical ejecta, which is larger for asymmetric binary
neutron stars than very-low-mass black hole--neutron star binaries due
probably to the presence of the stellar surface \citep[see also
discussions in][]{Most_PTR2021}. In addition, because the maximum
velocity of the dynamical ejecta is higher for binary neutron stars than
for black hole--neutron star binaries, the kilonova/macronova remnant
may be brighter from an early epoch for the former
\citep{Kyutoku_IOST2015,Brege_DFDCHKOPS2018,Most_PTR2021}. Possible
characteristic observable features associated with nonsphericity of the
dynamical ejecta from black hole--neutron star binaries could also be
utilized for distinction \citep{Kyutoku_Ioka_Shibata2013}, and further
investigation of this topic will be valuable.

\begin{acknowledgements}
 We are grateful to M.~D.~Duez, F.~Foucart, S.~Fujibayashi, and K.~Ioka
 for providing helpful comments. We also thank K.~Kawaguchi and
 K.~Kiuchi for allowing us to use numerical data of gravitational waves
 obtained in an ongoing project, in which numerical computations are
 performed on Cray XC50 at CfCA of National Astronomical Observatory of
 Japan and Cray XC40 at Yukawa Institute for Theoretical Physics of
 Kyoto University. The LS220 equation of state is taken from
 \url{https://stellarcollapse.org} \citep{OConnor_Ott2010}. This work was
 supported by JSPS KAKENHI Grant-in-Aid (Grants No.~JP16H02183,
 No.~JP17H01131, No.~JP18H05236, No.~JP19K14720, and No.~JP20H00158).
\end{acknowledgements}

%



\appendix
\normalsize

\section{Formulation of quasiequilibrium states} \label{app:init}

Minimal requirements for modeling black hole--neutron star binaries in
quasiequilibrium are to satisfy the Einstein constraint equations and
relativistic hydrostationary equations supplemented with an equation of
state. Sophisticated models of quasiequilibrium states can be obtained
by solving a part of the Einstein evolution equations with imposing
further conditions, e.g., slicing condition. For this purpose, two
approaches have been proposed to construct black hole--neutron star
binaries in quasiequilibrium. The main differences between these two
approaches lie in the method for handling the physical and/or coordinate
singularity associated with the black hole. One is the excision
approach, in which the inside of the black-hole horizon is explicitly
eliminated from the computational domain by excising a coordinate sphere
and by imposing an appropriate boundary condition there \citep[see also
\citealt{Dreyer_KSS2003,Ashtekar_Krishnan2004,Gourgoulhon_Jaramillo2006}
for reviews of the isolated
horizon]{Cook2002,Cook_Pfeiffer2004,Jaramillo_Gourgoulhon_Marugan2004}. The
other is the puncture approach, in which the black hole is modeled by a
puncture and the metric quantities are decomposed into singular and
regular parts \citep{Brandt_Brugmann1997}. In this approach, the inside
of the black-hole horizon is eliminated implicitly. In both approaches,
the outer boundary is located exactly at spatial infinity or at a very
large distance \citep{Pfeiffer_KST2003} by compactifying the radial
coordinate.

In this appendix, we first describe gravitational-field equations both
for the excision and the puncture approaches, which derive overlapping
but different sets of equations. Next, we describe relativistic
hydrostationary equations. These equations are solved simultaneously by
iterative methods until a sufficient level of convergence is
achieved. We refer the readers to
\citet{Taniguchi_BFS2005,Taniguchi_BFS2006,Taniguchi_BFS2007} and
\citet{Shibata_Uryu2006,Shibata_Uryu2007,Kyutoku_Shibata_Taniguchi2009}
for the excision and the puncture approaches applied to black
hole--neutron star binaries, respectively. We also refer the readers to
\citet{Cook2000,Tichy2017} for reviews of the initial value problem in
numerical relativity and to
\citet{Gourgoulhon_GTMB2001,Uryu_LFGS2009,Taniguchi_Shibata2010} for
details of the relativistic hydrostationarity. Geometrical units in
which $G=c=1$ is adopted entirely in Appendix~\ref{app:init}.

\subsection{Gravitational field} \label{app:init_grav}

In both the excision and the puncture approaches, a part of the Einstein
equation is solved based on the extended conformal thin-sandwich
formalism \citep{York1999,Pfeiffer_York2003}. The line element is
written in the $3+1$ form as
\begin{align}
 \dd{s}^2 & = g_{\mu \nu} \dd{x}^\mu \dd{x}^\nu \notag \\
 & = - \alpha^2 \dd{t}^2 + \gamma_{ij} \pqty{\dd{x}^i +
 \beta_\mathrm{com}^i \dd{t}} \pqty{\dd{x}^j + \beta_\mathrm{com}^j
 \dd{t}} .
\end{align}
As we stated in Sect.~\ref{sec:eq_param}, quasiequilibrium states are
modeled assuming the existence of a helical Killing vector,
Eq.~\eqref{eq:helical}. Thus, numerical computations of quasiequilibrium
states are usually performed in the comoving frame, because the system
appears stationary there \citep[see,
e.g.,][]{Gourgoulhon_GTMB2001,Gourgoulhon_Grandclement_Bonazzola2002}.
The shift vector in the comoving frame is expressed as
\begin{align}
 \beta_\mathrm{com}^i & = \beta^i + \beta_\mathrm{rot}^i ,
 \label{eq:shiftcom} \\
 \beta_\mathrm{rot}^i & = \Omega ( \partial_\varphi )^i ,
  \label{eq:shiftrot}
\end{align}
where $\beta^i$ is the shift vector in the asymptotic inertial frame and
$\beta_\mathrm{rot}^i$ is the vector connecting the two frames with
$\Omega$ being the orbital angular velocity of the binary measured at
infinity, which also appears in the helical Killing vector. The method
to determine the value of $\Omega$ will be discussed later. If the
radial velocity should be introduced to reduce or control the orbital
eccentricity, an additional vector needs to be incorporated as we
discuss in Appendix~\ref{app:init_beyond_ecc}
\citep{Pfeiffer_BKLLS2007,Foucart_KPT2008,Henriksson_FKT2016,Kyutoku_KKST2021}.

The extrinsic curvature is defined by
\begin{equation}
 K_{ij} = - \frac{1}{2} \mathcal{L}_n \gamma_{ij} ,
\end{equation}
where $n^\mu$ is the future-directed unit timelike vector normal to the
constant-time hypersurface and $\mathcal{L}$ denotes the Lie
derivative. The extrinsic curvature is decomposed into the trace and the
traceless parts as
\begin{equation}
 K_{ij} = \frac{1}{3} K \gamma_{ij} + A_{ij} , \label{eq:defextr}
\end{equation}
where $\gamma^{ij} A_{ij} = 0$. The trace, $K$, is usually regarded as a
freely-specifiable variable that determines the constant-time
hypersurface \citep{York1972}.

The energy-momentum tensor is decomposed in a $3+1$ manner as
\begin{align}
 \rho_\mathrm{H} & := n_\mu n_\nu T^{\mu \nu} , \\
 J^i & := - \gamma^i{}_\mu n_\nu T^{\mu \nu} , \\
 S_{ij} & := \gamma_{i\mu} \gamma_{j\nu} T^{\mu \nu} , \\
 S & := \gamma^{ij} S_{ij} .
\end{align}
For the case of an ideal fluid, the energy-momentum tensor takes the
form of
\begin{align}
 T^{\mu \nu} & = ( \rho + \rho \varepsilon + P ) u^\mu u^\nu + P g^{\mu
 \nu} \notag \\
 & = \rho h u^\mu u^\nu + P g^{\mu \nu} \label{eq:emtensor_ideal}
\end{align}
and is decomposed into
\begin{align}
 \rho_\mathrm{H} & = \rho h \pqty{\alpha u^t}^2 - P , \\
 J_i & = \rho h \pqty{\alpha u^t} u_i , \\
 S_{ij} & = \rho h u_i u_j + P \gamma_{ij} .
\end{align}

For the purpose of constructing a solvable set of elliptic equations,
conformal transformation is applied to gravitational fields in the
solution of the Einstein constraint equations
\citep{York1979}.\footnote{Conformal transformation is sometimes
performed also for matter variables. This strategy is advantageous to
ensure the uniqueness of the solution \citep{York1979}, but it has not
been found essential for obtaining quasiequilibrium states of black
hole--neutron star binaries.} The induced metric is decomposed into the
conformal factor $\psi$ and the conformal (or ``background'') metric
$\hat{\gamma}_{ij}$ as
\begin{equation}
 \gamma_{ij} = \psi^4 \hat{\gamma}_{ij} . \label{eq:traceextr}
\end{equation}
The traceless part of the extrinsic curvature is also transformed as
\citep{OMurchadha_York1974}
\begin{equation}
 A^{ij} = \psi^{-10} \hat{A}^{ij} \; , \;A_{ij} = \psi^{-2} \hat{A}_{ij}
  , \label{eq:transfextr}
\end{equation}
where $\hat{A}_{ij} = \hat{\gamma}_{ik} \hat{\gamma}_{jl}
\hat{A}^{kl}$. The reason for adopting this transformation law is that
the divergence of the traceless part of the extrinsic curvature, which
is the central quantity in the momentum constraint, is also conformally
transformed as $D_j A^{ij} = \psi^{-10} \hat{D}_j \hat{A}^{ij}$, where
$D_i$ and $\hat{D}_i$ are the covariant derivatives associated with
$\gamma_{ij}$ and $\hat{\gamma}_{ij}$, respectively. The Hamiltonian
constraint is rewritten to an elliptic equation to determine the
conformal factor as
\begin{equation}
 \hat{\Delta} \psi = - 2\pi \psi^5 \rho_\mathrm{H} + \frac{1}{8} \psi
  \hat{R} + \frac{1}{12} \psi^5 K^2 - \frac{1}{8} \psi^{-7} \hat{A}_{ij}
  \hat{A}^{ij} , \label{eq:confham}
\end{equation}
where $\hat{\Delta} = \hat{\gamma}^{ij} \hat{D}_i \hat{D}_j$ and
$\hat{R}$ is the scalar curvature of $\hat{\gamma}_{ij}$.

The conformal traceless extrinsic curvature is given from
Eqs.~\eqref{eq:defextr}, \eqref{eq:traceextr}, and \eqref{eq:transfextr}
by
\begin{equation}
 \hat{A}^{ij} = \frac{\psi^6}{2\alpha} \bqty{\pqty{\hat{\mathbb{L}}
  \beta_\mathrm{com} }^{ij} - \hat{u}^{ij}} , \label{eq:ctsextr}
\end{equation}
where we defined the time derivative (in the comoving frame for this
problem) of the conformal metric,
\begin{align}
 \hat{u}_{ij} & := \partial_t \hat{\gamma}_{ij} , \\
 \hat{u}^{ij} & = \hat{\gamma}^{ik} \hat{\gamma}^{jl} \hat{u}_{kl} ,
\end{align}
and the longitudinal derivative, or conformal Killing operator,
\begin{equation}
 \pqty{\hat{\mathbb{L}} V}^{ij} := \hat{D}^i V^j + \hat{D}^j V^i -
  \frac{2}{3} \hat{\gamma}^{ij} \hat{D}_k V^k .
\end{equation}
In the conformal thin-sandwich formalism \citep{York1999}, the momentum
constraint is rewritten to an elliptic equation to determine the shift
vector for given $\hat{u}^{ij}$ as
\begin{equation}
 \hat{\Delta}_{\mathbb{L}} \beta_\mathrm{com}^i = 16\pi \alpha
  \psi^4 J^i + \frac{4}{3} \alpha \hat{D}^i K + \frac{\alpha}{\psi^6}
  \hat{D}_j \pqty{\frac{\psi^6}{\alpha} \hat{u}^{ij}} -
  \pqty{\mathbb{\hat{L}} \beta_\mathrm{com}}^{ij} \hat{D}_j \ln
  \pqty{\frac{\psi^6}{\alpha}} , \label{eq:ctsmom}
\end{equation}
where we defined the vector Laplacian by
\begin{equation}
 \hat{\Delta}_{\mathbb{L}} V^i := \hat{D}_j ( \mathbb{L} V )^{ij} =
  \hat{\gamma}^{jk} \hat{D}_j \hat{D}_k V^i + \frac{1}{3} \hat{D}^i
  \hat{D}_j V^j + \hat{R}^i{}_j V^j .
\end{equation}

In the extended conformal thin-sandwich formalism
\citep{Pfeiffer_York2003}, the evolution equation of the trace of the
extrinsic curvature,
\begin{equation}
 ( \partial_t - \mathcal{L}_\beta ) K = - \psi^{-4} \bqty{\hat{\Delta}
  \alpha + 2 \pqty{\hat{D}^i \ln \psi} \pqty{\hat{D}_i \alpha}} + \alpha
  \bqty{4\pi ( \rho_\mathrm{H} + S ) + \psi^{-12} \hat{A}_{ij}
  \hat{A}^{ij} + \frac{K^2}{3}} ,
\end{equation}
is employed to derive the equation for the lapse function for given
$\partial_t K$. It is often numerically advantageous to combine this
equation with the Hamiltonian constraint to derive an elliptic equation
for $\Phi := \alpha \psi$ as
\citep{Wilson_Mathews1995,Wilson_Mathews_Marronetti1996,Taniguchi_BFS2008}
\begin{equation}
 \hat{\Delta} \Phi = 2\pi \Phi \psi^4 ( \rho_\mathrm{H} + 2S ) +
  \frac{1}{8} \Phi \hat{R} + \frac{5}{12} \Phi \psi^{-4} K^2 +
  \frac{7}{8} \Phi \psi^{-8} \hat{A}_{ij} \hat{A}^{ij} - \psi^5 (
  \partial_t - \beta^i \hat{D}_i ) K . \label{eq:xcts}
\end{equation}

The set of equations, Eqs.~\eqref{eq:confham}, \eqref{eq:ctsextr},
\eqref{eq:ctsmom}, and \eqref{eq:xcts}, contains four freely specifiable
quantities, $\hat{\gamma}_{ij}$, $K$, $\hat{u}_{ij}$, and $\partial_t
K$. Because the direction of the time in the comoving frame agrees with
the helical Killing vector so that $\xi^\mu = ( \partial_t )^\mu$, we
require $\hat{u}_{ij} = 0 = \partial_t K$. We caution that these
conditions sometimes require modification in the far zone, because the
helical Killing vector becomes spacelike
\citep{Shibata_Uryu_Friedman2004}. The remaining variables,
$\hat{\gamma}_{ij}$ and $K$, need to be chosen by other conditions. The
simplest and popular choice is the conformally-flat and maximal-slicing
conditions,
\begin{equation}
 \hat{\gamma}_{ij} = f_{ij} \; , \; K = 0 .
\end{equation}
This is numerically advantageous, because the source terms of
Eqs.~\eqref{eq:confham}, \eqref{eq:ctsmom}, and \eqref{eq:xcts} become
simple and fall off rapidly enough at spatial infinity to obtain
accurate results. Furthermore, the elliptic operator involved in the
problem becomes the Laplacian for a flat three space, methods for
solving which are well developed. The maximal-slicing condition
constrains only a single degree of freedom and is usually justified as
the gauge condition on the time. However, the conformal flatness
condition is physically too restrictive. For example, it is known that a
Kerr black hole cannot be constructed with this choice
\citep{Garat_Price2000,Kroon2004,DeFelice_LMO2019}. Moreover, the set of
equations can lead to nonunique solutions
\citep{Pfeiffer_York2005,Baumgarte_OMurchadha_Pfeiffer2007,Walsh2007,CorderoCarrion_CDJNG2009}. A
spinning black hole exhibits two solutions with different values of the
physical spin magnitude for a given input parameter that specifies the
spin (which depends on the approach). Only one branch of the solutions
is connected to a Schwarzschild solution in the nonspinning limit, but
the dimensionless spin parameter is limited to $\chi \lesssim
0.85$. Another popular choice is to adopt those for the Kerr-Schild
metric as $( \hat{\gamma}_{ij} , K )$. The obvious advantage of this
choice is that we can construct nearly-extremally spinning black holes
with $\chi \approx 1$. However, with this choice, the equations to be
solved become complicated and involve terms with slow falloff at spatial
infinity. Accordingly, the accuracy of numerical solutions tends to be
degraded compared to the conformally-flat and maximal-slicing cases
\citep{Taniguchi_BFS2006}. Yet another choice of the free variables, a
modified Kerr-Schild background, combines the advantages of these two
choices. That is, the metric takes the Kerr-Schild form only in the
vicinity of the black hole and approaches exponentially to the
conformally-flat and maximal-slicing one at the distant region to
maintain numerical accuracy \citep{Lovelace_OPC2008,Foucart_KPT2008}.

\subsubsection{Excision approach} \label{app:init_grav_ex}

In this Appendix~\ref{app:init_grav_ex}, we describe the central idea of
the excision approach with restricting ourselves to the conformally-flat
($\hat{\gamma}_{ij} = f_{ij}$) and maximally-slicing ($K=0$) cases. The
excision approach is not restricted to these choices, and general cases
are handled in, e.g., \citet{Taniguchi_BFS2006,Foucart_KPT2008}.

The relevant differential equations are now given by
\begin{align}
 \underline{\Delta} \psi & = - 2\pi \psi^5 \rho_\mathrm{H} - \frac{1}{8}
 \psi^{-7} \hat{A}_{ij} \hat{A}^{ij} , \label{eq:flatham} \\
 \underline{\Delta} \beta^i + \frac{1}{3} \underline{D}^i
 \underline{D}_j \beta^j & = 16\pi \Phi \psi^3 J^i + 2\hat{A}^{ij}
 \underline{D}_j \pqty{\Phi \psi^{-7}} , \label{eq:flatmom} \\
 \underline{\Delta} \Phi & = 2\pi \Phi \psi^4 ( \rho_\mathrm{H} + 2S ) +
 \frac{7}{8} \Phi \psi^{-8} \hat{A}_{ij} \hat{A}^{ij} ,
 \label{eq:flatms}
\end{align}
where $\underline{D}_i$ and $\underline{\Delta} := f^{ij}
\underline{D}_i \underline{D}_j$ denote the covariant derivative
operator and Laplacian of the flat metric, respectively. This covariant
derivative becomes the partial derivative in Cartesian
coordinates. Here, the shift vector in the comoving frame,
$\beta_\mathrm{com}^i$, is replaced by that in the asymptotic inertial
frame, $\beta^i$, because the rotational part, $\beta_\mathrm{rot}^i$,
given by Eq.~\eqref{eq:shiftrot} is a (conformal) Killing vector of the
flat metric. Similarly, Eq.~\eqref{eq:ctsextr} becomes
\begin{equation}
 \hat{A}^{ij} = \frac{\psi^7}{2\Phi} \pqty{\underline{D}^i \beta^j +
  \underline{D}^j \beta^i - \frac{2}{3} f^{ij} \underline{D}_k \beta^k}
  . \label{eq:exaij}
\end{equation}

A solution for these elliptic equations requires appropriate boundary
conditions at spatial infinity and the black-hole horizon $\mathcal{S}$,
which is set to be a coordinate sphere. By assuming that the spacetime
is asymptotically flat, appropriate boundary conditions at spatial
infinity are given by
\begin{align}
 \psi ( r \to \infty ) & = 1 , \\
 \beta^i ( r \to \infty ) & = 0 , \\
 \Phi^i ( r \to \infty ) & = 1 .
\end{align}
At the black-hole horizon, the assumption that the black hole is in
equilibrium leads to a set of boundary conditions for the conformal
factor and the shift vector \citep[see also
\citealt{Cook2002,Jaramillo_Gourgoulhon_Marugan2004,Caudill_CGP2006}]{Cook_Pfeiffer2004}. Specifically,
the requirement that the surface is nonexpanding and thus becomes an
apparent horizon (more strictly, marginally outer trapped surface)
derives
\begin{equation}
 \eval{\underline{s}^i \underline{D}_i \ln \psi}_\mathcal{S} = \eval{-
  \frac{1}{4} \pqty{\underline{h}^{ij} \underline{D}_i \underline{s}_j -
  \psi^2 L}}_\mathcal{S} ,
\end{equation}
where $s^i$ is the outward-pointing unit vector to the excision surface
with $\underline{s}^i := \psi^2 s^i$, $h_{ij} := \gamma_{ij} - s_i s_j$
is the induced metric on $\mathcal{S}$ with $\underline{h}_{ij} :=
\psi^{-4} h_{ij}$, and $L := h^{ij} K_{ij}$. The shift vector is treated
separately for the normal and tangential components with respect to the
excision surface. The normal component in the comoving frame is
determined by the condition that the coordinate location of
$\mathcal{S}$ does not change in time as
\begin{equation}
 \eval{\beta_\perp}_\mathcal{S} = \eval{\alpha}_\mathcal{S} .
\end{equation}
Note that the radius of $\mathcal{S}$ is iteratively determined by
requiring the mass (or area) of the black hole to take a desired
value. The tangential component is required to be a conformal Killing
vector on $\mathcal{S}$ with its magnitude undetermined. For the
conformally-flat case, we can express it as
\begin{equation}
 \eval{\beta_\parallel^i}_\mathcal{S} = \underline{\epsilon}^i{}_{jk}
  \Omega_r^j x_\mathrm{BH}^k ,
\end{equation}
where $\underline{\epsilon}_{ijk}$ is the flat Levi-Civita tensor,
$\Omega_r^i$ is a freely-specifiable vector that controls the black-hole
spin, and $x_\mathrm{BH}^i$ is the coordinate vector measured from the
center of $\mathcal{S}$. The magnitude (but not the direction) of a
quasilocal spin angular momentum of the black hole may be defined in
terms of an approximate Killing vector $\xi_\mathcal{S}^i$ for
axisymmetry on $\mathcal{S}$ as
\begin{equation}
 S_{( \xi )} := \frac{1}{8\pi} \oint_\mathcal{S} ( K_{ij} - K
  \gamma_{ij} ) \xi_\mathcal{S}^j \dd{S}^i .
\end{equation}
An approximate Killing vector may be found by solving Killing transport
equations \citep{Dreyer_KSS2003,Caudill_CGP2006} or by minimizing an
appropriately-defined norm of the shear
\citep{Cook_Whiting2007,Lovelace_OPC2008}. The magnitude and the
direction of $\Omega_r^i$ are iteratively determined by requiring the
quasilocal spin to take a desired value. However, because no method is
known for defining the direction of the spin in a geometric manner, the
direction is estimated in a coordinate-dependent manner from the
components of $\Omega_r^i$. Finally, the lapse function on $\mathcal{S}$
can be chosen freely. For example, we may choose a Neumann boundary
condition
\begin{equation}
 \eval{\underline{s}^i \underline{D}_i \Phi}_\mathcal{S} = 0 ,
\end{equation}
or simply require $\alpha$ to take a constant value on $\mathcal{S}$
\citep{Caudill_CGP2006}.

\subsubsection{Puncture approach} \label{app:init_grav_pu}

The puncture approach was first proposed to construct initial data
containing multiple black holes with arbitrary linear and spin angular
momenta \citep{Brandt_Brugmann1997} extending work on time-symmetric
(i.e., without momenta) multiple black holes
\citep{Brill_Lindquist1963}. Because of the simplicity of
moving-puncture methods for evolving black-hole spacetimes
\citep{Brugmann_Tichy_Jansen2004,Campanelli_LMZ2006,Baker_CCKV2006},
initial data of this type are popularly adopted in dynamical simulations
of binary black holes \citep{Ansorg_Brugmann_Tichy2004}.

Here, we describe the puncture approach for quasiequilibrium black
hole--neutron star binaries
\citep{Shibata_Uryu2006,Shibata_Uryu2007,Kyutoku_Shibata_Taniguchi2009}. In
contrast to the excision approach, the conformally-flat and
maximal-slicing conditions are essential for this approach (see
\citealt{Liu_Etienne_Shapiro2009} for possible extension). Thus, we
again assume $\hat{\gamma}_{ij} = f_{ij}$ and $K=0$. As a by-product,
this approach for solving the initial value problem can naturally be
adopted as a method of dynamical simulations within the
conformal-flatness approximation \citep{Just_BAGJ2015}.

The basic equations for gravitational fields are
Eqs.~\eqref{eq:flatham}, \eqref{eq:flatmom}, and \eqref{eq:flatms} as in
the case of the excision approach. To avoid appearances of divergent
terms in $\psi$ and $\Phi$, they are decomposed into singular and
regular parts by
\begin{align}
 \psi & = 1 + \frac{M_\mathrm{P}}{2r_\mathrm{BH}} + \phi , \\
 \Phi & = 1 - \frac{M_\Phi}{2r_\mathrm{BH}} + \eta ,
\end{align}
where $M_\mathrm{P}$ and $M_\Phi$ are positive mass parameters and
$r_\mathrm{BH} = \abs{x^i - x_\mathrm{P}^i}$ is the Euclidean coordinate
distance from the puncture located at $x_\mathrm{P}^i$. Because
$1/r_\mathrm{BH}$ is a homogeneous solution of the flat Laplacian, only
the regular parts $\phi$ and $\eta$ require numerical integrations.

The puncture approach for black hole--neutron star binaries employs a
mixture of the conformal thin-sandwich formalism and the conformal
transverse-traceless decomposition to determine the conformal traceless
part of the extrinsic curvature, $\hat{A}^{ij}$. Specifically, it is
also decomposed into singular and regular parts as
\begin{equation}
 \hat{A}^{ij} = \underline{D}^i W^j + \underline{D}^j W^i - \frac{2}{3}
 f^{ij} \underline{D}_k W^k + K_\mathrm{P}^{ij} . \label{eq:puncaij}
\end{equation}
The singular part associated with the puncture is given in a Bowen-York
form \citep{Bowen_York1980} as
\begin{equation}
 K_\mathrm{P}^{ij} = \frac{3}{2r_\mathrm{BH}^2} \bqty{l^i P_\mathrm{BH}^j +
  l^j P_\mathrm{BH}^i - \pqty{f^{ij} - l^i l^j} l_k P_\mathrm{BH}^k} +
  \frac{3}{r_\mathrm{BH}^3} \pqty{l^i \underline{\epsilon}^{jkl} + l^j
  \underline{\epsilon}^{ikl}} S^\mathrm{P}_k l_l ,
\end{equation}
where $l^i := x_\mathrm{BH}^i / r_\mathrm{BH}$ and $l_i := f_{ij}
l^j$. Vectorial constants $P_\mathrm{BH}^i$ and $S^\mathrm{P}_i$ are
related to the linear and spin angular momenta of the black hole,
respectively. The vectorial auxiliary function $W^i$ is determined by
the momentum constraint.

The set of the basic equations in the puncture approach is given by
\begin{align}
 \underline{\Delta} \phi & = - 2\pi \psi^5 \rho_\mathrm{H} - \frac{1}{8}
 \psi^{-7} \hat{A}_{ij} \hat{A}^{ij} , \\
 \underline{\Delta} \beta^i + \frac{1}{3} \underline{D}^i
 \underline{D}_j \beta^j & = 16\pi \Phi \psi^3 J^i + 2\hat{A}^{ij}
 \underline{D}_j \pqty{\Phi \psi^{-7}} , \\
 \underline{\Delta} \eta & = 2\pi \Phi \psi^4 ( \rho_\mathrm{H} + 2S ) +
 \frac{7}{8} \Phi \psi^{-8} \hat{A}_{ij} \hat{A}^{ij} , \\
 \underline{\Delta} W^i + \frac{1}{3} \underline{D}^i \underline{D}_j
 W^j & = 8\pi \psi^{10} J^i .
\end{align}
Because the matter fields are localized in the neutron star and
$\hat{A}_{ij} \hat{A}^{ij} \propto \order{r_\mathrm{BH}^{-4}}$ and
$\propto \order{r_\mathrm{BH}^{-6}}$ in the absence and the presence of
the black-hole spin, respectively, it is readily found that no source
term diverges at the puncture. In this approach, the conformal traceless
part of the extrinsic curvature, $\hat{A}^{ij}$, is determined not by
Eq.~\eqref{eq:exaij} but by Eq.~\eqref{eq:puncaij}, because the
regularity at the point $\alpha = 0 = \Phi$ is not ensured (see also
\citealt{Gourgoulhon_Grandclement_Bonazzola2002,Grandclement_Gourgoulhon_Bonazzola2002,Grandclement2006}
for related issues). The shift vector still needs to be solved, because
it is required in the solution of hydrostationary equations. A solution
for these elliptic equations needs boundary conditions at spatial
infinity, which is derived from the asymptotic flatness as
\begin{equation}
 \eval{\phi , \beta^i , \eta , W^i}_{r \to \infty} = 0 .
\end{equation}
The lack of the inner boundary could be a drawback, because we cannot
impose physical equilibrium conditions on the black-hole
horizon. However, we may also regard this lack as a flexibility for
adjusting initial data to a desirable state
\citep{Shibata_Taniguchi2008,Kyutoku_Shibata_Taniguchi2009}.

Free parameters in this formulation have to be chosen appropriately. The
puncture or ``bare'' mass, $M_\mathrm{P}$, and the spin parameter,
$S^\mathrm{P}_i$, are determined by requiring the black hole to have
desired mass and spin, respectively. The values of the mass and the spin
magnitude should be evaluated on the apparent horizon, which needs to be
located numerically in the computation of quasiequilibrium states in the
puncture approach (see \citealt{Thornburg2007} for
reviews). Empirically, the Euclidean norm of $S^\mathrm{P}_i$ coincides
with $S_{( \xi )}$ with the fractional error of $\lesssim \num{e-5}$ for
$\chi \lesssim 0.75$. Again, the direction of the spin is estimated in a
coordinate-dependent manner from the components of
$S^\mathrm{P}_i$. Another mass parameter, $M_\Phi$, is determined by the
virial theorem, i.e., the condition that the ADM and Komar masses agree
\citep{Beig1978,Ashtekar_MagnonAshtekar1979,Friedman_Uryu_Shibata2002,Shibata_Uryu_Friedman2004},
\begin{equation}
 \int_{r \to \infty} \underline{D}_i \Phi \dd{S}^i = - \int_{r \to
  \infty} \underline{D}_i \psi \dd{S}^i = 2\pi M_\mathrm{ADM} .
\end{equation}
The linear momentum parameter, $P_\mathrm{BH}^i$, is determined by
requiring that the total linear momentum of the system vanishes as
\begin{equation}
 P_\mathrm{BH}^i = - \int J^i \psi^{10} \dd{V} ,
\end{equation}
where the integral on the right-hand side is interpreted as the linear
momentum of the neutron star.

\subsection{Hydrostationarity} \label{app:init_hydro}

Fluids in quasiequilibrium binaries are required to satisfy the
hydrostationary equations derived by the continuity and Euler's
equations. Several methods for solving these equations have been
proposed focusing particularly on the irrotational configurations
\citep[see also \citealt{Tacik_FPMKSS2016} and references therein for
neutron stars with arbitrary
spins]{Bonazzola_Gourgoulhon_Marck1997,Asada1998,Shibata1998,Teukolsky1998,Gourgoulhon_GTMB2001,Shibata_Uryu_Friedman2004}. In
this Appendix~\ref{app:init_hydro}, we review formulation proposed in
\citet{Shibata_Uryu_Friedman2004,Uryu_LFGS2009}.

The continuity equation,
\begin{equation}
 \nabla_\mu \pqty{\rho u^\mu} = 0 , \label{eq:cont}
\end{equation}
and the local energy-momentum conservation equation,
\begin{equation}
 \nabla_\nu T^\nu{}_\mu = 0 ,
\end{equation}
govern the motion of the fluid inside neutron stars. The latter for the
ideal fluid is rewritten to
\begin{equation}
 \rho u^\nu \nabla_\nu ( hu_\mu ) + hu_\mu \nabla_\nu ( \rho u^\nu  ) +
  \nabla_\mu P = 0 , \label{eq:emcons}
\end{equation}
and the second term vanishes due to Eq.~\eqref{eq:cont}. If the first
law of thermodynamics under (i) zero temperature or isentropy and (ii)
chemical equilibrium, i.e.,
\begin{equation}
 \dd{h} = \frac{1}{\rho} \dd{P} , \label{eq:cold}
\end{equation}
holds throughout the fluid, Eq.~\eqref{eq:emcons} becomes
\begin{equation}
 u^\nu \nabla_\nu ( hu_\mu ) + \nabla_\mu h = 0 . \label{eq:coldemcons}
\end{equation}
By defining the relativistic vorticity tensor
\begin{align}
 \omega_{\mu \nu} & := \pqty{g_\mu{}^\alpha + u_\mu u^\alpha}
 \pqty{g_\nu{}^\beta + u_\nu u^\beta} ( \nabla_\alpha u_\beta -
 \nabla_\beta u_\alpha ) \notag \\
 & = h^{-1} [ \nabla_\mu ( hu_\nu ) - \nabla_\nu ( hu_\mu ) ] ,
\end{align}
where we used Eq.~\eqref{eq:coldemcons} to derive the second line, the
local energy-momentum tensor conservation under Eq.~\eqref{eq:cold} is
shown to be equivalent to
\begin{equation}
 u^\nu \omega_{\mu \nu} = 0 . \label{eq:coldemconsv}
\end{equation}

This equation may be integrated along the Killing vector for the
solution of hydrostationarity. We define the spatial velocity $v^\mu$ of
the fluid in the comoving frame by
\begin{equation}
 u^\mu = u^t \pqty{\xi^\mu + v^\mu} , \label{eq:veldecomp}
\end{equation}
where $\xi^\mu$ is the helical Killing vector for the case in which the
radial velocity is absent (see
\citealt{Kyutoku_Shibata_Taniguchi2014,Kyutoku_KKST2021} for an
approximate hydrostationarity with the radial velocity) and $v^\mu n_\mu
= 0$ is satisfied. As we described in Sect.~\ref{sec:eq_param}, the
helical Killing vector $\xi^\mu$ is timelike in the near zone. By
substituting this equation into Eq.~\eqref{eq:coldemconsv}, we obtain
\begin{equation}
 u^t \bqty{ \mathcal{L}_\xi ( hu_\mu ) - \nabla_\mu ( hu_\nu \xi^\nu ) +
  v^\nu \omega_{\nu \mu} } = 0 .
\end{equation}
Because $\xi^\mu$ is a Killing vector, we require that
\begin{equation}
 \mathcal{L}_\xi ( hu_\mu ) = 0 .
\end{equation}
If the fluid motion is synchronized with the orbital motion, i.e., the
fluid is in a corotational state with $v^\mu = 0$, the integration of
the local-energy momentum conservation gives rise to
\begin{equation}
 hu_\mu \xi^\mu = \mathrm{const} . \label{eq:eulerco}
\end{equation}

If instead the fluid motion is irrotational and $\omega_{ij} = 0$ is
satisfied, we again deduce that $hu_\mu \xi^\mu$ is constant in space
\citep{Shibata1998}. The irrotational condition also implies that the
spatial part of the specific momentum, $hu_i$, can be expressed as the
gradient of the velocity potential $\Psi$ as
\begin{equation}
 hu_i = D_i \Psi . \label{eq:velpot}
\end{equation}
The velocity potential is determined by the continuity equation, which
is rewritten as
\begin{equation}
 \nabla_\mu ( \rho u^\mu ) = \frac{1}{\alpha} D_i \pqty{\rho \alpha u^t
  v^i} = 0 ,
\end{equation}
recalling that the time coordinate is taken to be the direction of
$\xi^\mu$. By inserting Eqs.~\eqref{eq:veldecomp} and \eqref{eq:velpot}
with the expression of the helical Killing vector
\begin{equation}
 \xi^\mu = \alpha n^\mu + \beta_\mathrm{com}^\mu ,
  \label{eq:helical_hydro}
\end{equation}
we obtain the elliptic equation for the velocity potential,
\begin{equation}
 D^i D_i \Psi + \pqty{D^i \Psi - hu^t \beta_\mathrm{com}^i} D_i \ln
  \pqty{\frac{\rho \alpha}{h}} - D_i \pqty{hu^t \beta_\mathrm{com}^i} =
  0 .
\end{equation}


As in the case of gravitational-field equations, freely-specifiable
constants need to be chosen appropriately. The integration constant,
$C_\mathrm{E} := - hu_\mu \xi^\mu$, of relativistic Euler's equation is
usually fixed to its value at the center of the star, which is defined
as the location of the maximum baryon rest-mass density (see also
Appendix~\ref{app:init_angu}). We have another constant $\Omega$
implicitly in the helical Killing vector, and we describe the method to
fix it separately in Appendix~\ref{app:init_angu}.

\subsection{Orbital angular velocity for a quasicircular orbit}
\label{app:init_angu}

The first integral of Euler's equation, $C_\mathrm{E} = - hu_\mu
\xi^\mu$, involves the orbital angular velocity, $\Omega$, via
Eqs.~\eqref{eq:shiftcom}, \eqref{eq:veldecomp}, and
\eqref{eq:helical_hydro}. Thus, $\Omega$ needs to be determined
appropriately. In particular, we require that the binary is in a
quasicircular orbit to obtain quasiequilibrium binaries without radial
approaching velocity. In the following, we describe two typical methods
for deriving quasicircular orbits, referring to the line connecting the
centers of the black hole and the neutron star as the $X$ axis. Note
that, when we incorporate the radial approaching velocity to reduce or
control the orbital eccentricity, the orbital angular velocity is
usually prescribed to obtain a desired value of the eccentricity
\citep{Foucart_KPT2008,Kyutoku_KKST2021}.

One typical method to determine $\Omega$ is to require the force balance
along the $X$ axis. The force-balance condition is equivalent to
vanishing of the enthalpy gradient at the stellar center
$\mathcal{O}_\mathrm{NS}$, which is defined to satisfy
\begin{equation}
 \eval{\pdv{\ln h}{X}}_{\mathcal{O}_\mathrm{NS}} = 0
  . \label{eq:forcebalance}
\end{equation}
Because the pressure gradient and self-gravitational force vanish at the
stellar center, this equation combined with the first integral of
Euler's equation may be regarded as the condition that the gravitational
force from the black hole and the centrifugal force associated with the
orbital motion are commensurate at the stellar center. Hence, this
condition can be used to determine $\Omega$ for a given set of
gravitational-field variables. In this case, the constant $C_\mathrm{E}$
is determined by specifying the rest-mass density at the stellar center
as explained in Appendix~\ref{app:init_hydro}.

The other typical method to determine $\Omega$ is to require that the
specific enthalpy becomes unity, i.e., $h=1$, at two points on the
stellar surface along the $X$ axis.\footnote{This condition does not
hold if we consider special equations of state, e.g., those for quark
stars, for which the density at the stellar surface does not vanish.}
Because the pressure and the internal energy vanish at the stellar
surface, the sum of the black-hole gravitational force, the stellar
self-gravitational force, and the centrifugal force associated with the
orbital motion is balanced there. The two conditions derived at the two
points may be used as the conditions to determine $C_\mathrm{E}$ and
$\Omega$ from a given set of gravitational-field variables.

It has been confirmed that both methods derive accurate numerical
results with the reasonable number of iterations and that the results
derived by the two methods agree within the convergence level of the
specific enthalpy
\citep{Taniguchi_BFS2006,Taniguchi_BFS2007,Taniguchi_BFS2008,Taniguchi_Shibata2010}. That
is, both methods work well.

\subsection{Center of mass of a binary} \label{app:init_com}

Equation~\eqref{eq:forcebalance} also depends on the location of the
center of the mass, with respect to which the rotational shift vector,
$\beta_\mathrm{rot}^i$, is defined via $( \partial_\varphi )^i$. Thus,
we need a method to determine the center of mass for given locations of
the black hole and the neutron star. Again, the procedures are different
for the excision and the puncture approaches.

In the excision approach, the center of mass is determined by requiring
that the total linear momentum of the system vanishes
\citep{Taniguchi_BFS2006,Grandclement2006,Taniguchi_BFS2007,Taniguchi_BFS2008,Taniguchi_Shibata2010},
i.e.,
\begin{equation}
 P^i = \frac{1}{8\pi} \oint_{r \to \infty} \pqty{K^{ij} - K \gamma^{ij}}
  \dd{S}_j = 0 .
\end{equation}
The key idea behind this condition is that the total linear momentum
depends on the location of the center of mass for a hypothetical angular
velocity, $\Omega$. Hence, the location of the center of mass can be
determined to satisfy this condition. Once it is determined at an
iteration step, the black hole and the neutron star are moved keeping
their separation unchanged so that the center of mass of the binary is
located on the $Z$ axis.

The situation is totally different in the puncture approach. The reason
is that the vanishing of the total linear momentum is used to determine
the linear-momentum parameter of the puncture, $P_\mathrm{BH}^i$. There
has been no known natural and physical condition for determining the
center of the mass in the puncture approach, and three conditions have
been proposed. The first method is to require vanishing of the dipole
moment of the conformal factor, $\psi$, defined at spatial infinity
\citep{Shibata_Uryu2006,Shibata_Uryu2007}. The second method is to
require the minus of the azimuthal component of the shift vector,
$-\beta^\varphi$, at the puncture is equal to the orbital angular
velocity, so that the position of the puncture is fixed in the comoving
frame \citep{Shibata_Taniguchi2008}. However, the angular momentum of
quasiequilibrium states derived with these conditions is found to be
smaller by $\sim 2\%$ for a close orbit of $m_0 \Omega \ge 0.03$ than
the third or fourth post-Newtonian approximation for $Q=3$, and the
deviation is typically larger for higher mass-ratio
systems. Accordingly, these conditions lead to eccentric orbits, whereas
the second condition works slightly better than the first one.

The third method determines the location of the center of mass in a
phenomenological manner by requiring the total angular momentum of the
binary to agree with that derived by post-Newtonian approximations for a
given value of the orbital angular velocity $\Omega$ or equivalently the
post-Newton parameter $m_0 \Omega$
\citep{Kyutoku_Shibata_Taniguchi2009}. This method allows us to obtain
orbits with moderately low eccentricity of $\sim 0.01$, which is smaller
than the values achieved with previous two conditions.

When the orbital eccentricity is reduced by modifying the orbital
angular velocity and incorporating the radial approaching velocity (see
Appendix~\ref{app:init_beyond_ecc}), the puncture approach usually
adopts the second condition for determining the center of mass
\citep{Kyutoku_KKST2021}. This is because the radial velocity of the
puncture is also identified via the minus of the shift vector at the
puncture. Performance of the eccentricity reduction combined with other
methods for determining the center of mass has not been investigated.

\subsection{Beyond the helical symmetry} \label{app:init_beyond}

\subsubsection{Spin misalignment}

Until Appendix~\ref{app:init_com}, we have discussed quasiequilibrium
states in the presence of a helical Killing vector. This setup is
justified for black holes whose spins are absent or (anti-)aligned with
the orbital angular momentum of the binary as far as the orbital period
is much shorter than the time scale of gravitational radiation
reaction. However, if the spin of the black hole is inclined with
respect to the orbital angular momentum, the orbital plane as well as
the black-hole spin precesses due to the spin-orbit, spin-spin, and
monopole-quadrupole couplings
\citep{Barker_OConnel1975,Apostolatos_CST1994,Racine2008}. This suggests
that the helical symmetry is acceptable only when the orbital period is
much shorter than not only the radiation reaction time scale but also
the precession period. The same applies to the spin of the neutron star.

Spin misalignment breaks the reflection symmetry with respect to the
orbital plane. Accordingly, the separation of binary components along
the rotational axis, say the $Z$ axis, needs to be fixed in addition to
the other free parameters. In the computation of initial data, it is
usually determined by requiring the force-balance condition at the
stellar center like Eq.~\eqref{eq:forcebalance} along the $Z$ direction
\citep{Foucart_DKT2011,Kawaguchi_KNOST2015,Henriksson_FKT2016}. Furthermore,
the $Z$ component of the linear momentum of a neutron star does not
vanish in general in the absence of the reflection symmetry. In the
puncture approach, this is canceled by adjusting the linear-momentum
parameter of the puncture, $P_\mathrm{BH}^i$. In the excision approach,
the $Z$ component of the linear momentum is eliminated by applying the
boost to the whole system via the boundary condition at spatial infinity
\citep{Foucart_DKT2011,Henriksson_FKT2016}. This induces coordinate
motion of the system and may not be favorable for longterm simulations
\citep{Foucart_etal2021}.

\subsubsection{Eccentricity reduction with approaching velocity}
\label{app:init_beyond_ecc}

Although the orbital period is shorter than the time scale of
gravitational radiation reaction for typical initial data, the radiation
reaction cannot be fully neglected if our purpose is to perform accurate
simulations of the inspiral phase. Quantitatively, the quadrupole
formula derives the radial approaching velocity of $\{ 64Q/[5(1+Q)^2] \}
(m_0 \Omega)^3$, which amounts to $\order{1\%}$ of the orbital velocity
for typical initial data. If the initial data do not contain approaching
velocity, radiation reaction is suddenly excited at the beginning of the
dynamical simulations and induces residual eccentricity of $\sim
\order{0.01}$, which is harmful for modeling gravitational waveforms
\citep{Favata2014}. To start simulations of the inspiral phase smoothly,
we need to introduce approaching velocity to the initial data by
modifying the helical symmetry. Typically, the excision approach adopts
a symmetry vector of the form
\citep{Pfeiffer_BKLLS2007,Henriksson_FKT2016}
\begin{equation}
 \bar{\xi}^\mu = \alpha n^\mu + \beta^\mu + \beta_\mathrm{rot}^\mu +
  \dot{a}_0 ( \partial_r )^\mu .
\end{equation}
The last, additional term modifies the set of equations via the inner
boundary condition and hydrostationary equations. The puncture approach
adopts a slightly different form \citep[see also
\citealt{Kyutoku_Shibata_Taniguchi2014} for discussions about these two
choices]{Kyutoku_KKST2021}
\begin{equation}
 \bar{\xi}^\mu = \alpha n^\mu + \beta^\mu + \beta_\mathrm{rot}^\mu +
  v_\mathrm{NS}^i ( \partial_i )^\mu
\end{equation}
to modify hydrostationary equations. Additional parameters, $\dot{a}_0$
and $v_\mathrm{NS}^i$, are free parameters to control the approaching
velocity of the binary, $v_\mathrm{app}$.

To achieve low eccentricity, the values of the orbital angular velocity
$\Omega$ and the approaching velocity $v_\mathrm{app}$ need to be chosen
appropriately for a given value of the orbital separation, $d$. In the
case of black hole--neutron star binaries, these parameters are adjusted
iteratively by performing dynamical simulations for a few orbits
\citep{Foucart_KPT2008,Kyutoku_KKST2021} following the technique
developed in binary-black-hole simulations
\citep{Pfeiffer_BKLLS2007,Buonanno_KMPT2011}. Corrections at each
iteration step are estimated by following steps. First, we determine the
time evolution of the orbital separation $d(t)$ or angular velocity
$\Omega (t)$. Here, we focus on the latter \citep[see
\citealt{Pfeiffer_BKLLS2007,Boyle_BKMPSCT2007} for methods based on the
separation]{Buonanno_KMPT2011}. Next, we fit the evolution of its time
derivative to a function of the form
\begin{equation}
 \dot{\Omega} (t) = S_\Omega (t) + B \sin ( \omega t + \phi_0 ) ,
\end{equation}
where $S_\Omega (t)$ is chosen to be a smooth function, e.g., a
polynomial in $t$, which represents smooth evolution driven by radiation
reaction. The sinusoidal term describes the undesirable modulation
caused by the residual eccentricity. Once the parameters $B , \omega$,
and $\phi_0$ are determined by the fitting, we correct the orbital
angular velocity and the approaching velocity of the binary by using
formulae derived for Keplerian orbits as
\begin{align}
 \delta \Omega & = - \frac{B \omega \sin \phi_0}{4 \Omega^2} , \\
 \delta v_\mathrm{app} & = \frac{B d_0 \cos \phi_0}{2 \Omega} ,
\end{align}
where $d_0 = d(0)$, aiming at removing the modulation term. In this
procedure, the orbital eccentricity of the initial data is also
estimated by
\begin{equation}
 e \approx \frac{\abs{B}}{2 \omega \Omega} .
\end{equation}
By adopting the updated values of $\Omega$ and $v_\mathrm{app}$, we
compute new initial data and perform dynamical simulations until the
eccentricity is sufficiently reduced. By beginning with quasiequilibrium
states computed with the helical symmetry, the orbital eccentricity is
typically reduced by an order of magnitude after a few iterations.

\section{Formulation of dynamical simulations} \label{app:sim}

Dynamical simulations of black hole--neutron star binaries are performed
by numerically solving the evolution of geometric and hydrodynamical
variables. The time evolution of the metric is obtained by solving the
Einstein evolution equations with partially incorporating the constraint
equations (see also \citealt{Bonazzola_GGN2004} for a fully constrained
scheme). In Appendix~\ref{app:sim_grav}, we will review formalisms
employed for solving gravitational fields of black hole--neutron star
binary coalescences in full general relativity.

Hydrodynamics equations can be solved to different levels of
sophistication. The simplest strategy is to solve ideal hydrodynamics
equation. This is adequate for the inspiral phase and brief periods
after the onset of merger. To perform realistic longterm simulations of
the remnant disk, neutrino-radiation hydrodynamics and
magnetohydrodynamics are essential. Viscous hydrodynamics is a possible
phenomenological alternative to magnetohydrodynamics, whose accurate
simulations require extremely large computational resources. We will
review these schemes and their implementations in Appendix
\ref{app:sim_hydro}.

We conclude this section by reviewing in Appendix~\ref{app:sim_mr} the
mesh refinement technique, which plays a key role in performing
systematic studies of compact binary coalescences. All the topics are
covered only briefly in this article, and we refer the readers to
standard textbooks, e.g.,
\citet{Alcubierre,Baumgarte_Shapiro,Gourgoulhon,Shibata}, for details.

Geometrical units in which $G=c=1$ is adopted in Appendix
\ref{app:sim_grav} and Appendix~\ref{app:sim_hydro}.

\subsection{Evolution of the metric} \label{app:sim_grav}

Two formalisms are currently adopted for evolving the geometry of black
hole--neutron star spacetimes. One is the
Baumgarte-Shapiro-Shibata-Nakamura (BSSN) formalism \citep[Part 1
of][]{Nakamura_Oohara_Kojima1987,Shibata_Nakamura1995,Baumgarte_Shapiro1999}
with the moving-puncture gauge condition
\citep{Campanelli_LMZ2006,Baker_CCKV2006,Brugmann_GHHST2008,Marronetti_TBGS2008}. Another
formalism is the generalized harmonic formalism
\citep{Friedrich1985,Garfinkle2002,Gundlach_CHM2005,Pretorius2005,Pretorius2006,Lindblom_SKOR2006}. Both
formalisms are different from the standard $3+1$ formalism
\citep{Arnowitt_Deser_Misner2008,York1979}, which is incapable of
conducting numerical-relativity simulations \citep[see,
e.g.,][Sect.~2.3]{Shibata}. A common feature is that the numbers of
variables and constraints are increased simultaneously.

\subsubsection{BSSN formalism}

The BSSN formalism, the original version of which was first proposed by
Nakamura in \citet{Nakamura_Oohara_Kojima1987}, is formulated as a
modified version of the $3+1$ formalism
\citep{Arnowitt_Deser_Misner2008,York1979}. The essence of this
formalism is to rewrite the evolution equations by changing the
variables from the three metric $\gamma_{ij}$ and the extrinsic
curvature $K_{ij}$ to, as a typical choice,
\begin{align}
 \tilde{\gamma}_{ij} & := \gamma^{-1/3} \gamma_{ij} , \\
 W & := \gamma^{-1/6} , \\
 \tilde{A}_{ij} & := \gamma^{-1/3} \pqty{K_{ij} - \frac{1}{3} K
 \gamma_{ij}} , \\
 K & := \gamma^{ij} K_{ij} , \\
 \tilde{\Gamma}^i & := - \partial_j \tilde{\gamma}^{ij} ,
\end{align}
where $\gamma$ is the determinant of $\gamma_{ij}$. The choice of the
conformal-factor variable, $W$, is not unique
\citep{Campanelli_LMZ2006,Baker_CCKV2006,Marronetti_TBGS2008}, but the
one used in the original formulation was not suitable for evolving the
black-hole spacetime (see below). The definition of the three auxiliary
variable, $\tilde{\Gamma}^i$, is also not unique, and $F_i := \sum_j
\partial_j \tilde{\gamma}_{ij}$ is another standard choice
\citep{Shibata_Nakamura1995}. Because of the increased number of
variables, $\det(\tilde{\gamma}_{ij}) = 1 , \tilde{\gamma}^{ij}
\tilde{A}_{ij} = 0$, and $\tilde{\Gamma}^i + \partial_j
\tilde{\gamma}^{ij} = 0$ or $F_i - \sum_j \partial_j \tilde{\gamma}_{ij}
= 0$ are introduced as new constraints.

This change of variables allows us to write evolution equations for the
conformal three metric, $\tilde{\gamma}_{ij}$, and the conformal
traceless extrinsic curvature, $\tilde{A}_{ij}$, in the form of simple
wave equations. This prescription prevents the spurious growth of
unphysical modes and enables us to perform stable and longterm evolution
for a variety of systems. Specifically, the basic equations in the BSSN
formalism with the choice of variables $\tilde{\gamma}_{ij}$, $W$,
$\tilde{A}_{ij}$, $K$, and $\tilde{\Gamma}^i$ are
\begin{align}
 \pqty{\partial_t - \beta^k \partial_k} \tilde{\gamma}_{ij} & = -2\alpha
 \tilde{A}_{ij} + \tilde{\gamma}_{ik} \partial_j \beta^k +
 \tilde{\gamma}_{jk} \partial_i \beta^k - \frac{2}{3}
 \tilde{\gamma}_{ij} \partial_k \beta^k , \label{eq:bssng} \\
 \pqty{\partial_t - \beta^k \partial_k} W & = \frac{W}{3} \pqty{\alpha K
 - \partial_k \beta^k} , \\
 \pqty{\partial_t - \beta^k \partial_k} \tilde{A}_{ij} & = W^2
 \bqty{\alpha \pqty{R_{ij} - \frac{1}{3} R \gamma_{ij}} - \pqty{D_i D_j
 \alpha - \frac{1}{3} D^2 \alpha \gamma_{ij}}} \notag \\
 & + \alpha \pqty{K \tilde{A}_{ij} - 2 \tilde{A}_{ik} \tilde{A}_j{}^k} +
 \tilde{A}_{ik} \partial_j \beta^k + \tilde{A}_{jk} \partial_i \beta^k -
 \frac{2}{3} \tilde{A}_{ij} \partial_k \beta^k - 8 \pi \alpha W^2
 \pqty{S_{ij} - \frac{1}{3} S \gamma_{ij}} , \label{eq:bssna} \\
 \pqty{\partial_t - \beta^k \partial_k} K & = \alpha
 \pqty{\tilde{A}_{ij} \tilde{A}^{ij} + \frac{1}{3} K^2} - D^2 \alpha + 4
 \pi \alpha ( \rho_\mathrm{H} + S ) , \label{eq:bssnk} \\
 \pqty{\partial_t - \beta^k \partial_k} \tilde{\Gamma}^i & = - 2
 \tilde{A}^{ij} \partial_j \alpha + 2 \alpha
 \pqty{\tilde{\Gamma}^i{}_{jk} \tilde{A}^{jk} - \frac{2}{3}
 \tilde{\gamma}^{ij} \partial_j K - 8 \pi \tilde{\gamma}^{ij} J_j -
 \frac{3}{W} \tilde{A}^{ij} \partial_j W} \notag \\
 & - \tilde{\Gamma}^j \partial_j \beta^i + \frac{2}{3} \tilde{\Gamma}^i
 \partial_j \beta^j + \frac{1}{3} \tilde{\gamma}^{ij} \partial_j
 \partial_k \beta^k + \tilde{\gamma}^{jk} \partial_j \partial_k \beta^i
 . \label{eq:bssncc}
\end{align}
Here, $D^2 := D^i D_i$, $\tilde{\Gamma}^i{}_{jk}$ is the Christoffel
symbol associated with $\tilde{\gamma}_{ij}$, and the matter variables
are decomposed as described in Appendix~\ref{app:init_grav}. An
important remark is that the momentum constraint is used to derive the
evolution equation for $\tilde{\Gamma}^i$.

A point in the BSSN formulation is to decompose the Ricci tensor of the
induced metric, $\gamma_{ij}$, as
\begin{equation}
 R_{ij} = \tilde{R}_{ij} + R^W_{ij} ,
\end{equation}
where $\tilde{R}_{ij}$ is the Ricci tensor of the conformal metric,
$\tilde{\gamma}_{ij}$, and
\begin{equation}
 R_{ij}^W = \frac{1}{W} \tilde{D}_i \tilde{D}_j W + \tilde{\gamma}_{ij}
  \bqty{\frac{1}{W} \tilde{D}^k \tilde{D}_k W - \frac{2}{W^2}
  \pqty{\tilde{D}^k W} \pqty{\tilde{D}_k W}} ,
\end{equation}
where $\tilde{D}_i$ is the covariant derivative associated with
$\tilde{\gamma}_{ij}$. For stable numerical computations,
$\tilde{R}_{ij}$ needs to be written using $\tilde{\Gamma}^i$ as
\begin{align}
 \tilde{R}_{ij} & = - \frac{1}{2} \tilde{\gamma}^{kl} \partial_k
 \partial_l \tilde{\gamma}_{ij} + \frac{1}{2} \pqty{\tilde{\gamma}_{ik}
 \partial_j \tilde{\Gamma}^k + \tilde{\gamma}_{jk} \partial_i
 \tilde{\Gamma}^k} \notag \\
 & - \frac{1}{2} \bqty{\pqty{\partial_k \tilde{\gamma}_{il}}
 \pqty{\partial_j \tilde{\gamma}^{kl}} + \pqty{\partial_k
 \tilde{\gamma}_{jl}} \pqty{\partial_i \tilde{\gamma}^{kl}} -
 \tilde{\Gamma}^l \partial_l \tilde{\gamma}_{ij}} -
 \tilde{\Gamma}^l{}_{ik} \tilde{\Gamma}^k{}_{jl} .
\end{align}
With this prescription, $\tilde{R}_{ij} \sim - (1/2) \Delta
\tilde{\gamma}_{ij}$ in the weak-field limit, and thus
Eqs.~\eqref{eq:bssng} and \eqref{eq:bssna} essentially constitute a wave
equation for $\tilde{\gamma}_{ij}$, i.e., the strong hyperbolicity is
guaranteed.

In the puncture-BSSN formalism, accurate and stable evolution of a black
hole is accomplished by redefining the conformal-factor variable
\citep{Campanelli_LMZ2006,Baker_CCKV2006}, adopting finite-differencing
schemes higher than or equal to the fourth order, and employing
appropriate moving-puncture gauge conditions
\citep{Bona_MSS1995,Alcubierre_BDKPST2003}. Typical gauge conditions are
\begin{align}
 \pqty{\partial_t - \beta^k \partial_k} \alpha & = - 2 \alpha K , \\
 \pqty{\partial_t - \beta^k \partial_k} \beta^i & = \frac{3}{4} B^i , \\
 \pqty{\partial_t - \beta^k \partial_k} B^i & = \pqty{\partial_t -
 \beta^k \partial_k} \tilde{\Gamma}^i - \eta_\mathrm{B} B^i ,
\end{align}
where $B^i$ is an auxiliary vectorial variable and $\eta_\mathrm{B}$ is
a constant chosen to be $\sim 1/m_0$, whose magnitude limits the size of
the time step via the Courant-Friedrichs-Levy conditions
\citep{Schnetter2010}. This formulation usually adopts puncture-type
initial data for the black-hole spacetime, in which the physical
singularity is absent \citep{Brandt_Brugmann1997}. Although the
coordinate singularity may appear inside the horizon of a black hole, it
is effectively excised in the moving-puncture gauge conditions
\citep{Hannam_HPBO2007,Hannam_HOBO2008}. Consequently, the whole
computational region can be evolved without any special treatment of the
interior of horizons such as the excision.

To mitigate violation of the Hamiltonian and also momentum constraints
further by introducing a mechanism for constraint propagation, various
extension to the BSSN formalism have been proposed
\citep{Bernuzzi_Hilditch2010,Alic_BBRP2012} based on the so-called Z4
formalism \citep{Bona_Ledvinka_Palenzuela_Zacek2003} with a constraint
damping mechanism \citep{Gundlach_CHM2005}. Among them, the Z4c
prescription \citep{Bernuzzi_Hilditch2010,Hilditch_BTCTB2013} is
occasionally adopted in black hole--neutron star binary coalescences
\citep{Kawaguchi_KNOST2015,Hayashi_KKKS2021,Most_PTR2021}. Formally,
this prescription introduces an additional variable $\Theta$, which
vanishes when the Hamiltonian constraint is satisfied. Its evolution
equation implemented in the SACRA code
\citep{Yamamoto_Shibata_Taniguchi2008,Kyutoku_Shibata_Taniguchi2014} is
given by
\begin{equation}
 ( \partial_t - \beta^k \partial_k ) \Theta = F(r) \bqty{\frac{1}{2}
  \alpha \pqty{R - \tilde{A}_{ij} \tilde{A}^{ij} + \frac{2}{3} K^2 -
  16\pi \rho_\mathrm{H}} - \alpha \kappa_1 ( 2 + \kappa_2 ) \Theta} ,
\end{equation}
where $\kappa_1$ and $\kappa_2$ are parameters that control the damping
of constraints. Their values are taken typically to be $\kappa_1 \sim
0.01/m_0$ and $\kappa_2 = 0$ \citep[see,
e.g.,][]{Hilditch_BTCTB2013}. An overall function, $F(r)$, has been
introduced in \citet{Kyutoku_Shibata_Taniguchi2014} to avoid possible
numerical problems that often occur near the outer boundary. The
equation used in original publications corresponds to $F(r)=1$
\citep{Bernuzzi_Hilditch2010,Hilditch_BTCTB2013}, and boundary issues
are handled by imposing sophisticated boundary conditions
\citep{Ruiz_Hilditch_Bernuzzi2011,Hilditch_Ruiz2018}. The evolution
equation for $K$, Eq.~\eqref{eq:bssnk}, is reinterpreted as an evolution
equation for a modified variable $\hat{K} := K - 2\Theta$ with adding a
term $F(r) \alpha \kappa_1 ( 1 - \kappa_2 ) \Theta$ on the right-hand
side. This requires us to regard $K$ on the right-hand sides of
equations as a dependent variable determined by $K = \hat{K} +
2\Theta$. The evolution equation for $\tilde{\Gamma}^i$,
Eq.~\eqref{eq:bssncc}, is modified to
\begin{align}
 \pqty{\partial_t - \beta^k \partial_k} \tilde{\Gamma}^i & = - 2
 \tilde{A}^{ij} \partial_j \alpha + 2 \alpha
 \bqty{\tilde{\Gamma}^i{}_{jk} \tilde{A}^{jk} - \frac{1}{3}
 \tilde{\gamma}^{ij} \partial_j \pqty{2K - 3\Theta} - 8 \pi
 \tilde{\gamma}^{ij} J_j - \frac{3}{W} \tilde{A}^{ij} \partial_j W}
 \notag \\
 & - \tilde{\Gamma}_\mathrm{d}^j \partial_j \beta^i + \frac{2}{3}
 \tilde{\Gamma}_\mathrm{d}^i \partial_j \beta^j + \frac{1}{3}
 \tilde{\gamma}^{ij} \partial_j \partial_k \beta^k + \tilde{\gamma}^{jk}
 \partial_j \partial_k \beta^i - 2 F(r) \alpha \kappa_1
 \pqty{\tilde{\Gamma}^i - \tilde{\Gamma}_\mathrm{d}^i} ,
\end{align}
where $\tilde{\Gamma}_\mathrm{d}^i := \tilde{\gamma}^{jk}
\tilde{\Gamma}^i{}_{jk}$ is computed from $\tilde{\gamma}_{ij}$. These
modification allows violation of the Hamiltonian and also momentum
constraints to propagate away. Similarly, the CCZ4 formalism
\citep{Alic_BBRP2012} has also been adopted recently
\citep{East_LLP2021}.

\subsubsection{Generalized harmonic formalism}

The generalized harmonic formalism is based on the Einstein equation in
a modified harmonic gauge condition. The Einstein equation in the
harmonic gauge is employed in various branches of general relativity,
particularly in the post-Newtonian approximation \citep{Blanchet2014},
because the basic equations are written in a hyperbolic form. In the
generalized harmonic formalism, the gauge condition is written in the
form of
\begin{equation}
 \nabla^\mu \nabla_\mu x^\alpha = g^{\alpha \beta} H_\beta ,
\end{equation}
where $x^\alpha$ is not a vector but a collection of four coordinate
functions and $H_\alpha$ is a set of arbitrary functions. With an
appropriate choice of $H_\alpha$, the Einstein equation is written to a
hyperbolic equation for the spacetime metric, $g_{\mu \nu}$. Although
the usual harmonic gauge condition, $H_\alpha = 0$, also derives a
hyperbolic equation, the time coordinate is not guaranteed to remain
timelike and also coordinate singularities may develop. Introduction of
carefully-chosen gauge source functions, $H_\alpha$, enables us to
perform stable and longterm simulations in an analogous manner to the
introduction of $\tilde{\Gamma}^i$ or $F_i$ in the BSSN formalism
\citep{Garfinkle2002}. This formulation simultaneously introduces new
constraints,
\begin{equation}
 H_\alpha + g^{\mu \nu} \Gamma_{\alpha \mu \nu} = 0 ,
\end{equation}
where $\Gamma_{\alpha \mu \nu}$ is the Christoffel symbol associated
with $g_{\mu \nu}$.

Another modification is to cast the evolution equations to a first-order
form by introducing derivatives of $g_{\mu \nu}$
\citep{Alvi2002,Lindblom_SKOR2006},
\begin{align}
 Q_{\mu \nu} & := - n^\alpha \partial_\alpha g_{\mu \nu} , \\
 D_{i\mu \nu} & := \partial_i g_{\mu \nu} .
\end{align}
The latter definition simultaneously introduces another constraint,
$D_{i\mu \nu} - \partial_i g_{\mu \nu} = 0$. In contrast to the BSSN
formulation, there are several options for the basic equations,
particularly in the choice of the constraint damping terms in a similar
manner to the Z4c formulation \citep{Gundlach_CHM2005}. Details of the
formulation employed for simulating black hole--neutron star binaries
are described in, e.g., \citet{Lindblom_SKOR2006,Anderson_HLLMNPT2008}.

Gauge conditions in the generalized harmonic formalism can be expressed
in terms of the $3+1$ variables as
\begin{align}
 \pqty{\partial_t - \beta^k \partial_k} \alpha & = - \alpha \pqty{H_t -
 \beta^k H_k + \alpha K} , \\
 \pqty{\partial_t - \beta^k \partial_k} \beta^i & = \alpha \gamma^{ij}
 \bqty{\alpha H_j + \alpha \gamma^{kl} \Gamma_{jkl} - \partial_j \alpha}
 .
\end{align}
Various choices of the gauge source functions, $H_\mu$, have been
adopted in simulations of black hole--neutron star binaries \citep[see,
e.g.,][]{Duez_FKPST2008,Foucart_BDGKMMPSS2013}. Recent simulations
typically have adopted damped harmonic gauge conditions
\citep{Szilagyi_Lindblom_Scheel2009},
\begin{equation}
 H_\mu = \mu_L \ln \pqty{\frac{\sqrt{\gamma}}{\alpha}} n_\mu - \mu_S
  \frac{\beta_\mu}{\alpha} ,
\end{equation}
where $\mu_L$ and $\mu_S$ are arbitrary positive functions.

The generalized harmonic formalism, which does not have a strong
singularity-avoidance property \citep{Bona_Masso1992}, needs to adopt
the excision technique, in which a region inside the apparent horizon is
removed from the computational domains and is replaced by an inner
boundary, for handling black holes. Because the interior of the excision
surface is causally disconnected from the exterior spacetime, no
physical boundary condition is required during the simulations as far as
unphysical propagation of information is avoided appropriately. The
first success in simulating orbiting binary black holes is achieved with
this method \citep{Pretorius2005,Pretorius2006}, and subsequently, a
wide variety of highly-accurate numerical simulations for binary black
holes have been performed by the SXS collaboration in the generalized
harmonic formulation
\citep{Boyle_BKMPSCT2007,Boyle_BKMPPS2008,Scheel_BCKMP2009}.

\subsection{Hydrodynamics} \label{app:sim_hydro}

In addition to the evolution equations for the gravitational fields, we
need to solve hydrodynamics equations to evolve the neutron star in the
inspiral phase and the disrupted material, including the remnant disk
surrounding the black hole, fallback material, and the ejecta, in the
merger and postmerger phases. Moreover, neutrino-radiation hydrodynamics
and magnetohydrodynamics are essential for describing realistic
evolution of the accretion disk. We first discuss ideal hydrodynamics
equations, which are sufficient for modeling the inspiral phase and a
short period after the onset of merger. Later, neutrino-radiation
hydrodynamics and magnetohydrodynamics equations are described in
Appendix~\ref{app:sim_hydro_r} and Appendix~\ref{app:sim_hydro_m},
respectively.

In the $3+1$ formalism, the local energy-momentum conservation equation,
$\nabla_\nu T^{\mu \nu} = 0$, is decomposed into the spatial and
temporal parts as
\begin{align}
 \gamma_{i\mu} \nabla_\nu T^{\mu \nu} & = 0 , \\
 n_\mu \nabla_\nu T^{\mu \nu} & = 0 .
\end{align}
By defining variables
\begin{align}
 S_0 & := \rho_\mathrm{H} \sqrt{\gamma} , \label{eq:conservedE} \\
 S_i & := J_i \sqrt{\gamma} , \label{eq:conservedP}
\end{align}
these equations are explicitly written as \citep[see,
e.g.,][Sect.~4.3]{Shibata}
\begin{align}
 \partial_t S_i + \partial_j ( \alpha \sqrt{\gamma} S_i{}^j - \beta^j
 S_i ) & = - S_0 \partial_i \alpha + S_j \partial_i \beta^j -
 \frac{1}{2} \alpha \sqrt{\gamma} S_{jk} \partial_i \gamma^{jk} ,
 \label{eq:emcons_s} \\
 \partial_t S_0 + \partial_i ( \alpha S^i - \beta^i S_0 ) & = - S^i
 \partial_i \alpha + \alpha \sqrt{\gamma} S_{ij} K^{ij}
 . \label{eq:emcons_t}
\end{align}

In ideal hydrodynamics, Eqs.~\eqref{eq:emcons_s} and \eqref{eq:emcons_t}
give us relativistic Euler's equation and the energy equation,
respectively. We also solve the continuity equation,
Eq.~\eqref{eq:cont}. They constitute a system of evolution equations in
a conservative form (see, e.g., \citealt{Font2008} for reviews),
\begin{align}
 \partial_t \rho_* + \partial_i ( \rho_* v^i ) & = 0 ,
 \label{eq:evol_rho} \\
 \partial_t ( \rho_* \hat{u}_i ) + \partial_j ( \rho_* \hat{u}_i v^j + P
 \alpha \sqrt{\gamma} \delta_i{}^j ) & = P \partial_i ( \alpha
 \sqrt{\gamma} ) - \rho_* \pqty{hw \partial_i \alpha - \hat{u}_j
 \partial_i \beta^j + \frac{\alpha}{2hw} \hat{u}_j \hat{u}_k \partial_i
 \gamma^{jk} } \label{eq:evol_p} \\
 \partial_t ( \rho_* \hat{e} ) + \partial_i \bqty{\rho_* \hat{e} v^i + P
 \sqrt{\gamma} ( v^i + \beta^i )} & = P \alpha \sqrt{\gamma} K - \rho_*
 \hat{u}_i \gamma^{ij} \partial_j \alpha + \frac{\rho_* \alpha}{hw}
 \hat{u}_i \hat{u}_j K^{ij} \label{eq:evol_e} ,
\end{align}
where $v^i := u^i / u^t$, $w := \alpha u^t$, and the conserved variables
in ideal hydrodynamics are defined by
\begin{align}
 \rho_* & := \rho w \sqrt{\gamma} , \label{eq:defcd} \\
 \hat{u}_i & := \frac{S_i}{\rho_*} = hu_i , \label{eq:defcm} \\
 \hat{e} & := \frac{S_0}{\rho_*} = hw - \frac{P}{\rho w}
 . \label{eq:defce}
\end{align}
In relativistic hydrodynamics, we also need to determine the Lorentz
factor, $w$, to recover primitive variables such as $\rho$ and $u_i$
after solving for the conserved variables. This is accomplished by
solving the normalization condition of the four velocity, $u^\mu u_\mu =
-1$, together with the adopted equation of state and
Eqs.~\eqref{eq:defcd}, \eqref{eq:defcm}, and \eqref{eq:defce}.

These hydrodynamics equations in the conservative form are schematically
written in the form of
\begin{equation}
 \pdv{\vb{U}}{t} + \pdv{\vb{F}^i ( \vb{U} )}{x^i} = \vb{S} ( \vb{U} ) ,
\end{equation}
where $\vb{U}$ represents the set of the evolved variables $(\rho_* ,
S_i , S_0)$ and $\vb{F}^i$ are associated transport fluxes. The source
term, $\vb{S} ( \vb{U} )$, in ideal hydrodynamics comes from the
influence of gravity and consists of the metric and its first
derivatives. The source term seldom causes numerical instabilities and
can be evaluated in a straightforward manner as long as the size of the
time step is appropriate.

As is often the case in computational fluid dynamics (see, e.g.,
\citealt{Font2008,Marti_Muller2015,Balsara2017} for reviews), careful
treatment is required for numerically handling the transport terms. In
addition to the scheme for computing transport terms at a cell surface
from the evolved variables on each side, we also need a method of
reconstruction, i.e., interpolation of evolved variables to derive their
values at the cell surface.

For the case in which composition-dependent equations of state are
adopted, we also need to solve the evolution equation of the lepton
fractions, e.g., the electron fraction, $Y_\mathrm{e}$. As these
variables are tightly related to neutrino transport, we will discuss
their evolution equations in Appendix~\ref{app:sim_hydro_r}.

\subsubsection{Neutrino-radiation hydrodynamics} \label{app:sim_hydro_r}

One of the key features of radiation such as neutrinos is that they obey
kinetic theory. By assuming that the four momentum satisfies $p^\alpha
p_\alpha = \mathrm{const.}$, the distribution function $f (t,x^i,p^i)$
for a given species of neutrinos evolves according to Boltzmann's
equation \citep[see, e.g.,][]{Lindquist1966},
\begin{equation}
 p^\alpha \pdv{f}{x^\alpha} - \Gamma^i{}_{\alpha \beta} p^\alpha p^\beta
  \pdv{f}{p^i} = \mathcal{C} [f] ,
\end{equation}
where $\mathcal{C} [f]$ is the collision term. In typical numerical
simulations of compact binary coalescences, neutrinos are assumed to be
massless, i.e., $p^\alpha p_\alpha = 0$. For the collision term,
captures of electrons and positrons onto nucleons and nuclei,
electron-positron pair annihilation, nucleon Bremsstrahlung, and plasmon
decay are incorporated as emission processes
\citep{Deaton_DFOOKMSS2013,Foucart_DDOOHKPSS2014,Brege_DFDCHKOPS2018}. Some
simulations also incorporate captures of neutrinos onto nucleons as an
absorption process
\citep{Kyutoku_KSST2018,Fujibayashi_SWKKS2020,Fujibayashi_SWKKS2020-2}. Scattering
of neutrinos by nucleons and nuclei is also taken into account. Muon
neutrinos, muon antineutrinos, tau neutrinos, and tau antineutrinos are
usually treated collectively as ``x'' neutrinos in studies of compact
binary coalescences as well as supernova explosions. The reason for this
is that the matter temperature does not become as high as the mass of
muons and these neutrinos are produced and destroyed only via
neutral-current processes with the same rate.

Although it is desirable to solve Boltzmann's equation directly
\citep{Cardall_Endeve_Mezzacappa2013,Shibata_NSY2014}, this is
computationally extremely expensive because of the six-dimensional
nature of the phase space. A popular alternative in numerical
astrophysics is the truncated moment formalism, in which only the first
few moments of the distribution function are solved with imposing a
physically-motivated closure relation at some order \citep[see,
e.g.,][Sect.~6]{Mihalas_Mihalas}. A truncated moment formalism for
numerical relativity has been developed in \citet{Shibata_KKS2011} based
on \citet{Thorne1981}. Even if the dimension of the phase space is
reduced to four (three for spatial positions and one for the energy),
the time scale of weak interactions is much shorter than the dynamical
time scale in hot and dense regions such as the innermost region of the
accretion disk with $\rho \sim \SI{e12}{\gram\per\cubic\cm}$ and $kT
\sim \SI{10}{\mega\eV}$, and hence, a special treatment for solving
radiation-hydrodynamics and radiation-transfer equations are
necessary. This situation makes computational costs for radiation
hydrodynamics very high. To date, even the truncated moment formalism
has not been employed for studying black hole--neutron star binaries in
its full form.

All the neutrino-radiation-hydrodynamics simulations for black
hole--neutron star binaries have been performed solving the
energy-integrated moments of neutrinos combining some types of the
leakage scheme for handling neutrino interactions. For simplicity, we
describe equations neglecting frequency dependence in the following. We
also neglect neutrino absorption for a while and discuss it later. In
the leakage scheme, whose numerical-relativity version is first
formulated in \citet{Sekiguchi2010}, neutrinos are phenomenologically
separated into the trapped and streaming components (see, e.g.,
\citealt{Cooperstein1988} and references therein). The energy-momentum
tensor of the former and the latter is denoted by $T_{s
\mathrm{(T)}}^{\alpha \beta}$ and $T_{s \mathrm{(S)}}^{\alpha \beta}$,
respectively, for each species of neutrinos. Here, $s = \nu_e$,
$\bar{\nu}_e$, or $\nu_x$ denotes species of neutrinos. The trapped
neutrinos are assumed to couple tightly with the fluid, so that the
radiation hydrodynamics is fully applicable to the sum of these two
ingredients. The streaming neutrinos are assumed to stream out freely
from the system. It is also assumed that the four momentum of neutrinos
are transferred from the trapped component to the streaming component
with the leakage rate $Q_{s, \mathrm{leak}}^\alpha$ characterized not by
the time scale of weak interactions but by the diffusion time scale in
hot and dense environments. These assumptions lead to two sets of
equations to be solved,
\begin{align}
 \nabla_\beta \bqty{T_\mathrm{fluid}^{\alpha \beta} + \sum_s T_{s
 \mathrm{(T)}}^{\alpha \beta}} & = - \sum_s Q_{s,\mathrm{leak}}^\alpha ,
 \label{eq:leak_t} \\
 \nabla_\beta T_{s \mathrm{(S)}}^{\alpha \beta} & =
 Q_{s,\mathrm{leak}}^\alpha , \label{eq:leak_s}
\end{align}
where $T_\mathrm{fluid}^{\alpha \beta}$ is the energy-momentum tensor of
the fluid, e.g., Eq.~\eqref{eq:emtensor_ideal} for the ideal fluid.

The leakage rate, $Q_{s,\mathrm{leak}}^\alpha$, is determined
phenomenologically by interpolating the diffusion rate
$Q_{s,\mathrm{diff}}$ in the optically-thick region and the local
emission rate $Q_{s,\mathrm{local}}$ in the optically-thin region. The
latter is computed in a straightforward manner from microphysical
reaction rates. The former is determined from the optical depth of
neutrinos,
\begin{equation}
 \tau_s (r) := \int_r^\infty \bar{\kappa}_s (r') dr' ,
\end{equation}
with $\bar{\kappa}_s$ being the energy-averaged opacity, estimated by
combining the values along various paths on a constant-time slice. A
diffusion time scale is defined from $\tau_s$ as
\begin{equation}
 t_{s,\mathrm{diff}} := a_\mathrm{diff} \frac{\tau_s^2}{\bar{\kappa}_s}
  ,
\end{equation}
where $a_\mathrm{diff} \approx \order{1}$ is a free parameter, and then
the diffusion rate is estimated by (see \citealt{Sekiguchi2010} for a
detailed expression)
\begin{equation}
 Q_{s,\mathrm{diff}} \approx \frac{e_s}{t_{s,\mathrm{diff}}} ,
\end{equation}
with $e_s$ being the energy density of trapped neutrinos of species
$s$. Finally, the leakage rate is determined by
\begin{equation}
 Q_{s,\mathrm{leak}}^\alpha := u^\alpha \bqty{Q_{s,\mathrm{diff}} ( 1 -
  e^{-b \tau_s} ) + Q_{s,\mathrm{local}} e^{-b\tau_s}} ,
\end{equation}
where $b \approx \order{1}$ is also a free parameter. Although it has
been neglected in this explanation, neutrino absorption by nucleons may
also be incorporated in the leakage scheme as a heating term. A
prescription in the truncated moment formalism is described in
\citet{Fujibayashi_SKS2017}.

Equation \eqref{eq:leak_t} is solved in the same manner as
Eqs.~\eqref{eq:emcons_s} and \eqref{eq:emcons_t}, only by adding source
terms, $- \alpha \sqrt{\gamma} \gamma_{i\alpha}
Q_{s,\mathrm{leak}}^\alpha$ and $\alpha \sqrt{\gamma} n_\alpha
Q_{s,\mathrm{leak}}^\alpha$, respectively, to the right-hand sides. The
energy-momentum tensor of the streaming neutrinos may be decomposed in
two ways,
\begin{align}
 T_{s (S)}^{\alpha \beta} & = E_s n^\alpha n^\beta + F_s^\alpha n^\beta
 + F_s^\beta n^\alpha + P_s^{\alpha \beta} , \\
 & = J_s u^\alpha u^\beta + H_s^\alpha u^\beta + H_s^\beta u^\alpha +
 L_s^{\alpha \beta} ,
\end{align}
according to whether the decomposition is performed by the Eulerian
observer or in the rest frame of the fluid. The quantities in the former
expression are evolved in a conservative form as
\begin{align}
 \partial_t \mathcal{E}_s + \partial_i ( \alpha \mathcal{F}_s^i -
 \beta^i \mathcal{E}_s ) & = \alpha \mathcal{P}_s^{ij} K_{ij} -
 \mathcal{F}_s^i \partial_i \alpha - \alpha \sqrt{\gamma}
 Q_{s,\mathrm{leak}}^\mu n_\mu , \\
 \partial_t \mathcal{F}_{s,i} + \partial_j ( \alpha
 \mathcal{P}_{s,i}{}^j - \beta^j \mathcal{F}_{s,i} ) & = - \mathcal{E}_s
 \partial_i \alpha + \mathcal{F}_{s,j} \partial_i \beta^j +
 \frac{\alpha}{2} \mathcal{P}_s^{jk} \partial_i \gamma_{jk} + \alpha
 \sqrt{\gamma} Q_{s,\mathrm{leak}}^\mu \gamma_{\mu i} ,
\end{align}
where $\mathcal{E}_s := \sqrt{\gamma} E_s$, $\mathcal{F}_{s,i} :=
\sqrt{\gamma} F_{s,i}$, and $\mathcal{P}_{s,ij} := \sqrt{\gamma}
P_{s,ij}$.

In the truncated moment formalism, $P_{s,ij}$ is determined by a closure
relation derived by interpolating the form in the optically-thick and
optically-thin limits. Specifically, \citet{Shibata_KKS2011} proposed an
expression in terms of the so-called variable Eddington factor
\citep{Levermore1984},
\begin{align}
 \chi_{s,\mathrm{E}} & := \frac{3+4F_{s,\mathrm{E}}^2}{5+2\sqrt{4 -
 3F_{s,\mathrm{E}}^2}} , \\
 F_{s,\mathrm{E}} & := \frac{( g_{\alpha \beta} + u_\alpha u_\beta )
 H_s^\alpha H_s^\beta}{J_s^2} ,
\end{align}
given by
\begin{align}
 P_s^{ij} & = \frac{3\chi_{s,\mathrm{E}} - 1}{2}
 P_{s,\mathrm{thin}}^{ij} + \frac{3( 1 - \chi_{s,\mathrm{E}} )}{2}
 P_{s,\mathrm{thick}}^{ij} , \\
 P_{s,\mathrm{thick}}^{ij} & = \frac{E_s}{2w^2 + 1} [ (2w^2 - 1)
 \gamma^{ij} - 4V^i V^j] + \frac{1}{w} ( F_s^i V^j + F_s^j V^i ) +
 \frac{2F_s^k u_k}{w(2w^2 + 1)} ( - w^2 \gamma^{ij} + V^i V^j ) , \\
 P_{s,\mathrm{thin}}^{ij} & = E_s \frac{F_s^i F_s^j}{\gamma_{kl} F_s^k
 F_s^l} ,
\end{align}
where $V^i := \gamma^{ij} u_j$.

To handle the neutrino reactions, we also need to solve the chemical
composition of the matter in an appropriate manner. The electron
fraction evolves according to
\begin{equation}
 u^\alpha \nabla_\alpha Y_\mathrm{e} = \gamma_\mathrm{e} ,
\end{equation}
where $\gamma_\mathrm{e}$ is the net production rate (i.e., the creation
rate subtracted for the destruction rate) of electrons in the rest frame
of the fluid. This equation is rewritten to a conservative form by using
the continuity equation as
\begin{equation}
 \partial_t ( \rho_* Y_\mathrm{e} ) + \partial_i ( \rho_* Y_\mathrm{e}
  v^i ) = \frac{\rho_*}{u^t} \gamma_\mathrm{e} .
\end{equation}
After solving this evolution equation, $Y_\mathrm{e}$ is recovered
directly by $\rho_* Y_\mathrm{e} / \rho_*$. The fractions of trapped
neutrinos of each species obey similar evolution equations, and their
production rates are evaluated in the leakage scheme as (see
\citealt{Sekiguchi2010} for details)
\begin{equation}
 \gamma^\mathrm{leak}_s = ( 1 - e^{-b \tau_s} )
  \gamma^\mathrm{diff}_s + e^{-b \tau_s} \gamma^\mathrm{local}_s
\end{equation}
for each species $s$. We do not need to solve the fraction of streaming
neutrinos, because they are decoupled from the fluid.

We note that the leakage scheme can take a various form. Actually,
simulations performed with SpEC
\citep[e.g.,][]{Deaton_DFOOKMSS2013,Foucart_DDOOHKPSS2014,Brege_DFDCHKOPS2018}
adopted a scheme developed in \citet{OConnor_Ott2010}, which closely
resembles preceding formulation in Newtonian simulations
\citep{Ruffert_Janka_Schafer1996,Rosswog_Liebendorfer2003}. \citet{Most_PTR2021,Most_PTR2021-2}
also take a similar approach. Accuracy of results derived by the leakage
scheme should be checked via comparisons with simulations performed with
a sophisticated neutrino-transport scheme \citep{Foucart_ORKPS2016}.

\subsubsection{Magnetohydrodynamics} \label{app:sim_hydro_m}

Assuming infinite conductivity, we often employ ideal
magnetohydrodynamics in general relativity for exploring black
hole--neutron star binary coalescences \citep[see also
\citealt{Palenzuela_LRR2009,Palenzuela2013,Dionysopoulou_APRG2013,Shibata_Fujibayashi_Sekiguchi2021}
for resistive
magnetohydrodynamics]{Baumgarte_Shapiro2003,Shibata_Sekiguchi2005}. Indeed,
electrical conductivity is fairly high in most of the region encountered
in compact binary coalescences. In the limit of infinite conductivity,
the finiteness of the electric current requires the electric field to
vanish in the rest frame of the fluid via Ohm's law. As a result, the
magnetic field becomes the only variable to be evolved in ideal
magnetohydrodynamics.

Evolution of the magnetic field is governed by the spatial components of
the source-free half of Maxwell's equations, or the induction equation,
\begin{equation}
 \gamma^i{}_\mu \epsilon^{\mu \nu \alpha \beta} \nabla_\nu F_{\alpha
  \beta} = 0 ,
\end{equation}
where $\epsilon^{\mu \nu \alpha \beta}$ is the Levi-Civita tensor of the
spacetime and $F_{\alpha \beta}$ is the Faraday tensor. Specifically,
this can be written in the conservative form as
\begin{equation}
 \partial_t \mathcal{B}^i + \partial_j ( \mathcal{B}^i v^j -
  \mathcal{B}^j v^i ) = 0 ,
\end{equation}
where
\begin{equation}
 B^\mu := \frac{1}{2} n_\nu \epsilon^{\nu \mu \alpha \beta} F_{\alpha
  \beta} ,
\end{equation}
is the magnetic field measured by an Eulerian observer\footnote{Although
this symbol overlaps with a variable in the moving-puncture gauge
condition, we expect that no confusion arises.} and $\mathcal{B}^i :=
\sqrt{\gamma} B^i$. We do not need to solve the charge continuity to
determine the electric current, which can instead be determined by
Amp{\'e}re-Maxwell's law from the magnetic field. The energy-momentum
tensor acquires a contribution from the electromagnetic field,
\begin{equation}
 T_\mathrm{EM}^{\mu \nu} = \frac{1}{4\pi} \bqty{b^\alpha b_\alpha
  \pqty{u^\mu u^\nu + \frac{1}{2} g^{\mu \nu}} - b^\mu b^\nu} ,
\end{equation}
where
\begin{align}
 b^\mu & := \frac{1}{2} u_\nu \epsilon^{\nu \mu \alpha \beta} F_{\alpha
 \beta} \notag \\
 & = \frac{1}{w} \pqty{\delta^\mu{}_\alpha + u^\mu u_\alpha} B^\alpha
\end{align}
is the magnetic field measured in the rest frame of the fluid. They
modify the matter terms of the Einstein equation, $\rho_\mathrm{H}$,
$J_i$, and $S_{ij}$, as well as variables in the local energy-momentum
conservation, Eqs.~\eqref{eq:conservedE} and \eqref{eq:conservedP}. The
electric field measured by the Eulerian observer is determined by the
magnetic field as
\begin{align}
 E^\mu & = - n_\alpha \epsilon^{\alpha \mu \nu \alpha \beta} u_\nu
 b_\beta \notag \\
 & = - \frac{1}{w} n_\alpha \epsilon^{\alpha \mu \nu \beta} u_\nu
 B_\beta .
\end{align}

It is a vital requirement for numerical magnetohydrodynamics to ensure
the divergenceless condition of the magnetic field, $\partial_i
\mathcal{B}^i = 0$. This condition is derived by the temporal component
of the source-free half of Maxwell's equations. Here, we should
emphasize that satisfying this condition is merely a necessary condition
in numerical magnetohydrodynamics, because $\partial_i \mathcal{B}^i =
0$ is satisfied even if, e.g., the magnetic field is spuriously
magnified by a constant factor. Another condition to be satisfied in
numerical ideal magnetohydrodynamics is the conservation of the magnetic
flux. One robust method to achieve this goal is to employ the so-called
constrained transport scheme
\citep{Evans_Hawley1988,Balsara_Spicer1999}. This scheme enables the
divergenceless condition and the magnetic-flux conservation to be
preserved within the machine precision by carefully locating the
electromagnetic fields measured by an Eulerian observer on the discrete
grids. This scheme is also compatible with the mesh refinement algorithm
\citep[see Appendix~\ref{app:sim_mr} for the mesh
refinement]{Balsara2001,Balsara2009}, and \citet{Kiuchi_SKSTW2015}
employed this method for simulating black hole--neutron star binary
coalescences using the code developed in
\citet{Kiuchi_Kyutoku_Shibata2012}. Another simple method to keep the
magnetic field divergenceless is to evolve the vector potential $A^\mu$
instead of the magnetic field by rewriting the induction
equation. Simulations of black hole--neutron star binaries with the mesh
refinement are demonstrated to be feasible with this method if the
Lorenz gauge condition, $\nabla_\mu A^\mu = 0$ \citep{Etienne_PLS2012},
or its generalization, $\nabla_\mu A^\mu = \xi n_\mu A^\mu$ with $\xi
\sim 1/m_0$ being a damping parameter \citep{Farris_GPES2012}, is
adopted. It should be cautioned that, however, the magnetic-flux
conservation is not guaranteed to hold in this scheme.

Because no realistic initial data with magnetic fields have been
constructed for black hole--neutron star binaries, dynamical simulations
are performed by artificially superposing the magnetic field at the
beginning or in the early stage of the evolution. A typical choice is
confined toroidal magnetic fields given by the vector potential of the
form \citep{Shibata_DLSS2006}
\begin{equation}
 A_\varphi = A_b \varpi_\mathrm{NS}^2 [ \max ( P - P_\mathrm{crit} , 0 )
  ]^2 ,
\end{equation}
where $A_b$ is a constant specifying the magnetic-field strength,
$\varpi_\mathrm{NS}^2 := (x - x_\mathrm{NS})^2 + (y - y_\mathrm{NS})^2$,
$x_\mathrm{NS}$ and $y_\mathrm{NS}$ are the coordinates of the stellar
center, and $P_\mathrm{crit}$ is critical pressure chosen to be, e.g.,
4\% of the maximum value in the system \citep{Kiuchi_SKSTW2015}. Another
choice is pulsar-like dipolar magnetic fields modeled by
\citep{Paschalidis_Etienne_Shapiro2013}
\begin{equation}
 A_\varphi = \frac{\pi r_0^2 I_0 \varpi_\mathrm{NS}^2}{( r_\mathrm{NS}^2
  + r_0^2 )^{3/2}} \bqty{1 + \frac{15 r_0^2 ( \varpi_\mathrm{NS}^2 +
  r_0^2 )}{8 ( r_\mathrm{NS}^2 + r_0^2 )^2}} ,
\end{equation}
where constant parameters $r_0$ and $I_0$ denote the current-loop radius
and the loop current, respectively, and $r_\mathrm{NS} := \abs{x^i -
x_\mathrm{NS}^i}$. If the adopted magnetic field extends outside the
neutron star like this model, a non-tenuous artificial atmosphere is
required for stable evolution. The reason for this is that the
determination of the Lorentz factor, $w$, becomes highly inaccurate and
sometimes fails to find physical solutions in regions where the energy
and momentum are dominated by the magnetic field.

Accurate magnetohydrodynamics simulations for the postmerger remnant of
black hole--neutron star binaries such as the black hole--accretion disk
system require high computational costs. The reason for this is that the
accretion disk is likely to be unstable to the magnetorotational
instability, for which the wavelength of the fastest growing mode is as
short as \citep{Balbus_Hawley1998}
\begin{equation}
 \lambda_\mathrm{MRI} \approx \frac{B}{\sqrt{4\pi \rho}}
  \frac{2\pi}{\Omega} \approx \SI{18}{\meter}
  \pqty{\frac{B}{\SI{e12}{G}}}
  \pqty{\frac{\rho}{\SI{e12}{\gram\per\cubic\cm}}}^{-1/2}
  \pqty{\frac{\Omega}{\SI{e3}{\per\second}}}^{-1} ,
\end{equation}
where $B$ is the strength of magnetic fields. This strongly suggests
that current global, longterm magnetohydrodynamics simulations cannot
resolve magnetic-field amplification due to this instability if the
initial strength is taken to be a plausibly realistic value of $\sim
\SI{e12}{G}$ for neutron stars before merger. To make the matter worse,
magnetohydrodynamical turbulence cannot be handled accurately in
two-dimensional, axisymmetric simulations due to the antidynamo theorem
\citep{Cowling1933}. Moreover, we need to perform longterm simulations
for the entire evolution of the accretion disks, because they evolve
primarily via the viscous effect resulting from the
magnetohydrodynamical turbulence driven by the magnetorotational
instability [see Eq.~\eqref{eq:tvis}]. One method to overcome these
issues is to choose initial magnetic fields to be as strong as $\sim
\num{e16}$--\SI{e17}{G}. To incorporate effects of magnetic fields on
the evolution of postmerger remnants in an approximate manner with a
reasonable computational cost, viscous hydrodynamics discussed below
serves as a phenomenological alternative for the postmerger evolution
\citep{Fujibayashi_SWKKS2020,Fujibayashi_SWKKS2020-2}.

\subsubsection{Viscous hydrodynamics} \label{app:sim_hydro_v}

In contrast to ideal hydrodynamics, nonrelativistic viscous
hydrodynamics cannot be extended to relativistic cases in a
straightforward manner. The reason is that the Navier-Stokes equation
for viscous hydrodynamics is a diffusion equation for the velocity field
and does not respect causality. Practically, inappropriate formulation
of relativistic viscous hydrodynamics is known to result in extremely
rapid instabilities \citep{Hiscock_Lindblom1985}.

\citet{Shibata_Kiuchi_Sekiguchi2017} developed a scheme for relativistic
viscous hydrodynamics by adopting a simplified version of theory of
\citet{Israel_Stewart1979}, in which the viscosity respects the
causality. The energy-momentum tensor of the fluid is extended by
incorporating a viscous stress tensor $\tau^0_{\mu \nu}$ that satisfies
$\tau^0_{\mu \nu} u^\nu = 0$ as
\begin{equation}
 T_{\mu \nu} = \rho h u_\mu u_\nu + P g_{\mu \nu} - \rho h \nu
  \tau^0_{\mu \nu} ,
\end{equation}
where $\nu$ is the shear kinematic viscosity. The viscous stress tensor
evolves as \citep{Israel_Stewart1979,Hiscock_Lindblom1983}
\begin{equation}
 \mathcal{L}_u \tau^0_{\mu \nu} = - \zeta ( \tau^0_{\mu \nu} -
  \sigma_{\mu \nu} ) ,
\end{equation}
where $\zeta$ is a constant. \citet{Shibata_Kiuchi_Sekiguchi2017} define
the relativistic shear tensor $\sigma_{\mu \nu}$ by
\begin{align}
 \sigma_{\mu \nu} & := ( \delta_\mu{}^\alpha + u_\mu u^\alpha ) (
 \delta_\nu{}^\beta + u_\nu u^\beta ) ( \nabla_\alpha u_\beta +
 \nabla_\beta u_\alpha ) \notag \\
 & = \mathcal{L}_u ( g_{\mu \nu} + u_\mu u_\nu ) .
\end{align}
The value of $\zeta$ is chosen phenomenologically to be moderately
larger than the inverse of the dynamical time scale of the system,
rather than determined from microphysical considerations. This choice
allows $\tau^0_{\mu \nu}$ to approach $\sigma_{\mu \nu}$ in a
sufficiently short time. The value of $\zeta$ also needs to be much
smaller than the inverse of time steps in numerical simulations to avoid
a very stiff evolution equation. This evolution equation may be
rewritten in a form of
\begin{align}
 \mathcal{L}_u \tau_{\mu \nu} & = - \zeta \tau^0_{\mu \nu} , \\
 \tau_{\mu \nu} & := \tau^0_{\mu \nu} - \zeta ( g_{\mu \nu} + u_\mu
 u_\nu ) ,
\end{align}
to derive the equation in a conservative form for practical
computations,
\begin{equation}
 \partial_t ( \rho_* \tau_{ij} ) + \partial_k ( \rho_* \tau_{ij} v^k ) =
  - \rho_* ( \tau_{ik} \partial_j v^k + \tau_{jk} \partial_i v^k ) -
  \zeta \rho_* \frac{\alpha}{w}\tau^0_{ij} ,
\end{equation}
which contains only the first derivative of the velocity. In this
scheme, the source terms for the Einstein equation, $\rho_\mathrm{H}$,
$J_i$, and $S_{ij}$, contain the contribution from $\tau^0_{\mu \nu}$,
and other hydrodynamics equations are modified accordingly.

In the simulation of an accretion disk, the shear kinematic viscosity
$\nu$ is typically given according to the prescription of
\citet{Shakura_Sunyaev1973} as
\begin{equation}
 \nu = \alpha_\nu c_s H ,
\end{equation}
where $\alpha_\nu \approx 0.01$--$0.1$ is a free parameter
characterizing the strength of viscosity, $c_s$ is the sound speed, and
$H$ is another parameter denoting a typical scale height of the system,
which governs the size of the largest turbulent eddy.

\subsection{Fixed and adaptive mesh refinement} \label{app:sim_mr}

Compact binary coalescences are characterized by three length scales.
The shortest is the size of compact objects, $\sim GM_\mathrm{BH} / c^2$
for black holes and $R_\mathrm{NS} \sim 3$--$9 GM_\mathrm{NS} / c^2$ for
neutron stars, which may be collectively approximated by $Gm_0 /
c^2$. The middle is the orbital separation, $d$, over which compact
objects move throughout the space. This is approximately given in terms
of the orbital angular velocity $\Omega$ by
\begin{equation}
 d \approx \pqty{\frac{G m_0}{\Omega^2}}^{-1/3} = 10 \frac{G m_0}{c^2}
  \pqty{\frac{G m_0 \Omega / c^3}{0.03}}^{-2/3} .
\end{equation}
The longest is the wavelength of gravitational waves, $\lambda$. For a
binary in a quasicircular orbit with the orbital angular velocity
$\Omega$, the dominant quadrupole mode has
\begin{equation}
 \lambda \approx \frac{\pi c}{\Omega} = 105 \frac{Gm_0}{c^2}
  \pqty{\frac{G m_0 \Omega / c^3}{0.03}}^{-1} .
\end{equation}
Choices of the initial orbital angular velocity in numerical simulations
depend on the parameters of a binary and the purpose of the
simulations. Typical initial values for simulations of black
hole--neutron star binaries are $G m_0 \Omega / c^3 \sim 0.03$. Thus,
the longest wavelength $\lambda$ is larger by a factor of $\gtrsim 100$
than the length scale of compact objects, $Gm_0 / c^2$.

This hierarchy in the length scales makes computations with a uniform
grid extremely inefficient. First, since structures of each compact
object need to be finely resolved for accurate evolution, a grid spacing
has to be smaller than $\Delta_0 \sim \min [ GM_\mathrm{BH} /(40 c^2) ,
R_\mathrm{NS} / 100]$, where the numerical factors are taken as
representative choices. Second, the distance to the outer boundary along
each coordinate axis needs to be larger than the wavelength of
gravitational waves for accurate incorporation of radiation reaction and
accurate extraction of gravitational waves in the wave zone. If a
uniform grid with the spacing $\Delta_0$ is adopted, the total number of
grid points in one direction should be larger than
\begin{align}
 \frac{\lambda}{\Delta_0} \approx \max \bqty{4200 \pqty{1 + Q^{-1}}
 \pqty{\frac{G m_0 \Omega / c^3}{0.03}}^{-1} , 1750 (1+Q) \pqty{\frac{G
 M_\mathrm{NS} / ( c^2 R_\mathrm{NS} )}{1/6}} \pqty{\frac{G m_0 \Omega /
 c^3}{0.03}}^{-1}} .
\end{align}
In a three-dimensional space without any symmetry, the number of
required grid points is larger than $( 2 \lambda / \Delta_0 )^3 \gtrsim
10^{10}$. This is a prohibitively large number with current
computational resources, and systematic surveys over the parameter space
for black hole--neutron star binaries are far from feasible with a
uniform grid. Early simulations of black hole--neutron star binaries
have been performed with a nonuniform grid
\citep{Shibata_Uryu2006,Shibata_Uryu2007,Etienne_FLSTB2008,Shibata_Taniguchi2008}. However,
this is not an efficient way for the simulations in three spatial
dimensions.

A mesh refinement technique is indispensable for performing simulations
with finite differencing methods in an efficient manner. In computations
with a mesh refinement, we prepare several refinement levels of
Cartesian boxes (the geometry can be generalized) with different grid
resolutions; the smaller boxes usually have higher resolutions, and vice
versa. The black hole and the neutron star are resolved at the finest
refinement level with a sufficiently small grid spacing of $\lesssim
\Delta_0$. At the same time, propagation of gravitational waves is
followed in coarse but large boxes. While the grid spacings at large
boxes are much larger than $\Delta_0$, the wavelength of gravitational
radiation may still be covered by more than, e.g., $100$ grid
points. The mesh refinement with the grid structure fixed throughout the
simulation is called the fixed mesh refinement \citep[FMR; see,
e.g.,][]{Imbiriba_BCCFBVO2004,Schnetter_Hawley_Hawke2004}. The FMR is
adopted in simulations of black hole--neutron star binary coalescences
which aim at studying the merger and postmerger phases
\citep{Kiuchi_SKSTW2015,Kyutoku_KSST2018}. Still, simulations with the
FMR technique need to resolve the length scale of the orbital separation
$d$ by $\Delta_0$. Thus, the FMR is not satisfactory for performing
longterm simulations of the inspiral phase in a systematic manner.

Recent longterm simulations of compact binary coalescences with finite
differencing are frequently performed by employing an adaptive mesh
refinement (AMR) technique \citep[see,
e.g.,][]{Anderson_HLN2006,Brugmann_GHHST2008,Yamamoto_Shibata_Taniguchi2008}. The
AMR technique improves the efficiency of the FMR by allowing the grid
structure to change during the evolution so that the small grid spacing
is assigned only to the region requiring high resolution \citep[see,
e.g.,][]{Berger_Oliger1984}. For the specific application to compact
binary coalescences, boxes at the fine levels are allowed to move
following the orbital motion of each binary component. SpEC solves the
gravitational field based on the generalized harmonic formalism with
multidomain spectral methods, and the number of basis functions can be
chosen adaptively by the spectral AMR technique
\citep{Lovelace_Scheel_Szilagyi2011,Szilagyi2014}.

The mesh refinement is important not only for resolving gravitational
fields but also for matter fields. While the structure of a neutron star
has to be resolved during the inspiral phase, properties of the ejected
material can be analyzed accurately only in a distant region after the
motion of the ejecta is settled to homologous
expansion. Numerical-relativity codes based on the puncture-BSSN
formalism, such as SACRA, typically adopt the same AMR technique as that
for gravitational fields to evolve matter fields. In SpEC, hydrodynamics
equations are solved on another grid based on finite volume methods
\citep{Duez_FKPST2008}, and an AMR technique is implemented separately
from spectral grids for the gravitational fields
\citep{Foucart_DBDKHKPS2017}.

\section{Analytic estimate} \label{app:ae}

In this section, we present analytic estimates for stable mass transfer
and opening angles of the dynamical ejecta in Appendix~\ref{app:ae_smt}
and Appendix~\ref{app:ae_ej}, respectively.

\subsection{Stable mass transfer} \label{app:ae_smt}

Whether the stable mass transfer is possible in black hole--neutron star
binaries can be examined by analyzing conservation of the angular
momentum. In this Appendix~\ref{app:ae_smt}, we employ the Newtonian
equation of motion together with gravitational radiation reaction as the
only general-relativistic effect \citep{Cameron_Iben1986,Benz_BCP1990}
and approximately show that the stable mass transfer is not very likely
to occur in black hole--neutron star binaries as long as the
neutron-star radius is not very large.

Assuming that the point-particle approximation holds, the orbital
angular momentum of the binary $J_\mathrm{orb}$ with an orbital
separation $r$ is given by
\begin{equation}
 J_\mathrm{orb} = G M_\mathrm{BH} M_\mathrm{NS} \sqrt{\frac{r}{Gm_0}}
  . \label{eq:jorb}
\end{equation}
In the following, we suppose $M_\mathrm{BH} > M_\mathrm{NS}$ and thus $Q
= M_\mathrm{BH} / M_\mathrm{NS} > 1$. Equation~\eqref{eq:jorb} derives
\begin{equation}
 \frac{\dot{J}_\mathrm{orb}}{J_\mathrm{orb}} =
  \frac{\dot{M}_\mathrm{BH}}{M_\mathrm{BH}} +
  \frac{\dot{M}_\mathrm{NS}}{M_\mathrm{NS}} - \frac{\dot{m}_0}{2m_0} +
  \frac{\dot{r}}{2r} , \label{eq:smt_rel}
\end{equation}
where the overdot denotes the time derivative. The mass transfer occurs
only when the orbital separation is small enough to induce mass
shedding. Thus, we focus on close orbits with $r \lesssim
10Gm_0/c^2$. Because mass ejection is unlikely to occur during the phase
of mass transfer in such close orbits, we assume that the total mass is
conserved, i.e., $\dot{m}_0 = 0$, in the following. This implies that
$\dot{M}_\mathrm{BH} = -\dot{M}_\mathrm{NS} > 0$. If the mass is ejected
from the system, the angular momentum is removed, and hence, the stable
mass transfer is less likely to occur.

In close orbits with $r \lesssim 10Gm_0/c^2$, the orbital angular
velocity is so high that the gravitational radiation reaction plays a
central role in determining the orbital evolution. After the onset of
mass transfer, the spin-up of the black hole and the formation of a disk
around it are also important. That is, the orbital angular momentum is
transferred to the spin angular momentum of the black hole or disk, the
sum of which are denoted by $S$. The orbital angular momentum of the
binary evolves according to
\begin{equation}
 \dot{J}_\mathrm{orb} = - \dot{S} - \dot{J}_\mathrm{GW} ,
  \label{eq:smt_evol}
\end{equation}
where $\dot{J}_\mathrm{GW}$ is the dissipation rate of the angular
momentum due to the gravitational radiation reaction, which is written
in the quadrupole approximation as \citep{Peters1964}
\begin{equation}
 \dot{J}_\mathrm{GW} = \frac{32}{5} \frac{G^{7/2} m_0^{1/2}
  M_\mathrm{BH}^2 M_\mathrm{NS}^2}{c^5 r^{7/2}} . \label{eq:smt_gw}
\end{equation}
Because $\dot{S}$ should be proportional to $\dot{M}_\mathrm{NS}$, we
may express it as
\begin{equation}
 \dot{S} = \epsilon_\mathrm{mt} \dot{M}_\mathrm{NS} \sqrt{G
  M_\mathrm{BH} R_\mathrm{mt}} , \label{eq:smt_bh}
\end{equation}
where $0 \le \epsilon_\mathrm{mt} \le 1$ represents efficiency of the
spin-up of the black hole and the surrounding disk, and $R_\mathrm{mt}$
denotes an average radius of the material orbiting the black hole. The
remaining fraction $1 - \epsilon_\mathrm{mt}$ of the accreted angular
momentum is assumed to add to the orbital angular momentum of the black
hole. For the material swallowed by the black hole, $R_\mathrm{mt}$
should reflect the specific angular momentum at the innermost stable
circular orbit around the black hole. Hence, we suppose $R_\mathrm{mt}
\ge GM_\mathrm{BH}/c^2$, recalling that the specific angular momentum at
the innermost stable circular orbit is larger than
$GM_\mathrm{BH}/c^2$. In the following, we use a dimensionless quantity
$\hat{R}_\mathrm{mt} := R_\mathrm{mt} / (GM_\mathrm{BH}/c^2)$.

Substituting Eqs.~\eqref{eq:smt_gw} and \eqref{eq:smt_bh} into
Eq.~\eqref{eq:smt_evol}, we obtain
\begin{equation}
 \frac{\dot{J}_\mathrm{orb}}{J_\mathrm{orb}} = - \epsilon_\mathrm{mt}
  \frac{\dot{M}_\mathrm{NS}}{M_\mathrm{NS}} \sqrt{\frac{G m_0}{c^2 r}
  \hat{R}_\mathrm{mt}} - \frac{32}{5} \frac{G^3 m_0 M_\mathrm{BH}
  M_\mathrm{NS}}{c^5 r^4} ,
\end{equation}
and thus Eq.~\eqref{eq:smt_rel} is rewritten to
\begin{equation}
 \frac{\dot{r}}{2r} = \abs{\frac{\dot{M}_\mathrm{NS}}{M_\mathrm{NS}}}
  \pqty{1 - Q^{-1} - \epsilon_\mathrm{mt} \sqrt{\frac{Gm_0}{c^2 r}
  \hat{R}_\mathrm{mt}}} - \frac{32}{5} \frac{G^3 m_0 M_\mathrm{BH}
  M_\mathrm{NS}}{c^5 r^4} .
\end{equation}
Multiplying the orbital period $P = 2 \pi \sqrt{r^3/(Gm_0)}$, we obtain
\begin{equation}
 P \frac{\dot{r}}{2r} = \abs{\frac{\Delta M_\mathrm{NS}}{M_\mathrm{NS}}}
  \pqty{1 - Q^{-1} - \epsilon_\mathrm{mt} \sqrt{\frac{Gm_0}{c^2 r}
  \hat{R}_\mathrm{mt}}} - \frac{64\pi}{5} \frac{G^{5/2} m_0^{1/2}
  M_\mathrm{BH} M_\mathrm{NS}}{c^5 r^{5/2}} ,
\end{equation}
where $\Delta M_\mathrm{NS} := P \dot{M}_\mathrm{NS}$ denotes the mass
of the neutron star lost in a single orbital period.

For the stable mass transfer to continue longer than a couple of orbits,
$\Delta M_\mathrm{NS} / M_\mathrm{NS}$ should be smaller than $\sim
0.1$. Otherwise, tidal disruption occurs in a couple of orbits, because
the radius of the neutron star is increased significantly for
significant mass loss. As already shown in
Sect.~\ref{sec:intro_tidal_ms}, the mass shedding can set in only for
close orbits with $r \lesssim (2Q)^{1/3} c_\mathrm{R} R_\mathrm{NS}$
[see Eq.~\eqref{eq:msrad}]. This condition derives
\begin{align}
 \frac{64\pi}{5} \frac{G^{1/2} m_0^{1/2} M_\mathrm{BH}
 M_\mathrm{NS}}{c^5 r^{5/2}} & \gtrsim \frac{64\pi}{5} \frac{(1+Q)^{1/2}
 Q^{1/6}}{2^{5/6}} \pqty{\frac{\mathcal{C}}{c_\mathrm{R}}}^{5/2}
 \notag \\
 & \approx 0.12 \pqty{1 + \frac{Q}{3}}^{1/2} \pqty{\frac{Q}{3}}^{1/6}
 \pqty{\frac{c_\mathrm{R}}{1.9}}^{-5/2}
 \pqty{\frac{\mathcal{C}}{1/6}}^{5/2} ,
\end{align}
and thus, for a plausible value of $c_\mathrm{R} \approx 1.9$ (see
Sect.~\ref{sec:eq_end_ms}) and $\mathcal{C} = GM_\mathrm{NS} / (c^2
R_\mathrm{NS}) \ge 0.145$, we obtain
\begin{equation}
 \frac{64}{5} \frac{G^{5/2} m_0^{1/2} M_\mathrm{BH} M_\mathrm{NS}}{c^5
  r^{5/2}} \pqty{1 - Q^{-1}}^{-1} \gtrsim 0.13
\end{equation}
irrespective of the value of $Q$. Here, the minimum of this function
occurs at $Q = 2.5$--$3.5$ depending on the compactness of the neutron
star, $\mathcal{C}$. Therefore, as long as the compactness is in a
plausible range of $\mathcal{C} \gtrsim 0.145$, $\dot{r}$ could be
positive only for the case in which more than $\sim 15\%$ of the
neutron-star material are stripped toward the black hole in a short time
scale and that the efficiency of the spin-up, $\epsilon_\mathrm{mt}$, is
much smaller than unity. However, for the case in which the value of
$\epsilon_\mathrm{mt}$ is appreciable, say a realistic value of $0.5$,
with $\hat{R}_\mathrm{mt} \geq 1.5$, we have
\begin{equation}
 \frac{64\pi}{5} \frac{G^{5/2} m_0^{1/2} M_\mathrm{BH}
  M_\mathrm{NS}}{c^5 r^{5/2}} \pqty{1 - Q^{-1} - \epsilon_\mathrm{mt}
  \sqrt{\frac{Gm_0}{c^2 r} \hat{R}_\mathrm{mt}}}^{-1} \gtrsim 0.18
\end{equation}
(note that $\dot{r}>0$ is realized only if the sum of the terms in the
parenthesis is positive). Thus, the condition for $\dot{r} >0$ is
restricted further. This analysis indicates that the orbital separation
can increase only if a substantial amount of the neutron-star material
is stripped by the tidal force of the black hole. Hence, a steady and
gradual increase of the orbital separation is unlikely to occur for
black hole--neutron star binaries.

The values of $\epsilon_\mathrm{mt}$ and $\hat{R}_\mathrm{mt}$ are
nontrivial and can be determined only by numerical-relativity
simulations. Many numerical-relativity simulations with small values of
$Q$ have been done in the last decade. However, no simulation has shown
the evidence of the stable mass transfer. Thus, we may safely state that
the stable mass transfer is unlikely to occur in black hole--neutron
star binaries, if we suppose neutron stars with a typical mass in our
Galaxy of $M_\mathrm{NS} \approx 1.4\,M_\odot$ and $R_\mathrm{NS} \approx
10$--\SI{14}{\km}.

\subsection{Opening angle of the dynamical ejecta} \label{app:ae_ej}

As we discussed in Sect.~\ref{sec:dis_impl}, the nonspherical morphology
of the dynamical ejecta may be characterized by the opening angle in the
equatorial plane $\varphi$ and that in the meridional plane $\theta$,
where the latter is defined to refer only to the material above or below
the orbital plane. In the following, we reproduce analytic estimation of
these quantities by \citet{Kyutoku_IOST2015} to demonstrate weak
dependence on hypothetical equations of state.

Allowing more than one revolution, the opening angle of the dynamical
ejecta in the equatorial plane should be given by
\begin{equation}
 \varphi \approx 2 \pi \frac{t_\mathrm{td}}{P_\mathrm{td}} ,
\end{equation}
where $t_\mathrm{td}$ is the time scale of tidal disruption and
$P_\mathrm{td}$ is the orbital period at the tidal-disruption radius,
$r_\mathrm{td} \propto r_\mathrm{ms}$. First, $t_\mathrm{td}$ may be
given approximately by the sound crossing time $t_\mathrm{sc}$, which is
comparable to the dynamical time scale for a stellar configuration, of
the neutron star as
\begin{equation}
 t_\mathrm{td} \approx t_\mathrm{sc} \propto \frac{1}{\sqrt{\bar{\rho}}}
  \; ,
\end{equation}
where $\bar{\rho}$ is the average rest-mass density of the neutron star,
which is determined by the equation of state. Next, $P_\mathrm{td}$
should approximately be given by
\begin{equation}
 P_\mathrm{td} \approx 2 \pi \sqrt{\frac{r_\mathrm{td}^3}{Gm_0}} \propto
  \sqrt{\frac{Q}{(1+Q) G \bar{\rho}}} \; ,
\end{equation}
where Eq.~\eqref{eq:msrad} was used to derive the last expression. This
suggests that the dependence of $\varphi$ on the equation of state is
weak, because $\varphi$ is independent of $\bar{\rho}$. This expression
also suggests that $\varphi$ is smaller for a larger mass ratio, but the
expected variation is less than 10\% for $3 \le Q \le 7$. Prograde
black-hole spins will decrease $\varphi$, because the orbital frequency
of circular geodesic motion around a Kerr black hole is given by
\citep{Bardeen_Press_Teukolsky1972}
\begin{equation}
 \Omega_\mathrm{K} = \frac{(GM_\mathrm{BH})^{1/2}}{r^{3/2} + \chi
  (GM_\mathrm{BH}/c^2)^{3/2}} ,
\end{equation}
and thus $P_\mathrm{td}$ increases as $\chi$ increases.

The opening angle of the dynamical ejecta in the meridional plane,
$\theta$, is determined by the ratio of the velocity perpendicular to
the orbital plane, $v_\perp$, to that in the equatorial direction,
$v_\parallel$, as $\theta \approx \arctan ( v_\perp / v_\parallel )
\approx v_\perp / v_\parallel$. This value should be given by the ratio
of the neutron-star radius perpendicular to the orbital plane to the
tidal disruption radius, $r_\mathrm{td}$. Thus, the dependence of
$\theta$ on the equation of state will be weak again, because both
$v_\parallel$ and $v_\perp$ should scale linearly with
$R_\mathrm{NS}$. Dependence on the mass ratio is expected to be $\theta
\propto Q^{-1/3}$, inherited from that of $r_\mathrm{td} \propto
r_\mathrm{ms}$, but the expected change is only 25\% between $3 \le Q
\le 7$. The spin of the black hole will not modify the value of
$\theta$.

\newpage

\small
\phantomsection
\addcontentsline{toc}{section}{References}

\begin{thebibliography}{651}
\providecommand{\natexlab}[1]{#1}
\providecommand{\url}[1]{{#1}}
\providecommand{\urlprefix}{URL }
\expandafter\ifx\csname urlstyle\endcsname\relax
  \providecommand{\doi}[1]{DOI~\discretionary{}{}{}#1}\else
  \providecommand{\doi}{DOI~\discretionary{}{}{}\begingroup
  \urlstyle{rm}\Url}\fi
\providecommand{\eprint}[2][]{\url{#2}}

\bibitem[{{Abbott} et~al.(2016{\natexlab{a}}){Abbott}, {Abbott}, {Abbott},
  {Abernathy}, {Acernese}, {Ackley}, {Adams}, {Adams}, {Addesso}, {Adhikari},
  and et~al.}]{GW150914_2}
{Abbott} BP, {Abbott} R, {Abbott} TD, {Abernathy} MR, {Acernese} F, {Ackley} K,
  {Adams} C, {Adams} T, {Addesso} P, {Adhikari} RX, et~al (2016{\natexlab{a}})
  {Astrophysical Implications of the Binary Black-hole Merger GW150914}. \apj
  818:L22, \doi{10.3847/2041-8205/818/2/L22}, \eprint{1602.03846}

\bibitem[{{Abbott} et~al.(2016{\natexlab{b}}){Abbott}, {Abbott}, {Abbott},
  {Abernathy}, {Acernese}, {Ackley}, {Adams}, {Adams}, {Addesso}, {Adhikari},
  and et~al.}]{GW150914}
{Abbott} BP, {Abbott} R, {Abbott} TD, {Abernathy} MR, {Acernese} F, {Ackley} K,
  {Adams} C, {Adams} T, {Addesso} P, {Adhikari} RX, et~al (2016{\natexlab{b}})
  {Observation of Gravitational Waves from a Binary Black Hole Merger}. \prl
  116:061102, \doi{10.1103/PhysRevLett.116.061102}, \eprint{1602.03837}

\bibitem[{{Abbott} et~al.(2016{\natexlab{c}}){Abbott}, {Abbott}, {Abbott},
  {Abernathy}, {Acernese}, {Ackley}, {Adams}, {Adams}, {Addesso}, {Adhikari},
  and et~al.}]{GW150914GR}
{Abbott} BP, {Abbott} R, {Abbott} TD, {Abernathy} MR, {Acernese} F, {Ackley} K,
  {Adams} C, {Adams} T, {Addesso} P, {Adhikari} RX, et~al (2016{\natexlab{c}})
  {Tests of General Relativity with GW150914}. \prl 116:221101,
  \doi{10.1103/PhysRevLett.116.221101}, \eprint{1602.03841}

\bibitem[{{Abbott} et~al.(2017{\natexlab{a}}){Abbott}, {Abbott}, {Abbott},
  {Abernathy}, {Ackley}, {Adams}, {Addesso}, {Adhikari}, {Adya}, {Affeldt}, and
  et~al.}]{CosmicExplorer}
{Abbott} BP, {Abbott} R, {Abbott} TD, {Abernathy} MR, {Ackley} K, {Adams} C,
  {Addesso} P, {Adhikari} RX, {Adya} VB, {Affeldt} C, et~al
  (2017{\natexlab{a}}) {Exploring the sensitivity of next generation
  gravitational wave detectors}. Class Quantum Grav 34:044001,
  \doi{10.1088/1361-6382/aa51f4}, \eprint{1607.08697}

\bibitem[{{Abbott} et~al.(2017{\natexlab{b}}){Abbott}, {Abbott}, {Abbott},
  {Acernese}, {Ackley}, {Adams}, {Adams}, {Addesso}, {Adhikari}, {Adya}, and
  et~al.}]{GW170817Hubble}
{Abbott} BP, {Abbott} R, {Abbott} TD, {Acernese} F, {Ackley} K, {Adams} C,
  {Adams} T, {Addesso} P, {Adhikari} RX, {Adya} VB, et~al (2017{\natexlab{b}})
  {A gravitational-wave standard siren measurement of the Hubble constant}.
  \nat 551:85--88, \doi{10.1038/nature24471}, \eprint{1710.05835}

\bibitem[{{Abbott} et~al.(2017{\natexlab{c}}){Abbott}, {Abbott}, {Abbott},
  {Acernese}, {Ackley}, {Adams}, {Adams}, {Addesso}, {Adhikari}, {Adya}, and
  et~al.}]{GRB170817A}
{Abbott} BP, {Abbott} R, {Abbott} TD, {Acernese} F, {Ackley} K, {Adams} C,
  {Adams} T, {Addesso} P, {Adhikari} RX, {Adya} VB, et~al (2017{\natexlab{c}})
  {Gravitational Waves and Gamma-Rays from a Binary Neutron Star Merger:
  GW170817 and GRB 170817A}. \apj 848:L13, \doi{10.3847/2041-8213/aa920c},
  \eprint{1710.05834}

\bibitem[{{Abbott} et~al.(2017{\natexlab{d}}){Abbott}, {Abbott}, {Abbott},
  {Acernese}, {Ackley}, {Adams}, {Adams}, {Addesso}, {Adhikari}, {Adya}, and
  et~al.}]{GW170817}
{Abbott} BP, {Abbott} R, {Abbott} TD, {Acernese} F, {Ackley} K, {Adams} C,
  {Adams} T, {Addesso} P, {Adhikari} RX, {Adya} VB, et~al (2017{\natexlab{d}})
  {GW170817: Observation of Gravitational Waves from a Binary Neutron Star
  Inspiral}. \prl 119:161101, \doi{10.1103/PhysRevLett.119.161101},
  \eprint{1710.05832}

\bibitem[{{Abbott} et~al.(2017{\natexlab{e}}){Abbott}, {Abbott}, {Abbott},
  {Acernese}, {Ackley}, {Adams}, {Adams}, {Addesso}, {Adhikari}, {Adya}, and
  et~al.}]{EM170817}
{Abbott} BP, {Abbott} R, {Abbott} TD, {Acernese} F, {Ackley} K, {Adams} C,
  {Adams} T, {Addesso} P, {Adhikari} RX, {Adya} VB, et~al (2017{\natexlab{e}})
  {Multi-messenger Observations of a Binary Neutron Star Merger}. \apj 848:L12,
  \doi{10.3847/2041-8213/aa91c9}, \eprint{1710.05833}

\bibitem[{{Abbott} et~al.(2018){Abbott}, {Abbott}, {Abbott}, {Acernese},
  {Ackley}, {Adams}, {Adams}, {Addesso}, {Adhikari}, {Adya}, and
  et~al.}]{GW170817EOS}
{Abbott} BP, {Abbott} R, {Abbott} TD, {Acernese} F, {Ackley} K, {Adams} C,
  {Adams} T, {Addesso} P, {Adhikari} RX, {Adya} VB, et~al (2018) {GW170817:
  Measurements of Neutron Star Radii and Equation of State}. \prl 121:161101,
  \doi{10.1103/PhysRevLett.121.161101}, \eprint{1805.11581}

\bibitem[{{Abbott} et~al.(2019{\natexlab{a}}){Abbott}, {Abbott}, {Abbott},
  {Abraham}, {Acernese}, {Ackley}, {Adams}, {Adhikari}, {Adya}, {Affeldt}, and
  et~al.}]{GWTC1}
{Abbott} BP, {Abbott} R, {Abbott} TD, {Abraham} S, {Acernese} F, {Ackley} K,
  {Adams} C, {Adhikari} RX, {Adya} VB, {Affeldt} C, et~al (2019{\natexlab{a}})
  {GWTC-1: A Gravitational-Wave Transient Catalog of Compact Binary Mergers
  Observed by LIGO and Virgo during the First and Second Observing Runs}.
  Physical Review X 9:031040, \doi{10.1103/PhysRevX.9.031040},
  \eprint{1811.12907}

\bibitem[{{Abbott} et~al.(2019{\natexlab{b}}){Abbott}, {Abbott}, {Abbott},
  {Abraham}, {Acernese}, {Ackley}, {Adams}, {Adhikari}, {Adya}, {Affeldt}, and
  et~al.}]{GWTC1GR}
{Abbott} BP, {Abbott} R, {Abbott} TD, {Abraham} S, {Acernese} F, {Ackley} K,
  {Adams} C, {Adhikari} RX, {Adya} VB, {Affeldt} C, et~al (2019{\natexlab{b}})
  {Tests of general relativity with the binary black hole signals from the
  LIGO-Virgo catalog GWTC-1}. \prd 100:104036,
  \doi{10.1103/PhysRevD.100.104036}, \eprint{1903.04467}

\bibitem[{{Abbott} et~al.(2019{\natexlab{c}}){Abbott}, {Abbott}, {Abbott},
  {Acernese}, {Ackley}, {Adams}, {Adams}, {Addesso}, {Adhikari}, {Adya}, and
  et~al.}]{GW170817_2}
{Abbott} BP, {Abbott} R, {Abbott} TD, {Acernese} F, {Ackley} K, {Adams} C,
  {Adams} T, {Addesso} P, {Adhikari} RX, {Adya} VB, et~al (2019{\natexlab{c}})
  {Properties of the Binary Neutron Star Merger GW170817}. Physical Review X
  9:011001, \doi{10.1103/PhysRevX.9.011001}, \eprint{1805.11579}

\bibitem[{{Abbott} et~al.(2020{\natexlab{a}}){Abbott}, {Abbott}, {Abbott},
  {Abraham}, {Acernese}, {Ackley}, {Adams}, {Adhikari}, {Adya}, {Affeldt}, and
  et~al.}]{GW190425}
{Abbott} BP, {Abbott} R, {Abbott} TD, {Abraham} S, {Acernese} F, {Ackley} K,
  {Adams} C, {Adhikari} RX, {Adya} VB, {Affeldt} C, et~al (2020{\natexlab{a}})
  {GW190425: Observation of a Compact Binary Coalescence with Total Mass
  {\ensuremath{\sim}} 3.4 M$_{\odot}$}. \apjl 892:L3,
  \doi{10.3847/2041-8213/ab75f5}, \eprint{2001.01761}

\bibitem[{{Abbott} et~al.(2020{\natexlab{b}}){Abbott}, {Abbott}, {Abbott},
  {Abraham}, {Acernese}, {Ackley}, {Adams}, {Adya}, {Affeldt}, {Agathos}, and
  et~al.}]{KLV2020}
{Abbott} BP, {Abbott} R, {Abbott} TD, {Abraham} S, {Acernese} F, {Ackley} K,
  {Adams} C, {Adya} VB, {Affeldt} C, {Agathos} M, et~al (2020{\natexlab{b}})
  {Prospects for observing and localizing gravitational-wave transients with
  Advanced LIGO, Advanced Virgo and KAGRA}. Living Rev Relativ 23:3,
  \doi{10.1007/s41114-020-00026-9}

\bibitem[{{Abbott} et~al.(2020{\natexlab{c}}){Abbott}, {Abbott}, {Abraham},
  {Acernese}, {Ackley}, {Adams}, {Adhikari}, {Adya}, {Affeldt}, {Agathos}, and
  et~al.}]{GW190814}
{Abbott} R, {Abbott} TD, {Abraham} S, {Acernese} F, {Ackley} K, {Adams} C,
  {Adhikari} RX, {Adya} VB, {Affeldt} C, {Agathos} M, et~al
  (2020{\natexlab{c}}) {GW190814: Gravitational Waves from the Coalescence of a
  23 Solar Mass Black Hole with a 2.6 Solar Mass Compact Object}. \apjl
  896:L44, \doi{10.3847/2041-8213/ab960f}, \eprint{2006.12611}

\bibitem[{{Abbott} et~al.(2021{\natexlab{a}}){Abbott}, {Abbott}, {Abraham},
  {Acernese}, {Ackley}, {Adams}, {Adams}, {Adhikari}, {Adya}, {Affeldt}, and
  et~al.}]{GWTC2}
{Abbott} R, {Abbott} TD, {Abraham} S, {Acernese} F, {Ackley} K, {Adams} A,
  {Adams} C, {Adhikari} RX, {Adya} VB, {Affeldt} C, et~al (2021{\natexlab{a}})
  {GWTC-2: Compact Binary Coalescences Observed by LIGO and Virgo during the
  First Half of the Third Observing Run}. Phys Rev X 11:021053,
  \doi{10.1103/PhysRevX.11.021053}, \eprint{2010.14527}

\bibitem[{{Abbott} et~al.(2021{\natexlab{b}}){Abbott}, {Abbott}, {Abraham},
  {Acernese}, {Ackley}, {Adams}, {Adams}, {Adhikari}, {Adya}, {Affeldt}, and
  et~al.}]{GW200105200115}
{Abbott} R, {Abbott} TD, {Abraham} S, {Acernese} F, {Ackley} K, {Adams} A,
  {Adams} C, {Adhikari} RX, {Adya} VB, {Affeldt} C, et~al (2021{\natexlab{b}})
  {Observation of Gravitational Waves from Two Neutron Star-Black Hole
  Coalescences}. \apjl 915:L5, \doi{10.3847/2041-8213/ac082e},
  \eprint{2106.15163}

\bibitem[{{Abbott} et~al.(2021{\natexlab{c}}){Abbott}, {Abbott}, {Abraham},
  {Acernese}, {Ackley}, {Adams}, {Adams}, {Adhikari}, {Adya}, {Affeldt}, and
  et~al.}]{GWTC2GR}
{Abbott} R, {Abbott} TD, {Abraham} S, {Acernese} F, {Ackley} K, {Adams} A,
  {Adams} C, {Adhikari} RX, {Adya} VB, {Affeldt} C, et~al (2021{\natexlab{c}})
  {Tests of general relativity with binary black holes from the second
  LIGO-Virgo gravitational-wave transient catalog}. \prd 103:122002,
  \doi{10.1103/PhysRevD.103.122002}, \eprint{2010.14529}

\bibitem[{{Ajith} et~al.(2008){Ajith}, {Babak}, {Chen}, {Hewitson}, {Krishnan},
  {Sintes}, {Whelan}, {Br{\"u}gmann}, {Diener}, {Dorband}, and
  et~al.}]{Ajith_etal2008}
{Ajith} P, {Babak} S, {Chen} Y, {Hewitson} M, {Krishnan} B, {Sintes} AM,
  {Whelan} JT, {Br{\"u}gmann} B, {Diener} P, {Dorband} N, et~al (2008)
  {Template bank for gravitational waveforms from coalescing binary black
  holes: Nonspinning binaries}. \prd 77:104017,
  \doi{10.1103/PhysRevD.77.104017}, \eprint{0710.2335}

\bibitem[{{Akmal} et~al.(1998){Akmal}, {Pandharipande}, and
  {Ravenhall}}]{Akmal_Pandharipande_Ravehnall1998}
{Akmal} A, {Pandharipande} VR, {Ravenhall} DG (1998) {Equation of state of
  nucleon matter and neutron star structure}. \prc 58:1804--1828,
  \doi{10.1103/PhysRevC.58.1804}, \eprint{nucl-th/9804027}

\bibitem[{{Alcubierre}(2008)}]{Alcubierre}
{Alcubierre} M (2008) {Introduction to 3+1 Numerical Relativity}. Oxford
  University Press, Oxford, UK, \doi{10.1093/acprof:oso/9780199205677.001.0001}

\bibitem[{{Alcubierre} et~al.(2003){Alcubierre}, {Br{\"u}gmann}, {Diener},
  {Koppitz}, {Pollney}, {Seidel}, and {Takahashi}}]{Alcubierre_BDKPST2003}
{Alcubierre} M, {Br{\"u}gmann} B, {Diener} P, {Koppitz} M, {Pollney} D,
  {Seidel} E, {Takahashi} R (2003) {Gauge conditions for long-term numerical
  black hole evolutions without excision}. \prd 67:084023,
  \doi{10.1103/PhysRevD.67.084023}, \eprint{gr-qc/0206072}

\bibitem[{{Alexander} et~al.(2018){Alexander}, {Margutti}, {Blanchard}, {Fong},
  {Berger}, {Hajela}, {Eftekhari}, {Chornock}, {Cowperthwaite}, {Giannios}, and
  et~al.}]{Alexander_etal2018}
{Alexander} KD, {Margutti} R, {Blanchard} PK, {Fong} W, {Berger} E, {Hajela} A,
  {Eftekhari} T, {Chornock} R, {Cowperthwaite} PS, {Giannios} D, et~al (2018)
  {A Decline in the X-Ray through Radio Emission from GW170817 Continues to
  Support an Off-axis Structured Jet}. \apj 863:L18,
  \doi{10.3847/2041-8213/aad637}, \eprint{1805.02870}

\bibitem[{{Alford} et~al.(2005){Alford}, {Braby}, {Paris}, and
  {Reddy}}]{Alford_BPR2005}
{Alford} M, {Braby} M, {Paris} M, {Reddy} S (2005) {Hybrid Stars that
  Masquerade as Neutron Stars}. \apj 629:969--978, \doi{10.1086/430902},
  \eprint{nucl-th/0411016}

\bibitem[{{Alic} et~al.(2012){Alic}, {Bona-Casas}, {Bona}, {Rezzolla}, and
  {Palenzuela}}]{Alic_BBRP2012}
{Alic} D, {Bona-Casas} C, {Bona} C, {Rezzolla} L, {Palenzuela} C (2012)
  {Conformal and covariant formulation of the Z4 system with
  constraint-violation damping}. \prd 85:064040,
  \doi{10.1103/PhysRevD.85.064040}, \eprint{1106.2254}

\bibitem[{{Alvi}(2001)}]{Alvi2001}
{Alvi} K (2001) {Energy and angular momentum flow into a black hole in a
  binary}. \prd 64:104020, \doi{10.1103/PhysRevD.64.104020},
  \eprint{gr-qc/0107080}

\bibitem[{{Alvi}(2002)}]{Alvi2002}
{Alvi} K (2002) {First-order symmetrizable hyperbolic formulations of
  Einstein's equations including lapse and shift as dynamical fields}. Class
  Quantum Grav 19:5153--5162, \doi{10.1088/0264-9381/19/20/309},
  \eprint{gr-qc/0204068}

\bibitem[{{Anderson} et~al.(2006){Anderson}, {Hirschmann}, {Liebling}, and
  {Neilsen}}]{Anderson_HLN2006}
{Anderson} M, {Hirschmann} EW, {Liebling} SL, {Neilsen} D (2006) {Relativistic
  MHD with adaptive mesh refinement}. Class Quantum Grav 23:6503--6524,
  \doi{10.1088/0264-9381/23/22/025}, \eprint{gr-qc/0605102}

\bibitem[{{Anderson} et~al.(2008){Anderson}, {Hirschmann}, {Lehner},
  {Liebling}, {Motl}, {Neilsen}, {Palenzuela}, and
  {Tohline}}]{Anderson_HLLMNPT2008}
{Anderson} M, {Hirschmann} EW, {Lehner} L, {Liebling} SL, {Motl} PM, {Neilsen}
  D, {Palenzuela} C, {Tohline} JE (2008) {Simulating binary neutron stars:
  Dynamics and gravitational waves}. \prd 77:024006,
  \doi{10.1103/PhysRevD.77.024006}, \eprint{0708.2720}

\bibitem[{{Ansorg} et~al.(2003){Ansorg}, {Kleinw{\"a}chter}, and
  {Meinel}}]{Ansorg_KleinWachter_Meinel2003}
{Ansorg} M, {Kleinw{\"a}chter} A, {Meinel} R (2003) {Highly accurate
  calculation of rotating neutron stars. Detailed description of the numerical
  methods}. \aap 405:711--721, \doi{10.1051/0004-6361:20030618},
  \eprint{astro-ph/0301173}

\bibitem[{{Ansorg} et~al.(2004){Ansorg}, {Br{\"u}gmann}, and
  {Tichy}}]{Ansorg_Brugmann_Tichy2004}
{Ansorg} M, {Br{\"u}gmann} B, {Tichy} W (2004) {Single-domain spectral method
  for black hole puncture data}. \prd 70:064011,
  \doi{10.1103/PhysRevD.70.064011}, \eprint{gr-qc/0404056}

\bibitem[{{Antoniadis} et~al.(2013){Antoniadis}, {Freire}, {Wex}, {Tauris},
  {Lynch}, {van Kerkwijk}, {Kramer}, {Bassa}, {Dhillon}, {Driebe}, and
  et~al.}]{Antoniadis_etal2013}
{Antoniadis} J, {Freire} PCC, {Wex} N, {Tauris} TM, {Lynch} RS, {van Kerkwijk}
  MH, {Kramer} M, {Bassa} C, {Dhillon} VS, {Driebe} T, et~al (2013) {A Massive
  Pulsar in a Compact Relativistic Binary}. Science 340(6131):448,
  \doi{10.1126/science.1233232}, \eprint{1304.6875}

\bibitem[{{Apostolatos} et~al.(1994){Apostolatos}, {Cutler}, {Sussman}, and
  {Thorne}}]{Apostolatos_CST1994}
{Apostolatos} TA, {Cutler} C, {Sussman} GJ, {Thorne} KS (1994) {Spin-induced
  orbital precession and its modulation of the gravitational waveforms from
  merging binaries}. \prd 49:6274--6297, \doi{10.1103/PhysRevD.49.6274}

\bibitem[{{Arcavi} et~al.(2017){Arcavi}, {Hosseinzadeh}, {Howell}, {McCully},
  {Poznanski}, {Kasen}, {Barnes}, {Zaltzman}, {Vasylyev}, {Maoz}, and
  et~al.}]{Arcavi_etal2017}
{Arcavi} I, {Hosseinzadeh} G, {Howell} DA, {McCully} C, {Poznanski} D, {Kasen}
  D, {Barnes} J, {Zaltzman} M, {Vasylyev} S, {Maoz} D, et~al (2017) {Optical
  emission from a kilonova following a gravitational-wave-detected neutron-star
  merger}. \nat 551:64--66, \doi{10.1038/nature24291}, \eprint{1710.05843}

\bibitem[{{Arnett}(1982)}]{Arnett1982}
{Arnett} WD (1982) {Type I supernovae. I - Analytic solutions for the early
  part of the light curve}. \apj 253:785--797, \doi{10.1086/159681}

\bibitem[{{Arnowitt} et~al.(2008){Arnowitt}, {Deser}, and
  {Misner}}]{Arnowitt_Deser_Misner2008}
{Arnowitt} R, {Deser} S, {Misner} CW (2008) {Republication of: The dynamics of
  general relativity}. General Relativity and Gravitation 40:1997--2027,
  \doi{10.1007/s10714-008-0661-1}, \eprint{gr-qc/0405109}

\bibitem[{{Artemova} et~al.(1996){Artemova}, {Bjoernsson}, and
  {Novikov}}]{Artemova_Bjornsson_Novikov1996}
{Artemova} IV, {Bjoernsson} G, {Novikov} ID (1996) {Modified Newtonian
  Potentials for the Description of Relativistic Effects in Accretion Disks
  around Black Holes}. \apj 461:565, \doi{10.1086/177084}

\bibitem[{{Arzoumanian} et~al.(2018){Arzoumanian}, {Brazier}, {Burke-Spolaor},
  {Chamberlin}, {Chatterjee}, {Christy}, {Cordes}, {Cornish}, {Crawford},
  {Thankful Cromartie}, and et~al.}]{Arzoumanian_etal2018}
{Arzoumanian} Z, {Brazier} A, {Burke-Spolaor} S, {Chamberlin} S, {Chatterjee}
  S, {Christy} B, {Cordes} JM, {Cornish} NJ, {Crawford} F, {Thankful Cromartie}
  H, et~al (2018) {The NANOGrav 11-year Data Set: High-precision Timing of 45
  Millisecond Pulsars}. \apjs 235:37, \doi{10.3847/1538-4365/aab5b0},
  \eprint{1801.01837}

\bibitem[{{Asada}(1998)}]{Asada1998}
{Asada} H (1998) {Formulation for the internal motion of quasiequilibrium
  configurations in general relativity}. \prd 57:7292--7298,
  \doi{10.1103/PhysRevD.57.7292}, \eprint{gr-qc/9804003}

\bibitem[{{Ashtekar} and {Krishnan}(2004)}]{Ashtekar_Krishnan2004}
{Ashtekar} A, {Krishnan} B (2004) {Isolated and Dynamical Horizons and Their
  Applications}. Living Rev Relativ 7:10, \doi{10.12942/lrr-2004-10},
  \eprint{gr-qc/0407042}

\bibitem[{{Ashtekar} and {Magnon-Ashtekar}(1979)}]{Ashtekar_MagnonAshtekar1979}
{Ashtekar} A, {Magnon-Ashtekar} A (1979) {On conserved quantities in general
  relativity}. J Math Phys 20:793--800, \doi{10.1063/1.524151}

\bibitem[{{Baiotti} and {Rezzolla}(2017)}]{Baiotti_Rezzolla2017}
{Baiotti} L, {Rezzolla} L (2017) {Binary neutron star mergers: a review of
  Einstein{\textquoteright}s richest laboratory}. Reports on Progress in
  Physics 80:096901, \doi{10.1088/1361-6633/aa67bb}, \eprint{1607.03540}

\bibitem[{{Baker} et~al.(2006){Baker}, {Centrella}, {Choi}, {Koppitz}, and {van
  Meter}}]{Baker_CCKV2006}
{Baker} JG, {Centrella} J, {Choi} DI, {Koppitz} M, {van Meter} J (2006)
  {Gravitational-Wave Extraction from an Inspiraling Configuration of Merging
  Black Holes}. \prl 96:111102, \doi{10.1103/PhysRevLett.96.111102},
  \eprint{gr-qc/0511103}

\bibitem[{{Balbus} and {Hawley}(1991)}]{Balbus_Hawley1991}
{Balbus} SA, {Hawley} JF (1991) {A Powerful Local Shear Instability in Weakly
  Magnetized Disks. I. Linear Analysis}. \apj 376:214, \doi{10.1086/170270}

\bibitem[{{Balbus} and {Hawley}(1998)}]{Balbus_Hawley1998}
{Balbus} SA, {Hawley} JF (1998) {Instability, turbulence, and enhanced
  transport in accretion disks}. Reviews of Modern Physics 70:1--53,
  \doi{10.1103/RevModPhys.70.1}

\bibitem[{{Balsara}(2001)}]{Balsara2001}
{Balsara} DS (2001) {Divergence-Free Adaptive Mesh Refinement for
  Magnetohydrodynamics}. J Comput Phys 174:614--648,
  \doi{10.1006/jcph.2001.6917}, \eprint{astro-ph/0112150}

\bibitem[{{Balsara}(2009)}]{Balsara2009}
{Balsara} DS (2009) {Divergence-free reconstruction of magnetic fields and WENO
  schemes for magnetohydrodynamics}. J Comput Phys 228:5040--5056,
  \doi{10.1016/j.jcp.2009.03.038}, \eprint{0811.2192}

\bibitem[{{Balsara}(2017)}]{Balsara2017}
{Balsara} DS (2017) {Higher-order accurate space-time schemes for computational
  astrophysics{\textemdash}Part I: finite volume methods}. Living Rev Comput
  Astrophys 3:2, \doi{10.1007/s41115-017-0002-8}, \eprint{1703.01241}

\bibitem[{{Balsara} and {Spicer}(1999)}]{Balsara_Spicer1999}
{Balsara} DS, {Spicer} DS (1999) {A Staggered Mesh Algorithm Using High Order
  Godunov Fluxes to Ensure Solenoidal Magnetic Fields in Magnetohydrodynamic
  Simulations}. J Comput Phys 149:270--292, \doi{10.1006/jcph.1998.6153}

\bibitem[{{Banerjee} et~al.(2020){Banerjee}, {Tanaka}, {Kawaguchi}, {Kato}, and
  {Gaigalas}}]{Banerjee_TKKG2020}
{Banerjee} S, {Tanaka} M, {Kawaguchi} K, {Kato} D, {Gaigalas} G (2020)
  {Simulations of Early Kilonova Emission from Neutron Star Mergers}. \apj
  901:29, \doi{10.3847/1538-4357/abae61}, \eprint{2008.05495}

\bibitem[{{Banik} et~al.(2014){Banik}, {Hempel}, and
  {Bandyopadhyay}}]{Banik_Hempel_Bandyopadhyay2014}
{Banik} S, {Hempel} M, {Bandyopadhyay} D (2014) {New Hyperon Equations of State
  for Supernovae and Neutron Stars in Density-dependent Hadron Field Theory}.
  \apjs 214:22, \doi{10.1088/0067-0049/214/2/22}, \eprint{1404.6173}

\bibitem[{{Barbieri} et~al.(2019){Barbieri}, {Salafia}, {Perego}, {Colpi}, and
  {Ghirlanda}}]{Barbieri_SPCG2019}
{Barbieri} C, {Salafia} OS, {Perego} A, {Colpi} M, {Ghirlanda} G (2019)
  {Light-curve models of black hole - neutron star mergers: steps towards a
  multi-messenger parameter estimation}. \aap 625:A152,
  \doi{10.1051/0004-6361/201935443}, \eprint{1903.04543}

\bibitem[{{Barbieri} et~al.(2020){Barbieri}, {Salafia}, {Perego}, {Colpi}, and
  {Ghirlanda}}]{Barbieri_SPCG2020}
{Barbieri} C, {Salafia} OS, {Perego} A, {Colpi} M, {Ghirlanda} G (2020)
  {Electromagnetic counterparts of black hole-neutron star mergers: dependence
  on the neutron star properties}. Eur Phys J A 56:8,
  \doi{10.1140/epja/s10050-019-00013-x}, \eprint{1908.08822}

\bibitem[{{Bardeen} and {Petterson}(1975)}]{Bardeen_Petterson1975}
{Bardeen} JM, {Petterson} JA (1975) {The Lense-Thirring Effect and Accretion
  Disks around Kerr Black Holes}. \apjl 195:L65, \doi{10.1086/181711}

\bibitem[{{Bardeen} et~al.(1972){Bardeen}, {Press}, and
  {Teukolsky}}]{Bardeen_Press_Teukolsky1972}
{Bardeen} JM, {Press} WH, {Teukolsky} SA (1972) {Rotating Black Holes: Locally
  Nonrotating Frames, Energy Extraction, and Scalar Synchrotron Radiation}.
  \apj 178:347--370, \doi{10.1086/151796}

\bibitem[{{Barker} and {O'Connell}(1975)}]{Barker_OConnel1975}
{Barker} BM, {O'Connell} RF (1975) {Gravitational two-body problem with
  arbitrary masses, spins, and quadrupole moments}. \prd 12:329--335,
  \doi{10.1103/PhysRevD.12.329}

\bibitem[{{Barnes} et~al.(2016){Barnes}, {Kasen}, {Wu}, and
  {Mart{\'\i}nez-Pinedo}}]{Barnes_KWM2016}
{Barnes} J, {Kasen} D, {Wu} MR, {Mart{\'\i}nez-Pinedo} G (2016) {Radioactivity
  and Thermalization in the Ejecta of Compact Object Mergers and Their Impact
  on Kilonova Light Curves}. \apj 829:110, \doi{10.3847/0004-637X/829/2/110},
  \eprint{1605.07218}

\bibitem[{{Baumgarte} and {Shapiro}(1999)}]{Baumgarte_Shapiro1999}
{Baumgarte} TW, {Shapiro} SL (1999) {Numerical integration of Einstein's field
  equations}. \prd 59:024007, \doi{10.1103/PhysRevD.59.024007},
  \eprint{gr-qc/9810065}

\bibitem[{{Baumgarte} and {Shapiro}(2003)}]{Baumgarte_Shapiro2003}
{Baumgarte} TW, {Shapiro} SL (2003) {General Relativistic Magnetohydrodynamics
  for the Numerical Construction of Dynamical Spacetimes}. \apj 585:921--929,
  \doi{10.1086/346103}, \eprint{astro-ph/0211340}

\bibitem[{{Baumgarte} and {Shapiro}(2010)}]{Baumgarte_Shapiro}
{Baumgarte} TW, {Shapiro} SL (2010) {Numerical Relativity: Solving Einstein's
  Equations on the Computer}. Cambridge University Press

\bibitem[{{Baumgarte} et~al.(2004){Baumgarte}, {Skoge}, and
  {Shapiro}}]{Baumgarte_Skoge_Shapiro2004}
{Baumgarte} TW, {Skoge} ML, {Shapiro} SL (2004) {Black hole-neutron star
  binaries in general relativity: Quasiequilibrium formulation}. \prd
  70:064040, \doi{10.1103/PhysRevD.70.064040}, \eprint{gr-qc/0405077}

\bibitem[{{Baumgarte} et~al.(2007){Baumgarte}, {Murchadha}, and
  {Pfeiffer}}]{Baumgarte_OMurchadha_Pfeiffer2007}
{Baumgarte} TW, {Murchadha} N{\'O}, {Pfeiffer} HP (2007) {Einstein constraints:
  Uniqueness and nonuniqueness in the conformal thin sandwich approach}. \prd
  75:044009, \doi{10.1103/PhysRevD.75.044009}, \eprint{gr-qc/0610120}

\bibitem[{{Bauswein} et~al.(2010){Bauswein}, {Janka}, and
  {Oechslin}}]{Bauswein_Janka_Oechslin2010}
{Bauswein} A, {Janka} HT, {Oechslin} R (2010) {Testing approximations of
  thermal effects in neutron star merger simulations}. \prd 82:084043,
  \doi{10.1103/PhysRevD.82.084043}, \eprint{1006.3315}

\bibitem[{{Baym} et~al.(2018){Baym}, {Hatsuda}, {Kojo}, {Powell}, {Song}, and
  {Takatsuka}}]{Baym_HKPST2018}
{Baym} G, {Hatsuda} T, {Kojo} T, {Powell} PD, {Song} Y, {Takatsuka} T (2018)
  {From hadrons to quarks in neutron stars: a review}. Reports on Progress in
  Physics 81:056902, \doi{10.1088/1361-6633/aaae14}, \eprint{1707.04966}

\bibitem[{{Beig}(1978)}]{Beig1978}
{Beig} R (1978) {Arnowitt-Deser-Misner energy and $g_{00}$}. Phys Lett A
  69:153--155, \doi{10.1016/0375-9601(78)90198-6}

\bibitem[{{Bell}(1978)}]{Bell1978}
{Bell} AR (1978) {The acceleration of cosmic rays in shock fronts - I.} \mnras
  182:147--156, \doi{10.1093/mnras/182.2.147}

\bibitem[{{Bell}(2004)}]{Bell2004}
{Bell} AR (2004) {Turbulent amplification of magnetic field and diffusive shock
  acceleration of cosmic rays}. \mnras 353:550--558,
  \doi{10.1111/j.1365-2966.2004.08097.x}

\bibitem[{{Benacquista} and {Downing}(2013)}]{Benacquista_Downing2013}
{Benacquista} MJ, {Downing} JMB (2013) {Relativistic Binaries in Globular
  Clusters}. Living Rev Relativ 16:4, \doi{10.12942/lrr-2013-4},
  \eprint{1110.4423}

\bibitem[{{Benz} et~al.(1990){Benz}, {Bowers}, {Cameron}, and
  {Press}}]{Benz_BCP1990}
{Benz} W, {Bowers} RL, {Cameron} AGW, {Press} WH (1990) {Dynamic Mass Exchange
  in Doubly Degenerate Binaries. I. 0.9 and 1.2 $M_{\odot}$ Stars}. \apj
  348:647, \doi{10.1086/168273}

\bibitem[{{Berger}(2014)}]{Berger2014}
{Berger} E (2014) {Short-Duration Gamma-Ray Bursts}. \araa 52:43--105,
  \doi{10.1146/annurev-astro-081913-035926}, \eprint{1311.2603}

\bibitem[{{Berger} and {Oliger}(1984)}]{Berger_Oliger1984}
{Berger} MJ, {Oliger} J (1984) {Adaptive Mesh Refinement for Hyperbolic Partial
  Differential Equations}. J Comput Phys 53:484--512,
  \doi{10.1016/0021-9991(84)90073-1}

\bibitem[{{Bernuzzi} and {Hilditch}(2010)}]{Bernuzzi_Hilditch2010}
{Bernuzzi} S, {Hilditch} D (2010) {Constraint violation in free evolution
  schemes: Comparing the BSSNOK formulation with a conformal decomposition of
  the Z4 formulation}. \prd 81:084003, \doi{10.1103/PhysRevD.81.084003},
  \eprint{0912.2920}

\bibitem[{{Berti} et~al.(2009){Berti}, {Cardoso}, and
  {Starinets}}]{Berti_Cardoso_Starinets2009}
{Berti} E, {Cardoso} V, {Starinets} AO (2009) {TOPICAL REVIEW: Quasinormal
  modes of black holes and black branes}. Class Quantum Grav 26:163001,
  \doi{10.1088/0264-9381/26/16/163001}, \eprint{0905.2975}

\bibitem[{{Bildsten} and {Cutler}(1992)}]{Bildsten_Cutler1992}
{Bildsten} L, {Cutler} C (1992) {Tidal Interactions of Inspiraling Compact
  Binaries}. \apj 400:175, \doi{10.1086/171983}

\bibitem[{{Binnington} and {Poisson}(2009)}]{Binnington_Poisson2009}
{Binnington} T, {Poisson} E (2009) {Relativistic theory of tidal Love numbers}.
  \prd 80:084018, \doi{10.1103/PhysRevD.80.084018}, \eprint{0906.1366}

\bibitem[{{Birkl} et~al.(2007){Birkl}, {Aloy}, {Janka}, and
  {M{\"u}ller}}]{Birkl_AJM2007}
{Birkl} R, {Aloy} MA, {Janka} HT, {M{\"u}ller} E (2007) {Neutrino pair
  annihilation near accreting, stellar-mass black holes}. \aap 463:51--67,
  \doi{10.1051/0004-6361:20066293}, \eprint{astro-ph/0608543}

\bibitem[{{Blackburn} and {Detweiler}(1992)}]{Blackburn_Detweiler1992}
{Blackburn} JK, {Detweiler} S (1992) {Close black-hole binary systems}. \prd
  46:2318--2333, \doi{10.1103/PhysRevD.46.2318}

\bibitem[{{Blanchet}(2002)}]{Blanchet2002}
{Blanchet} L (2002) {Innermost circular orbit of binary black holes at the
  third post-Newtonian approximation}. \prd 65:124009,
  \doi{10.1103/PhysRevD.65.124009}, \eprint{gr-qc/0112056}

\bibitem[{{Blanchet}(2014)}]{Blanchet2014}
{Blanchet} L (2014) {Gravitational Radiation from Post-Newtonian Sources and
  Inspiralling Compact Binaries}. Living Rev Relativ 17:2,
  \doi{10.12942/lrr-2014-2}, \eprint{1310.1528}

\bibitem[{{Blanchet} and {Iyer}(2003)}]{Blanchet_Iyer2003}
{Blanchet} L, {Iyer} BR (2003) {Third post-Newtonian dynamics of compact
  binaries: equations of motion in the centre-of-mass frame}. Class Quantum
  Grav 20:755--776, \doi{10.1088/0264-9381/20/4/309}, \eprint{gr-qc/0209089}

\bibitem[{{Blanchet} et~al.(2005){Blanchet}, {Qusailah}, and
  {Will}}]{Blanchet_Qusailah_Will2005}
{Blanchet} L, {Qusailah} MSS, {Will} CM (2005) {Gravitational Recoil of
  Inspiraling Black Hole Binaries to Second Post-Newtonian Order}. \apj
  635:508--515, \doi{10.1086/497332}, \eprint{astro-ph/0507692}

\bibitem[{{Blandford} and {Ostriker}(1978)}]{Blandford_Ostriker1978}
{Blandford} RD, {Ostriker} JP (1978) {Particle acceleration by astrophysical
  shocks.} \apjl 221:L29--L32, \doi{10.1086/182658}

\bibitem[{{Blandford} and {Payne}(1982)}]{Blandford_Payne1982}
{Blandford} RD, {Payne} DG (1982) {Hydromagnetic flows from accretion disks and
  the production of radio jets}. \mnras 199:883--903,
  \doi{10.1093/mnras/199.4.883}

\bibitem[{{Blandford} and {Znajek}(1977)}]{Blandford_Znajek1977}
{Blandford} RD, {Znajek} RL (1977) {Electromagnetic extraction of energy from
  Kerr black holes}. \mnras 179:433--456, \doi{10.1093/mnras/179.3.433}

\bibitem[{{Blinnikov} et~al.(1984){Blinnikov}, {Novikov}, {Perevodchikova}, and
  {Polnarev}}]{Blinnikov_NPP1984}
{Blinnikov} SI, {Novikov} ID, {Perevodchikova} TV, {Polnarev} AG (1984)
  {Exploding Neutron Stars in Close Binaries}. Sov Astron Lett 10:177--179,
  \eprint{1808.05287}

\bibitem[{{Boh{\'e}} et~al.(2017){Boh{\'e}}, {Shao}, {Taracchini}, {Buonanno},
  {Babak}, {Harry}, {Hinder}, {Ossokine}, {P{\"u}rrer}, {Raymond}, and
  et~al.}]{Bohe_etal2017}
{Boh{\'e}} A, {Shao} L, {Taracchini} A, {Buonanno} A, {Babak} S, {Harry} IW,
  {Hinder} I, {Ossokine} S, {P{\"u}rrer} M, {Raymond} V, et~al (2017) {Improved
  effective-one-body model of spinning, nonprecessing binary black holes for
  the era of gravitational-wave astrophysics with advanced detectors}. \prd
  95:044028, \doi{10.1103/PhysRevD.95.044028}, \eprint{1611.03703}

\bibitem[{{Bona} and {Mass{\'o}}(1992)}]{Bona_Masso1992}
{Bona} C, {Mass{\'o}} J (1992) {Hyperbolic evolution system for numerical
  relativity}. \prl 68:1097--1099, \doi{10.1103/PhysRevLett.68.1097}

\bibitem[{{Bona} et~al.(1995){Bona}, {Mass{\'o}}, {Seidel}, and
  {Stela}}]{Bona_MSS1995}
{Bona} C, {Mass{\'o}} J, {Seidel} E, {Stela} J (1995) {New Formalism for
  Numerical Relativity}. \prl 75:600--603, \doi{10.1103/PhysRevLett.75.600},
  \eprint{gr-qc/9412071}

\bibitem[{{Bona} et~al.(2003){Bona}, {Ledvinka}, {Palenzuela}, and
  {{\v{Z}}{\'a}{\v{c}}ek}}]{Bona_Ledvinka_Palenzuela_Zacek2003}
{Bona} C, {Ledvinka} T, {Palenzuela} C, {{\v{Z}}{\'a}{\v{c}}ek} M (2003)
  {General-covariant evolution formalism for numerical relativity}. \prd
  67:104005, \doi{10.1103/PhysRevD.67.104005}, \eprint{gr-qc/0302083}

\bibitem[{{Bonazzola} et~al.(1997){Bonazzola}, {Gourgoulhon}, and
  {Marck}}]{Bonazzola_Gourgoulhon_Marck1997}
{Bonazzola} S, {Gourgoulhon} E, {Marck} JA (1997) {Relativistic formalism to
  compute quasiequilibrium configurations of nonsynchronized neutron star
  binaries}. \prd 56:7740--7749, \doi{10.1103/PhysRevD.56.7740},
  \eprint{gr-qc/9710031}

\bibitem[{{Bonazzola} et~al.(2004){Bonazzola}, {Gourgoulhon},
  {Grandcl{\'e}ment}, and {Novak}}]{Bonazzola_GGN2004}
{Bonazzola} S, {Gourgoulhon} E, {Grandcl{\'e}ment} P, {Novak} J (2004)
  {Constrained scheme for the Einstein equations based on the Dirac gauge and
  spherical coordinates}. \prd 70:104007, \doi{10.1103/PhysRevD.70.104007},
  \eprint{gr-qc/0307082}

\bibitem[{{Bovard} et~al.(2017){Bovard}, {Martin}, {Guercilena}, {Arcones},
  {Rezzolla}, and {Korobkin}}]{Bovard_MGARK2017}
{Bovard} L, {Martin} D, {Guercilena} F, {Arcones} A, {Rezzolla} L, {Korobkin} O
  (2017) {r -process nucleosynthesis from matter ejected in binary neutron star
  mergers}. \prd 96:124005, \doi{10.1103/PhysRevD.96.124005},
  \eprint{1709.09630}

\bibitem[{{Bowen} and {York}(1980)}]{Bowen_York1980}
{Bowen} JM, {York} J James~W (1980) {Time-asymmetric initial data for black
  holes and black-hole collisions}. \prd 21:2047--2056,
  \doi{10.1103/PhysRevD.21.2047}

\bibitem[{{Boyle} et~al.(2007){Boyle}, {Brown}, {Kidder}, {Mrou{\'e}},
  {Pfeiffer}, {Scheel}, {Cook}, and {Teukolsky}}]{Boyle_BKMPSCT2007}
{Boyle} M, {Brown} DA, {Kidder} LE, {Mrou{\'e}} AH, {Pfeiffer} HP, {Scheel} MA,
  {Cook} GB, {Teukolsky} SA (2007) {High-accuracy comparison of numerical
  relativity simulations with post-Newtonian expansions}. \prd 76:124038,
  \doi{10.1103/PhysRevD.76.124038}, \eprint{0710.0158}

\bibitem[{{Boyle} et~al.(2008){Boyle}, {Buonanno}, {Kidder}, {Mrou{\'e}},
  {Pan}, {Pfeiffer}, and {Scheel}}]{Boyle_BKMPPS2008}
{Boyle} M, {Buonanno} A, {Kidder} LE, {Mrou{\'e}} AH, {Pan} Y, {Pfeiffer} HP,
  {Scheel} MA (2008) {High-accuracy numerical simulation of black-hole
  binaries: Computation of the gravitational-wave energy flux and comparisons
  with post-Newtonian approximants}. \prd 78:104020,
  \doi{10.1103/PhysRevD.78.104020}, \eprint{0804.4184}

\bibitem[{{Boyle} et~al.(2019){Boyle}, {Hemberger}, {Iozzo}, {Lovelace},
  {Ossokine}, {Pfeiffer}, {Scheel}, {Stein}, {Woodford}, {Zimmerman}, and
  et~al.}]{Boyle_etal2019}
{Boyle} M, {Hemberger} D, {Iozzo} DAB, {Lovelace} G, {Ossokine} S, {Pfeiffer}
  HP, {Scheel} MA, {Stein} LC, {Woodford} CJ, {Zimmerman} A, et~al (2019) {The
  SXS collaboration catalog of binary black hole simulations}. Class Quantum
  Grav 36:195006, \doi{10.1088/1361-6382/ab34e2}, \eprint{1904.04831}

\bibitem[{{Brandt} and {Br{\"u}gmann}(1997)}]{Brandt_Brugmann1997}
{Brandt} S, {Br{\"u}gmann} B (1997) {A Simple Construction of Initial Data for
  Multiple Black Holes}. \prl 78:3606--3609, \doi{10.1103/PhysRevLett.78.3606},
  \eprint{gr-qc/9703066}

\bibitem[{{Brege} et~al.(2018){Brege}, {Duez}, {Foucart}, {Deaton}, {Caro},
  {Hemberger}, {Kidder}, {O'Connor}, {Pfeiffer}, and
  {Scheel}}]{Brege_DFDCHKOPS2018}
{Brege} W, {Duez} MD, {Foucart} F, {Deaton} MB, {Caro} J, {Hemberger} DA,
  {Kidder} LE, {O'Connor} E, {Pfeiffer} HP, {Scheel} MA (2018) {Black
  hole-neutron star mergers using a survey of finite-temperature equations of
  state}. \prd 98:063009, \doi{10.1103/PhysRevD.98.063009}, \eprint{1804.09823}

\bibitem[{{Brill} and {Lindquist}(1963)}]{Brill_Lindquist1963}
{Brill} DR, {Lindquist} RW (1963) {Interaction Energy in Geometrostatics}.
  Physical Review 131:471--476, \doi{10.1103/PhysRev.131.471}

\bibitem[{{Bromberg} et~al.(2011){Bromberg}, {Nakar}, {Piran}, and
  {Sari}}]{Bromberg_NPS2011}
{Bromberg} O, {Nakar} E, {Piran} T, {Sari} R (2011) {The Propagation of
  Relativistic Jets in External Media}. \apj 740:100,
  \doi{10.1088/0004-637X/740/2/100}, \eprint{1107.1326}

\bibitem[{{Brown} et~al.(2007){Brown}, {Sarbach}, {Schnetter}, {Tiglio},
  {Diener}, {Hawke}, and {Pollney}}]{Brown_SSTDHP2007}
{Brown} D, {Sarbach} O, {Schnetter} E, {Tiglio} M, {Diener} P, {Hawke} I,
  {Pollney} D (2007) {Excision without excision}. \prd 76:081503,
  \doi{10.1103/PhysRevD.76.081503}, \eprint{0707.3101}

\bibitem[{{Br{\"u}gmann} et~al.(2004){Br{\"u}gmann}, {Tichy}, and
  {Jansen}}]{Brugmann_Tichy_Jansen2004}
{Br{\"u}gmann} B, {Tichy} W, {Jansen} N (2004) {Numerical Simulation of
  Orbiting Black Holes}. \prl 92:211101, \doi{10.1103/PhysRevLett.92.211101},
  \eprint{gr-qc/0312112}

\bibitem[{{Br{\"u}gmann} et~al.(2008){Br{\"u}gmann}, {Gonz{\'a}lez}, {Hannam},
  {Husa}, {Sperhake}, and {Tichy}}]{Brugmann_GHHST2008}
{Br{\"u}gmann} B, {Gonz{\'a}lez} JA, {Hannam} M, {Husa} S, {Sperhake} U,
  {Tichy} W (2008) {Calibration of moving puncture simulations}. \prd
  77:024027, \doi{10.1103/PhysRevD.77.024027}, \eprint{gr-qc/0610128}

\bibitem[{{Bulla} et~al.(2021){Bulla}, {Kyutoku}, {Tanaka}, {Covino}, {Bruten},
  {Matsumoto}, {Maund}, {Testa}, and {Wiersema}}]{Bulla_KTCBMMTW2021}
{Bulla} M, {Kyutoku} K, {Tanaka} M, {Covino} S, {Bruten} JR, {Matsumoto} T,
  {Maund} JR, {Testa} V, {Wiersema} K (2021) {Polarized kilonovae from black
  hole-neutron star mergers}. \mnras 501:1891--1899,
  \doi{10.1093/mnras/staa3796}, \eprint{2009.07279}

\bibitem[{{Buonanno} and {Damour}(1999)}]{Buonanno_Damour1999}
{Buonanno} A, {Damour} T (1999) {Effective one-body approach to general
  relativistic two-body dynamics}. \prd 59:084006,
  \doi{10.1103/PhysRevD.59.084006}, \eprint{gr-qc/9811091}

\bibitem[{{Buonanno} and {Damour}(2000)}]{Buonanno_Damour2000}
{Buonanno} A, {Damour} T (2000) {Transition from inspiral to plunge in binary
  black hole coalescences}. \prd 62:064015, \doi{10.1103/PhysRevD.62.064015},
  \eprint{gr-qc/0001013}

\bibitem[{{Buonanno} et~al.(2003){Buonanno}, {Chen}, and
  {Vallisneri}}]{Buonanno_Chen_Vallisneri2003}
{Buonanno} A, {Chen} Y, {Vallisneri} M (2003) {Detecting gravitational waves
  from precessing binaries of spinning compact objects: Adiabatic limit}. \prd
  67:104025, \doi{10.1103/PhysRevD.67.104025}, \eprint{gr-qc/0211087}

\bibitem[{{Buonanno} et~al.(2006){Buonanno}, {Chen}, and
  {Damour}}]{Buonanno_Chen_Damour2006}
{Buonanno} A, {Chen} Y, {Damour} T (2006) {Transition from inspiral to plunge
  in precessing binaries of spinning black holes}. \prd 74:104005,
  \doi{10.1103/PhysRevD.74.104005}, \eprint{gr-qc/0508067}

\bibitem[{{Buonanno} et~al.(2009){Buonanno}, {Iyer}, {Ochsner}, {Pan}, and
  {Sathyaprakash}}]{Buonanno_IOPS2009}
{Buonanno} A, {Iyer} BR, {Ochsner} E, {Pan} Y, {Sathyaprakash} BS (2009)
  {Comparison of post-Newtonian templates for compact binary inspiral signals
  in gravitational-wave detectors}. \prd 80:084043,
  \doi{10.1103/PhysRevD.80.084043}, \eprint{0907.0700}

\bibitem[{{Buonanno} et~al.(2011){Buonanno}, {Kidder}, {Mrou{\'e}}, {Pfeiffer},
  and {Taracchini}}]{Buonanno_KMPT2011}
{Buonanno} A, {Kidder} LE, {Mrou{\'e}} AH, {Pfeiffer} HP, {Taracchini} A (2011)
  {Reducing orbital eccentricity of precessing black-hole binaries}. \prd
  83:104034, \doi{10.1103/PhysRevD.83.104034}, \eprint{1012.1549}

\bibitem[{{Burbidge} et~al.(1957){Burbidge}, {Burbidge}, {Fowler}, and
  {Hoyle}}]{Burbidge_BFH1957}
{Burbidge} EM, {Burbidge} GR, {Fowler} WA, {Hoyle} F (1957) {Synthesis of the
  Elements in Stars}. Reviews of Modern Physics 29:547--650,
  \doi{10.1103/RevModPhys.29.547}

\bibitem[{{Cameron}(1957)}]{Cameron1957}
{Cameron} AGW (1957) {Nuclear Reactions in Stars and Nucleogenesis}. \pasp
  69:201, \doi{10.1086/127051}

\bibitem[{{Cameron} and {Iben}(1986)}]{Cameron_Iben1986}
{Cameron} AGW, {Iben} J I (1986) {On the Behavior of Double Degenerate Binaries
  Associated with Type I Supernovae}. \apj 305:228, \doi{10.1086/164242}

\bibitem[{{Campanelli} et~al.(2006){Campanelli}, {Lousto}, {Marronetti}, and
  {Zlochower}}]{Campanelli_LMZ2006}
{Campanelli} M, {Lousto} CO, {Marronetti} P, {Zlochower} Y (2006) {Accurate
  Evolutions of Orbiting Black-Hole Binaries without Excision}. \prl 96:111101,
  \doi{10.1103/PhysRevLett.96.111101}, \eprint{gr-qc/0511048}

\bibitem[{{Cannizzo} et~al.(2011){Cannizzo}, {Troja}, and
  {Gehrels}}]{Cannizzo_Troja_Gehrels2011}
{Cannizzo} JK, {Troja} E, {Gehrels} N (2011) {Fall-back Disks in Long and Short
  Gamma-Ray Bursts}. \apj 734:35, \doi{10.1088/0004-637X/734/1/35},
  \eprint{1104.0456}

\bibitem[{{Carbone} and {Schwenk}(2019)}]{Carbone_Schwenk2019}
{Carbone} A, {Schwenk} A (2019) {Ab initio constraints on thermal effects of
  the nuclear equation of state}. \prc 100:025805,
  \doi{10.1103/PhysRevC.100.025805}, \eprint{1904.00924}

\bibitem[{{Cardall} et~al.(2013){Cardall}, {Endeve}, and
  {Mezzacappa}}]{Cardall_Endeve_Mezzacappa2013}
{Cardall} CY, {Endeve} E, {Mezzacappa} A (2013) {Conservative 3+1 general
  relativistic Boltzmann equation}. \prd 88:023011,
  \doi{10.1103/PhysRevD.88.023011}, \eprint{1305.0037}

\bibitem[{{Carrasco} and {Shibata}(2020)}]{Carrasco_Shibata2020}
{Carrasco} F, {Shibata} M (2020) {Magnetosphere of an orbiting neutron star}.
  \prd 101:063017, \doi{10.1103/PhysRevD.101.063017}, \eprint{2001.04210}

\bibitem[{{Carrasco} et~al.(2021){Carrasco}, {Shibata}, and
  {Reula}}]{Carrasco_Shibata_Reula2021}
{Carrasco} F, {Shibata} M, {Reula} O (2021) {Magnetospheres of black
  hole-neutron star binaries}. \prd 104:063004,
  \doi{10.1103/PhysRevD.104.063004}, \eprint{2106.09081}

\bibitem[{{Caudill} et~al.(2006){Caudill}, {Cook}, {Grigsby}, and
  {Pfeiffer}}]{Caudill_CGP2006}
{Caudill} M, {Cook} GB, {Grigsby} JD, {Pfeiffer} HP (2006) {Circular orbits and
  spin in black-hole initial data}. \prd 74:064011,
  \doi{10.1103/PhysRevD.74.064011}, \eprint{gr-qc/0605053}

\bibitem[{{Centrella} et~al.(2010){Centrella}, {Baker}, {Kelly}, and {van
  Meter}}]{Centrella_BKM2010}
{Centrella} J, {Baker} JG, {Kelly} BJ, {van Meter} JR (2010) {Black-hole
  binaries, gravitational waves, and numerical relativity}. Reviews of Modern
  Physics 82:3069--3119, \doi{10.1103/RevModPhys.82.3069}, \eprint{1010.5260}

\bibitem[{{Chandrasekhar}(1969)}]{Chandrasekhar}
{Chandrasekhar} S (1969) {Ellipsoidal figures of equilibrium}. Yale University
  Press, New Haven

\bibitem[{{Chatziioannou} et~al.(2013){Chatziioannou}, {Poisson}, and
  {Yunes}}]{Chatziioannou_Poisson_Yunes2013}
{Chatziioannou} K, {Poisson} E, {Yunes} N (2013) {Tidal heating and torquing of
  a Kerr black hole to next-to-leading order in the tidal coupling}. \prd
  87:044022, \doi{10.1103/PhysRevD.87.044022}, \eprint{1211.1686}

\bibitem[{{Chatziioannou} et~al.(2015){Chatziioannou}, {Cornish}, {Klein}, and
  {Yunes}}]{Chatziioannou_CKY2015}
{Chatziioannou} K, {Cornish} N, {Klein} A, {Yunes} N (2015) {Spin-precession:
  Breaking the Black Hole-Neutron Star Degeneracy}. \apjl 798:L17,
  \doi{10.1088/2041-8205/798/1/L17}, \eprint{1402.3581}

\bibitem[{{Chawla} et~al.(2010){Chawla}, {Anderson}, {Besselman}, {Lehner},
  {Liebling}, {Motl}, and {Neilsen}}]{Chawla_ABLLMN2010}
{Chawla} S, {Anderson} M, {Besselman} M, {Lehner} L, {Liebling} SL, {Motl} PM,
  {Neilsen} D (2010) {Mergers of Magnetized Neutron Stars with Spinning Black
  Holes: Disruption, Accretion, and Fallback}. \prl 105:111101,
  \doi{10.1103/PhysRevLett.105.111101}, \eprint{1006.2839}

\bibitem[{{Chen} and {Beloborodov}(2007)}]{Chen_Beloborodov2007}
{Chen} WX, {Beloborodov} AM (2007) {Neutrino-cooled Accretion Disks around
  Spinning Black Holes}. \apj 657:383--399, \doi{10.1086/508923},
  \eprint{astro-ph/0607145}

\bibitem[{{Christie} et~al.(2019){Christie}, {Lalakos}, {Tchekhovskoy},
  {Fern{\'a}ndez}, {Foucart}, {Quataert}, and {Kasen}}]{Christie_LTFFQK2019}
{Christie} IM, {Lalakos} A, {Tchekhovskoy} A, {Fern{\'a}ndez} R, {Foucart} F,
  {Quataert} E, {Kasen} D (2019) {The role of magnetic field geometry in the
  evolution of neutron star merger accretion discs}. \mnras 490:4811--4825,
  \doi{10.1093/mnras/stz2552}, \eprint{1907.02079}

\bibitem[{{Christodoulou}(1970)}]{Christodoulou1970}
{Christodoulou} D (1970) {Reversible and Irreversible Transformations in
  Black-Hole Physics}. \prl 25:1596--1597, \doi{10.1103/PhysRevLett.25.1596}

\bibitem[{{Clark} and {Eardley}(1977)}]{Clark_Eardley1977}
{Clark} JPA, {Eardley} DM (1977) {Evolution of close neutron star binaries.}
  \apj 215:311--322, \doi{10.1086/155360}

\bibitem[{{Constantinou} et~al.(2015){Constantinou}, {Muccioli}, {Prakash}, and
  {Lattimer}}]{Constantinou_MPL2015}
{Constantinou} C, {Muccioli} B, {Prakash} M, {Lattimer} JM (2015) {Thermal
  properties of hot and dense matter with finite range interactions}. \prc
  92:025801, \doi{10.1103/PhysRevC.92.025801}, \eprint{1504.03982}

\bibitem[{{Cook}(2000)}]{Cook2000}
{Cook} GB (2000) {Initial Data for Numerical Relativity}. Living Rev Relativ
  3:5, \doi{10.12942/lrr-2000-5}, \eprint{gr-qc/0007085}

\bibitem[{{Cook}(2002)}]{Cook2002}
{Cook} GB (2002) {Corotating and irrotational binary black holes in
  quasicircular orbits}. \prd 65:084003, \doi{10.1103/PhysRevD.65.084003},
  \eprint{gr-qc/0108076}

\bibitem[{{Cook} and {Baumgarte}(2008)}]{Cook_Baumgarte2008}
{Cook} GB, {Baumgarte} TW (2008) {Excision boundary conditions for the
  conformal metric}. \prd 78:104016, \doi{10.1103/PhysRevD.78.104016},
  \eprint{0810.4493}

\bibitem[{{Cook} and {Pfeiffer}(2004)}]{Cook_Pfeiffer2004}
{Cook} GB, {Pfeiffer} HP (2004) {Excision boundary conditions for black-hole
  initial data}. \prd 70:104016, \doi{10.1103/PhysRevD.70.104016},
  \eprint{gr-qc/0407078}

\bibitem[{{Cook} and {Whiting}(2007)}]{Cook_Whiting2007}
{Cook} GB, {Whiting} BF (2007) {Approximate Killing vectors on S$^{2}$}. \prd
  76:041501, \doi{10.1103/PhysRevD.76.041501}, \eprint{0706.0199}

\bibitem[{{Cooperstein}(1988)}]{Cooperstein1988}
{Cooperstein} J (1988) {Neutrinos in supernovae}. \physrep 163:95--126,
  \doi{10.1016/0370-1573(88)90038-5}

\bibitem[{{Cordero-Carri{\'o}n} et~al.(2009){Cordero-Carri{\'o}n},
  {Cerd{\'a}-Dur{\'a}n}, {Dimmelmeier}, {Jaramillo}, {Novak}, and
  {Gourgoulhon}}]{CorderoCarrion_CDJNG2009}
{Cordero-Carri{\'o}n} I, {Cerd{\'a}-Dur{\'a}n} P, {Dimmelmeier} H, {Jaramillo}
  JL, {Novak} J, {Gourgoulhon} E (2009) {Improved constrained scheme for the
  Einstein equations: An approach to the uniqueness issue}. \prd 79:024017,
  \doi{10.1103/PhysRevD.79.024017}, \eprint{0809.2325}

\bibitem[{{Coulter} et~al.(2017){Coulter}, {Foley}, {Kilpatrick}, {Drout},
  {Piro}, {Shappee}, {Siebert}, {Simon}, {Ulloa}, {Kasen}, and
  et~al.}]{Coulter_etal2017}
{Coulter} DA, {Foley} RJ, {Kilpatrick} CD, {Drout} MR, {Piro} AL, {Shappee} BJ,
  {Siebert} MR, {Simon} JD, {Ulloa} N, {Kasen} D, et~al (2017) {Swope Supernova
  Survey 2017a (SSS17a), the optical counterpart to a gravitational wave
  source}. Science 358(6370):1556--1558, \doi{10.1126/science.aap9811},
  \eprint{1710.05452}

\bibitem[{{Cowling}(1933)}]{Cowling1933}
{Cowling} TG (1933) {The magnetic field of sunspots}. \mnras 94:39--48,
  \doi{10.1093/mnras/94.1.39}

\bibitem[{{Cromartie} et~al.(2020){Cromartie}, {Fonseca}, {Ransom}, {Demorest},
  {Arzoumanian}, {Blumer}, {Brook}, {DeCesar}, {Dolch}, {Ellis}, and
  et~al.}]{Cromartie_etal2020}
{Cromartie} HT, {Fonseca} E, {Ransom} SM, {Demorest} PB, {Arzoumanian} Z,
  {Blumer} H, {Brook} PR, {DeCesar} ME, {Dolch} T, {Ellis} JA, et~al (2020)
  {Relativistic Shapiro delay measurements of an extremely massive millisecond
  pulsar}. Nature Astronomy 4:72--76, \doi{10.1038/s41550-019-0880-2},
  \eprint{1904.06759}

\bibitem[{{Cutler} and {Flanagan}(1994)}]{Cutler_Flanagan1994}
{Cutler} C, {Flanagan} {\'E}E (1994) {Gravitational waves from merging compact
  binaries: How accurately can one extract the binary's parameters from the
  inspiral waveform\textbackslash?} \prd 49:2658--2697,
  \doi{10.1103/PhysRevD.49.2658}, \eprint{gr-qc/9402014}

\bibitem[{{Dai}(2019)}]{Dai2019}
{Dai} ZG (2019) {Inspiral of a Spinning Black Hole-Magnetized Neutron Star
  Binary: Increasing Charge and Electromagnetic Emission}. \apjl 873:L13,
  \doi{10.3847/2041-8213/ab0b45}, \eprint{1902.07939}

\bibitem[{{Damour} and {Nagar}(2009)}]{Damour_Nagar2009}
{Damour} T, {Nagar} A (2009) {Relativistic tidal properties of neutron stars}.
  \prd 80:084035, \doi{10.1103/PhysRevD.80.084035}, \eprint{0906.0096}

\bibitem[{{Damour} et~al.(2001){Damour}, {Iyer}, and
  {Sathyaprakash}}]{Damour_Iyer_Sathyaprakash2001}
{Damour} T, {Iyer} BR, {Sathyaprakash} BS (2001) {Comparison of search
  templates for gravitational waves from binary inspiral}. \prd 63:044023,
  \doi{10.1103/PhysRevD.63.044023}, \eprint{gr-qc/0010009}

\bibitem[{{Darbha} et~al.(2021){Darbha}, {Kasen}, {Foucart}, and
  {Price}}]{Darbha_KFP2021}
{Darbha} S, {Kasen} D, {Foucart} F, {Price} DJ (2021) {Electromagnetic
  Signatures from the Tidal Tail of a Black Hole-Neutron Star Merger}. \apj
  915:69, \doi{10.3847/1538-4357/abff5d}, \eprint{2103.03378}

\bibitem[{{De} et~al.(2018){De}, {Finstad}, {Lattimer}, {Brown}, {Berger}, and
  {Biwer}}]{De_FLBBB2018}
{De} S, {Finstad} D, {Lattimer} JM, {Brown} DA, {Berger} E, {Biwer} CM (2018)
  {Tidal Deformabilities and Radii of Neutron Stars from the Observation of
  GW170817}. \prl 121:091102, \doi{10.1103/PhysRevLett.121.091102},
  \eprint{1804.08583}

\bibitem[{{De Felice} et~al.(2019){De Felice}, {Larrouturou}, {Mukohyama}, and
  {Oliosi}}]{DeFelice_LMO2019}
{De Felice} A, {Larrouturou} F, {Mukohyama} S, {Oliosi} M (2019) {On the
  absence of conformally flat slicings of the Kerr spacetime}. \prd 100:124044,
  \doi{10.1103/PhysRevD.100.124044}, \eprint{1908.03456}

\bibitem[{{Deaton} et~al.(2013){Deaton}, {Duez}, {Foucart}, {O'Connor}, {Ott},
  {Kidder}, {Muhlberger}, {Scheel}, and {Szilagyi}}]{Deaton_DFOOKMSS2013}
{Deaton} MB, {Duez} MD, {Foucart} F, {O'Connor} E, {Ott} CD, {Kidder} LE,
  {Muhlberger} CD, {Scheel} MA, {Szilagyi} B (2013) {Black Hole-Neutron Star
  Mergers with a Hot Nuclear Equation of State: Outflow and Neutrino-cooled
  Disk for a Low-mass, High-spin Case}. \apj 776:47,
  \doi{10.1088/0004-637X/776/1/47}, \eprint{1304.3384}

\bibitem[{{Demorest} et~al.(2010){Demorest}, {Pennucci}, {Ransom}, {Roberts},
  and {Hessels}}]{Demorest_PRRH2010}
{Demorest} PB, {Pennucci} T, {Ransom} SM, {Roberts} MSE, {Hessels} JWT (2010)
  {A two-solar-mass neutron star measured using Shapiro delay}. \nat
  467:1081--1083, \doi{10.1038/nature09466}, \eprint{1010.5788}

\bibitem[{{Desai} et~al.(2019){Desai}, {Metzger}, and
  {Foucart}}]{Desai_Metzger_Foucart2019}
{Desai} D, {Metzger} BD, {Foucart} F (2019) {Imprints of r-process heating on
  fall-back accretion: distinguishing black hole-neutron star from double
  neutron star mergers}. \mnras 485:4404--4412, \doi{10.1093/mnras/stz644},
  \eprint{1812.04641}

\bibitem[{{Detweiler}(1994)}]{Detweiler1994}
{Detweiler} S (1994) {Periodic solutions of the Einstein equations for binary
  systems}. \prd 50:4929--4943, \doi{10.1103/PhysRevD.50.4929},
  \eprint{gr-qc/9312016}

\bibitem[{{Di Matteo} et~al.(2002){Di Matteo}, {Perna}, and
  {Narayan}}]{DiMatteo_Perna_Narayan2002}
{Di Matteo} T, {Perna} R, {Narayan} R (2002) {Neutrino Trapping and Accretion
  Models for Gamma-Ray Bursts}. \apj 579:706--715, \doi{10.1086/342832},
  \eprint{astro-ph/0207319}

\bibitem[{{Dietrich} et~al.(2015){Dietrich}, {Bernuzzi}, {Ujevic}, and
  {Br{\"u}gmann}}]{Dietrich_BUB2015}
{Dietrich} T, {Bernuzzi} S, {Ujevic} M, {Br{\"u}gmann} B (2015) {Numerical
  relativity simulations of neutron star merger remnants using conservative
  mesh refinement}. \prd 91:124041, \doi{10.1103/PhysRevD.91.124041},
  \eprint{1504.01266}

\bibitem[{{Dietrich} et~al.(2017{\natexlab{a}}){Dietrich}, {Bernuzzi}, and
  {Tichy}}]{Dietrich_Bernuzzi_Tichy2017}
{Dietrich} T, {Bernuzzi} S, {Tichy} W (2017{\natexlab{a}}) {Closed-form tidal
  approximants for binary neutron star gravitational waveforms constructed from
  high-resolution numerical relativity simulations}. \prd 96:121501,
  \doi{10.1103/PhysRevD.96.121501}, \eprint{1706.02969}

\bibitem[{{Dietrich} et~al.(2017{\natexlab{b}}){Dietrich}, {Ujevic}, {Tichy},
  {Bernuzzi}, and {Br{\"u}gmann}}]{Dietrich_UTBB2017}
{Dietrich} T, {Ujevic} M, {Tichy} W, {Bernuzzi} S, {Br{\"u}gmann} B
  (2017{\natexlab{b}}) {Gravitational waves and mass ejecta from binary neutron
  star mergers: Effect of the mass ratio}. \prd 95:024029,
  \doi{10.1103/PhysRevD.95.024029}, \eprint{1607.06636}

\bibitem[{{Dietrich} et~al.(2019){Dietrich}, {Samajdar}, {Khan},
  {Johnson-McDaniel}, {Dudi}, and {Tichy}}]{Dietrich_SKJDT2019}
{Dietrich} T, {Samajdar} A, {Khan} S, {Johnson-McDaniel} NK, {Dudi} R, {Tichy}
  W (2019) {Improving the NRTidal model for binary neutron star systems}. \prd
  100:044003, \doi{10.1103/PhysRevD.100.044003}, \eprint{1905.06011}

\bibitem[{{Dionysopoulou} et~al.(2013){Dionysopoulou}, {Alic}, {Palenzuela},
  {Rezzolla}, and {Giacomazzo}}]{Dionysopoulou_APRG2013}
{Dionysopoulou} K, {Alic} D, {Palenzuela} C, {Rezzolla} L, {Giacomazzo} B
  (2013) {General-relativistic resistive magnetohydrodynamics in three
  dimensions: Formulation and tests}. \prd 88:044020,
  \doi{10.1103/PhysRevD.88.044020}, \eprint{1208.3487}

\bibitem[{{Dominik} et~al.(2015){Dominik}, {Berti}, {O'Shaughnessy}, {Mandel},
  {Belczynski}, {Fryer}, {Holz}, {Bulik}, and
  {Pannarale}}]{Dominik_BOMBFHBP2015}
{Dominik} M, {Berti} E, {O'Shaughnessy} R, {Mandel} I, {Belczynski} K, {Fryer}
  C, {Holz} DE, {Bulik} T, {Pannarale} F (2015) {Double Compact Objects III:
  Gravitational-wave Detection Rates}. \apj 806:263,
  \doi{10.1088/0004-637X/806/2/263}, \eprint{1405.7016}

\bibitem[{{D'Orazio} et~al.(2016){D'Orazio}, {Levin}, {Murray}, and
  {Price}}]{DOrazio_LMP2016}
{D'Orazio} DJ, {Levin} J, {Murray} NW, {Price} L (2016) {Bright transients from
  strongly-magnetized neutron star-black hole mergers}. \prd 94:023001,
  \doi{10.1103/PhysRevD.94.023001}, \eprint{1601.00017}

\bibitem[{{Douchin} and {Haensel}(2001)}]{Douchin_Haensel2001}
{Douchin} F, {Haensel} P (2001) {A unified equation of state of dense matter
  and neutron star structure}. \aap 380:151--167,
  \doi{10.1051/0004-6361:20011402}, \eprint{astro-ph/0111092}

\bibitem[{{Dreyer} et~al.(2003){Dreyer}, {Krishnan}, {Shoemaker}, and
  {Schnetter}}]{Dreyer_KSS2003}
{Dreyer} O, {Krishnan} B, {Shoemaker} D, {Schnetter} E (2003) {Introduction to
  isolated horizons in numerical relativity}. \prd 67:024018,
  \doi{10.1103/PhysRevD.67.024018}, \eprint{gr-qc/0206008}

\bibitem[{{Duez} and {Zlochower}(2019)}]{Duez_Zlochower2019}
{Duez} MD, {Zlochower} Y (2019) {Numerical relativity of compact binaries in
  the 21st century}. Reports on Progress in Physics 82:016902,
  \doi{10.1088/1361-6633/aadb16}, \eprint{1808.06011}

\bibitem[{{Duez} et~al.(2003){Duez}, {Marronetti}, {Shapiro}, and
  {Baumgarte}}]{Duez_MSB2003}
{Duez} MD, {Marronetti} P, {Shapiro} SL, {Baumgarte} TW (2003) {Hydrodynamic
  simulations in 3+1 general relativity}. \prd 67:024004,
  \doi{10.1103/PhysRevD.67.024004}, \eprint{gr-qc/0209102}

\bibitem[{{Duez} et~al.(2008){Duez}, {Foucart}, {Kidder}, {Pfeiffer}, {Scheel},
  and {Teukolsky}}]{Duez_FKPST2008}
{Duez} MD, {Foucart} F, {Kidder} LE, {Pfeiffer} HP, {Scheel} MA, {Teukolsky} SA
  (2008) {Evolving black hole-neutron star binaries in general relativity using
  pseudospectral and finite difference methods}. \prd 78:104015,
  \doi{10.1103/PhysRevD.78.104015}, \eprint{0809.0002}

\bibitem[{{Duez} et~al.(2010){Duez}, {Foucart}, {Kidder}, {Ott}, and
  {Teukolsky}}]{Duez_FKOT2010}
{Duez} MD, {Foucart} F, {Kidder} LE, {Ott} CD, {Teukolsky} SA (2010) {Equation
  of state effects in black hole-neutron star mergers}. Class Quantum Grav
  27:114106, \doi{10.1088/0264-9381/27/11/114106}, \eprint{0912.3528}

\bibitem[{{Duffell} et~al.(2015){Duffell}, {Quataert}, and
  {MacFadyen}}]{Duffell_Quataert_MacFadyen2015}
{Duffell} PC, {Quataert} E, {MacFadyen} AI (2015) {A Narrow Short-duration GRB
  Jet from a Wide Central Engine}. \apj 813:64,
  \doi{10.1088/0004-637X/813/1/64}, \eprint{1505.05538}

\bibitem[{{East} and {Lehner}(2019)}]{East_Lehner2019}
{East} WE, {Lehner} L (2019) {Fate of a neutron star with an endoparasitic
  black hole and implications for dark matter}. \prd 100:124026,
  \doi{10.1103/PhysRevD.100.124026}, \eprint{1909.07968}

\bibitem[{{East} et~al.(2012){East}, {Pretorius}, and
  {Stephens}}]{East_Pretorius_Stephens2012}
{East} WE, {Pretorius} F, {Stephens} BC (2012) {Eccentric black hole-neutron
  star mergers: Effects of black hole spin and equation of state}. \prd
  85:124009, \doi{10.1103/PhysRevD.85.124009}, \eprint{1111.3055}

\bibitem[{{East} et~al.(2015){East}, {Paschalidis}, and
  {Pretorius}}]{East_Paschalidis_Pretorius2015}
{East} WE, {Paschalidis} V, {Pretorius} F (2015) {Eccentric Mergers of Black
  Holes with Spinning Neutron Stars}. \apjl 807:L3,
  \doi{10.1088/2041-8205/807/1/L3}, \eprint{1503.07171}

\bibitem[{{East} et~al.(2021){East}, {Lehner}, {Liebling}, and
  {Palenzuela}}]{East_LLP2021}
{East} WE, {Lehner} L, {Liebling} SL, {Palenzuela} C (2021) {Multimessenger
  Signals from Black Hole-Neutron Star Mergers without Significant Tidal
  Disruption}. \apjl 912:L18, \doi{10.3847/2041-8213/abf566},
  \eprint{2101.12214}

\bibitem[{{Eastman} and {Pinto}(1993)}]{Eastman_Pinto1993}
{Eastman} RG, {Pinto} PA (1993) {Spectrum Formation in Supernovae: Numerical
  Techniques}. \apj 412:731, \doi{10.1086/172957}

\bibitem[{{Eggleton}(1983)}]{Eggleton1983}
{Eggleton} PP (1983) {Aproximations to the radii of Roche lobes.} \apj
  268:368--369, \doi{10.1086/160960}

\bibitem[{{Eichler} et~al.(1989){Eichler}, {Livio}, {Piran}, and
  {Schramm}}]{Eichler_LPS1989}
{Eichler} D, {Livio} M, {Piran} T, {Schramm} DN (1989) {Nucleosynthesis,
  neutrino bursts and {\ensuremath{\gamma}}-rays from coalescing neutron
  stars}. \nat 340:126--128, \doi{10.1038/340126a0}

\bibitem[{{Etienne} et~al.(2007){Etienne}, {Faber}, {Liu}, {Shapiro}, and
  {Baumgarte}}]{Etienne_FLSB2007}
{Etienne} ZB, {Faber} JA, {Liu} YT, {Shapiro} SL, {Baumgarte} TW (2007)
  {Filling the holes: Evolving excised binary black hole initial data with
  puncture techniques}. \prd 76:101503, \doi{10.1103/PhysRevD.76.101503},
  \eprint{0707.2083}

\bibitem[{{Etienne} et~al.(2008){Etienne}, {Faber}, {Liu}, {Shapiro},
  {Taniguchi}, and {Baumgarte}}]{Etienne_FLSTB2008}
{Etienne} ZB, {Faber} JA, {Liu} YT, {Shapiro} SL, {Taniguchi} K, {Baumgarte} TW
  (2008) {Fully general relativistic simulations of black hole-neutron star
  mergers}. \prd 77:084002, \doi{10.1103/PhysRevD.77.084002},
  \eprint{0712.2460}

\bibitem[{{Etienne} et~al.(2009){Etienne}, {Liu}, {Shapiro}, and
  {Baumgarte}}]{Etienne_LSB2009}
{Etienne} ZB, {Liu} YT, {Shapiro} SL, {Baumgarte} TW (2009) {General
  relativistic simulations of black-hole-neutron-star mergers: Effects of
  black-hole spin}. \prd 79:044024, \doi{10.1103/PhysRevD.79.044024},
  \eprint{0812.2245}

\bibitem[{{Etienne} et~al.(2012{\natexlab{a}}){Etienne}, {Liu}, {Paschalidis},
  and {Shapiro}}]{Etienne_LPS2012}
{Etienne} ZB, {Liu} YT, {Paschalidis} V, {Shapiro} SL (2012{\natexlab{a}})
  {General relativistic simulations of black-hole-neutron-star mergers: Effects
  of magnetic fields}. \prd 85:064029, \doi{10.1103/PhysRevD.85.064029},
  \eprint{1112.0568}

\bibitem[{{Etienne} et~al.(2012{\natexlab{b}}){Etienne}, {Paschalidis}, {Liu},
  and {Shapiro}}]{Etienne_PLS2012}
{Etienne} ZB, {Paschalidis} V, {Liu} YT, {Shapiro} SL (2012{\natexlab{b}})
  {Relativistic magnetohydrodynamics in dynamical spacetimes: Improved
  electromagnetic gauge condition for adaptive mesh refinement grids}. \prd
  85:024013, \doi{10.1103/PhysRevD.85.024013}, \eprint{1110.4633}

\bibitem[{{Etienne} et~al.(2012{\natexlab{c}}){Etienne}, {Paschalidis}, and
  {Shapiro}}]{Etienne_Paschalidis_Shapiro2012}
{Etienne} ZB, {Paschalidis} V, {Shapiro} SL (2012{\natexlab{c}})
  {General-relativistic simulations of black-hole-neutron-star mergers: Effects
  of tilted magnetic fields}. \prd 86:084026, \doi{10.1103/PhysRevD.86.084026},
  \eprint{1209.1632}

\bibitem[{{Evans} and {Hawley}(1988)}]{Evans_Hawley1988}
{Evans} CR, {Hawley} JF (1988) {Simulation of Magnetohydrodynamic Flows: A
  Constrained Transport Model}. \apj 332:659, \doi{10.1086/166684}

\bibitem[{{Faber} et~al.(2006{\natexlab{a}}){Faber}, {Baumgarte}, {Shapiro},
  and {Taniguchi}}]{Faber_BST2006}
{Faber} JA, {Baumgarte} TW, {Shapiro} SL, {Taniguchi} K (2006{\natexlab{a}})
  {General Relativistic Binary Merger Simulations and Short Gamma-Ray Bursts}.
  \apjl 641:L93--L96, \doi{10.1086/504111}, \eprint{astro-ph/0603277}

\bibitem[{{Faber} et~al.(2006{\natexlab{b}}){Faber}, {Baumgarte}, {Shapiro},
  {Taniguchi}, and {Rasio}}]{Faber_BSTR2006}
{Faber} JA, {Baumgarte} TW, {Shapiro} SL, {Taniguchi} K, {Rasio} FA
  (2006{\natexlab{b}}) {Dynamical evolution of black hole-neutron star binaries
  in general relativity: Simulations of tidal disruption}. \prd 73:024012,
  \doi{10.1103/PhysRevD.73.024012}, \eprint{astro-ph/0511366}

\bibitem[{{Farris} et~al.(2012){Farris}, {Gold}, {Paschalidis}, {Etienne}, and
  {Shapiro}}]{Farris_GPES2012}
{Farris} BD, {Gold} R, {Paschalidis} V, {Etienne} ZB, {Shapiro} SL (2012)
  {Binary Black-Hole Mergers in Magnetized Disks: Simulations in Full General
  Relativity}. \prl 109:221102, \doi{10.1103/PhysRevLett.109.221102},
  \eprint{1207.3354}

\bibitem[{{Farrow} et~al.(2019){Farrow}, {Zhu}, and
  {Thrane}}]{Farrow_Zhu_Thrane2019}
{Farrow} N, {Zhu} XJ, {Thrane} E (2019) {The Mass Distribution of Galactic
  Double Neutron Stars}. \apj 876:18, \doi{10.3847/1538-4357/ab12e3},
  \eprint{1902.03300}

\bibitem[{{Favata}(2014)}]{Favata2014}
{Favata} M (2014) {Systematic Parameter Errors in Inspiraling Neutron Star
  Binaries}. \prl 112:101101, \doi{10.1103/PhysRevLett.112.101101},
  \eprint{1310.8288}

\bibitem[{{Fern{\'a}ndez} and {Metzger}(2013)}]{Fernandez_Metzger2013}
{Fern{\'a}ndez} R, {Metzger} BD (2013) {Delayed outflows from black hole
  accretion tori following neutron star binary coalescence}. \mnras
  435:502--517, \doi{10.1093/mnras/stt1312}, \eprint{1304.6720}

\bibitem[{{Fern{\'a}ndez} et~al.(2017){Fern{\'a}ndez}, {Foucart}, {Kasen},
  {Lippuner}, {Desai}, and {Roberts}}]{Fernandez_FKLDR2017}
{Fern{\'a}ndez} R, {Foucart} F, {Kasen} D, {Lippuner} J, {Desai} D, {Roberts}
  LF (2017) {Dynamics, nucleosynthesis, and kilonova signature of black
  hole{\textemdash}neutron star merger ejecta}. Class Quantum Grav 34:154001,
  \doi{10.1088/1361-6382/aa7a77}, \eprint{1612.04829}

\bibitem[{{Fern{\'a}ndez} et~al.(2019){Fern{\'a}ndez}, {Tchekhovskoy},
  {Quataert}, {Foucart}, and {Kasen}}]{Fernandez_TQFK2019}
{Fern{\'a}ndez} R, {Tchekhovskoy} A, {Quataert} E, {Foucart} F, {Kasen} D
  (2019) {Long-term GRMHD simulations of neutron star merger accretion discs:
  implications for electromagnetic counterparts}. \mnras 482:3373--3393,
  \doi{10.1093/mnras/sty2932}, \eprint{1808.00461}

\bibitem[{{Fern{\'a}ndez} et~al.(2020){Fern{\'a}ndez}, {Foucart}, and
  {Lippuner}}]{Fernandez_Foucart_Lippuner2020}
{Fern{\'a}ndez} R, {Foucart} F, {Lippuner} J (2020) {The landscape of disc
  outflows from black hole-neutron star mergers}. \mnras 497:3221--3233,
  \doi{10.1093/mnras/staa2209}, \eprint{2005.14208}

\bibitem[{{Ferrari} et~al.(2009){Ferrari}, {Gualtieri}, and
  {Pannarale}}]{Ferrari_Gualtieri_Pannarale2009}
{Ferrari} V, {Gualtieri} L, {Pannarale} F (2009) {A semi-relativistic model for
  tidal interactions in BH-NS coalescing binaries}. Class Quantum Grav
  26:125004, \doi{10.1088/0264-9381/26/12/125004}, \eprint{0801.2911}

\bibitem[{{Ferrari} et~al.(2010){Ferrari}, {Gualtieri}, and
  {Pannarale}}]{Ferrari_Gualtieri_Pannarale2010}
{Ferrari} V, {Gualtieri} L, {Pannarale} F (2010) {Neutron star tidal disruption
  in mixed binaries: The imprint of the equation of state}. \prd 81:064026,
  \doi{10.1103/PhysRevD.81.064026}, \eprint{0912.3692}

\bibitem[{{Ferrari} et~al.(2012){Ferrari}, {Gualtieri}, and
  {Maselli}}]{Ferrari_Gualtieri_Maselli2012}
{Ferrari} V, {Gualtieri} L, {Maselli} A (2012) {Tidal interaction in compact
  binaries: A post-Newtonian affine framework}. \prd 85:044045,
  \doi{10.1103/PhysRevD.85.044045}, \eprint{1111.6607}

\bibitem[{{Figura} et~al.(2020){Figura}, {Lu}, {Burgio}, {Li}, and
  {Schulze}}]{Figura_LBLS2020}
{Figura} A, {Lu} JJ, {Burgio} GF, {Li} ZH, {Schulze} HJ (2020) {Hybrid equation
  of state approach in binary neutron-star merger simulations}. \prd
  102:043006, \doi{10.1103/PhysRevD.102.043006}, \eprint{2005.08691}

\bibitem[{{Finn} and {Chernoff}(1993)}]{Finn_Chernoff1993}
{Finn} LS, {Chernoff} DF (1993) {Observing binary inspiral in gravitational
  radiation: One interferometer}. \prd 47:2198--2219,
  \doi{10.1103/PhysRevD.47.2198}, \eprint{gr-qc/9301003}

\bibitem[{{Fishbone}(1973)}]{Fishbone1973}
{Fishbone} LG (1973) {The Relativistic Roche Problem. I. Equilibrium Theory for
  a Body in Equatorial, Circular Orbit around a Kerr Black Hole}. \apj
  185:43--68, \doi{10.1086/152395}

\bibitem[{{Flanagan} and {Hinderer}(2008)}]{Flanagan_Hinderer2008}
{Flanagan} {\'E}{\'E}, {Hinderer} T (2008) {Constraining neutron-star tidal
  Love numbers with gravitational-wave detectors}. \prd 77:021502,
  \doi{10.1103/PhysRevD.77.021502}, \eprint{0709.1915}

\bibitem[{{Fong} et~al.(2015){Fong}, {Berger}, {Margutti}, and
  {Zauderer}}]{Fong_BMZ2015}
{Fong} W, {Berger} E, {Margutti} R, {Zauderer} BA (2015) {A Decade of
  Short-duration Gamma-Ray Burst Broadband Afterglows: Energetics, Circumburst
  Densities, and Jet Opening Angles}. \apj 815:102,
  \doi{10.1088/0004-637X/815/2/102}, \eprint{1509.02922}

\bibitem[{{Fonseca} et~al.(2021){Fonseca}, {Cromartie}, {Pennucci}, {Ray},
  {Kirichenko}, {Ransom}, {Demorest}, {Stairs}, {Arzoumanian}, {Guillemot}, and
  et~al.}]{Fonseca_etal2021}
{Fonseca} E, {Cromartie} HT, {Pennucci} TT, {Ray} PS, {Kirichenko} AY, {Ransom}
  SM, {Demorest} PB, {Stairs} IH, {Arzoumanian} Z, {Guillemot} L, et~al (2021)
  {Refined Mass and Geometric Measurements of the High-mass PSR J0740+6620}.
  \apjl 915:L12, \doi{10.3847/2041-8213/ac03b8}, \eprint{2104.00880}

\bibitem[{{Font}(2008)}]{Font2008}
{Font} JA (2008) {Numerical Hydrodynamics and Magnetohydrodynamics in General
  Relativity}. Living Rev Relativ 11:7, \doi{10.12942/lrr-2008-7}

\bibitem[{{Foucart}(2012)}]{Foucart2012}
{Foucart} F (2012) {Black-hole-neutron-star mergers: Disk mass predictions}.
  \prd 86:124007, \doi{10.1103/PhysRevD.86.124007}, \eprint{1207.6304}

\bibitem[{{Foucart}(2020)}]{Foucart2020}
{Foucart} F (2020) {A brief overview of black hole-neutron star mergers}.
  Frontiers in Astronomy and Space Sciences 7:46,
  \doi{10.3389/fspas.2020.00046}, \eprint{2006.10570}

\bibitem[{{Foucart} et~al.(2008){Foucart}, {Kidder}, {Pfeiffer}, and
  {Teukolsky}}]{Foucart_KPT2008}
{Foucart} F, {Kidder} LE, {Pfeiffer} HP, {Teukolsky} SA (2008) {Initial data
  for black hole neutron star binaries: A flexible, high-accuracy spectral
  method}. \prd 77:124051, \doi{10.1103/PhysRevD.77.124051}, \eprint{0804.3787}

\bibitem[{{Foucart} et~al.(2011){Foucart}, {Duez}, {Kidder}, and
  {Teukolsky}}]{Foucart_DKT2011}
{Foucart} F, {Duez} MD, {Kidder} LE, {Teukolsky} SA (2011) {Black hole-neutron
  star mergers: Effects of the orientation of the black hole spin}. \prd
  83:024005, \doi{10.1103/PhysRevD.83.024005}, \eprint{1007.4203}

\bibitem[{{Foucart} et~al.(2012){Foucart}, {Duez}, {Kidder}, {Scheel},
  {Szilagyi}, and {Teukolsky}}]{Foucart_DKSST2012}
{Foucart} F, {Duez} MD, {Kidder} LE, {Scheel} MA, {Szilagyi} B, {Teukolsky} SA
  (2012) {Black hole-neutron star mergers for 10 $M_{\odot}$ black holes}. \prd
  85:044015, \doi{10.1103/PhysRevD.85.044015}, \eprint{1111.1677}

\bibitem[{{Foucart} et~al.(2013{\natexlab{a}}){Foucart}, {Buchman}, {Duez},
  {Grudich}, {Kidder}, {MacDonald}, {Mroue}, {Pfeiffer}, {Scheel}, and
  {Szilagyi}}]{Foucart_BDGKMMPSS2013}
{Foucart} F, {Buchman} L, {Duez} MD, {Grudich} M, {Kidder} LE, {MacDonald} I,
  {Mroue} A, {Pfeiffer} HP, {Scheel} MA, {Szilagyi} B (2013{\natexlab{a}})
  {First direct comparison of nondisrupting neutron star-black hole and binary
  black hole merger simulations}. \prd 88:064017,
  \doi{10.1103/PhysRevD.88.064017}, \eprint{1307.7685}

\bibitem[{{Foucart} et~al.(2013{\natexlab{b}}){Foucart}, {Deaton}, {Duez},
  {Kidder}, {MacDonald}, {Ott}, {Pfeiffer}, {Scheel}, {Szilagyi}, and
  {Teukolsky}}]{Foucart_DDKMOPSST2013}
{Foucart} F, {Deaton} MB, {Duez} MD, {Kidder} LE, {MacDonald} I, {Ott} CD,
  {Pfeiffer} HP, {Scheel} MA, {Szilagyi} B, {Teukolsky} SA (2013{\natexlab{b}})
  {Black-hole-neutron-star mergers at realistic mass ratios: Equation of state
  and spin orientation effects}. \prd 87:084006,
  \doi{10.1103/PhysRevD.87.084006}, \eprint{1212.4810}

\bibitem[{{Foucart} et~al.(2014){Foucart}, {Deaton}, {Duez}, {O'Connor}, {Ott},
  {Haas}, {Kidder}, {Pfeiffer}, {Scheel}, and
  {Szilagyi}}]{Foucart_DDOOHKPSS2014}
{Foucart} F, {Deaton} MB, {Duez} MD, {O'Connor} E, {Ott} CD, {Haas} R, {Kidder}
  LE, {Pfeiffer} HP, {Scheel} MA, {Szilagyi} B (2014) {Neutron star-black hole
  mergers with a nuclear equation of state and neutrino cooling: Dependence in
  the binary parameters}. \prd 90:024026, \doi{10.1103/PhysRevD.90.024026},
  \eprint{1405.1121}

\bibitem[{{Foucart} et~al.(2015){Foucart}, {O'Connor}, {Roberts}, {Duez},
  {Haas}, {Kidder}, {Ott}, {Pfeiffer}, {Scheel}, and
  {Szilagyi}}]{Foucart_ORDHKOPSS2015}
{Foucart} F, {O'Connor} E, {Roberts} L, {Duez} MD, {Haas} R, {Kidder} LE, {Ott}
  CD, {Pfeiffer} HP, {Scheel} MA, {Szilagyi} B (2015) {Post-merger evolution of
  a neutron star-black hole binary with neutrino transport}. \prd 91:124021,
  \doi{10.1103/PhysRevD.91.124021}, \eprint{1502.04146}

\bibitem[{{Foucart} et~al.(2016{\natexlab{a}}){Foucart}, {Haas}, {Duez},
  {O'Connor}, {Ott}, {Roberts}, {Kidder}, {Lippuner}, {Pfeiffer}, and
  {Scheel}}]{Foucart_HDOORKLPS2016}
{Foucart} F, {Haas} R, {Duez} MD, {O'Connor} E, {Ott} CD, {Roberts} L, {Kidder}
  LE, {Lippuner} J, {Pfeiffer} HP, {Scheel} MA (2016{\natexlab{a}}) {Low mass
  binary neutron star mergers: Gravitational waves and neutrino emission}. \prd
  93:044019, \doi{10.1103/PhysRevD.93.044019}, \eprint{1510.06398}

\bibitem[{{Foucart} et~al.(2016{\natexlab{b}}){Foucart}, {O'Connor}, {Roberts},
  {Kidder}, {Pfeiffer}, and {Scheel}}]{Foucart_ORKPS2016}
{Foucart} F, {O'Connor} E, {Roberts} L, {Kidder} LE, {Pfeiffer} HP, {Scheel} MA
  (2016{\natexlab{b}}) {Impact of an improved neutrino energy estimate on
  outflows in neutron star merger simulations}. \prd 94:123016,
  \doi{10.1103/PhysRevD.94.123016}, \eprint{1607.07450}

\bibitem[{{Foucart} et~al.(2017){Foucart}, {Desai}, {Brege}, {Duez}, {Kasen},
  {Hemberger}, {Kidder}, {Pfeiffer}, and {Scheel}}]{Foucart_DBDKHKPS2017}
{Foucart} F, {Desai} D, {Brege} W, {Duez} MD, {Kasen} D, {Hemberger} DA,
  {Kidder} LE, {Pfeiffer} HP, {Scheel} MA (2017) {Dynamical ejecta from
  precessing neutron star-black hole mergers with a hot, nuclear-theory based
  equation of state}. Class Quantum Grav 34:044002,
  \doi{10.1088/1361-6382/aa573b}, \eprint{1611.01159}

\bibitem[{{Foucart} et~al.(2018){Foucart}, {Hinderer}, and
  {Nissanke}}]{Foucart_Hinderer_Nissanke2018}
{Foucart} F, {Hinderer} T, {Nissanke} S (2018) {Remnant baryon mass in neutron
  star-black hole mergers: Predictions for binary neutron star mimickers and
  rapidly spinning black holes}. \prd 98:081501,
  \doi{10.1103/PhysRevD.98.081501}, \eprint{1807.00011}

\bibitem[{{Foucart} et~al.(2019{\natexlab{a}}){Foucart}, {Duez}, {Hinderer},
  {Caro}, {Williamson}, {Boyle}, {Buonanno}, {Haas}, {Hemberger}, {Kidder}, and
  et~al.}]{Foucart_etal2019}
{Foucart} F, {Duez} MD, {Hinderer} T, {Caro} J, {Williamson} AR, {Boyle} M,
  {Buonanno} A, {Haas} R, {Hemberger} DA, {Kidder} LE, et~al
  (2019{\natexlab{a}}) {Gravitational waveforms from spectral Einstein code
  simulations: Neutron star-neutron star and low-mass black hole-neutron star
  binaries}. \prd 99:044008, \doi{10.1103/PhysRevD.99.044008},
  \eprint{1812.06988}

\bibitem[{{Foucart} et~al.(2019{\natexlab{b}}){Foucart}, {Duez}, {Kidder},
  {Nissanke}, {Pfeiffer}, and {Scheel}}]{Foucart_DKNPS2019}
{Foucart} F, {Duez} MD, {Kidder} LE, {Nissanke} SM, {Pfeiffer} HP, {Scheel} MA
  (2019{\natexlab{b}}) {Numerical simulations of neutron star-black hole
  binaries in the near-equal-mass regime}. \prd 99:103025,
  \doi{10.1103/PhysRevD.99.103025}, \eprint{1903.09166}

\bibitem[{{Foucart} et~al.(2020){Foucart}, {Duez}, {Hebert}, {Kidder},
  {Pfeiffer}, and {Scheel}}]{Foucart_DHKPS2020}
{Foucart} F, {Duez} MD, {Hebert} F, {Kidder} LE, {Pfeiffer} HP, {Scheel} MA
  (2020) {Monte-Carlo Neutrino Transport in Neutron Star Merger Simulations}.
  \apjl 902:L27, \doi{10.3847/2041-8213/abbb87}, \eprint{2008.08089}

\bibitem[{{Foucart} et~al.(2021){Foucart}, {Chernoglazov}, {Boyle}, {Hinderer},
  {Miller}, {Moxon}, {Scheel}, {Deppe}, {Duez}, {H{\'e}bert}, and
  et~al.}]{Foucart_etal2021}
{Foucart} F, {Chernoglazov} A, {Boyle} M, {Hinderer} T, {Miller} M, {Moxon} J,
  {Scheel} MA, {Deppe} N, {Duez} MD, {H{\'e}bert} F, et~al (2021)
  {High-accuracy waveforms for black hole-neutron star systems with spinning
  black holes}. \prd 103:064007, \doi{10.1103/PhysRevD.103.064007},
  \eprint{2010.14518}

\bibitem[{{Fragile} et~al.(2007){Fragile}, {Blaes}, {Anninos}, and
  {Salmonson}}]{Fragile_BAS2007}
{Fragile} PC, {Blaes} OM, {Anninos} P, {Salmonson} JD (2007) {Global General
  Relativistic Magnetohydrodynamic Simulation of a Tilted Black Hole Accretion
  Disk}. \apj 668:417--429, \doi{10.1086/521092}, \eprint{0706.4303}

\bibitem[{{Fragione} et~al.(2019){Fragione}, {Grishin}, {Leigh}, {Perets}, and
  {Perna}}]{Fragione_GLPP2019}
{Fragione} G, {Grishin} E, {Leigh} NWC, {Perets} HB, {Perna} R (2019) {Black
  hole and neutron star mergers in galactic nuclei}. \mnras 488:47--63,
  \doi{10.1093/mnras/stz1651}, \eprint{1811.10627}

\bibitem[{{Friedman} and {Stergioulas}(2013)}]{Friedman_Stergioulas}
{Friedman} JL, {Stergioulas} N (2013) {Rotating Relativistic Stars}. Cambridge
  University Press, Cambridge, UK

\bibitem[{{Friedman} et~al.(2002){Friedman}, {Ury{\={u}}}, and
  {Shibata}}]{Friedman_Uryu_Shibata2002}
{Friedman} JL, {Ury{\={u}}} K, {Shibata} M (2002) {Thermodynamics of binary
  black holes and neutron stars}. \prd 65:064035,
  \doi{10.1103/PhysRevD.65.064035}, \eprint{gr-qc/0108070}

\bibitem[{{Friedrich}(1985)}]{Friedrich1985}
{Friedrich} H (1985) {On the hyperbolicity of Einstein's and other gauge field
  equations}. Communications in Mathematical Physics 100:525--543,
  \doi{10.1007/BF01217728}

\bibitem[{{Fujibayashi} et~al.(2017){Fujibayashi}, {Sekiguchi}, {Kiuchi}, and
  {Shibata}}]{Fujibayashi_SKS2017}
{Fujibayashi} S, {Sekiguchi} Y, {Kiuchi} K, {Shibata} M (2017) {Properties of
  Neutrino-driven Ejecta from the Remnant of a Binary Neutron Star Merger: Pure
  Radiation Hydrodynamics Case}. \apj 846:114, \doi{10.3847/1538-4357/aa8039},
  \eprint{1703.10191}

\bibitem[{{Fujibayashi} et~al.(2018){Fujibayashi}, {Kiuchi}, {Nishimura},
  {Sekiguchi}, and {Shibata}}]{Fujibayashi_KNSS2018}
{Fujibayashi} S, {Kiuchi} K, {Nishimura} N, {Sekiguchi} Y, {Shibata} M (2018)
  {Mass Ejection from the Remnant of a Binary Neutron Star Merger:
  Viscous-radiation Hydrodynamics Study}. \apj 860:64,
  \doi{10.3847/1538-4357/aabafd}, \eprint{1711.02093}

\bibitem[{{Fujibayashi} et~al.(2020{\natexlab{a}}){Fujibayashi}, {Shibata},
  {Wanajo}, {Kiuchi}, {Kyutoku}, and {Sekiguchi}}]{Fujibayashi_SWKKS2020}
{Fujibayashi} S, {Shibata} M, {Wanajo} S, {Kiuchi} K, {Kyutoku} K, {Sekiguchi}
  Y (2020{\natexlab{a}}) {Mass ejection from disks surrounding a low-mass black
  hole: Viscous neutrino-radiation hydrodynamics simulation in full general
  relativity}. \prd 101:083029, \doi{10.1103/PhysRevD.101.083029},
  \eprint{2001.04467}

\bibitem[{{Fujibayashi} et~al.(2020{\natexlab{b}}){Fujibayashi}, {Shibata},
  {Wanajo}, {Kiuchi}, {Kyutoku}, and {Sekiguchi}}]{Fujibayashi_SWKKS2020-2}
{Fujibayashi} S, {Shibata} M, {Wanajo} S, {Kiuchi} K, {Kyutoku} K, {Sekiguchi}
  Y (2020{\natexlab{b}}) {Viscous evolution of a massive disk surrounding
  stellar-mass black holes in full general relativity}. \prd 102:123014,
  \doi{10.1103/PhysRevD.102.123014}, \eprint{2009.03895}

\bibitem[{{Fuller} et~al.(1985){Fuller}, {Fowler}, and
  {Newman}}]{Fuller_Fowler_Newman1985}
{Fuller} GM, {Fowler} WA, {Newman} MJ (1985) {Stellar weak interaction rates
  for intermediate-mass nuclei. IV - Interpolation procedures for rapidly
  varying lepton capture rates using effective log (ft)-values}. \apj
  293:1--16, \doi{10.1086/163208}

\bibitem[{{Garat} and {Price}(2000)}]{Garat_Price2000}
{Garat} A, {Price} RH (2000) {Nonexistence of conformally flat slices of the
  Kerr spacetime}. \prd 61:124011, \doi{10.1103/PhysRevD.61.124011},
  \eprint{gr-qc/0002013}

\bibitem[{{Garfinkle}(2002)}]{Garfinkle2002}
{Garfinkle} D (2002) {Harmonic coordinate method for simulating generic
  singularities}. \prd 65:044029, \doi{10.1103/PhysRevD.65.044029},
  \eprint{gr-qc/0110013}

\bibitem[{{Glendenning} and {Moszkowski}(1991)}]{Glendenning_Moszkowski1991}
{Glendenning} NK, {Moszkowski} SA (1991) {Reconciliation of neutron-star masses
  and binding of the Lambda in hypernuclei}. \prl 67:2414--1417,
  \doi{10.1103/PhysRevLett.67.2414}

\bibitem[{{Goldstein} et~al.(2017){Goldstein}, {Veres}, {Burns}, {Briggs},
  {Hamburg}, {Kocevski}, {Wilson-Hodge}, {Preece}, {Poolakkil}, {Roberts}, and
  et~al.}]{Goldstein_etal2017}
{Goldstein} A, {Veres} P, {Burns} E, {Briggs} MS, {Hamburg} R, {Kocevski} D,
  {Wilson-Hodge} CA, {Preece} RD, {Poolakkil} S, {Roberts} OJ, et~al (2017) {An
  Ordinary Short Gamma-Ray Burst with Extraordinary Implications: Fermi-GBM
  Detection of GRB 170817A}. \apjl 848:L14, \doi{10.3847/2041-8213/aa8f41},
  \eprint{1710.05446}

\bibitem[{{Gompertz} et~al.(2013){Gompertz}, {O'Brien}, {Wynn}, and
  {Rowlinson}}]{Gompertz_OWR2013}
{Gompertz} BP, {O'Brien} PT, {Wynn} GA, {Rowlinson} A (2013) {Can magnetar
  spin-down power extended emission in some short GRBs?} \mnras 431:1745--1751,
  \doi{10.1093/mnras/stt293}, \eprint{1302.3643}

\bibitem[{{Goodman}(1986)}]{Goodman1986}
{Goodman} J (1986) {Are gamma-ray bursts optically thick?} \apjl 308:L47,
  \doi{10.1086/184741}

\bibitem[{{Gottlieb} et~al.(2018){Gottlieb}, {Nakar}, {Piran}, and
  {Hotokezaka}}]{Gottlieb_NPH2018}
{Gottlieb} O, {Nakar} E, {Piran} T, {Hotokezaka} K (2018) {A cocoon shock
  breakout as the origin of the {\ensuremath{\gamma}}-ray emission in
  GW170817}. \mnras 479:588--600, \doi{10.1093/mnras/sty1462},
  \eprint{1710.05896}

\bibitem[{{Gourgoulhon}(2012)}]{Gourgoulhon}
{Gourgoulhon} E (2012) {3+1 Formalism in General Relativity}, Lecture Notes in
  Physics, vol 846. Springer, Berlin, Heidelberg,
  \doi{10.1007/978-3-642-24525-1}

\bibitem[{{Gourgoulhon} and {Jaramillo}(2006)}]{Gourgoulhon_Jaramillo2006}
{Gourgoulhon} E, {Jaramillo} JL (2006) {A 3+1 perspective on null hypersurfaces
  and isolated horizons}. \physrep 423:159--294,
  \doi{10.1016/j.physrep.2005.10.005}, \eprint{gr-qc/0503113}

\bibitem[{{Gourgoulhon} et~al.(2001){Gourgoulhon}, {Grandcl{\'e}ment},
  {Taniguchi}, {Marck}, and {Bonazzola}}]{Gourgoulhon_GTMB2001}
{Gourgoulhon} E, {Grandcl{\'e}ment} P, {Taniguchi} K, {Marck} JA, {Bonazzola} S
  (2001) {Quasiequilibrium sequences of synchronized and irrotational binary
  neutron stars in general relativity: Method and tests}. \prd 63:064029,
  \doi{10.1103/PhysRevD.63.064029}, \eprint{gr-qc/0007028}

\bibitem[{{Gourgoulhon} et~al.(2002){Gourgoulhon}, {Grandcl{\'e}ment}, and
  {Bonazzola}}]{Gourgoulhon_Grandclement_Bonazzola2002}
{Gourgoulhon} E, {Grandcl{\'e}ment} P, {Bonazzola} S (2002) {Binary black holes
  in circular orbits. I. A global spacetime approach}. \prd 65:044020,
  \doi{10.1103/PhysRevD.65.044020}, \eprint{gr-qc/0106015}

\bibitem[{{Grandcl{\'e}ment}(2006)}]{Grandclement2006}
{Grandcl{\'e}ment} P (2006) {Accurate and realistic initial data for black hole
  neutron star binaries}. \prd 74:124002, \doi{10.1103/PhysRevD.74.124002},
  \eprint{gr-qc/0609044}

\bibitem[{{Grandcl{\'e}ment}(2007)}]{Grandclement2006e}
{Grandcl{\'e}ment} P (2007) {Erratum: Accurate and realistic initial data for
  black hole-neutron star binaries [Phys. Rev. D 74, 124002 (2006)]}. \prd
  75:129903, \doi{10.1103/PhysRevD.75.129903}

\bibitem[{{Grandcl{\'e}ment}(2010)}]{Grandclement2010}
{Grandcl{\'e}ment} P (2010) {KADATH: A spectral solver for theoretical
  physics}. J Comput Phys 229:3334--3357, \doi{10.1016/j.jcp.2010.01.005},
  \eprint{0909.1228}

\bibitem[{{Grandcl{\'e}ment} and {Novak}(2009)}]{Grandclement_Novak2009}
{Grandcl{\'e}ment} P, {Novak} J (2009) {Spectral Methods for Numerical
  Relativity}. Living Rev Relativ 12:1, \doi{10.12942/lrr-2009-1},
  \eprint{0706.2286}

\bibitem[{{Grandcl{\'e}ment} et~al.(2002){Grandcl{\'e}ment}, {Gourgoulhon}, and
  {Bonazzola}}]{Grandclement_Gourgoulhon_Bonazzola2002}
{Grandcl{\'e}ment} P, {Gourgoulhon} E, {Bonazzola} S (2002) {Binary black holes
  in circular orbits. II. Numerical methods and first results}. \prd 65:044021,
  \doi{10.1103/PhysRevD.65.044021}, \eprint{gr-qc/0106016}

\bibitem[{{Gundlach} et~al.(2005){Gundlach}, {Calabrese}, {Hinder}, and
  {Mart{\'\i}n-Garc{\'\i}a}}]{Gundlach_CHM2005}
{Gundlach} C, {Calabrese} G, {Hinder} I, {Mart{\'\i}n-Garc{\'\i}a} JM (2005)
  {Constraint damping in the Z4 formulation and harmonic gauge}. Class Quantum
  Grav 22:3767-3773, \doi{10.1088/0264-9381/22/17/025}, \eprint{gr-qc/0504114}

\bibitem[{{Haensel} and {Potekhin}(2004)}]{Haensel_Potekhin2004}
{Haensel} P, {Potekhin} AY (2004) {Analytical representations of unified
  equations of state of neutron-star matter}. \aap 428:191--197,
  \doi{10.1051/0004-6361:20041722}, \eprint{astro-ph/0408324}

\bibitem[{{Hamidani} and {Ioka}(2021)}]{Hamidani_Ioka2021}
{Hamidani} H, {Ioka} K (2021) {Jet propagation in expanding medium for
  gamma-ray bursts}. \mnras 500:627--642, \doi{10.1093/mnras/staa3276},
  \eprint{2007.10690}

\bibitem[{{Han} et~al.(2020){Han}, {Tang}, {Hu}, {Li}, {Jiang}, {Jin}, {Fan},
  and {Wei}}]{Han_THLJJFW2020}
{Han} MZ, {Tang} SP, {Hu} YM, {Li} YJ, {Jiang} JL, {Jin} ZP, {Fan} YZ, {Wei} DM
  (2020) {Is GW190425 Consistent with Being a Neutron Star-Black Hole Merger?}
  \apjl 891:L5, \doi{10.3847/2041-8213/ab745a}, \eprint{2001.07882}

\bibitem[{{Hannam} et~al.(2007){Hannam}, {Husa}, {Pollney}, {Br{\"u}gmann}, and
  {Murchadha}}]{Hannam_HPBO2007}
{Hannam} M, {Husa} S, {Pollney} D, {Br{\"u}gmann} B, {Murchadha} N{\'O} (2007)
  {Geometry and Regularity of Moving Punctures}. \prl 99:241102,
  \doi{10.1103/PhysRevLett.99.241102}, \eprint{gr-qc/0606099}

\bibitem[{{Hannam} et~al.(2008){Hannam}, {Husa}, {Ohme}, {Br{\"u}gmann}, and
  {{\'O} Murchadha}}]{Hannam_HOBO2008}
{Hannam} M, {Husa} S, {Ohme} F, {Br{\"u}gmann} B, {{\'O} Murchadha} N (2008)
  {Wormholes and trumpets: Schwarzschild spacetime for the moving-puncture
  generation}. \prd 78:064020, \doi{10.1103/PhysRevD.78.064020},
  \eprint{0804.0628}

\bibitem[{{Hannam} et~al.(2013){Hannam}, {Brown}, {Fairhurst}, {Fryer}, and
  {Harry}}]{Hannam_BFFH2013}
{Hannam} M, {Brown} DA, {Fairhurst} S, {Fryer} CL, {Harry} IW (2013) {When can
  Gravitational-wave Observations Distinguish between Black Holes and Neutron
  Stars?} \apjl 766:L14, \doi{10.1088/2041-8205/766/1/L14}, \eprint{1301.5616}

\bibitem[{{Hansen} and {Lyutikov}(2001)}]{Hansen_Lyutikov2001}
{Hansen} BMS, {Lyutikov} M (2001) {Radio and X-ray signatures of merging
  neutron stars}. \mnras 322:695--701, \doi{10.1046/j.1365-8711.2001.04103.x},
  \eprint{astro-ph/0003218}

\bibitem[{{Harada}(2001)}]{Harada2001}
{Harada} T (2001) {Reconstructing the equation of state for cold nuclear matter
  from the relationship of any two properties of neutron stars}. \prc
  64:048801, \doi{10.1103/PhysRevC.64.048801}, \eprint{astro-ph/0108410}

\bibitem[{{Harrison} et~al.(2018){Harrison}, {Gottlieb}, and
  {Nakar}}]{Harrison_Gottlieb_Nakar2018}
{Harrison} R, {Gottlieb} O, {Nakar} E (2018) {Numerically calibrated model for
  propagation of a relativistic unmagnetized jet in dense media}. \mnras
  477:2128--2140, \doi{10.1093/mnras/sty760}, \eprint{1707.06234}

\bibitem[{{Hartle}(1967)}]{Hartle1967}
{Hartle} JB (1967) {Slowly Rotating Relativistic Stars. I. Equations of
  Structure}. \apj 150:1005, \doi{10.1086/149400}

\bibitem[{{Hawking} and {Ellis}(1973)}]{Hawking_Ellis}
{Hawking} SW, {Ellis} GFR (1973) {The large scale structure of space-time}.
  Cambridge University Press, Cambridge, UK

\bibitem[{{Hawley}(1991)}]{Hawley1991}
{Hawley} JF (1991) {Three-dimensional Simulations of Black Hole Tori}. \apj
  381:496, \doi{10.1086/170673}

\bibitem[{{Hayashi} et~al.(2021){Hayashi}, {Kawaguchi}, {Kiuchi}, {Kyutoku},
  and {Shibata}}]{Hayashi_KKKS2021}
{Hayashi} K, {Kawaguchi} K, {Kiuchi} K, {Kyutoku} K, {Shibata} M (2021)
  {Properties of the remnant disk and the dynamical ejecta produced in low-mass
  black hole-neutron star mergers}. \prd 103:043007,
  \doi{10.1103/PhysRevD.103.043007}, \eprint{2010.02563}

\bibitem[{{Healy} et~al.(2019){Healy}, {Lousto}, {Lange}, {O'Shaughnessy},
  {Zlochower}, and {Campanelli}}]{Healy_LLOZC2019}
{Healy} J, {Lousto} CO, {Lange} J, {O'Shaughnessy} R, {Zlochower} Y,
  {Campanelli} M (2019) {Second RIT binary black hole simulations catalog and
  its application to gravitational waves parameter estimation}. \prd
  100:024021, \doi{10.1103/PhysRevD.100.024021}, \eprint{1901.02553}

\bibitem[{{Hempel} et~al.(2012){Hempel}, {Fischer}, {Schaffner-Bielich}, and
  {Liebend{\"o}rfer}}]{Hempel_FSL2012}
{Hempel} M, {Fischer} T, {Schaffner-Bielich} J, {Liebend{\"o}rfer} M (2012)
  {New Equations of State in Simulations of Core-collapse Supernovae}. \apj
  748:70, \doi{10.1088/0004-637X/748/1/70}, \eprint{1108.0848}

\bibitem[{{Henriksson} et~al.(2016){Henriksson}, {Foucart}, {Kidder}, and
  {Teukolsky}}]{Henriksson_FKT2016}
{Henriksson} K, {Foucart} F, {Kidder} LE, {Teukolsky} SA (2016) {Initial data
  for high-compactness black hole-neutron star binaries}. Class Quantum Grav
  33:105009, \doi{10.1088/0264-9381/33/10/105009}, \eprint{1409.7159}

\bibitem[{{Hilditch} and {Ruiz}(2018)}]{Hilditch_Ruiz2018}
{Hilditch} D, {Ruiz} M (2018) {The initial boundary value problem for
  free-evolution formulations of general relativity}. Class Quantum Grav
  35:015006, \doi{10.1088/1361-6382/aa96c6}, \eprint{1609.06925}

\bibitem[{{Hilditch} et~al.(2013){Hilditch}, {Bernuzzi}, {Thierfelder}, {Cao},
  {Tichy}, and {Br{\"u}gmann}}]{Hilditch_BTCTB2013}
{Hilditch} D, {Bernuzzi} S, {Thierfelder} M, {Cao} Z, {Tichy} W, {Br{\"u}gmann}
  B (2013) {Compact binary evolutions with the Z4c formulation}. \prd
  88:084057, \doi{10.1103/PhysRevD.88.084057}, \eprint{1212.2901}

\bibitem[{{Hinderer}(2008)}]{Hinderer2008}
{Hinderer} T (2008) {Tidal Love Numbers of Neutron Stars}. \apj 677:1216--1220,
  \doi{10.1086/533487}, \eprint{0711.2420}

\bibitem[{{Hinderer} et~al.(2019){Hinderer}, {Nissanke}, {Foucart},
  {Hotokezaka}, {Vincent}, {Kasliwal}, {Schmidt}, {Williamson}, {Nichols},
  {Duez}, and et~al.}]{Hinderer_etal2019}
{Hinderer} T, {Nissanke} S, {Foucart} F, {Hotokezaka} K, {Vincent} T,
  {Kasliwal} M, {Schmidt} P, {Williamson} AR, {Nichols} DA, {Duez} MD, et~al
  (2019) {Distinguishing the nature of comparable-mass neutron star binary
  systems with multimessenger observations: GW170817 case study}. \prd
  100:063021, \doi{10.1103/PhysRevD.100.063021}, \eprint{1808.03836}

\bibitem[{{Hiscock} and {Lindblom}(1983)}]{Hiscock_Lindblom1983}
{Hiscock} WA, {Lindblom} L (1983) {Stability and causality in dissipative
  relativistic fluids.} Annals of Physics 151:466--496,
  \doi{10.1016/0003-4916(83)90288-9}

\bibitem[{{Hiscock} and {Lindblom}(1985)}]{Hiscock_Lindblom1985}
{Hiscock} WA, {Lindblom} L (1985) {Generic instabilities in first-order
  dissipative relativistic fluid theories}. \prd 31:725--733,
  \doi{10.1103/PhysRevD.31.725}

\bibitem[{{Hoffman} et~al.(1997){Hoffman}, {Woosley}, and
  {Qian}}]{Hoffman_Woosley_Qian1997}
{Hoffman} RD, {Woosley} SE, {Qian} YZ (1997) {Nucleosynthesis in
  Neutrino-driven Winds. II. Implications for Heavy Element Synthesis}. \apj
  482:951--962, \doi{10.1086/304181}, \eprint{astro-ph/9611097}

\bibitem[{{Hossein Nouri} et~al.(2018){Hossein Nouri}, {Duez}, {Foucart},
  {Deaton}, {Haas}, {Haddadi}, {Kidder}, {Ott}, {Pfeiffer}, {Scheel}, and
  et~al.}]{Nouri_etal2018}
{Hossein Nouri} F, {Duez} MD, {Foucart} F, {Deaton} MB, {Haas} R, {Haddadi} M,
  {Kidder} LE, {Ott} CD, {Pfeiffer} HP, {Scheel} MA, et~al (2018) {Evolution of
  the magnetized, neutrino-cooled accretion disk in the aftermath of a black
  hole-neutron star binary merger}. \prd 97:083014,
  \doi{10.1103/PhysRevD.97.083014}

\bibitem[{{Hotokezaka} and {Nakar}(2020)}]{Hotokezaka_Nakar2020}
{Hotokezaka} K, {Nakar} E (2020) {Radioactive Heating Rate of r-process
  Elements and Macronova Light Curve}. \apj 891:152,
  \doi{10.3847/1538-4357/ab6a98}, \eprint{1909.02581}

\bibitem[{{Hotokezaka} et~al.(2011){Hotokezaka}, {Kyutoku}, {Okawa}, {Shibata},
  and {Kiuchi}}]{Hotokezaka_KOSK2011}
{Hotokezaka} K, {Kyutoku} K, {Okawa} H, {Shibata} M, {Kiuchi} K (2011) {Binary
  neutron star mergers: Dependence on the nuclear equation of state}. \prd
  83:124008, \doi{10.1103/PhysRevD.83.124008}, \eprint{1105.4370}

\bibitem[{{Hotokezaka} et~al.(2013{\natexlab{a}}){Hotokezaka}, {Kiuchi},
  {Kyutoku}, {Muranushi}, {Sekiguchi}, {Shibata}, and
  {Taniguchi}}]{Hotokezaka_KKMSST2013}
{Hotokezaka} K, {Kiuchi} K, {Kyutoku} K, {Muranushi} T, {Sekiguchi} Yi,
  {Shibata} M, {Taniguchi} K (2013{\natexlab{a}}) {Remnant massive neutron
  stars of binary neutron star mergers: Evolution process and gravitational
  waveform}. \prd 88:044026, \doi{10.1103/PhysRevD.88.044026},
  \eprint{1307.5888}

\bibitem[{{Hotokezaka} et~al.(2013{\natexlab{b}}){Hotokezaka}, {Kiuchi},
  {Kyutoku}, {Okawa}, {Sekiguchi}, {Shibata}, and
  {Taniguchi}}]{Hotokezaka_KKOSST2013}
{Hotokezaka} K, {Kiuchi} K, {Kyutoku} K, {Okawa} H, {Sekiguchi} Yi, {Shibata}
  M, {Taniguchi} K (2013{\natexlab{b}}) {Mass ejection from the merger of
  binary neutron stars}. \prd 87:024001, \doi{10.1103/PhysRevD.87.024001},
  \eprint{1212.0905}

\bibitem[{{Hotokezaka} et~al.(2016{\natexlab{a}}){Hotokezaka}, {Kyutoku},
  {Sekiguchi}, and {Shibata}}]{Hotokezaka_KSS2016}
{Hotokezaka} K, {Kyutoku} K, {Sekiguchi} Yi, {Shibata} M (2016{\natexlab{a}})
  {Measurability of the tidal deformability by gravitational waves from
  coalescing binary neutron stars}. \prd 93:064082,
  \doi{10.1103/PhysRevD.93.064082}, \eprint{1603.01286}

\bibitem[{{Hotokezaka} et~al.(2016{\natexlab{b}}){Hotokezaka}, {Wanajo},
  {Tanaka}, {Bamba}, {Terada}, and {Piran}}]{Hotokezaka_WTBTP2016}
{Hotokezaka} K, {Wanajo} S, {Tanaka} M, {Bamba} A, {Terada} Y, {Piran} T
  (2016{\natexlab{b}}) {Radioactive decay products in neutron star merger
  ejecta: heating efficiency and {\ensuremath{\gamma}}-ray emission}. \mnras
  459:35--43, \doi{10.1093/mnras/stw404}, \eprint{1511.05580}

\bibitem[{{Hotokezaka} et~al.(2017){Hotokezaka}, {Sari}, and
  {Piran}}]{Hotokezaka_Sari_Piran2017}
{Hotokezaka} K, {Sari} R, {Piran} T (2017) {Analytic heating rate of neutron
  star merger ejecta derived from Fermi's theory of beta decay}. \mnras
  468:91--96, \doi{10.1093/mnras/stx411}, \eprint{1701.02785}

\bibitem[{{Hotokezaka} et~al.(2018){Hotokezaka}, {Kiuchi}, {Shibata}, {Nakar},
  and {Piran}}]{Hotokezaka_KSNP2018}
{Hotokezaka} K, {Kiuchi} K, {Shibata} M, {Nakar} E, {Piran} T (2018)
  {Synchrotron Radiation from the Fast Tail of Dynamical Ejecta of Neutron Star
  Mergers}. \apj 867:95, \doi{10.3847/1538-4357/aadf92}, \eprint{1803.00599}

\bibitem[{{Hughes}(2001)}]{Hughes2001}
{Hughes} SA (2001) {Evolution of circular, nonequatorial orbits of Kerr black
  holes due to gravitational-wave emission. II. Inspiral trajectories and
  gravitational waveforms}. \prd 64:064004, \doi{10.1103/PhysRevD.64.064004},
  \eprint{gr-qc/0104041}

\bibitem[{{Imbiriba} et~al.(2004){Imbiriba}, {Baker}, {Choi}, {Centrella},
  {Fiske}, {Brown}, {van Meter}, and {Olson}}]{Imbiriba_BCCFBVO2004}
{Imbiriba} B, {Baker} J, {Choi} DI, {Centrella} J, {Fiske} DR, {Brown} JD, {van
  Meter} JR, {Olson} K (2004) {Evolving a puncture black hole with fixed mesh
  refinement}. \prd 70:124025, \doi{10.1103/PhysRevD.70.124025},
  \eprint{gr-qc/0403048}

\bibitem[{{Ioka} and {Taniguchi}(2000)}]{Ioka_Taniguchi2000}
{Ioka} K, {Taniguchi} K (2000) {Gravitational Waves from Inspiraling Compact
  Binaries with Magnetic Dipole Moments}. \apj 537:327--333,
  \doi{10.1086/309004}, \eprint{astro-ph/0001218}

\bibitem[{{Ishii} et~al.(2005){Ishii}, {Shibata}, and
  {Mino}}]{Ishii_Shibata_Mino2005}
{Ishii} M, {Shibata} M, {Mino} Y (2005) {Black hole tidal problem in the Fermi
  normal coordinates}. \prd 71:044017, \doi{10.1103/PhysRevD.71.044017},
  \eprint{gr-qc/0501084}

\bibitem[{{Ishizaki} et~al.(2021){Ishizaki}, {Kiuchi}, {Ioka}, and
  {Wanajo}}]{2021arXiv210404708I}
{Ishizaki} W, {Kiuchi} K, {Ioka} K, {Wanajo} S (2021) {Fallback Accretion
  Halted by R-process Heating in Neutron Star Mergers and Gamma-Ray Bursts}.
  arXiv e-prints arXiv:2104.04708, \eprint{2104.04708}

\bibitem[{{Israel} and {Stewart}(1979)}]{Israel_Stewart1979}
{Israel} W, {Stewart} JM (1979) {Transient relativistic thermodynamics and
  kinetic theory}. Annals of Physics 118:341--372,
  \doi{10.1016/0003-4916(79)90130-1}

\bibitem[{{Jani} et~al.(2016){Jani}, {Healy}, {Clark}, {London}, {Laguna}, and
  {Shoemaker}}]{Jani_HCLLS2016}
{Jani} K, {Healy} J, {Clark} JA, {London} L, {Laguna} P, {Shoemaker} D (2016)
  {Georgia tech catalog of gravitational waveforms}. Class Quantum Grav
  33:204001, \doi{10.1088/0264-9381/33/20/204001}, \eprint{1605.03204}

\bibitem[{{Janka} et~al.(1993){Janka}, {Zwerger}, and
  {Moenchmeyer}}]{Janka_Zwerger_Moenchmeyer1993}
{Janka} HT, {Zwerger} T, {Moenchmeyer} R (1993) {Does artificial viscosity
  destroy prompt type-II supernova explosions?} \aap 268:360--368

\bibitem[{{Janka} et~al.(1999){Janka}, {Eberl}, {Ruffert}, and
  {Fryer}}]{Janka_ERF1999}
{Janka} HT, {Eberl} T, {Ruffert} M, {Fryer} CL (1999) {Black Hole-Neutron Star
  Mergers as Central Engines of Gamma-Ray Bursts}. \apjl 527:L39--L42,
  \doi{10.1086/312397}, \eprint{astro-ph/9908290}

\bibitem[{{Jaramillo} et~al.(2004){Jaramillo}, {Gourgoulhon}, and
  {Marug{\'a}n}}]{Jaramillo_Gourgoulhon_Marugan2004}
{Jaramillo} JL, {Gourgoulhon} E, {Marug{\'a}n} GA (2004) {Inner boundary
  conditions for black hole initial data derived from isolated horizons}. \prd
  70:124036, \doi{10.1103/PhysRevD.70.124036}, \eprint{gr-qc/0407063}

\bibitem[{{Jaranowski} and {Krolak}(1994)}]{Jaranowski_Krolak1994}
{Jaranowski} P, {Krolak} A (1994) {Optimal solution to the inverse problem for
  the gravitational wave signal of a coalescing compact binary}. \prd
  49:1723--1739, \doi{10.1103/PhysRevD.49.1723}

\bibitem[{{Just} et~al.(2015){Just}, {Bauswein}, {Ardevol Pulpillo}, {Goriely},
  and {Janka}}]{Just_BAGJ2015}
{Just} O, {Bauswein} A, {Ardevol Pulpillo} R, {Goriely} S, {Janka} HT (2015)
  {Comprehensive nucleosynthesis analysis for ejecta of compact binary
  mergers}. \mnras 448:541--567, \doi{10.1093/mnras/stv009}, \eprint{1406.2687}

\bibitem[{{Just} et~al.(2016){Just}, {Obergaulinger}, {Janka}, {Bauswein}, and
  {Schwarz}}]{Just_OJBS2016}
{Just} O, {Obergaulinger} M, {Janka} HT, {Bauswein} A, {Schwarz} N (2016)
  {Neutron-star Merger Ejecta as Obstacles to Neutrino-powered Jets of
  Gamma-Ray Bursts}. \apjl 816:L30, \doi{10.3847/2041-8205/816/2/L30},
  \eprint{1510.04288}

\bibitem[{{Just} et~al.(2021){Just}, {Goriely}, {Janka}, {Nagataki}, and
  {Bauswein}}]{2021arXiv210208387J}
{Just} O, {Goriely} S, {Janka} HT, {Nagataki} S, {Bauswein} A (2021) {Neutrino
  absorption and other physics dependencies in neutrino-cooled black-hole
  accretion disks}. arXiv e-prints arXiv:2102.08387, \eprint{2102.08387}

\bibitem[{{Karp} et~al.(1977){Karp}, {Lasher}, {Chan}, and
  {Salpeter}}]{Karp_LCS1977}
{Karp} AH, {Lasher} G, {Chan} KL, {Salpeter} EE (1977) {The opacity of
  expanding media: the effect of spectral lines.} \apj 214:161--178,
  \doi{10.1086/155241}

\bibitem[{{Kasen} and {Barnes}(2019)}]{Kasen_Barnes2019}
{Kasen} D, {Barnes} J (2019) {Radioactive Heating and Late Time Kilonova Light
  Curves}. \apj 876:128, \doi{10.3847/1538-4357/ab06c2}, \eprint{1807.03319}

\bibitem[{{Kasen} et~al.(2013){Kasen}, {Badnell}, and
  {Barnes}}]{Kasen_Badnell_Barnes2013}
{Kasen} D, {Badnell} NR, {Barnes} J (2013) {Opacities and Spectra of the
  r-process Ejecta from Neutron Star Mergers}. \apj 774:25,
  \doi{10.1088/0004-637X/774/1/25}, \eprint{1303.5788}

\bibitem[{{Kasen} et~al.(2015){Kasen}, {Fern{\'a}ndez}, and
  {Metzger}}]{Kasen_Fernandez_Metzger2015}
{Kasen} D, {Fern{\'a}ndez} R, {Metzger} BD (2015) {Kilonova light curves from
  the disc wind outflows of compact object mergers}. \mnras 450:1777--1786,
  \doi{10.1093/mnras/stv721}, \eprint{1411.3726}

\bibitem[{{Kasen} et~al.(2017){Kasen}, {Metzger}, {Barnes}, {Quataert}, and
  {Ramirez-Ruiz}}]{Kasen_MBQR2017}
{Kasen} D, {Metzger} B, {Barnes} J, {Quataert} E, {Ramirez-Ruiz} E (2017)
  {Origin of the heavy elements in binary neutron-star mergers from a
  gravitational-wave event}. \nat 551:80--84, \doi{10.1038/nature24453},
  \eprint{1710.05463}

\bibitem[{{Kawaguchi} et~al.(2015){Kawaguchi}, {Kyutoku}, {Nakano}, {Okawa},
  {Shibata}, and {Taniguchi}}]{Kawaguchi_KNOST2015}
{Kawaguchi} K, {Kyutoku} K, {Nakano} H, {Okawa} H, {Shibata} M, {Taniguchi} K
  (2015) {Black hole-neutron star binary merger: Dependence on black hole spin
  orientation and equation of state}. \prd 92:024014,
  \doi{10.1103/PhysRevD.92.024014}, \eprint{1506.05473}

\bibitem[{{Kawaguchi} et~al.(2016){Kawaguchi}, {Kyutoku}, {Shibata}, and
  {Tanaka}}]{Kawaguchi_KST2016}
{Kawaguchi} K, {Kyutoku} K, {Shibata} M, {Tanaka} M (2016) {Models of
  Kilonova/Macronova Emission from Black Hole-Neutron Star Mergers}. \apj
  825:52, \doi{10.3847/0004-637X/825/1/52}, \eprint{1601.07711}

\bibitem[{{Kawaguchi} et~al.(2017){Kawaguchi}, {Kyutoku}, {Nakano}, and
  {Shibata}}]{Kawaguchi_KNS2017}
{Kawaguchi} K, {Kyutoku} K, {Nakano} H, {Shibata} M (2017) {Extracting the
  cutoff frequency in the gravitational-wave spectrum of black hole-neutron
  star mergers}. arXiv e-prints arXiv:1709.02754, \eprint{1709.02754}

\bibitem[{{Kawaguchi} et~al.(2018){Kawaguchi}, {Shibata}, and
  {Tanaka}}]{Kawaguchi_Shibata_Tanaka2018}
{Kawaguchi} K, {Shibata} M, {Tanaka} M (2018) {Radiative Transfer Simulation
  for the Optical and Near-infrared Electromagnetic Counterparts to GW170817}.
  \apjl 865:L21, \doi{10.3847/2041-8213/aade02}, \eprint{1806.04088}

\bibitem[{{Kawaguchi} et~al.(2020{\natexlab{a}}){Kawaguchi}, {Shibata}, and
  {Tanaka}}]{Kawaguchi_Shibata_Tanaka2020-2}
{Kawaguchi} K, {Shibata} M, {Tanaka} M (2020{\natexlab{a}}) {Constraint on the
  Ejecta Mass for Black Hole-Neutron Star Merger Event Candidate S190814bv}.
  \apj 893:153, \doi{10.3847/1538-4357/ab8309}, \eprint{2002.01662}

\bibitem[{{Kawaguchi} et~al.(2020{\natexlab{b}}){Kawaguchi}, {Shibata}, and
  {Tanaka}}]{Kawaguchi_Shibata_Tanaka2020}
{Kawaguchi} K, {Shibata} M, {Tanaka} M (2020{\natexlab{b}}) {Diversity of
  Kilonova Light Curves}. \apj 889:171, \doi{10.3847/1538-4357/ab61f6},
  \eprint{1908.05815}

\bibitem[{{Kawanaka} and {Mineshige}(2007)}]{Kawanaka_Mineshige2007}
{Kawanaka} N, {Mineshige} S (2007) {Neutrino-cooled Accretion Disk and Its
  Stability}. \apj 662:1156--1166, \doi{10.1086/517985},
  \eprint{astro-ph/0702630}

\bibitem[{{Kennel} and {Coroniti}(1984)}]{Kennel_Coroniti1984}
{Kennel} CF, {Coroniti} FV (1984) {Confinement of the Crab pulsar's wind by its
  supernova remnant.} \apj 283:694--709, \doi{10.1086/162356}

\bibitem[{{Khan} et~al.(2016){Khan}, {Husa}, {Hannam}, {Ohme}, {P{\"u}rrer},
  {Forteza}, and {Boh{\'e}}}]{Khan_HHOPFB2016}
{Khan} S, {Husa} S, {Hannam} M, {Ohme} F, {P{\"u}rrer} M, {Forteza} XJ,
  {Boh{\'e}} A (2016) {Frequency-domain gravitational waves from nonprecessing
  black-hole binaries. II. A phenomenological model for the advanced detector
  era}. \prd 93:044007, \doi{10.1103/PhysRevD.93.044007}, \eprint{1508.07253}

\bibitem[{{Kidder}(1995)}]{Kidder1995}
{Kidder} LE (1995) {Coalescing binary systems of compact objects to
  (post)$^{5/2}$-Newtonian order. V. Spin effects}. \prd 52:821--847,
  \doi{10.1103/PhysRevD.52.821}, \eprint{gr-qc/9506022}

\bibitem[{{Kidder} et~al.(1993){Kidder}, {Will}, and
  {Wiseman}}]{Kidder_Will_Wiseman1993}
{Kidder} LE, {Will} CM, {Wiseman} AG (1993) {Spin effects in the inspiral of
  coalescing compact binaries}. \prd 47:R4183--R4187,
  \doi{10.1103/PhysRevD.47.R4183}, \eprint{gr-qc/9211025}

\bibitem[{{Kisaka} and {Ioka}(2015)}]{Kisaka_Ioka2015}
{Kisaka} S, {Ioka} K (2015) {Long-lasting Black Hole Jets in Short Gamma-Ray
  Bursts}. \apjl 804:L16, \doi{10.1088/2041-8205/804/1/L16},
  \eprint{1503.06791}

\bibitem[{{Kisaka} et~al.(2017){Kisaka}, {Ioka}, and
  {Sakamoto}}]{Kisaka_Ioka_Sakamoto2017}
{Kisaka} S, {Ioka} K, {Sakamoto} T (2017) {Bimodal Long-lasting Components in
  Short Gamma-Ray Bursts: Promising Electromagnetic Counterparts to Neutron
  Star Binary Mergers}. \apj 846:142, \doi{10.3847/1538-4357/aa8775},
  \eprint{1707.00675}

\bibitem[{{Kiuchi} et~al.(2011){Kiuchi}, {Shibata}, {Montero}, and
  {Font}}]{Kiuchi_SMF2011}
{Kiuchi} K, {Shibata} M, {Montero} PJ, {Font} JA (2011) {Gravitational Waves
  from the Papaloizou-Pringle Instability in Black-Hole-Torus Systems}. \prl
  106:251102, \doi{10.1103/PhysRevLett.106.251102}, \eprint{1105.5035}

\bibitem[{{Kiuchi} et~al.(2012){Kiuchi}, {Kyutoku}, and
  {Shibata}}]{Kiuchi_Kyutoku_Shibata2012}
{Kiuchi} K, {Kyutoku} K, {Shibata} M (2012) {Three-dimensional evolution of
  differentially rotating magnetized neutron stars}. \prd 86:064008,
  \doi{10.1103/PhysRevD.86.064008}, \eprint{1207.6444}

\bibitem[{{Kiuchi} et~al.(2015{\natexlab{a}}){Kiuchi}, {Cerd{\'a}-Dur{\'a}n},
  {Kyutoku}, {Sekiguchi}, and {Shibata}}]{Kiuchi_CKSS2015}
{Kiuchi} K, {Cerd{\'a}-Dur{\'a}n} P, {Kyutoku} K, {Sekiguchi} Y, {Shibata} M
  (2015{\natexlab{a}}) {Efficient magnetic-field amplification due to the
  Kelvin-Helmholtz instability in binary neutron star mergers}. \prd 92:124034,
  \doi{10.1103/PhysRevD.92.124034}, \eprint{1509.09205}

\bibitem[{{Kiuchi} et~al.(2015{\natexlab{b}}){Kiuchi}, {Sekiguchi}, {Kyutoku},
  {Shibata}, {Taniguchi}, and {Wada}}]{Kiuchi_SKSTW2015}
{Kiuchi} K, {Sekiguchi} Y, {Kyutoku} K, {Shibata} M, {Taniguchi} K, {Wada} T
  (2015{\natexlab{b}}) {High resolution magnetohydrodynamic simulation of black
  hole-neutron star merger: Mass ejection and short gamma ray bursts}. \prd
  92:064034, \doi{10.1103/PhysRevD.92.064034}, \eprint{1506.06811}

\bibitem[{{Kiuchi} et~al.(2018){Kiuchi}, {Kyutoku}, {Sekiguchi}, and
  {Shibata}}]{Kiuchi_KSS2018}
{Kiuchi} K, {Kyutoku} K, {Sekiguchi} Y, {Shibata} M (2018) {Global simulations
  of strongly magnetized remnant massive neutron stars formed in binary neutron
  star mergers}. \prd 97:124039, \doi{10.1103/PhysRevD.97.124039},
  \eprint{1710.01311}

\bibitem[{{Kiuchi} et~al.(2019){Kiuchi}, {Kyutoku}, {Shibata}, and
  {Taniguchi}}]{Kiuchi_KST2019}
{Kiuchi} K, {Kyutoku} K, {Shibata} M, {Taniguchi} K (2019) {Revisiting the
  Lower Bound on Tidal Deformability Derived by AT 2017gfo}. \apjl 876:L31,
  \doi{10.3847/2041-8213/ab1e45}, \eprint{1903.01466}

\bibitem[{{Klu{\'z}niak} and {Lee}(1998)}]{Kluzniak_Lee1998}
{Klu{\'z}niak} W, {Lee} WH (1998) {Simulations of Binary Coalescence of a
  Neutron Star and a Black Hole}. \apjl 494:L53--L55, \doi{10.1086/311151},
  \eprint{astro-ph/9712019}

\bibitem[{{Kochanek}(1992)}]{Kochanek1992}
{Kochanek} CS (1992) {Coalescing Binary Neutron Stars}. \apj 398:234,
  \doi{10.1086/171851}

\bibitem[{{Kohri} and {Mineshige}(2002)}]{Kohri_Mineshige2002}
{Kohri} K, {Mineshige} S (2002) {Can Neutrino-cooled Accretion Disks Be an
  Origin of Gamma-Ray Bursts?} \apj 577:311--321, \doi{10.1086/342166},
  \eprint{astro-ph/0203177}

\bibitem[{{Kohri} et~al.(2005){Kohri}, {Narayan}, and
  {Piran}}]{Kohri_Narayan_Piran2005}
{Kohri} K, {Narayan} R, {Piran} T (2005) {Neutrino-dominated Accretion and
  Supernovae}. \apj 629:341--361, \doi{10.1086/431354},
  \eprint{astro-ph/0502470}

\bibitem[{{Komar}(1959)}]{Komar1959}
{Komar} A (1959) {Covariant Conservation Laws in General Relativity}. Physical
  Review 113:934--936, \doi{10.1103/PhysRev.113.934}

\bibitem[{{Korobkin} et~al.(2011){Korobkin}, {Abdikamalov}, {Schnetter},
  {Stergioulas}, and {Zink}}]{Korobkin_ASSZ2011}
{Korobkin} O, {Abdikamalov} EB, {Schnetter} E, {Stergioulas} N, {Zink} B (2011)
  {Stability of general-relativistic accretion disks}. \prd 83:043007,
  \doi{10.1103/PhysRevD.83.043007}, \eprint{1011.3010}

\bibitem[{{Kreidberg} et~al.(2012){Kreidberg}, {Bailyn}, {Farr}, and
  {Kalogera}}]{Kreidberg_BFK2012}
{Kreidberg} L, {Bailyn} CD, {Farr} WM, {Kalogera} V (2012) {Mass Measurements
  of Black Holes in X-Ray Transients: Is There a Mass Gap?} \apj 757:36,
  \doi{10.1088/0004-637X/757/1/36}, \eprint{1205.1805}

\bibitem[{{Kroon}(2004)}]{Kroon2004}
{Kroon} JA (2004) {Nonexistence of Conformally Flat Slices in Kerr and Other
  Stationary Spacetimes}. \prl 92:041101, \doi{10.1103/PhysRevLett.92.041101},
  \eprint{gr-qc/0310048}

\bibitem[{{Kruckow} et~al.(2018){Kruckow}, {Tauris}, {Langer}, {Kramer}, and
  {Izzard}}]{Kruckow_TLKI2018}
{Kruckow} MU, {Tauris} TM, {Langer} N, {Kramer} M, {Izzard} RG (2018)
  {Progenitors of gravitational wave mergers: binary evolution with the stellar
  grid-based code COMBINE}. \mnras 481:1908--1949, \doi{10.1093/mnras/sty2190},
  \eprint{1801.05433}

\bibitem[{{Kr{\"u}ger} and {Foucart}(2020)}]{Kruger_Foucart2020}
{Kr{\"u}ger} CJ, {Foucart} F (2020) {Estimates for disk and ejecta masses
  produced in compact binary mergers}. \prd 101:103002,
  \doi{10.1103/PhysRevD.101.103002}, \eprint{2002.07728}

\bibitem[{{Kulkarni}(2005)}]{Kulkarni2005}
{Kulkarni} SR (2005) {Modeling Supernova-like Explosions Associated with
  Gamma-ray Bursts with Short Durations}. arXiv e-prints astro-ph/0510256,
  \eprint{astro-ph/0510256}

\bibitem[{{Kyutoku} et~al.(2009){Kyutoku}, {Shibata}, and
  {Taniguchi}}]{Kyutoku_Shibata_Taniguchi2009}
{Kyutoku} K, {Shibata} M, {Taniguchi} K (2009) {Quasiequilibrium states of
  black hole-neutron star binaries in the moving-puncture framework}. \prd
  79:124018, \doi{10.1103/PhysRevD.79.124018}, \eprint{0906.0889}

\bibitem[{{Kyutoku} et~al.(2010){Kyutoku}, {Shibata}, and
  {Taniguchi}}]{Kyutoku_Shibata_Taniguchi2010}
{Kyutoku} K, {Shibata} M, {Taniguchi} K (2010) {Gravitational waves from
  nonspinning black hole-neutron star binaries: Dependence on equations of
  state}. \prd 82:044049, \doi{10.1103/PhysRevD.82.044049}, \eprint{1008.1460}

\bibitem[{{Kyutoku} et~al.(2011{\natexlab{a}}){Kyutoku}, {Okawa}, {Shibata},
  and {Taniguchi}}]{Kyutoku_OST2011}
{Kyutoku} K, {Okawa} H, {Shibata} M, {Taniguchi} K (2011{\natexlab{a}})
  {Gravitational waves from spinning black hole-neutron star binaries:
  dependence on black hole spins and on neutron star equations of state}. \prd
  84:064018, \doi{10.1103/PhysRevD.84.064018}, \eprint{1108.1189}

\bibitem[{{Kyutoku} et~al.(2011{\natexlab{b}}){Kyutoku}, {Shibata}, and
  {Taniguchi}}]{Kyutoku_Shibata_Taniguchi2010e}
{Kyutoku} K, {Shibata} M, {Taniguchi} K (2011{\natexlab{b}}) {Erratum:
  Gravitational waves from nonspinning black hole-neutron star binaries:
  Dependence on equations of state [Phys. Rev. DPRVDAQ1550-7998 82, 044049
  (2010)10.1103/PhysRevD.82.044049]}. \prd 84:049902,
  \doi{10.1103/PhysRevD.84.049902}

\bibitem[{{Kyutoku} et~al.(2013){Kyutoku}, {Ioka}, and
  {Shibata}}]{Kyutoku_Ioka_Shibata2013}
{Kyutoku} K, {Ioka} K, {Shibata} M (2013) {Anisotropic mass ejection from black
  hole-neutron star binaries: Diversity of electromagnetic counterparts}. \prd
  88:041503, \doi{10.1103/PhysRevD.88.041503}, \eprint{1305.6309}

\bibitem[{{Kyutoku} et~al.(2014){Kyutoku}, {Shibata}, and
  {Taniguchi}}]{Kyutoku_Shibata_Taniguchi2014}
{Kyutoku} K, {Shibata} M, {Taniguchi} K (2014) {Reducing orbital eccentricity
  in initial data of binary neutron stars}. \prd 90:064006,
  \doi{10.1103/PhysRevD.90.064006}, \eprint{1405.6207}

\bibitem[{{Kyutoku} et~al.(2015){Kyutoku}, {Ioka}, {Okawa}, {Shibata}, and
  {Taniguchi}}]{Kyutoku_IOST2015}
{Kyutoku} K, {Ioka} K, {Okawa} H, {Shibata} M, {Taniguchi} K (2015) {Dynamical
  mass ejection from black hole-neutron star binaries}. \prd 92:044028,
  \doi{10.1103/PhysRevD.92.044028}, \eprint{1502.05402}

\bibitem[{{Kyutoku} et~al.(2018){Kyutoku}, {Kiuchi}, {Sekiguchi}, {Shibata},
  and {Taniguchi}}]{Kyutoku_KSST2018}
{Kyutoku} K, {Kiuchi} K, {Sekiguchi} Y, {Shibata} M, {Taniguchi} K (2018)
  {Neutrino transport in black hole-neutron star binaries: Neutrino emission
  and dynamical mass ejection}. \prd 97:023009,
  \doi{10.1103/PhysRevD.97.023009}, \eprint{1710.00827}

\bibitem[{{Kyutoku} et~al.(2020){Kyutoku}, {Fujibayashi}, {Hayashi},
  {Kawaguchi}, {Kiuchi}, {Shibata}, and {Tanaka}}]{Kyutoku_FHKKST2020}
{Kyutoku} K, {Fujibayashi} S, {Hayashi} K, {Kawaguchi} K, {Kiuchi} K, {Shibata}
  M, {Tanaka} M (2020) {On the Possibility of GW190425 Being a Black
  Hole--Neutron Star Binary Merger}. \apjl 890:L4,
  \doi{10.3847/2041-8213/ab6e70}, \eprint{2001.04474}

\bibitem[{{Kyutoku} et~al.(2021){Kyutoku}, {Kawaguchi}, {Kiuchi}, {Shibata},
  and {Taniguchi}}]{Kyutoku_KKST2021}
{Kyutoku} K, {Kawaguchi} K, {Kiuchi} K, {Shibata} M, {Taniguchi} K (2021)
  {Reducing orbital eccentricity in initial data of black hole-neutron star
  binaries in the puncture framework}. \prd 103:023002,
  \doi{10.1103/PhysRevD.103.023002}, \eprint{2009.03896}

\bibitem[{{Lackey} et~al.(2006){Lackey}, {Nayyar}, and
  {Owen}}]{Lackey_Nayyar_Owen2006}
{Lackey} BD, {Nayyar} M, {Owen} BJ (2006) {Observational constraints on
  hyperons in neutron stars}. \prd 73:024021, \doi{10.1103/PhysRevD.73.024021},
  \eprint{astro-ph/0507312}

\bibitem[{{Lackey} et~al.(2012){Lackey}, {Kyutoku}, {Shibata}, {Brady}, and
  {Friedman}}]{Lackey_KSBF2012}
{Lackey} BD, {Kyutoku} K, {Shibata} M, {Brady} PR, {Friedman} JL (2012)
  {Extracting equation of state parameters from black hole-neutron star
  mergers: Nonspinning black holes}. \prd 85:044061,
  \doi{10.1103/PhysRevD.85.044061}, \eprint{1109.3402}

\bibitem[{{Lackey} et~al.(2014){Lackey}, {Kyutoku}, {Shibata}, {Brady}, and
  {Friedman}}]{Lackey_KSBF2014}
{Lackey} BD, {Kyutoku} K, {Shibata} M, {Brady} PR, {Friedman} JL (2014)
  {Extracting equation of state parameters from black hole-neutron star
  mergers: Aligned-spin black holes and a preliminary waveform model}. \prd
  89:043009, \doi{10.1103/PhysRevD.89.043009}, \eprint{1303.6298}

\bibitem[{{Lai}(2012)}]{Lai2012}
{Lai} D (2012) {DC Circuit Powered by Orbital Motion: Magnetic Interactions in
  Compact Object Binaries and Exoplanetary Systems}. \apjl 757:L3,
  \doi{10.1088/2041-8205/757/1/L3}, \eprint{1206.3723}

\bibitem[{{Lai} and {Wiseman}(1996)}]{Lai_Wiseman1996}
{Lai} D, {Wiseman} AG (1996) {Innermost stable circular orbit of inspiraling
  neutron-star binaries: Tidal effects, post-Newtonian effects, and the
  neutron-star equation of state}. \prd 54:3958--3964,
  \doi{10.1103/PhysRevD.54.3958}, \eprint{gr-qc/9609014}

\bibitem[{{Lai} et~al.(1993{\natexlab{a}}){Lai}, {Rasio}, and
  {Shapiro}}]{Lai_Rasio_Shapiro1993_2}
{Lai} D, {Rasio} FA, {Shapiro} SL (1993{\natexlab{a}}) {Ellipsoidal Figures of
  Equilibrium: Compressible Models}. \apjs 88:205, \doi{10.1086/191822}

\bibitem[{{Lai} et~al.(1993{\natexlab{b}}){Lai}, {Rasio}, and
  {Shapiro}}]{Lai_Rasio_Shapiro1993}
{Lai} D, {Rasio} FA, {Shapiro} SL (1993{\natexlab{b}}) {Hydrodynamic
  Instability and Coalescence of Close Binary Systems}. \apjl 406:L63,
  \doi{10.1086/186787}

\bibitem[{{Lai} et~al.(1994{\natexlab{a}}){Lai}, {Rasio}, and
  {Shapiro}}]{Lai_Rasio_Shapiro1994_2}
{Lai} D, {Rasio} FA, {Shapiro} SL (1994{\natexlab{a}}) {Equilibrium, Stability,
  and Orbital Evolution of Close Binary Systems}. \apj 423:344,
  \doi{10.1086/173812}, \eprint{astro-ph/9307032}

\bibitem[{{Lai} et~al.(1994{\natexlab{b}}){Lai}, {Rasio}, and
  {Shapiro}}]{Lai_Rasio_Shapiro1994}
{Lai} D, {Rasio} FA, {Shapiro} SL (1994{\natexlab{b}}) {Hydrodynamic
  Instability and Coalescence of Binary Neutron Stars}. \apj 420:811,
  \doi{10.1086/173606}, \eprint{astro-ph/9304027}

\bibitem[{{Lamb} et~al.(2019){Lamb}, {Lyman}, {Levan}, {Tanvir}, {Kangas},
  {Fruchter}, {Gompertz}, {Hjorth}, {Mandel}, {Oates}, and
  et~al.}]{Lamb_etal2019}
{Lamb} GP, {Lyman} JD, {Levan} AJ, {Tanvir} NR, {Kangas} T, {Fruchter} AS,
  {Gompertz} B, {Hjorth} J, {Mandel} I, {Oates} SR, et~al (2019) {The Optical
  Afterglow of GW170817 at One Year Post-merger}. \apj 870:L15,
  \doi{10.3847/2041-8213/aaf96b}, \eprint{1811.11491}

\bibitem[{{Lattimer}(2019)}]{Lattimer2019}
{Lattimer} JM (2019) {Impact of GW170817 for the nuclear physics of the EOS and
  the r-process}. Annals of Physics 411:167963, \doi{10.1016/j.aop.2019.167963}

\bibitem[{{Lattimer} and {Prakash}(2001)}]{Lattimer_Prakash2001}
{Lattimer} JM, {Prakash} M (2001) {Neutron Star Structure and the Equation of
  State}. \apj 550:426--442, \doi{10.1086/319702}, \eprint{astro-ph/0002232}

\bibitem[{{Lattimer} and {Prakash}(2016)}]{Lattimer_Prakash2016}
{Lattimer} JM, {Prakash} M (2016) {The equation of state of hot, dense matter
  and neutron stars}. \physrep 621:127--164,
  \doi{10.1016/j.physrep.2015.12.005}, \eprint{1512.07820}

\bibitem[{{Lattimer} and {Schramm}(1974)}]{Lattimer_Schramm1974}
{Lattimer} JM, {Schramm} DN (1974) {Black-Hole-Neutron-Star Collisions}. \apjl
  192:L145, \doi{10.1086/181612}

\bibitem[{{Lattimer} and {Schramm}(1976)}]{Lattimer_Schramm1976}
{Lattimer} JM, {Schramm} DN (1976) {The tidal disruption of neutron stars by
  black holes in close binaries.} \apj 210:549--567, \doi{10.1086/154860}

\bibitem[{{Lattimer} and {Swesty}(1991)}]{Lattimer_Swesty1991}
{Lattimer} JM, {Swesty} DF (1991) {A generalized equation of state for hot,
  dense matter}. \nphysa 535:331--376, \doi{10.1016/0375-9474(91)90452-C}

\bibitem[{{Lazzati} et~al.(2017){Lazzati}, {L{\'o}pez-C{\'a}mara}, {Cantiello},
  {Morsony}, {Perna}, and {Workman}}]{Lazzati_LCMPW2017}
{Lazzati} D, {L{\'o}pez-C{\'a}mara} D, {Cantiello} M, {Morsony} BJ, {Perna} R,
  {Workman} JC (2017) {Off-axis Prompt X-Ray Transients from the Cocoon of
  Short Gamma-Ray Bursts}. \apjl 848:L6, \doi{10.3847/2041-8213/aa8f3d},
  \eprint{1709.01468}

\bibitem[{{Lee}(2000)}]{Lee2000}
{Lee} WH (2000) {Newtonian hydrodynamics of the coalescence of black holes with
  neutron stars - III. Irrotational binaries with a stiff equation of state}.
  \mnras 318:606--624, \doi{10.1046/j.1365-8711.2000.03870.x},
  \eprint{astro-ph/0007206}

\bibitem[{{Lee}(2001)}]{Lee2001}
{Lee} WH (2001) {Newtonian hydrodynamics of the coalescence of black holes with
  neutron stars -- IV. Irrotational binaries with a soft equation of state}.
  \mnras 328:583--600, \doi{10.1046/j.1365-8711.2001.04898.x},
  \eprint{astro-ph/0108236}

\bibitem[{{Lee} and {Klu{\'z}niak}(1999{\natexlab{a}})}]{Lee_Kluzniak1999-2}
{Lee} WH, {Klu{\'z}niak} W (1999{\natexlab{a}}) {Newtonian hydrodynamics of the
  coalescence of black holes with neutron stars - II. Tidally locked binaries
  with a soft equation of state}. \mnras 308:780--794,
  \doi{10.1046/j.1365-8711.1999.02734.x}, \eprint{astro-ph/9904328}

\bibitem[{{Lee} and {Klu{\'z}niak}(1999{\natexlab{b}})}]{Lee_Kluzniak1999}
{Lee} WH, {Klu{\'z}niak} W (1999{\natexlab{b}}) {Newtonian Hydrodynamics of the
  Coalescence of Black Holes with Neutron Stars. I. Tidally Locked Binaries
  with a Stiff Equation of State}. \apj 526:178--199, \doi{10.1086/307958},
  \eprint{astro-ph/9808185}

\bibitem[{{Lee} et~al.(2004){Lee}, {Ramirez-Ruiz}, and
  {Page}}]{Lee_RamirezRuiz_Page2004}
{Lee} WH, {Ramirez-Ruiz} E, {Page} D (2004) {Opaque or Transparent? A Link
  between Neutrino Optical Depths and the Characteristic Duration of Short
  Gamma-Ray Bursts}. \apjl 608:L5--L8, \doi{10.1086/422217},
  \eprint{astro-ph/0404566}

\bibitem[{{Lee} et~al.(2005){Lee}, {Ramirez-Ruiz}, and
  {Page}}]{Lee_RamirezRuiz_Page2005}
{Lee} WH, {Ramirez-Ruiz} E, {Page} D (2005) {Dynamical Evolution of
  Neutrino-cooled Accretion Disks: Detailed Microphysics, Lepton-driven
  Convection, and Global Energetics}. \apj 632:421--437, \doi{10.1086/432373},
  \eprint{astro-ph/0506121}

\bibitem[{{Lee} et~al.(2009){Lee}, {Ramirez-Ruiz}, and
  {L{\'o}pez-C{\'a}mara}}]{Lee_RamirezRuiz_LopezCamara2009}
{Lee} WH, {Ramirez-Ruiz} E, {L{\'o}pez-C{\'a}mara} D (2009) {Phase Transitions
  and He-Synthesis-Driven Winds in Neutrino Cooled Accretion Disks: Prospects
  for Late Flares in Short Gamma-Ray Bursts}. \apjl 699:L93--L96,
  \doi{10.1088/0004-637X/699/2/L93}, \eprint{0904.3752}

\bibitem[{{Lehner} and {Pretorius}(2014)}]{Lehner_Pretorius2014}
{Lehner} L, {Pretorius} F (2014) {Numerical Relativity and Astrophysics}. \araa
  52:661--694, \doi{10.1146/annurev-astro-081913-040031}, \eprint{1405.4840}

\bibitem[{{Lehner} et~al.(2016){Lehner}, {Liebling}, {Palenzuela}, {Caballero},
  {O'Connor}, {Anderson}, and {Neilsen}}]{Lehner_LPCOAN2016}
{Lehner} L, {Liebling} SL, {Palenzuela} C, {Caballero} OL, {O'Connor} E,
  {Anderson} M, {Neilsen} D (2016) {Unequal mass binary neutron star mergers
  and multimessenger signals}. Class Quantum Grav 33:184002,
  \doi{10.1088/0264-9381/33/18/184002}, \eprint{1603.00501}

\bibitem[{{Levermore}(1984)}]{Levermore1984}
{Levermore} CD (1984) {Relating Eddington factors to flux limiters.} \jqsrt
  31:149--160, \doi{10.1016/0022-4073(84)90112-2}

\bibitem[{{Levin} et~al.(2018){Levin}, {D'Orazio}, and
  {Garcia-Saenz}}]{Levin_DOrazio_GarciaSaenz2018}
{Levin} J, {D'Orazio} DJ, {Garcia-Saenz} S (2018) {Black hole pulsar}. \prd
  98:123002, \doi{10.1103/PhysRevD.98.123002}, \eprint{1808.07887}

\bibitem[{{Li} and {Paczy{\'n}ski}(1998)}]{Li_Paczynski1998}
{Li} LX, {Paczy{\'n}ski} B (1998) {Transient Events from Neutron Star Mergers}.
  \apjl 507:L59--L62, \doi{10.1086/311680}, \eprint{astro-ph/9807272}

\bibitem[{{Li} and {Siegel}(2021)}]{Li_Siegel2021}
{Li} X, {Siegel} DM (2021) {Neutrino Fast Flavor Conversions in Neutron-Star
  Postmerger Accretion Disks}. \prl 126:251101,
  \doi{10.1103/PhysRevLett.126.251101}, \eprint{2103.02616}

\bibitem[{{Li} and {Shen}(2019)}]{Li_Shen2019}
{Li} Y, {Shen} RF (2019) {Polarization of Kilonova Emission from a Black
  Hole-Neutron Star Merger}. \apj 879:31, \doi{10.3847/1538-4357/ab2387},
  \eprint{1904.11841}

\bibitem[{{Lindblom}(1992)}]{Lindblom1992}
{Lindblom} L (1992) {Determining the Nuclear Equation of State from
  Neutron-Star Masses and Radii}. \apj 398:569, \doi{10.1086/171882}

\bibitem[{{Lindblom}(2010)}]{Lindblom2010}
{Lindblom} L (2010) {Spectral representations of neutron-star equations of
  state}. \prd 82:103011, \doi{10.1103/PhysRevD.82.103011}, \eprint{1009.0738}

\bibitem[{{Lindblom} et~al.(2006){Lindblom}, {Scheel}, {Kidder}, {Owen}, and
  {Rinne}}]{Lindblom_SKOR2006}
{Lindblom} L, {Scheel} MA, {Kidder} LE, {Owen} R, {Rinne} O (2006) {A new
  generalized harmonic evolution system}. Class Quantum Grav 23:S447--S462,
  \doi{10.1088/0264-9381/23/16/S09}, \eprint{gr-qc/0512093}

\bibitem[{{Lindquist}(1966)}]{Lindquist1966}
{Lindquist} RW (1966) {Relativistic transport theory}. Annals of Physics
  37:487--518, \doi{10.1016/0003-4916(66)90207-7}

\bibitem[{{Lipunov} et~al.(2017){Lipunov}, {Gorbovskoy}, {Kornilov}, {Tyurina},
  {Balanutsa}, {Kuznetsov}, {Vlasenko}, {Kuvshinov}, {Gorbunov}, {Buckley}, and
  et~al.}]{Lipunov_etal2017}
{Lipunov} VM, {Gorbovskoy} E, {Kornilov} VG, {Tyurina} N, {Balanutsa} P,
  {Kuznetsov} A, {Vlasenko} D, {Kuvshinov} D, {Gorbunov} I, {Buckley} DAH,
  et~al (2017) {MASTER Optical Detection of the First LIGO/Virgo Neutron Star
  Binary Merger GW170817}. \apjl 850:L1, \doi{10.3847/2041-8213/aa92c0},
  \eprint{1710.05461}

\bibitem[{{Liska} et~al.(2020){Liska}, {Tchekhovskoy}, and
  {Quataert}}]{Liska_Tchekhovskoy_QuataertQ2020}
{Liska} M, {Tchekhovskoy} A, {Quataert} E (2020) {Large-scale poloidal magnetic
  field dynamo leads to powerful jets in GRMHD simulations of black hole
  accretion with toroidal field}. \mnras 494:3656--3662,
  \doi{10.1093/mnras/staa955}, \eprint{1809.04608}

\bibitem[{{Littenberg} et~al.(2015){Littenberg}, {Farr}, {Coughlin},
  {Kalogera}, and {Holz}}]{Littenberg_FCKH2015}
{Littenberg} TB, {Farr} B, {Coughlin} S, {Kalogera} V, {Holz} DE (2015)
  {Neutron Stars versus Black Holes: Probing the Mass Gap with LIGO/Virgo}.
  \apjl 807:L24, \doi{10.1088/2041-8205/807/2/L24}, \eprint{1503.03179}

\bibitem[{{Liu} et~al.(2009){Liu}, {Etienne}, and
  {Shapiro}}]{Liu_Etienne_Shapiro2009}
{Liu} YT, {Etienne} ZB, {Shapiro} SL (2009) {Evolution of near-extremal-spin
  black holes using the moving puncture technique}. \prd 80:121503,
  \doi{10.1103/PhysRevD.80.121503}, \eprint{1001.4077}

\bibitem[{{Lodato} et~al.(2009){Lodato}, {King}, and
  {Pringle}}]{Lodato_King_Pringle2009}
{Lodato} G, {King} AR, {Pringle} JE (2009) {Stellar disruption by a
  supermassive black hole: is the light curve really proportional to
  t$^{-5/3}$?} \mnras 392:332--340, \doi{10.1111/j.1365-2966.2008.14049.x},
  \eprint{0810.1288}

\bibitem[{{L{\"o}ffler} et~al.(2006){L{\"o}ffler}, {Rezzolla}, and
  {Ansorg}}]{Loffler_Rezzolla_Ansorg2006}
{L{\"o}ffler} F, {Rezzolla} L, {Ansorg} M (2006) {Numerical evolutions of a
  black hole-neutron star system in full general relativity: Head-on
  collision}. \prd 74:104018, \doi{10.1103/PhysRevD.74.104018},
  \eprint{gr-qc/0606104}

\bibitem[{{Lovelace} et~al.(2008){Lovelace}, {Owen}, {Pfeiffer}, and
  {Chu}}]{Lovelace_OPC2008}
{Lovelace} G, {Owen} R, {Pfeiffer} HP, {Chu} T (2008) {Binary-black-hole
  initial data with nearly extremal spins}. \prd 78:084017,
  \doi{10.1103/PhysRevD.78.084017}, \eprint{0805.4192}

\bibitem[{{Lovelace} et~al.(2011){Lovelace}, {Scheel}, and
  {Szil{\'a}gyi}}]{Lovelace_Scheel_Szilagyi2011}
{Lovelace} G, {Scheel} MA, {Szil{\'a}gyi} B (2011) {Simulating merging binary
  black holes with nearly extremal spins}. \prd 83:024010,
  \doi{10.1103/PhysRevD.83.024010}, \eprint{1010.2777}

\bibitem[{{Lovelace} et~al.(2012){Lovelace}, {Boyle}, {Scheel}, and
  {Szil{\'a}gyi}}]{Lovelace_BSS2012}
{Lovelace} G, {Boyle} M, {Scheel} MA, {Szil{\'a}gyi} B (2012) {High-accuracy
  gravitational waveforms for binary black hole mergers with nearly extremal
  spins}. Class Quantum Grav 29:045003, \doi{10.1088/0264-9381/29/4/045003},
  \eprint{1110.2229}

\bibitem[{{Lovelace} et~al.(2013){Lovelace}, {Duez}, {Foucart}, {Kidder},
  {Pfeiffer}, {Scheel}, and {Szil{\'a}gyi}}]{Lovelace_DFKPSS2013}
{Lovelace} G, {Duez} MD, {Foucart} F, {Kidder} LE, {Pfeiffer} HP, {Scheel} MA,
  {Szil{\'a}gyi} B (2013) {Massive disc formation in the tidal disruption of a
  neutron star by a nearly extremal black hole}. Class Quantum Grav 30:135004,
  \doi{10.1088/0264-9381/30/13/135004}, \eprint{1302.6297}

\bibitem[{{Maggiore} et~al.(2020){Maggiore}, {Van Den Broeck}, {Bartolo},
  {Belgacem}, {Bertacca}, {Bizouard}, {Branchesi}, {Clesse}, {Foffa},
  {Garc{\'\i}a-Bellido}, and et~al.}]{EinsteinTelescope}
{Maggiore} M, {Van Den Broeck} C, {Bartolo} N, {Belgacem} E, {Bertacca} D,
  {Bizouard} MA, {Branchesi} M, {Clesse} S, {Foffa} S, {Garc{\'\i}a-Bellido} J,
  et~al (2020) {Science case for the Einstein telescope}. \jcap 2020:050,
  \doi{10.1088/1475-7516/2020/03/050}, \eprint{1912.02622}

\bibitem[{{Malkus} et~al.(2012){Malkus}, {Kneller}, {McLaughlin}, and
  {Surman}}]{Malkus_KMS2012}
{Malkus} A, {Kneller} JP, {McLaughlin} GC, {Surman} R (2012) {Neutrino
  oscillations above black hole accretion disks: Disks with electron-flavor
  emission}. \prd 86:085015, \doi{10.1103/PhysRevD.86.085015},
  \eprint{1207.6648}

\bibitem[{{Malkus} et~al.(2016){Malkus}, {McLaughlin}, and
  {Surman}}]{Malkus_McLaughlin_Surman2016}
{Malkus} A, {McLaughlin} GC, {Surman} R (2016) {Symmetric and standard matter
  neutrino resonances above merging compact objects}. \prd 93:045021,
  \doi{10.1103/PhysRevD.93.045021}, \eprint{1507.00946}

\bibitem[{{Manasse} and {Misner}(1963)}]{Manasse_Misner1963}
{Manasse} FK, {Misner} CW (1963) {Fermi Normal Coordinates and Some Basic
  Concepts in Differential Geometry}. J Math Phys 4:735--745,
  \doi{10.1063/1.1724316}

\bibitem[{{Marck}(1983)}]{Marck1983}
{Marck} JA (1983) {Solution to the Equations of Parallel Transport in Kerr
  Geometry; Tidal Tensor}. Proceedings of the Royal Society of London Series A
  385:431--438, \doi{10.1098/rspa.1983.0021}

\bibitem[{{Margalit} and {Piran}(2015)}]{Margalit_Piran2015}
{Margalit} B, {Piran} T (2015) {Radio flares of compact binary mergers: the
  effect of non-trivial outflow geometry}. \mnras 452:3419--3434,
  \doi{10.1093/mnras/stv1550}, \eprint{1503.06218}

\bibitem[{{Marronetti} et~al.(2008){Marronetti}, {Tichy}, {Br{\"u}gmann},
  {Gonz{\'a}lez}, and {Sperhake}}]{Marronetti_TBGS2008}
{Marronetti} P, {Tichy} W, {Br{\"u}gmann} B, {Gonz{\'a}lez} J, {Sperhake} U
  (2008) {High-spin binary black hole mergers}. \prd 77:064010,
  \doi{10.1103/PhysRevD.77.064010}, \eprint{0709.2160}

\bibitem[{{Mart{\'\i}} and {M{\"u}ller}(2015)}]{Marti_Muller2015}
{Mart{\'\i}} JM, {M{\"u}ller} E (2015) {Grid-based Methods in Relativistic
  Hydrodynamics and Magnetohydrodynamics}. Living Rev Comput Astrophys 1:3,
  \doi{10.1007/lrca-2015-3}

\bibitem[{{Maselli} et~al.(2012){Maselli}, {Gualtieri}, {Pannarale}, and
  {Ferrari}}]{Maselli_GPF2012}
{Maselli} A, {Gualtieri} L, {Pannarale} F, {Ferrari} V (2012) {On the validity
  of the adiabatic approximation in compact binary inspirals}. \prd 86:044032,
  \doi{10.1103/PhysRevD.86.044032}, \eprint{1205.7006}

\bibitem[{{Maselli} et~al.(2013){Maselli}, {Gualtieri}, and
  {Ferrari}}]{Maselli_Gualtieri_Ferrari2013}
{Maselli} A, {Gualtieri} L, {Ferrari} V (2013) {Constraining the equation of
  state of nuclear matter with gravitational wave observations: Tidal
  deformability and tidal disruption}. \prd 88:104040,
  \doi{10.1103/PhysRevD.88.104040}, \eprint{1310.5381}

\bibitem[{{Mashhoon}(1975)}]{Mashhoon1975}
{Mashhoon} B (1975) {On tidal phenomena in a strong gravitational field.} \apj
  197:705--716, \doi{10.1086/153560}

\bibitem[{{Mashonkina} et~al.(2014){Mashonkina}, {Christlieb}, and
  {Eriksson}}]{Mashonkina_Christlieb_Eriksson2014}
{Mashonkina} L, {Christlieb} N, {Eriksson} K (2014) {The Hamburg/ESO R-process
  Enhanced Star survey (HERES). X. HE 2252-4225, one more r-process enhanced
  and actinide-boost halo star}. \aap 569:A43,
  \doi{10.1051/0004-6361/201424017}, \eprint{1407.5379}

\bibitem[{{Matas} et~al.(2020){Matas}, {Dietrich}, {Buonanno}, {Hinderer},
  {P{\"u}rrer}, {Foucart}, {Boyle}, {Duez}, {Kidder}, {Pfeiffer}, and
  et~al.}]{Matas_etal2020}
{Matas} A, {Dietrich} T, {Buonanno} A, {Hinderer} T, {P{\"u}rrer} M, {Foucart}
  F, {Boyle} M, {Duez} MD, {Kidder} LE, {Pfeiffer} HP, et~al (2020)
  {Aligned-spin neutron-star-black-hole waveform model based on the
  effective-one-body approach and numerical-relativity simulations}. \prd
  102:043023, \doi{10.1103/PhysRevD.102.043023}, \eprint{2004.10001}

\bibitem[{{Matzner}(2003)}]{Matzner2003}
{Matzner} CD (2003) {Supernova hosts for gamma-ray burst jets: dynamical
  constraints}. \mnras 345:575--589, \doi{10.1046/j.1365-8711.2003.06969.x},
  \eprint{astro-ph/0203085}

\bibitem[{{McKinney} et~al.(2013){McKinney}, {Tchekhovskoy}, and {Bland
  ford}}]{McKinney_Tchekhovskoy_Blandford2013}
{McKinney} JC, {Tchekhovskoy} A, {Bland ford} RD (2013) {Alignment of
  Magnetized Accretion Disks and Relativistic Jets with Spinning Black Holes}.
  Science 339(6115):49, \doi{10.1126/science.1230811}, \eprint{1211.3651}

\bibitem[{{McWilliams} and {Levin}(2011)}]{McWilliams_Levin2011}
{McWilliams} ST, {Levin} J (2011) {Electromagnetic Extraction of Energy from
  Black-hole-Neutron-star Binaries}. \apj 742:90,
  \doi{10.1088/0004-637X/742/2/90}, \eprint{1101.1969}

\bibitem[{{Mendoza-Temis} et~al.(2015){Mendoza-Temis}, {Wu}, {Langanke},
  {Mart{\'\i}nez-Pinedo}, {Bauswein}, and {Janka}}]{MendozaTemis_WLMBJ2015}
{Mendoza-Temis} JdJ, {Wu} MR, {Langanke} K, {Mart{\'\i}nez-Pinedo} G,
  {Bauswein} A, {Janka} HT (2015) {Nuclear robustness of the r process in
  neutron-star mergers}. \prc 92:055805, \doi{10.1103/PhysRevC.92.055805},
  \eprint{1409.6135}

\bibitem[{{M{\'e}sz{\'a}ros} and {Rees}(1997)}]{Meszaros_Rees1997}
{M{\'e}sz{\'a}ros} P, {Rees} MJ (1997) {Poynting Jets from Black Holes and
  Cosmological Gamma-Ray Bursts}. \apjl 482:L29--L32, \doi{10.1086/310692},
  \eprint{astro-ph/9609065}

\bibitem[{{M{\'e}sz{\'a}ros} and {Rees}(2000)}]{Meszaros_Rees2000}
{M{\'e}sz{\'a}ros} P, {Rees} MJ (2000) {Steep Slopes and Preferred Breaks in
  Gamma-Ray Burst Spectra: The Role of Photospheres and Comptonization}. \apj
  530:292--298, \doi{10.1086/308371}, \eprint{astro-ph/9908126}

\bibitem[{{Metzger}(2019)}]{Metzger2019}
{Metzger} BD (2019) {Kilonovae}. Living Rev Relativ 23:1,
  \doi{10.1007/s41114-019-0024-0}, \eprint{1910.01617}

\bibitem[{{Metzger} et~al.(2010{\natexlab{a}}){Metzger}, {Arcones}, {Quataert},
  and {Mart{\'\i}nez-Pinedo}}]{Metzger_AQM2010}
{Metzger} BD, {Arcones} A, {Quataert} E, {Mart{\'\i}nez-Pinedo} G
  (2010{\natexlab{a}}) {The effects of r-process heating on fallback accretion
  in compact object mergers}. \mnras 402:2771--2777,
  \doi{10.1111/j.1365-2966.2009.16107.x}, \eprint{0908.0530}

\bibitem[{{Metzger} et~al.(2010{\natexlab{b}}){Metzger},
  {Mart{\'\i}nez-Pinedo}, {Darbha}, {Quataert}, {Arcones}, {Kasen}, {Thomas},
  {Nugent}, {Panov}, and {Zinner}}]{Metzger_MDQAKTNPZ2010}
{Metzger} BD, {Mart{\'\i}nez-Pinedo} G, {Darbha} S, {Quataert} E, {Arcones} A,
  {Kasen} D, {Thomas} R, {Nugent} P, {Panov} IV, {Zinner} NT
  (2010{\natexlab{b}}) {Electromagnetic counterparts of compact object mergers
  powered by the radioactive decay of r-process nuclei}. \mnras 406:2650--2662,
  \doi{10.1111/j.1365-2966.2010.16864.x}, \eprint{1001.5029}

\bibitem[{{Mewes} et~al.(2016){Mewes}, {Font}, {Galeazzi}, {Montero}, and
  {Stergioulas}}]{Mewes_FGMS2016}
{Mewes} V, {Font} JA, {Galeazzi} F, {Montero} PJ, {Stergioulas} N (2016)
  {Numerical relativity simulations of thick accretion disks around tilted Kerr
  black holes}. \prd 93:064055, \doi{10.1103/PhysRevD.93.064055},
  \eprint{1506.04056}

\bibitem[{{Mihalas} and {Mihalas}(1984)}]{Mihalas_Mihalas}
{Mihalas} D, {Mihalas} BW (1984) {Foundations of radiation hydrodynamics}.
  Oxford University Press, New York

\bibitem[{{Miller} et~al.(2019{\natexlab{a}}){Miller}, {Ryan}, {Dolence},
  {Burrows}, {Fontes}, {Fryer}, {Korobkin}, {Lippuner}, {Mumpower}, and
  {Wollaeger}}]{Miller_RDBFFKLMW2019}
{Miller} JM, {Ryan} BR, {Dolence} JC, {Burrows} A, {Fontes} CJ, {Fryer} CL,
  {Korobkin} O, {Lippuner} J, {Mumpower} MR, {Wollaeger} RT
  (2019{\natexlab{a}}) {Full transport model of GW170817-like disk produces a
  blue kilonova}. \prd 100:023008, \doi{10.1103/PhysRevD.100.023008},
  \eprint{1905.07477}

\bibitem[{{Miller}(2001)}]{Miller2001}
{Miller} M (2001) {General Relativistic Initial Data for the Binary Black Hole
  / Neutron Star System in Quasicircular Orbit}. arXiv e-prints gr-qc/0106017,
  \eprint{gr-qc/0106017}

\bibitem[{{Miller} et~al.(2019{\natexlab{b}}){Miller}, {Lamb}, {Dittmann},
  {Bogdanov}, {Arzoumanian}, {Gendreau}, {Guillot}, {Harding}, {Ho},
  {Lattimer}, and et~al.}]{Miller_etal2019}
{Miller} MC, {Lamb} FK, {Dittmann} AJ, {Bogdanov} S, {Arzoumanian} Z,
  {Gendreau} KC, {Guillot} S, {Harding} AK, {Ho} WCG, {Lattimer} JM, et~al
  (2019{\natexlab{b}}) {PSR J0030+0451 Mass and Radius from NICER Data and
  Implications for the Properties of Neutron Star Matter}. \apjl 887:L24,
  \doi{10.3847/2041-8213/ab50c5}, \eprint{1912.05705}

\bibitem[{{Mitman} et~al.(2020){Mitman}, {Moxon}, {Scheel}, {Teukolsky},
  {Boyle}, {Deppe}, {Kidder}, and {Throwe}}]{Mitman_MSTBDKT2020}
{Mitman} K, {Moxon} J, {Scheel} MA, {Teukolsky} SA, {Boyle} M, {Deppe} N,
  {Kidder} LE, {Throwe} W (2020) {Computation of displacement and spin
  gravitational memory in numerical relativity}. \prd 102:104007,
  \doi{10.1103/PhysRevD.102.104007}, \eprint{2007.11562}

\bibitem[{{Mizuta} and {Ioka}(2013)}]{Mizuta_Ioka2013}
{Mizuta} A, {Ioka} K (2013) {Opening Angles of Collapsar Jets}. \apj 777:162,
  \doi{10.1088/0004-637X/777/2/162}, \eprint{1304.0163}

\bibitem[{{Mochkovitch} et~al.(1993){Mochkovitch}, {Hernanz}, {Isern}, and
  {Martin}}]{Mochkovitch_HIM1993}
{Mochkovitch} R, {Hernanz} M, {Isern} J, {Martin} X (1993) {Gamma-ray bursts as
  collimated jets from neutron star/black hole mergers}. \nat 361:236--238,
  \doi{10.1038/361236a0}

\bibitem[{{Mooley} et~al.(2018{\natexlab{a}}){Mooley}, {Deller}, {Gottlieb},
  {Nakar}, {Hallinan}, {Bourke}, {Frail}, {Horesh}, {Corsi}, and
  {Hotokezaka}}]{Mooley_DGNHBFHCH2018}
{Mooley} KP, {Deller} AT, {Gottlieb} O, {Nakar} E, {Hallinan} G, {Bourke} S,
  {Frail} DA, {Horesh} A, {Corsi} A, {Hotokezaka} K (2018{\natexlab{a}})
  {Superluminal motion of a relativistic jet in the neutron-star merger
  GW170817}. \nat 561:355--359, \doi{10.1038/s41586-018-0486-3},
  \eprint{1806.09693}

\bibitem[{{Mooley} et~al.(2018{\natexlab{b}}){Mooley}, {Frail}, {Dobie},
  {Lenc}, {Corsi}, {De}, {Nayana}, {Makhathini}, {Heywood}, {Murphy}, and
  et~al.}]{Mooley_etal2018}
{Mooley} KP, {Frail} DA, {Dobie} D, {Lenc} E, {Corsi} A, {De} K, {Nayana} AJ,
  {Makhathini} S, {Heywood} I, {Murphy} T, et~al (2018{\natexlab{b}}) {A Strong
  Jet Signature in the Late-time Light Curve of GW170817}. \apj 868:L11,
  \doi{10.3847/2041-8213/aaeda7}, \eprint{1810.12927}

\bibitem[{{Most} et~al.(2021{\natexlab{a}}){Most}, {Papenfort}, {Tootle}, and
  {Rezzolla}}]{Most_PTR2021}
{Most} ER, {Papenfort} LJ, {Tootle} SD, {Rezzolla} L (2021{\natexlab{a}}) {Fast
  Ejecta as a Potential Way to Distinguish Black Holes from Neutron Stars in
  High-mass Gravitational-wave Events}. \apj 912:80,
  \doi{10.3847/1538-4357/abf0a5}, \eprint{2012.03896}

\bibitem[{{Most} et~al.(2021{\natexlab{b}}){Most}, {Papenfort}, {Tootle}, and
  {Rezzolla}}]{Most_PTR2021-2}
{Most} ER, {Papenfort} LJ, {Tootle} SD, {Rezzolla} L (2021{\natexlab{b}}) {On
  accretion disks formed in MHD simulations of black hole-neutron star mergers
  with accurate microphysics}. \mnras \doi{10.1093/mnras/stab1824},
  \eprint{2106.06391}

\bibitem[{{M{\"u}ller} and {Serot}(1996)}]{Muller_Serot1996}
{M{\"u}ller} H, {Serot} BD (1996) {Relativistic mean-field theory and the
  high-density nuclear equation of state}. \nphysa 606:508--537,
  \doi{10.1016/0375-9474(96)00187-X}, \eprint{nucl-th/9603037}

\bibitem[{{Mumpower} et~al.(2016){Mumpower}, {Surman}, {McLaughlin}, and
  {Aprahamian}}]{Mumpower_SMA2016}
{Mumpower} MR, {Surman} R, {McLaughlin} GC, {Aprahamian} A (2016) {The impact
  of individual nuclear properties on r-process nucleosynthesis}. Progress in
  Particle and Nuclear Physics 86:86--126, \doi{10.1016/j.ppnp.2015.09.001},
  \eprint{1508.07352}

\bibitem[{{Murguia-Berthier} et~al.(2014){Murguia-Berthier}, {Montes},
  {Ramirez-Ruiz}, {De Colle}, and {Lee}}]{MurguiaBerthier_MRDL2014}
{Murguia-Berthier} A, {Montes} G, {Ramirez-Ruiz} E, {De Colle} F, {Lee} WH
  (2014) {Necessary Conditions for Short Gamma-Ray Burst Production in Binary
  Neutron Star Mergers}. \apjl 788:L8, \doi{10.1088/2041-8205/788/1/L8},
  \eprint{1404.0383}

\bibitem[{{Murguia-Berthier} et~al.(2017){Murguia-Berthier}, {Ramirez-Ruiz},
  {Montes}, {De Colle}, {Rezzolla}, {Rosswog}, {Takami}, {Perego}, and
  {Lee}}]{MurguiaBerthier_RMDRRTPL2017}
{Murguia-Berthier} A, {Ramirez-Ruiz} E, {Montes} G, {De Colle} F, {Rezzolla} L,
  {Rosswog} S, {Takami} K, {Perego} A, {Lee} WH (2017) {The Properties of Short
  Gamma-Ray Burst Jets Triggered by Neutron Star Mergers}. \apjl 835:L34,
  \doi{10.3847/2041-8213/aa5b9e}, \eprint{1609.04828}

\bibitem[{{Nagakura} et~al.(2014){Nagakura}, {Hotokezaka}, {Sekiguchi},
  {Shibata}, and {Ioka}}]{Nagakura_HSSI2014}
{Nagakura} H, {Hotokezaka} K, {Sekiguchi} Y, {Shibata} M, {Ioka} K (2014) {Jet
  Collimation in the Ejecta of Double Neutron Star Mergers: A New Canonical
  Picture of Short Gamma-Ray Bursts}. \apjl 784:L28,
  \doi{10.1088/2041-8205/784/2/L28}, \eprint{1403.0956}

\bibitem[{{Nakamura} and {Oohara}(1983)}]{Nakamura_Oohara1983}
{Nakamura} T, {Oohara} KI (1983) {Gravitational radiation emitted by N
  particles in circular orbits}. Phys Lett A 98:403--406,
  \doi{10.1016/0375-9601(83)90248-7}

\bibitem[{{Nakamura} and {Sasaki}(1981)}]{Nakamura_Sasaki1981}
{Nakamura} T, {Sasaki} M (1981) {Is collapse of a deformed star always
  effectual for gravitational radiation?} Phys Lett B 106:69--72,
  \doi{10.1016/0370-2693(81)91082-0}

\bibitem[{{Nakamura} et~al.(1987){Nakamura}, {Oohara}, and
  {Kojima}}]{Nakamura_Oohara_Kojima1987}
{Nakamura} T, {Oohara} K, {Kojima} Y (1987) {General Relativistic Collapse to
  Black Holes and Gravitational Waves from Black Holes}. Progress of
  Theoretical Physics Supplement 90:1--218, \doi{10.1143/PTPS.90.1}

\bibitem[{{Nakar}(2007)}]{Nakar2007}
{Nakar} E (2007) {Short-hard gamma-ray bursts}. \physrep 442:166--236,
  \doi{10.1016/j.physrep.2007.02.005}, \eprint{astro-ph/0701748}

\bibitem[{{Nakar} and {Piran}(2011)}]{Nakar_Piran2011}
{Nakar} E, {Piran} T (2011) {Detectable radio flares following gravitational
  waves from mergers of binary neutron stars}. \nat 478:82--84,
  \doi{10.1038/nature10365}, \eprint{1102.1020}

\bibitem[{{Nakar} et~al.(2018){Nakar}, {Gottlieb}, {Piran}, {Kasliwal}, and
  {Hallinan}}]{Nakar_GPKH2018}
{Nakar} E, {Gottlieb} O, {Piran} T, {Kasliwal} MM, {Hallinan} G (2018) {From
  {\ensuremath{\gamma}} to Radio: The Electromagnetic Counterpart of GW170817}.
  \apj 867:18, \doi{10.3847/1538-4357/aae205}, \eprint{1803.07595}

\bibitem[{{Narayan} et~al.(1991){Narayan}, {Piran}, and
  {Shemi}}]{Narayan_Piran_Shemi1991}
{Narayan} R, {Piran} T, {Shemi} A (1991) {Neutron Star and Black Hole Binaries
  in the Galaxy}. \apjl 379:L17, \doi{10.1086/186143}

\bibitem[{{Narayan} et~al.(1992){Narayan}, {Paczynski}, and
  {Piran}}]{Narayan_Paczynski_Piran1992}
{Narayan} R, {Paczynski} B, {Piran} T (1992) {Gamma-Ray Bursts as the Death
  Throes of Massive Binary Stars}. \apjl 395:L83, \doi{10.1086/186493},
  \eprint{astro-ph/9204001}

\bibitem[{{Narayan} et~al.(2001){Narayan}, {Piran}, and
  {Kumar}}]{Narayan_Piran_Kumar2001}
{Narayan} R, {Piran} T, {Kumar} P (2001) {Accretion Models of Gamma-Ray
  Bursts}. \apj 557:949--957, \doi{10.1086/322267}, \eprint{astro-ph/0103360}

\bibitem[{{Narikawa} et~al.(2020){Narikawa}, {Uchikata}, {Kawaguchi}, {Kiuchi},
  {Kyutoku}, {Shibata}, and {Tagoshi}}]{Narikawa_UKKKST2020}
{Narikawa} T, {Uchikata} N, {Kawaguchi} K, {Kiuchi} K, {Kyutoku} K, {Shibata}
  M, {Tagoshi} H (2020) {Reanalysis of the binary neutron star mergers GW170817
  and GW190425 using numerical-relativity calibrated waveform models}. Physical
  Review Research 2:043039, \doi{10.1103/PhysRevResearch.2.043039},
  \eprint{1910.08971}

\bibitem[{{Neijssel} et~al.(2019){Neijssel}, {Vigna-G{\'o}mez}, {Stevenson},
  {Barrett}, {Gaebel}, {Broekgaarden}, {de Mink}, {Sz{\'e}csi}, {Vinciguerra},
  and {Mandel}}]{Neijssel_VSBGBMSVM2019}
{Neijssel} CJ, {Vigna-G{\'o}mez} A, {Stevenson} S, {Barrett} JW, {Gaebel} SM,
  {Broekgaarden} FS, {de Mink} SE, {Sz{\'e}csi} D, {Vinciguerra} S, {Mandel} I
  (2019) {The effect of the metallicity-specific star formation history on
  double compact object mergers}. \mnras 490:3740--3759,
  \doi{10.1093/mnras/stz2840}, \eprint{1906.08136}

\bibitem[{{Norris} and {Bonnell}(2006)}]{Norris_Bonnell2006}
{Norris} JP, {Bonnell} JT (2006) {Short Gamma-Ray Bursts with Extended
  Emission}. \apj 643:266--275, \doi{10.1086/502796}, \eprint{astro-ph/0601190}

\bibitem[{{{\'O} Murchadha} and {York}(1974)}]{OMurchadha_York1974}
{{\'O} Murchadha} N, {York} JW (1974) {Initial-value problem of general
  relativity. I. General formulation and physical interpretation}. \prd
  10:428--436, \doi{10.1103/PhysRevD.10.428}

\bibitem[{{O'Boyle} et~al.(2020){O'Boyle}, {Markakis}, {Stergioulas}, and
  {Read}}]{OBoyle_MSR2020}
{O'Boyle} MF, {Markakis} C, {Stergioulas} N, {Read} JS (2020) {Parametrized
  equation of state for neutron star matter with continuous sound speed}. \prd
  102:083027, \doi{10.1103/PhysRevD.102.083027}, \eprint{2008.03342}

\bibitem[{{O'Connor} and {Ott}(2010)}]{OConnor_Ott2010}
{O'Connor} E, {Ott} CD (2010) {A new open-source code for spherically symmetric
  stellar collapse to neutron stars and black holes}. Class Quantum Grav
  27:114103, \doi{10.1088/0264-9381/27/11/114103}, \eprint{0912.2393}

\bibitem[{{Oertel} et~al.(2017){Oertel}, {Hempel}, {Kl{\"a}hn}, and
  {Typel}}]{Oertel_HKT2017}
{Oertel} M, {Hempel} M, {Kl{\"a}hn} T, {Typel} S (2017) {Equations of state for
  supernovae and compact stars}. Reviews of Modern Physics 89:015007,
  \doi{10.1103/RevModPhys.89.015007}, \eprint{1610.03361}

\bibitem[{{Oppenheimer} and {Volkoff}(1939)}]{Oppenheimer_Volkoff1939}
{Oppenheimer} JR, {Volkoff} GM (1939) {On Massive Neutron Cores}. Physical
  Review 55:374--381, \doi{10.1103/PhysRev.55.374}

\bibitem[{{O'Shaughnessy} et~al.(2014{\natexlab{a}}){O'Shaughnessy}, {Farr},
  {Ochsner}, {Cho}, {Kim}, and {Lee}}]{OShaughnessy_FOCKL2014}
{O'Shaughnessy} R, {Farr} B, {Ochsner} E, {Cho} HS, {Kim} C, {Lee} CH
  (2014{\natexlab{a}}) {Parameter estimation of gravitational waves from
  nonprecessing black hole-neutron star inspirals with higher harmonics:
  Comparing Markov-chain Monte Carlo posteriors to an effective Fisher matrix}.
  \prd 89:064048, \doi{10.1103/PhysRevD.89.064048}, \eprint{1308.4704}

\bibitem[{{O'Shaughnessy} et~al.(2014{\natexlab{b}}){O'Shaughnessy}, {Farr},
  {Ochsner}, {Cho}, {Raymond}, {Kim}, and {Lee}}]{OShaughnessy_FOCRKL2014}
{O'Shaughnessy} R, {Farr} B, {Ochsner} E, {Cho} HS, {Raymond} V, {Kim} C, {Lee}
  CH (2014{\natexlab{b}}) {Parameter estimation of gravitational waves from
  precessing black hole-neutron star inspirals with higher harmonics}. \prd
  89:102005, \doi{10.1103/PhysRevD.89.102005}, \eprint{1403.0544}

\bibitem[{{{\"O}zel} and {Freire}(2016)}]{Ozel_Freire2016}
{{\"O}zel} F, {Freire} P (2016) {Masses, Radii, and the Equation of State of
  Neutron Stars}. \araa 54:401--440, \doi{10.1146/annurev-astro-081915-023322},
  \eprint{1603.02698}

\bibitem[{{{\"O}zel} and {Psaltis}(2009)}]{Ozel_Psaltis2009}
{{\"O}zel} F, {Psaltis} D (2009) {Reconstructing the neutron-star equation of
  state from astrophysical measurements}. \prd 80:103003,
  \doi{10.1103/PhysRevD.80.103003}, \eprint{0905.1959}

\bibitem[{{{\"O}zel} et~al.(2010){{\"O}zel}, {Psaltis}, {Narayan}, and
  {McClintock}}]{Ozel_PNM2010}
{{\"O}zel} F, {Psaltis} D, {Narayan} R, {McClintock} JE (2010) {The Black Hole
  Mass Distribution in the Galaxy}. \apj 725:1918--1927,
  \doi{10.1088/0004-637X/725/2/1918}, \eprint{1006.2834}

\bibitem[{{Paczy{\'n}ski}(1971)}]{Paczynski1971}
{Paczy{\'n}ski} B (1971) {Evolutionary Processes in Close Binary Systems}.
  \araa 9:183, \doi{10.1146/annurev.aa.09.090171.001151}

\bibitem[{{Paczynski}(1986)}]{Paczynski1986}
{Paczynski} B (1986) {Gamma-ray bursters at cosmological distances}. \apjl
  308:L43--L46, \doi{10.1086/184740}

\bibitem[{{Paczynski}(1991)}]{Paczynski1991}
{Paczynski} B (1991) {Cosmological gamma-ray bursts.} \actaa 41:257--267

\bibitem[{{Paczy{\'n}sky} and {Wiita}(1980)}]{Paczynski_Wiita1980}
{Paczy{\'n}sky} B, {Wiita} PJ (1980) {Thick accretion disks and supercritical
  luminosities.} \aap 500:203--211

\bibitem[{{Padilla-Gay} et~al.(2021){Padilla-Gay}, {Shalgar}, and
  {Tamborra}}]{PadillaGay_Shalgar_Tamborra2021}
{Padilla-Gay} I, {Shalgar} S, {Tamborra} I (2021) {Multi-dimensional solution
  of fast neutrino conversions in binary neutron star merger remnants}. \jcap
  2021:017, \doi{10.1088/1475-7516/2021/01/017}, \eprint{2009.01843}

\bibitem[{{Palenzuela}(2013)}]{Palenzuela2013}
{Palenzuela} C (2013) {Modelling magnetized neutron stars using resistive
  magnetohydrodynamics}. \mnras 431:1853--1865, \doi{10.1093/mnras/stt311},
  \eprint{1212.0130}

\bibitem[{{Palenzuela} et~al.(2009){Palenzuela}, {Lehner}, {Reula}, and
  {Rezzolla}}]{Palenzuela_LRR2009}
{Palenzuela} C, {Lehner} L, {Reula} O, {Rezzolla} L (2009) {Beyond ideal MHD:
  towards a more realistic modelling of relativistic astrophysical plasmas}.
  \mnras 394:1727--1740, \doi{10.1111/j.1365-2966.2009.14454.x},
  \eprint{0810.1838}

\bibitem[{{Palenzuela} et~al.(2015){Palenzuela}, {Liebling}, {Neilsen},
  {Lehner}, {Caballero}, {O'Connor}, and {Anderson}}]{Palenzuela_LNLCOA2015}
{Palenzuela} C, {Liebling} SL, {Neilsen} D, {Lehner} L, {Caballero} OL,
  {O'Connor} E, {Anderson} M (2015) {Effects of the microphysical equation of
  state in the mergers of magnetized neutron stars with neutrino cooling}. \prd
  92:044045, \doi{10.1103/PhysRevD.92.044045}, \eprint{1505.01607}

\bibitem[{{Pan} and {Yang}(2019)}]{Pan_Yang2019}
{Pan} Z, {Yang} H (2019) {Black hole discharge: Very-high-energy gamma rays
  from black hole-neutron star mergers}. \prd 100:043025,
  \doi{10.1103/PhysRevD.100.043025}, \eprint{1905.04775}

\bibitem[{{Pannarale}(2013)}]{Pannarale2013}
{Pannarale} F (2013) {Black hole remnant of black hole-neutron star coalescing
  binaries}. \prd 88:104025, \doi{10.1103/PhysRevD.88.104025},
  \eprint{1208.5869}

\bibitem[{{Pannarale}(2014)}]{Pannarale2014}
{Pannarale} F (2014) {Black hole remnant of black hole-neutron star coalescing
  binaries with arbitrary black hole spin}. \prd 89:044045,
  \doi{10.1103/PhysRevD.89.044045}, \eprint{1311.5931}

\bibitem[{{Pannarale} et~al.(2011){Pannarale}, {Tonita}, and
  {Rezzolla}}]{Pannarale_Tonita_Rezzolla2011}
{Pannarale} F, {Tonita} A, {Rezzolla} L (2011) {Black Hole-Neutron Star Mergers
  and Short Gamma-ray Bursts: A Relativistic Toy Model to Estimate the Mass of
  the Torus}. \apj 727:95, \doi{10.1088/0004-637X/727/2/95}, \eprint{1007.4160}

\bibitem[{{Pannarale} et~al.(2013){Pannarale}, {Berti}, {Kyutoku}, and
  {Shibata}}]{Pannarale_BKS2013}
{Pannarale} F, {Berti} E, {Kyutoku} K, {Shibata} M (2013) {Nonspinning black
  hole-neutron star mergers: A model for the amplitude of gravitational
  waveforms}. \prd 88:084011, \doi{10.1103/PhysRevD.88.084011},
  \eprint{1307.5111}

\bibitem[{{Pannarale} et~al.(2015{\natexlab{a}}){Pannarale}, {Berti},
  {Kyutoku}, {Lackey}, and {Shibata}}]{Pannarale_BKLS2015_2}
{Pannarale} F, {Berti} E, {Kyutoku} K, {Lackey} BD, {Shibata} M
  (2015{\natexlab{a}}) {Aligned spin neutron star-black hole mergers: A
  gravitational waveform amplitude model}. \prd 92:084050,
  \doi{10.1103/PhysRevD.92.084050}, \eprint{1509.00512}

\bibitem[{{Pannarale} et~al.(2015{\natexlab{b}}){Pannarale}, {Berti},
  {Kyutoku}, {Lackey}, and {Shibata}}]{Pannarale_BKLS2015}
{Pannarale} F, {Berti} E, {Kyutoku} K, {Lackey} BD, {Shibata} M
  (2015{\natexlab{b}}) {Gravitational-wave cutoff frequencies of tidally
  disruptive neutron star-black hole binary mergers}. \prd 92:081504,
  \doi{10.1103/PhysRevD.92.081504}, \eprint{1509.06209}

\bibitem[{{Papaloizou} and {Pringle}(1977)}]{Papaloizou_Pringle1977}
{Papaloizou} J, {Pringle} JE (1977) {Tidal torques on accretion discs in close
  binary systems}. \mnras 181:441--454, \doi{10.1093/mnras/181.3.441}

\bibitem[{{Papaloizou} and {Pringle}(1983)}]{Papaloizou_Pringle1983}
{Papaloizou} JCB, {Pringle} JE (1983) {The time-dependence of non-planar
  accretion discs}. \mnras 202:1181--1194, \doi{10.1093/mnras/202.4.1181}

\bibitem[{{Papenfort} et~al.(2021){Papenfort}, {Tootle}, {Grandcl{\'e}ment},
  {Most}, and {Rezzolla}}]{Papenfort_TGMR2021}
{Papenfort} LJ, {Tootle} SD, {Grandcl{\'e}ment} P, {Most} ER, {Rezzolla} L
  (2021) {New public code for initial data of unequal-mass, spinning
  compact-object binaries}. \prd 104:024057, \doi{10.1103/PhysRevD.104.024057},
  \eprint{2103.09911}

\bibitem[{{Paschalidis}(2017)}]{Paschalidis2017}
{Paschalidis} V (2017) {General relativistic simulations of compact binary
  mergers as engines for short gamma-ray bursts}. Class Quantum Grav 34:084002,
  \doi{10.1088/1361-6382/aa61ce}, \eprint{1611.01519}

\bibitem[{{Paschalidis} et~al.(2013){Paschalidis}, {Etienne}, and
  {Shapiro}}]{Paschalidis_Etienne_Shapiro2013}
{Paschalidis} V, {Etienne} ZB, {Shapiro} SL (2013) {General-relativistic
  simulations of binary black hole-neutron stars: Precursor electromagnetic
  signals}. \prd 88:021504, \doi{10.1103/PhysRevD.88.021504},
  \eprint{1304.1805}

\bibitem[{{Paschalidis} et~al.(2015){Paschalidis}, {Ruiz}, and
  {Shapiro}}]{Paschalidis_Ruiz_Shapiro2015}
{Paschalidis} V, {Ruiz} M, {Shapiro} SL (2015) {Relativistic Simulations of
  Black Hole-Neutron Star Coalescence: The Jet Emerges}. \apjl 806:L14,
  \doi{10.1088/2041-8205/806/1/L14}, \eprint{1410.7392}

\bibitem[{{Penrose}(1969)}]{Penrose1969}
{Penrose} R (1969) {Gravitational Collapse: the Role of General Relativity}.
  Nuovo Cimento Rivista Serie 1:252

\bibitem[{{Penrose}(2002)}]{Penrose2002}
{Penrose} R (2002) {``Golden Oldie'': Gravitational Collapse: The Role of
  General Relativity}. General Relativity and Gravitation 7:1141--1165,
  \doi{10.1023/A:1016578408204}

\bibitem[{{Perego} et~al.(2017){Perego}, {Radice}, and
  {Bernuzzi}}]{Perego_Radice_Bernuzzi2017}
{Perego} A, {Radice} D, {Bernuzzi} S (2017) {AT 2017gfo: An Anisotropic and
  Three-component Kilonova Counterpart of GW170817}. \apjl 850:L37,
  \doi{10.3847/2041-8213/aa9ab9}, \eprint{1711.03982}

\bibitem[{{Peters}(1964)}]{Peters1964}
{Peters} PC (1964) {Gravitational Radiation and the Motion of Two Point
  Masses}. Physical Review 136:1224--1232, \doi{10.1103/PhysRev.136.B1224}

\bibitem[{{Peters} and {Mathews}(1963)}]{Peters_Mathews1963}
{Peters} PC, {Mathews} J (1963) {Gravitational Radiation from Point Masses in a
  Keplerian Orbit}. Physical Review 131:435--440, \doi{10.1103/PhysRev.131.435}

\bibitem[{{Pfeiffer} and {York}(2003)}]{Pfeiffer_York2003}
{Pfeiffer} HP, {York} JW (2003) {Extrinsic curvature and the Einstein
  constraints}. \prd 67:044022, \doi{10.1103/PhysRevD.67.044022},
  \eprint{gr-qc/0207095}

\bibitem[{{Pfeiffer} and {York}(2005)}]{Pfeiffer_York2005}
{Pfeiffer} HP, {York} JW (2005) {Uniqueness and Nonuniqueness in the Einstein
  Constraints}. \prl 95:091101, \doi{10.1103/PhysRevLett.95.091101},
  \eprint{gr-qc/0504142}

\bibitem[{{Pfeiffer} et~al.(2003){Pfeiffer}, {Kidder}, {Scheel}, and
  {Teukolsky}}]{Pfeiffer_KST2003}
{Pfeiffer} HP, {Kidder} LE, {Scheel} MA, {Teukolsky} SA (2003) {A multidomain
  spectral method for solving elliptic equations}. Computer Physics
  Communications 152:253--273, \doi{10.1016/S0010-4655(02)00847-0},
  \eprint{gr-qc/0202096}

\bibitem[{{Pfeiffer} et~al.(2007){Pfeiffer}, {Brown}, {Kidder}, {Lindblom},
  {Lovelace}, and {Scheel}}]{Pfeiffer_BKLLS2007}
{Pfeiffer} HP, {Brown} DA, {Kidder} LE, {Lindblom} L, {Lovelace} G, {Scheel} MA
  (2007) {Reducing orbital eccentricity in binary black hole simulations}.
  Class Quantum Grav 24:S59--S81, \doi{10.1088/0264-9381/24/12/S06},
  \eprint{gr-qc/0702106}

\bibitem[{{Phinney}(1989)}]{Phinney1989}
{Phinney} ES (1989) {Manifestations of a Massive Black Hole in the Galactic
  Center}. In: {Morris} M (ed) The Center of the Galaxy, IAU Symposium, vol
  136, p 543

\bibitem[{{Phinney}(1991)}]{Phinney1991}
{Phinney} ES (1991) {The Rate of Neutron Star Binary Mergers in the Universe:
  Minimal Predictions for Gravity Wave Detectors}. \apjl 380:L17,
  \doi{10.1086/186163}

\bibitem[{{Piran} et~al.(2013){Piran}, {Nakar}, and
  {Rosswog}}]{Piran_Nakar_Rosswog2013}
{Piran} T, {Nakar} E, {Rosswog} S (2013) {The electromagnetic signals of
  compact binary mergers}. \mnras 430:2121--2136, \doi{10.1093/mnras/stt037},
  \eprint{1204.6242}

\bibitem[{{Poisson}(2004)}]{Poisson2004}
{Poisson} E (2004) {Absorption of mass and angular momentum by a black hole:
  Time-domain formalisms for gravitational perturbations, and the small-hole or
  slow-motion approximation}. \prd 70:084044, \doi{10.1103/PhysRevD.70.084044},
  \eprint{gr-qc/0407050}

\bibitem[{{Poisson} and {Sasaki}(1995)}]{Poisson_Sasaki1995}
{Poisson} E, {Sasaki} M (1995) {Gravitational radiation from a particle in
  circular orbit around a black hole. V. Black-hole absorption and tail
  corrections}. \prd 51:5753--5767, \doi{10.1103/PhysRevD.51.5753},
  \eprint{gr-qc/9412027}

\bibitem[{{Poisson} and {Will}(1995)}]{Poisson_Will1995}
{Poisson} E, {Will} CM (1995) {Gravitational waves from inspiraling compact
  binaries: Parameter estimation using second-post-Newtonian waveforms}. \prd
  52:848--855, \doi{10.1103/PhysRevD.52.848}, \eprint{gr-qc/9502040}

\bibitem[{{Poisson} and {Will}(2014)}]{Poisson_Will}
{Poisson} E, {Will} CM (2014) {Gravity}. Cambridge University Press, Cambridge,
  UK

\bibitem[{{Popham} et~al.(1999){Popham}, {Woosley}, and
  {Fryer}}]{Popham_Woosley_Fryer1999}
{Popham} R, {Woosley} SE, {Fryer} C (1999) {Hyperaccreting Black Holes and
  Gamma-Ray Bursts}. \apj 518:356--374, \doi{10.1086/307259},
  \eprint{astro-ph/9807028}

\bibitem[{{Postnov} and {Yungelson}(2014)}]{Postnov_Yungelson2014}
{Postnov} KA, {Yungelson} LR (2014) {The Evolution of Compact Binary Star
  Systems}. Living Rev Relativ 17:3, \doi{10.12942/lrr-2014-3},
  \eprint{1403.4754}

\bibitem[{{Potekhin} et~al.(2013){Potekhin}, {Fantina}, {Chamel}, {Pearson},
  and {Goriely}}]{Potekhin_FCPG2013}
{Potekhin} AY, {Fantina} AF, {Chamel} N, {Pearson} JM, {Goriely} S (2013)
  {Analytical representations of unified equations of state for neutron-star
  matter}. \aap 560:A48, \doi{10.1051/0004-6361/201321697}, \eprint{1310.0049}

\bibitem[{{Potekhin} et~al.(2015){Potekhin}, {Pons}, and
  {Page}}]{Potekhin_Pons_Page2015}
{Potekhin} AY, {Pons} JA, {Page} D (2015) {Neutron Stars{\textemdash}Cooling
  and Transport}. \ssr 191:239--291, \doi{10.1007/s11214-015-0180-9},
  \eprint{1507.06186}

\bibitem[{{Pretorius}(2005)}]{Pretorius2005}
{Pretorius} F (2005) {Evolution of Binary Black-Hole Spacetimes}. \prl
  95:121101, \doi{10.1103/PhysRevLett.95.121101}, \eprint{gr-qc/0507014}

\bibitem[{{Pretorius}(2006)}]{Pretorius2006}
{Pretorius} F (2006) {Simulation of binary black hole spacetimes with a
  harmonic evolution scheme}. Class Quantum Grav 23:S529--S552,
  \doi{10.1088/0264-9381/23/16/S13}, \eprint{gr-qc/0602115}

\bibitem[{{Raaijmakers} et~al.(2019){Raaijmakers}, {Riley}, {Watts}, {Greif},
  {Morsink}, {Hebeler}, {Schwenk}, {Hinderer}, {Nissanke}, {Guillot}, and
  et~al.}]{Raaijmakers_etal2019}
{Raaijmakers} G, {Riley} TE, {Watts} AL, {Greif} SK, {Morsink} SM, {Hebeler} K,
  {Schwenk} A, {Hinderer} T, {Nissanke} S, {Guillot} S, et~al (2019) {A Nicer
  View of PSR J0030+0451: Implications for the Dense Matter Equation of State}.
  \apjl 887:L22, \doi{10.3847/2041-8213/ab451a}, \eprint{1912.05703}

\bibitem[{{Raaijmakers} et~al.(2021){Raaijmakers}, {Nissanke}, {Foucart},
  {Kasliwal}, {Bulla}, {Fernandez}, {Henkel}, {Hinderer}, {Hotokezaka},
  {Luko{\v{s}}i{\={u}}t{\\.{e}}}, and et~al.}]{2021arXiv210211569R}
{Raaijmakers} G, {Nissanke} S, {Foucart} F, {Kasliwal} MM, {Bulla} M,
  {Fernandez} R, {Henkel} A, {Hinderer} T, {Hotokezaka} K,
  {Luko{\v{s}}i{\={u}}t{\\{e}}} K, et~al (2021) {The Challenges Ahead for
  Multimessenger Analyses of Gravitational Waves and Kilonova: a Case Study on
  GW190425}. arXiv e-prints arXiv:2102.11569, \eprint{2102.11569}

\bibitem[{{Racine}(2008)}]{Racine2008}
{Racine} {\'E} (2008) {Analysis of spin precession in binary black hole systems
  including quadrupole-monopole interaction}. \prd 78:044021,
  \doi{10.1103/PhysRevD.78.044021}, \eprint{0803.1820}

\bibitem[{{Radice} et~al.(2016){Radice}, {Galeazzi}, {Lippuner}, {Roberts},
  {Ott}, and {Rezzolla}}]{Radice_GLROR2016}
{Radice} D, {Galeazzi} F, {Lippuner} J, {Roberts} LF, {Ott} CD, {Rezzolla} L
  (2016) {Dynamical mass ejection from binary neutron star mergers}. \mnras
  460:3255--3271, \doi{10.1093/mnras/stw1227}, \eprint{1601.02426}

\bibitem[{{Radice} et~al.(2018){Radice}, {Perego}, {Hotokezaka}, {Fromm},
  {Bernuzzi}, and {Roberts}}]{Radice_PHFBR2018}
{Radice} D, {Perego} A, {Hotokezaka} K, {Fromm} SA, {Bernuzzi} S, {Roberts} LF
  (2018) {Binary Neutron Star Mergers: Mass Ejection, Electromagnetic
  Counterparts, and Nucleosynthesis}. \apj 869:130,
  \doi{10.3847/1538-4357/aaf054}, \eprint{1809.11161}

\bibitem[{{Raithel} et~al.(2019){Raithel}, {{\"O}zel}, and
  {Psaltis}}]{Raithel_Ozel_Psaltis2019}
{Raithel} CA, {{\"O}zel} F, {Psaltis} D (2019) {Finite-temperature Extension
  for Cold Neutron Star Equations of State}. \apj 875:12,
  \doi{10.3847/1538-4357/ab08ea}, \eprint{1902.10735}

\bibitem[{{Rantsiou} et~al.(2008){Rantsiou}, {Kobayashi}, {Laguna}, and
  {Rasio}}]{Rantsiou_KLR2008}
{Rantsiou} E, {Kobayashi} S, {Laguna} P, {Rasio} FA (2008) {Mergers of Black
  Hole-Neutron Star Binaries. I. Methods and First Results}. \apj
  680:1326--1349, \doi{10.1086/587858}, \eprint{astro-ph/0703599}

\bibitem[{{Rasio} and {Shapiro}(1992)}]{Rasio_Shapiro1992}
{Rasio} FA, {Shapiro} SL (1992) {Hydrodynamical Evolution of Coalescing Binary
  Neutron Stars}. \apj 401:226, \doi{10.1086/172055}

\bibitem[{{Rasio} and {Shapiro}(1994)}]{Rasio_Shapiro1994}
{Rasio} FA, {Shapiro} SL (1994) {Hydrodynamics of Binary Coalescence. I.
  Polytropes with Stiff Equations of State}. \apj 432:242,
  \doi{10.1086/174566}, \eprint{astro-ph/9401027}

\bibitem[{{Read} et~al.(2009{\natexlab{a}}){Read}, {Lackey}, {Owen}, and
  {Friedman}}]{Read_LOF2009}
{Read} JS, {Lackey} BD, {Owen} BJ, {Friedman} JL (2009{\natexlab{a}})
  {Constraints on a phenomenologically parametrized neutron-star equation of
  state}. \prd 79:124032, \doi{10.1103/PhysRevD.79.124032}, \eprint{0812.2163}

\bibitem[{{Read} et~al.(2009{\natexlab{b}}){Read}, {Markakis}, {Shibata},
  {Ury{\={u}}}, {Creighton}, and {Friedman}}]{Read_MSUCF2009}
{Read} JS, {Markakis} C, {Shibata} M, {Ury{\={u}}} K, {Creighton} JDE,
  {Friedman} JL (2009{\natexlab{b}}) {Measuring the neutron star equation of
  state with gravitational wave observations}. \prd 79:124033,
  \doi{10.1103/PhysRevD.79.124033}, \eprint{0901.3258}

\bibitem[{{Rees}(1988)}]{Rees1988}
{Rees} MJ (1988) {Tidal disruption of stars by black holes of
  {}10$^{6}$-{}10$^{8}$ solar masses in nearby galaxies}. \nat 333:523--528,
  \doi{10.1038/333523a0}

\bibitem[{{Rees} and {Gunn}(1974)}]{Rees_Gunn1974}
{Rees} MJ, {Gunn} JE (1974) {The origin of the magnetic field and relativistic
  particles in the Crab Nebula}. \mnras 167:1--12, \doi{10.1093/mnras/167.1.1}

\bibitem[{{Rees} and {Meszaros}(1992)}]{Rees_Meszaros1992}
{Rees} MJ, {Meszaros} P (1992) {Relativistic fireballs - Energy conversion and
  time-scales.} \mnras 258:41, \doi{10.1093/mnras/258.1.41P}

\bibitem[{{Richards} et~al.(2021){Richards}, {Baumgarte}, and
  {Shapiro}}]{Richards_Baumgarte_Shapiro2021}
{Richards} CB, {Baumgarte} TW, {Shapiro} SL (2021) {Accretion onto a small
  black hole at the center of a neutron star}. \prd 103:104009,
  \doi{10.1103/PhysRevD.103.104009}, \eprint{2102.09574}

\bibitem[{{Richers} et~al.(2019){Richers}, {McLaughlin}, {Kneller}, and
  {Vlasenko}}]{Richers_MKV2019}
{Richers} SA, {McLaughlin} GC, {Kneller} JP, {Vlasenko} A (2019) {Neutrino
  quantum kinetics in compact objects}. \prd 99:123014,
  \doi{10.1103/PhysRevD.99.123014}, \eprint{1903.00022}

\bibitem[{{Roberts} et~al.(2017){Roberts}, {Lippuner}, {Duez}, {Faber},
  {Foucart}, {Lombardi}, {Ning}, {Ott}, and {Ponce}}]{Roberts_LDFFLNOP2017}
{Roberts} LF, {Lippuner} J, {Duez} MD, {Faber} JA, {Foucart} F, {Lombardi} J
  James~C, {Ning} S, {Ott} CD, {Ponce} M (2017) {The influence of neutrinos on
  r-process nucleosynthesis in the ejecta of black hole-neutron star mergers}.
  \mnras 464:3907--3919, \doi{10.1093/mnras/stw2622}, \eprint{1601.07942}

\bibitem[{{Rossi} and {Begelman}(2009)}]{Rossi_Begelman2009}
{Rossi} EM, {Begelman} MC (2009) {Delayed X-ray emission from fallback in
  compact-object mergers}. \mnras 392:1451--1455,
  \doi{10.1111/j.1365-2966.2008.14139.x}, \eprint{0808.1284}

\bibitem[{{Rosswog}(2005)}]{Rosswog2005}
{Rosswog} S (2005) {Mergers of Neutron Star-Black Hole Binaries with Small Mass
  Ratios: Nucleosynthesis, Gamma-Ray Bursts, and Electromagnetic Transients}.
  \apj 634:1202--1213, \doi{10.1086/497062}, \eprint{astro-ph/0508138}

\bibitem[{{Rosswog}(2007)}]{Rosswog2007}
{Rosswog} S (2007) {Fallback accretion in the aftermath of a compact binary
  merger}. \mnras 376:L48--L51, \doi{10.1111/j.1745-3933.2007.00284.x},
  \eprint{astro-ph/0611440}

\bibitem[{{Rosswog} and {Diener}(2021)}]{Rosswog_Diener2021}
{Rosswog} S, {Diener} P (2021) {SPHINCS\_BSSN: a general relativistic smooth
  particle hydrodynamics code for dynamical spacetimes}. Class Quantum Grav
  38:115002, \doi{10.1088/1361-6382/abee65}, \eprint{2012.13954}

\bibitem[{{Rosswog} and {Liebend{\"o}rfer}(2003)}]{Rosswog_Liebendorfer2003}
{Rosswog} S, {Liebend{\"o}rfer} M (2003) {High-resolution calculations of
  merging neutron stars -- II. Neutrino emission}. \mnras 342:673--689,
  \doi{10.1046/j.1365-8711.2003.06579.x}, \eprint{astro-ph/0302301}

\bibitem[{{Rosswog} et~al.(2000){Rosswog}, {Davies}, {Thielemann}, and
  {Piran}}]{Rosswog_DTP2000}
{Rosswog} S, {Davies} MB, {Thielemann} FK, {Piran} T (2000) {Merging neutron
  stars: asymmetric systems}. \aap 360:171--184, \eprint{astro-ph/0005550}

\bibitem[{{Rosswog} et~al.(2004){Rosswog}, {Speith}, and
  {Wynn}}]{Rosswog_Speith_Wynn2004}
{Rosswog} S, {Speith} R, {Wynn} GA (2004) {Accretion dynamics in neutron
  star-black hole binaries}. \mnras 351:1121--1133,
  \doi{10.1111/j.1365-2966.2004.07865.x}, \eprint{astro-ph/0403500}

\bibitem[{{Rosswog} et~al.(2013){Rosswog}, {Piran}, and
  {Nakar}}]{Rosswog_Piran_Nakar2013}
{Rosswog} S, {Piran} T, {Nakar} E (2013) {The multimessenger picture of compact
  object encounters: binary mergers versus dynamical collisions}. \mnras
  430:2585--2604, \doi{10.1093/mnras/sts708}, \eprint{1204.6240}

\bibitem[{{Rosswog} et~al.(2014){Rosswog}, {Korobkin}, {Arcones}, {Thielemann},
  and {Piran}}]{Rosswog_KATP2014}
{Rosswog} S, {Korobkin} O, {Arcones} A, {Thielemann} FK, {Piran} T (2014) {The
  long-term evolution of neutron star merger remnants - I. The impact of
  r-process nucleosynthesis}. \mnras 439:744--756, \doi{10.1093/mnras/stt2502},
  \eprint{1307.2939}

\bibitem[{{Rowlinson} et~al.(2013){Rowlinson}, {O'Brien}, {Metzger}, {Tanvir},
  and {Levan}}]{Rowlinson_OMTL2013}
{Rowlinson} A, {O'Brien} PT, {Metzger} BD, {Tanvir} NR, {Levan} AJ (2013)
  {Signatures of magnetar central engines in short GRB light curves}. \mnras
  430:1061--1087, \doi{10.1093/mnras/sts683}, \eprint{1301.0629}

\bibitem[{{Ruchlin} et~al.(2017){Ruchlin}, {Healy}, {Lousto}, and
  {Zlochower}}]{Ruchlin_Healy_Lousto_Zlochower2017}
{Ruchlin} I, {Healy} J, {Lousto} CO, {Zlochower} Y (2017) {Puncture initial
  data for black-hole binaries with high spins and high boosts}. \prd
  95:024033, \doi{10.1103/PhysRevD.95.024033}

\bibitem[{{Ruffert} and {Janka}(2010)}]{Ruffert_Janka2010}
{Ruffert} M, {Janka} HT (2010) {Polytropic neutron star - black hole merger
  simulations with a Paczy{\'n}ski-Wiita potential}. \aap 514:A66,
  \doi{10.1051/0004-6361/200912738}, \eprint{0906.3998}

\bibitem[{{Ruffert} et~al.(1996){Ruffert}, {Janka}, and
  {Schaefer}}]{Ruffert_Janka_Schafer1996}
{Ruffert} M, {Janka} HT, {Schaefer} G (1996) {Coalescing neutron stars: a step
  towards physical models. I. Hydrodynamic evolution and gravitational-wave
  emission.} \aap 311:532--566, \eprint{astro-ph/9509006}

\bibitem[{{Ruffert} et~al.(1997){Ruffert}, {Janka}, {Takahashi}, and
  {Schaefer}}]{Ruffert_JTS1997}
{Ruffert} M, {Janka} HT, {Takahashi} K, {Schaefer} G (1997) {Coalescing neutron
  stars: a step towards physical models. II. Neutrino emission, neutron tori,
  and gamma-ray bursts.} \aap 319:122--153, \eprint{astro-ph/9606181}

\bibitem[{{Ruiz} et~al.(2011){Ruiz}, {Hilditch}, and
  {Bernuzzi}}]{Ruiz_Hilditch_Bernuzzi2011}
{Ruiz} M, {Hilditch} D, {Bernuzzi} S (2011) {Constraint preserving boundary
  conditions for the Z4c formulation of general relativity}. \prd 83:024025,
  \doi{10.1103/PhysRevD.83.024025}, \eprint{1010.0523}

\bibitem[{{Ruiz} et~al.(2018){Ruiz}, {Shapiro}, and
  {Tsokaros}}]{Ruiz_Shapiro_Tsokaros2018}
{Ruiz} M, {Shapiro} SL, {Tsokaros} A (2018) {Jet launching from binary black
  hole-neutron star mergers: Dependence on black hole spin, binary mass ratio,
  and magnetic field orientation}. \prd 98:123017,
  \doi{10.1103/PhysRevD.98.123017}, \eprint{1810.08618}

\bibitem[{{Ruiz} et~al.(2020){Ruiz}, {Paschalidis}, {Tsokaros}, and
  {Shapiro}}]{Ruiz_PTS2020}
{Ruiz} M, {Paschalidis} V, {Tsokaros} A, {Shapiro} SL (2020) {Black
  hole-neutron star coalescence: Effects of the neutron star spin on jet
  launching and dynamical ejecta mass}. \prd 102:124077,
  \doi{10.1103/PhysRevD.102.124077}, \eprint{2011.08863}

\bibitem[{{Saijo} and {Nakamura}(2000)}]{Saijo_Nakamura2000}
{Saijo} M, {Nakamura} T (2000) {Possible Direct Method to Determine the Radius
  of a Star from the Spectrum of Gravitational Wave Signals}. \prl
  85:2665--2668, \doi{10.1103/PhysRevLett.85.2665}, \eprint{astro-ph/0008309}

\bibitem[{{Saijo} and {Nakamura}(2001)}]{Saijo_Nakamura2001}
{Saijo} M, {Nakamura} T (2001) {Possible direct method to determine the radius
  of a star from the spectrum of gravitational wave signals. II. Spectra for
  various cases}. \prd 63:064004, \doi{10.1103/PhysRevD.63.064004},
  \eprint{astro-ph/0012061}

\bibitem[{{Santamar{\'\i}a} et~al.(2010){Santamar{\'\i}a}, {Ohme}, {Ajith},
  {Br{\"u}gmann}, {Dorband}, {Hannam}, {Husa}, {M{\"o}sta}, {Pollney},
  {Reisswig}, and et~al.}]{Santamaria_etal2010}
{Santamar{\'\i}a} L, {Ohme} F, {Ajith} P, {Br{\"u}gmann} B, {Dorband} N,
  {Hannam} M, {Husa} S, {M{\"o}sta} P, {Pollney} D, {Reisswig} C, et~al (2010)
  {Matching post-Newtonian and numerical relativity waveforms: Systematic
  errors and a new phenomenological model for nonprecessing black hole
  binaries}. \prd 82:064016, \doi{10.1103/PhysRevD.82.064016},
  \eprint{1005.3306}

\bibitem[{{Santoliquido} et~al.(2021){Santoliquido}, {Mapelli}, {Giacobbo},
  {Bouffanais}, and {Artale}}]{Santoilquido_MGBA2021}
{Santoliquido} F, {Mapelli} M, {Giacobbo} N, {Bouffanais} Y, {Artale} MC (2021)
  {The cosmic merger rate density of compact objects: impact of star formation,
  metallicity, initial mass function, and binary evolution}. \mnras
  502:4877--4889, \doi{10.1093/mnras/stab280}, \eprint{2009.03911}

\bibitem[{{Sathyaprakash} and
  {Dhurandhar}(1991)}]{Sathyaprakash_Dhurandhar1991}
{Sathyaprakash} BS, {Dhurandhar} SV (1991) {Choice of filters for the detection
  of gravitational waves from coalescing binaries}. \prd 44:3819--3834,
  \doi{10.1103/PhysRevD.44.3819}

\bibitem[{{Savchenko} et~al.(2017){Savchenko}, {Ferrigno}, {Kuulkers},
  {Bazzano}, {Bozzo}, {Brandt}, {Chenevez}, {Courvoisier}, {Diehl}, {Domingo},
  and et~al.}]{Savchenko_etal2017}
{Savchenko} V, {Ferrigno} C, {Kuulkers} E, {Bazzano} A, {Bozzo} E, {Brandt} S,
  {Chenevez} J, {Courvoisier} TJL, {Diehl} R, {Domingo} A, et~al (2017)
  {INTEGRAL Detection of the First Prompt Gamma-Ray Signal Coincident with the
  Gravitational-wave Event GW170817}. \apjl 848:L15,
  \doi{10.3847/2041-8213/aa8f94}, \eprint{1710.05449}

\bibitem[{{Scheel} et~al.(2009){Scheel}, {Boyle}, {Chu}, {Kidder}, {Matthews},
  and {Pfeiffer}}]{Scheel_BCKMP2009}
{Scheel} MA, {Boyle} M, {Chu} T, {Kidder} LE, {Matthews} KD, {Pfeiffer} HP
  (2009) {High-accuracy waveforms for binary black hole inspiral, merger, and
  ringdown}. \prd 79:024003, \doi{10.1103/PhysRevD.79.024003},
  \eprint{0810.1767}

\bibitem[{{Scheel} et~al.(2015){Scheel}, {Giesler}, {Hemberger}, {Lovelace},
  {Kuper}, {Boyle}, {Szil{\'a}gyi}, and {Kidder}}]{Scheel_GHLKBSK2015}
{Scheel} MA, {Giesler} M, {Hemberger} DA, {Lovelace} G, {Kuper} K, {Boyle} M,
  {Szil{\'a}gyi} B, {Kidder} LE (2015) {Improved methods for simulating nearly
  extremal binary black holes}. Class Quantum Grav 32:105009,
  \doi{10.1088/0264-9381/32/10/105009}, \eprint{1412.1803}

\bibitem[{{Schnetter}(2010)}]{Schnetter2010}
{Schnetter} E (2010) {Time step size limitation introduced by the BSSN Gamma
  Driver}. Class Quantum Grav 27:167001, \doi{10.1088/0264-9381/27/16/167001},
  \eprint{1003.0859}

\bibitem[{{Schnetter} et~al.(2004){Schnetter}, {Hawley}, and
  {Hawke}}]{Schnetter_Hawley_Hawke2004}
{Schnetter} E, {Hawley} SH, {Hawke} I (2004) {Evolutions in 3D numerical
  relativity using fixed mesh refinement}. Class Quantum Grav 21:1465--1488,
  \doi{10.1088/0264-9381/21/6/014}, \eprint{gr-qc/0310042}

\bibitem[{{Schutz}(2011)}]{Schutz2011}
{Schutz} BF (2011) {Networks of gravitational wave detectors and three figures
  of merit}. Class Quantum Grav 28:125023,
  \doi{10.1088/0264-9381/28/12/125023}, \eprint{1102.5421}

\bibitem[{{Sekiguchi}(2010)}]{Sekiguchi2010}
{Sekiguchi} Y (2010) {Stellar Core Collapse in Full General Relativity with
  Microphysics -- Formulation and Spherical Collapse Test}. Progress of
  Theoretical Physics 124:331--379, \doi{10.1143/PTP.124.331},
  \eprint{1009.3320}

\bibitem[{{Sekiguchi} and {Shibata}(2011)}]{Sekiguchi_Shibata2011}
{Sekiguchi} Y, {Shibata} M (2011) {Formation of Black Hole and Accretion Disk
  in a Massive High-entropy Stellar Core Collapse}. \apj 737:6,
  \doi{10.1088/0004-637X/737/1/6}, \eprint{1009.5303}

\bibitem[{{Sekiguchi} et~al.(2011{\natexlab{a}}){Sekiguchi}, {Kiuchi},
  {Kyutoku}, and {Shibata}}]{Sekiguchi_KKS2011-2}
{Sekiguchi} Y, {Kiuchi} K, {Kyutoku} K, {Shibata} M (2011{\natexlab{a}})
  {Effects of Hyperons in Binary Neutron Star Mergers}. \prl 107:211101,
  \doi{10.1103/PhysRevLett.107.211101}, \eprint{1110.4442}

\bibitem[{{Sekiguchi} et~al.(2011{\natexlab{b}}){Sekiguchi}, {Kiuchi},
  {Kyutoku}, and {Shibata}}]{Sekiguchi_KKS2011}
{Sekiguchi} Y, {Kiuchi} K, {Kyutoku} K, {Shibata} M (2011{\natexlab{b}})
  {Gravitational Waves and Neutrino Emission from the Merger of Binary Neutron
  Stars}. \prl 107:051102, \doi{10.1103/PhysRevLett.107.051102},
  \eprint{1105.2125}

\bibitem[{{Sekiguchi} et~al.(2015){Sekiguchi}, {Kiuchi}, {Kyutoku}, and
  {Shibata}}]{Sekiguchi_KKS2015}
{Sekiguchi} Y, {Kiuchi} K, {Kyutoku} K, {Shibata} M (2015) {Dynamical mass
  ejection from binary neutron star mergers: Radiation-hydrodynamics study in
  general relativity}. \prd 91:064059, \doi{10.1103/PhysRevD.91.064059},
  \eprint{1502.06660}

\bibitem[{{Sekiguchi} et~al.(2016){Sekiguchi}, {Kiuchi}, {Kyutoku}, {Shibata},
  and {Taniguchi}}]{Sekiguchi_KKST2016}
{Sekiguchi} Y, {Kiuchi} K, {Kyutoku} K, {Shibata} M, {Taniguchi} K (2016)
  {Dynamical mass ejection from the merger of asymmetric binary neutron stars:
  Radiation-hydrodynamics study in general relativity}. \prd 93:124046,
  \doi{10.1103/PhysRevD.93.124046}, \eprint{1603.01918}

\bibitem[{{Setiawan} et~al.(2004){Setiawan}, {Ruffert}, and
  {Janka}}]{Setiawan_Ruffert_Janka2004}
{Setiawan} S, {Ruffert} M, {Janka} HT (2004) {Non-stationary hyperaccretion of
  stellar-mass black holes in three dimensions: torus evolution and neutrino
  emission}. \mnras 352:753--758, \doi{10.1111/j.1365-2966.2004.07974.x},
  \eprint{astro-ph/0402481}

\bibitem[{{Setiawan} et~al.(2006){Setiawan}, {Ruffert}, and
  {Janka}}]{Setiawan_Ruffert_Janka2006}
{Setiawan} S, {Ruffert} M, {Janka} HT (2006) {Three-dimensional simulations of
  non-stationary accretion by remnant black holes of compact object mergers}.
  \aap 458:553--567, \doi{10.1051/0004-6361:20054193},
  \eprint{astro-ph/0509300}

\bibitem[{{Shakura} and {Sunyaev}(1973)}]{Shakura_Sunyaev1973}
{Shakura} NI, {Sunyaev} RA (1973) {Reprint of 1973A\&A....24..337S. Black holes
  in binary systems. Observational appearance}. \aap 500:33--51

\bibitem[{{Shapiro} and {Wasserman}(1982)}]{Shapiro_Wasserman1982}
{Shapiro} SL, {Wasserman} I (1982) {Gravitational radiation from nonspherical
  infall into black holes}. \apj 260:838--848, \doi{10.1086/160302}

\bibitem[{{Shen} et~al.(2011){Shen}, {Horowitz}, and
  {O'Connor}}]{Shen_Horowitz_OConner2011}
{Shen} G, {Horowitz} CJ, {O'Connor} E (2011) {Second relativistic mean field
  and virial equation of state for astrophysical simulations}. \prc 83:065808,
  \doi{10.1103/PhysRevC.83.065808}, \eprint{1103.5174}

\bibitem[{{Shen} et~al.(1998){Shen}, {Toki}, {Oyamatsu}, and
  {Sumiyoshi}}]{Shen_TOS1998}
{Shen} H, {Toki} H, {Oyamatsu} K, {Sumiyoshi} K (1998) {Relativistic equation
  of state of nuclear matter for supernova and neutron star}. \nphysa
  637:435--450, \doi{10.1016/S0375-9474(98)00236-X}, \eprint{nucl-th/9805035}

\bibitem[{{Shibata}(1996)}]{Shibata1996}
{Shibata} M (1996) {Relativistic Roche-Riemann Problems around a Black Hole}.
  Progress of Theoretical Physics 96:917--932, \doi{10.1143/PTP.96.917}

\bibitem[{{Shibata}(1998)}]{Shibata1998}
{Shibata} M (1998) {Relativistic formalism for computation of irrotational
  binary stars in quasiequilibrium states}. \prd 58:024012,
  \doi{10.1103/PhysRevD.58.024012}, \eprint{gr-qc/9803085}

\bibitem[{{Shibata}(1999)}]{Shibata1999}
{Shibata} M (1999) {Fully general relativistic simulation of coalescing binary
  neutron stars: Preparatory tests}. \prd 60:104052,
  \doi{10.1103/PhysRevD.60.104052}, \eprint{gr-qc/9908027}

\bibitem[{{Shibata}(2005)}]{Shibata2005}
{Shibata} M (2005) {Constraining Nuclear Equations of State Using Gravitational
  Waves from Hypermassive Neutron Stars}. \prl 94:201101,
  \doi{10.1103/PhysRevLett.94.201101}, \eprint{gr-qc/0504082}

\bibitem[{{Shibata}(2007)}]{Shibata2007}
{Shibata} M (2007) {Rotating black hole surrounded by self-gravitating torus in
  the puncture framework}. \prd 76:064035, \doi{10.1103/PhysRevD.76.064035}

\bibitem[{{Shibata}(2016)}]{Shibata}
{Shibata} M (2016) {100 Years of General Relativity, Volume 1: Numerical
  Relativity}. World Scientific, Singapore, \doi{10.1142/9692}

\bibitem[{{Shibata} and {Hotokezaka}(2019)}]{Shibata_Hotokezaka2019}
{Shibata} M, {Hotokezaka} K (2019) {Merger and Mass Ejection of Neutron Star
  Binaries}. Annu Rev Nucl Part Sci 69:41--64,
  \doi{10.1146/annurev-nucl-101918-023625}, \eprint{1908.02350}

\bibitem[{{Shibata} and {Nakamura}(1995)}]{Shibata_Nakamura1995}
{Shibata} M, {Nakamura} T (1995) {Evolution of three-dimensional gravitational
  waves: Harmonic slicing case}. \prd 52:5428--5444,
  \doi{10.1103/PhysRevD.52.5428}

\bibitem[{{Shibata} and {Sekiguchi}(2012)}]{Shibata_Sekiguchi2012}
{Shibata} M, {Sekiguchi} Y (2012) {Radiation Magnetohydrodynamics for Black
  Hole-Torus System in Full General Relativity: A Step toward Physical
  Simulation}. Progress of Theoretical Physics 127:535--559,
  \doi{10.1143/PTP.127.535}, \eprint{1206.5911}

\bibitem[{{Shibata} and {Sekiguchi}(2005)}]{Shibata_Sekiguchi2005}
{Shibata} M, {Sekiguchi} YI (2005) {Magnetohydrodynamics in full general
  relativity: Formulation and tests}. \prd 72:044014,
  \doi{10.1103/PhysRevD.72.044014}, \eprint{astro-ph/0507383}

\bibitem[{{Shibata} and {Taniguchi}(2006)}]{Shibata_Taniguchi2006}
{Shibata} M, {Taniguchi} K (2006) {Merger of binary neutron stars to a black
  hole: Disk mass, short gamma-ray bursts, and quasinormal mode ringing}. \prd
  73:064027, \doi{10.1103/PhysRevD.73.064027}, \eprint{astro-ph/0603145}

\bibitem[{{Shibata} and {Taniguchi}(2008)}]{Shibata_Taniguchi2008}
{Shibata} M, {Taniguchi} K (2008) {Merger of black hole and neutron star in
  general relativity: Tidal disruption, torus mass, and gravitational waves}.
  \prd 77:084015, \doi{10.1103/PhysRevD.77.084015}, \eprint{0711.1410}

\bibitem[{{Shibata} and {Taniguchi}(2011)}]{1st}
{Shibata} M, {Taniguchi} K (2011) {Coalescence of Black Hole-Neutron Star
  Binaries}. Living Rev Relativ 14:6, \doi{10.12942/lrr-2011-6}

\bibitem[{{Shibata} and {Ury{\={u}}}(2002)}]{Shibata_Uryu2002}
{Shibata} M, {Ury{\={u}}} K (2002) {Gravitational Waves from Merger of Binary
  Neutron Stars in Fully General Relativistic Simulation}. Progress of
  Theoretical Physics 107:265--303, \doi{10.1143/PTP.107.265},
  \eprint{gr-qc/0203037}

\bibitem[{{Shibata} and {Ury{\={u}}}(2006)}]{Shibata_Uryu2006}
{Shibata} M, {Ury{\={u}}} K (2006) {Merger of black hole-neutron star binaries:
  Nonspinning black hole case}. \prd 74:121503,
  \doi{10.1103/PhysRevD.74.121503}, \eprint{gr-qc/0612142}

\bibitem[{{Shibata} and {Ury{\={u}}}(2007)}]{Shibata_Uryu2007}
{Shibata} M, {Ury{\={u}}} K (2007) {Merger of black hole neutron star binaries
  in full general relativity}. Class Quantum Grav 24:S125--S137,
  \doi{10.1088/0264-9381/24/12/S09}, \eprint{astro-ph/0611522}

\bibitem[{{Shibata} and {Ury{\={u}}}(2000)}]{Shibata_Uryu2000}
{Shibata} M, {Ury{\={u}}} K{\={o}} (2000) {Simulation of merging binary neutron
  stars in full general relativity: {\ensuremath{\Gamma}}=2 case}. \prd
  61:064001, \doi{10.1103/PhysRevD.61.064001}, \eprint{gr-qc/9911058}

\bibitem[{{Shibata} et~al.(2003){Shibata}, {Taniguchi}, and
  {Ury{\={u}}}}]{Shibata_Taniguchi_Uryu2003}
{Shibata} M, {Taniguchi} K, {Ury{\={u}}} K (2003) {Merger of binary neutron
  stars of unequal mass in full general relativity}. \prd 68:084020,
  \doi{10.1103/PhysRevD.68.084020}, \eprint{gr-qc/0310030}

\bibitem[{{Shibata} et~al.(2004){Shibata}, {Ury{\={u}}}, and
  {Friedman}}]{Shibata_Uryu_Friedman2004}
{Shibata} M, {Ury{\={u}}} K, {Friedman} JL (2004) {Deriving formulations for
  numerical computation of binary neutron stars in quasicircular orbits}. \prd
  70:044044, \doi{10.1103/PhysRevD.70.044044}, \eprint{gr-qc/0407036}

\bibitem[{{Shibata} et~al.(2005){Shibata}, {Taniguchi}, and
  {Ury{\={u}}}}]{Shibata_Taniguchi_Uryu2005}
{Shibata} M, {Taniguchi} K, {Ury{\={u}}} K (2005) {Merger of binary neutron
  stars with realistic equations of state in full general relativity}. \prd
  71:084021, \doi{10.1103/PhysRevD.71.084021}, \eprint{gr-qc/0503119}

\bibitem[{{Shibata} et~al.(2006){Shibata}, {Duez}, {Liu}, {Shapiro}, and
  {Stephens}}]{Shibata_DLSS2006}
{Shibata} M, {Duez} MD, {Liu} YT, {Shapiro} SL, {Stephens} BC (2006)
  {Magnetized Hypermassive Neutron-Star Collapse: A Central Engine for Short
  Gamma-Ray Bursts}. \prl 96:031102, \doi{10.1103/PhysRevLett.96.031102},
  \eprint{astro-ph/0511142}

\bibitem[{{Shibata} et~al.(2007){Shibata}, {Sekiguchi}, and
  {Takahashi}}]{Shibata_Sekiguchi_Takahashi2007}
{Shibata} M, {Sekiguchi} YI, {Takahashi} R (2007) {Magnetohydrodynamics of
  Neutrino-Cooled Accretion Tori around a Rotating Black Hole in General
  Relativity}. Progress of Theoretical Physics 118:257--302,
  \doi{10.1143/PTP.118.257}, \eprint{0709.1766}

\bibitem[{{Shibata} et~al.(2009){Shibata}, {Kyutoku}, {Yamamoto}, and
  {Taniguchi}}]{Shibata_KYT2009}
{Shibata} M, {Kyutoku} K, {Yamamoto} T, {Taniguchi} K (2009) {Gravitational
  waves from black hole-neutron star binaries: Classification of waveforms}.
  \prd 79:044030, \doi{10.1103/PhysRevD.79.044030}, \eprint{0902.0416}

\bibitem[{{Shibata} et~al.(2011){Shibata}, {Kiuchi}, {Sekiguchi}, and
  {Suwa}}]{Shibata_KKS2011}
{Shibata} M, {Kiuchi} K, {Sekiguchi} Y, {Suwa} Y (2011) {Truncated Moment
  Formalism for Radiation Hydrodynamics in Numerical Relativity}. Progress of
  Theoretical Physics 125:1255--1287, \doi{10.1143/PTP.125.1255},
  \eprint{1104.3937}

\bibitem[{{Shibata} et~al.(2012){Shibata}, {Kyutoku}, {Yamamoto}, and
  {Taniguchi}}]{Shibata_KYT2009e}
{Shibata} M, {Kyutoku} K, {Yamamoto} T, {Taniguchi} K (2012) {Erratum and
  Addendum: Gravitational waves from black hole-neutron star binaries:
  Classification of waveforms}. \prd 85:127502,
  \doi{10.1103/PhysRevD.85.127502}

\bibitem[{{Shibata} et~al.(2014){Shibata}, {Nagakura}, {Sekiguchi}, and
  {Yamada}}]{Shibata_NSY2014}
{Shibata} M, {Nagakura} H, {Sekiguchi} Y, {Yamada} S (2014) {Conservative form
  of Boltzmann's equation in general relativity}. \prd 89:084073,
  \doi{10.1103/PhysRevD.89.084073}

\bibitem[{{Shibata} et~al.(2017{\natexlab{a}}){Shibata}, {Fujibayashi},
  {Hotokezaka}, {Kiuchi}, {Kyutoku}, {Sekiguchi}, and
  {Tanaka}}]{Shibata_FHKKST2017}
{Shibata} M, {Fujibayashi} S, {Hotokezaka} K, {Kiuchi} K, {Kyutoku} K,
  {Sekiguchi} Y, {Tanaka} M (2017{\natexlab{a}}) {Modeling GW170817 based on
  numerical relativity and its implications}. \prd 96:123012,
  \doi{10.1103/PhysRevD.96.123012}, \eprint{1710.07579}

\bibitem[{{Shibata} et~al.(2017{\natexlab{b}}){Shibata}, {Kiuchi}, and
  {Sekiguchi}}]{Shibata_Kiuchi_Sekiguchi2017}
{Shibata} M, {Kiuchi} K, {Sekiguchi} Yi (2017{\natexlab{b}}) {General
  relativistic viscous hydrodynamics of differentially rotating neutron stars}.
  \prd 95:083005, \doi{10.1103/PhysRevD.95.083005}, \eprint{1703.10303}

\bibitem[{{Shibata} et~al.(2021){Shibata}, {Fujibayashi}, and
  {Sekiguchi}}]{Shibata_Fujibayashi_Sekiguchi2021}
{Shibata} M, {Fujibayashi} S, {Sekiguchi} Y (2021) {Long-term evolution of a
  merger-remnant neutron star in general relativistic magnetohydrodynamics:
  Effect of magnetic winding}. \prd 103:043022,
  \doi{10.1103/PhysRevD.103.043022}, \eprint{2102.01346}

\bibitem[{{Siegel} and {Metzger}(2017)}]{Siegel_Metzger2017}
{Siegel} DM, {Metzger} BD (2017) {Three-Dimensional General-Relativistic
  Magnetohydrodynamic Simulations of Remnant Accretion Disks from Neutron Star
  Mergers: Outflows and r -Process Nucleosynthesis}. \prl 119:231102,
  \doi{10.1103/PhysRevLett.119.231102}, \eprint{1705.05473}

\bibitem[{{Siegel} and {Metzger}(2018)}]{Siegel_Metzger2018}
{Siegel} DM, {Metzger} BD (2018) {Three-dimensional GRMHD Simulations of
  Neutrino-cooled Accretion Disks from Neutron Star Mergers}. \apj 858:52,
  \doi{10.3847/1538-4357/aabaec}, \eprint{1711.00868}

\bibitem[{{Soares-Santos} et~al.(2017){Soares-Santos}, {Holz}, {Annis},
  {Chornock}, {Herner}, {Berger}, {Brout}, {Chen}, {Kessler}, {Sako}, and
  et~al.}]{SoaresSantos_etal2017}
{Soares-Santos} M, {Holz} DE, {Annis} J, {Chornock} R, {Herner} K, {Berger} E,
  {Brout} D, {Chen} HY, {Kessler} R, {Sako} M, et~al (2017) {The
  Electromagnetic Counterpart of the Binary Neutron Star Merger LIGO/Virgo
  GW170817. I. Discovery of the Optical Counterpart Using the Dark Energy
  Camera}. \apjl 848:L16, \doi{10.3847/2041-8213/aa9059}, \eprint{1710.05459}

\bibitem[{{Steiner} et~al.(2013){Steiner}, {Hempel}, and
  {Fischer}}]{Steiner_Hempel_Fisher2013}
{Steiner} AW, {Hempel} M, {Fischer} T (2013) {Core-collapse Supernova Equations
  of State Based on Neutron Star Observations}. \apj 774:17,
  \doi{10.1088/0004-637X/774/1/17}, \eprint{1207.2184}

\bibitem[{{Stephens} et~al.(2011){Stephens}, {East}, and
  {Pretorius}}]{Stephens_East_Pretorius2011}
{Stephens} BC, {East} WE, {Pretorius} F (2011) {Eccentric
  Black-hole-Neutron-star Mergers}. \apjl 737:L5,
  \doi{10.1088/2041-8205/737/1/L5}, \eprint{1105.3175}

\bibitem[{{Stone} et~al.(2013){Stone}, {Loeb}, and
  {Berger}}]{Stone_Loeb_Berger2013}
{Stone} N, {Loeb} A, {Berger} E (2013) {Pulsations in short gamma ray bursts
  from black hole-neutron star mergers}. \prd 87:084053,
  \doi{10.1103/PhysRevD.87.084053}, \eprint{1209.4097}

\bibitem[{{Stovall} et~al.(2018){Stovall}, {Freire}, {Chatterjee}, {Demorest},
  {Lorimer}, {McLaughlin}, {Pol}, {van Leeuwen}, {Wharton}, {Allen}, and
  et~al.}]{Stovall_etal2018}
{Stovall} K, {Freire} PCC, {Chatterjee} S, {Demorest} PB, {Lorimer} DR,
  {McLaughlin} MA, {Pol} N, {van Leeuwen} J, {Wharton} RS, {Allen} B, et~al
  (2018) {PALFA Discovery of a Highly Relativistic Double Neutron Star Binary}.
  \apjl 854:L22, \doi{10.3847/2041-8213/aaad06}, \eprint{1802.01707}

\bibitem[{{Szil{\'a}gyi}(2014)}]{Szilagyi2014}
{Szil{\'a}gyi} B (2014) {Key elements of robustness in binary black hole
  evolutions using spectral methods}. Int J Mod Phys D 23:1430014,
  \doi{10.1142/S0218271814300146}, \eprint{1405.3693}

\bibitem[{{Szil{\'a}gyi} et~al.(2009){Szil{\'a}gyi}, {Lindblom}, and
  {Scheel}}]{Szilagyi_Lindblom_Scheel2009}
{Szil{\'a}gyi} B, {Lindblom} L, {Scheel} MA (2009) {Simulations of binary black
  hole mergers using spectral methods}. \prd 80:124010,
  \doi{10.1103/PhysRevD.80.124010}, \eprint{0909.3557}

\bibitem[{{Tacik} et~al.(2016){Tacik}, {Foucart}, {Pfeiffer}, {Muhlberger},
  {Kidder}, {Scheel}, and {Szil{\'a}gyi}}]{Tacik_FPMKSS2016}
{Tacik} N, {Foucart} F, {Pfeiffer} HP, {Muhlberger} C, {Kidder} LE, {Scheel}
  MA, {Szil{\'a}gyi} B (2016) {Initial data for black hole-neutron star
  binaries, with rotating stars}. Class Quantum Grav 33:225012,
  \doi{10.1088/0264-9381/33/22/225012}, \eprint{1607.07962}

\bibitem[{{Takami} et~al.(2014){Takami}, {Kyutoku}, and
  {Ioka}}]{Takami_Kyutoku_Ioka2014}
{Takami} H, {Kyutoku} K, {Ioka} K (2014) {High-energy radiation from remnants
  of neutron star binary mergers}. \prd 89:063006,
  \doi{10.1103/PhysRevD.89.063006}, \eprint{1307.6805}

\bibitem[{{Takami} et~al.(2015){Takami}, {Rezzolla}, and
  {Baiotti}}]{Takami_Rezzolla_Baiotti2015}
{Takami} K, {Rezzolla} L, {Baiotti} L (2015) {Spectral properties of the
  post-merger gravitational-wave signal from binary neutron stars}. \prd
  91:064001, \doi{10.1103/PhysRevD.91.064001}, \eprint{1412.3240}

\bibitem[{{Tanaka} and {Hotokezaka}(2013)}]{Tanaka_Hotokezaka2013}
{Tanaka} M, {Hotokezaka} K (2013) {Radiative Transfer Simulations of Neutron
  Star Merger Ejecta}. \apj 775:113, \doi{10.1088/0004-637X/775/2/113},
  \eprint{1306.3742}

\bibitem[{{Tanaka} et~al.(2014){Tanaka}, {Hotokezaka}, {Kyutoku}, {Wanajo},
  {Kiuchi}, {Sekiguchi}, and {Shibata}}]{Tanaka_HKWKSS2014}
{Tanaka} M, {Hotokezaka} K, {Kyutoku} K, {Wanajo} S, {Kiuchi} K, {Sekiguchi} Y,
  {Shibata} M (2014) {Radioactively Powered Emission from Black Hole-Neutron
  Star Mergers}. \apj 780:31, \doi{10.1088/0004-637X/780/1/31},
  \eprint{1310.2774}

\bibitem[{{Tanaka} et~al.(2017){Tanaka}, {Utsumi}, {Mazzali}, {Tominaga},
  {Yoshida}, {Sekiguchi}, {Morokuma}, {Motohara}, {Ohta}, {Kawabata}, and
  et~al.}]{Tanaka_etal2017}
{Tanaka} M, {Utsumi} Y, {Mazzali} PA, {Tominaga} N, {Yoshida} M, {Sekiguchi} Y,
  {Morokuma} T, {Motohara} K, {Ohta} K, {Kawabata} KS, et~al (2017) {Kilonova
  from post-merger ejecta as an optical and near-Infrared counterpart of
  GW170817}. \pasj 69:102, \doi{10.1093/pasj/psx121}, \eprint{1710.05850}

\bibitem[{{Tanaka} et~al.(2018){Tanaka}, {Kato}, {Gaigalas}, {Rynkun},
  {Rad{\v{z}}i{\={u}}t$\dot{\textrm{e}}$}, {Wanajo}, {Sekiguchi}, {Nakamura},
  {Tanuma}, {Murakami}, and et~al.}]{Tanaka_etal2018}
{Tanaka} M, {Kato} D, {Gaigalas} G, {Rynkun} P,
  {Rad{\v{z}}i{\={u}}t$\dot{\textrm{e}}$} L, {Wanajo} S, {Sekiguchi} Y,
  {Nakamura} N, {Tanuma} H, {Murakami} I, et~al (2018) {Properties of Kilonovae
  from Dynamical and Post-merger Ejecta of Neutron Star Mergers}. \apj 852:109,
  \doi{10.3847/1538-4357/aaa0cb}, \eprint{1708.09101}

\bibitem[{{Tanaka} et~al.(2020){Tanaka}, {Kato}, {Gaigalas}, and
  {Kawaguchi}}]{Tanaka_KGK2020}
{Tanaka} M, {Kato} D, {Gaigalas} G, {Kawaguchi} K (2020) {Systematic opacity
  calculations for kilonovae}. \mnras 496:1369--1392,
  \doi{10.1093/mnras/staa1576}, \eprint{1906.08914}

\bibitem[{{Taniguchi} and {Gourgoulhon}(2002)}]{Taniguchi_Gourgoulhon2002}
{Taniguchi} K, {Gourgoulhon} E (2002) {Quasiequilibrium sequences of
  synchronized and irrotational binary neutron stars in general relativity.
  III. Identical and different mass stars with {\ensuremath{\gamma}}=2}. \prd
  66:104019, \doi{10.1103/PhysRevD.66.104019}, \eprint{gr-qc/0207098}

\bibitem[{{Taniguchi} and {Gourgoulhon}(2003)}]{Taniguchi_Gourgoulhon2003}
{Taniguchi} K, {Gourgoulhon} E (2003) {Various features of quasiequilibrium
  sequences of binary neutron stars in general relativity}. \prd 68:124025,
  \doi{10.1103/PhysRevD.68.124025}, \eprint{gr-qc/0309045}

\bibitem[{{Taniguchi} and {Nakamura}(1996)}]{Taniguchi_Nakamura1996}
{Taniguchi} K, {Nakamura} T (1996) {Innermost Stable Circular Orbit of
  Coalescing Neutron Star-Black Hole Binary --- Generalized Pseudo-Newtonian
  Potential Approach ---}. Progress of Theoretical Physics 96:693--712,
  \doi{10.1143/PTP.96.693}, \eprint{astro-ph/9609009}

\bibitem[{{Taniguchi} and {Shibata}(2010)}]{Taniguchi_Shibata2010}
{Taniguchi} K, {Shibata} M (2010) {Binary Neutron Stars in Quasi-equilibrium}.
  \apjs 188:187--208, \doi{10.1088/0067-0049/188/1/187}, \eprint{1005.0958}

\bibitem[{{Taniguchi} et~al.(2005){Taniguchi}, {Baumgarte}, {Faber}, and
  {Shapiro}}]{Taniguchi_BFS2005}
{Taniguchi} K, {Baumgarte} TW, {Faber} JA, {Shapiro} SL (2005) {Black
  hole-neutron star binaries in general relativity: Effects of neutron star
  spin}. \prd 72:044008, \doi{10.1103/PhysRevD.72.044008},
  \eprint{astro-ph/0505450}

\bibitem[{{Taniguchi} et~al.(2006){Taniguchi}, {Baumgarte}, {Faber}, and
  {Shapiro}}]{Taniguchi_BFS2006}
{Taniguchi} K, {Baumgarte} TW, {Faber} JA, {Shapiro} SL (2006)
  {Quasiequilibrium sequences of black-hole neutron-star binaries in general
  relativity}. \prd 74:041502, \doi{10.1103/PhysRevD.74.041502},
  \eprint{gr-qc/0609053}

\bibitem[{{Taniguchi} et~al.(2007){Taniguchi}, {Baumgarte}, {Faber}, and
  {Shapiro}}]{Taniguchi_BFS2007}
{Taniguchi} K, {Baumgarte} TW, {Faber} JA, {Shapiro} SL (2007)
  {Quasiequilibrium black hole-neutron star binaries in general relativity}.
  \prd 75:084005, \doi{10.1103/PhysRevD.75.084005}, \eprint{gr-qc/0701110}

\bibitem[{{Taniguchi} et~al.(2008){Taniguchi}, {Baumgarte}, {Faber}, and
  {Shapiro}}]{Taniguchi_BFS2008}
{Taniguchi} K, {Baumgarte} TW, {Faber} JA, {Shapiro} SL (2008) {Relativistic
  black hole-neutron star binaries in quasiequilibrium: Effects of the black
  hole excision boundary condition}. \prd 77:044003,
  \doi{10.1103/PhysRevD.77.044003}, \eprint{0710.5169}

\bibitem[{{Tanvir} et~al.(2017){Tanvir}, {Levan}, {Gonz{\'a}lez-Fern{\'a}ndez},
  {Korobkin}, {Mandel}, {Rosswog}, {Hjorth}, {D'Avanzo}, {Fruchter}, {Fryer},
  and et~al.}]{Tanvir_etal2017}
{Tanvir} NR, {Levan} AJ, {Gonz{\'a}lez-Fern{\'a}ndez} C, {Korobkin} O, {Mandel}
  I, {Rosswog} S, {Hjorth} J, {D'Avanzo} P, {Fruchter} AS, {Fryer} CL, et~al
  (2017) {The Emergence of a Lanthanide-rich Kilonova Following the Merger of
  Two Neutron Stars}. \apjl 848:L27, \doi{10.3847/2041-8213/aa90b6},
  \eprint{1710.05455}

\bibitem[{{Taracchini} et~al.(2012){Taracchini}, {Pan}, {Buonanno}, {Barausse},
  {Boyle}, {Chu}, {Lovelace}, {Pfeiffer}, and
  {Scheel}}]{Taracchini_PBBBCLPS2012}
{Taracchini} A, {Pan} Y, {Buonanno} A, {Barausse} E, {Boyle} M, {Chu} T,
  {Lovelace} G, {Pfeiffer} HP, {Scheel} MA (2012) {Prototype effective-one-body
  model for nonprecessing spinning inspiral-merger-ringdown waveforms}. \prd
  86:024011, \doi{10.1103/PhysRevD.86.024011}, \eprint{1202.0790}

\bibitem[{{Tauris} et~al.(2017){Tauris}, {Kramer}, {Freire}, {Wex}, {Janka},
  {Langer}, {Podsiadlowski}, {Bozzo}, {Chaty}, {Kruckow}, and
  et~al.}]{Tauris_etal2017}
{Tauris} TM, {Kramer} M, {Freire} PCC, {Wex} N, {Janka} HT, {Langer} N,
  {Podsiadlowski} P, {Bozzo} E, {Chaty} S, {Kruckow} MU, et~al (2017)
  {Formation of Double Neutron Star Systems}. \apj 846:170,
  \doi{10.3847/1538-4357/aa7e89}, \eprint{1706.09438}

\bibitem[{{Teukolsky}(1998)}]{Teukolsky1998}
{Teukolsky} SA (1998) {Irrotational Binary Neutron Stars in Quasi-Equilibrium
  in General Relativity}. \apj 504:442--449, \doi{10.1086/306082},
  \eprint{gr-qc/9803082}

\bibitem[{{Thompson} et~al.(2020){Thompson}, {Fauchon-Jones}, {Khan},
  {Nitoglia}, {Pannarale}, {Dietrich}, and {Hannam}}]{Thompson_FKNPDH2020}
{Thompson} JE, {Fauchon-Jones} E, {Khan} S, {Nitoglia} E, {Pannarale} F,
  {Dietrich} T, {Hannam} M (2020) {Modeling the gravitational wave signature of
  neutron star black hole coalescences}. \prd 101:124059,
  \doi{10.1103/PhysRevD.101.124059}, \eprint{2002.08383}

\bibitem[{{Thompson} et~al.(2019){Thompson}, {Kochanek}, {Stanek}, {Badenes},
  {Post}, {Jayasinghe}, {Latham}, {Bieryla}, {Esquerdo}, {Berlind}, and
  et~al.}]{Thompson_etal2019}
{Thompson} TA, {Kochanek} CS, {Stanek} KZ, {Badenes} C, {Post} RS, {Jayasinghe}
  T, {Latham} DW, {Bieryla} A, {Esquerdo} GA, {Berlind} P, et~al (2019) {A
  noninteracting low-mass black hole-giant star binary system}. Science
  366(6465):637--640, \doi{10.1126/science.aau4005}, \eprint{1806.02751}

\bibitem[{{Thornburg}(2007)}]{Thornburg2007}
{Thornburg} J (2007) {Event and Apparent Horizon Finders for 3 + 1 Numerical
  Relativity}. Living Rev Relativ 10:3, \doi{10.12942/lrr-2007-3}

\bibitem[{{Thorne}(1981)}]{Thorne1981}
{Thorne} KS (1981) {Relativistic radiative transfer - Moment formalisms}.
  \mnras 194:439--473, \doi{10.1093/mnras/194.2.439}

\bibitem[{{Thorne}(1987)}]{Thorne}
{Thorne} KS (1987) {Gravitational radiation}. In: Hawking SW, Israel W (eds)
  Three Hundred Years of Gravitation, Cambridge University Press, pp 330--458

\bibitem[{{Tichy}(2017)}]{Tichy2017}
{Tichy} W (2017) {The initial value problem as it relates to numerical
  relativity}. Reports on Progress in Physics 80:026901,
  \doi{10.1088/1361-6633/80/2/026901}, \eprint{1610.03805}

\bibitem[{{Timmes} and {Swesty}(2000)}]{Timmes_Swesty2000}
{Timmes} FX, {Swesty} FD (2000) {The Accuracy, Consistency, and Speed of an
  Electron-Positron Equation of State Based on Table Interpolation of the
  Helmholtz Free Energy}. \apjs 126:501--516, \doi{10.1086/313304}

\bibitem[{{Todd-Rutel} and {Piekarewicz}(2005)}]{ToddRutel_Piekarewicz2005}
{Todd-Rutel} BG, {Piekarewicz} J (2005) {Neutron-Rich Nuclei and Neutron Stars:
  A New Accurately Calibrated Interaction for the Study of Neutron-Rich
  Matter}. \prl 95:122501, \doi{10.1103/PhysRevLett.95.122501},
  \eprint{nucl-th/0504034}

\bibitem[{{Tolman}(1939)}]{Tolman1939}
{Tolman} RC (1939) {Static Solutions of Einstein's Field Equations for Spheres
  of Fluid}. Physical Review 55:364--373, \doi{10.1103/PhysRev.55.364}

\bibitem[{{Tsang} et~al.(2012){Tsang}, {Read}, {Hinderer}, {Piro}, and
  {Bondarescu}}]{Tsang_RHPB2012}
{Tsang} D, {Read} JS, {Hinderer} T, {Piro} AL, {Bondarescu} R (2012) {Resonant
  Shattering of Neutron Star Crusts}. \prl 108:011102,
  \doi{10.1103/PhysRevLett.108.011102}, \eprint{1110.0467}

\bibitem[{{Ury{\={u}}} and {Eriguchi}(1998)}]{Uryu_Eriguchi1998}
{Ury{\={u}}} K, {Eriguchi} Y (1998) {Stationary Structures of Irrotational
  Binary Systems: Models for Close Binary Systems of Compact Stars}. \apjs
  118:563--587, \doi{10.1086/313146}, \eprint{astro-ph/9808118}

\bibitem[{{Ury{\={u}}} and {Eriguchi}(1999)}]{Uryu_Eriguchi1999}
{Ury{\={u}}} K, {Eriguchi} Y (1999) {Newtonian models for black hole-gaseous
  star close binary systems}. \mnras 303:329--342,
  \doi{10.1046/j.1365-8711.1999.02224.x}, \eprint{astro-ph/9808120}

\bibitem[{{Ury{\={u}}} et~al.(2006){Ury{\={u}}}, {Limousin}, {Friedman},
  {Gourgoulhon}, and {Shibata}}]{Uryu_LFGS2006}
{Ury{\={u}}} K, {Limousin} F, {Friedman} JL, {Gourgoulhon} E, {Shibata} M
  (2006) {Binary Neutron Stars: Equilibrium Models beyond Spatial Conformal
  Flatness}. \prl 97:171101, \doi{10.1103/PhysRevLett.97.171101},
  \eprint{gr-qc/0511136}

\bibitem[{{Ury{\={u}}} et~al.(2009){Ury{\={u}}}, {Limousin}, {Friedman},
  {Gourgoulhon}, and {Shibata}}]{Uryu_LFGS2009}
{Ury{\={u}}} K, {Limousin} F, {Friedman} JL, {Gourgoulhon} E, {Shibata} M
  (2009) {Nonconformally flat initial data for binary compact objects}. \prd
  80:124004, \doi{10.1103/PhysRevD.80.124004}, \eprint{0908.0579}

\bibitem[{{Valenti} et~al.(2017){Valenti}, {Sand}, {Yang}, {Cappellaro},
  {Tartaglia}, {Corsi}, {Jha}, {Reichart}, {Haislip}, and
  {Kouprianov}}]{Valenti_SYCTCJRHK2017}
{Valenti} S, {Sand} DJ, {Yang} S, {Cappellaro} E, {Tartaglia} L, {Corsi} A,
  {Jha} SW, {Reichart} DE, {Haislip} J, {Kouprianov} V (2017) {The Discovery of
  the Electromagnetic Counterpart of GW170817: Kilonova AT 2017gfo/DLT17ck}.
  \apjl 848:L24, \doi{10.3847/2041-8213/aa8edf}, \eprint{1710.05854}

\bibitem[{{Vallisneri}(2000)}]{Vallisneri2000}
{Vallisneri} M (2000) {Prospects for Gravitational-Wave Observations of
  Neutron-Star Tidal Disruption in Neutron-Star-Black-Hole Binaries}. \prl
  84:3519--3522, \doi{10.1103/PhysRevLett.84.3519}, \eprint{gr-qc/9912026}

\bibitem[{{van de Meent}(2020)}]{vandeMeent2020}
{van de Meent} M (2020) {Analytic solutions for parallel transport along
  generic bound geodesics in Kerr spacetime}. Class Quantum Grav 37:145007,
  \doi{10.1088/1361-6382/ab79d5}, \eprint{1906.05090}

\bibitem[{{Villar} et~al.(2017){Villar}, {Guillochon}, {Berger}, {Metzger},
  {Cowperthwaite}, {Nicholl}, {Alexander}, {Blanchard}, {Chornock},
  {Eftekhari}, and et~al.}]{Villar_etal2017}
{Villar} VA, {Guillochon} J, {Berger} E, {Metzger} BD, {Cowperthwaite} PS,
  {Nicholl} M, {Alexander} KD, {Blanchard} PK, {Chornock} R, {Eftekhari} T,
  et~al (2017) {The Combined Ultraviolet, Optical, and Near-infrared Light
  Curves of the Kilonova Associated with the Binary Neutron Star Merger
  GW170817: Unified Data Set, Analytic Models, and Physical Implications}.
  \apjl 851:L21, \doi{10.3847/2041-8213/aa9c84}, \eprint{1710.11576}

\bibitem[{{Vincent} et~al.(2020){Vincent}, {Foucart}, {Duez}, {Haas}, {Kidder},
  {Pfeiffer}, and {Scheel}}]{Vincent_FDHKPS2020}
{Vincent} T, {Foucart} F, {Duez} MD, {Haas} R, {Kidder} LE, {Pfeiffer} HP,
  {Scheel} MA (2020) {Unequal mass binary neutron star simulations with
  neutrino transport: Ejecta and neutrino emission}. \prd 101:044053,
  \doi{10.1103/PhysRevD.101.044053}, \eprint{1908.00655}

\bibitem[{{Wada} et~al.(2020){Wada}, {Shibata}, and
  {Ioka}}]{Wada_Shibata_Ioka2020}
{Wada} T, {Shibata} M, {Ioka} K (2020) {Analytic properties of the
  electromagnetic field of binary compact stars and electromagnetic precursors
  to gravitational waves}. Progress of Theoretical and Experimental Physics
  2020:103E01, \doi{10.1093/ptep/ptaa126}, \eprint{2008.04661}

\bibitem[{{Wade} et~al.(2014){Wade}, {Creighton}, {Ochsner}, {Lackey}, {Farr},
  {Littenberg}, and {Raymond}}]{Wade_COLFLR2014}
{Wade} L, {Creighton} JDE, {Ochsner} E, {Lackey} BD, {Farr} BF, {Littenberg}
  TB, {Raymond} V (2014) {Systematic and statistical errors in a Bayesian
  approach to the estimation of the neutron-star equation of state using
  advanced gravitational wave detectors}. \prd 89:103012,
  \doi{10.1103/PhysRevD.89.103012}, \eprint{1402.5156}

\bibitem[{{Walsh}(2007)}]{Walsh2007}
{Walsh} DM (2007) {Non-uniqueness in conformal formulations of the Einstein
  constraints}. Class Quantum Grav 24:1911--1925,
  \doi{10.1088/0264-9381/24/8/002}, \eprint{gr-qc/0610129}

\bibitem[{{Wan}(2017)}]{Wan2017}
{Wan} MB (2017) {Effects of magnetic field topology in black hole-neutron star
  mergers: Long-term simulations}. \prd 95:104013,
  \doi{10.1103/PhysRevD.95.104013}, \eprint{1606.09090}

\bibitem[{{Wanajo}(2018)}]{Wanajo2018}
{Wanajo} S (2018) {Physical Conditions for the r-process. I. Radioactive Energy
  Sources of Kilonovae}. \apj 868:65, \doi{10.3847/1538-4357/aae0f2},
  \eprint{1808.03763}

\bibitem[{{Wanajo} et~al.(2014){Wanajo}, {Sekiguchi}, {Nishimura}, {Kiuchi},
  {Kyutoku}, and {Shibata}}]{Wanajo_SNKKS2014}
{Wanajo} S, {Sekiguchi} Y, {Nishimura} N, {Kiuchi} K, {Kyutoku} K, {Shibata} M
  (2014) {Production of All the r-process Nuclides in the Dynamical Ejecta of
  Neutron Star Mergers}. \apjl 789:L39, \doi{10.1088/2041-8205/789/2/L39},
  \eprint{1402.7317}

\bibitem[{{Watson} et~al.(2019){Watson}, {Hansen}, {Selsing}, {Koch},
  {Malesani}, {Andersen}, {Fynbo}, {Arcones}, {Bauswein}, {Covino}, and
  et~al.}]{Watson_etal2019}
{Watson} D, {Hansen} CJ, {Selsing} J, {Koch} A, {Malesani} DB, {Andersen} AC,
  {Fynbo} JPU, {Arcones} A, {Bauswein} A, {Covino} S, et~al (2019)
  {Identification of strontium in the merger of two neutron stars}. \nat
  574:497--500, \doi{10.1038/s41586-019-1676-3}, \eprint{1910.10510}

\bibitem[{{Waxman} et~al.(2018){Waxman}, {Ofek}, {Kushnir}, and
  {Gal-Yam}}]{Waxman_OKG2018}
{Waxman} E, {Ofek} EO, {Kushnir} D, {Gal-Yam} A (2018) {Constraints on the
  ejecta of the GW170817 neutron star merger from its electromagnetic
  emission}. \mnras 481:3423--3441, \doi{10.1093/mnras/sty2441},
  \eprint{1711.09638}

\bibitem[{{Waxman} et~al.(2019){Waxman}, {Ofek}, and
  {Kushnir}}]{Waxman_Ofek_Kushnir2019}
{Waxman} E, {Ofek} EO, {Kushnir} D (2019) {Late-time Kilonova Light Curves and
  Implications to GW170817}. \apj 878:93, \doi{10.3847/1538-4357/ab1f71},
  \eprint{1902.01197}

\bibitem[{{Weibel}(1959)}]{Weibel1959}
{Weibel} ES (1959) {Spontaneously Growing Transverse Waves in a Plasma Due to
  an Anisotropic Velocity Distribution}. \prl 2:83--84,
  \doi{10.1103/PhysRevLett.2.83}

\bibitem[{{Wessel} et~al.(2021){Wessel}, {Paschalidis}, {Tsokaros}, {Ruiz}, and
  {Shapiro}}]{Wessel_PTRS2021}
{Wessel} E, {Paschalidis} V, {Tsokaros} A, {Ruiz} M, {Shapiro} SL (2021)
  {Gravitational waves from disks around spinning black holes: Simulations in
  full general relativity}. \prd 103:043013, \doi{10.1103/PhysRevD.103.043013},
  \eprint{2011.04077}

\bibitem[{{Wiggins} and {Lai}(2000)}]{Wiggins_Lai2000}
{Wiggins} P, {Lai} D (2000) {Tidal Interaction between a Fluid Star and a Kerr
  Black Hole in Circular Orbit}. \apj 532:530--539, \doi{10.1086/308565},
  \eprint{astro-ph/9907365}

\bibitem[{{Wilson} and {Mathews}(1995)}]{Wilson_Mathews1995}
{Wilson} JR, {Mathews} GJ (1995) {Instabilities in Close Neutron Star
  Binaries}. \prl 75:4161--4164, \doi{10.1103/PhysRevLett.75.4161}

\bibitem[{{Wilson} et~al.(1996){Wilson}, {Mathews}, and
  {Marronetti}}]{Wilson_Mathews_Marronetti1996}
{Wilson} JR, {Mathews} GJ, {Marronetti} P (1996) {Relativistic numerical model
  for close neutron-star binaries}. \prd 54:1317--1331,
  \doi{10.1103/PhysRevD.54.1317}, \eprint{gr-qc/9601017}

\bibitem[{{Wiseman}(1992)}]{Wiseman1992}
{Wiseman} AG (1992) {Coalescing binary systems of compact objects to
  (post)$^{5/2}$-Newtonian order. II. Higher-order wave forms and radiation
  recoil}. \prd 46:1517--1539, \doi{10.1103/PhysRevD.46.1517}

\bibitem[{{Wu} and {Tamborra}(2017)}]{Wu_Tamborra2017}
{Wu} MR, {Tamborra} I (2017) {Fast neutrino conversions: Ubiquitous in compact
  binary merger remnants}. \prd 95:103007, \doi{10.1103/PhysRevD.95.103007},
  \eprint{1701.06580}

\bibitem[{{Wu} et~al.(2016){Wu}, {Fern{\'a}ndez}, {Mart{\'\i}nez-Pinedo}, and
  {Metzger}}]{Wu_FMM2016}
{Wu} MR, {Fern{\'a}ndez} R, {Mart{\'\i}nez-Pinedo} G, {Metzger} BD (2016)
  {Production of the entire range of r-process nuclides by black hole accretion
  disc outflows from neutron star mergers}. \mnras 463:2323--2334,
  \doi{10.1093/mnras/stw2156}, \eprint{1607.05290}

\bibitem[{{Wu} et~al.(2017){Wu}, {Tamborra}, {Just}, and {Janka}}]{Wu_TJJ2017}
{Wu} MR, {Tamborra} I, {Just} O, {Janka} HT (2017) {Imprints of neutrino-pair
  flavor conversions on nucleosynthesis in ejecta from neutron-star merger
  remnants}. \prd 96:123015, \doi{10.1103/PhysRevD.96.123015},
  \eprint{1711.00477}

\bibitem[{{Wu} et~al.(2019){Wu}, {Barnes}, {Mart{\'\i}nez-Pinedo}, and
  {Metzger}}]{Wu_BMM2019}
{Wu} MR, {Barnes} J, {Mart{\'\i}nez-Pinedo} G, {Metzger} BD (2019)
  {Fingerprints of Heavy-Element Nucleosynthesis in the Late-Time Lightcurves
  of Kilonovae}. \prl 122:062701, \doi{10.1103/PhysRevLett.122.062701},
  \eprint{1808.10459}

\bibitem[{{Yagi} and {Yunes}(2014)}]{Yagi_Yunes2014}
{Yagi} K, {Yunes} N (2014) {Love number can be hard to measure}. \prd
  89:021303, \doi{10.1103/PhysRevD.89.021303}, \eprint{1310.8358}

\bibitem[{{Yamamoto} et~al.(2008){Yamamoto}, {Shibata}, and
  {Taniguchi}}]{Yamamoto_Shibata_Taniguchi2008}
{Yamamoto} T, {Shibata} M, {Taniguchi} K (2008) {Simulating coalescing compact
  binaries by a new code (SACRA)}. \prd 78:064054,
  \doi{10.1103/PhysRevD.78.064054}, \eprint{0806.4007}

\bibitem[{{Yang} et~al.(2018){Yang}, {East}, and
  {Lehner}}]{Yang_East_Lehner2018}
{Yang} H, {East} WE, {Lehner} L (2018) {Can We Distinguish Low-mass Black Holes
  in Neutron Star Binaries?} \apj 856:110, \doi{10.3847/1538-4357/aab2b0},
  \eprint{1710.05891}

\bibitem[{{Ye} et~al.(2020){Ye}, {Fong}, {Kremer}, {Rodriguez}, {Chatterjee},
  {Fragione}, and {Rasio}}]{Ye_FKRCFR2020}
{Ye} CS, {Fong} Wf, {Kremer} K, {Rodriguez} CL, {Chatterjee} S, {Fragione} G,
  {Rasio} FA (2020) {On the Rate of Neutron Star Binary Mergers from Globular
  Clusters}. \apjl 888:L10, \doi{10.3847/2041-8213/ab5dc5}, \eprint{1910.10740}

\bibitem[{{York}(1979)}]{York1979}
{York} J J~W (1979) {Kinematics and dynamics of general relativity}. In:
  {Smarr} LL (ed) Sources of Gravitational Radiation, pp 83--126

\bibitem[{{York}(1999)}]{York1999}
{York} J James~W (1999) {Conformal ``Thin-Sandwich'' Data for the Initial-Value
  Problem of General Relativity}. \prl 82:1350--1353,
  \doi{10.1103/PhysRevLett.82.1350}, \eprint{gr-qc/9810051}

\bibitem[{{York}(1972)}]{York1972}
{York} JW (1972) {Role of Conformal Three-Geometry in the Dynamics of
  Gravitation}. \prl 28:1082--1085, \doi{10.1103/PhysRevLett.28.1082}

\bibitem[{{Yunes} et~al.(2016){Yunes}, {Yagi}, and
  {Pretorius}}]{Yunes_Yagi_Pretorius2016}
{Yunes} N, {Yagi} K, {Pretorius} F (2016) {Theoretical physics implications of
  the binary black-hole mergers GW150914 and GW151226}. \prd 94:084002,
  \doi{10.1103/PhysRevD.94.084002}, \eprint{1603.08955}

\bibitem[{{Zalamea} and {Beloborodov}(2011)}]{Zalamea_Beloborodov2011}
{Zalamea} I, {Beloborodov} AM (2011) {Neutrino heating near hyper-accreting
  black holes}. \mnras 410:2302--2308, \doi{10.1111/j.1365-2966.2010.17600.x},
  \eprint{1003.0710}

\bibitem[{{Zevin} et~al.(2020){Zevin}, {Spera}, {Berry}, and
  {Kalogera}}]{Zevin_SBK2020}
{Zevin} M, {Spera} M, {Berry} CPL, {Kalogera} V (2020) {Exploring the Lower
  Mass Gap and Unequal Mass Regime in Compact Binary Evolution}. \apjl 899:L1,
  \doi{10.3847/2041-8213/aba74e}, \eprint{2006.14573}

\bibitem[{{Zhang}(2019)}]{Zhang2019}
{Zhang} B (2019) {Charged Compact Binary Coalescence Signal and Electromagnetic
  Counterpart of Plunging Black Hole-Neutron Star Mergers}. \apjl 873:L9,
  \doi{10.3847/2041-8213/ab0ae8}, \eprint{1901.11177}

\bibitem[{{Zhong} et~al.(2019){Zhong}, {Dai}, and {Deng}}]{Zhong_Dai_Deng2019}
{Zhong} SQ, {Dai} ZG, {Deng} CM (2019) {Electromagnetic Emission Post Spinning
  Black Hole Magnetized Neutron Star Mergers}. \apjl 883:L19,
  \doi{10.3847/2041-8213/ab40c5}, \eprint{1909.00494}

\bibitem[{{Zhu} et~al.(2020){Zhu}, {Yang}, {Liu}, {Huang}, {Zhang}, {Li}, {Yu},
  and {Gao}}]{Zhu_YLHZLYG2020}
{Zhu} JP, {Yang} YP, {Liu} LD, {Huang} Y, {Zhang} B, {Li} Z, {Yu} YW, {Gao} H
  (2020) {Kilonova Emission from Black Hole-Neutron Star Mergers. I.
  Viewing-angle-dependent Lightcurves}. \apj 897:20,
  \doi{10.3847/1538-4357/ab93bf}, \eprint{2003.06733}

\bibitem[{{Zhu} et~al.(2018){Zhu}, {Wollaeger}, {Vassh}, {Surman}, {Sprouse},
  {Mumpower}, {M{\"o}ller}, {McLaughlin}, {Korobkin}, {Kawano}, and
  et~al.}]{Zhu_etal2018}
{Zhu} Y, {Wollaeger} RT, {Vassh} N, {Surman} R, {Sprouse} TM, {Mumpower} MR,
  {M{\"o}ller} P, {McLaughlin} GC, {Korobkin} O, {Kawano} T, et~al (2018)
  {Californium-254 and Kilonova Light Curves}. \apjl 863:L23,
  \doi{10.3847/2041-8213/aad5de}, \eprint{1806.09724}

\bibitem[{{Zhu} et~al.(2021){Zhu}, {Lund}, {Barnes}, {Sprouse}, {Vassh},
  {McLaughlin}, {Mumpower}, and {Surman}}]{Zhu_LBSVMMS2021}
{Zhu} YL, {Lund} KA, {Barnes} J, {Sprouse} TM, {Vassh} N, {McLaughlin} GC,
  {Mumpower} MR, {Surman} R (2021) {Modeling Kilonova Light Curves: Dependence
  on Nuclear Inputs}. \apj 906:94, \doi{10.3847/1538-4357/abc69e},
  \eprint{2010.03668}

\end{thebibliography}

\end{document}